\documentclass[usenatbib,usegraphicx]{mn2e}
\input epsf
\usepackage{graphics}
\usepackage{epsfig}
\usepackage{url}

\newcommand{\Ha} {H$\alpha$}
\newcommand{\ha} {H$\alpha$}

\newcommand{\Deg}{${}^{\circ}$}

\newcommand{\kms}{~km~s$^{-1}$}
\newcommand{\kmsMpc}{km~s$^{-1}~$Mpc$^{-1}$}
\def\degr{\hbox{$^\circ$}}
\def\deg{\hbox{$^\circ$}}
\newcommand{\PVM}{position-velocity diagram}

\newcommand{\PPVMs}{Position-velocity diagrams}
\newcommand{\pvm}{position-velocity diagram}
\newcommand{\PVMs}{position-velocity diagrams}

\newcommand{\pa}{position angle}
\newcommand{\PA}{position angle}
\newcommand{\pas}{position angles}
\newcommand{\PAs}{position angles}

\newcommand{\vf}{velocity field}
\newcommand{\VF}{velocity field}
\newcommand{\VFs}{velocity fields}
\newcommand{\rc}{rotation curve}
\newcommand{\SNR}{signal-to-noise ratio}
\newcommand{\snr}{signal-to-noise ratio}
\newcommand{\RC}{rotation curve}
\newcommand{\rcs}{rotation curves}
\newcommand{\RCs}{rotation curves}
\newcommand{\FOV}{field-of-view}
\newcommand{\TF}{Tully-Fisher}
\title[GHASP : An \Ha~kinematic survey of 203 spiral and irregular galaxies - VII.]
      {GHASP : An \Ha~kinematic survey of 203 spiral and irregular galaxies
- VII. Revisiting the analysis of \ha~data cubes for 97 galaxies.}

\author[B. Epinat et al.]
{Epinat, B., Amram, P. \& Marcelin M. \\
Laboratoire d'Astrophysique de Marseille, OAMP, Universit\'e
Aix-Marseille \& CNRS, 38 rue Fr\'ed\'eric Joliot-Curie,\\
13388 Marseille Cedex 13, France\\}
\date{Accepted. Received; in original form }

\pubyear{2008}
\begin{document}
\maketitle

\label{firstpage}

\begin{abstract}

The GHASP survey (Gassendi HAlpha survey of SPirals) consists of
3D \ha~data cubes for 203 spiral and irregular galaxies, covering
a large range in morphological types and absolute magnitudes, for
kinematics analysis. It is the largest sample of Fabry-Perot data
published up to now.
In order to provide an homogenous sample, reduced and analyzed
using the same procedure, we present in this paper the new
reduction and analysis for a set of 97 galaxies already published
in previous papers but now using the new data reduction procedure adopted for the whole sample.
The GHASP survey is now achieved and the whole sample is reduced
using adaptive binning techniques based on Voronoi tessellations.
We have derived \ha~data cubes from which are computed \ha~maps,
radial \VFs~as well as residual \VFs, \PVMs, \RCs~and
kinematical parameters for almost all galaxies. The \RCs, the kinematical parameters and their
uncertainties are computed homogeneously using the new method based
on the power spectrum of the residual \VF. This paper provides the
kinematical parameters for the whole sample.
For the first time, the integrated \Ha~profiles have been computed
and are presented for the whole sample. The total \Ha~fluxes
deduced from these profiles have been used in order to provide a
flux calibration for the 203 GHASP galaxies.
This paper confirms the conclusions already drawn from half the sample
concerning (i) the increased accuracy of \PAs~measurements using
kinematical data, (ii) the difficulty to have robust
determinations of both morphological and kinematical inclinations
in particular for low inclination galaxies and (iii) the very good
agreement between the \TF~relationship derived from our data and
previous determinations found in the literature.

\end{abstract}


\section{Introduction}
\label{intro}

The GHASP survey consists of a large sample of spiral and
irregular galaxies observed with a scanning Fabry-Perot for
studying their kinematical and dynamical properties through the
ionized hydrogen component. The goals of this study have been
described in \citet{Epinat:2008}.
The GHASP sample is by now the largest homogenous sample of
Fabry-Perot data ever published, comprising 3D data for 203
galaxies.
This paper is the seventh of a series called hereafter Paper I to
VI
\citep{Garrido:2002,Garrido:2003,Garrido:2004,Garrido:2005,Spano:2008,
Epinat:2008} presenting the data obtained in the frame of the
GHASP survey.  The observations were lead during fourteen observing
runs at the ``Observatoire de Haute Provence (OHP)'', France, from
1998 to 2004. The survey is now achieved.
The observing runs number 8 to 14 have been presented
in Paper VI, which relies on a set of 108 galaxies, providing 106
\VFs~and 93 \RCs.  Those data have been reduced with the new data
reduction procedure (see Paper VI and references therein).
The data presented in this paper are those of the seven first
observing runs (already presented in Papers I to IV) that have
been re-reduced using the same method as in Paper VI. It also
contains data for UGC 3382 and UGC 11300 that have been improved
by adding new data (runs 3, 5 and 6) to the ones already published
in Paper VI (runs 10 and 13). Thus, the data presented here
consist of a set of 97 galaxies, providing 96 \VFs~and 82 \RCs.

To be clear on the goals and limits of this present work, we
summarize hereafter what we present and what we do not in this
paper.
We present:
\begin{itemize}
    \item the new individual maps and \PVMs~in Appendix \ref{maps} (on line data
    only),
    \item the new \rcs, in Appendix \ref{rc} and the new corresponding tables in Appendix \ref{rc_tables}.
\end{itemize}
In this paper, we lead the same kind of analysis as the one
presented in Paper VI concerning:
\begin{itemize}
    \item the study of the parameters of the kinematical models,
    \item the study of the residual \VFs,
    \item the \TF~relation.
\end{itemize}
Because it is useful to display and
analyze all the data together, we put here in the same tables (in Appendix \ref{tables}) the
new parameters and the results already published in Paper VI so that the reader does not have to compile tables coming from different publications. With respect to Papers I to IV, some distances and absolute magnitudes have been recomputed (using better estimations).
%Furthermore, all the parameters for the whole GHASP sample are in
%a single table and the reader does not have to compile tables
%coming from different publications.
%Thus we provide:
%\begin{itemize}
%    \item all the parameters of the kinematical models (in section \ref{xxx} and \ref{tables}).
%    \item all the residual \VFs~(in section \ref{xxx}),
%    \item the \TF~relation.
%\end{itemize}
For the whole GHASP sample we make a new analysis on an absolute flux calibration made using the data calibrated by \citet{James:2004} and the integrated \ha~profiles deduced from our data cubes.

Because this has been discussed in previous papers, we do not
present any more:
\begin{itemize}
    \item the morphological types and luminosity distributions of the whole GHASP sample (see Paper
    VI),
    \item the data reduction procedure used here, including the computation of the \RCs, the determination of the kinematical parameters and the determination of the uncertainties (see Paper VI),
    \item the instrumental set-up of the instrument for the data re-reduced in this paper (see Papers I to
    IV),
%    \item the log of the observation (see Papers I to IV) xxxpourtant on remet des info recapitulatives en table B1: que dire donc?xxx,
    \item the individual comments for each galaxy (see Papers I to IV),
    except when the new reduction procedure leads to new comments or to conclusions noticeably
    different from the previous ones (see Appendix \ref{notes}).
\end{itemize}

In section \ref{calibration} we make an indirect flux calibration
of the \ha~profiles. In section \ref{analysis} we present the data
and in section \ref{tullyfisher} we compute the \TF~relation.
In section \ref{conclusion} we give the summary and conclusions.
When the distances of the galaxies are not known, a Hubble
constant H$_{0}$=75\kms~Mpc$^{-1}$ is used throughout this paper.

\section{Calibration and \ha~profiles}
\label{calibration}

\begin{figure}
\begin{center}
\includegraphics[width=7.5cm]{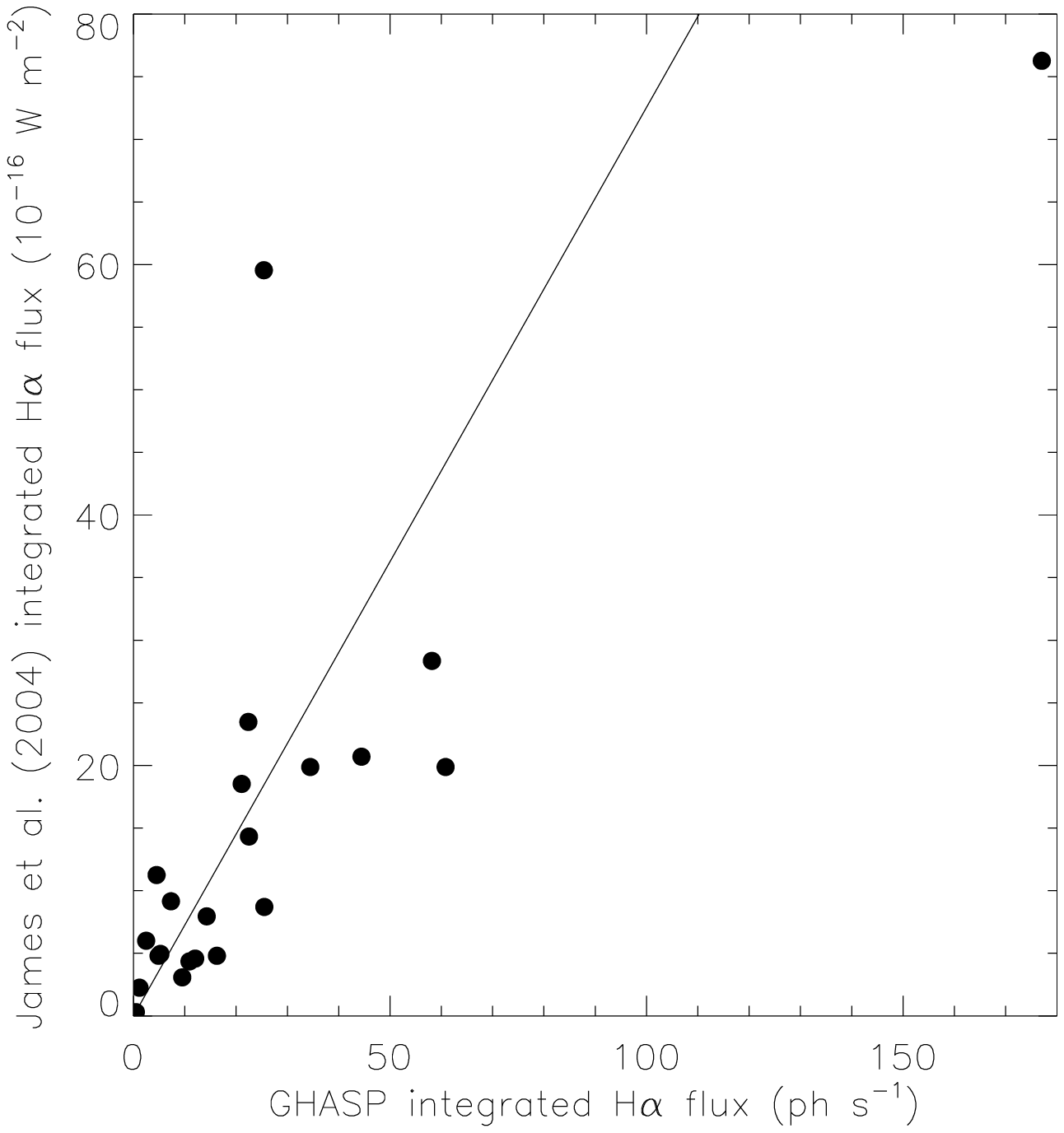}
\includegraphics[width=7.5cm]{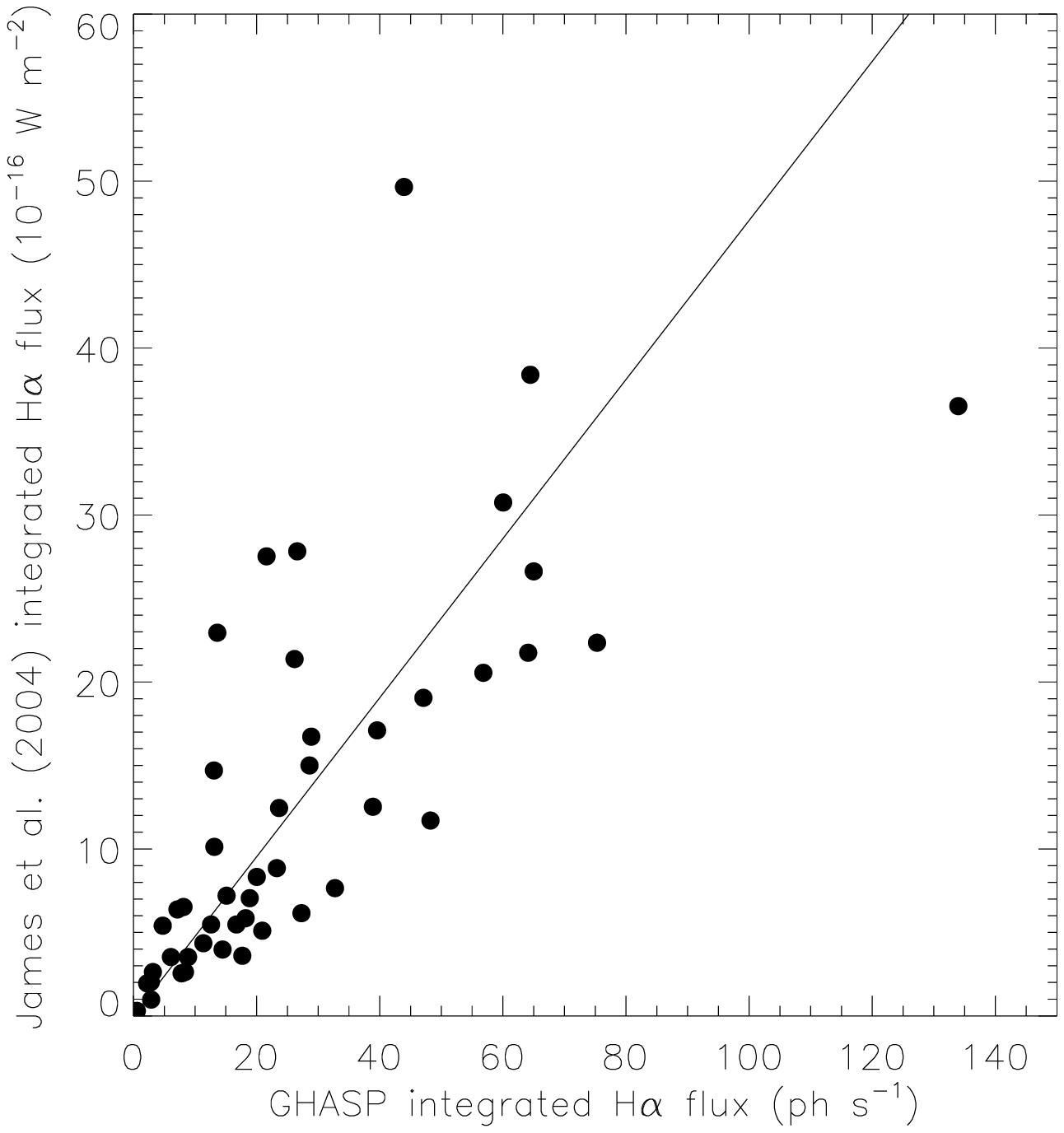}
\end{center}
\caption{\Ha~flux measured by GHASP~versus \Ha~flux from
\citet{James:2004}. The dashed line represents the linear
regression on the data from which results our calibration.
\textbf{Top:} calibration for the IPCS 512x512. \textbf{Bottom:}
calibration for the IPCS 256x256.} \label{calib}
\end{figure}

Even if direct flux calibration is always possible using well
calibrated and extended \ha~emitters like planetary nebulae
\citep{Dopita:1997}, during the observations, we decided not to
calibrate our data, thus saving observing time. Indeed, our major
scientific goal was not to use the Fabry-Perot technique to make
photometric studies but kinematic ones. We estimated that \ha~flux
calibrations require additional observing times ranging from 25 to
33\%. Nevertheless, an indirect calibration of the total \Ha~flux
of the 201/203 galaxies from the GHASP sample has been made using
%22+47 (pourquoi 22+47 et pas la somme)
69 of the 71 galaxies we have in common with \citet{James:2004}.
From their study using narrow band filters including \ha~and
[NII], \citet{James:2004} provided calibrated fluxes for their
sample of 334 galaxies. Our spectral resolution allows us to
resolve the \ha~line, moreover the full width half maximum (FWHM) of the narrow band filters
are narrow enought to reject [NII].  This is not the case for
\citet{James:2004} who do not separate \Ha~from [NII] lines. We
have corrected this effect assuming a mean and constant
spectral ratio \Ha/[NII]=3. Distinct calibrations have been made
depending on the observing setup. Indeed, the GHASP sample has
been obtained using a 256$\times$256 IPCS (Imaging Photon Counting
System) until March 2000 (runs number 1 to 4) and with a new
512$\times$512 IPCS since October 2000 (runs number 5 to 14), with
respective pixel scales of 0.96\arcsec~and 0.68\arcsec~(given in
Table \ref{table_calib}). We reject from the calibration the galaxies
for which we have added data observed with both detectors.
%However, we are then able to compute their flux
Their flux was computed \textit{a posteriori} by taking into
account the response of each detector and the corresponding
exposure time. To compute the total \Ha~flux for each galaxy, we
compute the \ha~profile by summing the flux in the spectrum of
each spatial element (see Appendix \ref{profiles}). To minimize the foreground sky
contamination, only the spatial elements used to compute the
different moments (monochromatic images, velocity fields, ...)
have been added. We use the \VF~in order to disentangle
free spectral range overlaps.
%For each pixel, we
%center the line in the spectrum and then enlarge the spectrum by a
%factor three for each pixel with the continuum value.
%At that point we then only need to shift
%replace the line to its initial position and then sum the
%contribution of each pixel containing signal from the galaxy.
%To improve the accuracy of the profiles (normalised with the
%exposure time),
We correct the fluxes from the interference filter response,
taking into account the aperture of the instrument, the
inclination of the filter and its temperature during the
observations. We subtract the continuum taking into account the
periodicity of the Fabry-Perot transmission (difference between
the free spectral range of the interferometer and the FWHM of the
interference filter). The systemic velocity computed from the kinematical model (vertical dashed line in Figure \ref{plot_profiles}) is globally well centered on the integrated \Ha~profile.
%-free profile for this filter response. The correction
%applied to the continuum is different. Indeed, for a given spacing
%of the Fabry-Perot plates periodical wavelengths are allowed to
%get out from it. Making the assumption that the continuum is flat,
%for each channel (Fabry-Perot spacing), we estimate how much
%continuum is transmitted by the interference filter, rescale
%accordingly the continuum and add it to the corrected
%continuum-free profile.
%The flux are computed as the sum of the corrected continuum-free
%profile.
Figure \ref{calib} shows the comparison between
\citet{James:2004} fluxes and our estimated fluxes for the old
IPCS (top) and the new IPCS (bottom). The linear regressions are
plotted with the dashed lines and their coefficients are
respectively 0.73$\pm$0.13 $10^{-16}~W~m^{-2}~ph^{-1}~s$ and 0.48$\pm$0.06
$10^{-16}~W~m^{-2}~ph^{-1}~s$. The fluxes estimated from this calibration
are presented in Table \ref{table_calib}.

\section{Data Analysis}
\label{analysis}

%\subsection{Maps from the 3D data cubes}
%\label{maps}

As detailed in Paper VI, the same procedure (adaptive binning, sky subtraction, ghost removals, ...) has been used to compute the data cubes and the maps deduced from these cubes.
For each galaxy, in Appendix D (available online only), from Figure
D1 to D96, when possible, we present
five frames per figure: the XDSS blue (or red) image (top/left),
the \ha~\VF~(top/right), the \ha~monochromatic image (middle/left)
and, when a model fits the \VF, the \ha~residual
\VF~(middle/right) and finally, the \PVM~along the major axis
(bottom). The white and black double crosses indicate the centers used for the kinematical analysis
(given in Table \ref{table_calib}, see Paper VI for details) while
the black line traces the kinematical major axis deduced from the
\VF~analysis (see Paper VI) or the morphological one (taken from
HyperLeda) when no position angle of the kinematical major axis
could be derived using the kinematics (e.g. Table \ref{tablemod}).
This line ends at the radius D$_{25}$/2 corresponding to the
isophotal level 25 mag arcsec$^{-2}$ in the B-band (given in Table
\ref{tabletf}) in order to compare the \VF~extent with the optical
disk of the galaxies. \PPVMs~are computed along the axis defined
by this black line, using a virtual slit width of seven pixels.
The red line superimposed on the \PVM~is the \RC~deduced from the
model \VF~along this virtual slit (see Paper VI). When no fit is
satisfactory (generally because of poor \snr), we used the real
\VF~instead of the model (see individual captions in Figures
D1 to D96). The \RCs~are given in Appendix
\ref{rc} (figures) and \ref{rc_tables} (tables). They are computed
and displayed following the method described in Paper VI. These
figures are also available on the Fabry-Perot data base:
\url{http://FabryPerot.oamp.fr/}. Appendix
\ref{rc_tables} (that contains the tables corresponding to the
\RCs) is available online only.
The curves are plotted with both sides superimposed in the same
quadrant, using different symbols for the receding (crosses) and
approaching (dots) side (with respect to the center). The black
vertical arrow on the X-axis represents the radius $D_{25}/2$
while the smaller grey arrow on the X-axis represents the
transition radius (defined in Paper VI), always smaller than $D_{25}/2$ by definition.
\par
For galaxies seen almost edge-on (inclination higher than
75$\degr$) our model does not describe accurately the rotation of
a galaxy (see Paper VI). Furthermore, for UGC 1249, UGC 2082, UGC
3851, UGC 4278, UGC 5272, UGC 5935 and UGC 11909, neither \RC~nor
residual \VFs~have been plotted. For them, the \pvm~gives more
suitable information than the \RC~and allows the
peak-to-peak or peak-to-valley velocity distribution along the
major axis to be followed.

The \RCs~recomputed in this paper may be different from the ones published in Papers I to V
since: (i) the adaptive binning gives different weights to low
\SNR~regions in the \VF~from which is computed the \RC; (ii) the
exclusion sector around the minor axis is always set at
22.5\degr~(in the galaxy plane) contrarily to what had been done
in previous papers where the exclusion sector varied from one
galaxy to the other; (iii) the inclination and major axis \PA~may
be different; (iv) the center may have changed. Indeed, as has been done in Paper VI, the \vf~center chosen to compute the \RC~matches the morphological center (nucleus) when it is unambiguously defined from the morphology.
No comment is given in Appendix \ref{notes} unless the differences between the previous and
the new \VFs~and \RCs~lead to inconsistent results.

%\subsection{Residual velocity fields}
%\label{section_residuals}

\begin{figure}
\begin{center}
\includegraphics[width=7.5cm]{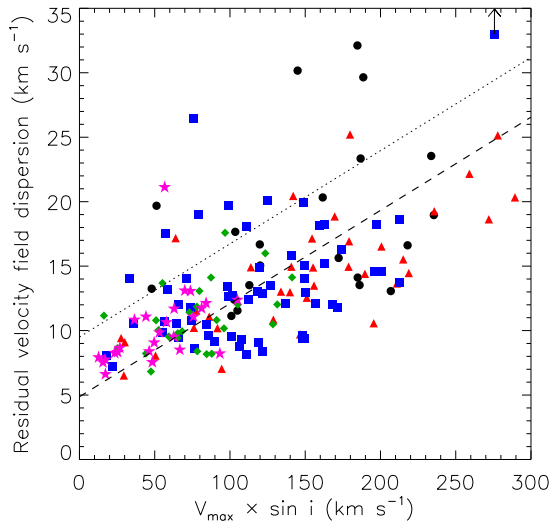}
%\centerline{\epsfxsize=7.5truecm\epsfbox{dispersion_res_vs_vsini}}
\end{center}
\caption{Dispersion in residual \VF~versus maximum velocity,
subdivided by Hubble morphological type: black circles 0$\leq$t$<$2,
red triangles 2$\leq$t$<$4, blue squares 4$\leq$t$<$6, green
rhombuses 6$\leq$t$<$8 and pink stars 8$\leq$t$<$10. The dashed
line represents the linear regression on the data. The points
above the dotted line are discussed in section
\ref{analysis}. UGC 3334 labelled with an arrow has
actually a hudge residual velocity dispersion of 54\kms~(see Table
\ref{tablemod}).} \label{resi}
\end{figure}

The mean velocity dispersion on each residual \VF~has been
computed for each galaxy and tabulated in Table
\ref{tablemod}.
Details on the computation and on the analysis of residual \VFs~are
given in Paper VI.
%
%they range from 4 to 54\kms~with a mean value around 13\kms.
%
Taking into account the whole sample, the plot presented in Paper
VI has been updated in Figure \ref{resi}.  It still shows that
the residual velocity dispersion is correlated with the maximum
amplitude of the \VF~(shown by the dashed linear regression).
We observe a set of galaxies with a high residual velocity
dispersion (points above the dotted line in Figure \ref{resi}).
These points correspond to galaxies: (i) dominated by strong bars
(UGC 89, UGC 3013, UGC 11283 and UGC 11407), or strong spiral
structures (UGC 5786 and UGC 3334) and not correctly described by
our model which does not take into account non-axisymmetric
motions; (ii) having \VFs~of lower quality (UGC 1655, UGC 1736, UGC
3382, UGC 3528, IC 476, UGC 4256, UGC 4456, UGC 4543, IC 2542, UGC
6277, UGC 6523, UGC 9406, UGC 10502, UGC 11269, UGC 11891 and UGC
12276), all these galaxies present a mean size of the bins greater
than 25 pixels and an integrated total \Ha~flux on average lower
than 2.7$\pm$2.6 $10^{-16}~W~m^{-2}$.
We also confirm that there is no evidence for correlations between the residual velocity dispersion and the morphological type or the presence of a bar (when it is not dominating the potential of the galaxy).

%We also confirm the other conclusions already reached and
%discussed with half the sample (Paper VI).

%\subsection{Kinematical parameters}

The morphological parameters (input parameters of the fits) and the results of the
fits (kinematical parameters, $\chi^2$, and parameters of the residual
maps) are given in Table \ref{tablemod}.
Morphological types, distances, absolute magnitudes M$_B$, optical
radii D$_{25}/2$, axis ratios and references for HI \VFs~compiled
from the literature are given in Table \ref{tabletf}, together
with maximum velocity parameters computed from the \RCs~(V$_{max}$
and quality flag on V$_{max}$).
The 25 galaxies larger than the field of view of the instrument
are flagged in Col. 8 of Table \ref{tabletf}.
%
%For some galaxies for which the \SNR~or the spatial coverage is
%too low, the fit could not converge correctly and one or two
%parameters ($i$ and $PA$) were usually fixed to the morphological
%values to achieve the fit.
%
%The galaxies for which it was necessary to decrease the degrees of
%freedom of the model are flagged with an asterisk ($^{*}$) in Col. 5 and/or Col. 7
%Table \ref{tablemod} (see Paper VI for more details).
The galaxies for which it was necessary to decrease the degrees of
freedom of the model have their fixed parameters flagged with an asterisk ($^{*}$) in Table \ref{tablemod} (see Paper VI for more details).

%When it is not the case, parameter determinations are discussed in
%Appendix \ref{notes}. For some extreme cases, even when $i$ and
%$PA$ were fixed, the fit does not converge. In particular for
%galaxies having high inclinations, then no model was computed (see
%section \ref{maps}).

%See Paper VI and references therein for a discussion on the
%absolute systemic velocities.

\begin{figure}
\begin{center}
\includegraphics[width=7.5cm]{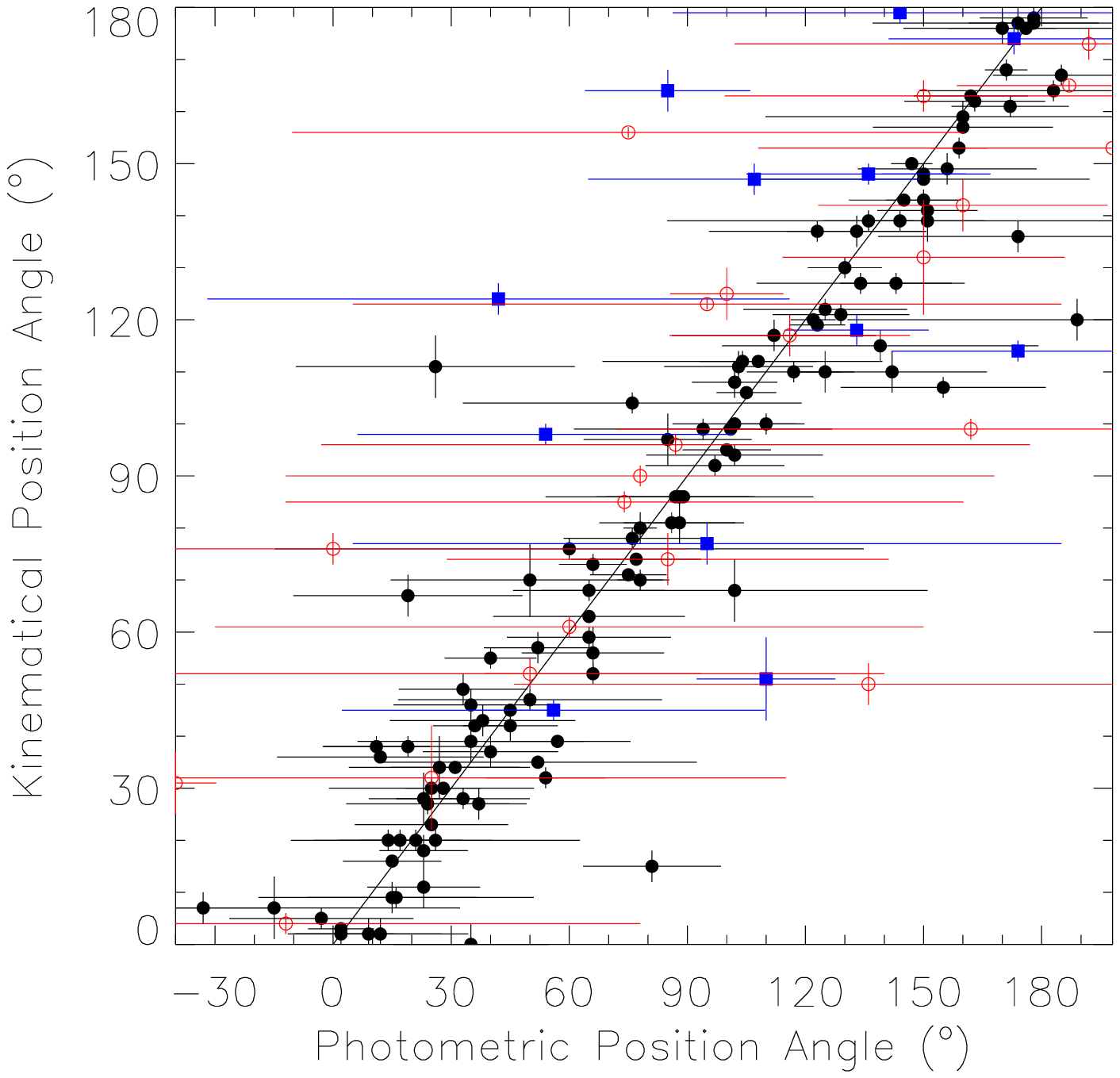}
\includegraphics[width=7.5cm]{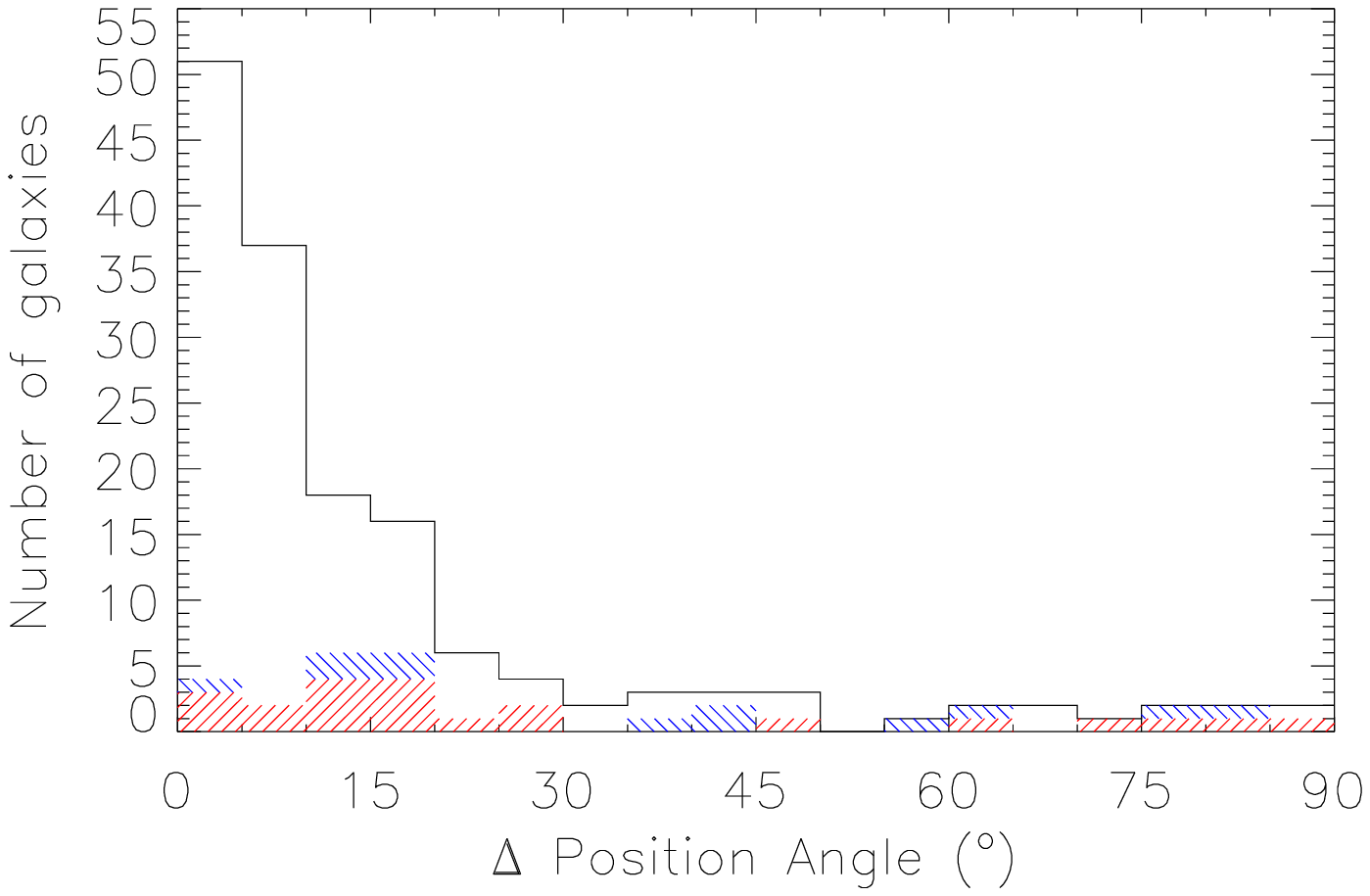}
\end{center}
\caption{\textbf{Top:} kinematical versus morphological (HyperLeda)
\PAs~of the major axis. Galaxies for which no accurate
morphological \PA~has been computed are shown by red open circles;
galaxies having an inclination lower than 25\degr~are displayed by
blue squares; the other galaxies are represented by black circles.
\textbf{Bottom:} histogram of the variation between kinematical and
morphological \PAs. The red hash, blue hash and residual white
represent respectively the galaxies for which no accurate \PA~has
been measured, for which inclination is lower than 25\degr~and the
other galaxies of the sample.} \label{pa}
\end{figure}

The kinematical \PAs~obtained by GHASP are compared with the
photometric \PAs~(found in HyperLeda) and plotted in Figure
\ref{pa}. The error bar on the morphological \PA~has been
estimated using the axis ratio and optical radius uncertainties;
for clarity, only one morphological \PA~is plotted (see Paper VI).
In Figure \ref{pa}, we have used special symbols for galaxies
with no accurate morphological \PA~(red open circles) and with an inclination
lower than 25\degr~(blue squares).
Most of the galaxies showing a disagreement in \PAs~larger than
20\degr~present: (i) a bad morphological determination of the
\PA; or (ii) a kinematical inclination lower than 25\deg; or (iii)
are specific cases discussed in Appendix \ref{notes} or in
Appendix B of Paper VI (namely these galaxies are: UGC 3013, UGC
3740, IC 476, UGC 4256, UGC 4273, UGC 4422, UGC 4543, UGC 5931,
UGC 10310, UGC 10359, UGC 10470, UGC 10445, UGC 10897, UGC 11283,
UGC 11861 and UGC 12060).
Morphological \PAs~of low inclination galaxies have systematically
higher uncertainties than kinematical ones (see Figure \ref{pa}).
For kinematical inclinations greater than 25\degr, the mean error
on morphological \PAs~is $\sim$20\degr~and the mean error on
kinematical \PAs~is $\sim$2\degr.
For inclinations lower than 25\degr~the difference is larger: the
mean error on morphological \PAs~is $\sim$40\degr~while the mean
error on kinematical \PAs~is $\sim$3\degr.
For the whole sample, the histogram of the variation between
kinematical and morphological \PAs~given in Figure \ref{pa}
(bottom) indicates that (i) for 57\% of the galaxies, the
agreement is better than 10\degr; (ii) for 79\%, the agreement is
better than 20\degr; (iii) the disagreement is larger than
30\degr~for 15\% of the galaxies.
%
%(i) for more than 67\% of the galaxies, the agreement is better
%than 10\degr; (ii) for more than 85\%, the agreement is better
%than 20\degr; (iii) the disagreement is larger than 30\degr~for
%8\% of these galaxies.
%
The conclusion addressed in Paper VI remains valid, i.e.
integral field spectroscopy constitutes the best technique to
determine \PAs~and as a consequence, \RCs.

\begin{figure}
\begin{center}
\includegraphics[width=7.5cm]{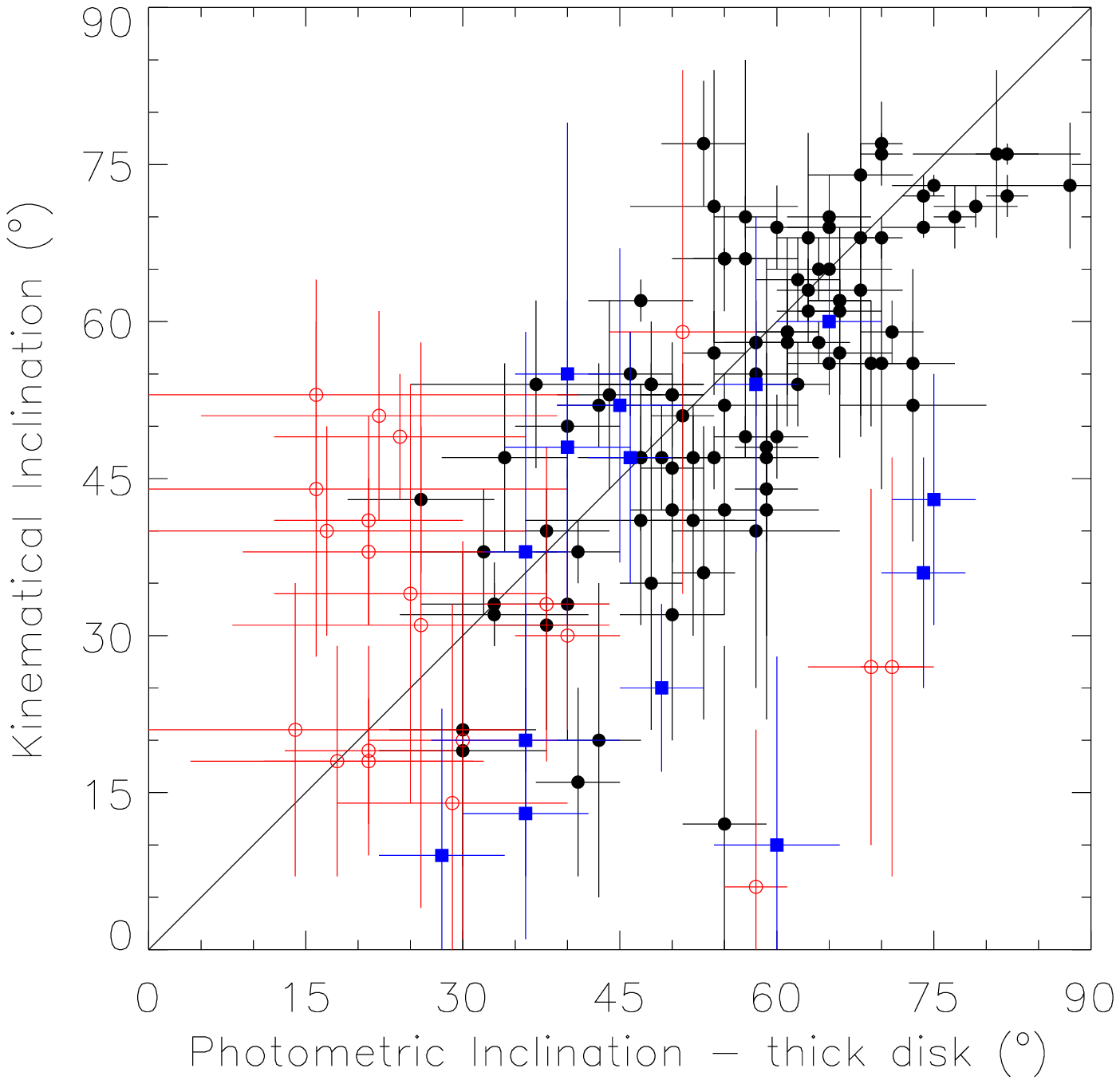}
\includegraphics[width=7.5cm]{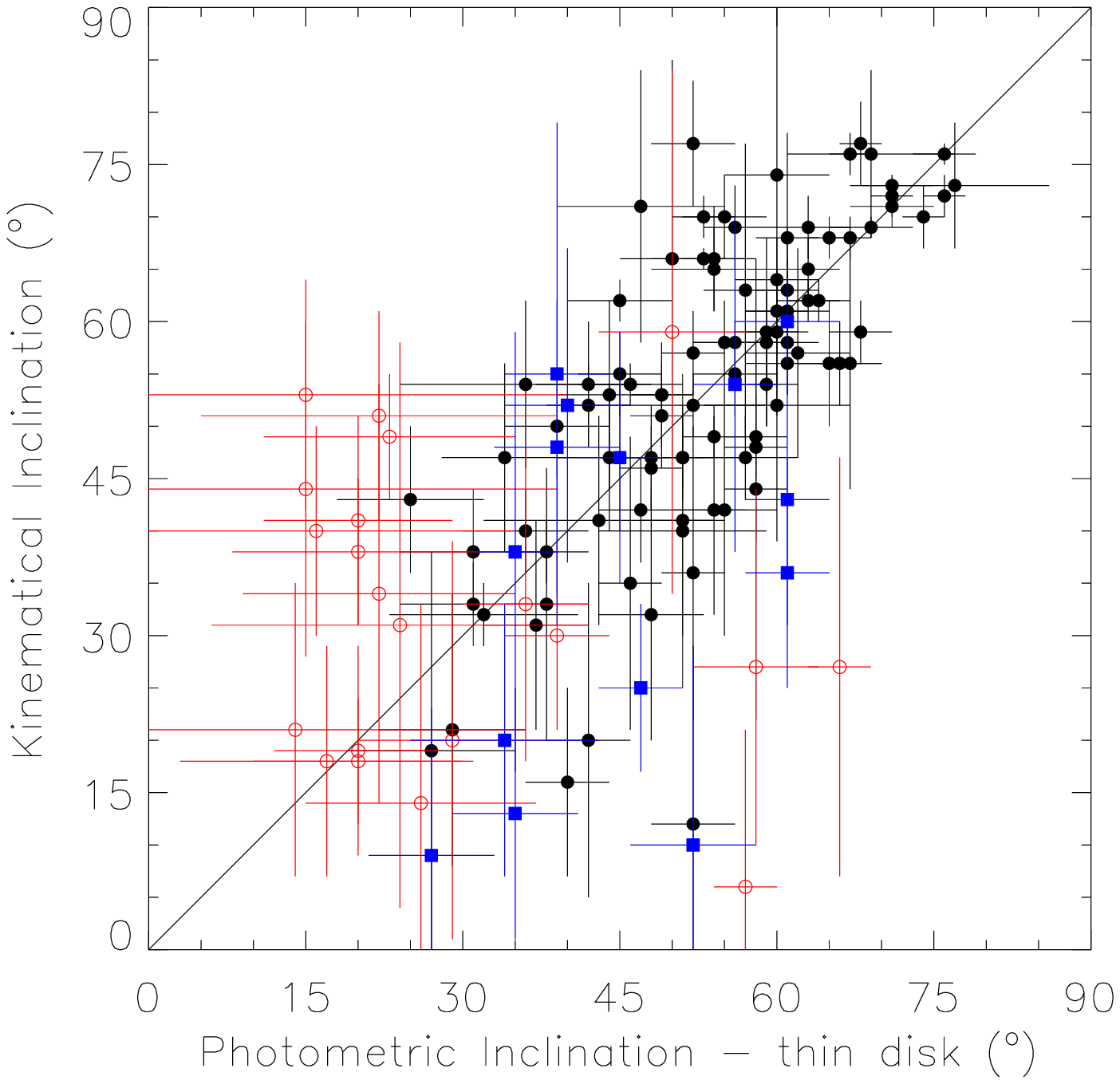}
\includegraphics[width=7.5cm]{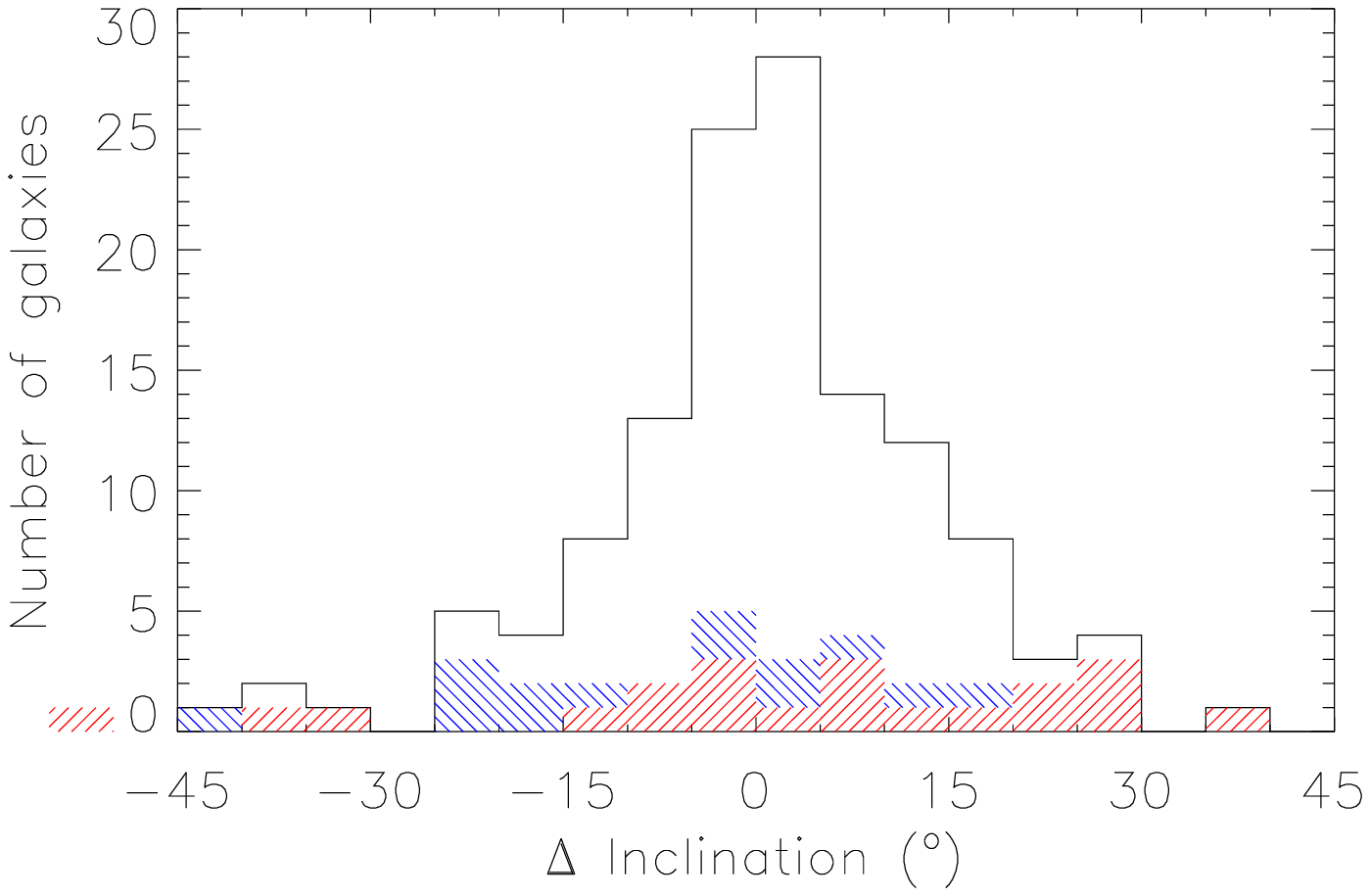}
\end{center}
\caption{\textbf{Top:} kinematical versus thick disk morphological inclinations.
\textbf{Middle:} kinematical versus thin disk morphological inclinations.
\textbf{Top and Middle:} Galaxies for which no accurate morphological
\PA~has been computed are shown by red open circles; galaxies with
a difference between the kinematical and morphological \PAs~larger
than 20\degr~are displayed with blue squares; the other galaxies are
represented by black circles. \textbf{Bottom:} histogram of the variation
between kinematical and morphological inclinations. The red hash,
blue hash and residual white represent respectively the galaxies
for which no accurate \PA~has been measured, for which the
difference between the kinematical and morphological \PAs~is
larger than 20\degr~and the other galaxies of the sample.}
\label{inclination}
\end{figure}

The kinematical and morphological inclinations are compared in
Figure \ref{inclination}. On the top panel, the photometric
inclination is computed using a correction factor depending on the
morphological type (see Paper VI).
On the middle panel the photometric inclination is derived from
the axis ratio. Galaxies for which the morphological \PA~could not
be determined accurately are represented by red open circles while
galaxies with a difference between morphological and kinematical
\PAs~larger than 20\degr~are displayed with blue squares.
The conclusions given in Paper VI are still valid: (i) the
agreement between photometric and kinematical inclinations is
better for high inclination galaxies; (ii) morphological
inclination determination is unreliable if the measure of the
\PA~is not reliable; (iii) the errors on morphological
inclinations and on kinematical inclinations are comparable; (iv)
kinematical methods may underestimate (or morphological method overestimate) the inclination; (v) the kinematical
inclinations are closer to the morphological inclinations when
the latter are computed without any correction for the morphological type (see Figure \ref{inclination} top
and middle).
The histogram of the difference between morphological and
kinematical inclinations (Figure \ref{inclination}, bottom) shows
that a difference of inclination larger than 10\degr~is found for
less than 40\% of the sample.  Taking into account the whole GHASP
sample, the number of galaxies by bin is higher and the histogram
is more symmetrical with respect to that computed in Paper VI.

\section{The Tully-Fisher relation}
\label{tullyfisher}

\begin{figure}
\begin{center}
\includegraphics[width=7.5cm]{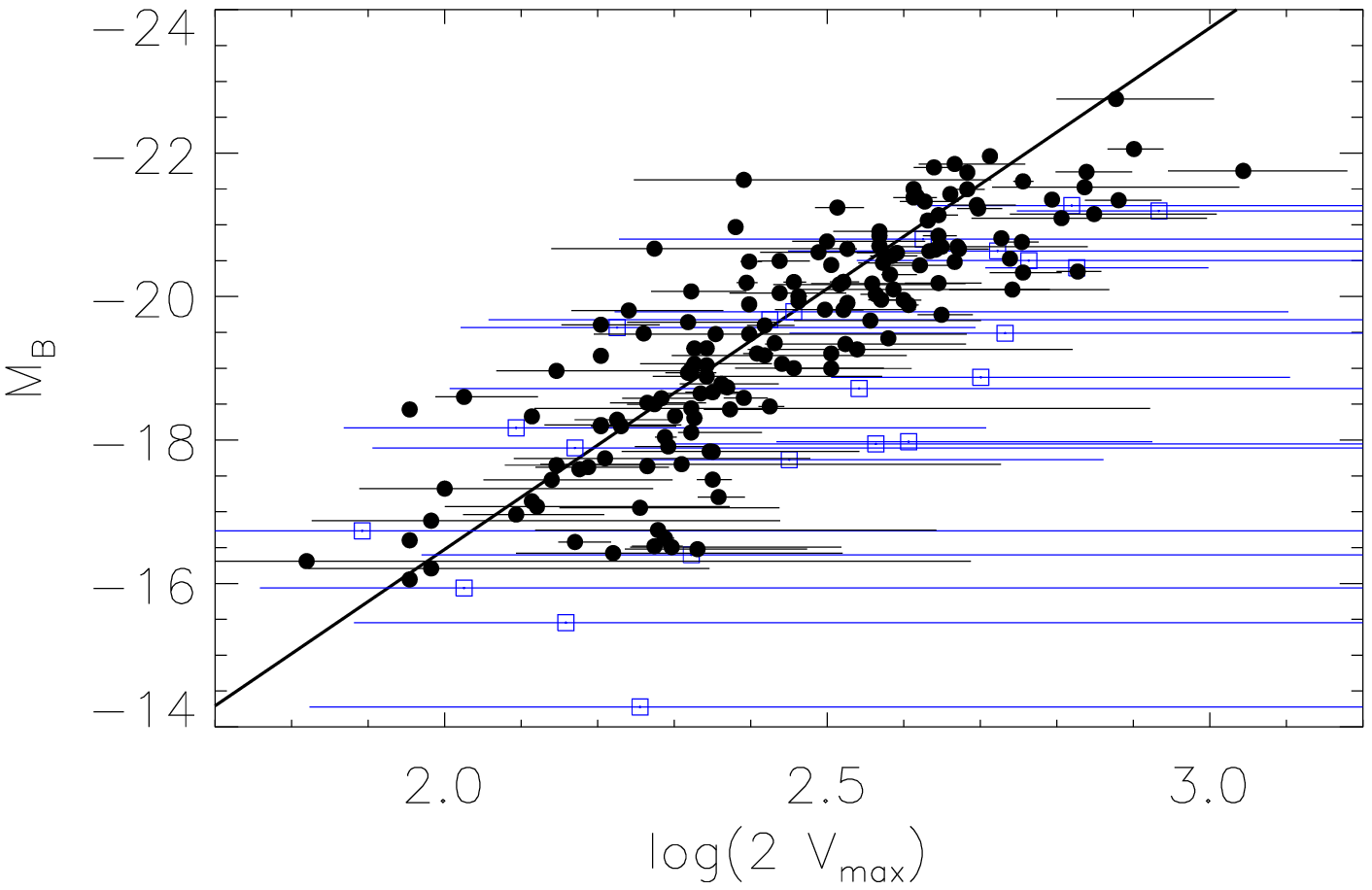}
\includegraphics[width=7.5cm]{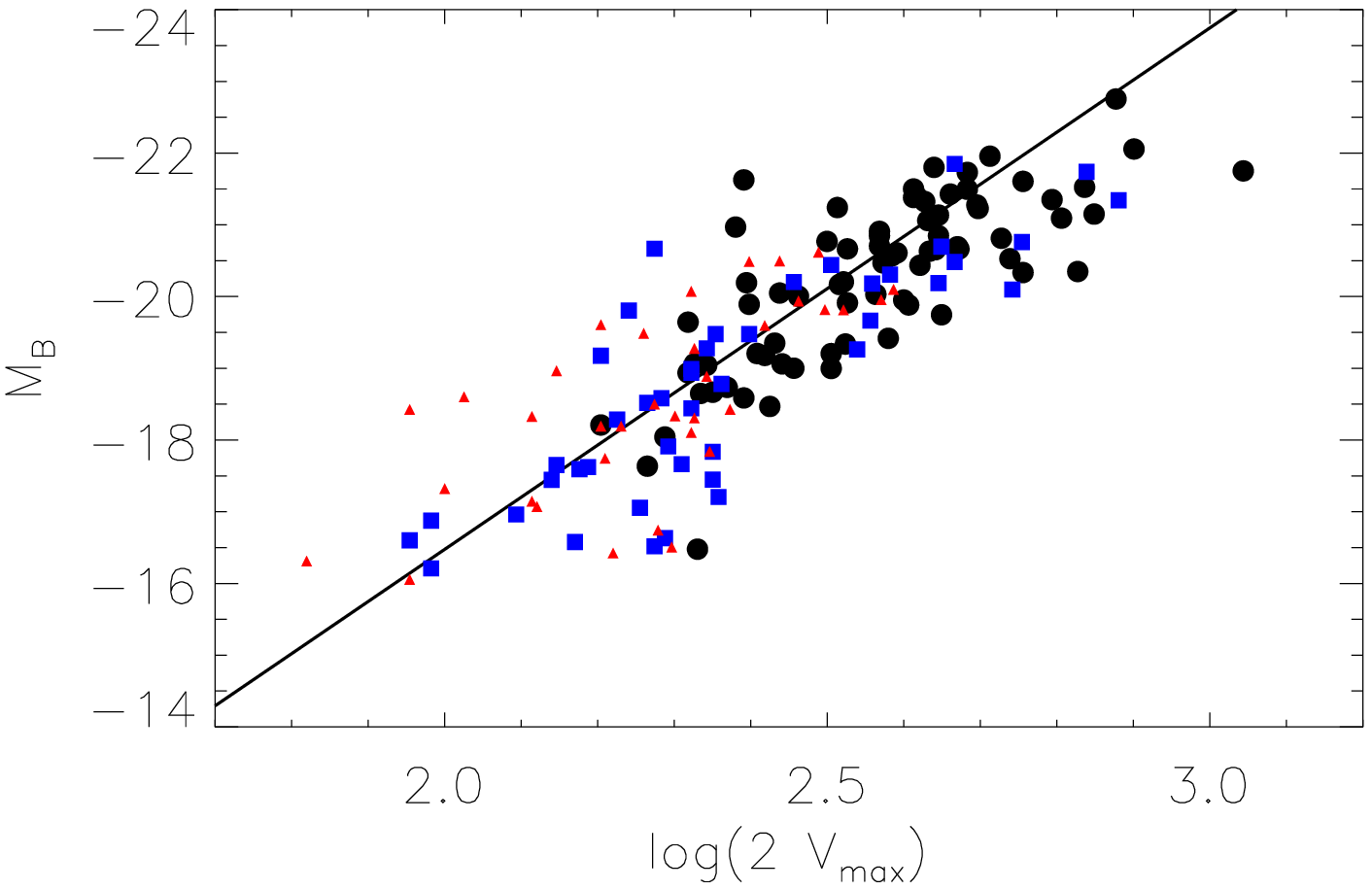}
\includegraphics[width=7.5cm]{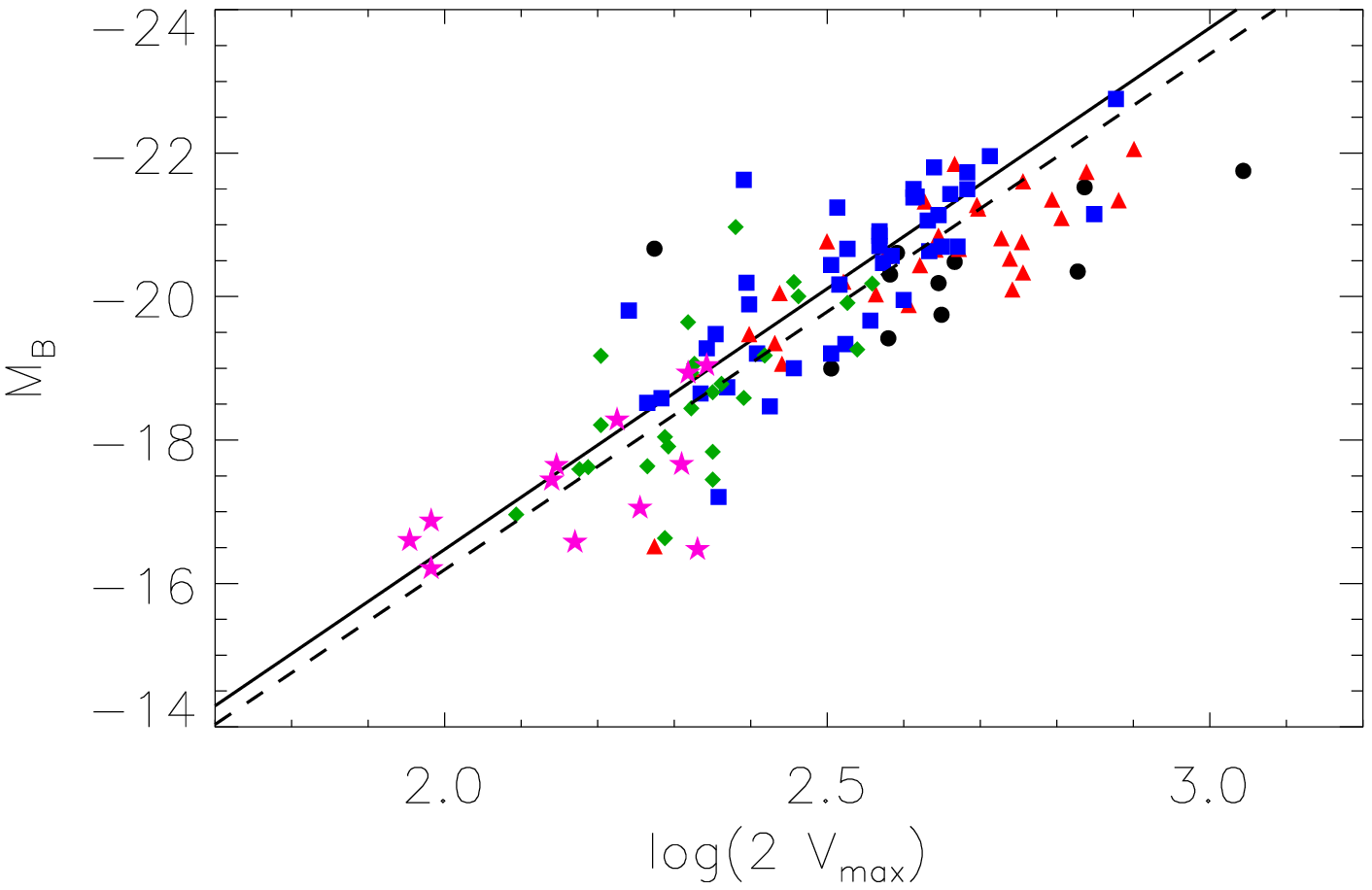}
\end{center}
\caption{TF relation for our sample of galaxies. The solid line
represents the B magnitude \TF~relation determined by
\citet{Tully:2000} from nearby galaxies in clusters (Ursa Major,
Pisces filament, Coma). \textbf{Top:} subdivided by inclination - low
inclination galaxies ($i<$25\degr): blue squares; other galaxies
($i$$\geq$25\degr): black circles. \textbf{Middle:} subdivided by
V$_{max}$ flags - V$_{max}$ reached: black dots, large size;
V$_{max}$ probably reached: blue squares, medium size; V$_{max}$
probably not reached: red triangles, small size. \textbf{Bottom:}
subdivided by morphological type - black circles from 0 to 2; red
triangles from 2 to 4; blue squares from 4 to 6; green rhombuses
from 6 to 8; pink stars from 8 to 10; the dashed line represents
the best linear fit to the data.}
\label{tullyfisher_plot}      % for cross-references
\end{figure}

Among the whole sample of 203 galaxies, we have plotted the
\TF~relation (\citealp{Tully:1977}, M$_{B}$ as a function of
$\log{2 V_{max}}$) for a sub-sample of 177 galaxies in Figure
\ref{tullyfisher_plot}. The 26 other galaxies are not considered
in the present discussion because (i) for seven galaxies the
\RC~does not reach the maximum rotation velocity (UGC 1117, UGC
1655, UGC 2455, UGC 4393, UGC 6523, UGC 8898 and UGC 9406); (ii)
no B magnitude is available for three galaxies (UGC 2800, UGC 11496,
UGC 12212) and (iii) no velocity measurement, either on the
\RC~or on the \PVM~is possible for 16 other galaxies (see Table
\ref{tabletf}).
The maximum velocity V$_{max}$ and its error have been obtained
from the fit to the \VF~and the solid line in Figure
\ref{tullyfisher_plot} is the relation found by \citet{Tully:2000}
(see Paper VI).

In Figure \ref{tullyfisher_plot} (Top), the error bars on the
velocity are displayed and galaxies with inclination lower than
25\degr~are distinguished (blue open squares). As already noticed
in Paper VI, these galaxies have statistically higher velocities
than expected from the \citet{Tully:2000} relation and have large
error bars. Considering this effect, we choose to exclude the 22
galaxies with inclinations lower than 25\degr~from the
\TF~analysis (see Paper VI).
Among the 155 remaining galaxies, the maximum velocity V$_{max}$
is reached for 76 of them (black dots, large size), probably
reached for 44 of them (blue squares, medium size) and probably
not reached for 35 of them (red triangles, small size). They are
distinguished in Figure \ref{tullyfisher_plot} (Middle) and
flagged in Table \ref{tabletf}.
The quality flag on the maximum velocity is given in Table
\ref{tabletf} (see Paper VI).

%It appears from this last point that the HI line width at 20\%~has
%most often the best agreement with the \ha~\VF~amplitude (better
%than the line width at 50\%) xxx.

Figure \ref{tullyfisher_plot} (Middle) confirms the two
classifications ``V$_{max}$ probably reached'' and``V$_{max}$
probably not reached'' since for the majority of each class the
points are respectively in agreement and above the
\citet{Tully:2000} relation. From the two classes ``V$_{max}$
reached'' and ``V$_{max}$ probably reached'', we find the following
relation (see Paper VI for additional details):
\begin{equation}
%\mathbf{M_{B}=(-6.9\pm 1.6)[\log{2 V_{max}} - 2.5] -(19.8\pm0.1)}
M_{B}=(-7.2\pm 1.2)[\log{2 V_{max}} - 2.5] -(19.8\pm0.1)
\label{tfghasp}
\end{equation}

This relation is displayed as a dashed line in Figure
\ref{tullyfisher_plot}, in which morphological types are
distinguished for the two best classes (black circles from 0 to 2,
red triangles from 2 to 4, blue squares from 4 to 6, green
rhombuses from 6 to 8 and pink stars from 8 to 10).
The slope of the linear regression computed here from the whole
GHASP sample is now exactly the same as the one computed by
\citet{Tully:2000} (this slope was found to be somewhat lower in Paper
VI), and its uncertainty has been reduced by a factor 0.75.
%-
For the \TF~relation, we note that fast rotators
(V$_{max}>300$\kms: UGC 89, UGC 508, UGC 3429, UGC 4422, UGC 4820,
UGC 5532, UGC 8900, UGC 8937, UGC 9969 and UGC 11470) are less
luminous than expected. This trend has already been noticed and
discussed in Paper VI.

\section{Summary and conclusions}
\label{conclusion}

The knowledge of the links between the kinematical and dynamical
state of galaxies helps us to increase our understanding of the
physics and evolution of galaxies. The GHASP sample, which
consists of 203 spiral and irregular galaxies, covering a wide
range of morphological types and absolute magnitudes, has been
constituted in order to provide a kinematical reference sample of
nearby galaxies. The galaxies have been observed in the \ha~line
using Fabry-Perot techniques, leading to the construction of data
cubes.  This sample is by now the largest set of galaxies ever
homogeneously observed with Fabry-Perot techniques.  Major
improvements in the reduction (adaptive binning techniques, ghost
suppression, treatment of faint outskirts regions, etc) and in the
analysis (determination of the \RC~and of the kinematical
parameters and their uncertainties, etc) have been developed and
implemented in the data reduction procedure and homogeneously
applied to the whole GHASP sample (see Paper VI for additional
details).

In this paper, 97 galaxies have been re-reduced using adaptive
binning techniques in order to provide homogeneous data for the
whole sample. For each galaxy, we have presented the \ha~\VF, the
\ha~monochromatic image and eventually the \ha~residual \VF, the
\PVM~along the major axis and the \RC, when available, leading for
the whole sample to 200 \VFs~and 177 \RCs.

From the data cubes, integrated \Ha~profiles have also been
produced. A post calibration has allowed to compute indirect
absolute \Ha~flux for all the galaxies belonging to the GHASP
sample.  This post calibration has been done using fluxes for 69
galaxies found in the literature \citep{James:2004}.

We confirm and strengthen most of the results already obtained from half
the sample:

\begin{itemize}

    \item A high quality model has been achieved to represent the
axi-symmetric rotational component of the galaxies since no
typical signatures for biases are observed in the
residual \VFs. This means that the residuals observed in the
residual \VF~are due to actual non circular motions and not to an
uncorrect determination of the kinematical parameters (position of
the center, \PA, inclination and rotation velocity). In addition,
the \PVMs~confirm the validity of the \RCs.

    \item The mean residual velocity dispersion is strongly
correlated with the maximum amplitude of the \VF. For a given
velocity range, this correlation does not clearly depend on the
morphological type. However strongly barred galaxies have a higher
residual velocity dispersion than mild-barred or unbarred
galaxies. Peculiar galaxies also show a high residual velocity
dispersion.

    \item The determinations of kinematical \pas~are robust whatever the inclination
of the galaxy whereas morphological \pas~are poorly determined for
low inclination systems. Moreover, morphological \PAs~have
systematically higher uncertainties than kinematical ones. This is
a major argument for deriving \rcs~from integral field
spectroscopy rather than long slit spectroscopy instruments that
could lead to incorrect positioning of the slit (a difference
between the morphological and kinematical \PAs~larger than
30\degr~is found for $\sim$15$\%$ of the GHASP sample). This may
strongly bias mass distribution models and \TF~studies.
In order to build a mass model, the stellar mass distribution derived from the surface brightness profile is combined with the \rc~deduced from the \vf. The \pas~of the major axis deduced from the surface brightness image and from the \vf~should be identical.
Important inconsistencies may appear if these \PAs~are misaligned.

    \item Galaxies with poor determination of their morphological
\pas~have usually unreliable and overestimated morphological
inclinations. The agreement between kinematical and morphological
inclinations is better when assuming a thin disk in particular for high
inclination galaxies. For galaxies with intermediate disk
inclinations (higher than 25\degr~and lower than 75\degr), to
improve the quality of the \RC, it is possible to reduce the
degrees of freedom in kinematical models by fixing the inclination to the morphological value. This is specially true when only
low quality kinematical data are available as it is the case for
high redshift galaxies.

    \item The use of the whole GHASP sample leads to a \TF~relationship in perfect agreement
with \citet{Tully:2000}, despite important differences in the
selection of both samples. With respect to the result presented in
Paper VI, the use of the whole sample increases the agreement with
\citet{Tully:2000}. Three comments should be underlined: (i)
galaxies with inclination lower than 25\degr~are inappropriate for
\TF~relation determination since their estimated velocities are
easily overestimated; (ii) fast rotators (V$_{max}>$300\kms) are
maybe less luminous (than expected from the \TF~relation); (iii)
for fast rotators and high luminosity galaxies, the agreement with
the \TF~relation is better when the morphological inclination of
the galaxy is computed without taking into account the increasing
thickness of the disk when the morphological type of the galaxies
moves from early to late types.

\end{itemize}

From these data and analysis, it is now possible to adress the
scientific drivers on the whole GHASP sample in forthcoming
papers.

\section*{acknowledgements} {The authors warmly thank their collaborators:
Philippe Balard, Chantal Balkowski, Jacques Boulesteix, Olivier
Boissin, Claude Carignan, Laurent Chemin, Olivier Daigle, Jean-Luc
Gach, Olivia Garrido and Olivier Hernandez for having participated
to the previous works making possible the new analyze of the
observations. They thank Isabelle J\'egouzo and Christian Surace for building the Fabry-Perot Database. The authors wish to thank the referee Dr P. James who helped improving
the manuscript.
They also thank the Programme National Galaxies for supporting the
GHASP project in allocating continuously observing time during
several years, the Observatoire de Haute-Provence team for its
technical assistance during the observations.
This research has made use of the
NASA/IPAC Extragalactic Database (NED) which is operated by the
Jet Propulsion Laboratory, California Institute of Technology,
under contract with the National Aeronautics and Space
Administration. The authors have also made an extensive use of the
HyperLeda Database (\url{http://leda.univ-lyon1.fr}).
The Digitized Sky Surveys were produced at the Space Telescope
Science Institute under U.S. Government grant NAG W-2166. The
images of these surveys are based on photographic data obtained
using the Oschin Schmidt Telescope on Palomar Mountain and the UK
Schmidt Telescope. The plates were processed into the present
compressed digital form with the permission of these
institutions.}

\bibliographystyle{mn2e}
\bibliography{biblio}

\begin{thebibliography}{}

\bibitem[\protect\citeauthoryear{{de Vaucouleurs}}{{de
  Vaucouleurs}}{1979}]{de-Vaucouleurs:1979}
{de Vaucouleurs} G.,  1979, \apj, 227, 380

\bibitem[\protect\citeauthoryear{{Dopita} \& {Hua}}{{Dopita} \&
  {Hua}}{1997}]{Dopita:1997}
{Dopita} M.~A.,  {Hua} C.~T.,  1997, \apjs, 108, 515

\bibitem[\protect\citeauthoryear{{Epinat}, {Amram}, {Marcelin}, {Balkowski},
  {Daigle}, {Hernandez}, {Chemin}, {Carignan}, {Gach} \& {Balard}}{{Epinat}
  et~al.}{2008}]{Epinat:2008}
{Epinat} B.,  {Amram} P.,  {Marcelin} M.,  {Balkowski} C.,  {Daigle} O.,
  {Hernandez} O.,  {Chemin} L.,  {Carignan} C.,  {Gach} J.~.,    {Balard} P.,
  2008, ArXiv e-prints, 805

\bibitem[\protect\citeauthoryear{{Garrido}, {Marcelin} \& {Amram}}{{Garrido}
  et~al.}{2004}]{Garrido:2004}
{Garrido} O.,  {Marcelin} M.,    {Amram} P.,  2004, \mnras, 349, 225

\bibitem[\protect\citeauthoryear{{Garrido}, {Marcelin}, {Amram}, {Balkowski},
  {Gach} \& {Boulesteix}}{{Garrido} et~al.}{2005}]{Garrido:2005}
{Garrido} O.,  {Marcelin} M.,  {Amram} P.,  {Balkowski} C.,  {Gach} J.~L.,
  {Boulesteix} J.,  2005, \mnras, 362, 127

\bibitem[\protect\citeauthoryear{{Garrido}, {Marcelin}, {Amram} \&
  {Boissin}}{{Garrido} et~al.}{2003}]{Garrido:2003}
{Garrido} O.,  {Marcelin} M.,  {Amram} P.,    {Boissin} O.,  2003, \aap, 399,
  51

\bibitem[\protect\citeauthoryear{{Garrido}, {Marcelin}, {Amram} \&
  {Boulesteix}}{{Garrido} et~al.}{2002}]{Garrido:2002}
{Garrido} O.,  {Marcelin} M.,  {Amram} P.,    {Boulesteix} J.,  2002, \aap,
  387, 821

\bibitem[\protect\citeauthoryear{{Haynes}, {Giovanelli}, {Salzer}, {Wegner},
  {Freudling}, {da Costa}, {Herter} \& {Vogt}}{{Haynes}
  et~al.}{1999}]{Haynes:1999}
{Haynes} M.~P.,  {Giovanelli} R.,  {Salzer} J.~J.,  {Wegner} G.,  {Freudling}
  W.,  {da Costa} L.~N.,  {Herter} T.,    {Vogt} N.~P.,  1999, \aj, 117, 1668

\bibitem[\protect\citeauthoryear{{Irwin}}{{Irwin}}{1994}]{Irwin:1994}
{Irwin} J.~A.,  1994, \apj, 429, 618

\bibitem[\protect\citeauthoryear{{James}, {Shane}, {Beckman}, {Cardwell},
  {Collins}, {Etherton}, {de Jong}, {Fathi}, {Knapen}, {Peletier}, {Percival},
  {Pollacco}, {Seigar}, {Stedman} \& {Steele}}{{James}
  et~al.}{2004}]{James:2004}
{James} P.~A.,  {Shane} N.~S.,  {Beckman} J.~E.,  {Cardwell} A.,  {Collins}
  C.~A.,  {Etherton} J.,  {de Jong} R.~S.,  {Fathi} K.,  {Knapen} J.~H.,
  {Peletier} R.~F.,  {Percival} S.~M.,  {Pollacco} D.~L.,  {Seigar} M.~S.,
  {Stedman} S.,    {Steele} I.~A.,  2004, \aap, 414, 23

\bibitem[\protect\citeauthoryear{{Karachentsev}, {Karachentseva}, {Huchtmeier}
  \& {Makarov}}{{Karachentsev} et~al.}{2004}]{Karachentsev:2004}
{Karachentsev} I.~D.,  {Karachentseva} V.~E.,  {Huchtmeier} W.~K.,    {Makarov}
  D.~I.,  2004, \aj, 127, 2031

\bibitem[\protect\citeauthoryear{{Koopmann}, {Haynes} \&
  {Catinella}}{{Koopmann} et~al.}{2006}]{Koopmann:2006}
{Koopmann} R.~A.,  {Haynes} M.~P.,    {Catinella} B.,  2006, \aj, 131, 716

\bibitem[\protect\citeauthoryear{{Kornreich}, {Haynes}, {Lovelace} \& {van
  Zee}}{{Kornreich} et~al.}{2000}]{Kornreich:2000}
{Kornreich} D.~A.,  {Haynes} M.~P.,  {Lovelace} R.~V.~E.,    {van Zee} L.,
  2000, \aj, 120, 139

\bibitem[\protect\citeauthoryear{{Laine} \& {Gottesman}}{{Laine} \&
  {Gottesman}}{1998}]{Laine:1998}
{Laine} S.,  {Gottesman} S.~T.,  1998, \mnras, 297, 1041

\bibitem[\protect\citeauthoryear{{Marcum et al.}}{{Marcum et
  al.}}{2001}]{Marcum:2001}
{Marcum et al.} P.~M.,  2001, \apjs, 132, 129

\bibitem[\protect\citeauthoryear{{Moustakas} \& {Kennicutt} Jr.}{{Moustakas} \&
  {Kennicutt}}{2006}]{Moustakas:2006}
{Moustakas} J.,  {Kennicutt} Jr. R.~C.,  2006, \apjs, 164, 81

\bibitem[\protect\citeauthoryear{{Nilson}}{{Nilson}}{1973}]{Nilson:1973}
{Nilson} P.,  1973, Nova Acta Regiae Soc.~Sci.~Upsaliensis Ser.~V, pp~0--+

\bibitem[\protect\citeauthoryear{{Noordermeer}, {van der Hulst}, {Sancisi},
  {Swaters} \& {van Albada}}{{Noordermeer} et~al.}{2005}]{Noordermeer:2005}
{Noordermeer} E.,  {van der Hulst} J.~M.,  {Sancisi} R.,  {Swaters} R.~A.,
  {van Albada} T.~S.,  2005, \aap, 442, 137

\bibitem[\protect\citeauthoryear{{O'Connell}, {Gallagher} III \&
  {Hunter}}{{O'Connell} et~al.}{1994}]{OConnell:1994}
{O'Connell} R.~W.,  {Gallagher} III J.~S.,    {Hunter} D.~A.,  1994, \apj, 433,
  65

\bibitem[\protect\citeauthoryear{{Paturel}, {Andernach}, {Bottinelli}, {di
  Nella}, {Durand}, {Garnier}, {Gouguenheim}, {Lanoix}, {Marthinet}, {Petit},
  {Rousseau}, {Theureau} \& {Vauglin}}{{Paturel} et~al.}{1997}]{Paturel:1997}
{Paturel} G.,  {Andernach} H.,  {Bottinelli} L.,  {di Nella} H.,  {Durand} N.,
  {Garnier} R.,  {Gouguenheim} L.,  {Lanoix} P.,  {Marthinet} M.~C.,  {Petit}
  C.,  {Rousseau} J.,  {Theureau} G.,    {Vauglin} I.,  1997, \aaps, 124, 109

\bibitem[\protect\citeauthoryear{{Paturel}, {Fang}, {Petit}, {Garnier} \&
  {Rousseau}}{{Paturel} et~al.}{2000}]{Paturel:2000}
{Paturel} G.,  {Fang} Y.,  {Petit} C.,  {Garnier} R.,    {Rousseau} J.,  2000,
  \aaps, 146, 19

\bibitem[\protect\citeauthoryear{{Paturel}, {Garcia}, {Fouque} \&
  {Buta}}{{Paturel} et~al.}{1991}]{Paturel:1991}
{Paturel} G.,  {Garcia} A.~M.,  {Fouque} P.,    {Buta} R.,  1991, \aap, 243,
  319

\bibitem[\protect\citeauthoryear{{Rownd}, {Dickey} \& {Helou}}{{Rownd}
  et~al.}{1994}]{Rownd:1994}
{Rownd} B.~K.,  {Dickey} J.~M.,    {Helou} G.,  1994, \aj, 108, 1638

\bibitem[\protect\citeauthoryear{{Saha}, {Thim}, {Tammann}, {Reindl} \&
  {Sandage}}{{Saha} et~al.}{2006}]{Saha:2006}
{Saha} A.,  {Thim} F.,  {Tammann} G.~A.,  {Reindl} B.,    {Sandage} A.,  2006,
  \apjs, 165, 108

\bibitem[\protect\citeauthoryear{{Schulman}, {Bregman}, {Brinks} \&
  {Roberts}}{{Schulman} et~al.}{1996}]{Schulman:1996}
{Schulman} E.,  {Bregman} J.~N.,  {Brinks} E.,    {Roberts} M.~S.,  1996, \aj,
  112, 960

\bibitem[\protect\citeauthoryear{{Shapley}, {Fabbiano} \& {Eskridge}}{{Shapley}
  et~al.}{2001}]{Shapley:2001}
{Shapley} A.,  {Fabbiano} G.,    {Eskridge} P.~B.,  2001, \apjs, 137, 139

\bibitem[\protect\citeauthoryear{{Spano}, {Marcelin}, {Amram}, {Carignan},
  {Epinat} \& {Hernandez}}{{Spano} et~al.}{2007}]{Spano:2008}
{Spano} M.,  {Marcelin} M.,  {Amram} P.,  {Carignan} C.,  {Epinat} B.,
  {Hernandez} O.,  2007, \mnras, pp 1084--+

\bibitem[\protect\citeauthoryear{{Springob}, {Masters}, {Haynes}, {Giovanelli}
  \& {Marinoni}}{{Springob} et~al.}{2007}]{Springob:2007}
{Springob} C.~M.,  {Masters} K.~L.,  {Haynes} M.~P.,  {Giovanelli} R.,
  {Marinoni} C.,  2007, \apjs, 172, 599

\bibitem[\protect\citeauthoryear{{Swaters}}{{Swaters}}{1999}]{Swaters:phd}
{Swaters} R.~A.,  1999, PhD thesis, , Rijksuniversiteit Groningen, (1999)

\bibitem[\protect\citeauthoryear{{Swaters}, {Madore}, {van den Bosch} \&
  {Balcells}}{{Swaters} et~al.}{2003}]{Swaters:2003}
{Swaters} R.~A.,  {Madore} B.~F.,  {van den Bosch} F.~C.,    {Balcells} M.,
  2003, \apj, 583, 732

\bibitem[\protect\citeauthoryear{{Swaters}, {van Albada}, {van der Hulst} \&
  {Sancisi}}{{Swaters} et~al.}{2002}]{Swaters:2002}
{Swaters} R.~A.,  {van Albada} T.~S.,  {van der Hulst} J.~M.,    {Sancisi} R.,
  2002, \aap, 390, 829

\bibitem[\protect\citeauthoryear{{Tully} \& {Fisher}}{{Tully} \&
  {Fisher}}{1977}]{Tully:1977}
{Tully} R.~B.,  {Fisher} J.~R.,  1977, \aap, 54, 661

\bibitem[\protect\citeauthoryear{{Tully} \& {Pierce}}{{Tully} \&
  {Pierce}}{2000}]{Tully:2000}
{Tully} R.~B.,  {Pierce} M.~J.,  2000, \apj, 533, 744

\bibitem[\protect\citeauthoryear{{Tully}, {Verheijen}, {Pierce}, {Huang} \&
  {Wainscoat}}{{Tully} et~al.}{1996}]{Tully:1996}
{Tully} R.~B.,  {Verheijen} M.~A.~W.,  {Pierce} M.~J.,  {Huang} J.-S.,
  {Wainscoat} R.~J.,  1996, \aj, 112, 2471

\bibitem[\protect\citeauthoryear{{van der Kruit} \& {Allen}}{{van der Kruit} \&
  {Allen}}{1978}]{van-der-Kruit:1978}
{van der Kruit} P.~C.,  {Allen} R.~J.,  1978, \araa, 16, 103

\bibitem[\protect\citeauthoryear{{Vauglin}, {Paturel}, {Borsenberger},
  {Fouqu{\'e}}, {Epchtein}, {Kimeswenger}, {Tiph{\`e}ne}, {Lanoix} \&
  {Courtois}}{{Vauglin} et~al.}{1999}]{Vauglin:1999}
{Vauglin} I.,  {Paturel} G.,  {Borsenberger} J.,  {Fouqu{\'e}} P.,  {Epchtein}
  N.,  {Kimeswenger} S.,  {Tiph{\`e}ne} D.,  {Lanoix} P.,    {Courtois} H.,
  1999, \aaps, 135, 133

\bibitem[\protect\citeauthoryear{{Wilcots} \& {Prescott}}{{Wilcots} \&
  {Prescott}}{2004}]{Wilcots:2004}
{Wilcots} E.~M.,  {Prescott} M.~K.~M.,  2004, \aj, 127, 1900

\bibitem[\protect\citeauthoryear{{Wilcots}, {Turnbull} \& {Brinks}}{{Wilcots}
  et~al.}{2001}]{Wilcots:2001}
{Wilcots} E.~M.,  {Turnbull} M.~C.,    {Brinks} E.,  2001, \apj, 560, 110

\bibitem[\protect\citeauthoryear{{Williams}, {Yun} \&
  {Verdes-Montenegro}}{{Williams} et~al.}{2002}]{Williams:2002}
{Williams} B.~A.,  {Yun} M.~S.,    {Verdes-Montenegro} L.,  2002, \aj, 123,
  2417

\end{thebibliography}

\appendix

\clearpage
\section{Notes on individual galaxies}
\label{notes}
\noindent \textbf{UGC 508}. HI inclination (25\degr, \citealp{Noordermeer:2005}) as well as photometric inclination (14\degr, from an axis
ratio of 0.97) lead to a very high maximum rotation velocity
($\sim$550\kms). From \TF~relationship, its absolute magnitude
M$_B$ suggests a maximum velocity around 270\kms, leading to an
axis ratio of 0.5 (hence an inclination of 60\degr). This
strongly barred galaxy shows moreover clear evidence for
interaction, resulting in perturbed morphology and velocity field.
This biases the determination of the inclination by morphological
as well as kinematical methods, leading to an unrealistically high
maximum rotation velocity.
\\
\noindent \textbf{UGC 1117}. This galaxy is the famous M33.
Because of the limited \FOV~of GHASP we only observed the solid
body central part of the \RC.  The external round shape structures
in the different images are due to edge-effects of the
interference filter vignetting the \FOV.
\\
\noindent \textbf{UGC 1249}. No \RC~has been computed because of
its high inclination (90\degr).
\\
\noindent \textbf{UGC 1256}. Within $\sim$25\arcsec~($\sim$1 kpc),
the \RC~shows negative rotation velocity. This is due to the fact
that the bar is almost parallel to the major axis.
\\
\noindent \textbf{UGC 1736}. The kinematical center chosen in
Paper IV is different from the morphological center chosen here,
leading to a different \RC.
\\
\noindent \textbf{UGC 1913}. Same comment as for UGC 1736.
\\
\noindent \textbf{UGC 2023}. Despite the fact that this
galaxy has been observed again (2 hours exposure time on September
11th 2002) and that these data have been compared and added to the
data presented in Paper II, the \SNR~remains very weak.  From the
R-band image observed by \citet{James:2004}, it is now possible to
accurately determine the center of that very low surface
brightness object.  The major axis \PA~and the inclination have
been set to the values determined from HI data
\citet{Swaters:phd}. The kinematical center chosen in Paper II is
quite different from the one used here (13\arcsec~westward now,
which is 0.5 kpc), leading to a very different shape for the \RC.
Despite the fact that the systemic velocities are almost the same,
in Paper II the \RC~reaches a plateau at $\sim$20\arcsec whereas,
with the new rotation center, the \RC~now shows a solid body shape
up to 60\arcsec.
%
%Michel, trouver le
%centre d'olivia pour verifier que l'on a pas ecrit d'anerie.
%Retracer aussi la CR à partir des donnees d'olivia.  Verifier le
%PA car il y a aussi 10 degres d'écart.
%
%OK c'est fait, mais je suis tres embete car, en reprenant les VR d'Olivia
%je n'ai pas de courbe "solid body" mais toujours une courbe qui retombe,
%meme en decalant bien le centre de rotation. A moins de ne considerer
%qu'un tout petit secteur... et encore...
%J'ai le meme probleme en reprenant les donnees de la deuxieme observation
%avec le nouveau centre propose par Benoit.
%CONCLUSION : Que Benoit verifie bien son champ de VR et sa courbe de rotation !
%
% Par Benoit: le pa ayant ŽtŽ fixŽ, il est possible qu'il soit assez diffŽrent de celui proposŽ par Olivia... Mon champ de vitesse est tres proche de celui d'Olivia.
\\
\noindent \textbf{UGC 2034}. Despite the fact that this
galaxy has been observed again for 1.5 hours (on September
12th 2002) and that these
data have been compared and added to the data presented in Paper
II, the \SNR~remains very weak.  Due to the lack of rotation and
of spatial coverage, our model does not converge. Thus the
parameters have been set to HI values from \citet{Swaters:phd}.
%Par Benoit: inclinaison a 19 degres. comme Swater these.
\\
\noindent \textbf{UGC 2045}. A difference of 6\degr~is computed
between the major axis \PA~given in this paper and in Paper IV.
This is due to the warp which biases the automatic determination
of the major axis \PA~with respect to the morphological one. It
leads to little change however in the \RC. The maximum rotation
velocity is thus directly taken from the \PVM~plot.
\\
\noindent \textbf{UGC 2053}. The \SNR~and the total \ha~flux of
this galaxy are very low.  The new maps are not much different from the
ones published in Paper II, so that they are not presented here.
\\
\noindent \textbf{UGC 2080}.  The determination of the inclination
of this almost face-on galaxy leads to lower value when
using the kinematics than when using the morphology. A too low
inclination leads to a maximum rotation velocity too high with
respect to its magnitude and its optical radius. The distance of this
nearby galaxy is not accurately determined, but even if the distance is
underestimated by a factor two, the kinematical inclination is
still too low. Thus the inclination has been set to the
morphological value.
\\
\noindent \textbf{UGC 2082}. Using the rule defined in Paper VI,
we have not plotted the \RC~of this edge-on galaxy.  The maximum
rotation velocity may be not reached.
\\
\noindent \textbf{UGC 2141}. The signature of a strong bar is
clearly visible in the \VF.  It is almost aligned with the major
axis and may explain the difference between the value
of major axis \PA~found in this study and that published in
Paper IV (7\degr). The major axis \PA~probably changes with
radius within the optical limits for this galaxy. Because of
the resulting uncertainty on the major axis \PA, the maximum rotation
velocity of that galaxy has been determined directly from its \VF.
\\
\noindent \textbf{UGC 2183}.  The inclination computed here
(41$\pm$10\degr) is similar to the morphological one (47\degr) but
quite different from that found in HI by
\citet{Swaters:phd} and adopted in Paper IV (62\degr). The value 90\deg~suggested by \citet{Noordermeer:2005} from optical measurements does not seem realistic.
\\
%
%\noindent \textbf{UGC 2193}. no comment
%\\
%
\noindent \textbf{UGC 2455}. The \VF~of this faint low surface
brightness galaxy shows a small amplitude making difficult the
determination of the \RC~which is, moreover, affected by a strong
bar.
\\
%
%\noindent \textbf{UGC 2503}. no comment
%\\
%
%\noindent \textbf{UGC 2800}. no comment
%\\
%
%\noindent \textbf{UGC 2855}. no comment
%\\
%
\noindent \textbf{UGC 3013}. The determination of the morphological \pa~is biased by a strong bar and spiral arms.
\\
%
%\noindent \textbf{UGC 3273}. no comment
%\\
%
\noindent \textbf{UGC 3382}. This galaxy has been published in
Paper VI from data coming only from run 13.  Nevertheless, this
galaxy was already observed in run 5 but never published because
the \SNR~was too low. In this paper, the data from both runs
have been added, leading to higher \SNR~data and smaller bins
allowing a refinement in the kinematical parameters.
\\
%
%\noindent \textbf{UGC 3384}.  no comment.
%
\noindent \textbf{UGC 3429}. The nucleus of the galaxy is
probably hidden behind a dust lane, so that its true position is
hard to find on the continuum images because of strong absorption.
Thus, we use the center making the central part of the rotation
curve most symmetric. This leads to a satisfactory position for the
rotation center on the continuum image when assuming that the dust
lane is symmetrical with respect to the major axis. Beyond
65\arcsec, the \RC~is unsure, due to obvious strong non circular
motions in the \VF~of this postmerger candidate
\citep{Marcum:2001}.
\\
%
%Michel, verifier le centre d'og.
%OK c'est fait. A premiere vue le centre adopte par Olivia est tres
%proche. Sa RC semble pourtant bien meilleure au point de vue de la
%symetrie dans la partie montante. De fait, si je decale le centre
%d'a peine 2 pixels en X et 1 pixel en Y je retrouve l'allure de la
%RC de Benoit.
%
%CONCLUSION : Benoit doit decaler legerement son centre afin de
%trouver une RC plus symetrique (pour info Olivia avait comme
%centre Xc=258 et Yc=242)
% Par Benoit: c'etait fait... quel dommage de ne pas avoir eu le temps de donner les nouvelles cartes+rc a michel...
%
\noindent \textbf{UGC 3574}.  With respect to Paper I, the
inclination has been reduced from 30 to 19$\pm$10\degr, this new
value is more compatible with the morphological one (21\degr) but
leads to a very high maximum rotation velocity with respect to its
faint absolute magnitude (M$_B$=-18.0, \citealp{James:2004}). For this
nearby object (V$_{sys}$=1433\kms), the distance determined using the
Hubble relationship (corrected from Virgo infall) is nevertheless
unsure, as can be also suspected from its maximal \ha~extension
reaching $\sim$2.5 its optical radius.
\\
%
%\noindent \textbf{UGC 3691}.  no comment.
%
%\noindent \textbf{UGC 3734}.  no comment.
%Pb d'inclinaison aussi
%mais dans l'autre sens qui colle mieux avec TF du coup mais bon
%mais l'inclinaison cinématique ne semble pas trop réaliste...
%vérifier les chi2 pour voir si on est pas dans un minimum de la
%vallée... NON, PAS DANS UNE VALLEE... Sinon la CR est compatible.
%
%\\
%\noindent \textbf{UGC 3809}.  no comment.
%\\
%des points bleus de la CR avec des V trop grandes ont ete vires car 1pt par bin+mauvais SNR
%
\noindent \textbf{UGC 3851}.  The ghost on the data (located on
the northern side of the image) has been removed.  No \RC~has been
computed because of its high inclination and to the fact that the
\VF~corresponding to the very bright region south of the galaxy
may be an artefact due to the detector.
\\
\noindent \textbf{UGC 4273}. The determination of the morphological \pa~is biased by the bar and spiral arms.
\\
%\noindent \textbf{UGC 4274}.  no comment.
%
\noindent \textbf{UGC 4278}. No \RC~has been computed because of
its high inclination (90\degr).
\\
\noindent \textbf{UGC 4284}. The \RC~is more symmetrical using a kinematical center 5\arcsec~south from the morphological center. However, to have a consistent analysis with the rest of the sample, we decided to keep the morphological center.
\\
%Par Benoit: le centre morphologique semble tout de meme decalle du centre des isophotes externes
%Michel, peux-tu vérifier pourquoi OG n'a pas le petit plateau a
%moins de 10 arcsec du centre sur le chp de vit d'OG ? OK c'est
%fait. A premiere vue la difference s'explique principalement par
%un centre legerement different (quelques pixels plus au nord pour
%Benoit). Pour info, Olivia a pris Xc=131 Yc=135 (sur l'image
%d'origine, non tournee). Par ailleurs l'observation était entachee
%d'une forte derive en alpha (visible sur le site web de GHASP) qui
%a pu lisser certains details. Cette derive est corrigee dans
%l'analyse de Benoit.
%
%CONCLUSION : Je suggere quand meme a Benoit de voir ce que donne
%sa RC en decalant le centre de quelques pixels vers le sud (2 ou 3
%a peine) et en remontant la Vs en consequence. Je ne serais pas
%etonne de voir une RC plus symetrique.
%
%A faire uniquement si le centre est mal défini.  Sinon indiquer
%que la courbe est meilleure avec un autre centre mais ne pas le
%faire pour rester cohérent avec le reste. Ph.
%
\noindent \textbf{UGC 4305}. The center has been changed from Paper III, and is now more to the East. It has been determined from a 2MASS image in the infrared. This center still gives a fairly symmetric \RC. It coincides approximately with the HI kinematical center but absolutely not with the optical center of isophotes. The counter rotation seen in the \rc~of Paper III is not seen anymore in the new \rc~due to the new center. However peculiar motions are still seen in the \VF~near the optical center of isophotes. Depending on the center chosen for kinematical analysis, they can be interpreted as non regular motions or as a counter rotation.
\\
% Par Benoit: On prends le centre dŽterminŽ sur les images 2MASS (ˆ l'oeil). Il coincide assez bien avec le centre cinŽmatique HI. Il est un peu plus ˆ l'est que le pic du continu. On fixe l'inclinaison a la valeur HI (40deg, Swaters), mais on laisse le PA libre car un warp est possible.
%
%Vérifier comment OG a tracer la CR,
%quels sont les parametres, le centre en particulier et surtout
%comment le justifier ! Michel. OK c'est fait. Olivia a adopte
%comme centre de rotation le noyau vu sur l'image continuum (Xc=125
%Yc=138 sur image non tournee). Le PA est assez bien defini. J'ai
%repris son champ de VR d'origine et il n'y a guere de doute, la RC
%donnee sur le site web de GHASP semble realiste et fiable.
%CONCLUSION : Essayer de voir ce que donne le champ de VR de Benoit
%avec les parametres d'Olivia (pour le noyau voir
%
%\noindent \textbf{UGC 4325}. no comment.
%
\noindent \textbf{UGC 4499}. The inclination has been set to the
value computed by \citet{Swaters:2003}.
\\
\noindent \textbf{UGC 4543}. The morphological \pa~of this galaxy is difficult to estimate  due to its ``Magellanic'' structure.
\\
\noindent \textbf{UGC 4936}. The inclination found now is 17\degr~lower
than the one published in Paper III.  This value is more
compatible with the absolute magnitude of the galaxy as well as
its optical radius.
\\
%
%\noindent \textbf{UGC 5253}. no comment.
%
\noindent \textbf{UGC 5272}. The \PVM~displays a solid body
structure characteristic of a bar like structure or an edge-on
galaxy. The optical and \Ha~images both look like a bar rather than a disk galaxy, suggesting a morphological
inclination close to 90\degr. Thus no \RC~has been plotted. The
HI inclination is 59\degr, computed from a different major axis
\PA~(20\degr), this is an additional argument in favour of the bar-like structure seen in the \ha~image.
\\
%
%\noindent \textbf{UGC 5316}. no comment.
%
%\noindent \textbf{UGC 5414}. no comment.
%
\noindent \textbf{UGC 5721}. The uncommon \RC~results from the
very irregular \VF~perturbed by a strong bar.
\\
%
%\noindent \textbf{UGC 5789}. no comment.
%
%\noindent \textbf{UGC 5829}. no comment.
%
\noindent \textbf{UGC 5931}. The morphological \pa~of this galaxy is difficult to estimate due to the interaction with its close companion UGC 5935.
\\
\noindent \textbf{UGC 5935}. No \RC~has been computed because of
its high inclination, close to 90\degr.
\\
%
%\noindent \textbf{UGC 5982}. no comment.
%
\noindent \textbf{UGC 6537}. The rotation center has been assumed to be the
strong nucleus, leading to an asymmetric \RC~in the outer regions,
compatible with the asymmetric \VF. In Paper III, in order to get a
symmetric \RC, the rotation center had been shifted by 7\arcsec~($\sim$0.5
kpc) with respect to the nucleus, and the major axis \PA~rotated by 6\degr.
\\
%
%\noindent \textbf{UGC 6628}. no comment.
%(mais alors la
%difference de CR est completement incompréhensible, Michel jette
%éventuellement un oeil sur la CR d'OG qui devrait au moins avoir
%un point qui monte a 87 km/s, vu le chp de vit et l'inclinaison).
%Inclinaison Swaters phd
%
%OK c'est fait. Si je prends un PA a 0° (comme Benoit)
%au lieu de 18° (valeur adoptee par Olivia)
%avec le champ de VR d'Olivia je trouve des points qui montent
%au dessus de 100 km/s au bout de la RC, mais ça reste tres marginal et,
%dans l'ensemble, la RC reste en plateau vers 60 km/s.
%
%CONCLUSION : Meme en jouant sur les parametres (i, PA, Xc et Yc)
%j'ai toujours une RC qui reste en plateau avec le champ de VR d'Olivia...
%a l'exception, comme souligne ci-dessus, d'un ou deux points a l'extremite.
%Ceci dit, le champ de VR de Benoit me semble de meilleure qualite
%(notamment en terme de rapport S/B) et je pense donc qu'il faut le preferer
%a celui d'origine presente sur le site web de GHASP
%(sur lequel les barres d'erreur de la RC sont enormes !).
%De plus, la RC de Benoit me semble assez symetrique,
%ce qui conforte le nouveau resultat.
%
%\noindent \textbf{UGC 6702}. no comment.
%
%\noindent \textbf{UGC 6778}. no comment.
%
\noindent \textbf{UGC 7278}. This galaxy does not show any evidence for rotation. However, a velocity amplitude of about 10\kms~is observed on its velocity field.
\\
%
%\noindent \textbf{UGC 7323}. no comment.
%
\noindent \textbf{UGC 7524}. The \VF~only covers the bar of the
galaxy which is almost aligned with the HI major axis \PA, thus no
\RC~has been plotted. Furthermore, only the central part of this galaxy is seen in our \FOV.
\\
\noindent \textbf{UGC 7592}.  This galaxy does not show any evidence for rotation. However, a velocity amplitude of about 25\kms~is observed on its velocity field.
\\
%
%\noindent \textbf{UGC 7971}. no comment.
%
%\noindent \textbf{UGC 8490}. no comment.
%
%\noindent \textbf{UGC 9366}. no comment.
%
%\noindent \textbf{UGC 9649}. no comment.
%Par Benoit: inclinaison fixee a celle du site WHISP=incl NED (photom)=54deg
%\\
%
%\noindent \textbf{UGC 9753}. no comment.
%
%\noindent \textbf{UGC 9858}. no comment.
%
%\noindent \textbf{UGC 9969}. no comment.
%
%\noindent \textbf{UGC 9992}. no comment.
%
\noindent \textbf{UGC 10310}.  The morphological \pa~is difficult to estimate due to the nature of this barred Magellanic galaxy.
\\
\noindent \textbf{UGC 10359}.  The determination of the morphological \pa~is biased by a strong bar and spiral arms.
\\
%Inclination is HI value of 44deg from WHISP web site = incl from NED. Centre has been adjusted but is very close to the center of the bar (see maps)
%
\noindent \textbf{UGC 10445}. The presence of a bar and spiral arms make both morphological and kinematical \pas~determinations difficult.
\\
\noindent \textbf{UGC 10470}. A strong bar in this galaxy
biases the determination of the major axis \PA~and of the
inclination. For the latter we adopted the value suggested by
the morphological axis ratio (34\degr~from the NED).
\\
%Inclination is HI value of 34deg from WHISP web site = incl from NED. 
%HLEDA:45¡ LIBRE:15¡->340km/s pour
%Mag=-20.2
%
\noindent \textbf{UGC 10502}. The kinematical inclination is much
higher than the morphological one (24\degr~from the axis ratio).
Since this galaxy is interacting, its morphological inclination
may be underestimated because of its open spiral arms distorted by
streaming motions. The average morphological inclination given in
Hyperleda is instead 40\degr, still lower than the kinematical
one.
\\
%
%\noindent \textbf{UGC 10546}. no comment.
%
%\noindent \textbf{UGC 10564}. no comment.
%Prendre le vmax sur le
%champ de vitesse a cause de l'inclinaison, environ 150 km/s%
%\\
\noindent \textbf{UGC 10897}. The morphological \pa~is poorly determined for this low inclination galaxy.
\\
%Inclination is HI value of 31deg from WHISP web site = incl from NED. 
%
%Michel, il y a une
%grande difference d'amplitude sur le champ de vitesses entre og et
%be, ce qui conduit a des CRs différentes en amplitude.
%
%OK c'est fait.
%Effectivement Olivia a adopte une inclinaison de 30° seulement
%contre 55 pour Benoit. Cela explique la difference d'amplitude.
%L'inclinaison morphologique est de 31° ce qui explique sans doute
%le choix precedemment fait.
%
%CONCLUSION : Si la valeur trouvee par Benoit donne un meilleur
%resultat pour Tully-Fisher c'est parfait et on la garde.
%Dans le cas contraire on peut toujours justifier du choix
%de l'adoption de l'inclinaison morphologique...
%
% Par Beno"t en rŽponse ˆ MICHEL: j'ai fixŽ l'inclinaison ˆ 31deg et j'ai des vitesses plus grandes qu'Olivia. As-tu retracŽ la CR pour vŽrifier?
%
\noindent \textbf{UGC 11124}. The inclination is difficult to
determine because of a strong bar in this interacting galaxy.
The kinematical inclination (51\degr) is higher than
the morphological one determined from the axis ratio (22\degr).
Despite the fact that the external isocontours seem almost round,
the luminosity of this galaxy is nevertheless more consistent with
a high inclination than with a lower one with respect to its rotation velocity.
\\
%
%\noindent \textbf{UGC 11218}. no comment.
%
\noindent \textbf{UGC 11283}. The presence of a strong bar and spiral arms make both morphological and kinematical \pa~determinations difficult.
\\
%\noindent \textbf{UGC 11283c}. no comment.
%
%\noindent \textbf{UGC 11300}. no comment.
%
\noindent \textbf{UGC 11429}. The very peculiar \VF~of this
galaxy, which is in pair, displays a concave curvature towards the South on
both sides of the galaxy. Thus, the residual \VF~shows a very high
dispersion and a signature typical of a \RC~having a rotation
center abnormally displaced toward the South on the major axis
\citep{van-der-Kruit:1978}. This is obviously not the
case from the morphology.
%van der kruit and allen 78
\\
%inclination only is fixed
%
%\noindent \textbf{UGC 11557}. no comment.
%
%\noindent \textbf{UGC 11707}. no comment.
%
%\noindent \textbf{UGC 11852}. no comment.
%
\noindent \textbf{UGC 11861}. The morphological \pa~of this galaxy is difficult to estimate  due to its ``Magellanic'' structure.
\\
%\noindent \textbf{UGC 11891}. no comment.
%
\noindent \textbf{UGC 11909}. No \RC~has been computed because of
its high inclination (90\degr).
\\
%
%\noindent \textbf{UGC 11914}. no comment.
%
\noindent \textbf{UGC 11951}. The \RC~published in Paper II is
incorrect due to a typo in the value of the \PA~of the major axis.
\\
%no comment.  Michel, virer cette
%galaxie du site web d'olivia car la CR est incompatible avec le
%Chp de vitesse, verifie par toi meme (pb d'amplitude de la cr et
%du chp vit).
%
%OK c'est fait.
%Effectivement il y a une grosse couille avec le choix du PA !!!
%Il est dans les choux d'une quarantaine de degres...
%Pire, on l'a publiee comme ca...
%J'ai revu les parametres et, avec le champ de VR d'origine,
%je retrouve une allure de RC comparable a ce qu'obtient Benoit,
%cad que l'on semble atteindre un plateau vers 100 km au bout
%d'une minute d'arc.
%
%CONCLUSION : On oublie ce qui avait ete publie
%(il faudra peut-etre en dire un mot quand meme...)
%En attendant, et a titre provisoire, je vais mettre une RC correcte sur le site web de GHASP.
%Le champ de VR qu'on y a mis est correct.
%
\noindent \textbf{UGC 12060}. This galaxy is irregular, barred and has a low surface brightness. These features make the morphological \pa~determination difficult.
\\
\noindent \textbf{UGC 12276}. The previous data reduction (Paper
IV) missed a large part of the \VF.
\\
% Par Benoit: j'ai vŽrifiŽ la carte mono et les profils:pas de doutes, on voit bien la distribution plus intense de gaz le long des bras spiraux, et plus diffus sur tout le disque. Rien ˆ voir avec la carte mono d'Olivia. A noter que mon temps de pose total est de 4680s contre 7200 pour Olivia. Je soupconne donc qu'elle ait utilisŽ des cycles qui contenaient bcp de bruit.
%
%Michel, le chp de vit
%d'olivia est horrible par rapport a celui de benoit,
%incomprehensible.  tu peux jeter un oeil stp.
%
%OK c'est fait.
%J'ai repris les donnees d'observation a partir de zero.
%Je confirme que le champ de vitesses ne s'etend pas beaucoup plus loin
%que ce qu'on donne sur le site web de GHASP.
%On peut baisser un peu le seuil et gratter quelques points de VR mais ca n'est pas significatif...
%
%CONCLUSION : Je ne sais par quel prodige Benoit a pu obtenir un
%champ de vitesses aussi fourni ! J'aimerais qu'il verifie qu'il
%retrouve bien le champ de VR d'Olivia en montant son seuil pour ne
%garder que les points de poids eleve. Je trouve qu'il y a beaucoup
%de points tout a fait en dehors des contours de l'image du DSS...
%c'est louche... Mais, ce qui m'etonne le plus, c'est que Benoit
%obtienne une RC avec un plateau plus bas (80 km/s au lieu de 110
%km/s pour Olivia) alors qu'il a adopte une inclinaison plus basse
%(33° au lieu de 38° pour Olivia, qui est la valeur morphologique).
%Bref, c'est tout l'inverse de ce qu'on devrait avoir... Je note
%aussi que, pour le PA et le centre, il n'y a pas de pb (noyau bien
%visible et PA identique a deux degres pres).
%
%
\noindent \textbf{UGC 12276c}. The total \ha~diameter of the
galaxy ($\sim$8\arcsec) is not much larger than the seeing spot of
the observations ($\sim$5\arcsec). Thus no \RC~has been computed.
\\
%
%Remarque ajoutee par Michel :
%DOMMAGE !
%En effet, la RC que l'on a pu tracer (voir site web de GHASP) ne semble pas debile... loin de la !
%
%
%\noindent \textbf{UGC 12343}. no comment.
%
\noindent \textbf{UGC 12632}. The very high
velocity bump on the blue side of the \ha~\RC~is also visible in
the HI data \citep{Swaters:phd}.
\\
%Par benoit: Inclination is HI value of 44deg from WHISP Swaters. PA Sw=36deg.
%
%Michel, peux-tu vérifier le pa
%de og qui est de 162 sur la page web et qui semble bizarre.
%Prendre inclinaison HI
%car ca crache en HI.  34 degreweb=NED), 46 Swaters phd.
%
%OK c'est fait.
%La valeur de 162° pour le PA, indiquee sur la page web, est erronnee...
%La veritable valeur adoptee par Olivia est 30° (contre 40° pour Benoit).
%Elle est donnee correcte (210°, soit 30 + 180) dans le papier GHASP IV.
%Quant a l'inclinaison, on avait adopte 45°, quasiment comme Swaters.
%
%CONCLUSION : Je vais corriger la valeur du PA donnee sur le site
%web. En ce qui concerne l'inclinaison, celle trouvee par Benoit
%semble effectivement trop forte.
%Benoit, a partir du commentaire de michel on va vérifier TF.
% A
%part ça, je trouve que Benoit a binne un beu fort du cote
%"blueshifted". Quelle en est la raison ? (rapport S/B ?) Si c'est
%par manque de points il faudrait elargir le secteur, ou alors
%reduire le PA car la valeur 30° me semble plus realiste que 40°.
%
%\noindent \textbf{UGC 12754}. no comment.
% Par Benoit: on a decallŽ un poil le centre pour symetriser la RC, le nouveau centre est toujours bien compatible avec le centre morphologique.
%\\

\clearpage
\section{Tables}
\label{tables}

\begin{table*}
\caption{Calibration parameters.}
\begin{tabular}{cccccccc}
\noalign{\medskip} \hline
N\Deg & N\Deg &  $\alpha$    & $\delta$   &  Exposure time & Scale & Flux & Paper\\
 UGC        &  NGC       &  (2000) & (2000) & $s$ & " & $10^{-16}~W~m^{-2}$ &\\
 (1)&(2)&(3)&(4)&(5)&(6)&(7)&(8)\\
\hline
12893 &  & 00$\rm^{h}$00$\rm^{m}$28.0$\rm^{s}$ & 17$\degr$13'09" &  8640 & 0.68 & 0.3$\pm$0.1 & VI \\
89 & 0023 & 00$\rm^{h}$09$\rm^{m}$53.4$\rm^{s}$ & 25$\degr$55'24" &  8640 & 0.68 & 3.3$\pm$0.4 & VI \\
94 & 0026 & 00$\rm^{h}$10$\rm^{m}$25.9$\rm^{s}$ & 25$\degr$49'54" &  6240 & 0.68 & 2.2$\pm$0.3 & VI \\
508 & 0266 & 00$\rm^{h}$49$\rm^{m}$47.8$\rm^{s}$ & 32$\degr$16'40" &  9720 & 0.68 & 4.8$\pm$0.6 & IV  \\
528 & 0278 & 00$\rm^{h}$52$\rm^{m}$04.6$\rm^{s}$ & 47$\degr$33'01" &  3600 & 0.96 & 75.4$\pm$13.1 & II  \\
763 & 0428 & 01$\rm^{h}$12$\rm^{m}$55.7$\rm^{s}$ & 00$\degr$58'53" &  4680 & 0.68 & 27.8$\pm$3.3 & IV  \\
1013 & 0536 & \textit{01$\rm^{h}$26$\rm^{m}$21.9$\rm^{s}$} & \textit{34$\degr$42'11"} &  3840 & 0.68 &  & VI \\
 & 0542 & \textit{01$\rm^{h}$26$\rm^{m}$30.9$\rm^{s}$} & \textit{34$\degr$40'31"} &  3840 & 0.68 & 0.5$\pm$0.1 & VI \\
1117 & 0598 & 01$\rm^{h}$33$\rm^{m}$51.0$\rm^{s}$ & 30$\degr$39'35" &  5760 & 0.68 & $\geq$97.6$\pm$11.6 & IV  \\
1249 &  & \textit{01$\rm^{h}$47$\rm^{m}$29.9$\rm^{s}$} & \textit{27$\degr$19'60"} &  6000 & 0.96 & $\geq$17.8$\pm$3.1 & II  \\
1256 & 0672 & 01$\rm^{h}$47$\rm^{m}$54.7$\rm^{s}$ & 27$\degr$25'57" & 11040 & 0.96 & $\geq$25.9$\pm$4.5 & II  \\
1317 & 0697 & 01$\rm^{h}$51$\rm^{m}$17.6$\rm^{s}$ & 22$\degr$21'28" &  4800 & 0.68 & 8.3$\pm$1.0 & VI \\
1437 & 0753 & 01$\rm^{h}$57$\rm^{m}$42.2$\rm^{s}$ & 35$\degr$54'58" &  6720 & 0.68 & 7.9$\pm$0.9 & VI \\
1655 & 0828 & 02$\rm^{h}$10$\rm^{m}$09.7$\rm^{s}$ & 39$\degr$11'25" & 12720 & 0.68 & 0.3$\pm$0.1 & VI \\
1736 & 0864 & 02$\rm^{h}$15$\rm^{m}$27.7$\rm^{s}$ & 06$\degr$00'08" &  7560 & 0.68 & 6.2$\pm$0.7 & IV  \\
1810 &  & \textit{02$\rm^{h}$21$\rm^{m}$28.7$\rm^{s}$} & \textit{39$\degr$22'32"} &  7200 & 0.68 & 2.1$\pm$0.2 & VI \\
1886 &  & 02$\rm^{h}$26$\rm^{m}$00.6$\rm^{s}$ & 39$\degr$28'15" &  6120 & 0.68 & 3.2$\pm$0.4 & IV  \\
1913 & 0925 & 02$\rm^{h}$27$\rm^{m}$17.3$\rm^{s}$ & 33$\degr$34'44" &  3960 & 0.68 & $\geq$44.9$\pm$5.3 & IV  \\
2023 &  & 02$\rm^{h}$33$\rm^{m}$18.5$\rm^{s}$ & 33$\degr$29'30" & 14160 & 0.96 & 3.9$\pm$0.7 & II  \\
2034 &  & 02$\rm^{h}$33$\rm^{m}$42.9$\rm^{s}$ & 40$\degr$31'41" & 15840 & 0.96 & 1.5$\pm$0.3 & I  \\
2045 & 0972 & 02$\rm^{h}$34$\rm^{m}$13.4$\rm^{s}$ & 29$\degr$18'40" &  6120 & 0.68 & 22.4$\pm$2.7 & IV  \\
2053 &  & \textit{02$\rm^{h}$34$\rm^{m}$29.3$\rm^{s}$} & \textit{29$\degr$44'60"} &  9120 & 0.96 & 0.3$\pm$0.1 & II  \\
2082 &  & \textit{02$\rm^{h}$36$\rm^{m}$16.2$\rm^{s}$} & \textit{25$\degr$25'25"} &  7200 & 0.96 & 4.0$\pm$0.7 & II  \\
2080 &  & 02$\rm^{h}$36$\rm^{m}$27.9$\rm^{s}$ & 38$\degr$58'09" & 21840 & 0.96 & 12.9$\pm$2.2 & I  \\
2141 & 1012 & 02$\rm^{h}$39$\rm^{m}$14.9$\rm^{s}$ & 30$\degr$09'04" &  5040 & 0.68 & 23.0$\pm$2.7 & IV  \\
2183 & 1056 & 02$\rm^{h}$42$\rm^{m}$48.3$\rm^{s}$ & 28$\degr$34'27" &  5760 & 0.68 & 3.4$\pm$0.4 & IV  \\
2193 & 1058 & 02$\rm^{h}$43$\rm^{m}$30.1$\rm^{s}$ & 37$\degr$20'28" &  7920 & 0.68 & 6.5$\pm$0.8 & IV  \\
2455 & 1156 & 02$\rm^{h}$59$\rm^{m}$42.3$\rm^{s}$ & 25$\degr$14'13" &  3840 & 0.96 & 32.2$\pm$5.6 & I  \\
2503 & 1169 & 03$\rm^{h}$03$\rm^{m}$34.8$\rm^{s}$ & 46$\degr$23'10" &  7200 & 0.68 & 4.5$\pm$0.5 & IV  \\
2800 &  & 03$\rm^{h}$40$\rm^{m}$03.6$\rm^{s}$ & 71$\degr$24'19" &  8880 & 0.96 & 2.0$\pm$0.4 & I  \\
2855 &  & 03$\rm^{h}$48$\rm^{m}$20.7$\rm^{s}$ & 70$\degr$07'57" &  7200 & 0.96 & 25.0$\pm$4.3 & II  \\
3013 & 1530 & 04$\rm^{h}$23$\rm^{m}$26.7$\rm^{s}$ & 75$\degr$17'44" &  3360 & 0.68 & 4.8$\pm$0.6 & IV  \\
3056 & 1569 & \textit{04$\rm^{h}$30$\rm^{m}$49.2$\rm^{s}$} & \textit{64$\degr$50'52"} &  6000 & 0.68 & 270.5$\pm$32.0 & VI \\
3273 &  & 05$\rm^{h}$17$\rm^{m}$45.1$\rm^{s}$ & 53$\degr$33'01" &  4800 & 0.68 & 3.6$\pm$0.4 & IV  \\
3334 & 1961 & 05$\rm^{h}$42$\rm^{m}$04.6$\rm^{s}$ & 69$\degr$22'43" &  5040 & 0.68 & 20.2$\pm$2.4 & VI \\
3382 &  & 05$\rm^{h}$59$\rm^{m}$47.7$\rm^{s}$ & 62$\degr$09'28" & 19200 & 0.68 & 0.4$\pm$0.1 & VI \\
3384 &  & \textit{06$\rm^{h}$01$\rm^{m}$37.2$\rm^{s}$} & \textit{73$\degr$07'01"} &  4320 & 0.68 & 1.4$\pm$0.2 & IV  \\
3429 & 2146 & 06$\rm^{h}$18$\rm^{m}$38.4$\rm^{s}$ & 78$\degr$21'26" &  6480 & 0.68 & 30.7$\pm$3.6 & IV  \\
3463 &  & 06$\rm^{h}$26$\rm^{m}$55.8$\rm^{s}$ & 59$\degr$04'47" &  6960 & 0.68 & 11.1$\pm$1.3 & VI \\
3574 &  & 06$\rm^{h}$53$\rm^{m}$10.4$\rm^{s}$ & 57$\degr$10'40" & 10080 & 0.96 & $\geq$8.7$\pm$1.5 & I  \\
3521 &  & 06$\rm^{h}$55$\rm^{m}$00.1$\rm^{s}$ & 84$\degr$02'30" &  8400 & 0.68 & 0.7$\pm$0.1 & VI \\
3528 &  & 06$\rm^{h}$56$\rm^{m}$10.6$\rm^{s}$ & 84$\degr$04'44" &  8400 & 0.68 & 0.3$\pm$0.1 & VI \\
3618 & 2308 & \textit{06$\rm^{h}$58$\rm^{m}$37.6$\rm^{s}$} & \textit{45$\degr$12'38"} &  6000 & 0.68 &  & VI \\
3691 &  & 07$\rm^{h}$08$\rm^{m}$01.4$\rm^{s}$ & 15$\degr$10'39" &  3960 & 0.68 & 12.2$\pm$1.4 & IV  \\
3685 &  & 07$\rm^{h}$09$\rm^{m}$05.9$\rm^{s}$ & 61$\degr$35'44" &  5520 & 0.68 & 8.0$\pm$0.9 & VI \\
3708 & 2341 & 07$\rm^{h}$09$\rm^{m}$12.0$\rm^{s}$ & 20$\degr$36'11" &  6480 & 0.68 & 1.2$\pm$0.1 & VI \\
3709 & 2342 & 07$\rm^{h}$09$\rm^{m}$18.1$\rm^{s}$ & 20$\degr$38'10" &  6480 & 0.68 & 4.2$\pm$0.5 & VI \\
3734 & 2344 & 07$\rm^{h}$12$\rm^{m}$28.7$\rm^{s}$ & 47$\degr$10'00" & 11040 & 0.68 & $\geq$3.7$\pm$0.4 & IV  \\
3826 &  & 07$\rm^{h}$24$\rm^{m}$28.0$\rm^{s}$ & 61$\degr$41'38" &  4560 & 0.68 & 4.0$\pm$0.5 & VI \\
3809 & 2336 & 07$\rm^{h}$27$\rm^{m}$03.9$\rm^{s}$ & 80$\degr$10'42" &  9120 & 0.96 & $\geq$25.4$\pm$4.4 & III  \\
3740 & 2276 & 07$\rm^{h}$27$\rm^{m}$13.1$\rm^{s}$ & 85$\degr$45'16" &  4800 & 0.68 & 63.8$\pm$7.6 & VI \\
3851 & 2366 & \textit{07$\rm^{h}$28$\rm^{m}$51.9$\rm^{s}$} & \textit{69$\degr$12'31"} &  7200 & 0.96 & $\geq$128.5$\pm$22.3 & III  \\
3876 &  & 07$\rm^{h}$29$\rm^{m}$17.5$\rm^{s}$ & 27$\degr$54'00" & 15600 & 0.68 & 1.5$\pm$0.2 & VI \\
3915 &  & 07$\rm^{h}$34$\rm^{m}$55.8$\rm^{s}$ & 31$\degr$16'34" &  6000 & 0.68 & 9.3$\pm$1.1 & VI \\
IC476 &  & 07$\rm^{h}$47$\rm^{m}$16.5$\rm^{s}$ & 26$\degr$57'03" &  6240 & 0.68 & 0.2$\pm$0.1 & VI \\
4026 & 2449 & 07$\rm^{h}$47$\rm^{m}$20.4$\rm^{s}$ & 26$\degr$55'48" &  6240 & 0.68 & 0.5$\pm$0.1 & VI \\
4165 & 2500 & 08$\rm^{h}$01$\rm^{m}$53.2$\rm^{s}$ & 50$\degr$44'15" &  9600 & 0.68 & 13.6$\pm$1.6 & VI \\
4256 & 2532 & 08$\rm^{h}$10$\rm^{m}$15.2$\rm^{s}$ & 33$\degr$57'24" &  6960 & 0.68 & 2.9$\pm$0.3 & VI \\
4273 & 2543 & 08$\rm^{h}$12$\rm^{m}$57.9$\rm^{s}$ & 36$\degr$15'16" &  6240 & 0.68 & 9.0$\pm$1.1 & IV  \\
4274 & 2537 & 08$\rm^{h}$13$\rm^{m}$14.9$\rm^{s}$ & 45$\degr$59'31" &  2640 & 0.96 & 15.3$\pm$2.7 & III  \\
\hline
\label{table_calib}
\end{tabular}
\end{table*}
\begin{table*}
\contcaption{}
\begin{tabular}{cccccccc}
\noalign{\medskip} \hline
N\Deg & N\Deg &  $\alpha$    & $\delta$   &  Exposure time & Scale & Flux & Paper\\
 UGC        &  NGC       &  (2000) & (2000) & $s$ & " & $10^{-16}~W~m^{-2}$ &\\
 (1)&(2)&(3)&(4)&(5)&(6)&(7)&(8)\\
\hline
4278 &  & \textit{08$\rm^{h}$13$\rm^{m}$58.9$\rm^{s}$} & \textit{45$\degr$44'37"} &  7200 & 0.96 & $\geq$14.9$\pm$2.6 & III  \\
4284 & 2541 & 08$\rm^{h}$14$\rm^{m}$40.2$\rm^{s}$ & 49$\degr$03'42" &  7440 & 0.96 & $\geq$34.0$\pm$5.9 & III  \\
4305 &  & 08$\rm^{h}$19$\rm^{m}$18.4$\rm^{s}$ & 70$\degr$43'03" &  6240 & 0.96 & $\geq$94.2$\pm$16.4 & III  \\
4325 & 2552 & 08$\rm^{h}$19$\rm^{m}$20.0$\rm^{s}$ & 50$\degr$00'31" &  6240 & 0.96 & 3.8$\pm$0.7 & I  \\
4393 &  & 08$\rm^{h}$26$\rm^{m}$04.4$\rm^{s}$ & 45$\degr$58'02" &  6960 & 0.68 & 5.4$\pm$0.6 & VI \\
4422 & 2595 & 08$\rm^{h}$27$\rm^{m}$42.0$\rm^{s}$ & 21$\degr$28'45" &  7200 & 0.68 & 3.2$\pm$0.4 & VI \\
4456 &  & 08$\rm^{h}$32$\rm^{m}$03.5$\rm^{s}$ & 24$\degr$00'39" &  6720 & 0.68 & 2.0$\pm$0.2 & VI \\
4499 &  & 08$\rm^{h}$37$\rm^{m}$41.5$\rm^{s}$ & 51$\degr$39'10" & 10560 & 0.96 & 3.5$\pm$0.6 & I  \\
4543 &  & 08$\rm^{h}$43$\rm^{m}$21.4$\rm^{s}$ & 45$\degr$44'10" & 13680 & 0.96 & 2.8$\pm$0.5 & III  \\
4555 & 2649 & 08$\rm^{h}$44$\rm^{m}$08.4$\rm^{s}$ & 34$\degr$43'02" &  6960 & 0.68 & 1.5$\pm$0.2 & VI \\
4770 & 2746 & 09$\rm^{h}$05$\rm^{m}$59.4$\rm^{s}$ & 35$\degr$22'38" &  7920 & 0.68 & 0.6$\pm$0.1 & VI \\
4820 & 2775 & 09$\rm^{h}$10$\rm^{m}$20.1$\rm^{s}$ & 07$\degr$02'17" &  8400 & 0.68 & 9.5$\pm$1.1 & VI \\
4936 & 2805 & 09$\rm^{h}$20$\rm^{m}$20.4$\rm^{s}$ & 64$\degr$06'10" &  6960 & 0.96 & 16.6$\pm$2.9 & III  \\
5045 &  & 09$\rm^{h}$28$\rm^{m}$10.2$\rm^{s}$ & 44$\degr$39'52" &  6960 & 0.68 & 3.5$\pm$0.4 & VI \\
5175 & 2977 & 09$\rm^{h}$43$\rm^{m}$46.8$\rm^{s}$ & 74$\degr$51'35" &  5520 & 0.68 & 5.2$\pm$0.6 & VI \\
5228 &  & 09$\rm^{h}$46$\rm^{m}$03.8$\rm^{s}$ & 01$\degr$40'06" &  9120 & 0.68 & 7.3$\pm$0.9 & VI \\
5251 & 3003 & 09$\rm^{h}$48$\rm^{m}$36.4$\rm^{s}$ & 33$\degr$25'17" &  1680 & 0.68 & 23.1$\pm$2.7 & VI \\
5253 & 2985 & 09$\rm^{h}$50$\rm^{m}$21.9$\rm^{s}$ & 72$\degr$16'47" &  6960 & 0.96 & 9.5$\pm$1.7 & I  \\
5272 &  & \textit{09$\rm^{h}$50$\rm^{m}$22.4$\rm^{s}$} & \textit{31$\degr$29'16"} &  8160 & 0.96 & 6.9$\pm$1.2 & III  \\
5279 & 3026 & \textit{09$\rm^{h}$50$\rm^{m}$55.3$\rm^{s}$} & \textit{28$\degr$33'04"} &  6000 & 0.68 & 5.0$\pm$0.6 & VI \\
5316 & 3027 & 09$\rm^{h}$55$\rm^{m}$40.4$\rm^{s}$ & 72$\degr$12'13" &  6240 & 0.96 & 8.7$\pm$1.5 & I  \\
5319 & 3061 & 09$\rm^{h}$56$\rm^{m}$12.0$\rm^{s}$ & 75$\degr$51'59" &  7200 & 0.68 & 3.2$\pm$0.4 & VI \\
5351 & 3067 & \textit{09$\rm^{h}$58$\rm^{m}$21.2$\rm^{s}$} & \textit{32$\degr$22'12"} &  7200 & 0.68 & 8.2$\pm$1.0 & VI \\
5373 &  & 10$\rm^{h}$00$\rm^{m}$00.5$\rm^{s}$ & 05$\degr$19'58" &  7680 & 0.68 & 6.0$\pm$0.7 & VI \\
5398 & 3077 & \textit{10$\rm^{h}$03$\rm^{m}$20.0$\rm^{s}$} & \textit{68$\degr$44'01"} &  7920 & 0.68 & 20.9$\pm$2.5 & VI \\
5414 & 3104 & 10$\rm^{h}$03$\rm^{m}$57.1$\rm^{s}$ & 40$\degr$45'21" &  7200 & 0.96 & 11.8$\pm$2.1 & III  \\
IC2542 &  & 10$\rm^{h}$07$\rm^{m}$50.5$\rm^{s}$ & 34$\degr$18'55" &  7440 & 0.68 & 1.6$\pm$0.2 & VI \\
5510 & 3162 & 10$\rm^{h}$13$\rm^{m}$31.7$\rm^{s}$ & 22$\degr$44'14" &  6240 & 0.68 & 18.9$\pm$2.2 & VI \\
5532 & 3147 & 10$\rm^{h}$16$\rm^{m}$53.5$\rm^{s}$ & 73$\degr$24'03" &  6480 & 0.68 & 27.8$\pm$3.3 & VI \\
5556 & 3187 & \textit{10$\rm^{h}$17$\rm^{m}$47.9$\rm^{s}$} & \textit{21$\degr$52'24"} &  5760 & 0.68 & 3.2$\pm$0.4 & VI \\
5721 & 3274 & 10$\rm^{h}$32$\rm^{m}$16.9$\rm^{s}$ & 27$\degr$40'08" &  8160 & 0.96 & 5.3$\pm$0.9 & I  \\
5786 & 3310 & 10$\rm^{h}$38$\rm^{m}$45.9$\rm^{s}$ & 53$\degr$30'12" &  4080 & 0.68 & 183.5$\pm$21.7 & VI \\
5789 & 3319 & 10$\rm^{h}$39$\rm^{m}$09.4$\rm^{s}$ & 41$\degr$41'11" &  6300 & 0.96 & $\geq$16.0$\pm$2.8 & I  \\
5829 &  & 10$\rm^{h}$42$\rm^{m}$43.7$\rm^{s}$ & 34$\degr$26'54" & 11760 & 0.96 & $\geq$3.3$\pm$0.6 & I  \\
5840 & 3344 & 10$\rm^{h}$43$\rm^{m}$31.1$\rm^{s}$ & 24$\degr$55'21" &  5760 & 0.68 & $\geq$60.4$\pm$7.2 & VI \\
5842 & 3346 & 10$\rm^{h}$43$\rm^{m}$39.0$\rm^{s}$ & 14$\degr$52'18" &  6000 & 0.68 & 9.9$\pm$1.2 & VI \\
5931 & 3395 & 10$\rm^{h}$49$\rm^{m}$50.2$\rm^{s}$ & 32$\degr$58'59" & 10320 & 0.96 & 10.5$\pm$1.8 & I  \\
5935 & 3396 & \textit{10$\rm^{h}$49$\rm^{m}$55.2$\rm^{s}$} & \textit{32$\degr$59'27"} & 10320 & 0.96 & 10.2$\pm$1.8 & I  \\
5982 & 3430 & 10$\rm^{h}$52$\rm^{m}$11.5$\rm^{s}$ & 32$\degr$56'59" & 10080 & 0.96 & 8.3$\pm$1.4 & I  \\
6118 & 3504 & 11$\rm^{h}$03$\rm^{m}$11.3$\rm^{s}$ & 27$\degr$58'20" &  5760 & 0.68 & 27.4$\pm$3.2 & VI \\
6277 & 3596 & 11$\rm^{h}$15$\rm^{m}$06.2$\rm^{s}$ & 14$\degr$47'12" & 10320 & 0.68 & 9.4$\pm$1.1 & VI \\
6419 & 3664 & 11$\rm^{h}$24$\rm^{m}$24.6$\rm^{s}$ & 03$\degr$19'36" &  6960 & 0.68 & 12.2$\pm$1.4 & VI \\
6521 & 3719 & 11$\rm^{h}$32$\rm^{m}$13.4$\rm^{s}$ & 00$\degr$49'09" &  6960 & 0.68 & 3.6$\pm$0.4 & VI \\
6523 & 3720 & 11$\rm^{h}$32$\rm^{m}$21.6$\rm^{s}$ & 00$\degr$48'14" &  6960 & 0.68 & 1.4$\pm$0.2 & VI \\
6537 & 3726 & 11$\rm^{h}$33$\rm^{m}$21.2$\rm^{s}$ & 47$\degr$01'45" &  6240 & 0.96 & $\geq$73.8$\pm$12.8 & III  \\
6628 &  & 11$\rm^{h}$40$\rm^{m}$05.7$\rm^{s}$ & 45$\degr$56'33" &  5760 & 0.96 & 7.9$\pm$1.4 & III  \\
6702 & 3840 & 11$\rm^{h}$43$\rm^{m}$59.0$\rm^{s}$ & 20$\degr$04'37" &  5280 & 0.68 & 2.9$\pm$0.3 & IV  \\
6778 & 3893 & 11$\rm^{h}$48$\rm^{m}$38.4$\rm^{s}$ & 48$\degr$42'38" & 10080 & 0.96 & 18.4$\pm$3.2 & I  \\
6787 & 3898 & 11$\rm^{h}$49$\rm^{m}$15.6$\rm^{s}$ & 56$\degr$05'04" &  9360 & 0.68 & 2.0$\pm$0.2 & VI \\
7021 & 4045 & 12$\rm^{h}$02$\rm^{m}$42.3$\rm^{s}$ & 01$\degr$58'36" &  7200 & 0.68 & 4.6$\pm$0.5 & VI \\
7045 & 4062 & 12$\rm^{h}$04$\rm^{m}$03.8$\rm^{s}$ & 31$\degr$53'42" &  6000 & 0.68 & 13.8$\pm$1.6 & VI \\
7154 & 4145 & 12$\rm^{h}$10$\rm^{m}$01.4$\rm^{s}$ & 39$\degr$53'02" &  5520 & 0.68 & 30.3$\pm$3.6 & VI \\
7278 & 4214 & \textit{12$\rm^{h}$15$\rm^{m}$39.1$\rm^{s}$} & \textit{36$\degr$19'41"} &  5040 & 0.96 & 142.4$\pm$24.8 & III  \\
7323 & 4242 & 12$\rm^{h}$17$\rm^{m}$30.2$\rm^{s}$ & 45$\degr$37'12" &  7200 & 0.96 & 18.5$\pm$3.2 & III  \\
7429 & 4319 & \textit{12$\rm^{h}$21$\rm^{m}$43.1$\rm^{s}$} & \textit{75$\degr$19'22"} &  6960 & 0.68 & 0.1$\pm$0.1 & VI \\
7524 & 4395 & \textit{12$\rm^{h}$25$\rm^{m}$48.9$\rm^{s}$} & \textit{33$\degr$32'48"} &  8400 & 0.96 & $\geq$17.7$\pm$3.1 & I  \\
7592 & 4449 & \textit{12$\rm^{h}$28$\rm^{m}$10.9$\rm^{s}$} & \textit{44$\degr$05'33"} &  2880 & 0.96 & 429.6$\pm$74.7 & III  \\
7699 &  & \textit{12$\rm^{h}$32$\rm^{m}$48.1$\rm^{s}$} & \textit{37$\degr$37'20"} &  6720 & 0.68 & 8.3$\pm$1.0 & VI \\
7766 & 4559 & 12$\rm^{h}$35$\rm^{m}$57.3$\rm^{s}$ & 27$\degr$57'38" &  2400 & 0.68 & $\geq$50.4$\pm$6.0 & VI \\
7831 & 4605 & 12$\rm^{h}$39$\rm^{m}$59.7$\rm^{s}$ & 61$\degr$36'29" &  3840 & 0.68 & 62.1$\pm$7.4 & VI \\
\hline
\end{tabular}
\end{table*}
\begin{table*}
\contcaption{}
\begin{tabular}{cccccccc}
\noalign{\medskip} \hline
N\Deg & N\Deg &  $\alpha$    & $\delta$   &  Exposure time & Scale & Flux & Paper\\
 UGC        &  NGC       &  (2000) & (2000) & $s$ & " & $10^{-16}~W~m^{-2}$ &\\
 (1)&(2)&(3)&(4)&(5)&(6)&(7)&(8)\\
\hline
7853 & 4618 & 12$\rm^{h}$41$\rm^{m}$33.1$\rm^{s}$ & 41$\degr$09'05" &  5040 & 0.68 & 49.8$\pm$5.9 & VI \\
7861 & 4625 & 12$\rm^{h}$41$\rm^{m}$52.9$\rm^{s}$ & 41$\degr$16'25" &  3840 & 0.68 & 6.3$\pm$0.7 & VI \\
7876 & 4635 & 12$\rm^{h}$42$\rm^{m}$39.3$\rm^{s}$ & 19$\degr$56'44" &  6720 & 0.68 & 4.1$\pm$0.5 & VI \\
7901 & 4651 & 12$\rm^{h}$43$\rm^{m}$42.7$\rm^{s}$ & 16$\degr$23'35" &  8160 & 0.68 & 12.7$\pm$1.5 & VI \\
7971 & 4707 & 12$\rm^{h}$48$\rm^{m}$22.9$\rm^{s}$ & 51$\degr$09'57" & 13200 & 0.96 & 0.8$\pm$0.1 & I  \\
7985 & 4713 & 12$\rm^{h}$49$\rm^{m}$57.9$\rm^{s}$ & 05$\degr$18'42" &  6240 & 0.68 & 28.6$\pm$3.4 & VI \\
8334 & 5055 & 13$\rm^{h}$15$\rm^{m}$49.4$\rm^{s}$ & 42$\degr$01'46" &  4320 & 0.68 & $\geq$88.2$\pm$10.4 & VI \\
8403 & 5112 & 13$\rm^{h}$21$\rm^{m}$56.6$\rm^{s}$ & 38$\degr$44'05" &  4080 & 0.68 & 30.5$\pm$3.6 & VI \\
8490 & 5204 & 13$\rm^{h}$29$\rm^{m}$36.5$\rm^{s}$ & 58$\degr$25'09" & 13920 & 0.96 & $\geq$16.2$\pm$2.8 & I  \\
 & 5296 & 13$\rm^{h}$46$\rm^{m}$18.7$\rm^{s}$ & 43$\degr$51'04" &  6000 & 0.68 & 0.8$\pm$0.1 & VI \\
8709 & 5297 & 13$\rm^{h}$46$\rm^{m}$23.7$\rm^{s}$ & 43$\degr$52'20" &  6000 & 0.68 & 15.2$\pm$1.8 & VI \\
8852 & 5376 & 13$\rm^{h}$55$\rm^{m}$16.1$\rm^{s}$ & 59$\degr$30'25" &  7680 & 0.68 & 4.4$\pm$0.5 & VI \\
8863 & 5377 & 13$\rm^{h}$56$\rm^{m}$16.7$\rm^{s}$ & 47$\degr$14'08" &  5040 & 0.68 & 0.4$\pm$0.1 & VI \\
8898 & 5394 & 13$\rm^{h}$58$\rm^{m}$33.7$\rm^{s}$ & 37$\degr$27'12" &  4320 & 0.68 & 1.0$\pm$0.1 & VI \\
8900 & 5395 & 13$\rm^{h}$58$\rm^{m}$38.0$\rm^{s}$ & 37$\degr$25'28" &  4320 & 0.68 & 5.6$\pm$0.7 & VI \\
8937 & 5430 & 14$\rm^{h}$00$\rm^{m}$45.8$\rm^{s}$ & 59$\degr$19'43" &  8880 & 0.68 & 10.7$\pm$1.3 & VI \\
9013 & 5474 & 14$\rm^{h}$05$\rm^{m}$02.0$\rm^{s}$ & 53$\degr$39'08" &  5040 & 0.68 & $\geq$18.8$\pm$2.2 & VI \\
9179 & 5585 & 14$\rm^{h}$19$\rm^{m}$48.1$\rm^{s}$ & 56$\degr$43'45" &  8160 & 0.68 & 12.5$\pm$1.5 & VI \\
9219 & 5608 & \textit{14$\rm^{h}$23$\rm^{m}$17.5$\rm^{s}$} & \textit{41$\degr$46'34"} &  6480 & 0.68 & 2.9$\pm$0.3 & VI \\
9248 & 5622 & 14$\rm^{h}$26$\rm^{m}$12.2$\rm^{s}$ & 48$\degr$33'51" & 10800 & 0.68 & 1.8$\pm$0.2 & VI \\
9358 & 5678 & 14$\rm^{h}$32$\rm^{m}$05.6$\rm^{s}$ & 57$\degr$55'16" &  7680 & 0.68 & 8.8$\pm$1.0 & VI \\
9366 & 5676 & 14$\rm^{h}$32$\rm^{m}$46.8$\rm^{s}$ & 49$\degr$27'29" &  5040 & 0.68 & 34.7$\pm$4.1 & IV  \\
9363 & 5668 & 14$\rm^{h}$33$\rm^{m}$24.4$\rm^{s}$ & 04$\degr$27'02" &  8160 & 0.68 & $\geq$21.5$\pm$2.5 & VI \\
9406 & 5693 & 14$\rm^{h}$36$\rm^{m}$11.1$\rm^{s}$ & 48$\degr$35'06" &  6000 & 0.68 & 0.9$\pm$0.1 & VI \\
9465 & 5727 & 14$\rm^{h}$40$\rm^{m}$26.1$\rm^{s}$ & 33$\degr$59'23" &  6480 & 0.68 & 4.2$\pm$0.5 & VI \\
9576 & 5774 & 14$\rm^{h}$53$\rm^{m}$42.5$\rm^{s}$ & 03$\degr$34'57" &  6840 & 0.68 & 11.3$\pm$1.3 & VI \\
9649 & 5832 & 14$\rm^{h}$57$\rm^{m}$46.2$\rm^{s}$ & 71$\degr$40'53" &  5520 & 0.68 & 3.9$\pm$0.5 & IV  \\
9736 & 5874 & 15$\rm^{h}$07$\rm^{m}$51.9$\rm^{s}$ & 54$\degr$45'08" &  3840 & 0.68 & 2.8$\pm$0.3 & VI \\
9753 & 5879 & 15$\rm^{h}$09$\rm^{m}$46.8$\rm^{s}$ & 57$\degr$00'01" &  5040 & 0.68 & 18.7$\pm$2.2 & IV  \\
9858 &  & 15$\rm^{h}$26$\rm^{m}$41.6$\rm^{s}$ & 40$\degr$33'53" &  5280 & 0.68 & 6.9$\pm$0.8 & IV  \\
9866 & 5949 & 15$\rm^{h}$28$\rm^{m}$00.6$\rm^{s}$ & 64$\degr$45'46" &  5280 & 0.68 & 7.2$\pm$0.9 & VI \\
9943 & 5970 & 15$\rm^{h}$38$\rm^{m}$30.0$\rm^{s}$ & 12$\degr$11'11" &  8400 & 0.68 & 16.0$\pm$1.9 & VI \\
9969 & 5985 & 15$\rm^{h}$39$\rm^{m}$37.2$\rm^{s}$ & 59$\degr$19'54" & 18840 & 0.96 & 12.3$\pm$2.1 & I  \\
9992 &  & \textit{15$\rm^{h}$41$\rm^{m}$47.9$\rm^{s}$} & \textit{67$\degr$15'14"} &  5760 & 0.68 & 0.5$\pm$0.1 & IV  \\
10075 & 6015 & 15$\rm^{h}$51$\rm^{m}$25.3$\rm^{s}$ & 62$\degr$18'36" &  5760 & 0.68 & 27.1$\pm$3.2 & VI \\
10310 &  & 16$\rm^{h}$16$\rm^{m}$18.5$\rm^{s}$ & 47$\degr$02'45" & 12340 & 0.96 & 1.8$\pm$0.3 & I  \\
10359 & 6140 & 16$\rm^{h}$20$\rm^{m}$56.9$\rm^{s}$ & 65$\degr$23'23" &  4560 & 0.68 & $\geq$17.9$\pm$2.1 & IV  \\
10470 & 6217 & 16$\rm^{h}$32$\rm^{m}$39.1$\rm^{s}$ & 78$\degr$11'53" &  7920 & 0.68 & 35.9$\pm$4.2 & IV  \\
10445 &  & 16$\rm^{h}$33$\rm^{m}$47.7$\rm^{s}$ & 28$\degr$59'05" &  4320 & 0.68 & 8.7$\pm$1.0 & IV  \\
10502 &  & 16$\rm^{h}$37$\rm^{m}$37.8$\rm^{s}$ & 72$\degr$22'26" &  3120 & 0.68 & 6.9$\pm$0.8 & IV  \\
10521 & 6207 & 16$\rm^{h}$43$\rm^{m}$03.7$\rm^{s}$ & 36$\degr$49'55" &  7680 & 0.68 & 10.3$\pm$1.2 & VI \\
10546 & 6236 & 16$\rm^{h}$44$\rm^{m}$34.4$\rm^{s}$ & 70$\degr$46'47" &  5280 & 0.68 & 13.0$\pm$1.5 & IV  \\
10564 & 6237 & 16$\rm^{h}$46$\rm^{m}$22.5$\rm^{s}$ & 70$\degr$21'20" &  4560 & 0.68 & 6.9$\pm$0.8 & IV  \\
10652 & 6283 & 16$\rm^{h}$59$\rm^{m}$26.5$\rm^{s}$ & 49$\degr$55'20" &  6000 & 0.68 & 6.4$\pm$0.8 & VI \\
10713 &  & \textit{17$\rm^{h}$04$\rm^{m}$33.7$\rm^{s}$} & \textit{72$\degr$26'44"} &  5040 & 0.68 & 1.9$\pm$0.2 & VI \\
10757 &  & 17$\rm^{h}$10$\rm^{m}$13.4$\rm^{s}$ & 72$\degr$24'38" &  5280 & 0.68 & 2.8$\pm$0.3 & VI \\
10769 &  & \textit{17$\rm^{h}$11$\rm^{m}$33.5$\rm^{s}$} & \textit{72$\degr$24'07"} &  5280 & 0.68 & 0.1$\pm$0.1 & VI \\
10791 &  & 17$\rm^{h}$14$\rm^{m}$38.5$\rm^{s}$ & 72$\degr$23'56" &  6000 & 0.68 & 0.0$\pm$0.1 & VI \\
10897 & 6412 & 17$\rm^{h}$29$\rm^{m}$37.5$\rm^{s}$ & 75$\degr$42'16" &  6960 & 0.96 & 16.3$\pm$2.8 & II  \\
11012 & 6503 & 17$\rm^{h}$49$\rm^{m}$26.3$\rm^{s}$ & 70$\degr$08'40" &  5040 & 0.68 & 25.3$\pm$3.0 & VI \\
11124 &  & 18$\rm^{h}$07$\rm^{m}$27.6$\rm^{s}$ & 35$\degr$33'51" &  3600 & 0.68 & 8.4$\pm$1.0 & IV  \\
11218 & 6643 & 18$\rm^{h}$19$\rm^{m}$46.7$\rm^{s}$ & 74$\degr$34'06" &  7200 & 0.96 & 42.2$\pm$7.3 & II  \\
11269 & 6667 & 18$\rm^{h}$30$\rm^{m}$39.7$\rm^{s}$ & 67$\degr$59'14" &  3840 & 0.68 & 1.1$\pm$0.1 & VI \\
11283 &  & 18$\rm^{h}$33$\rm^{m}$52.4$\rm^{s}$ & 49$\degr$16'41" &  8640 & 0.96 & 10.4$\pm$1.8 & II  \\
11283c &  & \textit{18$\rm^{h}$34$\rm^{m}$00.7$\rm^{s}$} & \textit{49$\degr$22'21"} &  7200 & 0.96 & 0.6$\pm$0.1 & II  \\
11300 & 6689 & 18$\rm^{h}$34$\rm^{m}$49.9$\rm^{s}$ & 70$\degr$31'27" & 17520 & 0.96 & $\geq$9.3$\pm$1.6 & II IV VI \\
11332 & 6654A & \textit{18$\rm^{h}$39$\rm^{m}$25.2$\rm^{s}$} & \textit{73$\degr$34'48"} &  7200 & 0.68 & 15.6$\pm$1.8 & VI \\
11407 & 6764 & 19$\rm^{h}$08$\rm^{m}$16.4$\rm^{s}$ & 50$\degr$55'59" &  6480 & 0.68 & 6.6$\pm$0.8 & VI \\
11429 & 6792 & 19$\rm^{h}$20$\rm^{m}$57.4$\rm^{s}$ & 43$\degr$07'57" &  6000 & 0.68 & 4.4$\pm$0.5 & IV  \\
11466 &  & 19$\rm^{h}$42$\rm^{m}$59.1$\rm^{s}$ & 45$\degr$17'58" &  5040 & 0.68 & 14.2$\pm$1.7 & VI \\

\hline
\end{tabular}
\end{table*}
\begin{table*}
\contcaption{}
\begin{tabular}{cccccccc}
\noalign{\medskip} \hline
N\Deg & N\Deg &  $\alpha$    & $\delta$   &  Exposure time & Scale & Flux & Paper\\
 UGC        &  NGC       &  (2000) & (2000) & $s$ & " & $10^{-16}~W~m^{-2}$ &\\
 (1)&(2)&(3)&(4)&(5)&(6)&(7)&(8)\\
\hline11470 & 6824 & 19$\rm^{h}$43$\rm^{m}$40.8$\rm^{s}$ & 56$\degr$06'34" &  3600 & 0.68 & 2.9$\pm$0.3 & VI \\
11496 &  & 19$\rm^{h}$53$\rm^{m}$01.8$\rm^{s}$ & 67$\degr$39'54" &  5520 & 0.68 & 0.9$\pm$0.1 & VI \\
11498 &  & 19$\rm^{h}$57$\rm^{m}$15.1$\rm^{s}$ & 05$\degr$53'24" &  8400 & 0.68 & 1.9$\pm$0.2 & VI \\
11557 &  & 20$\rm^{h}$24$\rm^{m}$00.7$\rm^{s}$ & 60$\degr$11'41" &  4320 & 0.68 & 10.0$\pm$1.2 & IV  \\
11597 & 6946 & 20$\rm^{h}$34$\rm^{m}$52.5$\rm^{s}$ & 60$\degr$09'12" &  5280 & 0.68 & $\geq$103.2$\pm$12.2 & VI \\
11670 & 7013 & 21$\rm^{h}$03$\rm^{m}$33.7$\rm^{s}$ & 29$\degr$53'50" &  7440 & 0.68 & 2.9$\pm$0.3 & VI \\
11707 &  & 21$\rm^{h}$14$\rm^{m}$31.8$\rm^{s}$ & 26$\degr$44'05" &  4320 & 0.68 & 6.9$\pm$0.8 & IV  \\
11852 &  & 21$\rm^{h}$55$\rm^{m}$59.4$\rm^{s}$ & 27$\degr$53'53" &  6480 & 0.68 & 0.6$\pm$0.1 & IV  \\
11861 &  & 21$\rm^{h}$56$\rm^{m}$24.2$\rm^{s}$ & 73$\degr$15'39" &  4800 & 0.68 & 9.5$\pm$1.1 & IV  \\
11872 & 7177 & 22$\rm^{h}$00$\rm^{m}$41.2$\rm^{s}$ & 17$\degr$44'18" &  7200 & 0.68 & 6.3$\pm$0.7 & VI \\
11891 &  & 22$\rm^{h}$03$\rm^{m}$33.8$\rm^{s}$ & 43$\degr$44'57" & 10560 & 0.96 & 0.6$\pm$0.1 & II  \\
11909 &  & \textit{22$\rm^{h}$06$\rm^{m}$16.4$\rm^{s}$} & \textit{47$\degr$15'10"} &  4320 & 0.68 & 13.8$\pm$1.6 & IV  \\
11914 & 7217 & 22$\rm^{h}$07$\rm^{m}$52.5$\rm^{s}$ & 31$\degr$21'32" &  6120 & 0.68 & 20.2$\pm$2.4 & IV  \\
11951 & 7231 & 22$\rm^{h}$12$\rm^{m}$30.2$\rm^{s}$ & 45$\degr$19'42" &  7200 & 0.96 & 14.7$\pm$2.6 & II  \\
12060 &  & 22$\rm^{h}$30$\rm^{m}$34.0$\rm^{s}$ & 33$\degr$49'09" & 18000 & 0.96 & 4.1$\pm$0.7 & I  \\
12082 &  & 22$\rm^{h}$34$\rm^{m}$11.3$\rm^{s}$ & 32$\degr$51'42" &  7440 & 0.68 & 1.3$\pm$0.2 & VI \\
12101 & 7320 & 22$\rm^{h}$36$\rm^{m}$03.5$\rm^{s}$ & 33$\degr$56'52" &  4680 & 0.68 & 2.3$\pm$0.3 & IV  \\
12212 &  & 22$\rm^{h}$50$\rm^{m}$30.6$\rm^{s}$ & 29$\degr$08'20" &  9360 & 0.96 & 1.0$\pm$0.2 & II  \\
12276 & 7440 & 22$\rm^{h}$58$\rm^{m}$32.6$\rm^{s}$ & 35$\degr$48'08" &  4680 & 0.68 & 3.1$\pm$0.4 & IV  \\
12276c &  & \textit{22$\rm^{h}$58$\rm^{m}$41.4$\rm^{s}$} & \textit{35$\degr$48'33"} &  4680 & 0.68 & 0.2$\pm$0.1 & IV  \\
12343 & 7479 & 23$\rm^{h}$04$\rm^{m}$56.7$\rm^{s}$ & 12$\degr$19'21" &  5520 & 0.68 & 31.0$\pm$3.7 & IV  \\
12632 &  & 23$\rm^{h}$29$\rm^{m}$58.7$\rm^{s}$ & 40$\degr$59'25" &  5760 & 0.68 & 3.4$\pm$0.4 & IV  \\
12754 & 7741 & 23$\rm^{h}$43$\rm^{m}$54.4$\rm^{s}$ & 26$\degr$04'31" & 10800 & 0.96 & $\geq$44.1$\pm$7.7 & I  \\
\hline
\end{tabular}
\\(1) Name of the galaxy in the UGC catalog except for NGC 542, IC 476, IC 2542 and NGC 5296 that do not have UGC name. (2) Name in the NGC catalog when available. (3\&4) Coordinates (in 2000) of the center of the galaxy used for the kinematic study except those in italic (taken from HyperLeda). (5) Total exposure time in second. (6) Pixel scale in arcsec. (7) Flux deduced from the comparison with \citet{James:2004} data. When the galaxy is larger than the \FOV~(see Table \ref{tabletf}), we only have a lower limit on the integrated \Ha~flux. (8) Publication papers.
\end{table*}

\begin{table*}
\caption{Model parameters.}
\begin{tabular}{cccccccccc}
\noalign{\medskip} \hline 
N\Deg & V$_{sys\_Leda}$ & V$_{sys\_FP}$ & i$_{Morph}$ & i$_{Kin}$ & P.A.$_{Morph}$ & P.A.$_{Kin}$ & $\overline{Res}$ & $\sigma_{res}$ & $\chi^{2}_{red}$\\
 UGC & \kms & \kms & \Deg & \Deg & \Deg & \Deg & 10$^{-3}$\kms & \kms &\\
(1)&(2)&(3)&(4)&(5)&(6)&(7)&(8)&(9)&(10)\\
\hline
12893 & 1102$\pm$7 & 1097$\pm$2 & 30$\pm$8 & 19$\pm$19 & 95$\pm$90 & 77$^\#$$\pm$5 & -28.1 & 8 & 1.2 \\
89 & 4564$\pm$3 & 4510$\pm$5 & 40$\pm$4 & 33$\pm$13 & 174$\pm$37 & 177$\pm$4 & 42.5 & 23 & 8.6 \\
94 & 4592$\pm$4 & 4548$\pm$2 & 50$\pm$4 & 42$\pm$5 & 102$\pm$22 & 94$\pm$2 & -3.9 & 13 & 2.7 \\
508 & 4665$\pm$4 & 4641$\pm$2 & 15$\pm$27 & 25$\pm$7$^*$ & 95$^{2M}$$\pm$90 & 123$\pm$2 & -0.2 & 24 & 8.7 \\
528 & 639$\pm$7 & 628$\pm$1 & 14$\pm$22 & 21$\pm$14 & 50$^{2M}$/30$^{Ha}$$\pm$90 & 52$\pm$3 & 0.6 & 9 & 1.3 \\
763 & 1156$\pm$4 & 1148$\pm$2 & 48$\pm$5 & 54$\pm$6 & 112$\pm$26 & 117$\pm$3 & -0.4 & 12 & 2.3 \\
1013 & 5190$\pm$4 &  & 80$\pm$3 &  & 68$\pm$7 &  &  &  &  \\
NGC 542 & 4660$\pm$8 &  & 90$\pm$0 &  & 143$\pm$8 &  &  &  &  \\
1117 & -182$\pm$3 & -191$\pm$1 & 56$\pm$5 & 56$\pm$16$^*$ & 23$\pm$11 & 18$^\#$$\pm$3 & 2.1 & 9 & 1.4 \\
1249 & 340$\pm$14 &  & 90$\pm$0 &  & 150$\pm$8 &  &  &  &  \\
1256 & 424$\pm$6 & 428$\pm$1 & 70$\pm$2 & 76$\pm$2 & 66$\pm$9 & 73$\pm$2 & 1.0 & 13 & 2.5 \\
1317 & 3111$\pm$5 & 3090$\pm$2 & 75$\pm$4 & 73$\pm$1 & 105$\pm$8 & 106$\pm$1 & -1.3 & 15 & 3.3 \\
1437 & 4893$\pm$4 & 4858$\pm$2 & 52$\pm$3 & 47$\pm$4 & 134$\pm$26 & 127$^\#$$\pm$2 & -0.8 & 18 & 5.2 \\
1655 & 5340$\pm$16 & 5427$\pm$7 & 45$\pm$9 & 45$\pm$18$^*$ &  & 138$^\#$$\pm$6 & -2744.9 & 30 & 15.4 \\
1736 & 1561$\pm$2 & 1522$\pm$2 & 48$\pm$3 & 35$\pm$14 & 24$\pm$21 & 27$\pm$3 & -6.5 & 18 & 5.1 \\
1810 & 7556$\pm$21 &  & 75$\pm$3 &  & 42$\pm$9 &  &  &  &  \\
1886 & 4859$\pm$23 & 4836$\pm$3 & 47$\pm$5 & 62$\pm$2 & 52$\pm$40 & 35$\pm$2 & 0.8 & 19 & 5.9 \\
1913 & 554$\pm$4 & 536$\pm$2 & 59$\pm$3 & 48$\pm$9 & 102$\pm$11 & 108$^\#$$\pm$3 & -12.7 & 12 & 2.2 \\
2023 & 604$\pm$6 & 593$\pm$2 & 15$\pm$30 & 19$\pm$18$^*$ &  & 135$^\#$$\pm$9$^*$ & $<$0.1 & 7 & 0.7 \\
2034 & 579$\pm$6 & 567$\pm$2 & 37$\pm$12 & 19$\pm$23$^*$ & 170$\pm$57 & 162$^\#$$\pm$17$^*$ & -0.3 & 8 & 1.0 \\
2045 & 1546$\pm$3 & 1525$\pm$3 & 66$\pm$4 & 61$\pm$8 & 151$\pm$14 & 139$^\#$$\pm$4 & -3.7 & 17 & 4.4 \\
2053 & 1026$\pm$5 &  & 66$\pm$9 &  & 43$\pm$23 &  &  &  &  \\
2082 & 707$\pm$3 &  & 87$\pm$4 &  & 133$\pm$3 &  &  &  &  \\
2080 & 903$\pm$5 & 893$\pm$1 & 25$\pm$12 & 25$\pm$9$^*$ & 75$^{2M}$$\pm$85 & 156$^\#$$\pm$2 & -0.1 & 10 & 1.5 \\
2141 & 985$\pm$4 & 965$\pm$3 & 68$\pm$5 & 74$\pm$23 & 23$\pm$14 & 11$^\#$$\pm$5 & -11.6 & 11 & 1.9 \\
2183 & 1546$\pm$7 & 1475$\pm$4 & 47$\pm$11 & 41$\pm$10 & 160$\pm$50 & 159$\pm$3 & -2.0 & 12 & 2.1 \\
2193 & 517$\pm$6 & 519$\pm$1 & 58$\pm$3 & 6$\pm$15 & 100$\pm$14 & 125$^\#$$\pm$6 & $<$0.1 & 8 & 1.0 \\
2455 & 374$\pm$2 & 367$\pm$2 & 43$\pm$7 & 51$\pm$30$^*$ & 25$\pm$28 & 83$^\#$$\pm$21$^*$ &  &  &  \\
2503 & 2387$\pm$4 & 2377$\pm$2 & 57$\pm$3 & 49$\pm$2 & 31$\pm$16 & 34$\pm$1 & -8.4 & 16 & 3.8 \\
2800 & 1174$\pm$5 & 1174$\pm$2 & 73$\pm$7 & 52$\pm$13 & 103$\pm$19 & 111$^\#$$\pm$4 & 0.9 & 12 & 2.2 \\
2855 & 1202$\pm$4 & 1188$\pm$2 & 68$\pm$2 & 68$\pm$2 & 110$\pm$10 & 100$\pm$2 & -0.6 & 19 & 5.4 \\
3013 & 2468$\pm$14 & 2453$\pm$6 & 58$\pm$3 & 58$\pm$8$^*$ & 81$\pm$18 & 15$^\#$$\pm$4 & -3.5 & 25 & 10.0 \\
3056 & -100$\pm$7 &  & 65$\pm$7 &  & 120$\pm$17 &  &  &  &  \\
3273 & 616$\pm$7 & 615$\pm$2 & 90$\pm$0 & 82$\pm$7 & 45$\pm$12 & 42$\pm$4 & 4.0 & 12 & 2.4 \\
3334 & 3934$\pm$4 & 3952$\pm$13 & 47$\pm$7 & 47$\pm$14$^*$ & 85$\pm$21 & 97$^\#$$\pm$6 & 9.9 & 54 & 46.0 \\
3382 & 4497$\pm$6 & 4490$\pm$2 & 21$\pm$10 & 18$\pm$6 & 30$^{2M}$/168$^{Pa}$$\pm$90 & 4$^\#$$\pm$2 & -2.2 & 18 & 5.0 \\
3384 & 1088$\pm$3 &  & 45$\pm$7 &  & 62$\pm$59 &  &  &  &  \\
3429 & 898$\pm$6 & 868$\pm$4 & 37$\pm$12 & 54$\pm$8 & 123$\pm$28 & 137$^\#$$\pm$3 & -7.6 & 17 & 4.3 \\
3463 & 2692$\pm$4 & 2679$\pm$3 & 63$\pm$3 & 63$\pm$3 & 117$\pm$12 & 110$\pm$2 & -0.1 & 15 & 3.6 \\
3574 & 1441$\pm$3 & 1433$\pm$1 & 21$\pm$8 & 19$\pm$10 & 10$^{2M}$/121$^{Ha}$/162$^{Pa}$$\pm$90 & 99$\pm$3 & $<$0.1 & 12 & 2.3 \\
3521 & 4426$\pm$7 & 4415$\pm$2 & 61$\pm$3 & 58$\pm$5 & 76$\pm$17 & 78$^\#$$\pm$3 & 3.2 & 16 & 4.1 \\
3528 & 4421$\pm$18 & 4340$\pm$5 & 59$\pm$5 & 42$\pm$12 & 38$\pm$24 & 43$^\#$$\pm$4 & -17.8 & 32 & 17.6 \\
3618 & 5851$\pm$6 &  & 49$\pm$5 &  & 171$\pm$24 &  &  &  &  \\
3691 & 2202$\pm$4 & 2203$\pm$2 & 62$\pm$4 & 64$\pm$4 & 65$\pm$19 & 68$^\#$$\pm$3 & $<$0.1 & 10 & 1.7 \\
3685 & 1796$\pm$4 & 1795$\pm$1 & 55$\pm$4 & 12$\pm$17 & 133$\pm$18 & 118$^\#$$\pm$4 & -1.3 & 9 & 1.4 \\
3708 & 5201$\pm$26 & 5161$\pm$4 & 16$\pm$24 & 44$\pm$16 & 136$^{Ni}$$\pm$90 & 50$^\#$$\pm$4 & -11.6 & 18 & 5.4 \\
3709 & 5223$\pm$50 & 5292$\pm$4 & 46$\pm$4 & 55$\pm$4 & 66$\pm$27 & 52$^\#$$\pm$2 & -4.7 & 18 & 5.3 \\
3734 & 969$\pm$5 & 966$\pm$1 & 26$\pm$7 & 43$\pm$7 & 144$\pm$59 & 139$\pm$2 & 7.7 & 12 & 2.2 \\
3826 & 1733$\pm$3 & 1724$\pm$2 & 30$\pm$9 & 20$\pm$19 & 160$^{2M}$/85$^{Ni}$$\pm$56 & 74$^\#$$\pm$5 & 135.4 & 9 & 1.2 \\
3809 & 2202$\pm$4 & 2200$\pm$1 & 58$\pm$4 & 58$\pm$2 & 178$\pm$17 & 177$^\#$$\pm$1 & $<$0.1 & 14 & 3.3 \\
3740 & 2417$\pm$6 & 2416$\pm$2 & 40$\pm$6 & 48$\pm$14 & 19$\pm$29 & 67$^\#$$\pm$4 & $<$0.1 & 11 & 1.7 \\
3851 & 99$\pm$2 &  & 90$\pm$0 &  & 30$\pm$7 &  &  &  &  \\
3876 & 860$\pm$7 & 854$\pm$2 & 61$\pm$3 & 59$\pm$5 & 178$\pm$14 & 178$^\#$$\pm$3 & -0.2 & 10 & 1.6 \\
3915 & 4679$\pm$7 & 4659$\pm$3 & 59$\pm$5 & 47$\pm$4 & 25$\pm$26 & 30$\pm$2 & 6.8 & 13 & 2.7 \\
IC 476 & 4734$\pm$32 & 4767$\pm$3 & 40$\pm$5 & 55$\pm$24 & 102$\pm$49 & 68$\pm$6 & -5.9 & 18 & 5.5 \\
4026 & 4782$\pm$11 & 4892$\pm$3 & 73$\pm$4 & 56$\pm$4 & 136$\pm$12 & 139$\pm$2 & 4.0 & 19 & 5.7 \\
4165 & 515$\pm$4 & 504$\pm$1 & 21$\pm$9 & 41$\pm$10 & 74$^{Pa}$$\pm$86 & 85$^\#$$\pm$2 & -0.4 & 10 & 1.6 \\
4256 & 5252$\pm$5 & 5252$\pm$3 & 36$\pm$4 & 38$\pm$21 & 26$\pm$35 & 111$^\#$$\pm$6 & 15.9 & 26 & 11.1 \\
4273 & 2458$\pm$18 & 2398$\pm$2 & 65$\pm$5 & 60$\pm$4 & 54$\pm$15 & 32$^\#$$\pm$2 & -2.3 & 14 & 3.3 \\
4274 & 445$\pm$3 & 430$\pm$1 & 69$\pm$6 & 27$\pm$17 &  & 175$\pm$3 & $<$0.1 & 8 & 1.1 \\
\hline
\label{tablemod}
\end{tabular}
\end{table*}
\begin{table*}
\contcaption{}
\begin{tabular}{cccccccccc}
\noalign{\medskip} \hline 
N\Deg & V$_{sys\_Leda}$ & V$_{sys\_FP}$ & i$_{Morph}$ & i$_{Kin}$ & P.A.$_{Morph}$ & P.A.$_{Kin}$ & $\overline{Res}$ & $\sigma_{res}$ & $\chi^{2}_{red}$\\
 UGC & \kms & \kms & \Deg & \Deg & \Deg & \Deg & 10$^{-3}$\kms & \kms &\\
(1)&(2)&(3)&(4)&(5)&(6)&(7)&(8)&(9)&(10)\\
\hline
4278 & 558$\pm$4 &  & 90$\pm$0 &  & 172$\pm$3 &  &  &  &  \\
4284 & 559$\pm$3 & 536$\pm$2 & 61$\pm$3 & 59$\pm$9 & 170$\pm$14 & 176$\pm$3 & 0.5 & 10 & 1.4 \\
4305 & 158$\pm$2 & 139$\pm$1 & 51$\pm$6 & 40$\pm$27$^*$ & 15$\pm$21 & 9$^\#$$\pm$4 &  &  &  \\
4325 & 518$\pm$4 & 508$\pm$2 & 68$\pm$4 & 63$\pm$14 & 52$\pm$14 & 57$\pm$3 & -0.4 & 11 & 2.0 \\
4393 & 2126$\pm$5 & 2119$\pm$4 & 50$\pm$5 & 50$\pm$9$^*$ & 50$\pm$35 & 70$^\#$$\pm$7 & 0.6 & 11 & 1.7 \\
4422 & 4333$\pm$4 & 4321$\pm$2 & 49$\pm$4 & 25$\pm$8 & 12$\pm$26 & 36$\pm$2 & -18.3 & 20 & 6.2 \\
4456 & 5497$\pm$23 & 5470$\pm$1 & 28$\pm$6 & 9$\pm$14 & 42$\pm$74 & 124$\pm$3 & 7.3 & 14 & 3.1 \\
4499 & 692$\pm$4 & 682$\pm$1 & 80$\pm$6 & 50$\pm$14$^*$ & 151$\pm$13 & 141$\pm$4 & 0.5 & 7 & 0.7 \\
4543 & 1960$\pm$4 & 1948$\pm$2 & 45$\pm$6 & 52$\pm$15 & 174$\pm$36 & 136$^\#$$\pm$4 & 0.7 & 14 & 3.0 \\
4555 & 4235$\pm$6 & 4235$\pm$2 & 21$\pm$12 & 38$\pm$7 & 140$^{2M}$/78$^{SDSS}$$\pm$90 & 90$\pm$2 & 0.5 & 15 & 3.5 \\
4770 & 7063$\pm$9 & 7026$\pm$3 & 36$\pm$9 & 20$\pm$13 & 54$\pm$48 & 98$^\#$$\pm$3 & -36.8 & 14 & 3.0 \\
4820 & 1355$\pm$4 & 1350$\pm$2 & 41$\pm$4 & 38$\pm$3 & 160$\pm$23 & 157$\pm$2 & $<$0.1 & 13 & 2.7 \\
4936 & 1733$\pm$5 & 1728$\pm$1 & 36$\pm$6 & 13$\pm$12 & 174$\pm$32 & 114$^\#$$\pm$2 & 0.2 & 10 & 1.4 \\
5045 & 7716$\pm$23 & 7667$\pm$2 & 41$\pm$4 & 16$\pm$9 & 136$\pm$31 & 148$\pm$2 & -2.8 & 13 & 2.7 \\
5175 & 3052$\pm$11 & 3049$\pm$2 & 65$\pm$4 & 56$\pm$3 & 145$\pm$14 & 143$\pm$2 & -2.3 & 13 & 2.9 \\
5228 & 1873$\pm$7 & 1869$\pm$2 & 82$\pm$2 & 72$\pm$2 & 122$\pm$6 & 120$\pm$2 & -0.7 & 9 & 1.3 \\
5251 & 1481$\pm$3 & 1465$\pm$3 & 88$\pm$9 & 73$\pm$6 & 78$\pm$4 & 80$^\#$$\pm$3 & $<$0.1 & 15 & 3.5 \\
5253 & 1322$\pm$7 & 1322$\pm$2 & 38$\pm$6 & 40$\pm$4 & 176$\pm$31 & 176$^\#$$\pm$2 & -0.7 & 13 & 2.5 \\
5272 & 520$\pm$3 &  & 90$\pm$0 &  & 118$\pm$6 &  &  &  &  \\
5279 & 1488$\pm$4 &  & 90$\pm$0 &  & 83$\pm$6 &  &  &  &  \\
5316 & 1059$\pm$3 & 1031$\pm$2 & 70$\pm$2 & 77$\pm$4 & 130$\pm$9 & 130$\pm$3 & -5.4 & 14 & 3.2 \\
5319 & 2448$\pm$9 & 2439$\pm$1 & 40$\pm$5 & 30$\pm$9 & 125$^{2M}$/7$^{Pa}$$\pm$29 & 165$^\#$$\pm$2 & -21.0 & 9 & 1.3 \\
5351 & 1473$\pm$4 &  & 82$\pm$6 &  & 105$\pm$6 &  &  &  &  \\
5373 & 302$\pm$3 & 291$\pm$2 & 60$\pm$6 & 10$\pm$18 & 110$\pm$18 & 51$\pm$8 & -0.5 & 8 & 0.9 \\
5398 & 14$\pm$5 &  & 40$\pm$10 &  & 45$\pm$35 &  &  &  &  \\
5414 & 603$\pm$3 & 592$\pm$2 & 54$\pm$8 & 71$\pm$13 & 35$\pm$29 & 39$^\#$$\pm$4 & -0.4 & 13 & 2.7 \\
IC 2542 & 6113$\pm$20 & 6111$\pm$2 & 43$\pm$4 & 20$\pm$15 & 173$\pm$32 & 174$\pm$3 & -0.9 & 20 & 6.3 \\
5510 & 1301$\pm$3 & 1298$\pm$2 & 38$\pm$5 & 31$\pm$10 & 26$\pm$37 & 20$^\#$$\pm$3 & -0.3 & 10 & 1.4 \\
5532 & 2812$\pm$8 & 2802$\pm$1 & 33$\pm$9 & 32$\pm$3 & 150$\pm$42 & 147$\pm$1 & $<$0.1 & 14 & 2.9 \\
5556 & 1581$\pm$3 &  & 75$\pm$2 &  & 105$\pm$7 &  &  &  &  \\
5721 & 537$\pm$4 & 527$\pm$6 & 62$\pm$3 & 62$\pm$30$^*$ & 95$\pm$14 & 95$^\#$$\pm$21$^*$ & 0.2 & 14 & 3.1 \\
5786 & 990$\pm$3 & 992$\pm$4 & 16$\pm$25 & 53$\pm$11 & 18$^{SDSS}$$\pm$90 & 153$\pm$5 & 0.5 & 17 & 4.6 \\
5789 & 742$\pm$2 & 730$\pm$2 & 63$\pm$3 & 68$\pm$10 & 37$\pm$12 & 27$\pm$3 & 4.0 & 8 & 1.1 \\
5829 & 630$\pm$3 & 626$\pm$2 & 25$\pm$13 & 34$\pm$20 &  & 18$^\#$$\pm$6 & -0.5 & 9 & 1.2 \\
5840 & 582$\pm$4 & 580$\pm$1 & 18$\pm$14 & 18$\pm$11 &  & 153$^\#$$\pm$3 & $<$0.1 & 12 & 2.1 \\
5842 & 1261$\pm$10 & 1245$\pm$1 & 34$\pm$6 & 47$\pm$9 & 104$\pm$36 & 112$^\#$$\pm$2 & -1.5 & 10 & 1.7 \\
5931 & 1621$\pm$3 & 1604$\pm$3 & 58$\pm$4 & 54$\pm$16 & 35$\pm$20 & 180$^\#$$\pm$5 & 10.2 & 13 & 2.9 \\
5935 & 1656$\pm$7 &  & 90$\pm$0 &  & 99$\pm$14 &  &  &  &  \\
5982 & 1583$\pm$6 & 1573$\pm$2 & 58$\pm$4 & 55$\pm$4 & 33$\pm$17 & 28$\pm$2 & -1.5 & 15 & 3.6 \\
6118 & 1539$\pm$5 & 1525$\pm$2 & 27$\pm$7 & 39$\pm$8$^*$ & 150$^{2M}$/57$^{Pa}$$\pm$50 & 163$^\#$$\pm$3 & 16.3 & 11 & 1.9 \\
6277 & 1192$\pm$3 & 1191$\pm$3 & 17$\pm$16 & 17$\pm$17$^*$ & 0$^{2M}$$\pm$90 & 76$\pm$3 & -5.1 & 19 & 5.7 \\
6419 & 1365$\pm$24 & 1381$\pm$2 & 57$\pm$5 & 66$\pm$19 & 27$\pm$23 & 34$\pm$7 & -1.8 & 8 & 0.9 \\
6521 & 5879$\pm$5 & 5842$\pm$2 & 50$\pm$3 & 46$\pm$4 & 21$\pm$19 & 20$\pm$2 & -0.9 & 17 & 4.5 \\
6523 & 5913$\pm$13 & 5947$\pm$2 & 24$\pm$7 & 24$\pm$14$^*$ & 36$^{Va}$/12$^{Pa}$/51$^{Pa}$$\pm$90 & 173$^\#$$\pm$3 & -6.4 & 13 & 2.8 \\
6537 & 862$\pm$5 & 856$\pm$2 & 49$\pm$5 & 47$\pm$5 & 14$\pm$19 & 20$^\#$$\pm$2 & -0.2 & 12 & 2.3 \\
6628 & 850$\pm$5 & 863$\pm$1 & 33$\pm$10 & 20$\pm$20$^*$ & 144$\pm$58 & 179$\pm$2 & 1.4 & 10 & 1.4 \\
6702 & 7372$\pm$10 & 7332$\pm$2 & 32$\pm$7 & 38$\pm$6 & 60$\pm$75 & 76$^\#$$\pm$2 & 8.4 & 15 & 3.6 \\
6778 & 969$\pm$4 & 951$\pm$1 & 60$\pm$3 & 49$\pm$4 & 162$\pm$15 & 163$^\#$$\pm$2 & 2.7 & 12 & 2.3 \\
6787 & 1173$\pm$3 & 1157$\pm$3 & 57$\pm$3 & 70$\pm$2 & 108$\pm$15 & 112$\pm$2 & -0.7 & 17 & 4.4 \\
7021 & 1979$\pm$8 & 1976$\pm$3 & 56$\pm$4 & 56$\pm$7$^*$ & 89$\pm$17 & 86$^\#$$\pm$2 & -0.6 & 14 & 3.1 \\
7045 & 770$\pm$6 & 758$\pm$1 & 70$\pm$2 & 68$\pm$2 & 101$\pm$8 & 99$\pm$2 & 0.2 & 10 & 1.5 \\
7154 & 1011$\pm$28 & 1009$\pm$1 & 64$\pm$3 & 65$\pm$3 & 100$\pm$11 & 95$^\#$$\pm$2 & $<$0.1 & 12 & 2.3 \\
7278 & 291$\pm$2 &  & 44$\pm$9 &  &  &  &  &  &  \\
7323 & 517$\pm$4 & 505$\pm$1 & 51$\pm$4 & 51$\pm$11$^*$ & 19$\pm$21 & 38$\pm$3 & $<$0.1 & 10 & 1.5 \\
7429 & 1476$\pm$24 &  & 73$\pm$3 &  & 162$\pm$10 &  &  &  &  \\
7524 & 317$\pm$3 &  & 90$\pm$0 &  & 127$\pm$7 &  &  &  &  \\
7592 & 204$\pm$3 &  & 63$\pm$5 &  & 47$\pm$16 &  &  &  &  \\
7699 & 496$\pm$1 &  & 78$\pm$2 &  & 32$\pm$6 &  &  &  &  \\
7766 & 814$\pm$3 & 807$\pm$1 & 65$\pm$4 & 69$\pm$3 & 150$\pm$10 & 143$^\#$$\pm$2 & $<$0.1 & 12 & 2.4 \\
7831 & 147$\pm$4 & 136$\pm$3 & 70$\pm$3 & 56$\pm$12 & 125$\pm$7 & 110$^\#$$\pm$5 & $<$0.1 & 9 & 1.2 \\
\hline
\end{tabular}
\end{table*}
\begin{table*}
\contcaption{}
\begin{tabular}{cccccccccc}
\noalign{\medskip} \hline 
N\Deg & V$_{sys\_Leda}$ & V$_{sys\_FP}$ & i$_{Morph}$ & i$_{Kin}$ & P.A.$_{Morph}$ & P.A.$_{Kin}$ & $\overline{Res}$ & $\sigma_{res}$ & $\chi^{2}_{red}$\\
 UGC & \kms & \kms & \Deg & \Deg & \Deg & \Deg & 10$^{-3}$\kms & \kms &\\
(1)&(2)&(3)&(4)&(5)&(6)&(7)&(8)&(9)&(10)\\
\hline
7853 & 537$\pm$4 & 530$\pm$2 & 58$\pm$5 & 58$\pm$28$^*$ & 40$\pm$17 & 37$^\#$$\pm$4 & 12.8 & 8 & 1.1 \\
7861 & 611$\pm$4 & 598$\pm$1 & 47$\pm$6 & 47$\pm$24$^*$ & 116$^{Pa}$/30$^{SDSS}$$\pm$31 & 117$^\#$$\pm$4 & $<$0.1 & 11 & 1.9 \\
7876 & 955$\pm$8 & 944$\pm$1 & 44$\pm$5 & 53$\pm$9 & 3$\pm$34 & 164$^\#$$\pm$3 & 0.3 & 8 & 1.1 \\
7901 & 799$\pm$2 & 788$\pm$2 & 50$\pm$3 & 53$\pm$2 & 77$\pm$16 & 74$^\#$$\pm$2 & $<$0.1 & 12 & 2.2 \\
7971 & 467$\pm$6 & 457$\pm$2 & 26$\pm$18 & 31$\pm$27 & 25$^{Ni}$$\pm$90 & 32$\pm$11 & 0.3 & 8 & 1.0 \\
7985 & 653$\pm$3 & 642$\pm$2 & 24$\pm$12 & 49$\pm$6 & 110$^{2M}$/87$^{Ha}$/153$^{Pa}$/100$^{Ni}$/88$^{SDSS}$$\pm$90 & 96$^\#$$\pm$3 & -0.1 & 8 & 1.0 \\
8334 & 508$\pm$3 & 484$\pm$1 & 55$\pm$5 & 66$\pm$1 & 102$\pm$16 & 100$\pm$1 & -0.1 & 11 & 1.7 \\
8403 & 969$\pm$4 & 975$\pm$2 & 54$\pm$3 & 57$\pm$4 & 129$\pm$17 & 121$\pm$2 & -0.1 & 9 & 1.3 \\
8490 & 202$\pm$2 & 190$\pm$1 & 58$\pm$8 & 40$\pm$15 & 5$\pm$17 & 167$\pm$3 & -0.2 & 11 & 1.8 \\
NGC 5296 & 2243$\pm$3 & 2254$\pm$2 & 65$\pm$6 & 65$\pm$4 & 12$\pm$22 & 2$\pm$3 & -6.2 & 4 & 0.3 \\
8709 & 2407$\pm$13 & 2405$\pm$3 & 82$\pm$3 & 76$\pm$1 & 147$\pm$5 & 150$^\#$$\pm$2 & 0.1 & 15 & 3.3 \\
8852 & 2023$\pm$17 & 2075$\pm$1 & 55$\pm$7 & 52$\pm$3 & 65$\pm$24 & 63$\pm$2 & 0.9 & 10 & 1.5 \\
8863 & 1796$\pm$7 & 1789$\pm$4 & 77$\pm$4 & 77$\pm$13$^*$ & 38$\pm$8 & 38$^\#$$\pm$7$^*$ & -41.4 & 14 & 3.3 \\
8898 & 3464$\pm$10 & 3448$\pm$2 & 71$\pm$3 & 27$\pm$20 & 30$^{2M}$/140$^{Pa}$/117$^{SDSS}$$\pm$10 & 31$\pm$6 & 41.5 & 7 & 0.7 \\
8900 & 3466$\pm$11 & 3511$\pm$3 & 66$\pm$5 & 57$\pm$10 & 172$\pm$15 & 161$\pm$2 & 5.8 & 20 & 6.6 \\
8937 & 2968$\pm$9 & 2961$\pm$5 & 50$\pm$5 & 32$\pm$12 & 177$\pm$23 & 5$^\#$$\pm$3 & 2361.4 & 19 & 5.7 \\
9013 & 255$\pm$23 & 262$\pm$1 & 50$\pm$4 & 21$\pm$16$^*$ & 85$\pm$21 & 164$\pm$4 & 1.8 & 7 & 0.8 \\
9179 & 302$\pm$2 & 293$\pm$2 & 53$\pm$3 & 36$\pm$14 & 33$\pm$16 & 49$\pm$4 & $<$0.1 & 9 & 1.4 \\
9219 & 666$\pm$11 &  & 81$\pm$6 &  & 99$\pm$13 &  &  &  &  \\
9248 & 3867$\pm$6 & 3865$\pm$2 & 58$\pm$3 & 58$\pm$4 & 86$\pm$18 & 81$^\#$$\pm$2 & 3.5 & 15 & 3.6 \\
9358 & 1907$\pm$4 & 1912$\pm$3 & 62$\pm$3 & 54$\pm$4 & 2$\pm$14 & 2$^\#$$\pm$2 & 2.3 & 15 & 3.5 \\
9366 & 2102$\pm$3 & 2109$\pm$2 & 66$\pm$3 & 62$\pm$2 & 45$\pm$10 & 45$^\#$$\pm$2 & -0.6 & 14 & 2.9 \\
9363 & 1584$\pm$3 & 1577$\pm$1 & 33$\pm$6 & 18$\pm$14$^*$ & 107$\pm$42 & 147$\pm$3 & 3.2 & 8 & 1.1 \\
9406 & 2279$\pm$2 & 2281$\pm$2 & 51$\pm$7 & 59$\pm$25 & 60$^{2M}$/150$^{SDSS}$$\pm$36 & 132$\pm$12 & 78.0 & 11 & 2.0 \\
9465 & 1495$\pm$3 & 1485$\pm$2 & 90$\pm$0 & 65$\pm$4 & 143$\pm$14 & 127$\pm$3 & 0.3 & 8 & 1.1 \\
9576 & 1565$\pm$5 & 1555$\pm$2 & 52$\pm$4 & 41$\pm$11 & 125$\pm$21 & 122$\pm$3 & -0.4 & 10 & 1.6 \\
9649 & 447$\pm$4 & 440$\pm$1 & 70$\pm$3 & 54$\pm$6$^*$ & 40$\pm$12 & 55$^\#$$\pm$3 & $<$0.1 & 10 & 1.6 \\
9736 & 3128$\pm$4 & 3135$\pm$2 & 51$\pm$3 & 51$\pm$5 & 57$\pm$19 & 39$^\#$$\pm$2 & 2.9 & 14 & 3.2 \\
9753 & 771$\pm$3 & 764$\pm$2 & 74$\pm$4 & 69$\pm$1 & 2$\pm$8 & 3$\pm$2 & 0.5 & 11 & 1.8 \\
9858 & 2621$\pm$3 & 2638$\pm$3 & 90$\pm$0 & 75$\pm$2 & 78$\pm$6 & 70$\pm$2 & 0.6 & 17 & 4.6 \\
9866 & 427$\pm$4 & 430$\pm$1 & 69$\pm$3 & 56$\pm$6 & 150$\pm$11 & 148$\pm$2 & -5.1 & 7 & 0.8 \\
9943 & 1958$\pm$4 & 1946$\pm$1 & 48$\pm$5 & 54$\pm$2 & 87$\pm$20 & 86$^\#$$\pm$2 & -0.1 & 9 & 1.4 \\
9969 & 2519$\pm$3 & 2516$\pm$2 & 63$\pm$3 & 61$\pm$1 & 15$\pm$12 & 16$\pm$1 & 7.4 & 19 & 5.4 \\
9992 & 426$\pm$4 &  & 60$\pm$5 &  &  &  &  &  &  \\
10075 & 831$\pm$3 & 827$\pm$1 & 66$\pm$3 & 62$\pm$2 & 28$\pm$10 & 30$^\#$$\pm$1 & $<$0.1 & 9 & 1.4 \\
10310 & 716$\pm$3 & 702$\pm$2 & 42$\pm$11 & 42$\pm$20$^*$ & 165$\pm$47 & 7$^\#$$\pm$7 & $<$0.1 & 11 & 2.0 \\
10359 & 907$\pm$4 & 911$\pm$2 & 32$\pm$8 & 44$\pm$12$^*$ & 76$\pm$43 & 104$^\#$$\pm$3 & $<$0.1 & 13 & 2.5 \\
10470 & 1368$\pm$3 & 1354$\pm$2 & 45$\pm$5 & 34$\pm$9$^*$ & 155$\pm$26 & 107$^\#$$\pm$2 & -0.6 & 10 & 1.6 \\
10445 & 962$\pm$3 & 961$\pm$2 & 46$\pm$4 & 47$\pm$12 & 142$\pm$24 & 110$\pm$4 & -0.4 & 11 & 1.8 \\
10502 & 4297$\pm$5 & 4291$\pm$2 & 40$\pm$5 & 50$\pm$5 & 94$\pm$33 & 99$\pm$2 & -0.2 & 20 & 6.4 \\
10521 & 852$\pm$2 & 832$\pm$2 & 71$\pm$3 & 59$\pm$3 & 17$\pm$10 & 20$\pm$2 & 0.7 & 9 & 1.2 \\
10546 & 1280$\pm$4 & 1268$\pm$2 & 55$\pm$3 & 42$\pm$10 & 9$\pm$19 & 2$^\#$$\pm$3 & -2.9 & 14 & 3.1 \\
10564 & 1132$\pm$4 & 1120$\pm$2 & 53$\pm$4 & 77$\pm$6 & 156$\pm$23 & 149$\pm$3 & 0.9 & 11 & 2.1 \\
10652 & 1092$\pm$26 & 1089$\pm$1 & 30$\pm$7 & 21$\pm$13 & 56$\pm$54 & 45$^\#$$\pm$3 & 0.6 & 8 & 1.0 \\
10713 & 1073$\pm$4 &  & 90$\pm$0 &  & 8$\pm$7 &  &  &  &  \\
10757 & 1168$\pm$8 & 1210$\pm$2 & 59$\pm$3 & 44$\pm$22 & 66$\pm$18 & 56$\pm$6 & 0.7 & 11 & 1.8 \\
10769 & 1230$\pm$13 &  & 57$\pm$4 &  & 41$\pm$24 &  &  &  &  \\
10791 & 1328$\pm$6 & 1318$\pm$3 & 0$\pm$0 & 34$\pm$20$^*$ &  & 92$\pm$4 & 726.5 & 10 & 1.7 \\
10897 & 1334$\pm$3 & 1313$\pm$1 & 31$\pm$6 & 31$\pm$17$^*$ & 139$\pm$40 & 115$\pm$3 & 0.3 & 13 & 2.7 \\
11012 & 36$\pm$12 & 25$\pm$1 & 74$\pm$2 & 72$\pm$2 & 123$\pm$7 & 119$^\#$$\pm$2 & 0.4 & 8 & 1.0 \\
11124 & 1608$\pm$4 & 1606$\pm$2 & 22$\pm$17 & 51$\pm$10 &  & 2$^\#$$\pm$3 & -0.4 & 11 & 1.8 \\
11218 & 1483$\pm$2 & 1477$\pm$2 & 64$\pm$3 & 58$\pm$2 & 36$\pm$11 & 42$\pm$2 & -0.5 & 12 & 2.3 \\
11269 & 2590$\pm$6 & 2563$\pm$6 & 60$\pm$3 & 69$\pm$4 & 97$\pm$18 & 92$^\#$$\pm$3 & 3.8 & 30 & 14.4 \\
11283 & 1963$\pm$5 & 1944$\pm$4 & 34$\pm$7 & 34$\pm$17$^*$ & 9$\pm$58 & 120$\pm$5 & -3.7 & 18 & 4.9 \\
11283c & 1963$\pm$5 &  & 68$\pm$3 &  & 82$\pm$14 &  &  &  &  \\
11300 & 488$\pm$3 & 482$\pm$1 & 77$\pm$2 & 70$\pm$3 & 171$\pm$5 & 168$\pm$2 & -1.6 & 12 & 2.3 \\
11332 & 1569$\pm$25 &  & 82$\pm$2 &  & 65$\pm$5 &  &  &  &  \\
11407 & 2412$\pm$4 & 2402$\pm$8 & 64$\pm$3 & 64$\pm$22$^*$ & 65$\pm$13 & 65$\pm$10$^*$ & 1.0 & 20 & 6.6 \\
11429 & 4642$\pm$4 & 4679$\pm$7 & 61$\pm$3 & 61$\pm$16$^*$ & 23$\pm$14 & 28$^\#$$\pm$6 &  &  &  \\
11466 & 820$\pm$9 & 826$\pm$3 & 55$\pm$3 & 66$\pm$5 & 35$\pm$20 & 46$^\#$$\pm$3 & -0.3 & 13 & 2.6 \\
\hline
\end{tabular}
\end{table*}
\begin{table*}
\contcaption{}
\begin{tabular}{cccccccccc}
\noalign{\medskip} \hline 
N\Deg & V$_{sys\_Leda}$ & V$_{sys\_FP}$ & i$_{Morph}$ & i$_{Kin}$ & P.A.$_{Morph}$ & P.A.$_{Kin}$ & $\overline{Res}$ & $\sigma_{res}$ & $\chi^{2}_{red}$\\
 UGC & \kms & \kms & \Deg & \Deg & \Deg & \Deg & 10$^{-3}$\kms & \kms &\\
(1)&(2)&(3)&(4)&(5)&(6)&(7)&(8)&(9)&(10)\\
\hline
11470 & 3530$\pm$40 & 3546$\pm$5 & 47$\pm$6 & 47$\pm$7 & 50$\pm$33 & 47$\pm$3 & 258.0 & 25 & 10.7 \\
11496 & 2105$\pm$6 & 2115$\pm$2 & 0$\pm$0 & 44$\pm$16 &  & 167$\pm$4 & 135.8 & 9 & 1.2 \\
11498 & 3266$\pm$8 & 3284$\pm$4 & 79$\pm$4 & 71$\pm$2 & 75$\pm$10 & 71$^\#$$\pm$2 & -23.4 & 22 & 7.9 \\
11557 & 1388$\pm$4 & 1392$\pm$1 & 29$\pm$16 & 29$\pm$22$^*$ &  & 96$^\#$$\pm$3 & 0.1 & 11 & 1.8 \\
11597 & 46$\pm$3 & 40$\pm$2 & 17$\pm$19 & 40$\pm$10 & 60$^{Sp}$$\pm$90 & 61$^\#$$\pm$3 & 0.4 & 13 & 2.5 \\
11670 & 778$\pm$3 & 776$\pm$3 & 90$\pm$0 & 65$\pm$2 & 159$\pm$7 & 153$^\#$$\pm$2 & 0.8 & 16 & 3.9 \\
11707 & 906$\pm$4 & 897$\pm$2 & 65$\pm$4 & 70$\pm$4 & 65$\pm$21 & 59$\pm$3 & 0.3 & 11 & 1.8 \\
11852 & 5850$\pm$8 & 5821$\pm$3 & 47$\pm$5 & 47$\pm$7 & 16$\pm$35 & 9$^\#$$\pm$3 & 4.1 & 20 & 6.8 \\
11861 & 1477$\pm$4 & 1476$\pm$2 & 75$\pm$4 & 43$\pm$12 & 11$\pm$14 & 38$^\#$$\pm$3 & $<$0.1 & 16 & 4.0 \\
11872 & 1147$\pm$5 & 1140$\pm$1 & 54$\pm$6 & 47$\pm$3 & 88$\pm$19 & 86$\pm$2 & -0.1 & 13 & 2.7 \\
11891 & 461$\pm$4 & 466$\pm$6 & 43$\pm$10 & 43$\pm$23$^*$ &  & 119$\pm$10 & 11.1 & 21 & 7.4 \\
11909 & 1106$\pm$5 &  & 90$\pm$0 &  & 1$\pm$6 &  &  &  &  \\
11914 & 951$\pm$2 & 945$\pm$1 & 33$\pm$7 & 33$\pm$4 & 88$\pm$34 & 86$^\#$$\pm$2 & 0.2 & 15 & 3.5 \\
11951 & 1085$\pm$10 & 1085$\pm$2 & 81$\pm$8 & 76$\pm$8 & 88$\pm$14 & 81$^\#$$\pm$4 & 0.2 & 12 & 2.4 \\
12060 & 884$\pm$5 & 879$\pm$2 & 74$\pm$4 & 36$\pm$11 & 147$\pm$18 & 7$^\#$$\pm$3 & -0.4 & 12 & 2.2 \\
12082 & 803$\pm$2 & 792$\pm$2 & 29$\pm$11 & 14$\pm$19 &  & 143$\pm$5 & 0.6 & 8 & 1.1 \\
12101 & 777$\pm$3 & 770$\pm$2 & 58$\pm$5 & 58$\pm$9$^*$ & 133$\pm$18 & 137$^\#$$\pm$4 & 0.5 & 13 & 2.8 \\
12212 & 892$\pm$4 & 899$\pm$2 & 71$\pm$9 & 71$\pm$27$^*$ &  & 82$\pm$7 & 0.9 & 13 & 2.8 \\
12276 & 5664$\pm$4 & 5642$\pm$2 & 38$\pm$6 & 33$\pm$15 & 160$^{Pa}$/55$^{2M}$$\pm$37 & 142$^\#$$\pm$5 & 9.2 & 20 & 6.2 \\
12276c & 5664$\pm$4 &  & 24$\pm$10 &  & 41$^{Pa}$$\pm$90 &  &  &  &  \\
12343 & 2381$\pm$3 & 2371$\pm$2 & 43$\pm$4 & 52$\pm$4 & 25$\pm$19 & 23$^\#$$\pm$2 & -0.3 & 16 & 4.2 \\
12632 & 422$\pm$5 & 415$\pm$2 & 37$\pm$13 & 46$\pm$16$^*$ &  & 47$\pm$4 & -0.4 & 9 & 1.3 \\
12754 & 753$\pm$9 & 742$\pm$2 & 50$\pm$3 & 53$\pm$5 & 163$\pm$18 & 162$^\#$$\pm$2 & 0.2 & 13 & 2.8 \\
\hline
\end{tabular}
\\(1) Name in the UGC catalog (see table \ref{table_calib}). (2) Systemic velocity found in HyperLeda data base. (3) Systemic velocity deduced from our velocity field analysis. (4) Morphological inclination from HyperLeda \citep{Paturel:1997}. (5) Inclination deduced from the analysis of our \VF; those marked with an asterisk ($^*$) have been fixed equal to morphological value from HyperLeda, except UGC 9649, UGC 10359, UGC 10470 for which we used morphological inclinations from NED, and UGC 508, UGC 2023, UGC 2034, UGC 2455, UGC 4499, UGC 6118, UGC 6628, UGC 9013, UGC 9363, UGC 10791 and UGC 12632 for which we used inclinations determined from HI data (see table \ref{tabletf}). (6) Morphological position angle from HyperLeda, except for those marked ($Ha$: \citealp{Haynes:1999}; $Ni$: \citealp{Nilson:1973}; $Pa$: \citealp{Paturel:2000}; $SDSS$: 2006 Sloan Digital Sky Survey, DR5; $Sp$: \citealp{Springob:2007}; $2M$: Two Micron All Sky Survey team 2003, 2MASS extended objects; $Va$: \citealp{Vauglin:1999}). (7) Position angle deduced from our \VF; those marked with an asterisk ($^*$) have been fixed equal to morphological value. The symbol $^\#$ indicates that the position angle refers to the approaching side. (8) Mean residual velocity on the whole \VF. (9) Residual velocity dispersion on the whole \VF. (10) Reduced $\chi^{2}$ of the model.
\end{table*}

\begin{table*}
\caption{Galaxy parameters.}
\begin{tabular}{ccccccccccc}
\noalign{\medskip} \hline 
N\Deg & t & Type & D & M$_{b}$ & b/a & i$_{b/a}$ & D$_{25}$/2 & V$_{max}$ & V$_{max}$& HI data\\
UGC & & & Mpc & mag & & \Deg & "/kpc & \kms & flag & \\
(1)&(2)&(3)&(4)&(5)&(6)&(7)&(8)&(9)&(10)&(11)\\
\hline
12893 & 8.4$\pm$0.8 & Sd & 12.5$^{Ja}$ & -15.5 & 0.89$\pm$0.06 & 27$\pm$7 & 34$\pm$5/2.1$\pm$0.3 & 72$\pm$67 & 2 &  \\
89 & 1.2$\pm$0.6 & SBa & 64.2$^{Mo}$ & -21.5 & 0.79$\pm$0.04 & 38$\pm$3 & 46$\pm$4/14.5$\pm$1.4 & 343$\pm$117 & 1 & W$^{N05}$ \\
94 & 2.4$\pm$0.6 & S(r)ab & 64.2$^{Mo}$ & -20.4 & 0.68$\pm$0.04 & 47$\pm$3 & 34$\pm$3/10.5$\pm$0.8 & 209$\pm$21 & 1 & W$^{N05}$ \\
508 & 1.5$\pm$0.9 & SBab & 63.8 & -21.8 & 0.97$\pm$0.11 & 14$\pm$25 & 85$\pm$7/26.3$\pm$2.2 & 553$\pm$127 & 1 & W$^{N05}$ \\
528 & 2.9$\pm$1.1 & SABb & 12.1 & -19.6 & 0.97$\pm$0.09 & 14$\pm$21 & 72$\pm$5/4.2$\pm$0.3 & 84$\pm$52 & 1 & W$^{Web}$ \\
763 & 8.6$\pm$1.0 & SABm & 12.7$^{Ja}$ & -18.9 & 0.74$\pm$0.04 & 42$\pm$3 & 85$\pm$6/5.2$\pm$0.4 & 104$\pm$11 & 1 &  \\
1013 & 3.1$\pm$0.2 & SB(r)b pec & 70.8 & -22.0 & 0.31$\pm$0.03 & 72$\pm$2 & 88$\pm$6/30.1$\pm$2.0 &  &  & W$^{Web}$ \\
NGC 542 & 2.8$\pm$3.9 & Sb pec & 63.7 & -19.5 & 0.22$\pm$0.03 & 77$\pm$2 & 34$\pm$6/10.4$\pm$1.7 & 125$\pm$8$^{PV}$ & 2 &  \\
1117 & 6.0$\pm$0.4 & Sc & 0.9$^{Sa}$ & -18.9 & 0.58$\pm$0.06 & 54$\pm$4 & 1863$\pm$64/8.1$\pm$0.3$^*$ & 79$\pm$17 & 4 &  \\
1249 & 8.8$\pm$0.6 & SBm pec & 7.2$^{Ka}$ & -18.3 & 0.38$\pm$0.05 & 68$\pm$3 & 193$\pm$11/6.7$\pm$0.4$^*$ & 65$\pm$8$^{PV}$ & 3 & W$^{S02}$ \\
1256 & 6.0$\pm$0.3 & SBc pec & 7.2$^{Ka}$ & -18.9 & 0.39$\pm$0.02 & 67$\pm$1 & 210$\pm$12/7.3$\pm$0.4$^*$ & 105$\pm$9 & 2 & W$^{Web}$ \\
1317 & 4.9$\pm$0.7 & SAB(r)c & 42.2 & -21.5 & 0.33$\pm$0.04 & 71$\pm$2 & 114$\pm$7/23.4$\pm$1.5 & 205$\pm$9 & 1 & W$^{Web}$ \\
1437 & 4.9$\pm$1.0 & SABc & 66.8 & -21.8 & 0.63$\pm$0.03 & 51$\pm$2 & 43$\pm$5/13.8$\pm$1.7 & 218$\pm$15 & 1 & W$^{Web}$ \\
1655 & 1.0$\pm$0.5 & Sa & 73.0 & -21.6 & 0.75$\pm$0.09 & 42$\pm$8 & 86$\pm$10/30.3$\pm$3.5 & 205$\pm$64 & 4 &  \\
1736 & 5.1$\pm$0.6 & SABc & 17.6$^{Ja}$ & -20.1 & 0.69$\pm$0.03 & 46$\pm$3 & 112$\pm$7/9.5$\pm$0.6 & 193$\pm$68 & 3 &  \\
1810 & 3.1$\pm$0.6 & Sb pec & 102.4 & -22.2 & 0.36$\pm$0.02 & 69$\pm$1 & 52$\pm$4/26.0$\pm$2.0 &  &  & W$^{Web}$ \\
1886 & 3.6$\pm$0.6 & SABb & 66.5 & -20.8 & 0.70$\pm$0.06 & 45$\pm$5 & 16$\pm$3/5.1$\pm$0.8 & 267$\pm$8 & 1 & W$^{Web}$ \\
1913 & 7.0$\pm$0.4 & SBcd & 9.3$^{Ka}$ & -20.1 & 0.53$\pm$0.04 & 58$\pm$3 & 319$\pm$13/14.4$\pm$0.6$^*$ & 105$\pm$16 & 3 & W$^{Web}$ \\
2023 & 9.9$\pm$0.6 & I & 7.8$^{Ja}$ & -15.9 & 0.97$\pm$0.11 & 13$\pm$27 & 77$\pm$10/2.9$\pm$0.4 & 53$\pm$49 & 3 & W$^{S02}$ \\
2034 & 9.7$\pm$1.1 & IAB & 10.1 & -16.7 & 0.84$\pm$0.10 & 33$\pm$10 & 90$\pm$11/4.4$\pm$0.5 & 39$\pm$46 & 2 & W$^{S02}$ \\
2045 & 2.0$\pm$0.4 & Sab & 22.7$^{Mo}$ & -20.5 & 0.48$\pm$0.05 & 61$\pm$3 & 112$\pm$10/12.4$\pm$1.1 & 185$\pm$16$^{PV}$ & 3 & W$^{N05}$ \\
2053 & 9.9$\pm$0.4 & IB & 12.7 & -16.3 & 0.55$\pm$0.08 & 56$\pm$6 & 56$\pm$8/3.4$\pm$0.5 &  &  & W$^{S02}$ \\
2082 & 5.9$\pm$0.5 & Sc & 10.2 & -18.3 & 0.19$\pm$0.01 & 79$\pm$1 & 156$\pm$7/7.7$\pm$0.4 & 100$\pm$8$^{PV}$ & 3 & W$^{Web}$ \\
2080 & 6.0$\pm$0.3 & SABc & 13.7 & -19.2 & 0.91$\pm$0.08 & 24$\pm$11 & 127$\pm$11/8.5$\pm$0.7 & 131$\pm$42 & 1 & W$^{Web}$ \\
2141 & 0.4$\pm$1.5 & S0-a & 12.2$^{Ja}$ & -18.1 & 0.50$\pm$0.03 & 60$\pm$2 & 63$\pm$5/3.7$\pm$0.3 & 157$\pm$20$^{PV}$ & 3 & W$^{N05}$ \\
2183 & 1.0$\pm$0.4 & Sa & 18.6$^{Ja}$ & -19.0 & 0.73$\pm$0.11 & 43$\pm$9 & 56$\pm$11/5.1$\pm$1.0 & 160$\pm$32 & 1 & W$^{N05}$ \\
2193 & 5.3$\pm$0.6 & Sc & 9.8$^{Mo}$ & -18.7 & 0.55$\pm$0.04 & 57$\pm$2 & 74$\pm$5/3.5$\pm$0.2 & 174$\pm$423 & 2 & W$^{Web}$ \\
2455 & 9.8$\pm$0.6 & IB & 7.8$^{Ka}$ & -18.6 & 0.78$\pm$0.06 & 38$\pm$5 & 86$\pm$6/3.3$\pm$0.2 & 21$\pm$12 & 4 & W$^{S02}$ \\
2503 & 2.6$\pm$1.0 & SB(r)b & 34.4$^{Ko}$ & -21.6 & 0.59$\pm$0.03 & 54$\pm$2 & 97$\pm$6/16.2$\pm$1.1 & 285$\pm$12 & 1 & W$^{Web}$ \\
2800 & 9.9$\pm$0.7 & I & 20.6$^{Sw}$ &  & 0.50$\pm$0.06 & 60$\pm$4 & 70$\pm$9/7.0$\pm$0.9 & 103$\pm$20 & 2 & W$^{Web}$ \\
2855 & 5.1$\pm$0.6 & SABc & 17.5$^{Ja}$ & -21.4 & 0.41$\pm$0.03 & 65$\pm$2 & 106$\pm$7/9.0$\pm$0.6 & 229$\pm$9 & 1 & W$^{Web}$ \\
3013 & 3.1$\pm$0.5 & SB(r)b & 36.3 & -21.3 & 0.57$\pm$0.03 & 55$\pm$2 & 55$\pm$5/9.6$\pm$0.8 & 212$\pm$21 & 1 & W$^{Web}$ \\
3056 & 9.6$\pm$1.2 & IB & 2.5$^{Oc}$ & -18.7 & 0.57$\pm$0.06 & 55$\pm$4 & 119$\pm$9/1.4$\pm$0.1 &  &  &  \\
3273 & 8.8$\pm$1.2 & SAB(r)m & 12.2$^{Sw}$ & -18.3 & 0.33$\pm$0.04 & 71$\pm$3 & 77$\pm$11/4.6$\pm$0.7 & 106$\pm$7 & 3 & W$^{Web}$ \\
3334 & 4.2$\pm$1.0 & SABb & 55.6 & -22.8 & 0.70$\pm$0.07 & 45$\pm$6 & 132$\pm$8/35.7$\pm$2.2 & 377$\pm$85 & 1 & W$^{Web}$ \\
3382 & 1.0$\pm$0.4 & SB(r)a & 62.8 & -20.4 & 0.94$\pm$0.05 & 20$\pm$9 & 38$\pm$4/11.5$\pm$1.1 & 335$\pm$111 & 2 & W$^{N05}$ \\
3384 & 8.8$\pm$0.5 & Sm & 17.0 & -14.6 & 0.77$\pm$0.06 & 40$\pm$5 & 13$\pm$2/1.0$\pm$0.2 &  &  & W$^{Web}$ \\
3429 & 2.3$\pm$0.7 & SBab pec & 17.9$^{Mo}$ & -21.3 & 0.81$\pm$0.11 & 36$\pm$11 & 159$\pm$8/13.8$\pm$0.7 & 322$\pm$30 & 1 &  \\
3463 & 4.7$\pm$0.9 & SABc & 38.6 & -20.7 & 0.49$\pm$0.03 & 61$\pm$2 & 66$\pm$4/12.4$\pm$0.8 & 168$\pm$9 & 1 &  \\
3574 & 5.9$\pm$0.5 & Sc & 21.8$^{Ja}$ & -18.0 & 0.94$\pm$0.05 & 20$\pm$8 & 44$\pm$4/4.7$\pm$0.4$^*$ & 202$\pm$96 & 1 & W$^{Web}$ \\
3521 & 4.8$\pm$1.8 & Sc & 62.6 & -19.8 & 0.52$\pm$0.03 & 59$\pm$2 & 35$\pm$4/10.7$\pm$1.1 & 166$\pm$12 & 3 &  \\
3528 & 2.0$\pm$0.3 & SBab & 61.8 & -20.1 & 0.58$\pm$0.06 & 55$\pm$4 & 41$\pm$5/12.2$\pm$1.6 & 276$\pm$66 & 2 &  \\
3618 & 2.0$\pm$0.3 & Sab & 80.0 & -20.9 & 0.70$\pm$0.05 & 46$\pm$4 & 44$\pm$4/16.9$\pm$1.4 &  &  &  \\
3691 & 6.1$\pm$1.0 & S(r)c & 28.9$^{Sh}$ & -20.2 & 0.50$\pm$0.06 & 60$\pm$4 & 61$\pm$8/8.5$\pm$1.1 & 143$\pm$10 & 2 &  \\
3685 & 3.0$\pm$0.4 & SB(r)b & 26.3$^{Ja}$ & -19.7 & 0.61$\pm$0.04 & 52$\pm$3 & 57$\pm$4/7.3$\pm$0.5 & 133$\pm$177 & 3 & W$^{Web}$ \\
3708 & 4.5$\pm$1.7 & Sbc pec & 70.0 & -20.7 & 0.96$\pm$0.11 & 15$\pm$23 & 25$\pm$6/8.3$\pm$2.0 & 234$\pm$69 & 1 &  \\
3709 & 5.7$\pm$1.6 & Sc & 70.7 & -21.5 & 0.71$\pm$0.05 & 45$\pm$4 & 35$\pm$3/12.1$\pm$1.1 & 241$\pm$14 & 1 &  \\
3734 & 4.4$\pm$0.9 & SABb & 15.9$^{Ja}$ & -18.6 & 0.90$\pm$0.04 & 25$\pm$6 & 62$\pm$5/4.7$\pm$0.3$^*$ & 108$\pm$16 & 1 & W$^{Web}$ \\
3826 & 6.5$\pm$0.8 & SABc & 25.7$^{Ja}$ & -17.9 & 0.87$\pm$0.07 & 29$\pm$8 & 98$\pm$9/12.2$\pm$1.2 & 74$\pm$66 & 1 & W$^{Web}$ \\
3809 & 4.0$\pm$0.3 & SB(r)bc & 32.9 & -22.0 & 0.56$\pm$0.05 & 56$\pm$3 & 150$\pm$12/24.0$\pm$1.9$^*$ & 258$\pm$9 & 1 &  \\
3740 & 5.4$\pm$0.6 & SAB(r)c pec & 17.1$^{Sh}$ & -19.8 & 0.78$\pm$0.06 & 39$\pm$5 & 67$\pm$5/5.5$\pm$0.4 & 87$\pm$20 & 2 & W$^{Web}$ \\
3851 & 9.8$\pm$0.6 & IB & 3.4$^{Sh}$ & -17.1 & 0.33$\pm$0.02 & 71$\pm$1 & 132$\pm$8/2.2$\pm$0.1$^*$ & 65$\pm$8$^{PV}$ & 3 & W$^{S02}$ \\
3876 & 6.5$\pm$0.8 & Scd & 14.5$^{Ja}$ & -17.4 & 0.50$\pm$0.03 & 60$\pm$2 & 57$\pm$4/4.0$\pm$0.3 & 112$\pm$10 & 2 &  \\
3915 & 4.6$\pm$1.6 & SBc & 63.6 & -21.4 & 0.55$\pm$0.06 & 57$\pm$4 & 34$\pm$6/10.3$\pm$1.7 & 205$\pm$16 & 1 &  \\
IC 476 & 4.2$\pm$2.6 & SABb & 63.9 & -19.0 & 0.78$\pm$0.05 & 39$\pm$5 & 18$\pm$3/5.7$\pm$0.8 & 70$\pm$22 & 3 &  \\
4026 & 2.0$\pm$0.4 & Sab & 64.7 & -20.8 & 0.41$\pm$0.03 & 66$\pm$2 & 43$\pm$4/13.5$\pm$1.2 & 284$\pm$14 & 2 &  \\
4165 & 6.9$\pm$0.4 & SBcd & 11.0$^{Mo}$ & -18.2 & 0.94$\pm$0.05 & 20$\pm$8 & 74$\pm$5/3.9$\pm$0.2 & 80$\pm$18 & 1 & W$^{Web}$ \\
4256 & 5.2$\pm$0.6 & SABc & 71.7 & -21.6 & 0.82$\pm$0.04 & 35$\pm$4 & 50$\pm$4/17.3$\pm$1.3 & 123$\pm$59 & 1 & W$^{Web}$ \\
4273 & 3.1$\pm$0.4 & SBb pec & 35.4$^{Ja}$ & -20.7 & 0.48$\pm$0.05 & 61$\pm$3 & 65$\pm$6/11.2$\pm$1.1 & 219$\pm$11 & 1 & W$^{Web}$ \\
4274 & 8.4$\pm$1.6 & SBd & 6.9$^{Ka}$ & -17.7 & 0.53$\pm$0.05 & 58$\pm$4 & 62$\pm$8/2.1$\pm$0.3 & 102$\pm$57 & 2 & W$^{S02}$ \\
\hline
\label{tabletf}
\end{tabular}
\end{table*}
\begin{table*}
\contcaption{}
\begin{tabular}{ccccccccccc}
\noalign{\medskip} \hline 
N\Deg & t & Type & D & M$_{b}$ & b/a & i$_{b/a}$ & D$_{25}$/2 & V$_{max}$ & V$_{max}$& HI data\\
UGC & & & Mpc & mag & & \Deg & "/kpc & \kms & flag & \\
(1)&(2)&(3)&(4)&(5)&(6)&(7)&(8)&(9)&(10)&(11)\\
\hline
4278 & 6.5$\pm$0.8 & SBc & 9.6 & -19.2 & 0.14$\pm$0.01 & 82$\pm$1 & 87$\pm$6/4.0$\pm$0.3$^*$ & 80$\pm$8$^{PV}$ & 2 & W$^{S02}$ \\
4284 & 6.0$\pm$0.4 & SABc & 9.8 & -18.4 & 0.52$\pm$0.03 & 59$\pm$2 & 91$\pm$6/4.3$\pm$0.3$^*$ & 118$\pm$14 & 3 & W$^{Web}$ \\
4305 & 9.8$\pm$0.8 & I & 3.4$^{Sw}$ & -16.9 & 0.70$\pm$0.05 & 45$\pm$4 & 200$\pm$12/3.3$\pm$0.2$^*$ & 48$\pm$28 & 2 & W$^{S02}$ \\
4325 & 9.0$\pm$0.5 & SABm & 10.9$^{Mo}$ & -18.2 & 0.54$\pm$0.03 & 57$\pm$2 & 89$\pm$6/4.7$\pm$0.3 & 85$\pm$13 & 3 & W$^{S02}$ \\
4393 & 4.6$\pm$1.3 & SBc & 31.5$^{Ja}$ & -19.3 & 0.66$\pm$0.05 & 49$\pm$4 & 44$\pm$7/6.7$\pm$1.1 & 47$\pm$10 & 4 &  \\
4422 & 4.9$\pm$0.6 & SAB(r)c & 58.1 & -21.1 & 0.68$\pm$0.04 & 47$\pm$3 & 51$\pm$5/14.4$\pm$1.4 & 353$\pm$94 & 1 &  \\
4456 & 5.2$\pm$0.6 & S(r)c & 74.0 & -20.8 & 0.89$\pm$0.04 & 27$\pm$5 & 31$\pm$3/11.0$\pm$1.1 & 211$\pm$321 & 1 &  \\
4499 & 7.9$\pm$0.8 & SBd & 12.2$^{Ja}$ & -17.0 & 0.45$\pm$0.03 & 63$\pm$2 & 57$\pm$5/3.4$\pm$0.3 & 62$\pm$13 & 2 & W$^{S02}$ \\
4543 & 8.0$\pm$0.5 & Sd & 30.3$^{Sw}$ & -17.6 & 0.77$\pm$0.05 & 40$\pm$5 & 37$\pm$4/5.4$\pm$0.5 & 70$\pm$15 & 2 & W$^{S02}$ \\
4555 & 4.0$\pm$0.6 & SABb & 58.0 & -20.9 & 0.94$\pm$0.07 & 20$\pm$11 & 45$\pm$5/12.7$\pm$1.4 & 185$\pm$30 & 1 &  \\
4770 & 1.1$\pm$0.6 & SBa & 95.9 & -21.3 & 0.83$\pm$0.07 & 34$\pm$8 & 48$\pm$5/22.3$\pm$2.3 & 330$\pm$194 & 3 &  \\
4820 & 1.7$\pm$0.8 & S(r)ab & 17.1$^{Sh}$ & -20.3 & 0.79$\pm$0.04 & 38$\pm$4 & 127$\pm$6/10.6$\pm$0.5 & 336$\pm$20 & 1 &  \\
4936 & 6.9$\pm$0.3 & SAB(r)c & 25.6$^{Ko}$ & -20.6 & 0.81$\pm$0.06 & 35$\pm$6 & 102$\pm$7/12.6$\pm$0.8 & 264$\pm$227 & 2 &  \\
5045 & 5.0$\pm$0.5 & SAB(r)c & 105.1 & -21.2 & 0.76$\pm$0.04 & 40$\pm$4 & 35$\pm$3/18.0$\pm$1.5 & 429$\pm$228 & 1 &  \\
5175 & 3.2$\pm$0.7 & Sb & 44.1 & -20.6 & 0.48$\pm$0.05 & 61$\pm$3 & 62$\pm$5/13.2$\pm$1.1 & 188$\pm$10 & 1 &  \\
5228 & 4.9$\pm$0.5 & SBc & 24.7 & -19.9 & 0.25$\pm$0.02 & 76$\pm$1 & 68$\pm$5/8.2$\pm$0.6 & 125$\pm$9 & 1 &  \\
5251 & 4.3$\pm$0.8 & SBbc pec & 21.5 & -20.5 & 0.22$\pm$0.01 & 77$\pm$1 & 142$\pm$7/14.8$\pm$0.7 & 125$\pm$9 & 3 & W$^{Web}$ \\
5253 & 2.3$\pm$0.9 & Sab & 21.1 & -20.7 & 0.81$\pm$0.05 & 36$\pm$5 & 108$\pm$7/11.1$\pm$0.7 & 235$\pm$17 & 1 & W$^{N05}$ \\
5272 & 9.8$\pm$0.6 & IB & 7.1$^{Ka}$ & -16.1 & 0.23$\pm$0.02 & 76$\pm$1 & 48$\pm$4/1.7$\pm$0.2 & 45$\pm$8$^{PV}$ & 3 & W$^{S02}$ \\
5279 & 9.7$\pm$1.1 & IB & 21.3 & -19.0 & 0.27$\pm$0.02 & 74$\pm$1 & 67$\pm$5/6.9$\pm$0.5 & 110$\pm$8$^{PV}$ & 1 &  \\
5316 & 6.5$\pm$0.8 & SBc & 17.6 & -19.9 & 0.38$\pm$0.03 & 68$\pm$2 & 100$\pm$7/8.6$\pm$0.6 & 145$\pm$9 & 3 & W$^{Web}$ \\
5319 & 5.3$\pm$0.6 & SB(r)c & 35.8 & -19.7 & 0.77$\pm$0.05 & 39$\pm$4 & 47$\pm$3/8.2$\pm$0.6 & 180$\pm$47 & 2 &  \\
5351 & 2.1$\pm$0.6 & SABa & 19.3$^{Sh}$ & -19.4 & 0.32$\pm$0.04 & 71$\pm$2 & 62$\pm$3/5.8$\pm$0.3 & 135$\pm$8$^{PV}$ & 1 & W$^{N05}$ \\
5373 & 9.9$\pm$0.3 & IB & 1.4$^{Ka}$ & -14.3 & 0.62$\pm$0.05 & 52$\pm$4 & 148$\pm$10/1.0$\pm$0.1 & 90$\pm$162 & 2 &  \\
5398 & 7.9$\pm$3.8 & Sd & 3.8$^{Ka}$ & -17.8 & 0.81$\pm$0.08 & 36$\pm$8 & 162$\pm$13/3.0$\pm$0.2 &  &  &  \\
5414 & 9.9$\pm$0.2 & IAB(r) & 10.0$^{Sw}$ & -16.6 & 0.68$\pm$0.08 & 47$\pm$6 & 90$\pm$10/4.3$\pm$0.5 & 74$\pm$10 & 2 & W$^{S02}$ \\
IC 2542 & 4.6$\pm$1.3 & SBc & 83.4 & -20.5 & 0.75$\pm$0.04 & 42$\pm$4 & 31$\pm$3/12.4$\pm$1.2 & 290$\pm$192 & 2 &  \\
5510 & 4.6$\pm$1.0 & SAB(r)c & 18.6 & -19.3 & 0.80$\pm$0.05 & 37$\pm$5 & 64$\pm$6/5.8$\pm$0.5 & 167$\pm$44 & 1 &  \\
5532 & 3.9$\pm$0.6 & Sbc & 41.1 & -22.1 & 0.85$\pm$0.08 & 32$\pm$8 & 122$\pm$10/24.2$\pm$1.9 & 398$\pm$24 & 1 & W$^{Web}$ \\
5556 & 5.0$\pm$0.8 & SBc pec & 22.2 & -18.9 & 0.32$\pm$0.02 & 71$\pm$1 & 67$\pm$4/7.2$\pm$0.4 &  &  & W$^{Web}$ \\
5721 & 6.6$\pm$0.8 & SBcd & 6.5$^{Ka}$ & -16.5 & 0.50$\pm$0.03 & 60$\pm$2 & 49$\pm$4/1.5$\pm$0.1 & 99$\pm$29 & 3 & W$^{S02}$ \\
5786 & 4.0$\pm$0.1 & SAB(r)b & 14.2$^{Sh}$ & -19.6 & 0.96$\pm$0.11 & 15$\pm$23 & 54$\pm$6/3.7$\pm$0.4 & 80$\pm$15 & 3 & W$^{Web}$ \\
5789 & 5.9$\pm$0.4 & SBc & 14.1$^{Sa}$ & -19.6 & 0.49$\pm$0.04 & 61$\pm$2 & 109$\pm$7/7.4$\pm$0.5$^*$ & 131$\pm$10 & 3 & W$^{Web}$ \\
5829 & 9.7$\pm$0.9 & IB & 9.0$^{Sw}$ & -16.2 & 0.92$\pm$0.07 & 22$\pm$11 & 135$\pm$9/5.9$\pm$0.4$^*$ & 48$\pm$26 & 2 & W$^{S02}$ \\
5840 & 4.0$\pm$0.3 & SB(r)bc & 6.9$^{Ka}$ & -18.9 & 0.95$\pm$0.07 & 17$\pm$13 & 200$\pm$10/6.7$\pm$0.3$^*$ & 251$\pm$138 & 1 & W$^{Web}$ \\
5842 & 6.0$\pm$0.4 & SBc & 15.2$^{Sh}$ & -18.8 & 0.83$\pm$0.05 & 34$\pm$5 & 79$\pm$5/5.8$\pm$0.4 & 115$\pm$18 & 2 &  \\
5931 & 5.9$\pm$0.5 & SBc pec & 21.2$^{Sh}$ & -19.8 & 0.55$\pm$0.05 & 56$\pm$4 & 52$\pm$6/5.3$\pm$0.6 & 157$\pm$32 & 3 &  \\
5935 & 9.4$\pm$1.5 & SBm pec & 26.4$^{Sw}$ & -20.5 & 0.41$\pm$0.04 & 66$\pm$2 & 93$\pm$11/11.9$\pm$1.4 &  & 4 & W$^{S02}$ \\
5982 & 5.1$\pm$0.7 & SBc & 20.8$^{Sh}$ & -20.0 & 0.56$\pm$0.05 & 56$\pm$4 & 120$\pm$10/12.1$\pm$1.0 & 199$\pm$13 & 1 & W$^{Web}$ \\
6118 & 2.1$\pm$0.6 & SB(r)ab & 19.8$^{Sh}$ & -20.0 & 0.90$\pm$0.05 & 26$\pm$7 & 74$\pm$5/7.1$\pm$0.4 & 137$\pm$24 & 1 & W$^{N05}$ \\
6277 & 5.1$\pm$0.5 & SABc & 16.9 & -19.5 & 0.96$\pm$0.07 & 17$\pm$15 & 106$\pm$9/8.7$\pm$0.7 & 270$\pm$258 & 2 & V$^{K00}$ \\
6419 & 8.9$\pm$0.9 & SBm & 18.8 & -18.6 & 0.65$\pm$0.04 & 50$\pm$3 & 44$\pm$4/4.0$\pm$0.4 & 53$\pm$11 & 3 & V$^{W04}$ \\
6521 & 3.7$\pm$0.9 & S(r)bc & 78.6 & -21.2 & 0.67$\pm$0.03 & 48$\pm$2 & 50$\pm$3/19.1$\pm$1.2 & 249$\pm$18 & 1 &  \\
6523 & 1.4$\pm$1.1 & Sa & 80.0 & -21.0 & 0.92$\pm$0.04 & 23$\pm$7 & 32$\pm$3/12.5$\pm$1.3 & 118$\pm$63 & 4 &  \\
6537 & 5.1$\pm$0.5 & SB(r)c & 14.3$^{Tu}$ & -20.5 & 0.67$\pm$0.05 & 48$\pm$4 & 158$\pm$9/11.0$\pm$0.6$^*$ & 187$\pm$17 & 1 & W$^{Web}$ \\
6628 & 8.8$\pm$1.0 & SAB(r)m & 15.3$^{Sw}$ & -17.9 & 0.87$\pm$0.07 & 30$\pm$8 & 65$\pm$6/4.8$\pm$0.5 & 183$\pm$168 & 3 & W$^{S02}$ \\
6702 & 1.3$\pm$1.0 & Sa & 99.8 & -20.6 & 0.86$\pm$0.05 & 31$\pm$6 & 24$\pm$3/11.5$\pm$1.5 & 195$\pm$23 & 1 &  \\
6778 & 5.1$\pm$0.8 & SABc pec & 15.5$^{Sh}$ & -20.7 & 0.53$\pm$0.03 & 58$\pm$2 & 81$\pm$6/6.1$\pm$0.4 & 223$\pm$14 & 2 & W$^{Web}$ \\
6787 & 1.7$\pm$0.8 & Sab & 18.9 & -20.5 & 0.60$\pm$0.03 & 53$\pm$2 & 104$\pm$6/9.5$\pm$0.5 & 232$\pm$10 & 2 & W$^{N05}$ \\
7021 & 1.3$\pm$0.8 & SAB(r)a & 26.8 & -19.7 & 0.62$\pm$0.04 & 52$\pm$3 & 77$\pm$5/10.0$\pm$0.6 & 223$\pm$18 & 1 &  \\
7045 & 5.3$\pm$0.6 & SABc & 11.4$^{Mo}$ & -19.2 & 0.39$\pm$0.03 & 67$\pm$2 & 124$\pm$6/6.9$\pm$0.3 & 160$\pm$9 & 1 &  \\
7154 & 6.9$\pm$0.4 & SBcd & 16.2 & -20.0 & 0.46$\pm$0.03 & 63$\pm$2 & 139$\pm$9/10.9$\pm$0.7 & 145$\pm$9 & 1 & W$^{Web}$ \\
7278 & 9.8$\pm$0.5 & IB & 2.9$^{Sh}$ & -17.4 & 0.78$\pm$0.08 & 39$\pm$8 & 204$\pm$12/2.9$\pm$0.2 &  &  & W$^{S02}$ \\
7323 & 8.0$\pm$0.4 & SBd & 8.1$^{Sw}$ & -18.3 & 0.70$\pm$0.04 & 45$\pm$3 & 114$\pm$7/4.5$\pm$0.3 & 84$\pm$15 & 2 & W$^{S02}$ \\
7429 & 2.4$\pm$0.7 & SB(r)ab & 23.7 & -19.8 & 0.40$\pm$0.03 & 67$\pm$2 & 73$\pm$5/8.4$\pm$0.6 &  &  &  \\
7524 & 8.8$\pm$0.7 & SB(r)m & 4.6$^{Ko}$ & -18.4 & 0.33$\pm$0.02 & 71$\pm$1 & 125$\pm$8/2.8$\pm$0.2$^*$ & 45$\pm$8$^{PV}$ & 3 & W$^{S02}$ \\
7592 & 9.8$\pm$0.5 & IB & 2.9$^{Sh}$ & -18.3 & 0.58$\pm$0.04 & 54$\pm$3 & 140$\pm$9/2.0$\pm$0.1 &  &  & W$^{S02}$ \\
7699 & 6.0$\pm$0.6 & SBc & 9.3 & -17.6 & 0.28$\pm$0.01 & 74$\pm$1 & 108$\pm$6/4.9$\pm$0.3 & 92$\pm$8$^{PV}$ & 1 &  \\
7766 & 6.0$\pm$0.4 & SBc & 13.0 & -21.0 & 0.46$\pm$0.05 & 63$\pm$3 & 317$\pm$15/20.0$\pm$1.0$^*$ & 120$\pm$9 & 1 & W$^{Web}$ \\
7831 & 4.9$\pm$0.4 & SBc & 5.2$^{Ka}$ & -18.5 & 0.39$\pm$0.03 & 67$\pm$2 & 177$\pm$8/4.5$\pm$0.2 & 92$\pm$15 & 2 & W$^{Web}$ \\
\hline
\end{tabular}
\end{table*}
\begin{table*}
\contcaption{}
\begin{tabular}{ccccccccccc}
\noalign{\medskip} \hline 
N\Deg & t & Type & D & M$_{b}$ & b/a & i$_{b/a}$ & D$_{25}$/2 & V$_{max}$ & V$_{max}$& HI data\\
UGC & & & Mpc & mag & & \Deg & "/kpc & \kms & flag & \\
(1)&(2)&(3)&(4)&(5)&(6)&(7)&(8)&(9)&(10)&(11)\\
\hline
7853 & 8.6$\pm$1.1 & SBm & 8.9$^{Mo}$ & -18.9 & 0.64$\pm$0.05 & 50$\pm$3 & 106$\pm$6/4.6$\pm$0.3 & 110$\pm$35 & 3 & W$^{Web}$ \\
7861 & 8.8$\pm$0.7 & SAB(r)m pec & 10.2$^{Mo}$ & -17.3 & 0.75$\pm$0.05 & 41$\pm$4 & 41$\pm$4/2.0$\pm$0.2 & 50$\pm$21 & 3 & W$^{Web}$ \\
7876 & 6.5$\pm$0.9 & SABc & 14.5 & -17.9 & 0.72$\pm$0.05 & 44$\pm$4 & 58$\pm$7/4.1$\pm$0.5 & 98$\pm$14 & 2 &  \\
7901 & 5.2$\pm$0.6 & Sc pec & 20.7$^{Sh}$ & -20.6 & 0.66$\pm$0.03 & 49$\pm$3 & 115$\pm$6/11.6$\pm$0.6 & 215$\pm$10 & 1 &  \\
7971 & 8.7$\pm$0.5 & Sm & 8.4$^{Sw}$ & -16.3 & 0.92$\pm$0.11 & 24$\pm$15 & 66$\pm$8/2.7$\pm$0.3 & 33$\pm$27 & 3 & W$^{S02}$ \\
7985 & 6.9$\pm$0.5 & SBcd & 13.7$^{Mo}$ & -18.7 & 0.92$\pm$0.08 & 23$\pm$11 & 51$\pm$5/3.4$\pm$0.3 & 112$\pm$13 & 1 &  \\
8334 & 4.0$\pm$0.2 & Sbc & 9.8 & -21.1 & 0.61$\pm$0.06 & 53$\pm$4 & 356$\pm$20/16.9$\pm$1.0$^*$ & 214$\pm$9 & 1 &  \\
8403 & 5.8$\pm$0.6 & SBc & 19.1$^{Ja}$ & -19.2 & 0.61$\pm$0.04 & 52$\pm$3 & 90$\pm$6/8.3$\pm$0.6 & 128$\pm$10 & 1 & W$^{Web}$ \\
8490 & 8.9$\pm$0.4 & Sm & 4.7$^{Sh}$ & -17.1 & 0.64$\pm$0.07 & 51$\pm$5 & 135$\pm$8/3.1$\pm$0.2$^*$ & 90$\pm$29 & 2 & W$^{S02}$ \\
NGC 5296 & -1.1$\pm$0.8 & S0-a & 32.8 & -18.2 & 0.58$\pm$0.04 & 54$\pm$3 & 28$\pm$3/4.5$\pm$0.5 & 80$\pm$9 & 3 & W$^{Web}$ \\
8709 & 4.9$\pm$0.8 & SABc pec & 35.0 & -21.4 & 0.24$\pm$0.02 & 76$\pm$1 & 112$\pm$7/19.0$\pm$1.1 & 207$\pm$9 & 1 & W$^{Web}$ \\
8852 & 2.3$\pm$0.6 & SAB(r)a & 30.6 & -20.0 & 0.62$\pm$0.08 & 52$\pm$6 & 77$\pm$9/11.4$\pm$1.3 & 186$\pm$10 & 3 &  \\
8863 & 1.1$\pm$0.4 & SBa & 25.5$^{Ko}$ & -20.3 & 0.39$\pm$0.03 & 67$\pm$2 & 108$\pm$5/13.4$\pm$0.6 & 191$\pm$13 & 2 & W$^{N05}$ \\
8898 & 3.1$\pm$0.6 & SBb pec & 49.0 & -20.5 & 0.41$\pm$0.03 & 66$\pm$2 & 79$\pm$6/18.7$\pm$1.3 & 65$\pm$45 & 4 & W$^{Web}$ \\
8900 & 3.2$\pm$0.6 & Sb pec & 49.2 & -21.7 & 0.47$\pm$0.06 & 62$\pm$4 & 75$\pm$8/17.8$\pm$1.8 & 345$\pm$37 & 2 & W$^{Web}$ \\
8937 & 3.1$\pm$0.4 & SBb & 49.0$^{Mo}$ & -21.1 & 0.67$\pm$0.06 & 48$\pm$4 & 69$\pm$6/16.4$\pm$1.4 & 320$\pm$105 & 1 &  \\
9013 & 6.0$\pm$0.3 & Sc pec & 7.2$^{Ka}$ & -18.2 & 0.66$\pm$0.04 & 49$\pm$3 & 72$\pm$5/2.5$\pm$0.2$^*$ & 62$\pm$45 & 2 & V$^{R94}$ \\
9179 & 6.9$\pm$0.4 & SABc & 5.7$^{Ka}$ & -17.8 & 0.61$\pm$0.03 & 52$\pm$2 & 128$\pm$8/3.5$\pm$0.2 & 111$\pm$36 & 3 &  \\
9219 & 9.7$\pm$1.4 & IB & 10.2$^{Ja}$ & -16.6 & 0.44$\pm$0.03 & 64$\pm$2 & 49$\pm$4/2.4$\pm$0.2 & 45$\pm$8$^{PV}$ & 2 &  \\
9248 & 3.1$\pm$0.5 & Sb & 54.9 & -20.2 & 0.57$\pm$0.03 & 55$\pm$2 & 40$\pm$4/10.6$\pm$1.0 & 166$\pm$11 & 1 &  \\
9358 & 3.3$\pm$0.8 & SABb & 29.1 & -20.8 & 0.52$\pm$0.03 & 59$\pm$2 & 94$\pm$6/13.3$\pm$0.9 & 221$\pm$14 & 1 &  \\
9366 & 4.7$\pm$0.9 & Sc & 37.7$^{Mo}$ & -21.7 & 0.45$\pm$0.03 & 63$\pm$2 & 107$\pm$6/19.6$\pm$1.2 & 241$\pm$9 & 1 & W$^{Web}$ \\
9363 & 6.9$\pm$0.4 & S(r)cd & 22.3 & -19.8 & 0.84$\pm$0.05 & 33$\pm$6 & 57$\pm$5/6.2$\pm$0.5$^*$ & 143$\pm$105 & 1 & V$^{S96}$ \\
9406 & 6.9$\pm$0.4 & SB(r)cd & 33.8 & -19.0 & 0.64$\pm$0.08 & 50$\pm$6 & 44$\pm$8/7.3$\pm$1.3 & 19$\pm$10 & 4 &  \\
9465 & 7.9$\pm$0.9 & SABd & 26.4$^{Ja}$ & -18.0 & 0.40$\pm$0.03 & 67$\pm$2 & 23$\pm$3/3.0$\pm$0.4 & 97$\pm$9 & 1 &  \\
9576 & 6.9$\pm$0.4 & SABc pec & 27.4$^{Ja}$ & -19.6 & 0.63$\pm$0.05 & 51$\pm$4 & 51$\pm$4/6.8$\pm$0.6 & 104$\pm$25 & 1 & V$^{I94}$ \\
9649 & 3.1$\pm$0.7 & SBb & 7.7$^{Ja}$ & -16.5 & 0.42$\pm$0.03 & 65$\pm$2 & 56$\pm$5/2.1$\pm$0.2 & 94$\pm$11 & 2 & W$^{Web}$ \\
9736 & 5.0$\pm$0.7 & SABc & 45.4 & -20.6 & 0.65$\pm$0.03 & 49$\pm$2 & 71$\pm$4/15.7$\pm$1.0 & 192$\pm$16 & 1 &  \\
9753 & 3.5$\pm$0.7 & Sbc & 12.4$^{Sh}$ & -19.1 & 0.36$\pm$0.04 & 69$\pm$2 & 114$\pm$7/6.8$\pm$0.4 & 138$\pm$9 & 1 & W$^{Web}$ \\
9858 & 4.0$\pm$0.5 & SABb & 38.2 & -20.4 & 0.20$\pm$0.02 & 79$\pm$1 & 118$\pm$12/21.8$\pm$2.3 & 160$\pm$9 & 2 & W$^{Web}$ \\
9866 & 4.0$\pm$0.3 & S(r)bc & 7.4$^{Ja}$ & -17.2 & 0.41$\pm$0.03 & 65$\pm$2 & 55$\pm$4/2.0$\pm$0.1 & 114$\pm$11 & 2 &  \\
9943 & 5.0$\pm$0.6 & SB(r)c & 28.0 & -20.7 & 0.69$\pm$0.05 & 46$\pm$4 & 82$\pm$5/11.1$\pm$0.7 & 185$\pm$10 & 1 &  \\
9969 & 3.1$\pm$0.3 & SB(r)b & 36.0$^{Sh}$ & -21.4 & 0.50$\pm$0.03 & 60$\pm$2 & 119$\pm$7/20.8$\pm$1.3 & 311$\pm$9 & 1 & W$^{Web}$ \\
9992 & 9.8$\pm$0.6 & I & 10.4 & -15.3 & 0.61$\pm$0.05 & 52$\pm$4 & 44$\pm$6/2.2$\pm$0.3 &  &  & W$^{S02}$ \\
10075 & 6.0$\pm$0.4 & Sc & 14.7$^{Ja}$ & -19.9 & 0.44$\pm$0.04 & 64$\pm$3 & 174$\pm$9/12.4$\pm$0.7 & 168$\pm$9 & 1 &  \\
10310 & 9.2$\pm$0.6 & SBm & 12.7$^{Sw}$ & -17.1 & 0.80$\pm$0.09 & 37$\pm$9 & 77$\pm$10/4.7$\pm$0.6 & 66$\pm$27 & 3 & W$^{S02}$ \\
10359 & 5.6$\pm$0.6 & SBc pec & 16.0 & -19.0 & 0.85$\pm$0.07 & 32$\pm$8 & 63$\pm$5/4.9$\pm$0.4$^*$ & 143$\pm$30 & 1 & W$^{Web}$ \\
10470 & 4.0$\pm$0.2 & SB(r)bc & 21.2$^{Ja}$ & -20.2 & 0.73$\pm$0.05 & 43$\pm$4 & 67$\pm$5/6.9$\pm$0.5 & 164$\pm$39 & 1 & W$^{Web}$ \\
10445 & 6.0$\pm$0.9 & SB(r)c pec & 16.9$^{Ja}$ & -17.6 & 0.71$\pm$0.04 & 45$\pm$3 & 57$\pm$4/4.7$\pm$0.3 & 77$\pm$17 & 2 & W$^{Web}$ \\
10502 & 5.3$\pm$0.6 & Sc & 61.2 & -21.2 & 0.78$\pm$0.05 & 39$\pm$4 & 60$\pm$5/17.9$\pm$1.5 & 163$\pm$14 & 1 & W$^{Web}$ \\
10521 & 4.9$\pm$0.7 & Sc & 18.0$^{Mo}$ & -20.2 & 0.38$\pm$0.03 & 68$\pm$2 & 106$\pm$9/9.3$\pm$0.7 & 124$\pm$9 & 1 &  \\
10546 & 6.0$\pm$0.5 & SABc & 20.4$^{Ja}$ & -19.1 & 0.59$\pm$0.04 & 54$\pm$3 & 55$\pm$5/5.4$\pm$0.4 & 106$\pm$22 & 1 & W$^{Web}$ \\
10564 & 6.5$\pm$0.8 & SBc & 18.4$^{Ja}$ & -17.6 & 0.62$\pm$0.05 & 52$\pm$3 & 37$\pm$4/3.3$\pm$0.3 & 75$\pm$8 & 2 & W$^{Web}$ \\
10652 & 3.8$\pm$2.6 & S(r)bc & 18.2 & -17.7 & 0.87$\pm$0.05 & 29$\pm$6 & 33$\pm$3/2.9$\pm$0.3 & 141$\pm$82 & 2 &  \\
10713 & 3.0$\pm$0.4 & Sb & 18.3 & -19.0 & 0.19$\pm$0.02 & 79$\pm$1 & 54$\pm$7/4.8$\pm$0.7 & 105$\pm$8$^{PV}$ & 2 & W$^{Web}$ \\
10757 & 6.0$\pm$0.4 & Sc & 19.5 & -17.7 & 0.53$\pm$0.04 & 58$\pm$2 & 36$\pm$4/3.4$\pm$0.4 & 81$\pm$33 & 3 & W$^{Web}$ \\
10769 & 3.0$\pm$0.5 & SABb & 20.0 & -17.0 & 0.59$\pm$0.04 & 54$\pm$3 & 28$\pm$3/2.7$\pm$0.3 &  &  & W$^{Web}$ \\
10791 & 8.8$\pm$0.5 & SABm & 21.7 & -16.7 & 1.00$\pm$0.13 & 0$\pm$0 & 56$\pm$9/5.9$\pm$0.9 & 95$\pm$48 & 3 & W$^{Web}$ \\
10897 & 5.2$\pm$0.5 & SABc & 20.5$^{Ja}$ & -19.5 & 0.86$\pm$0.05 & 30$\pm$6 & 64$\pm$4/6.3$\pm$0.4 & 113$\pm$56 & 2 & W$^{Web}$ \\
11012 & 5.9$\pm$0.7 & Sc & 5.3$^{Ka}$ & -18.7 & 0.33$\pm$0.03 & 71$\pm$2 & 185$\pm$11/4.7$\pm$0.3 & 117$\pm$9 & 1 &  \\
11124 & 5.9$\pm$0.5 & SBc & 23.7$^{Ja}$ & -18.6 & 0.93$\pm$0.10 & 22$\pm$16 & 67$\pm$10/7.7$\pm$1.1 & 96$\pm$15 & 2 & W$^{Web}$ \\
11218 & 5.2$\pm$0.6 & Sc & 22.8$^{Ko}$ & -20.8 & 0.48$\pm$0.03 & 61$\pm$2 & 100$\pm$5/11.0$\pm$0.6 & 185$\pm$9 & 1 & W$^{Web}$ \\
11269 & 2.0$\pm$0.5 & SABa & 35.0$^{Ja}$ & -19.9 & 0.56$\pm$0.04 & 56$\pm$3 & 56$\pm$5/9.5$\pm$0.8 & 202$\pm$13 & 1 & W$^{N05}$ \\
11283 & 7.8$\pm$0.9 & SBd & 31.3$^{Sw}$ & -19.3 & 0.86$\pm$0.05 & 31$\pm$6 & 40$\pm$4/6.1$\pm$0.7 & 173$\pm$73 & 2 & W$^{Web}$ \\
11283c & 5.4$\pm$2.2 & Sc & 31.3$^{Sw}$ & -16.5 & 0.41$\pm$0.03 & 66$\pm$2 & 28$\pm$3/4.2$\pm$0.5 &  &  & W$^{Web}$ \\
11300 & 6.4$\pm$0.9 & SABc & 8.4$^{Ja}$ & -17.8 & 0.28$\pm$0.01 & 74$\pm$1 & 99$\pm$5/4.0$\pm$0.2$^*$ & 112$\pm$9 & 2 & W$^{Web}$ \\
11332 & 7.0$\pm$0.5 & SBcd & 23.0$^{Ja}$ & -19.5 & 0.21$\pm$0.01 & 78$\pm$1 & 63$\pm$5/7.1$\pm$0.5 & 91$\pm$8$^{PV}$ & 3 &  \\
11407 & 3.6$\pm$0.6 & SBbc & 35.8 & -20.8 & 0.49$\pm$0.03 & 61$\pm$2 & 75$\pm$6/13.0$\pm$1.0 & 158$\pm$30 & 1 & V$^{W01}$ \\
11429 & 3.1$\pm$0.5 & SBb & 65.2 & -21.8 & 0.53$\pm$0.03 & 58$\pm$2 & 63$\pm$4/19.8$\pm$1.3 & 232$\pm$35 & 2 & W$^{Web}$ \\
11466 & 4.8$\pm$1.9 & Sc & 14.2 & -18.5 & 0.59$\pm$0.03 & 54$\pm$2 & 45$\pm$4/3.1$\pm$0.3 & 133$\pm$10 & 1 & W$^{Web}$ \\
\hline
\end{tabular}
\end{table*}
\begin{table*}
\contcaption{}
\begin{tabular}{ccccccccccc}
\noalign{\medskip} \hline 
N\Deg & t & Type & D & M$_{b}$ & b/a & i$_{b/a}$ & D$_{25}$/2 & V$_{max}$ & V$_{max}$& HI data\\
UGC & & & Mpc & mag & & \Deg & "/kpc & \kms & flag & \\
(1)&(2)&(3)&(4)&(5)&(6)&(7)&(8)&(9)&(10)&(11)\\
\hline
11470 & 2.2$\pm$0.6 & Sab & 50.8 & -21.3 & 0.71$\pm$0.07 & 45$\pm$5 & 72$\pm$9/17.8$\pm$2.2 & 380$\pm$40 & 2 &  \\
11496 & 8.8$\pm$0.5 & Sm & 31.9 &  & 1.00$\pm$0.13 & 0$\pm$0 & 57$\pm$9/8.9$\pm$1.3 & 96$\pm$29 & 2 & W$^{Web}$ \\
11498 & 3.1$\pm$0.7 & SBb & 44.9 & -20.5 & 0.32$\pm$0.04 & 71$\pm$2 & 84$\pm$9/18.2$\pm$1.9 & 274$\pm$9 & 1 &  \\
11557 & 7.8$\pm$0.9 & SABd & 19.7$^{Sw}$ & -18.4 & 0.90$\pm$0.11 & 26$\pm$14 & 59$\pm$8/5.7$\pm$0.8 & 105$\pm$72 & 2 & W$^{S02}$ \\
11597 & 5.9$\pm$0.3 & SABc & 5.9$^{Ka}$ & -20.6 & 0.96$\pm$0.09 & 16$\pm$18 & 342$\pm$14/9.8$\pm$0.4$^*$ & 154$\pm$32 & 3 &  \\
11670 & 0.5$\pm$1.0 & S(r)a & 12.8 & -19.4 & 0.33$\pm$0.03 & 71$\pm$2 & 125$\pm$7/7.7$\pm$0.4 & 190$\pm$9 & 1 & W$^{N05}$ \\
11707 & 7.9$\pm$1.0 & Sd & 15.9$^{Sw}$ & -16.6 & 0.57$\pm$0.03 & 55$\pm$2 & 31$\pm$3/2.4$\pm$0.3 & 97$\pm$5 & 2 & W$^{S02}$ \\
11852 & 1.0$\pm$0.5 & SBa & 80.1 & -20.2 & 0.72$\pm$0.05 & 44$\pm$4 & 28$\pm$3/10.8$\pm$1.3 & 221$\pm$27 & 2 & W$^{N05}$ \\
11861 & 7.8$\pm$0.9 & SABd & 25.1$^{Sw}$ & -20.2 & 0.48$\pm$0.03 & 61$\pm$2 & 54$\pm$4/6.5$\pm$0.5 & 181$\pm$39 & 2 & W$^{S02}$ \\
11872 & 2.5$\pm$0.5 & SAB(r)b & 18.1$^{Ko}$ & -20.0 & 0.63$\pm$0.07 & 51$\pm$5 & 85$\pm$6/7.4$\pm$0.5 & 183$\pm$12 & 1 &  \\
11891 & 9.9$\pm$0.5 & I & 9.0 & -16.4 & 0.78$\pm$0.08 & 39$\pm$8 & 100$\pm$15/4.3$\pm$0.6 & 83$\pm$35 & 3 & W$^{Web}$ \\
11909 & 4.5$\pm$1.6 & Sbc & 17.7 & -19.3 & 0.20$\pm$0.01 & 78$\pm$1 & 61$\pm$6/5.2$\pm$0.5 & 110$\pm$8$^{PV}$ & 2 & W$^{Web}$ \\
11914 & 2.5$\pm$0.7 & S(r)ab & 15.0$^{Ko}$ & -20.3 & 0.86$\pm$0.06 & 31$\pm$6 & 135$\pm$7/9.8$\pm$0.5 & 285$\pm$26 & 1 & W$^{N05}$ \\
11951 & 1.1$\pm$0.8 & SBa & 17.4 & -19.3 & 0.36$\pm$0.04 & 69$\pm$3 & 51$\pm$8/4.3$\pm$0.7 & 106$\pm$7 & 3 & W$^{N05}$ \\
12060 & 9.9$\pm$0.5 & IB & 15.7$^{Sw}$ & -16.5 & 0.49$\pm$0.03 & 61$\pm$2 & 33$\pm$4/2.5$\pm$0.3 & 107$\pm$27 & 1 & W$^{S02}$ \\
12082 & 8.7$\pm$0.8 & SABm & 10.1$^{Ja}$ & -16.4 & 0.90$\pm$0.07 & 26$\pm$10 & 81$\pm$9/3.9$\pm$0.4 & 105$\pm$137 & 3 & W$^{Web}$ \\
12101 & 6.6$\pm$0.9 & Scd & 15.1$^{Sh}$ & -18.5 & 0.55$\pm$0.06 & 57$\pm$4 & 77$\pm$7/5.6$\pm$0.5 & 94$\pm$12 & 3 & V$^{W02}$ \\
12212 & 8.7$\pm$0.5 & Sm & 15.5$^{Sw}$ &  & 0.51$\pm$0.07 & 59$\pm$5 & 59$\pm$10/4.4$\pm$0.7 & 78$\pm$15 & 4 & W$^{Web}$ \\
12276 & 1.1$\pm$0.5 & SB(r)a & 77.8 & -20.7 & 0.81$\pm$0.06 & 36$\pm$5 & 34$\pm$3/12.8$\pm$1.1 & 94$\pm$37 & 2 & W$^{N05}$ \\
12276c & 5.1$\pm$5.0 & S? & 77.8 & -17.4 & 0.92$\pm$0.06 & 23$\pm$9 & 9$\pm$2/3.3$\pm$0.7 &  &  & W$^{N05}$ \\
12343 & 4.4$\pm$0.9 & SBbc & 26.9$^{Ja}$ & -21.1 & 0.75$\pm$0.03 & 42$\pm$3 & 109$\pm$5/14.2$\pm$0.6 & 221$\pm$14 & 1 & V$^{L98}$ \\
12632 & 8.7$\pm$0.5 & SABm & 8.0 & -17.4 & 0.83$\pm$0.11 & 34$\pm$11 & 127$\pm$14/4.9$\pm$0.6 & 69$\pm$19 & 2 & W$^{S02}$ \\
12754 & 6.0$\pm$0.4 & SBc & 8.9$^{Sw}$ & -18.6 & 0.65$\pm$0.03 & 49$\pm$2 & 109$\pm$6/4.7$\pm$0.3$^*$ & 123$\pm$11 & 1 & W$^{Web}$ \\
\hline
\end{tabular}
\\(1) Name of the galaxy in the UGC catalog (see table \ref{table_calib}). (2) Morphological type from the de Vaucouleurs classification \citep{de-Vaucouleurs:1979} in HyperLeda data base. (3) Morphological type from HyperLeda data base. (4) Distance D, deduced from the systemic velocity taken in NED corrected from Virgo infall, assuming H$_{o}$ = 75 \kmsMpc, except for those marked ($Ja$: \citealp{James:2004}; $Ka$: \citealp{Karachentsev:2004}; $Ko$: \citealp{Koopmann:2006}; $Mo$: \citealp{Moustakas:2006}; $Oc$: \citealp{OConnell:1994}; $Sa$: \citealp{Saha:2006}; $Sh$: \citealp{Shapley:2001}; $Tu$: \citealp{Tully:1996}). (5) Absolute B magnitude from D and apparent corrected B magnitude (HyperLeda). (6) Axis ratio from HyperLeda. (7) Inclination derived from the axis ratio ($\arccos{b/a}$). (8) Isophotal radius at the limiting surface brightness of 25 B mag/sq arcsec, from HyperLeda \citep{Paturel:1991} in arcsecond and kpc adopting the distance given in column 4; an asterisk ($^*$) indicates that the galaxy is larger than GHASP field of view. (9) Maximum velocity, V$_{max}$, derived from the fit of the velocity field or from the \PVM~(marked with $^{PV}$). (10) Quality flag on V$_{max}$ (1: reached; 2: probably reached; 3 probably not reached; 4: not reached). (11) Aperture synthesis HI data references: W for WHISP data ($S02$: \citealp{Swaters:2002}; $N05$: \citealp{Noordermeer:2005}; $web$: \url{http://www.astro.rug.nl/~whisp}); V for VLA data ($I94$: \citealp{Irwin:1994}; $R94$: \citealp{Rownd:1994}; $S96$: \citealp{Schulman:1996}; $L98$: \citealp{Laine:1998}; $K00$: \citealp{Kornreich:2000}; $W01$: \citealp{Wilcots:2001}; $W02$: \citealp{Williams:2002}; $W04$: \citealp{Wilcots:2004}).
%(8) Explanation about the reason why a
%galaxy has not been taken into account in the analysis: cut when
%the galaxy has been cut by the field of view or by a filter;
%peculiar when kinematics of the galaxy is very disturbed (e.g.
%counter-rotation motions).
\end{table*}

\clearpage
\section{\Ha~profiles}
\clearpage
\label{profiles}
\begin{figure}
\begin{minipage}{180mm}
\begin{center}
\includegraphics[width=3.5cm]{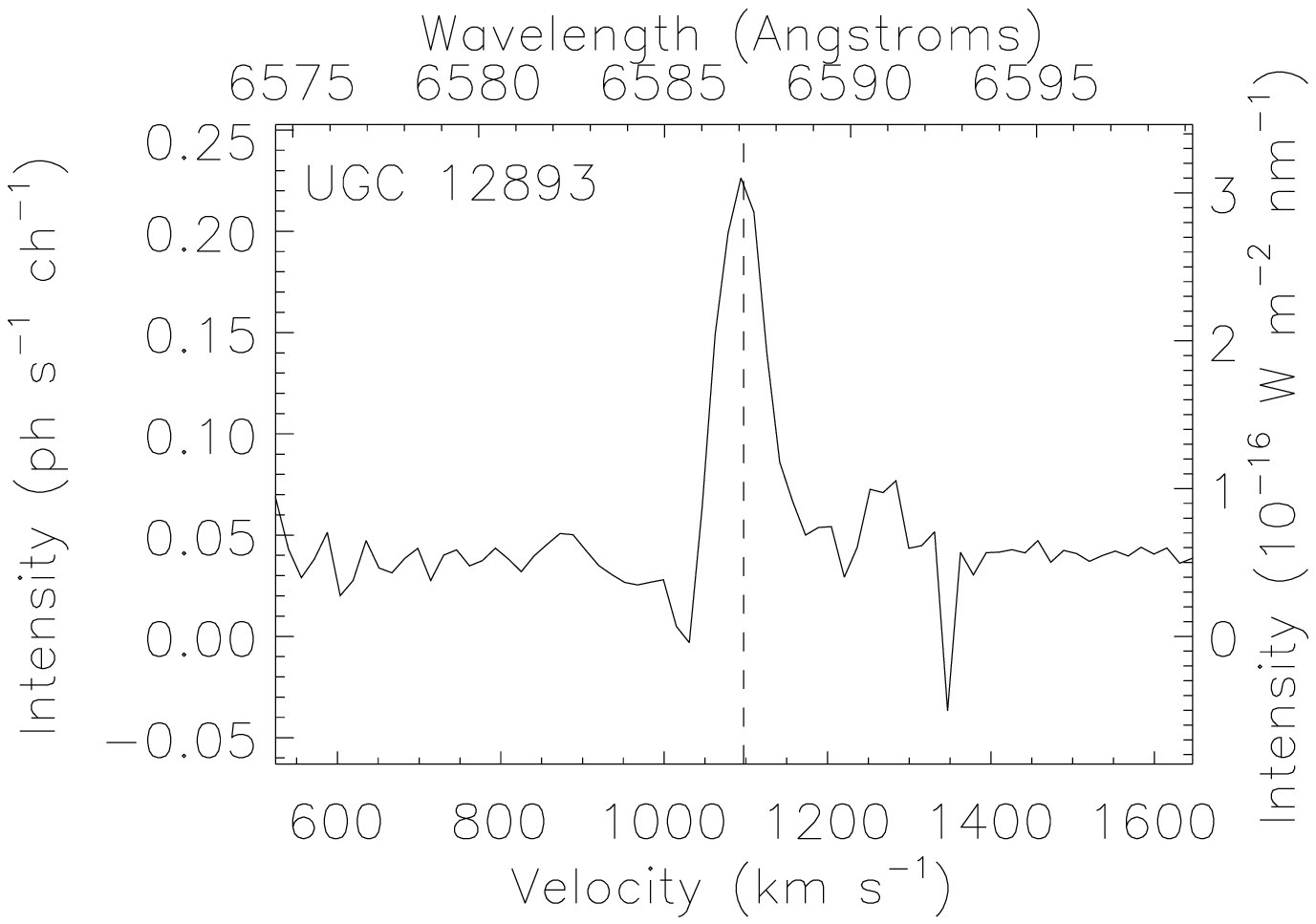}
\includegraphics[width=3.5cm]{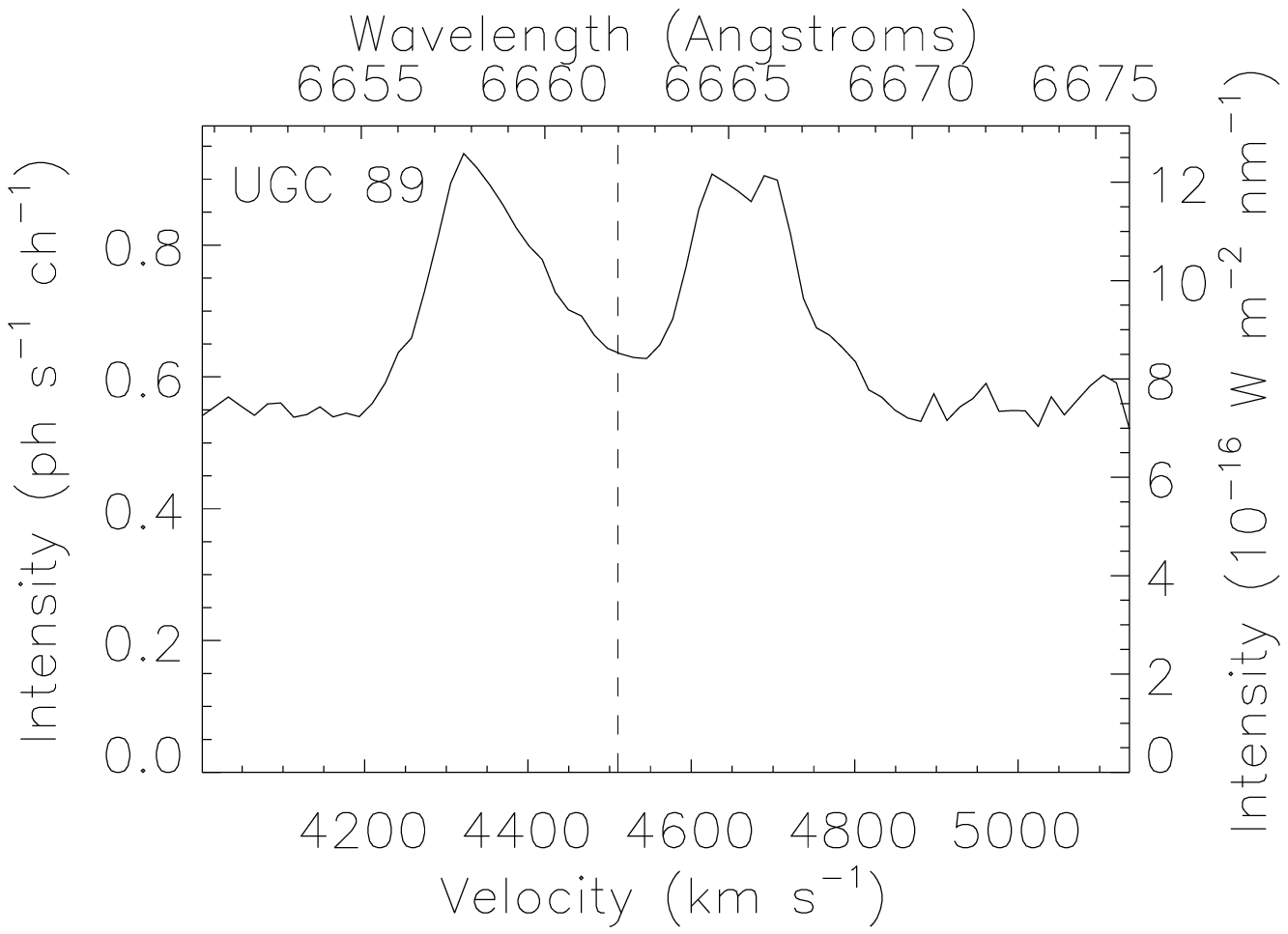}
\includegraphics[width=3.5cm]{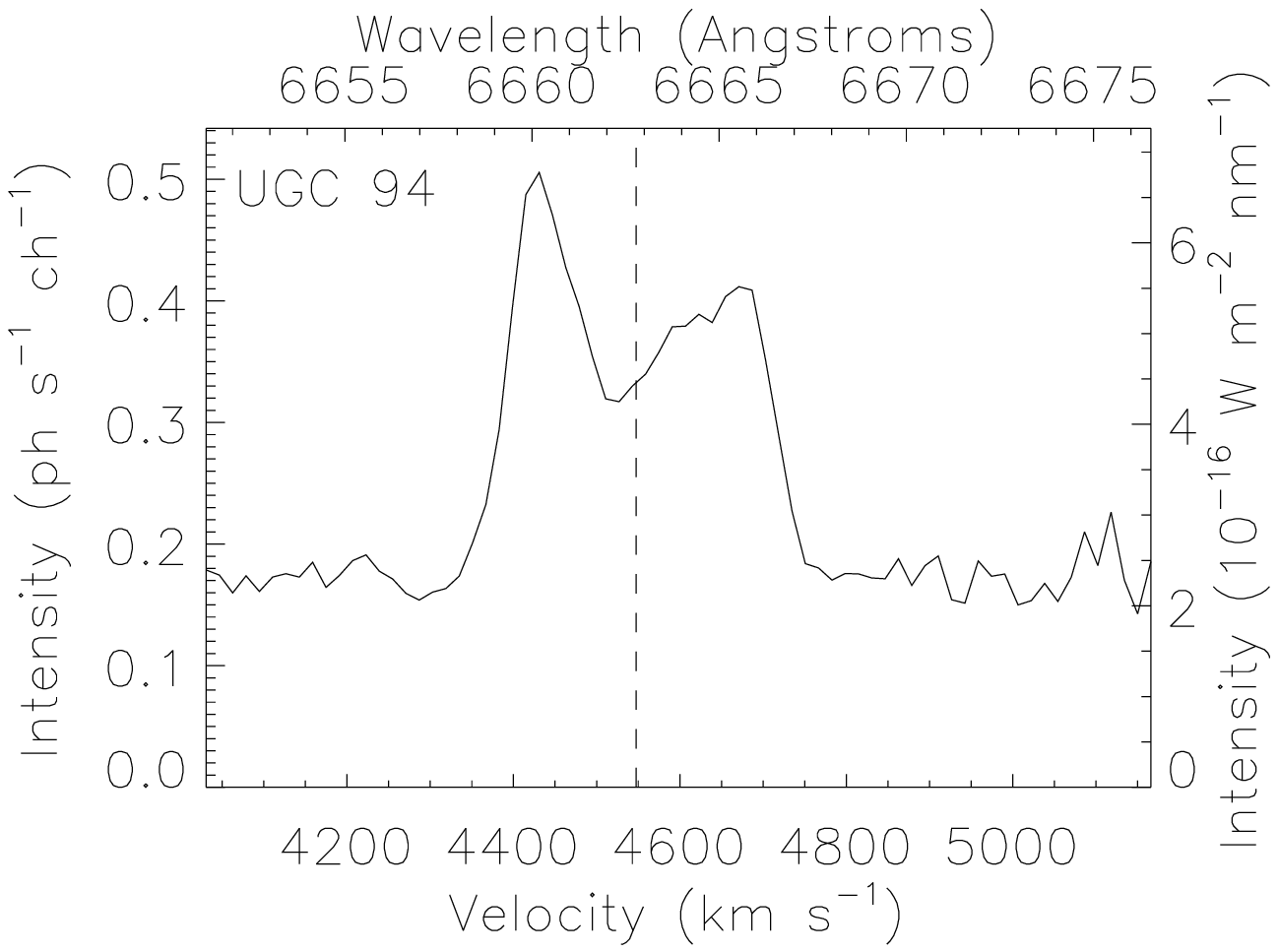}
\includegraphics[width=3.5cm]{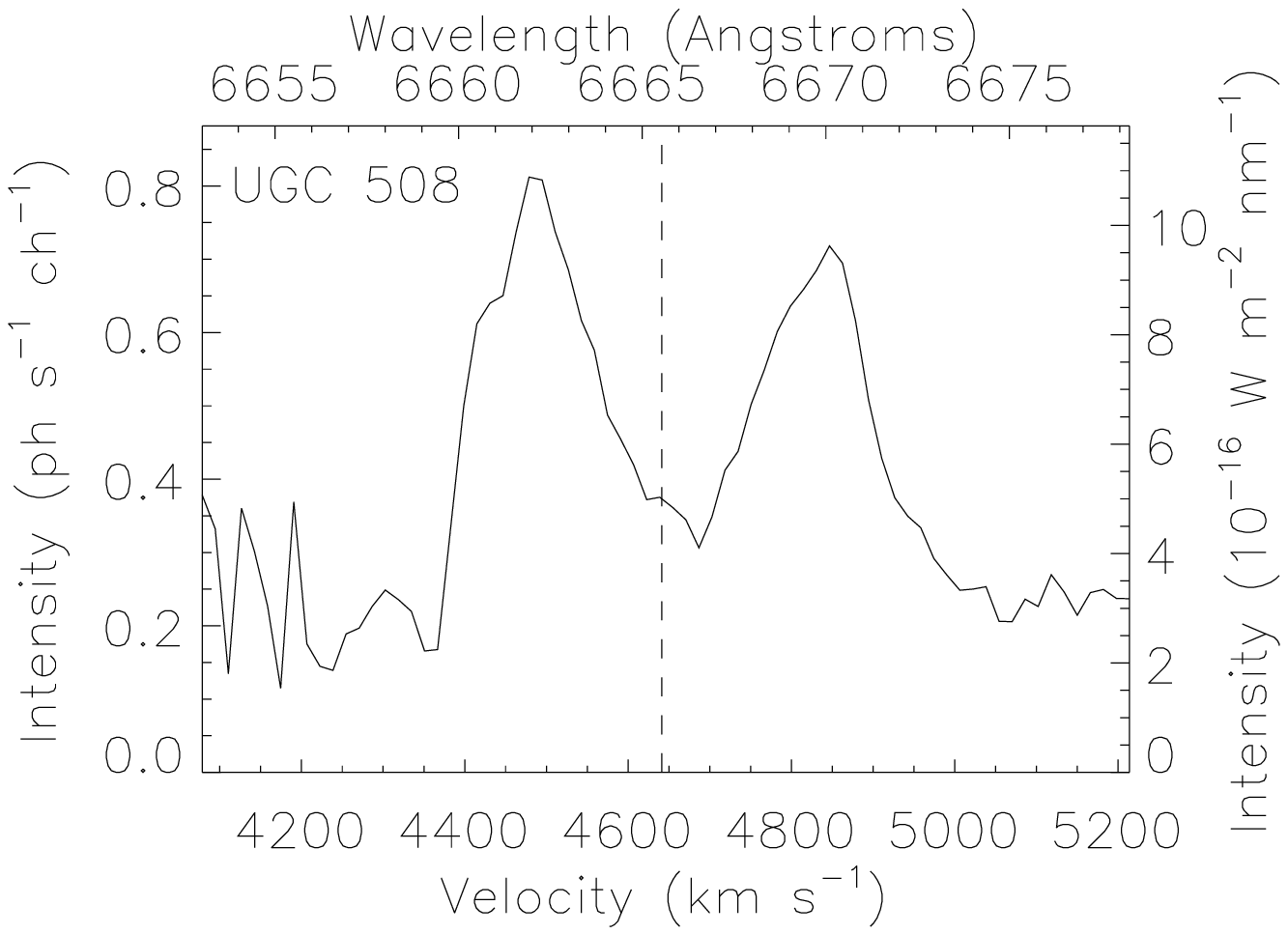}
\includegraphics[width=3.5cm]{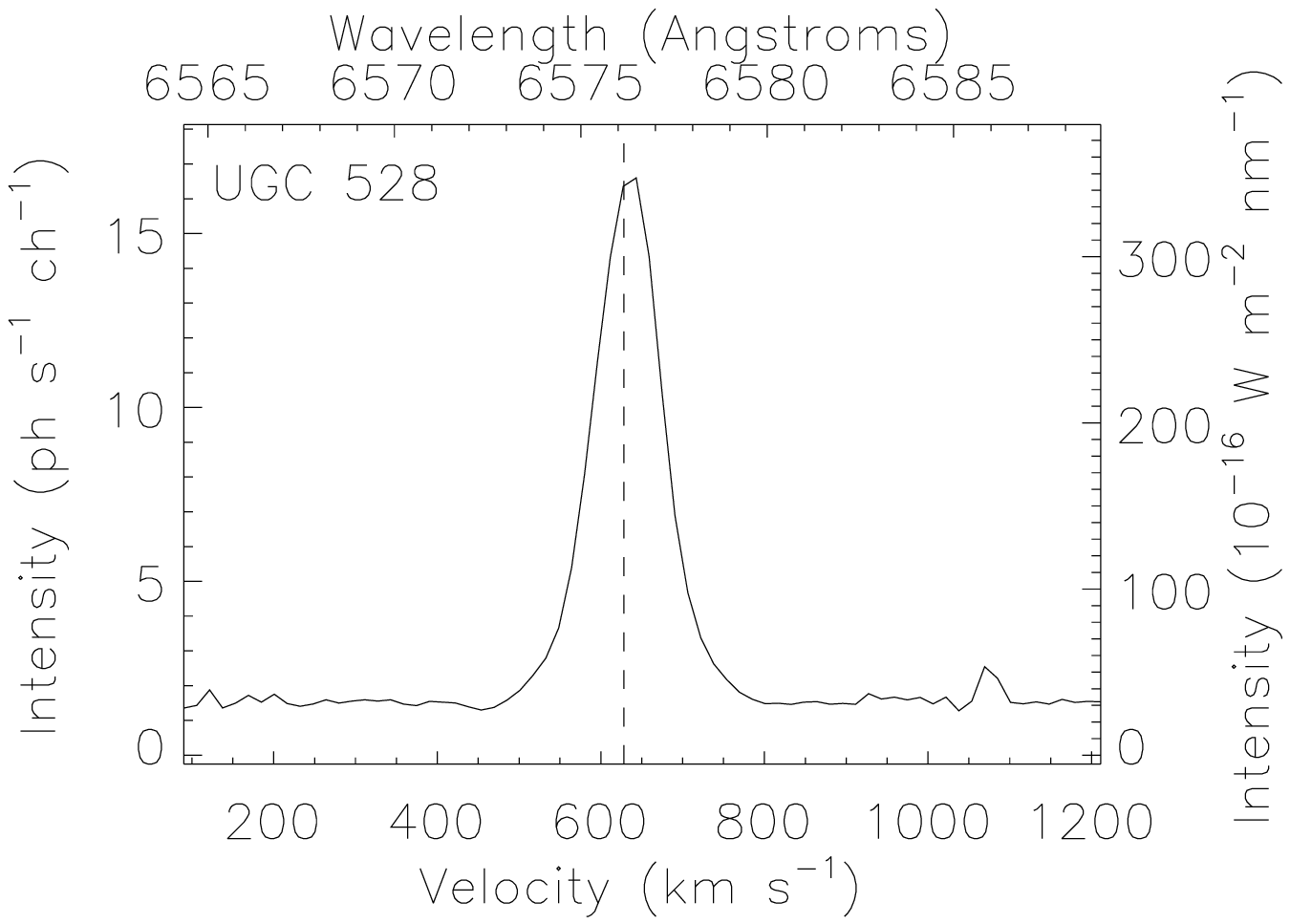}
\includegraphics[width=3.5cm]{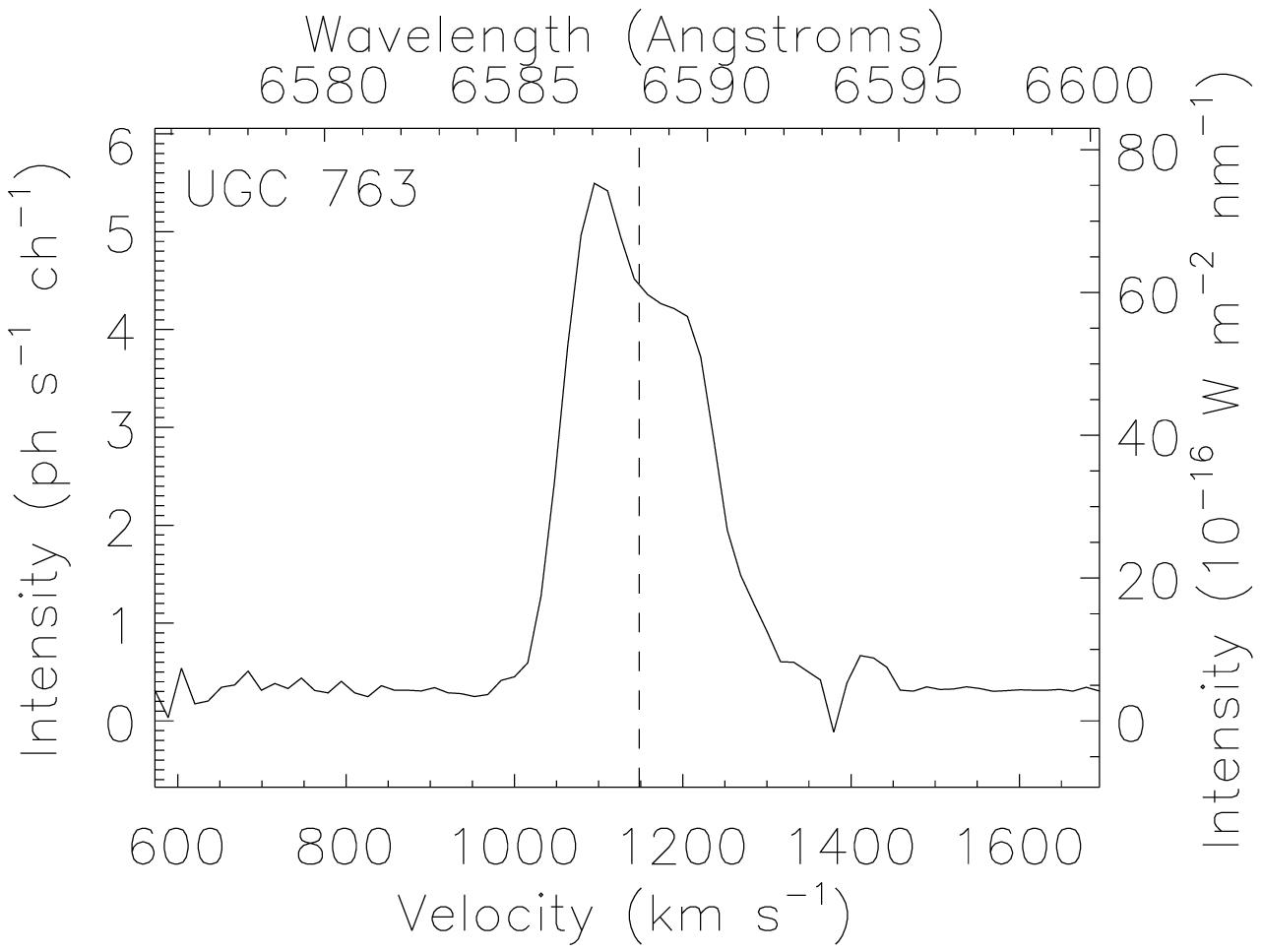}
\includegraphics[width=3.5cm]{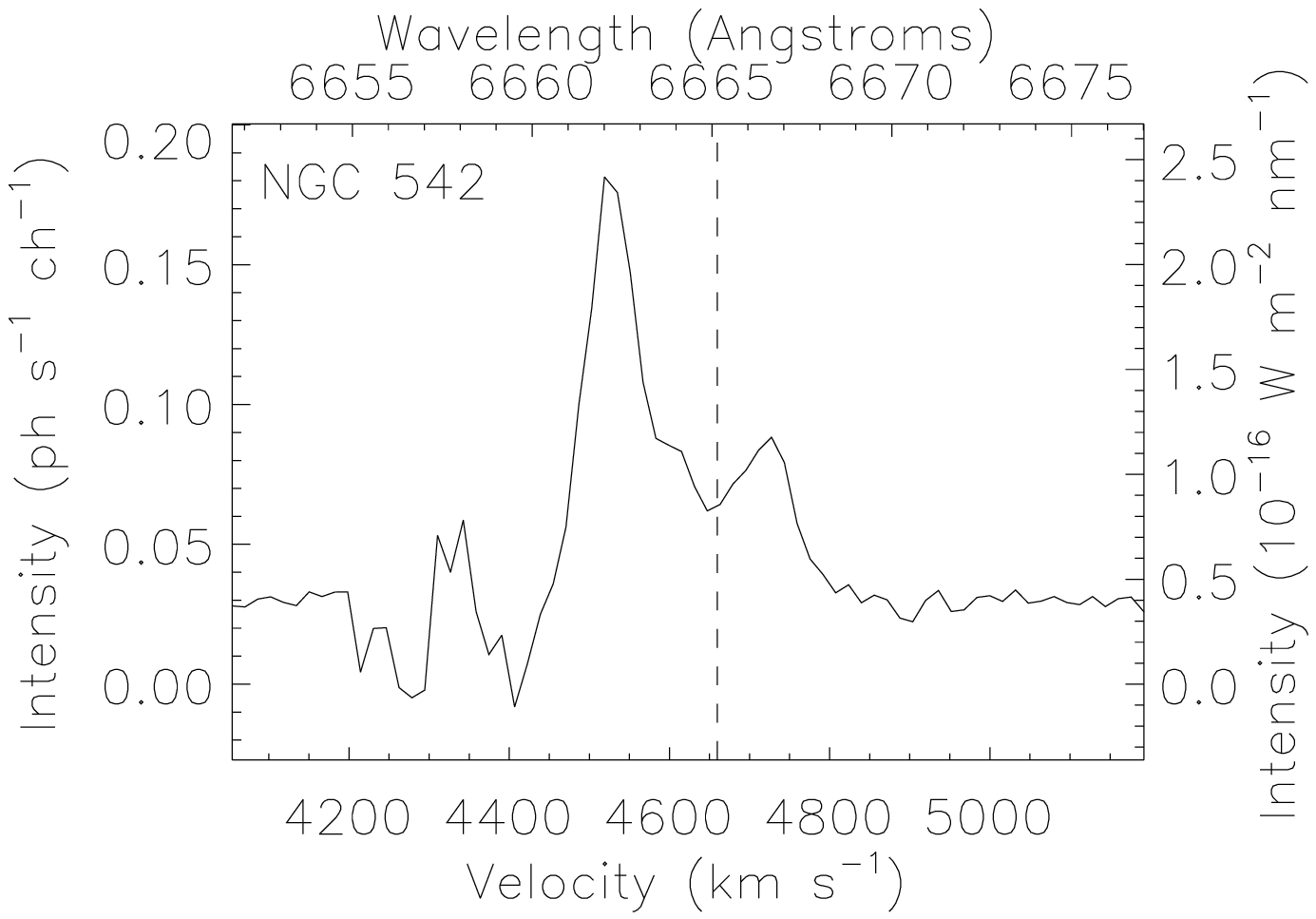}
\includegraphics[width=3.5cm]{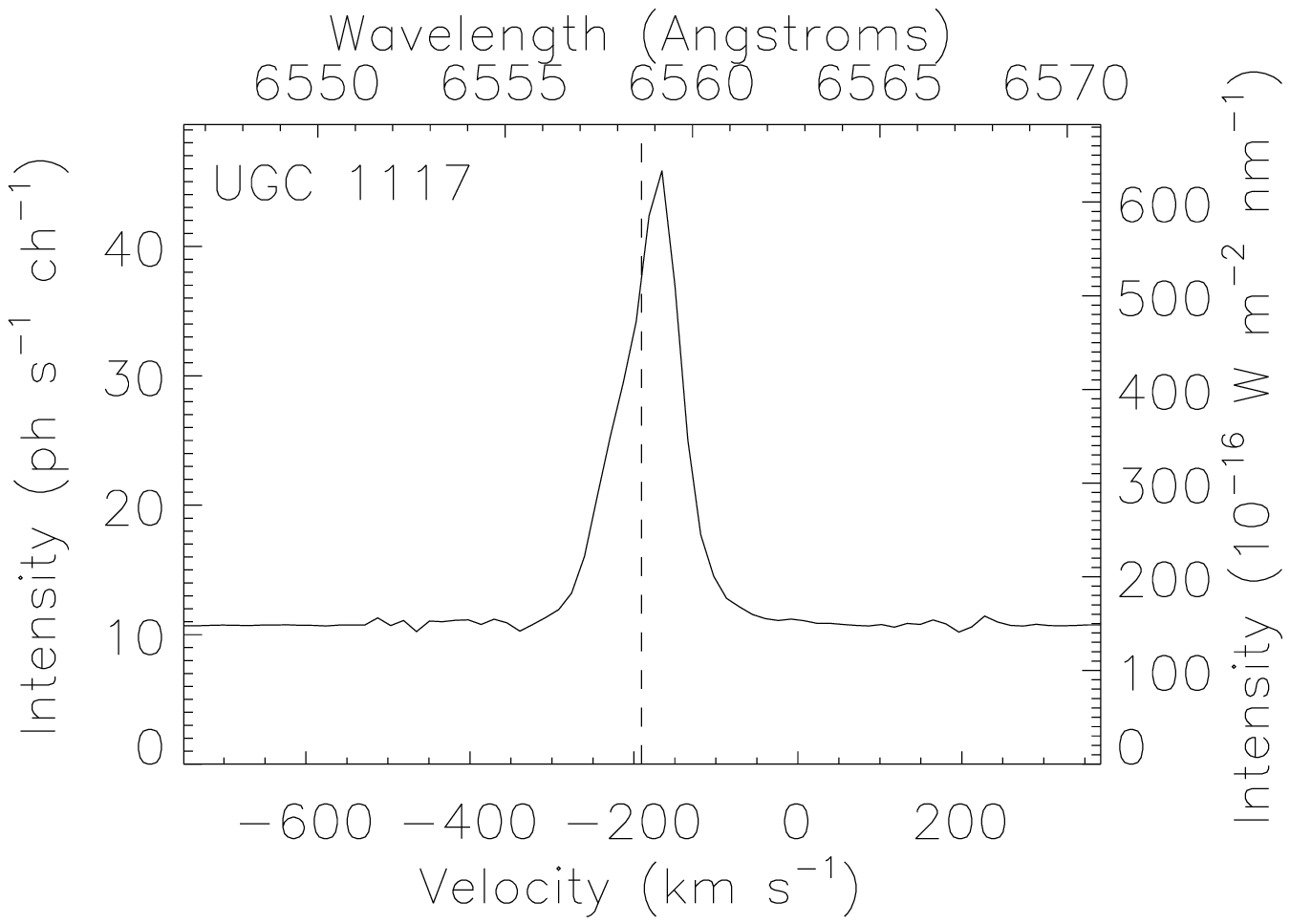}
\includegraphics[width=3.5cm]{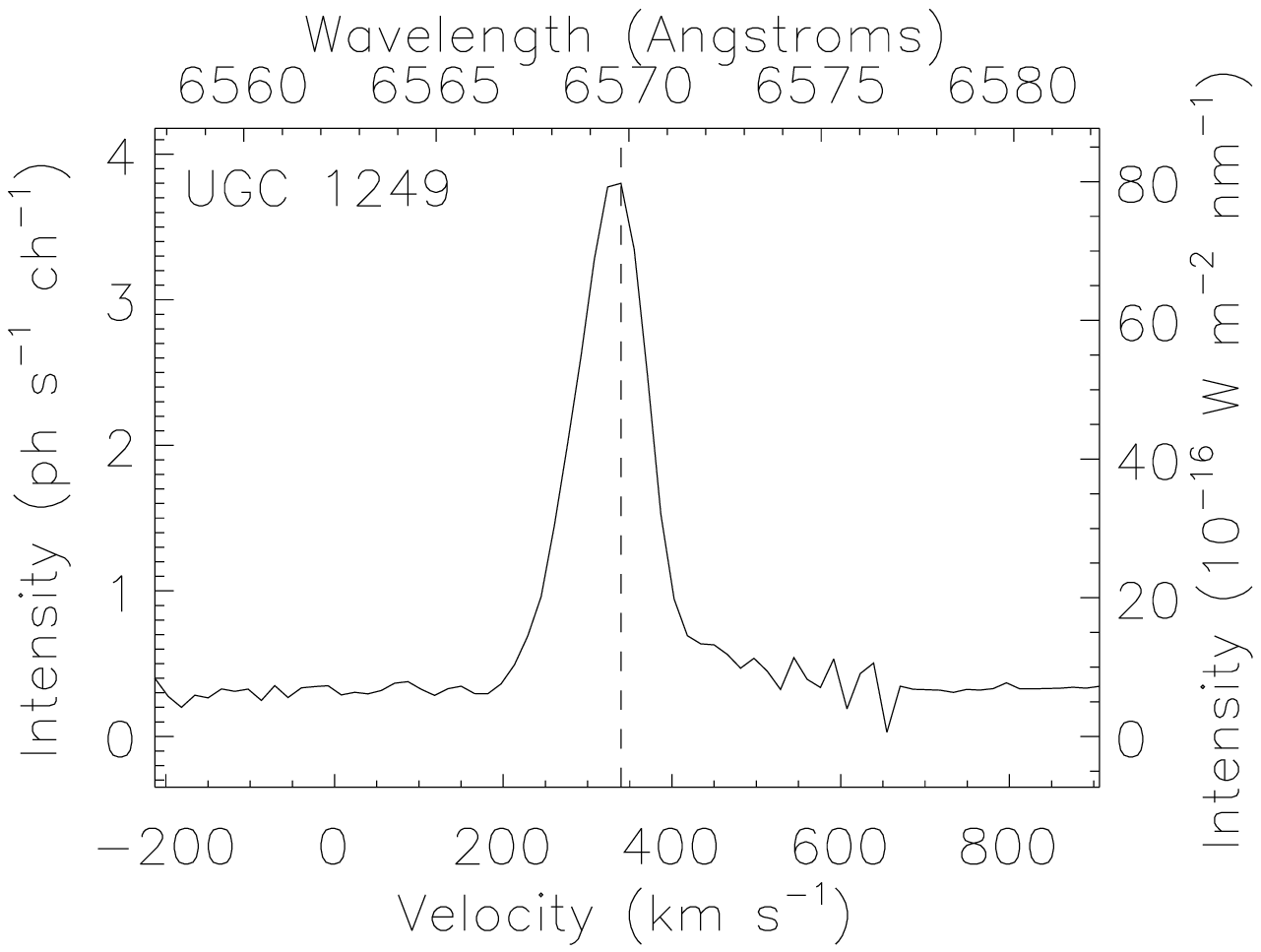}
\includegraphics[width=3.5cm]{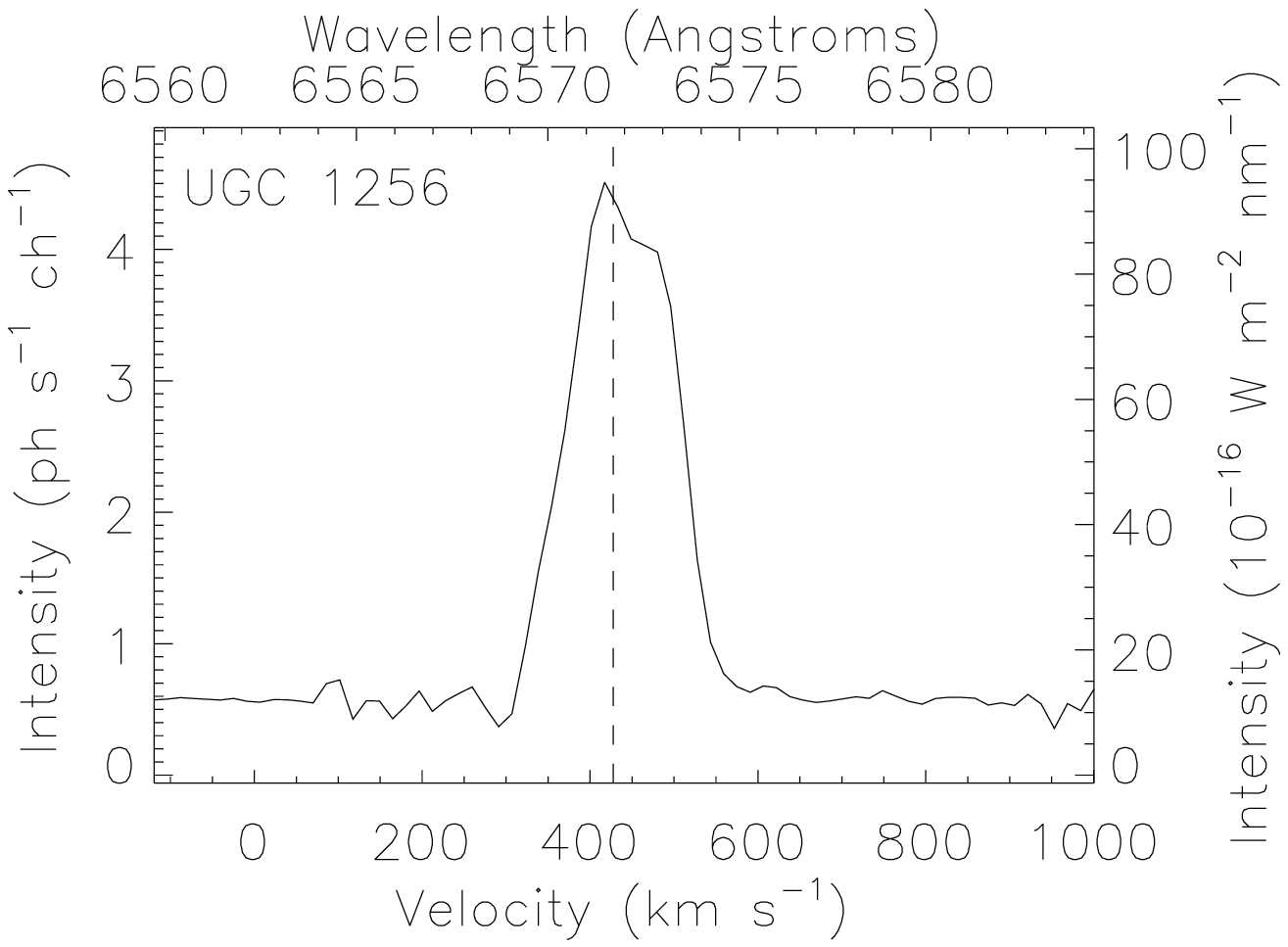}
\includegraphics[width=3.5cm]{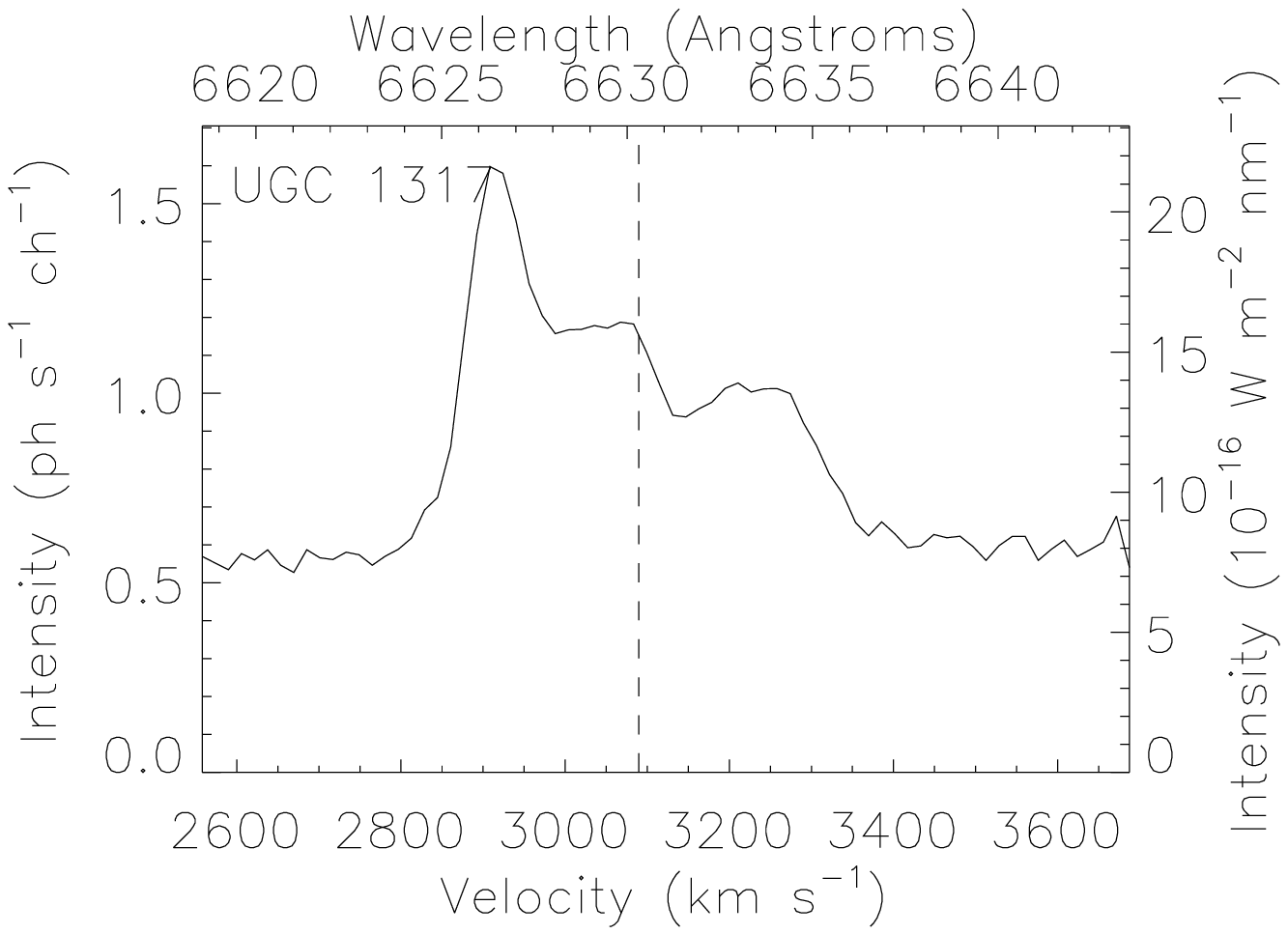}
\includegraphics[width=3.5cm]{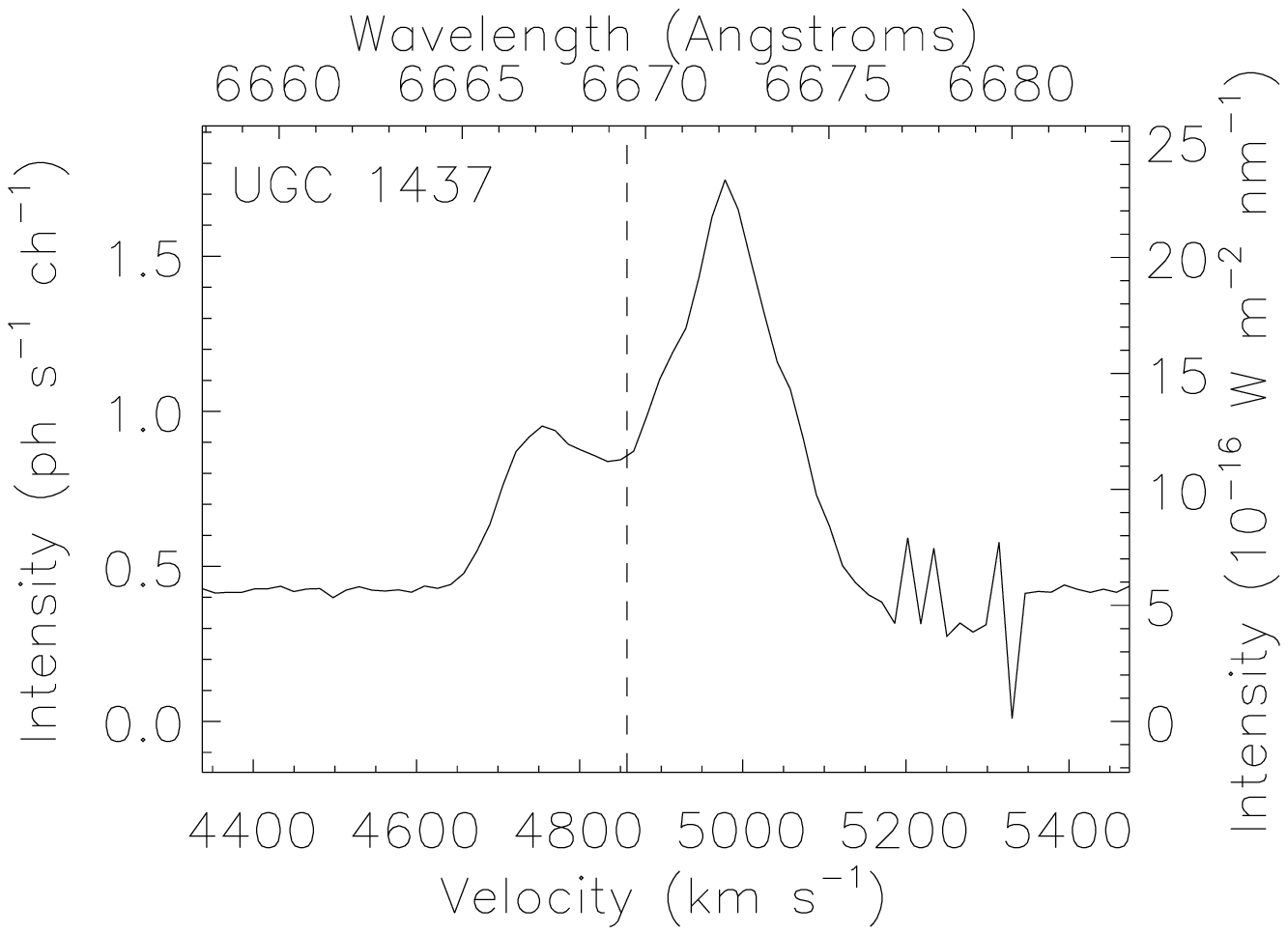}
\includegraphics[width=3.5cm]{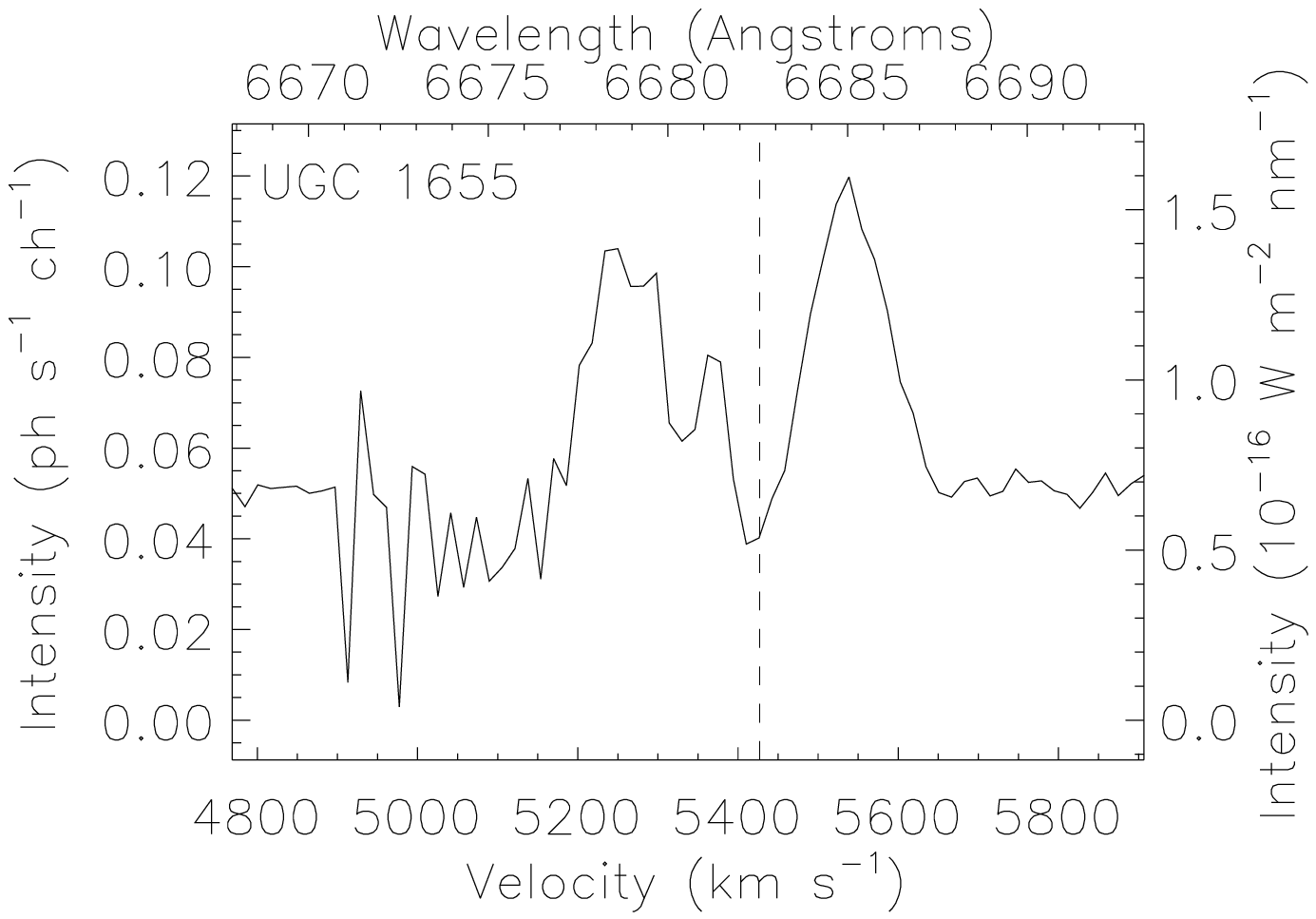}
\includegraphics[width=3.5cm]{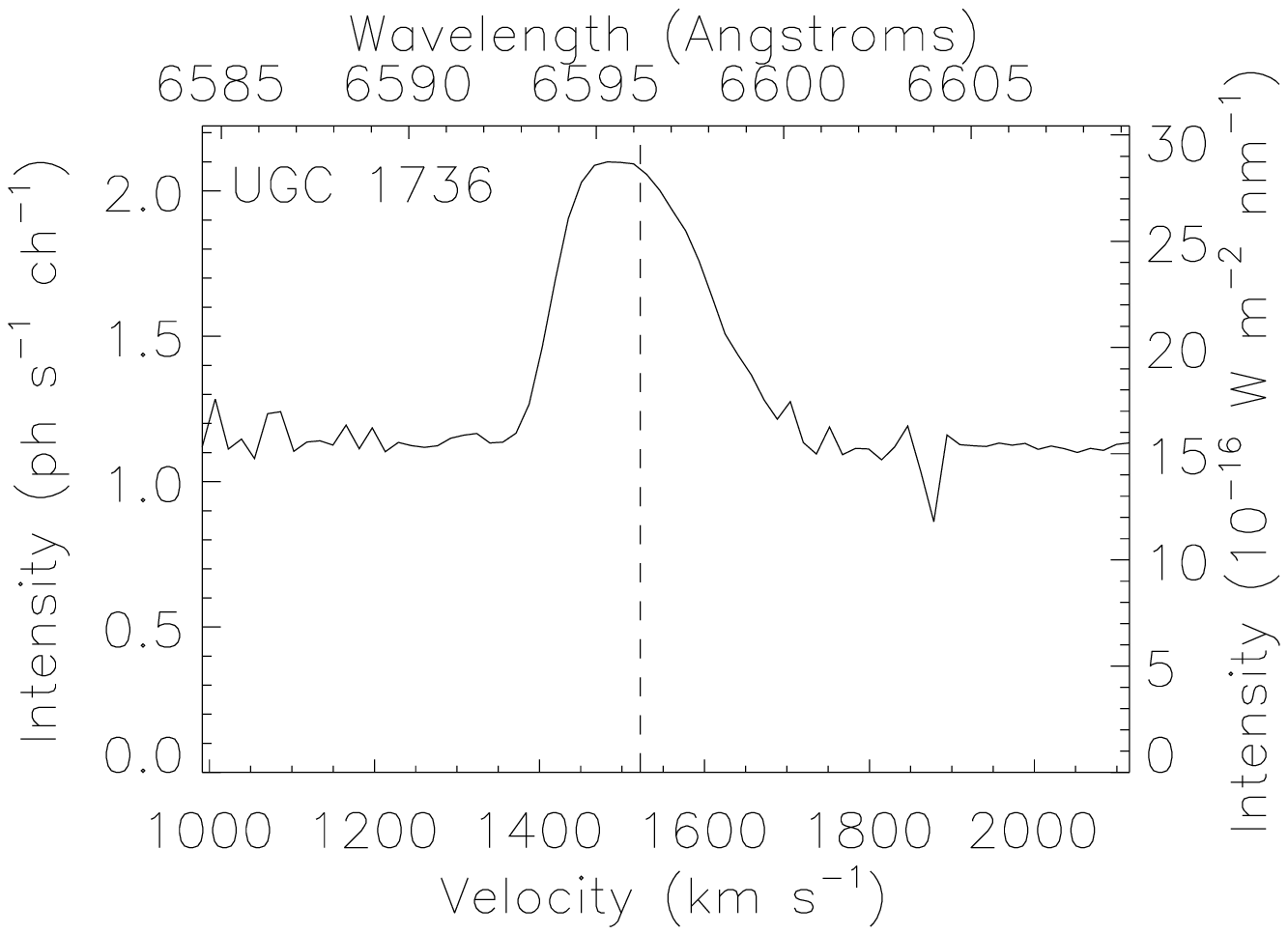}
\includegraphics[width=3.5cm]{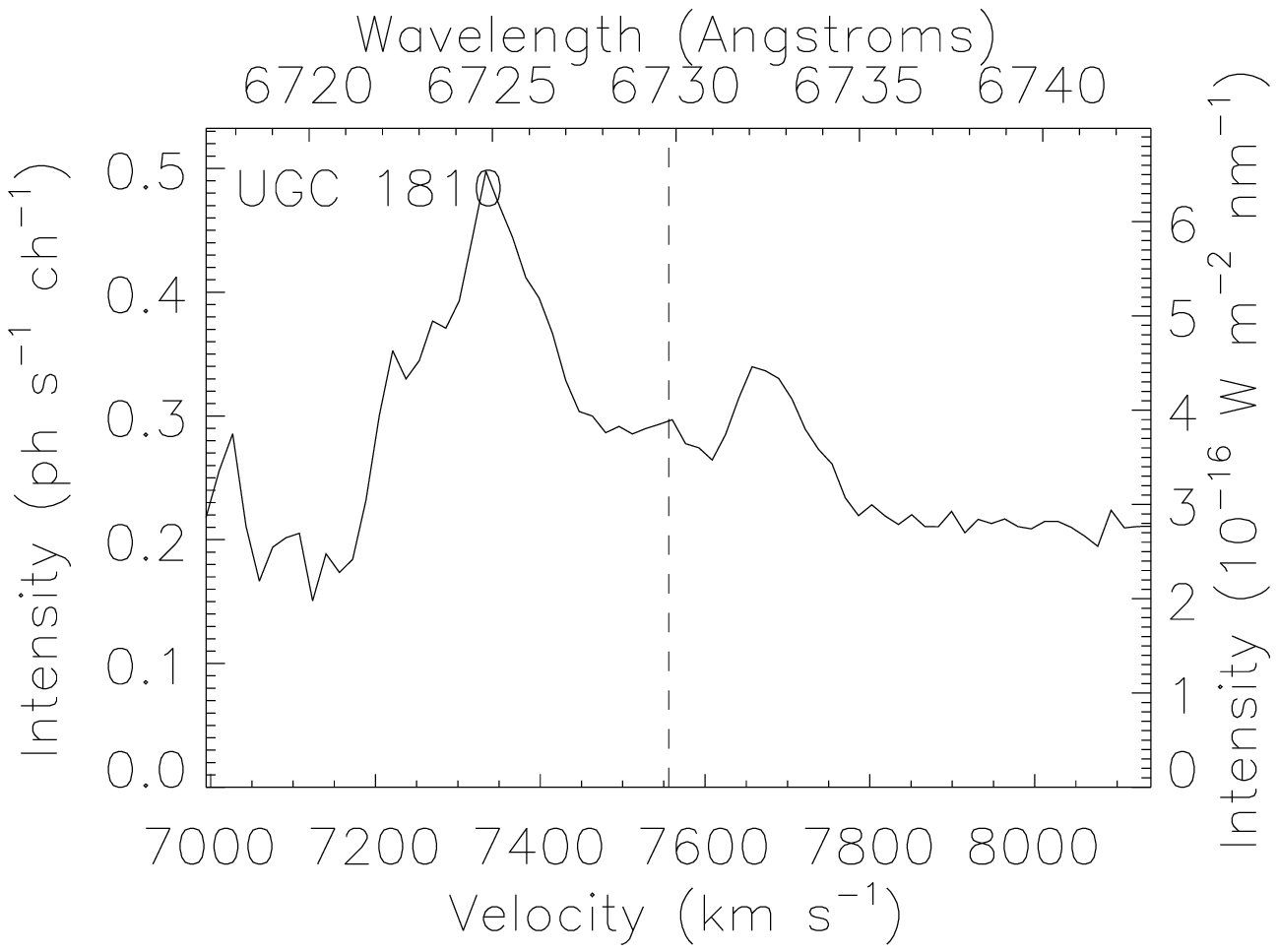}
\includegraphics[width=3.5cm]{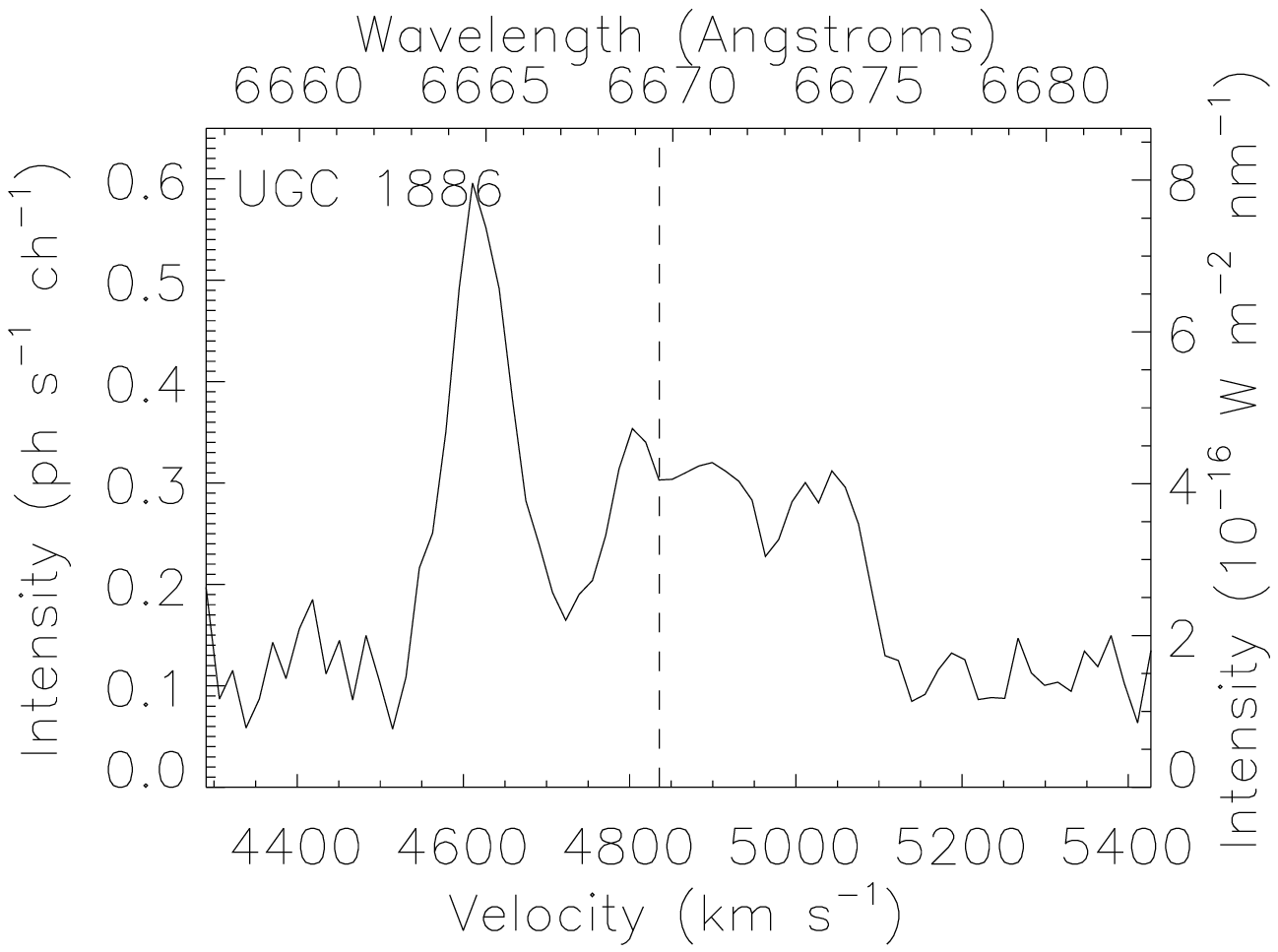}
\includegraphics[width=3.5cm]{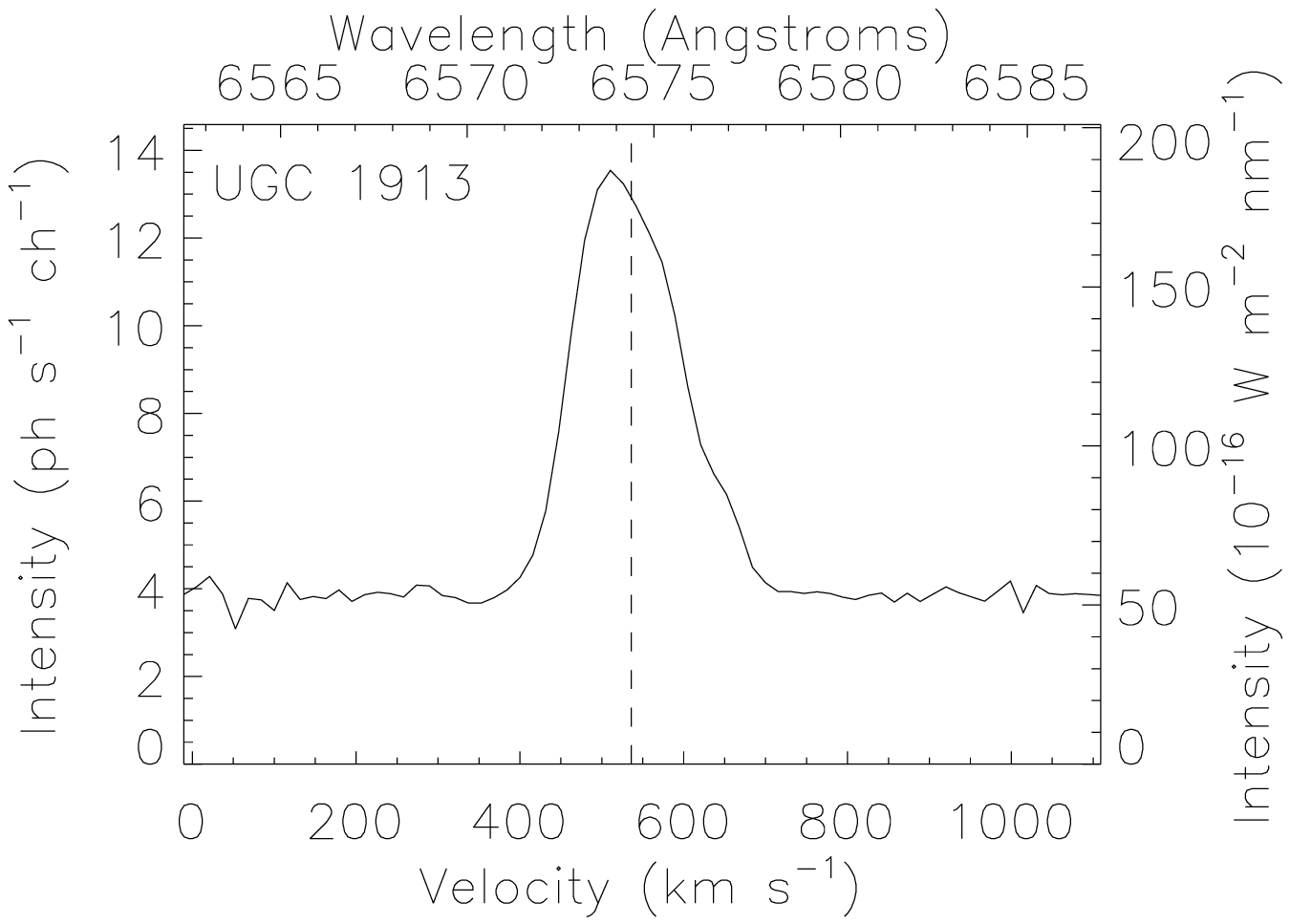}
\includegraphics[width=3.5cm]{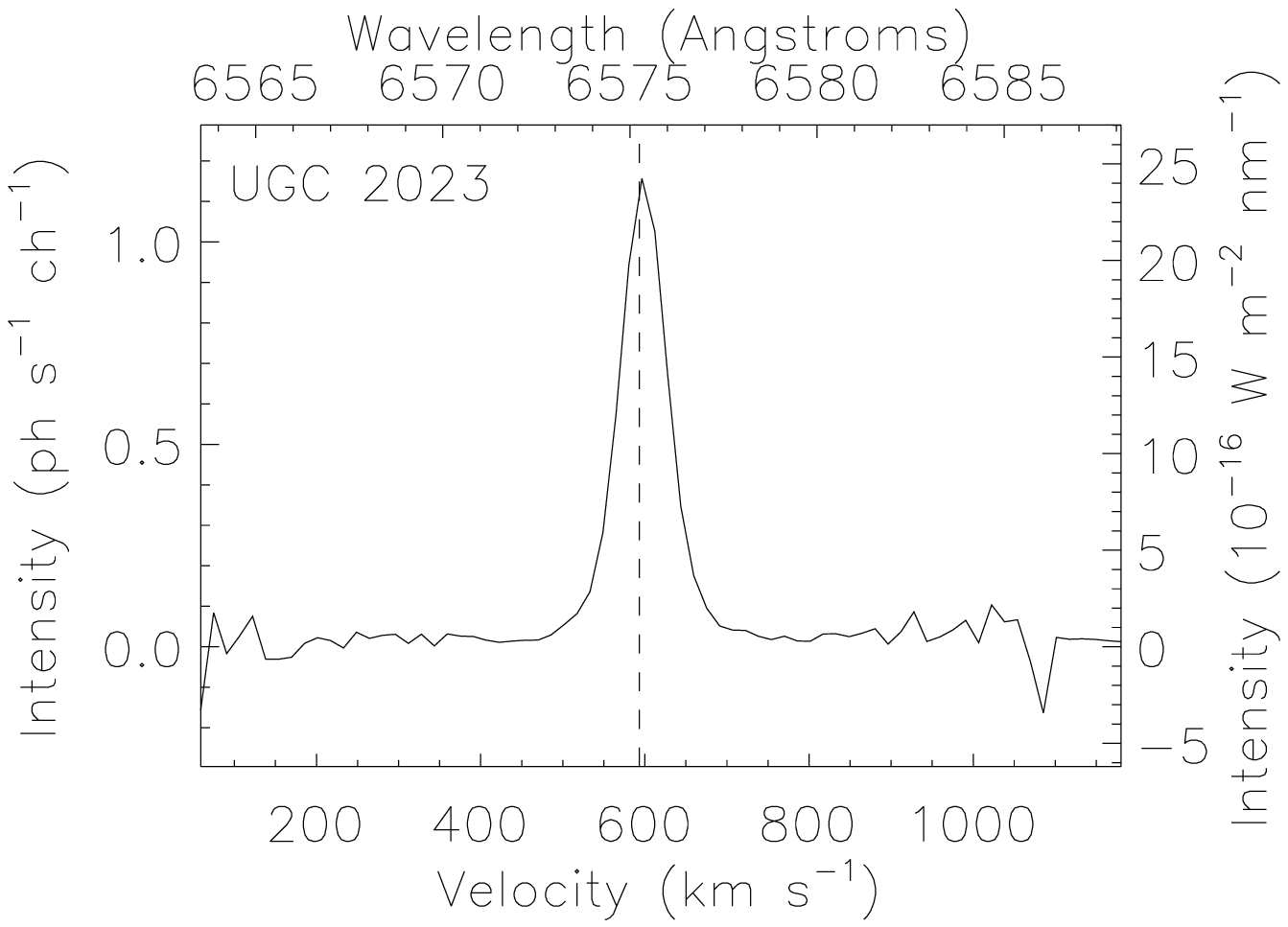}
\includegraphics[width=3.5cm]{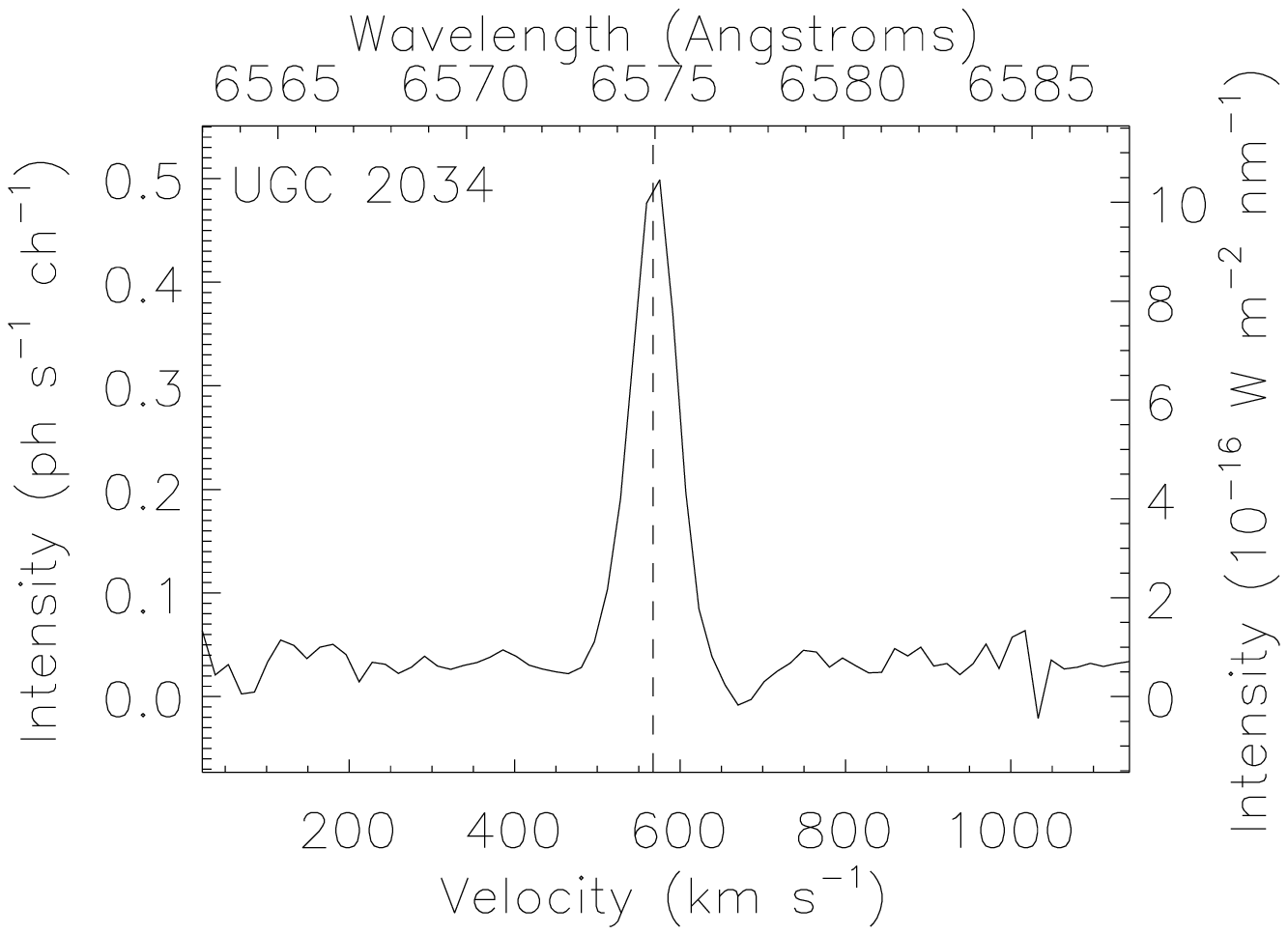}
\includegraphics[width=3.5cm]{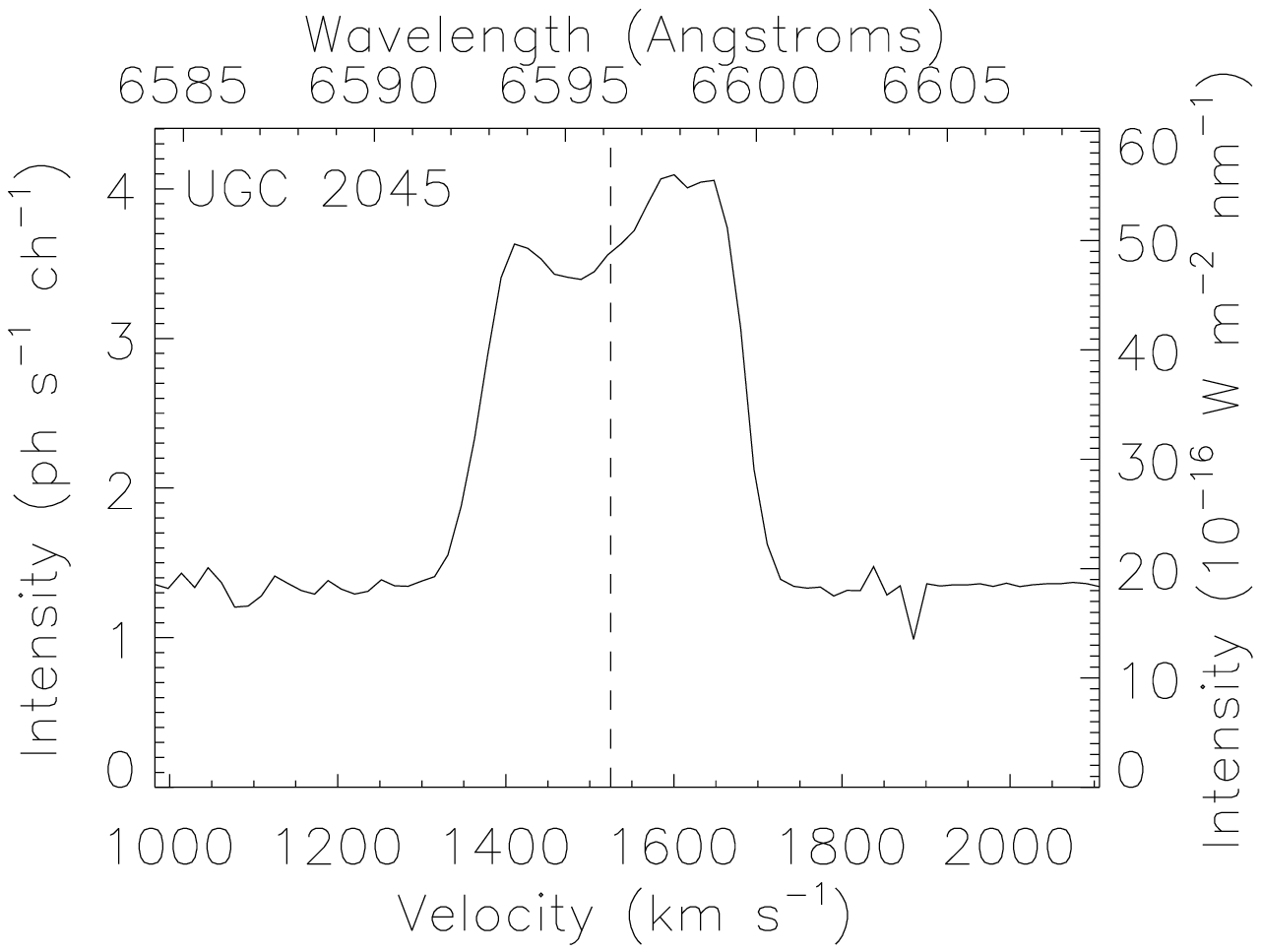}
\includegraphics[width=3.5cm]{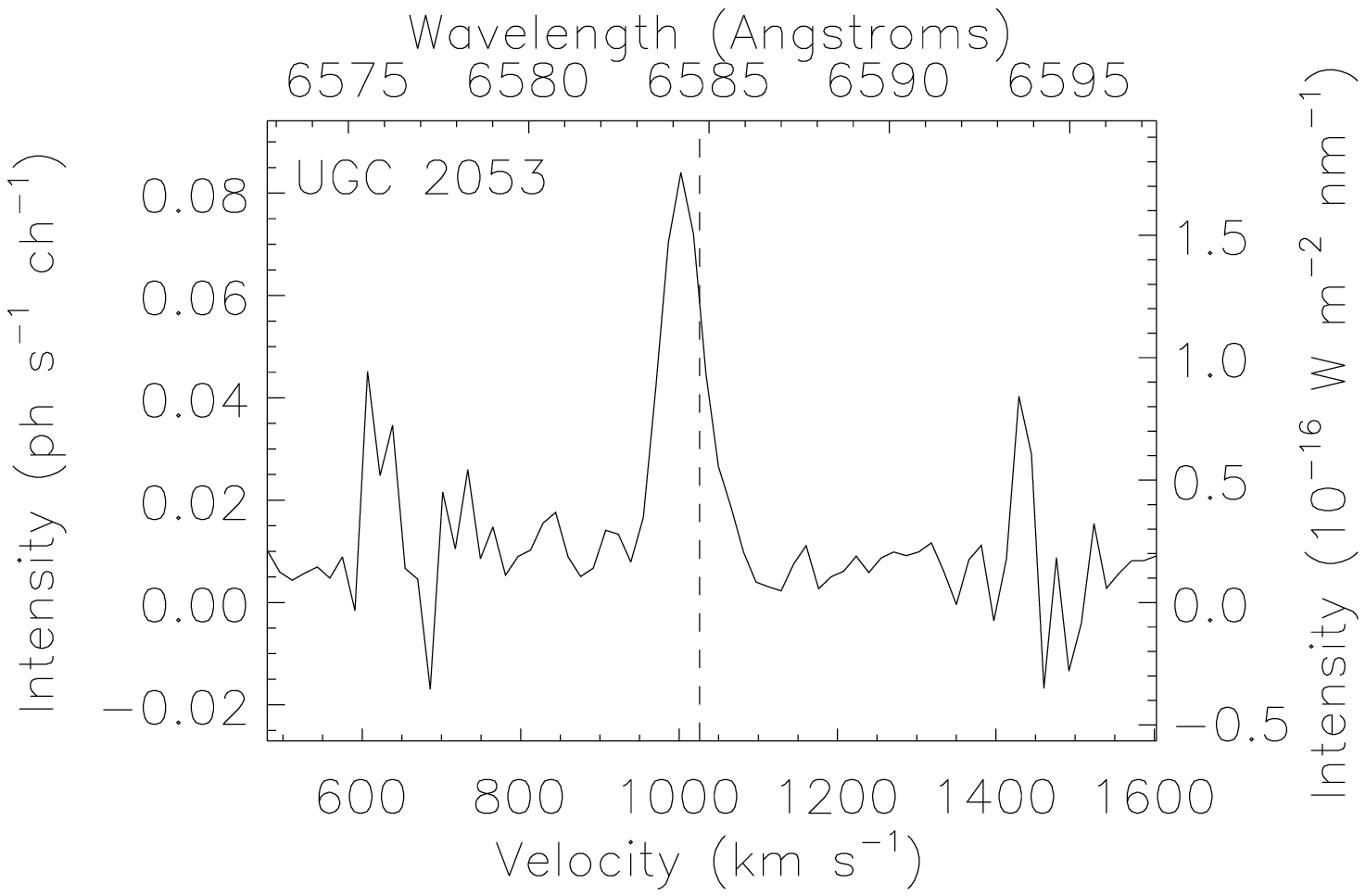}
\includegraphics[width=3.5cm]{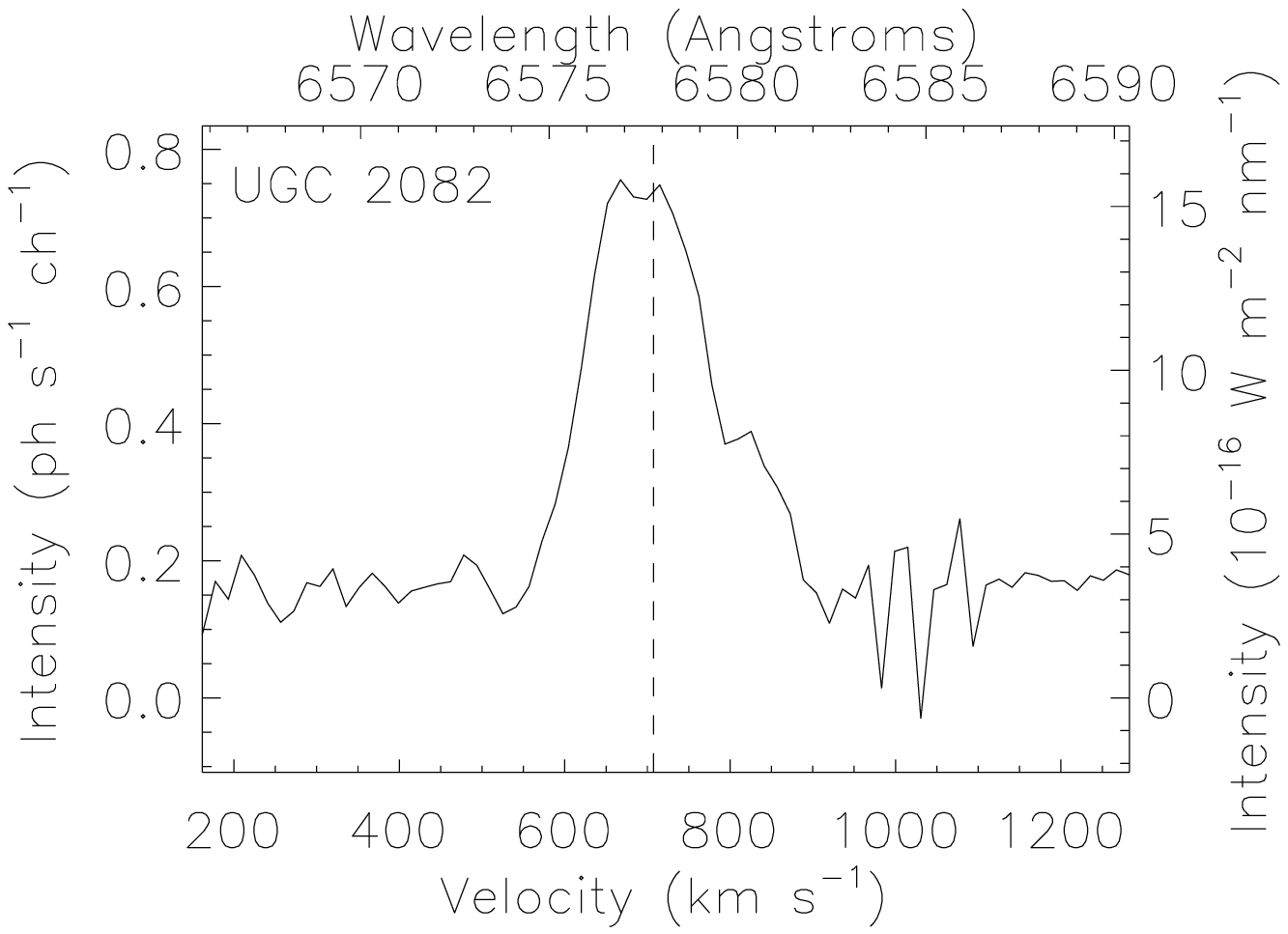}
\includegraphics[width=3.5cm]{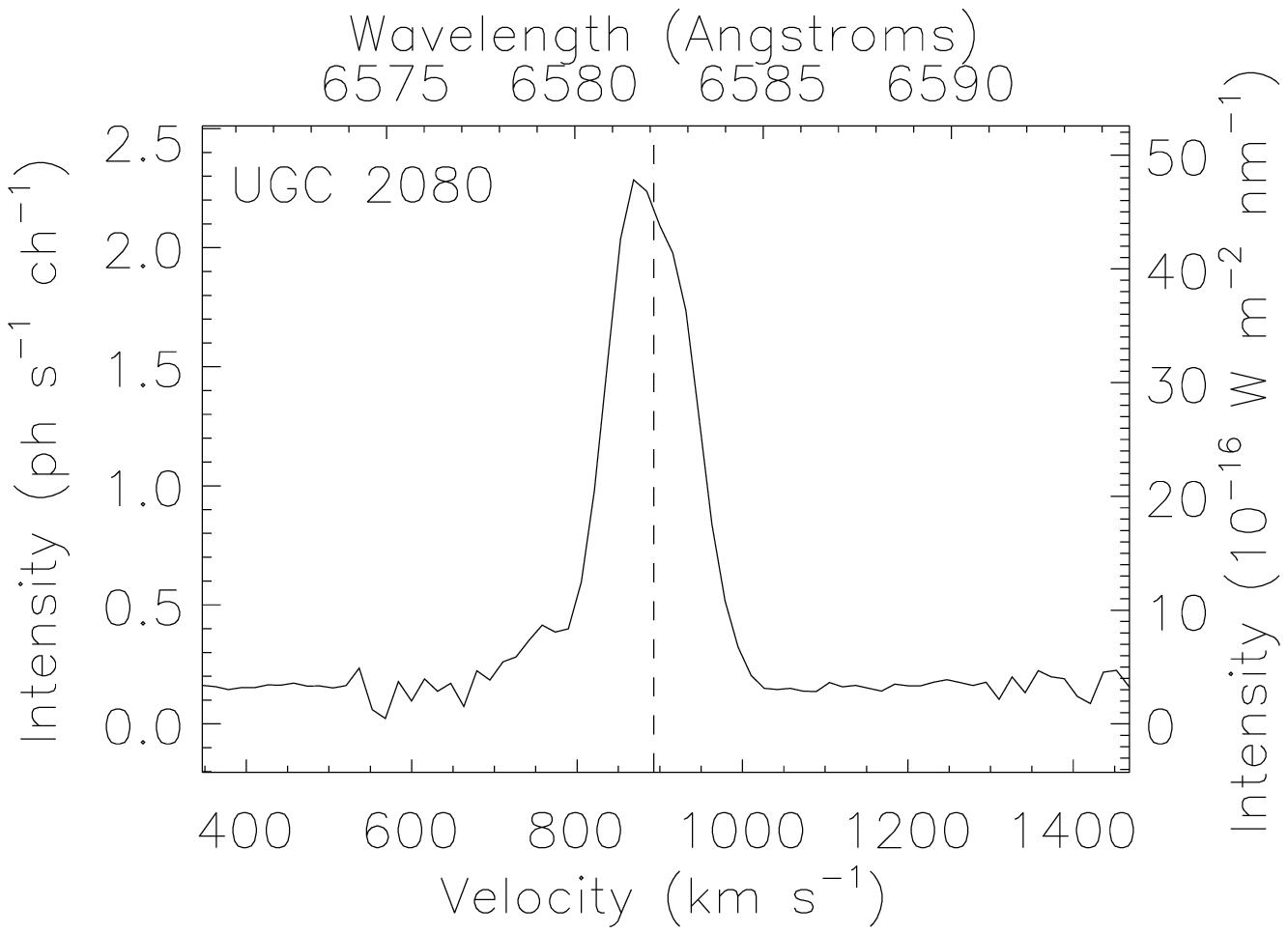}
\includegraphics[width=3.5cm]{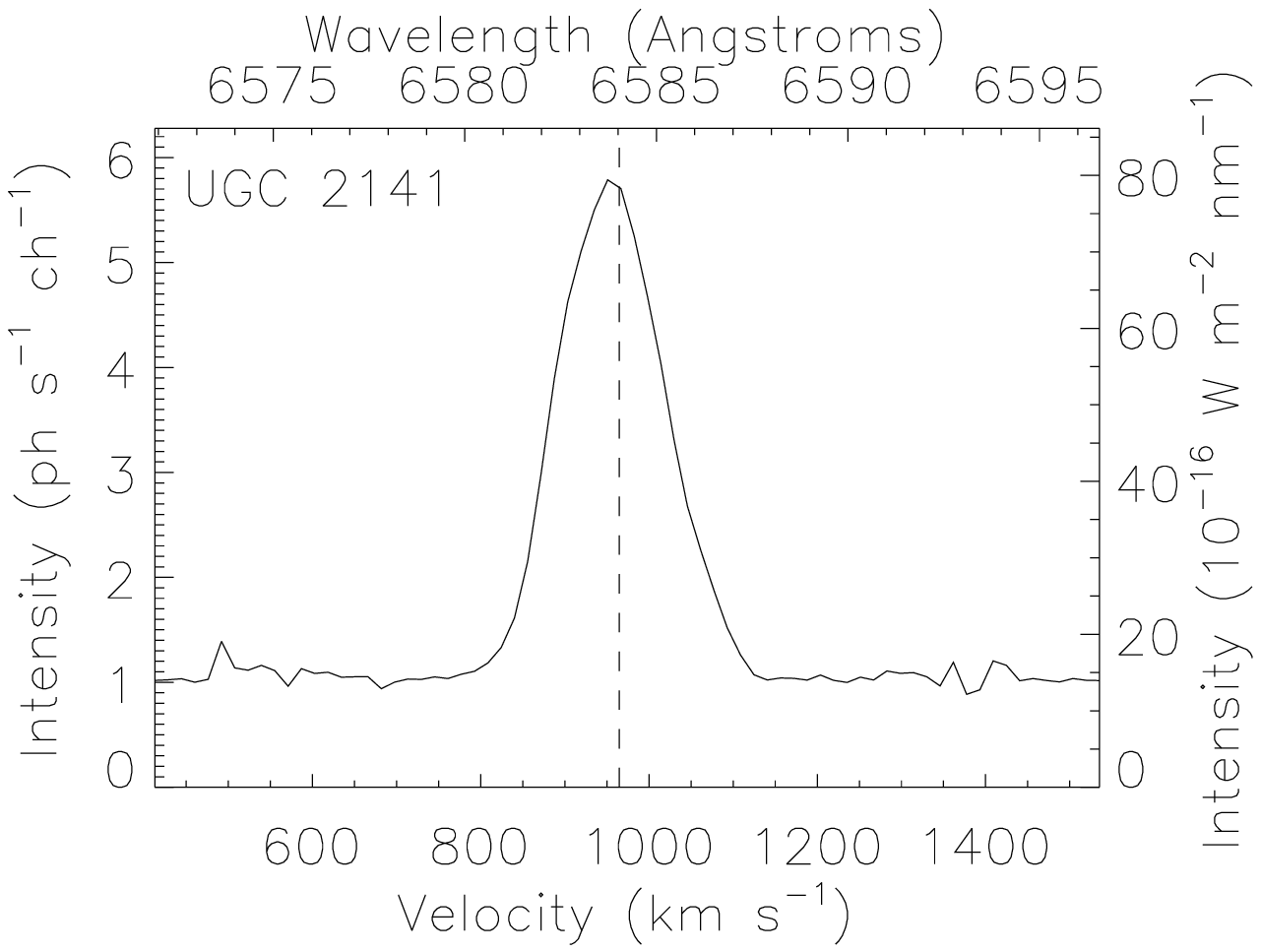}
\includegraphics[width=3.5cm]{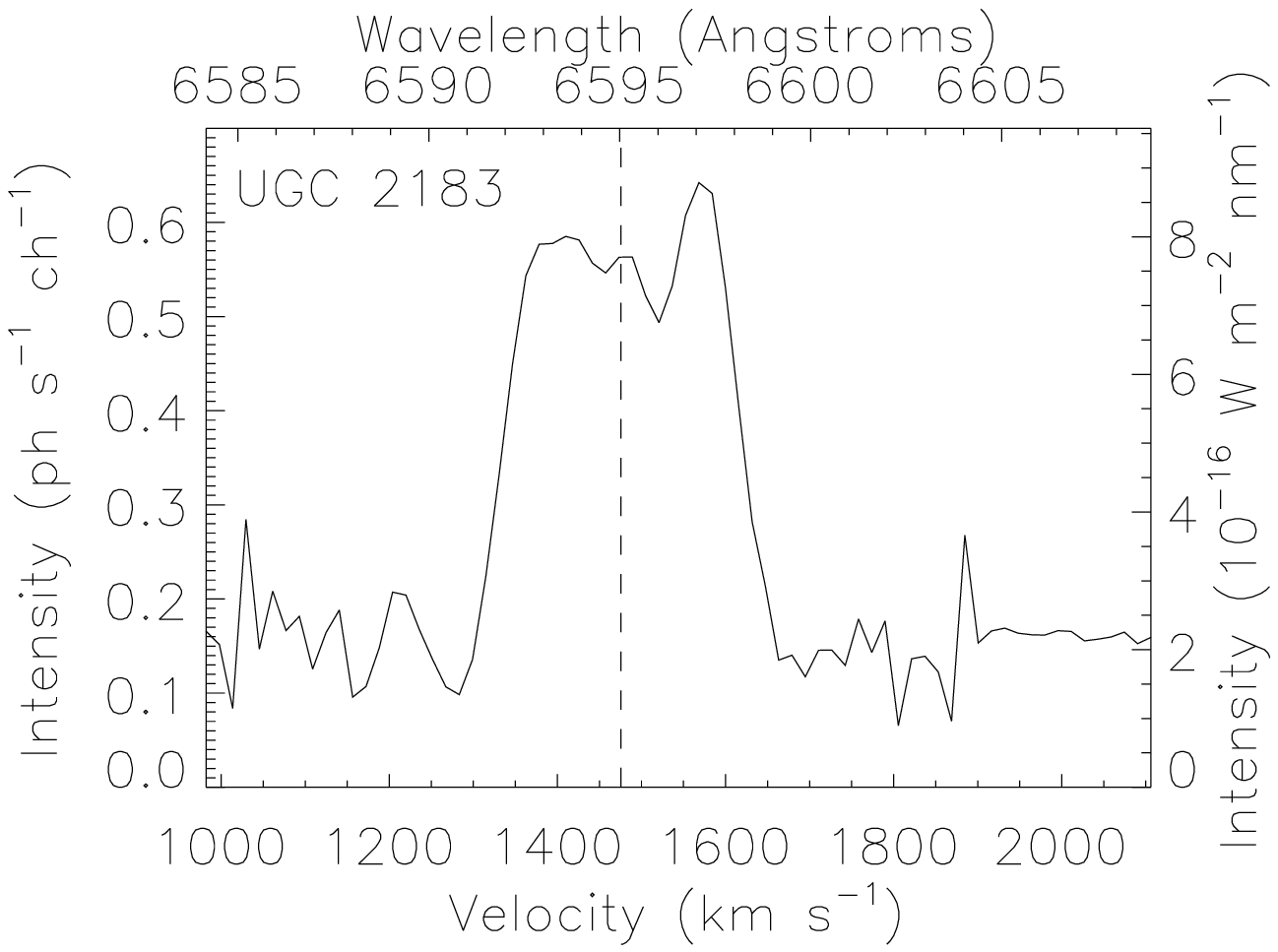}
\includegraphics[width=3.5cm]{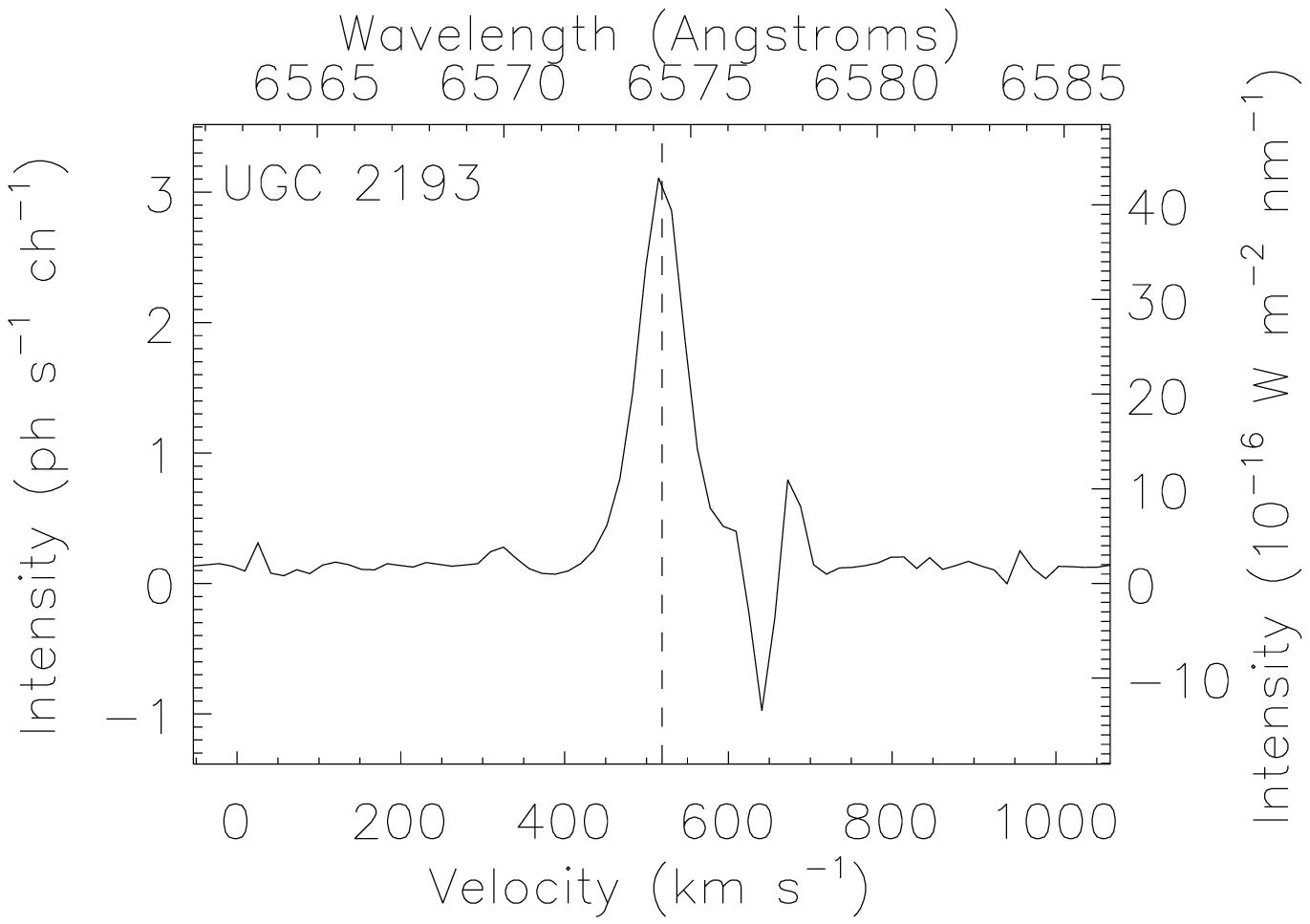}
\includegraphics[width=3.5cm]{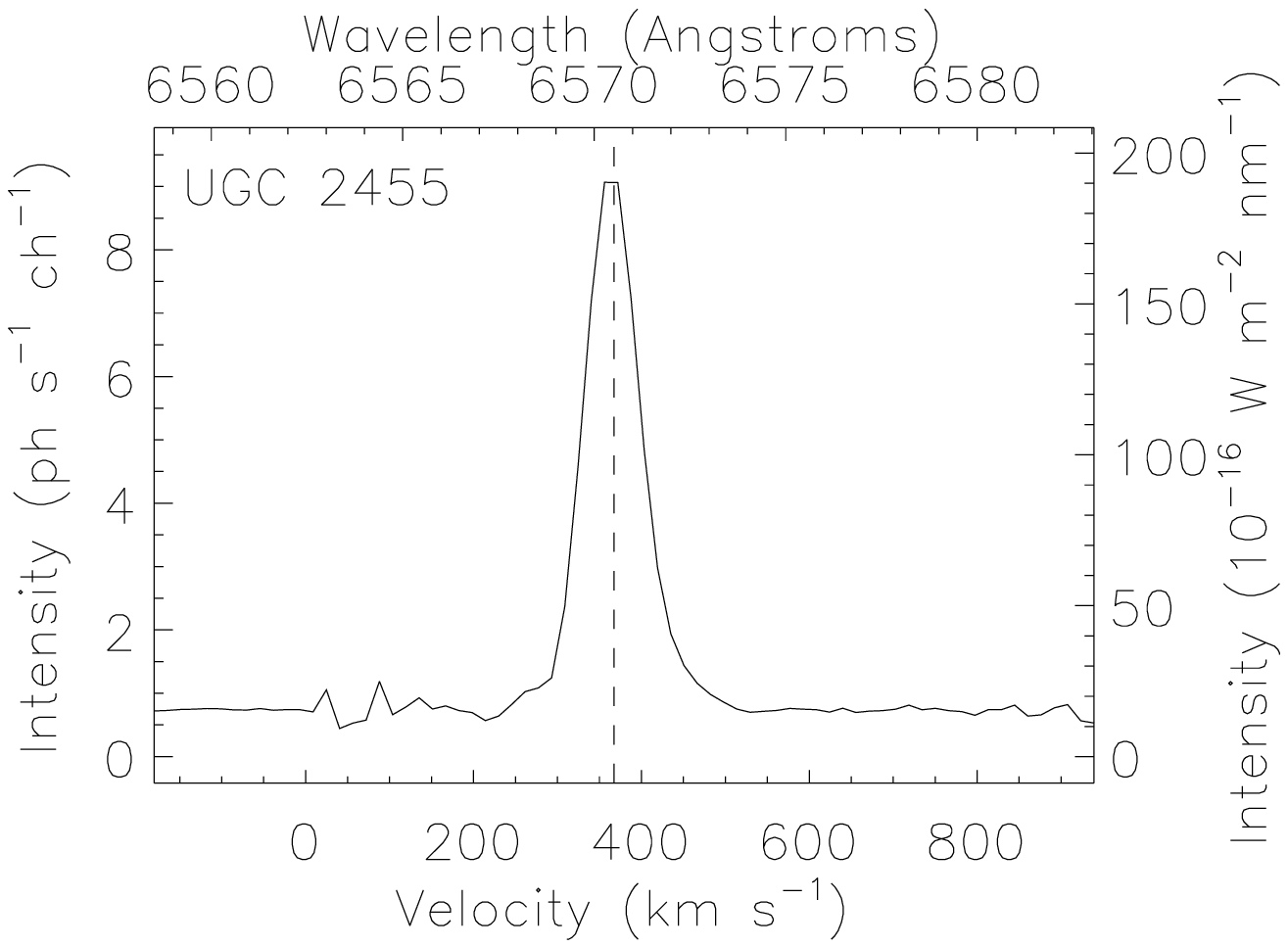}
\includegraphics[width=3.5cm]{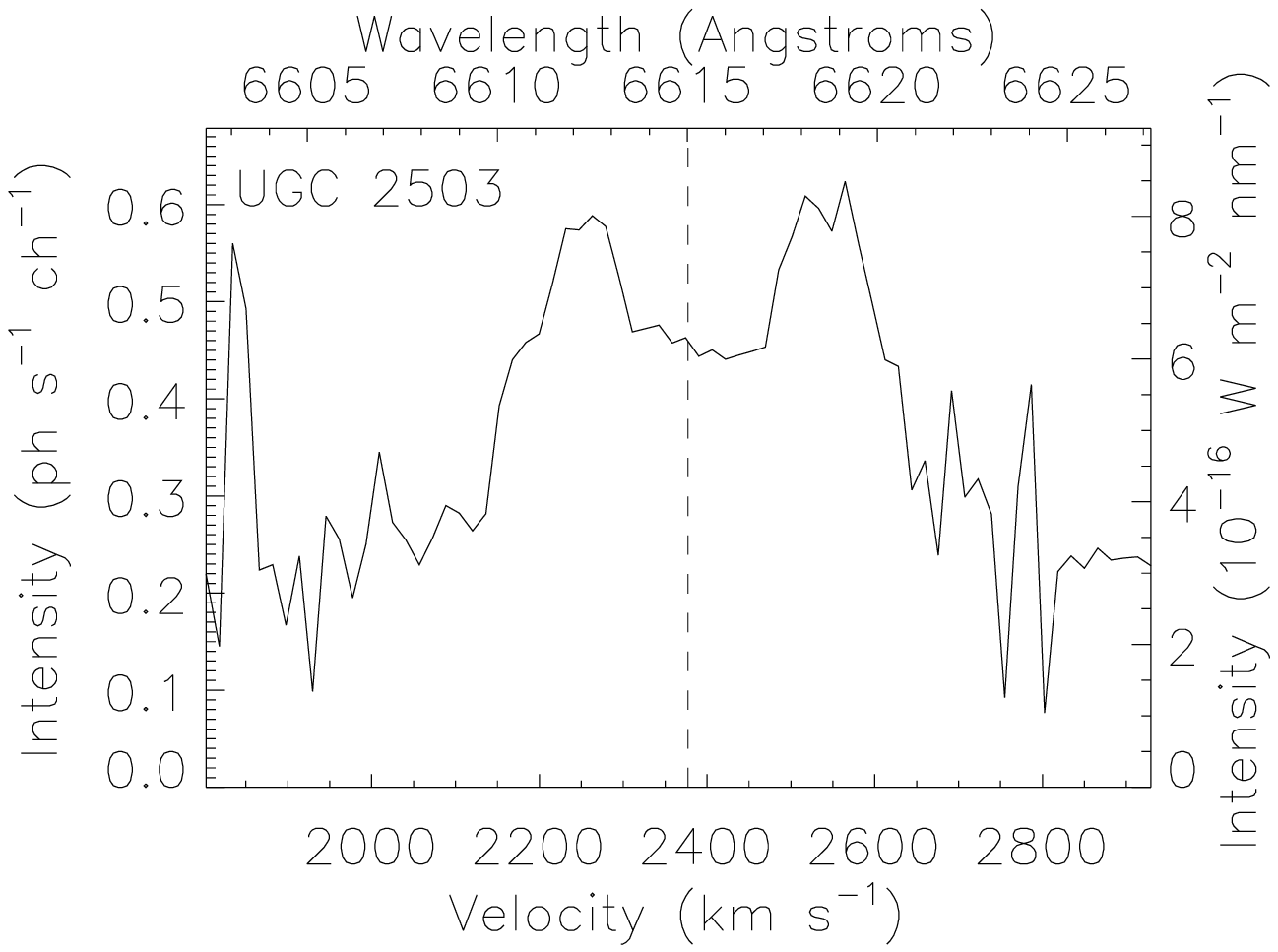}
\includegraphics[width=3.5cm]{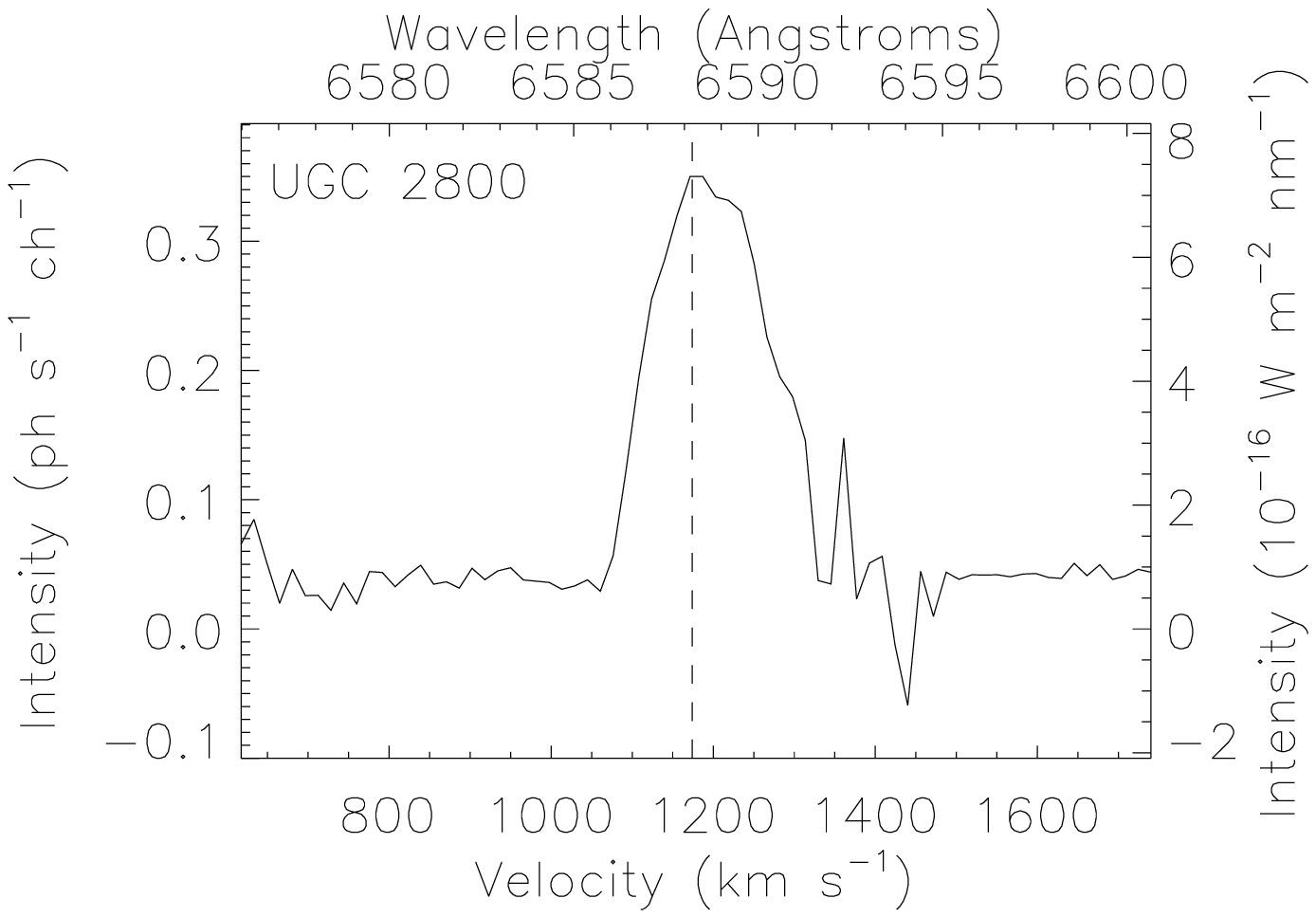}
\includegraphics[width=3.5cm]{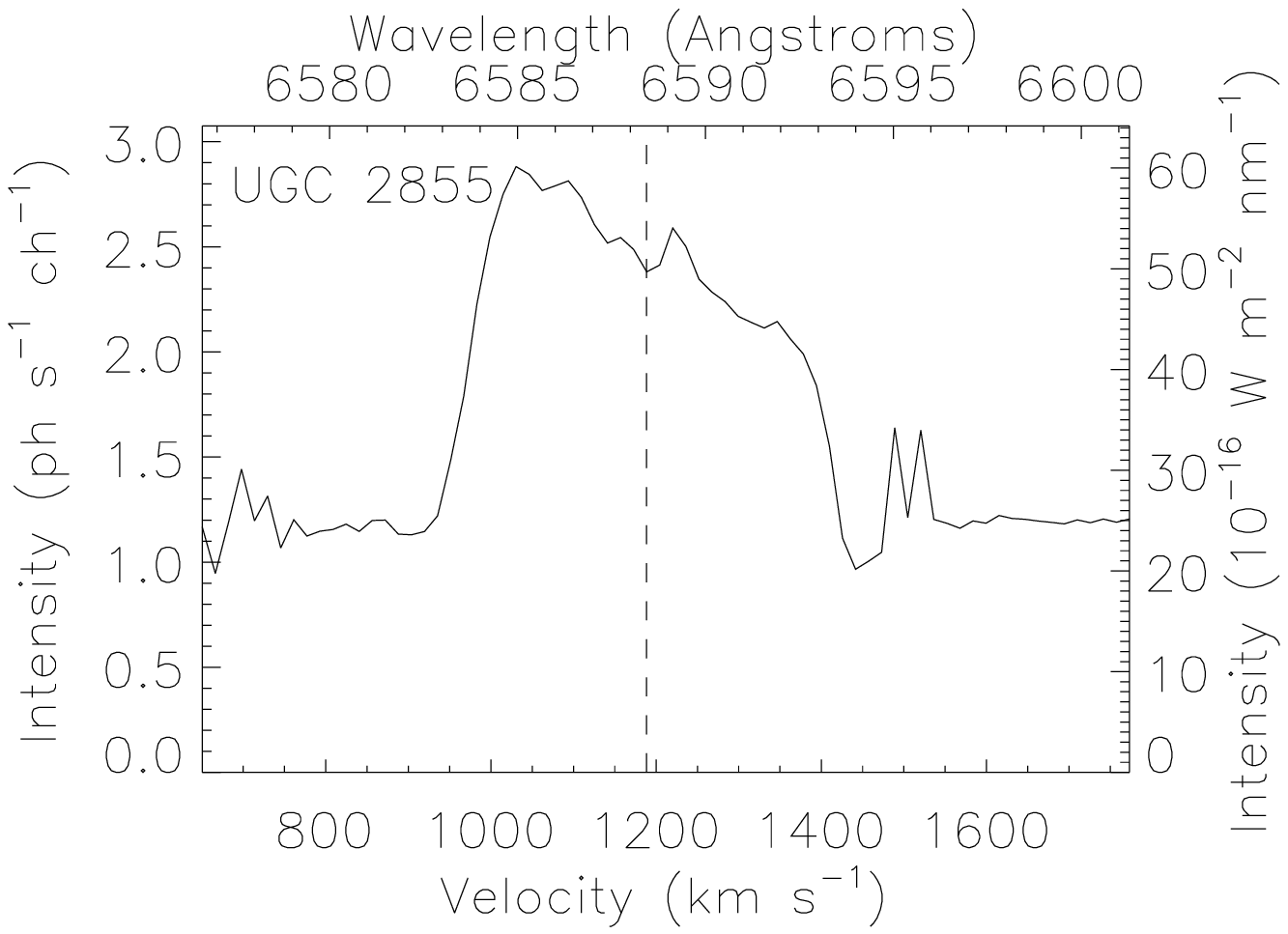}
\includegraphics[width=3.5cm]{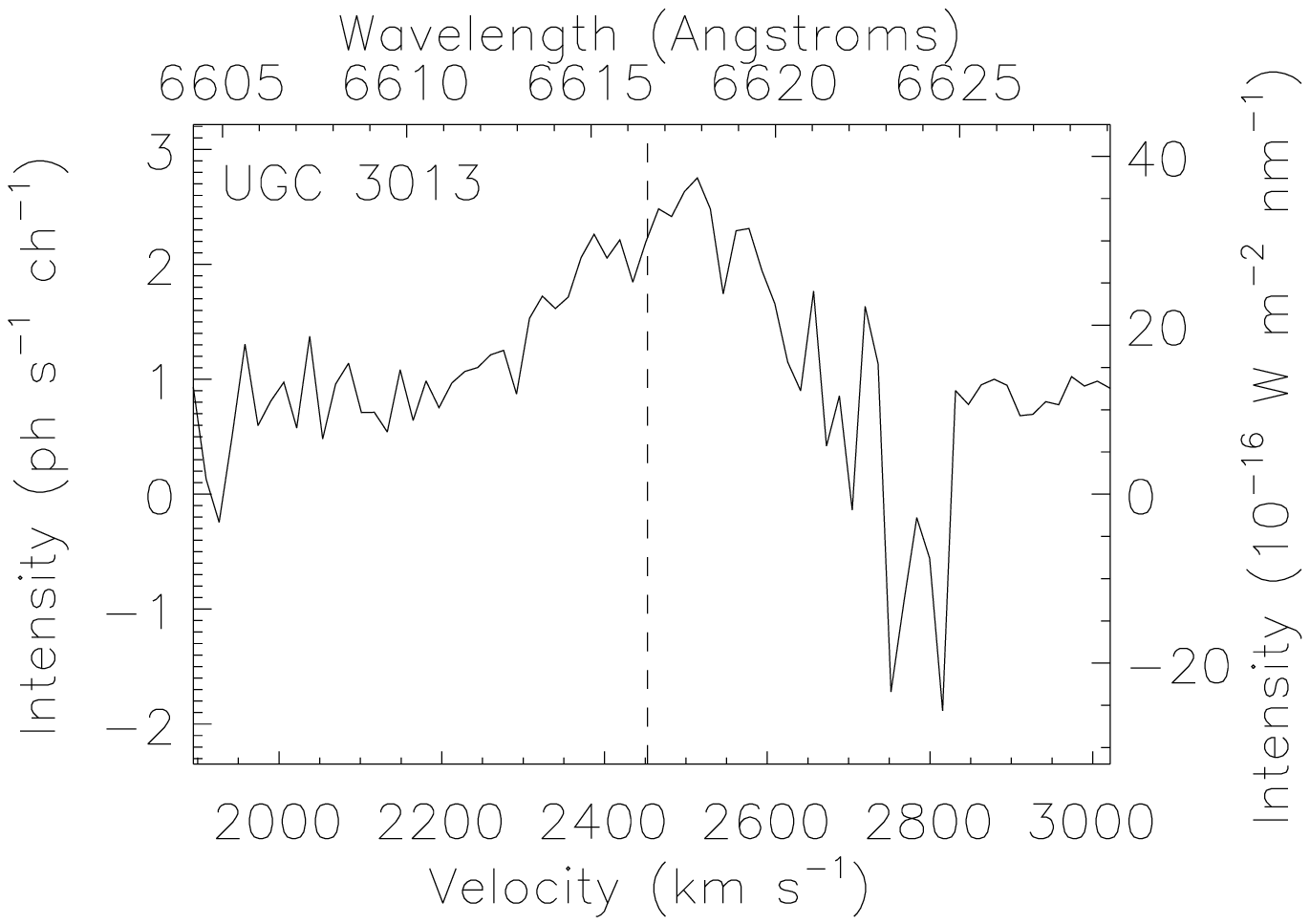}
\includegraphics[width=3.5cm]{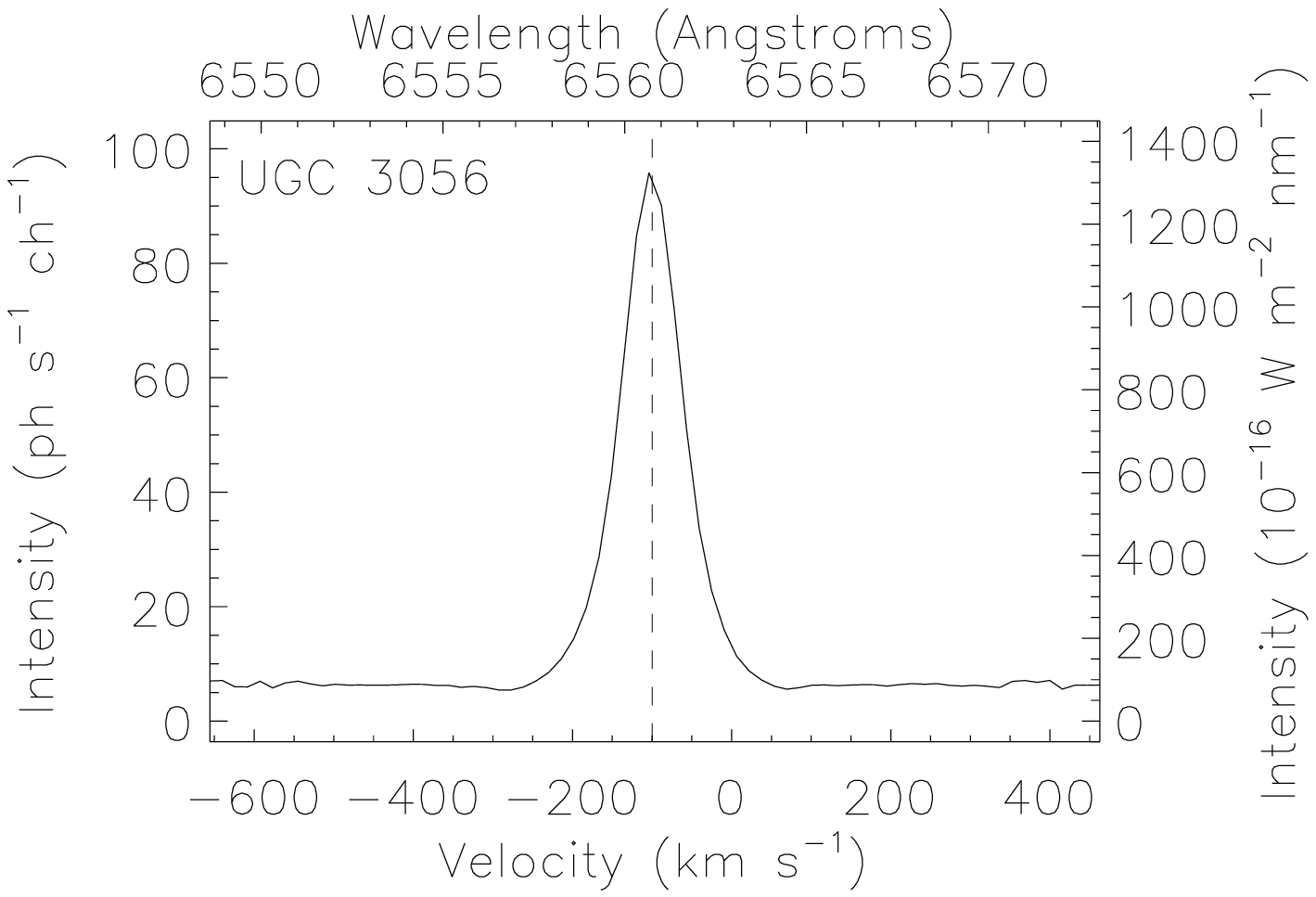}
\includegraphics[width=3.5cm]{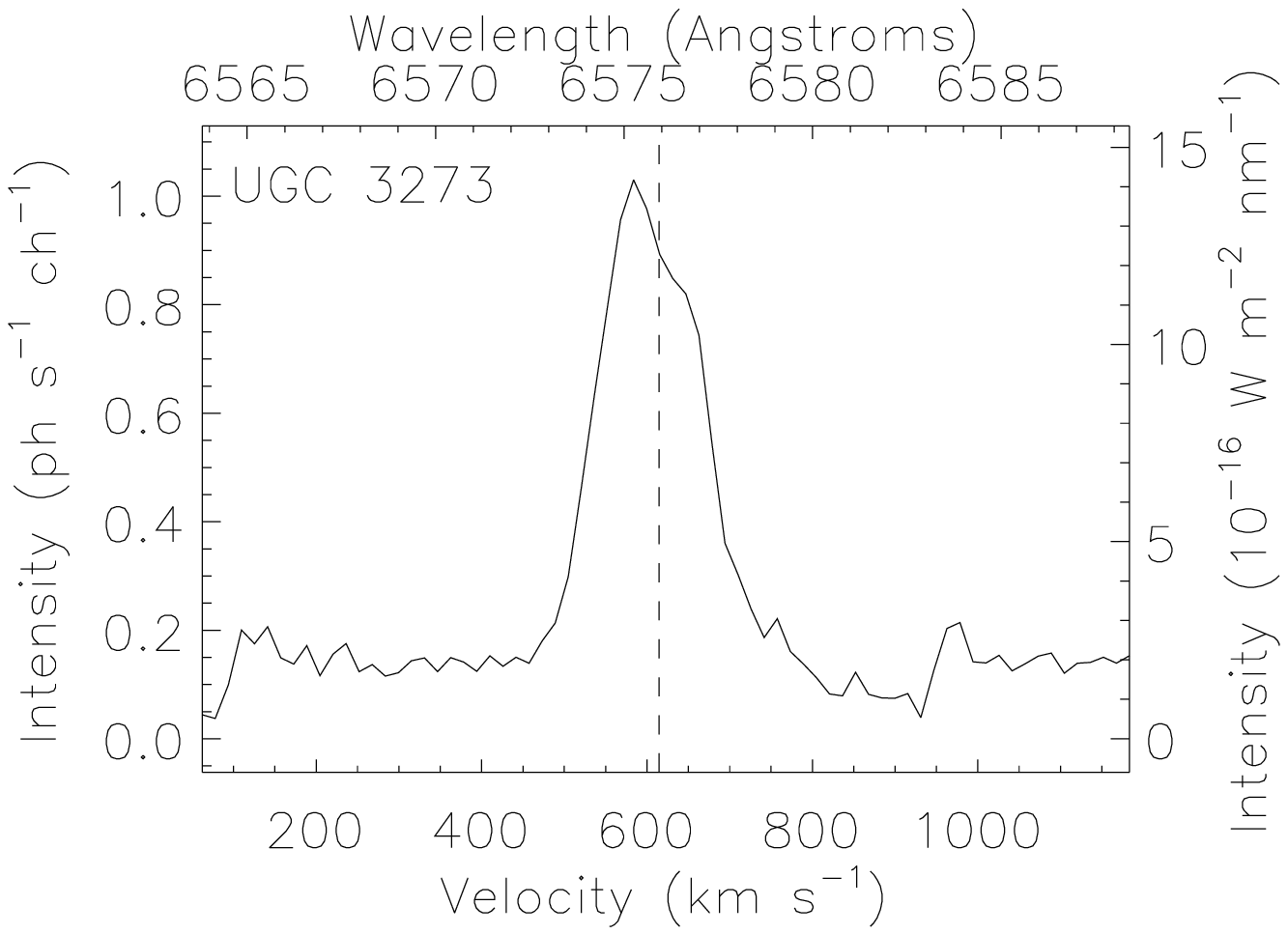}
\includegraphics[width=3.5cm]{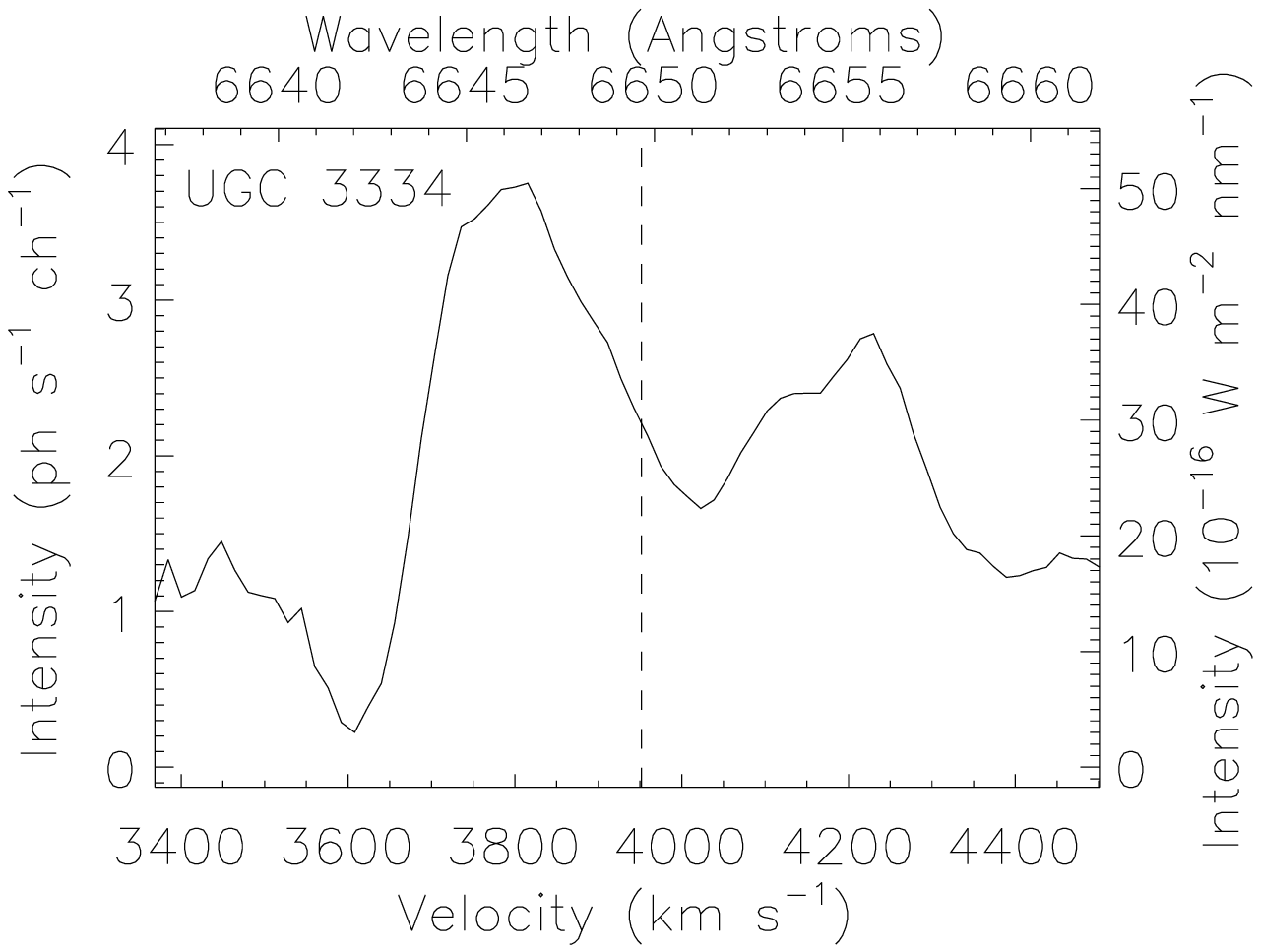}
\includegraphics[width=3.5cm]{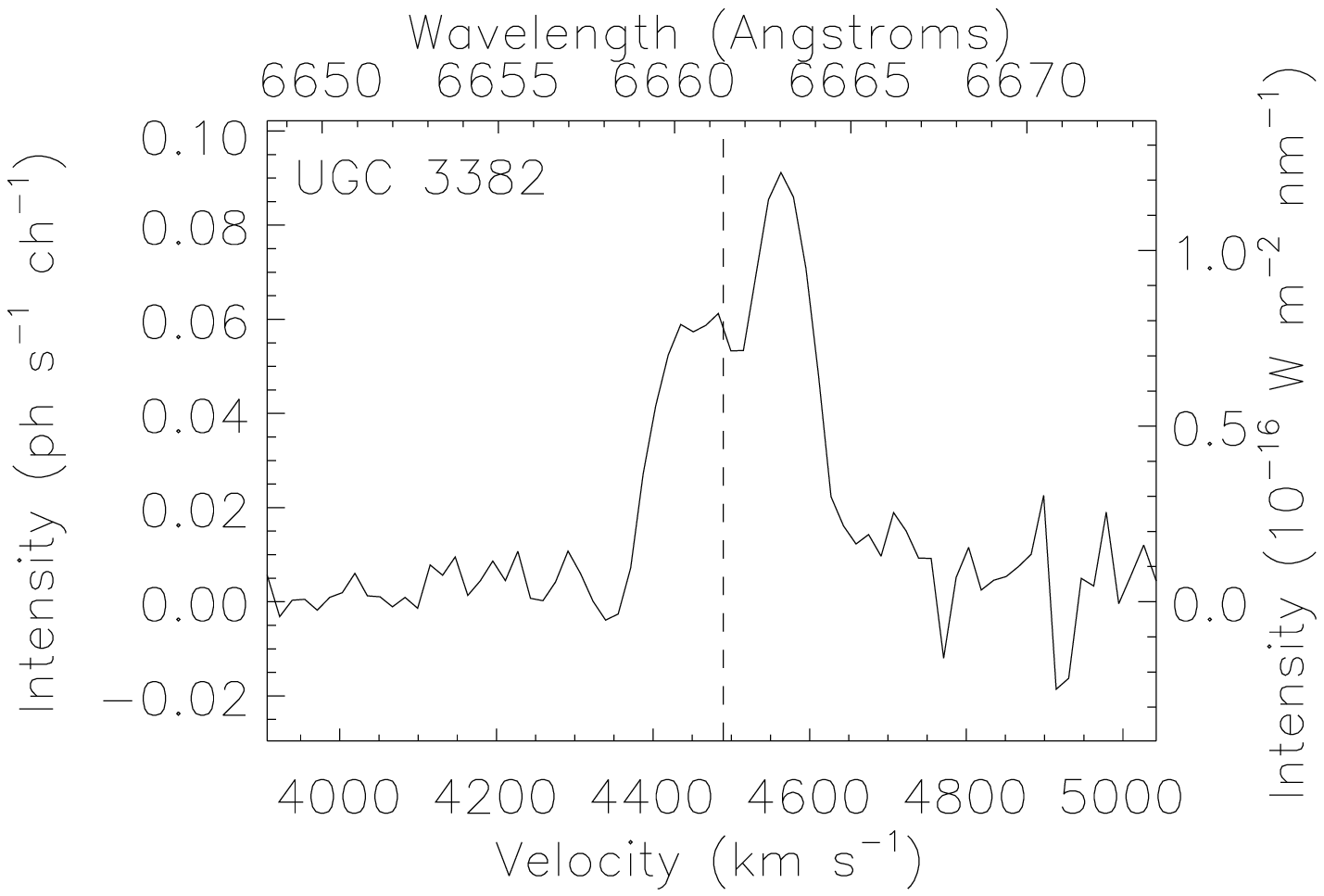}
\includegraphics[width=3.5cm]{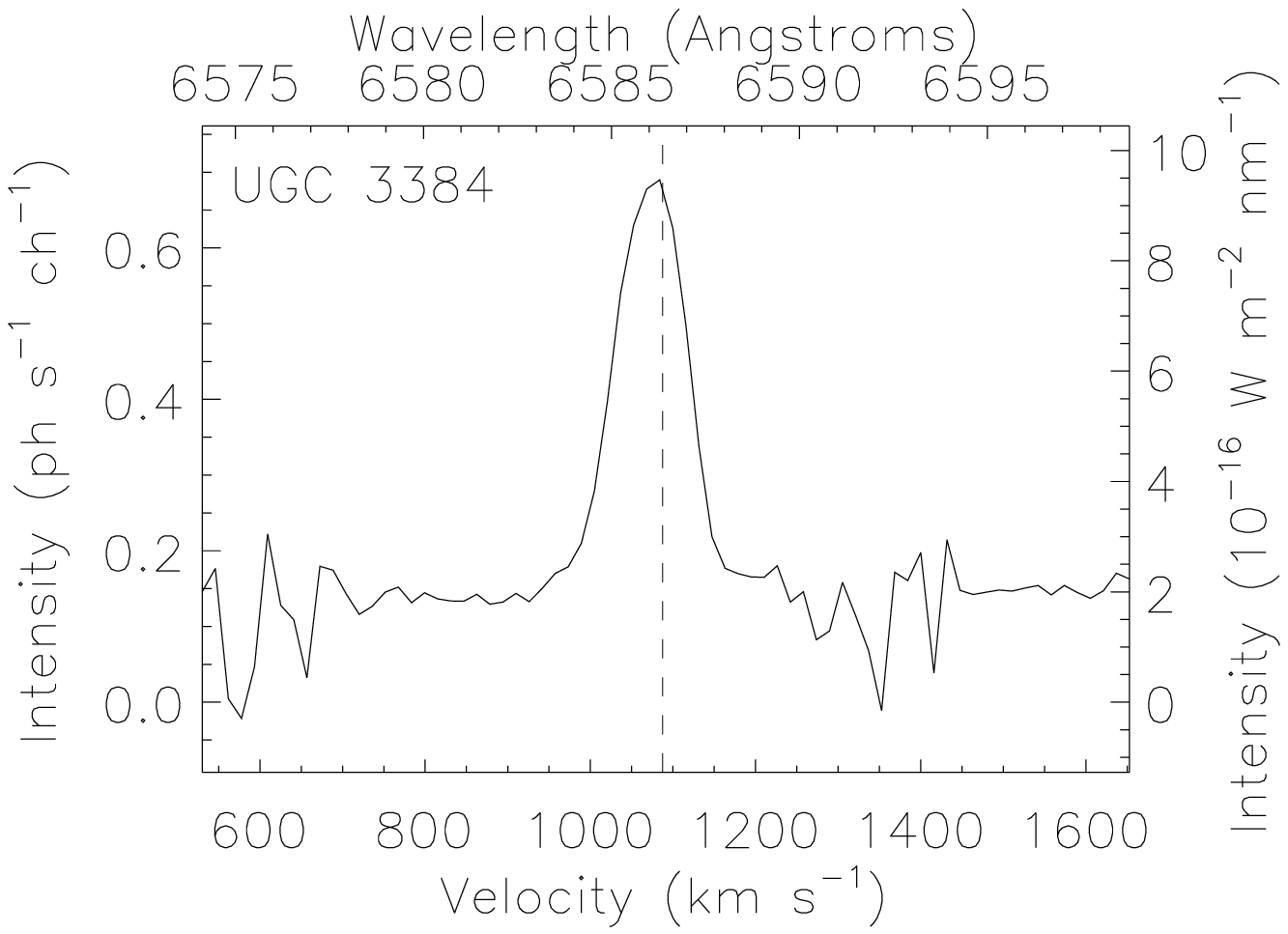}
\includegraphics[width=3.5cm]{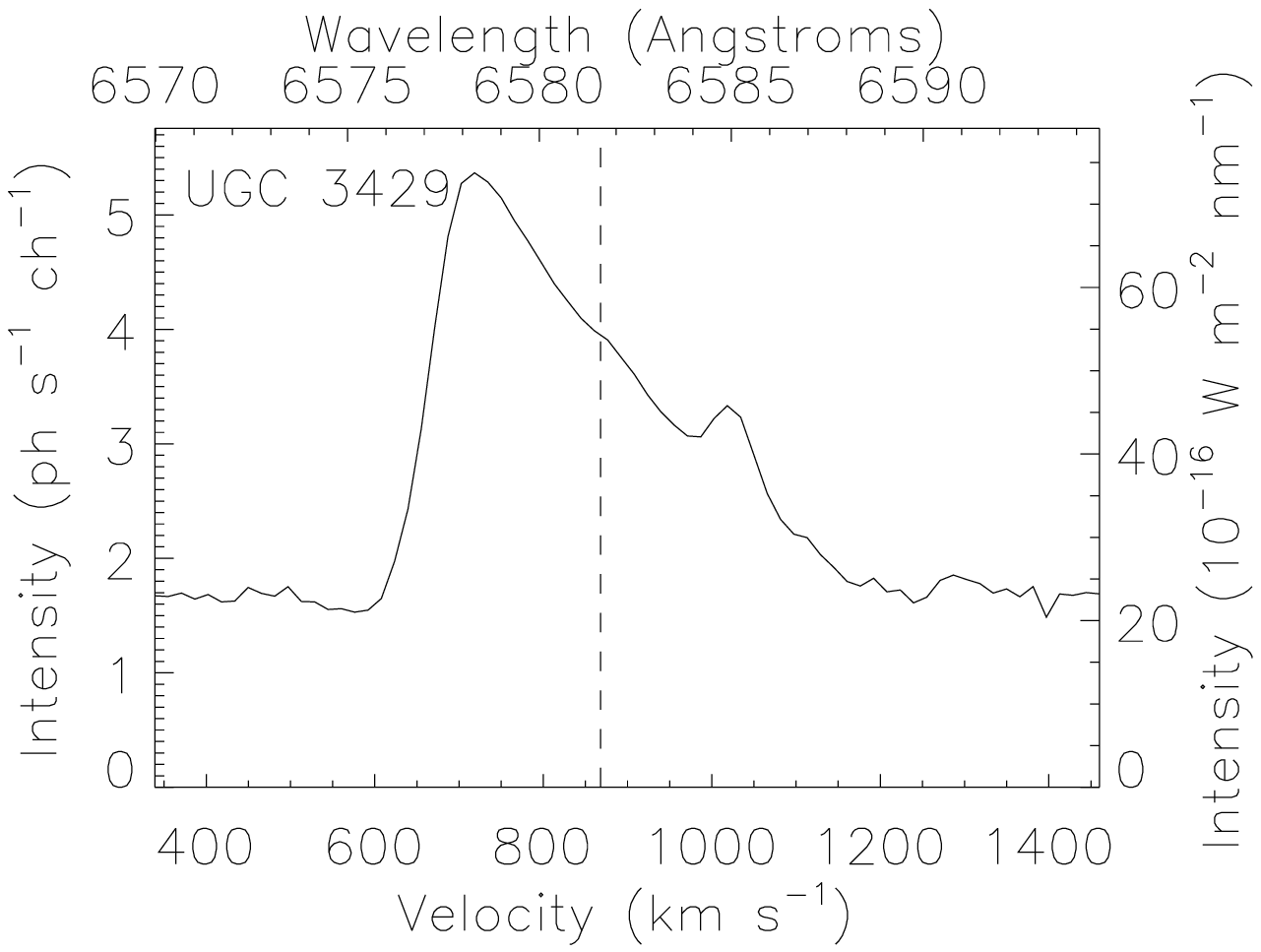}
\includegraphics[width=3.5cm]{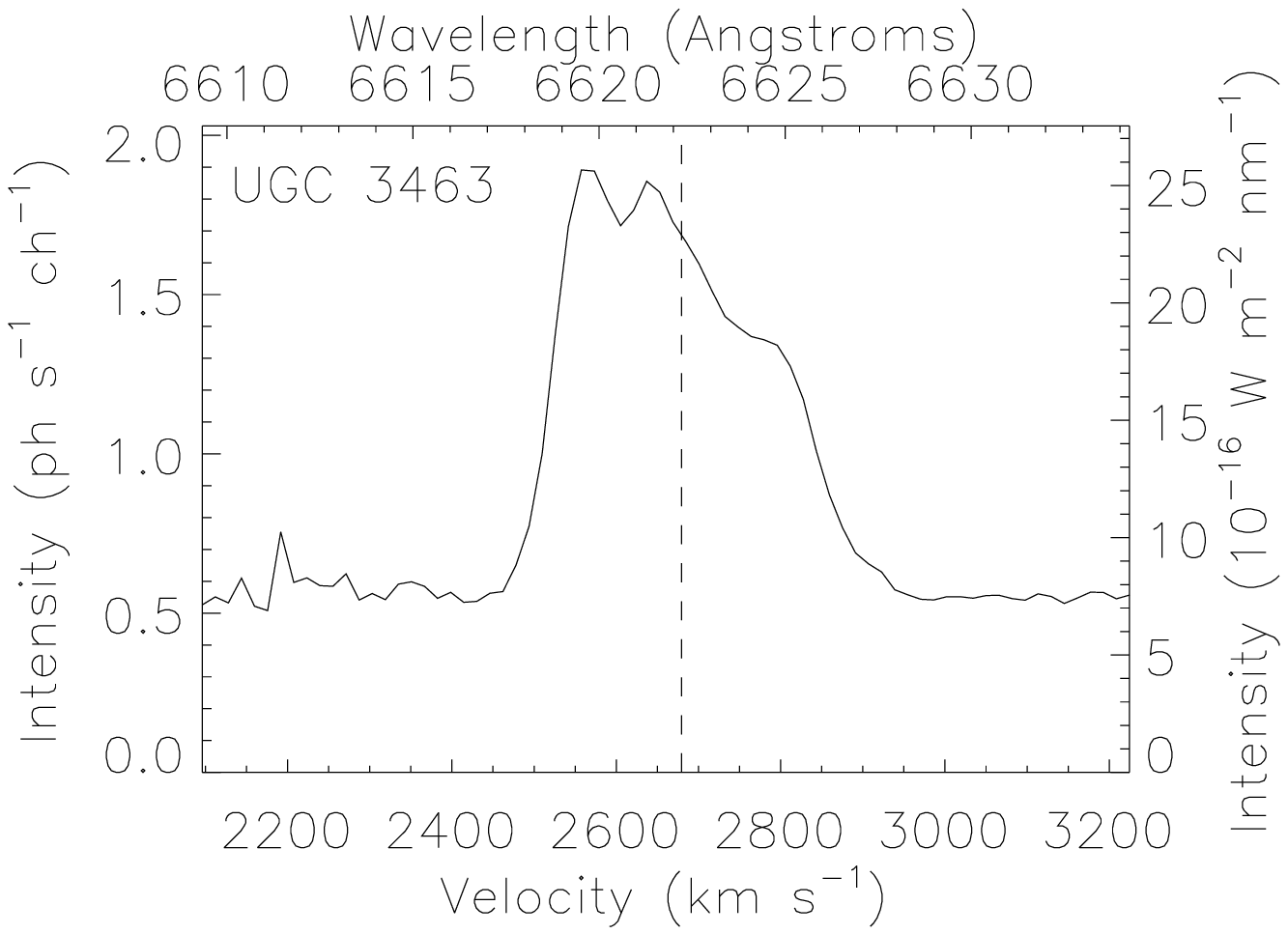}
\includegraphics[width=3.5cm]{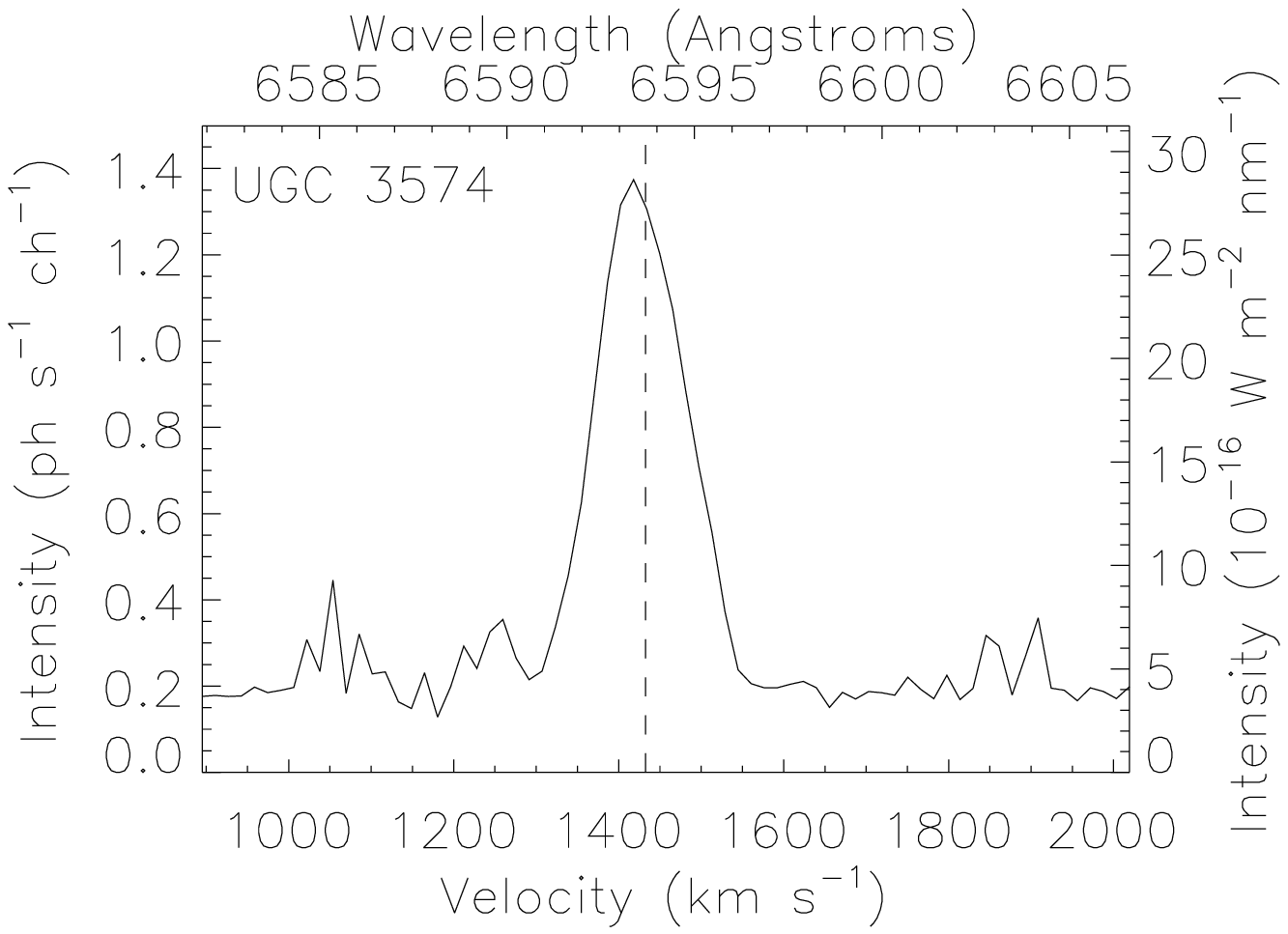}
\includegraphics[width=3.5cm]{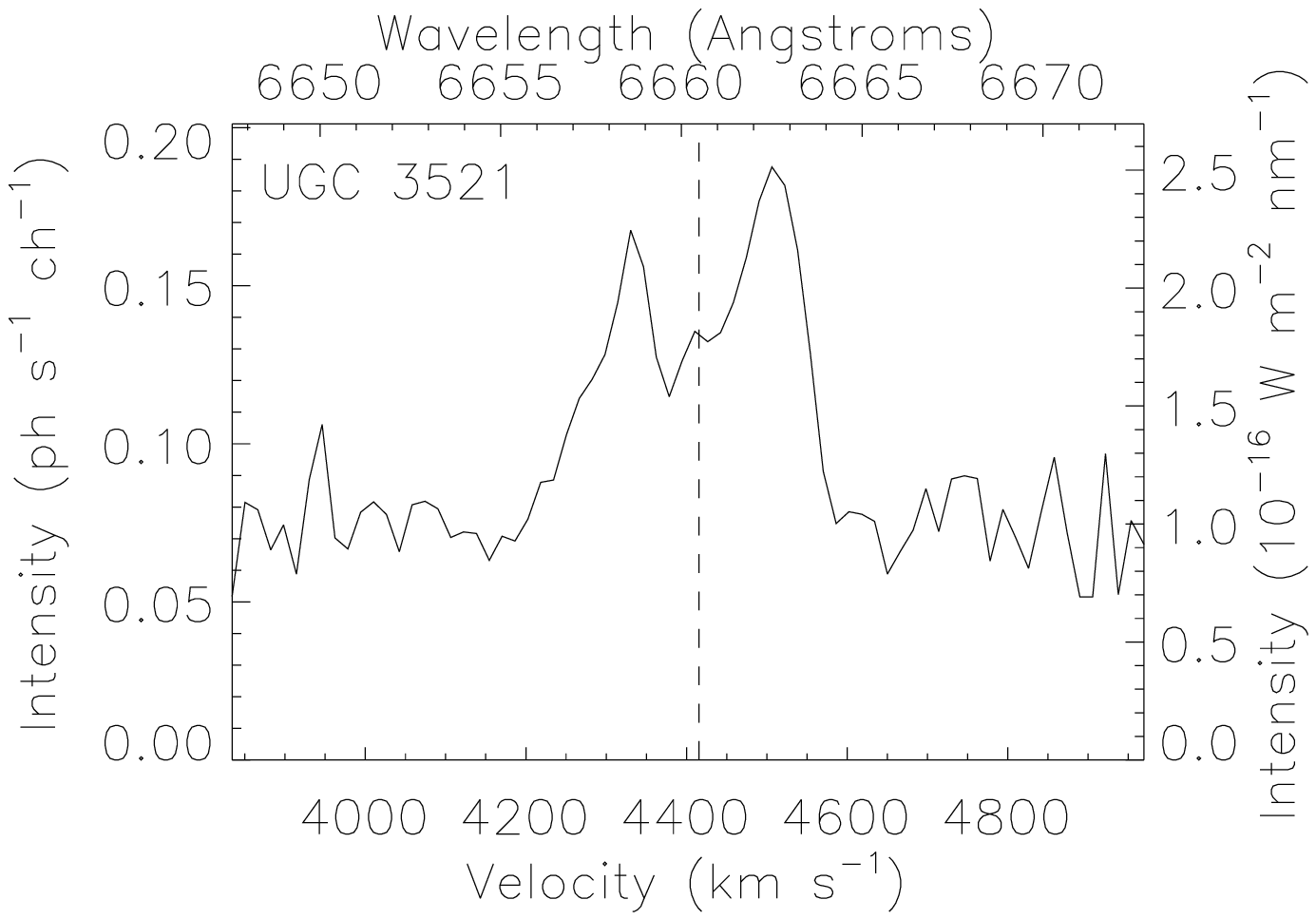}
\includegraphics[width=3.5cm]{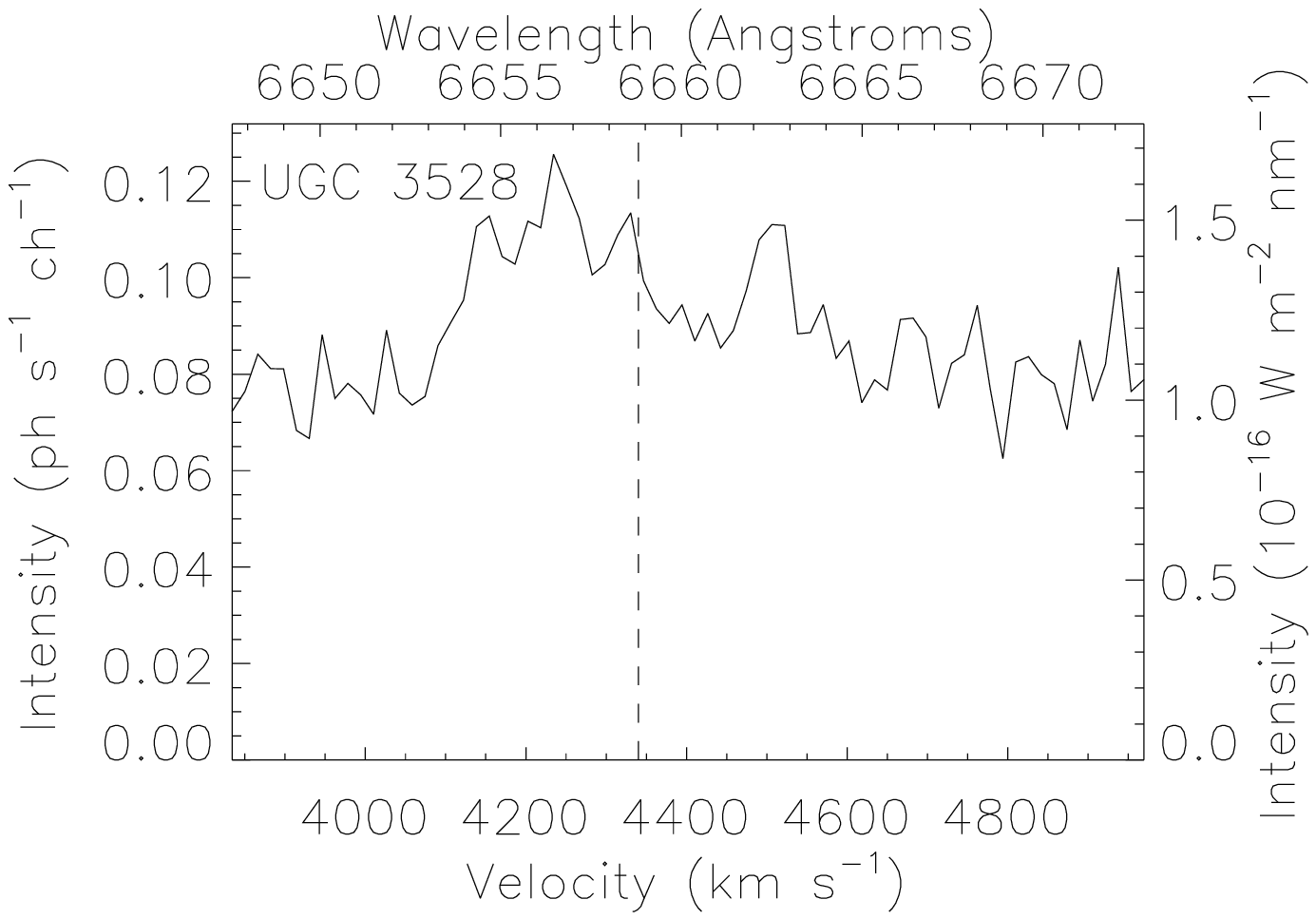}
\includegraphics[width=3.5cm]{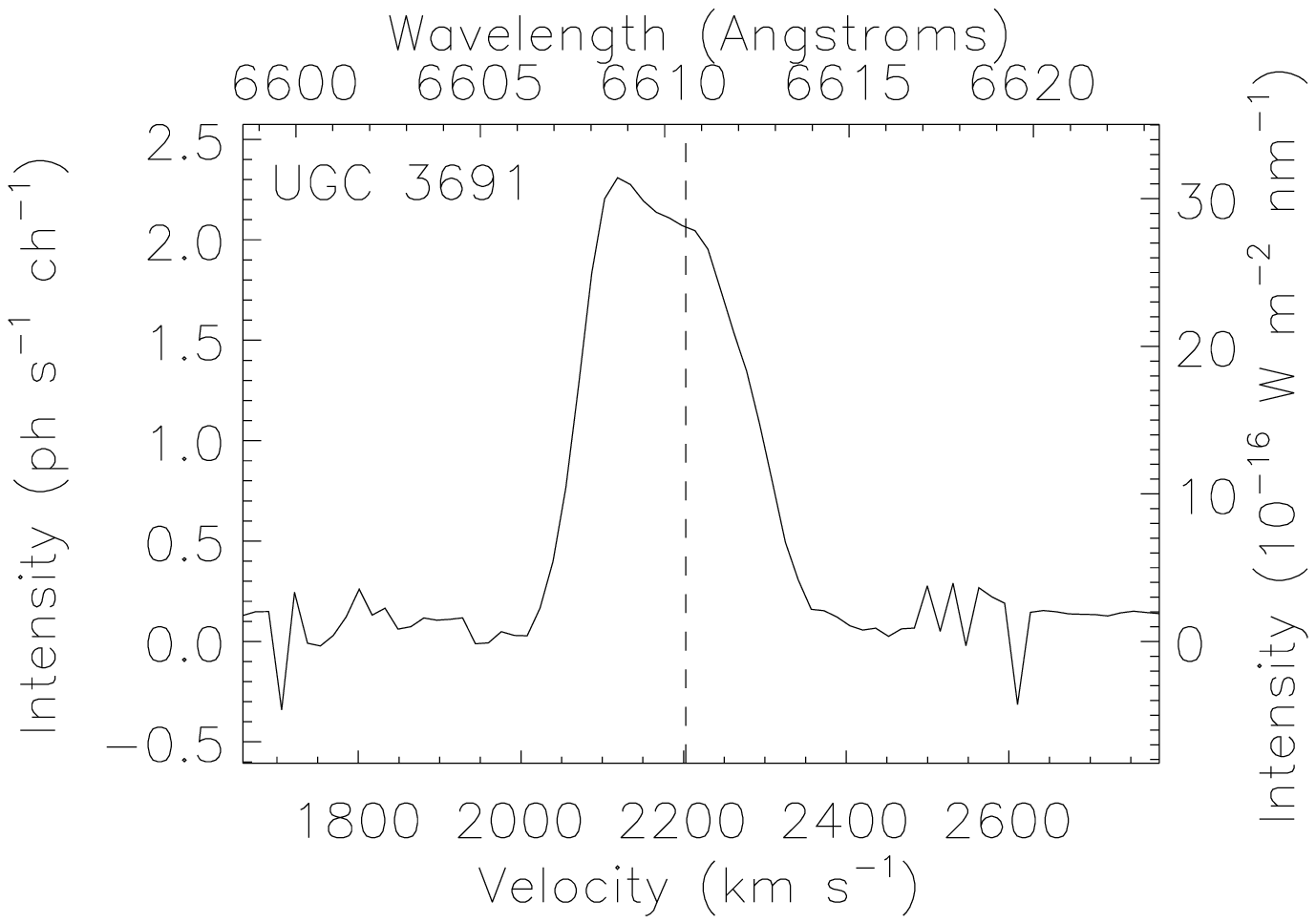}
\includegraphics[width=3.5cm]{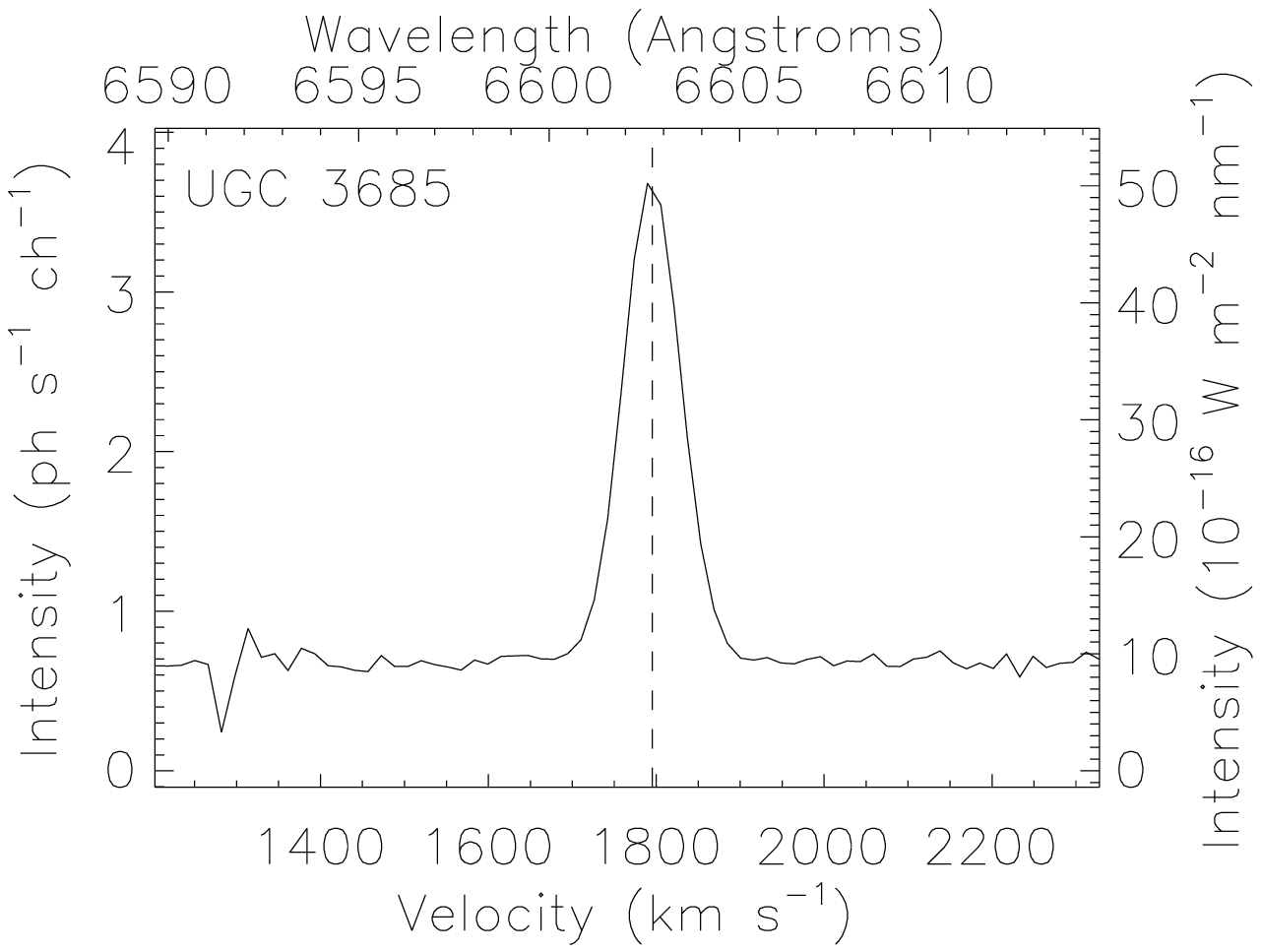}
\includegraphics[width=3.5cm]{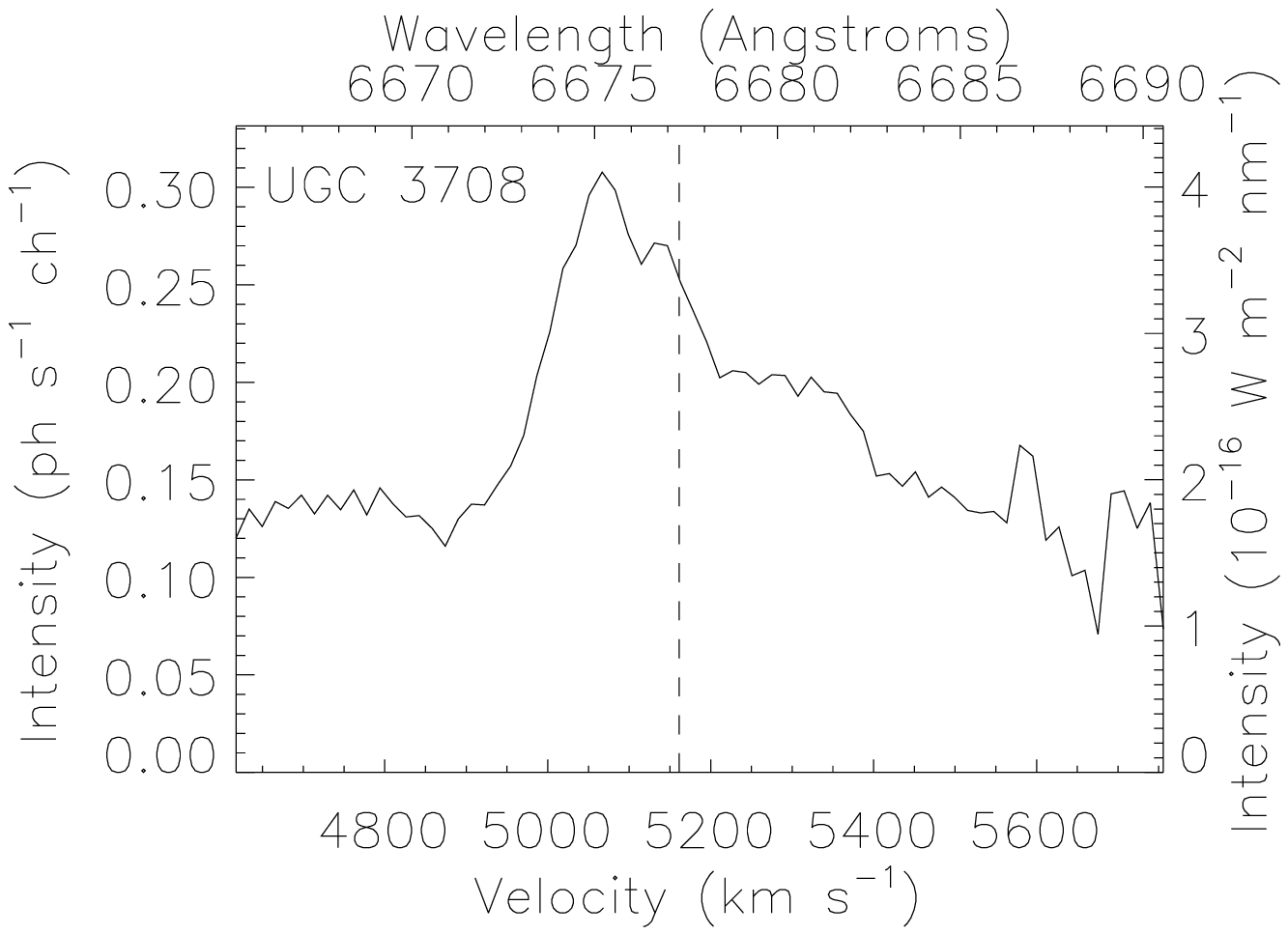}
\includegraphics[width=3.5cm]{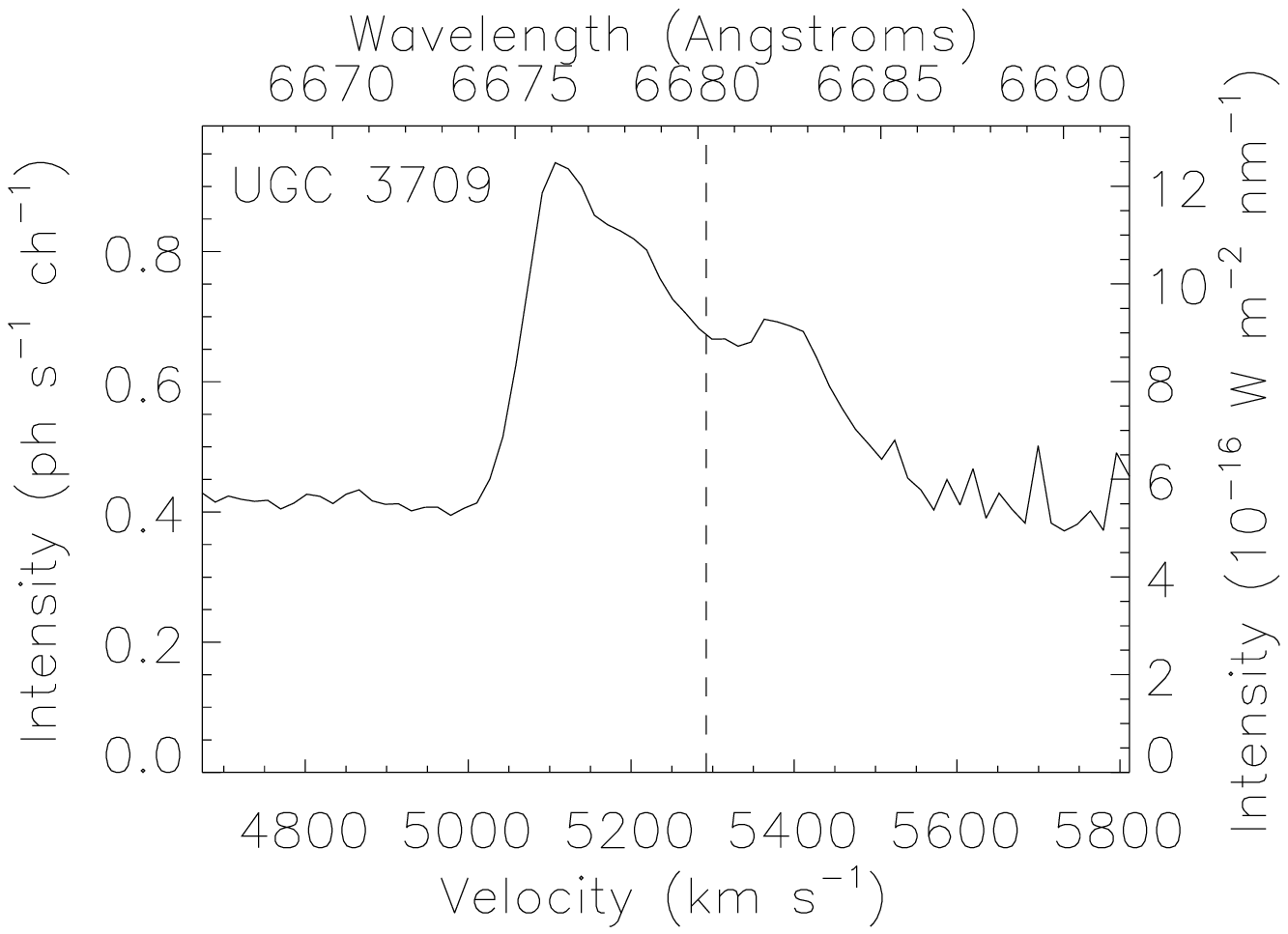}
\includegraphics[width=3.5cm]{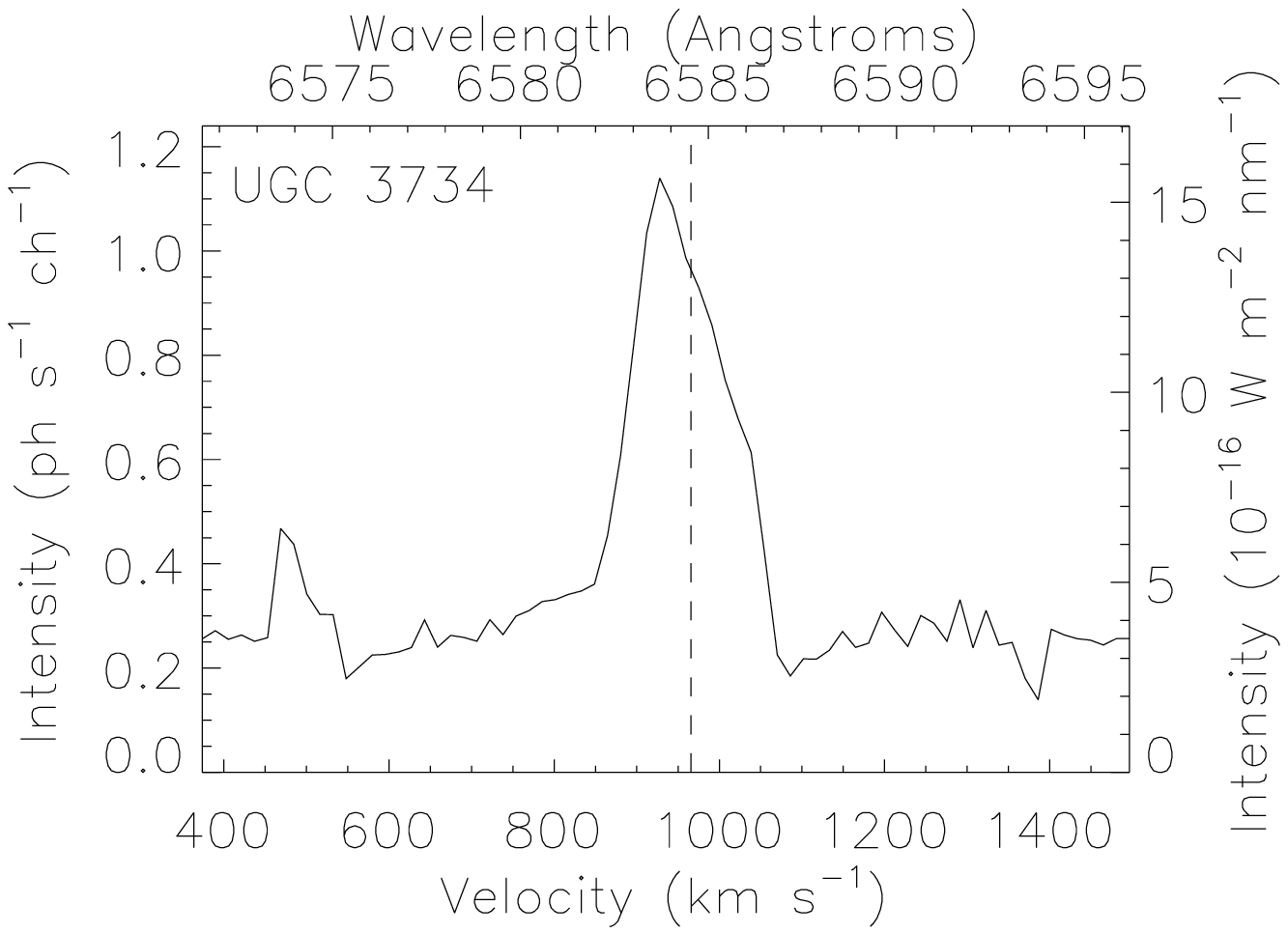}
\includegraphics[width=3.5cm]{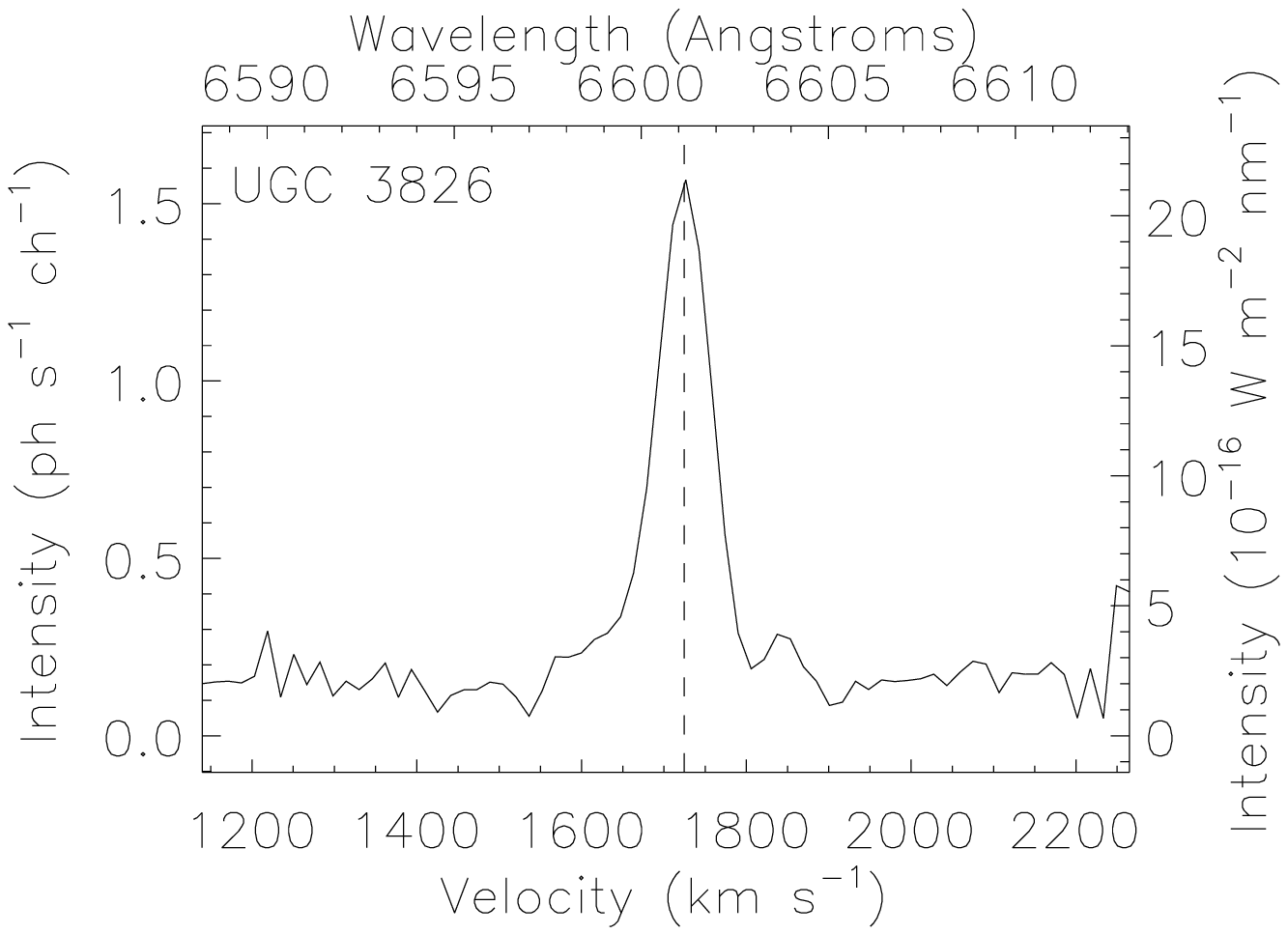}
\includegraphics[width=3.5cm]{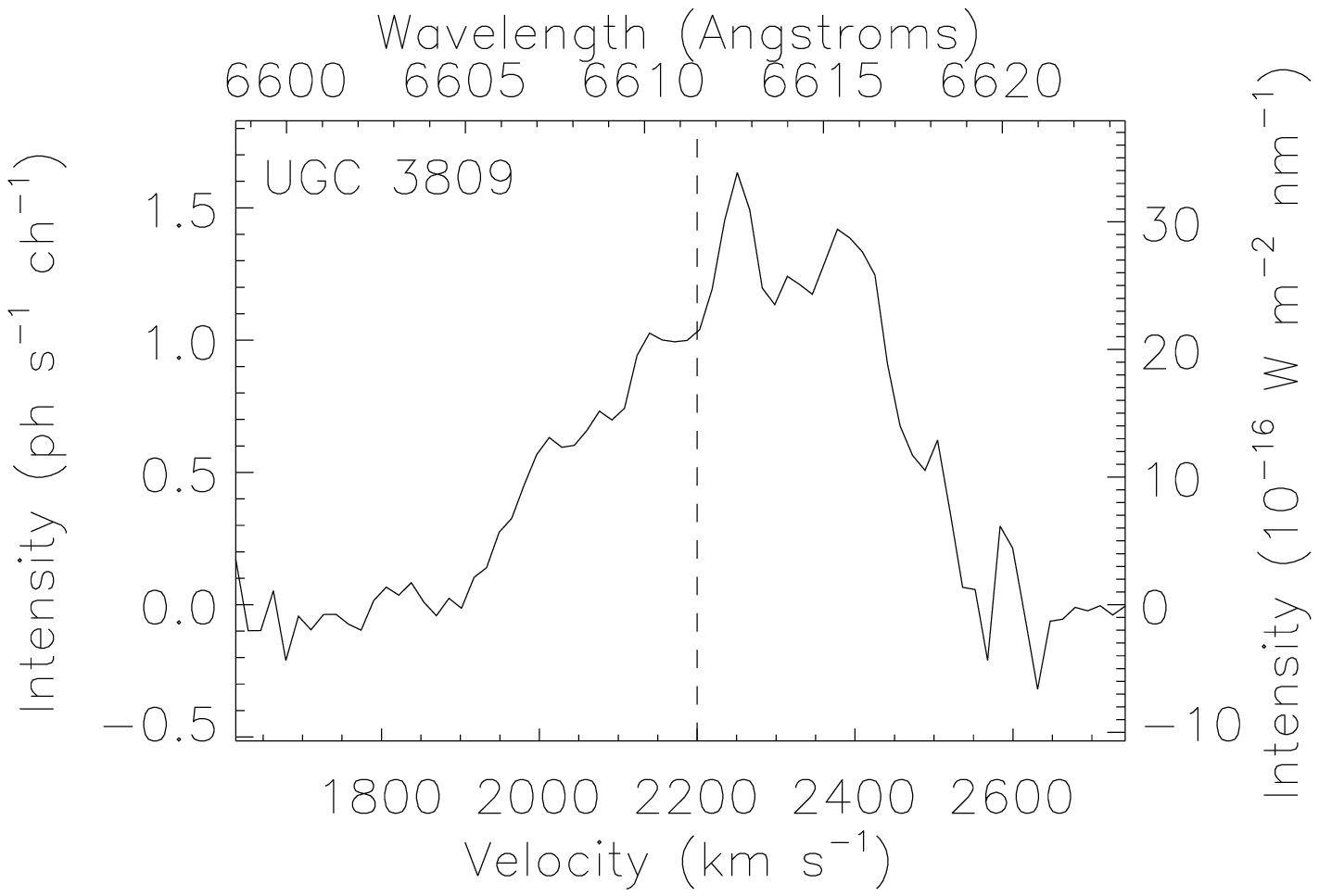}
\includegraphics[width=3.5cm]{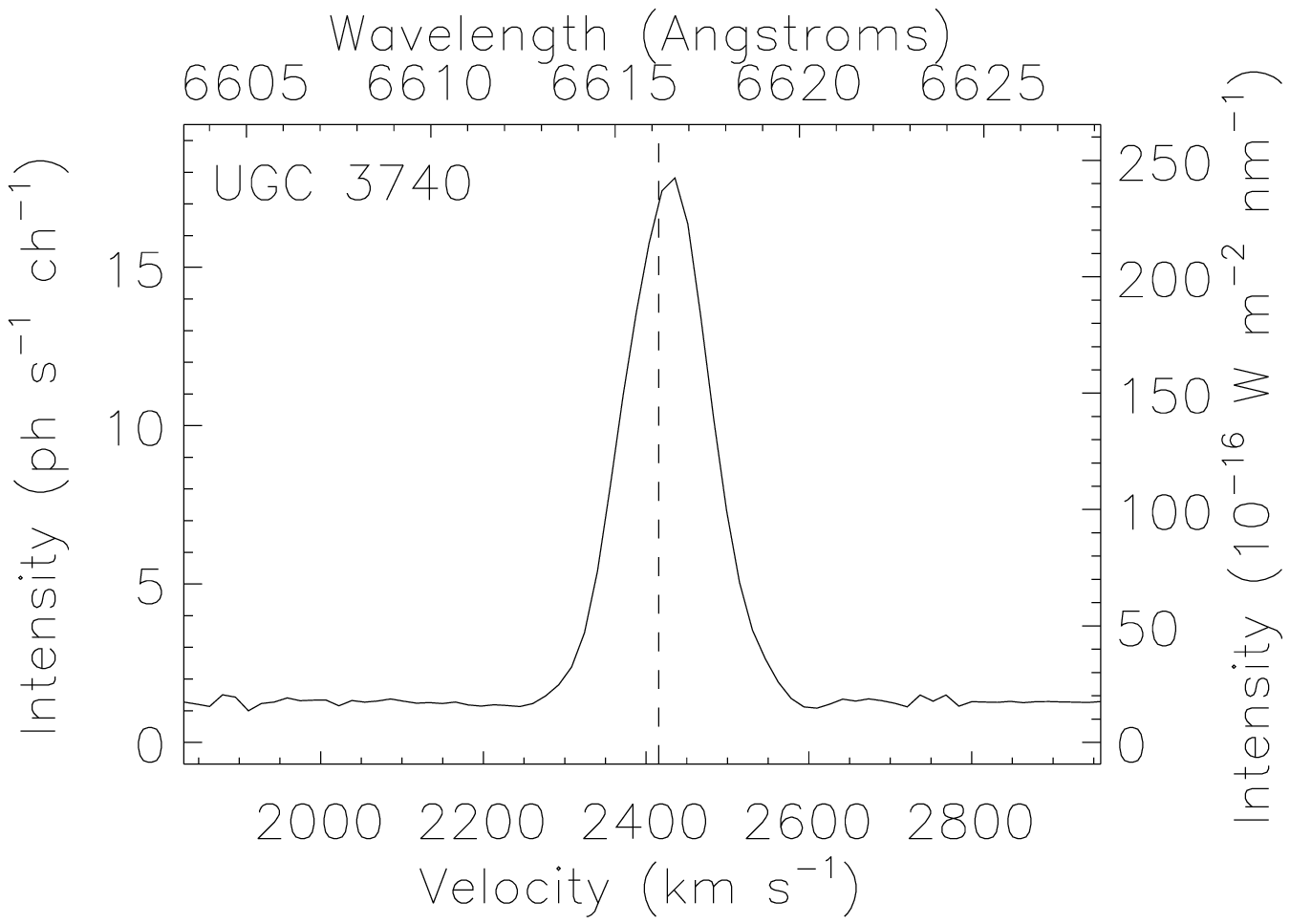}
\includegraphics[width=3.5cm]{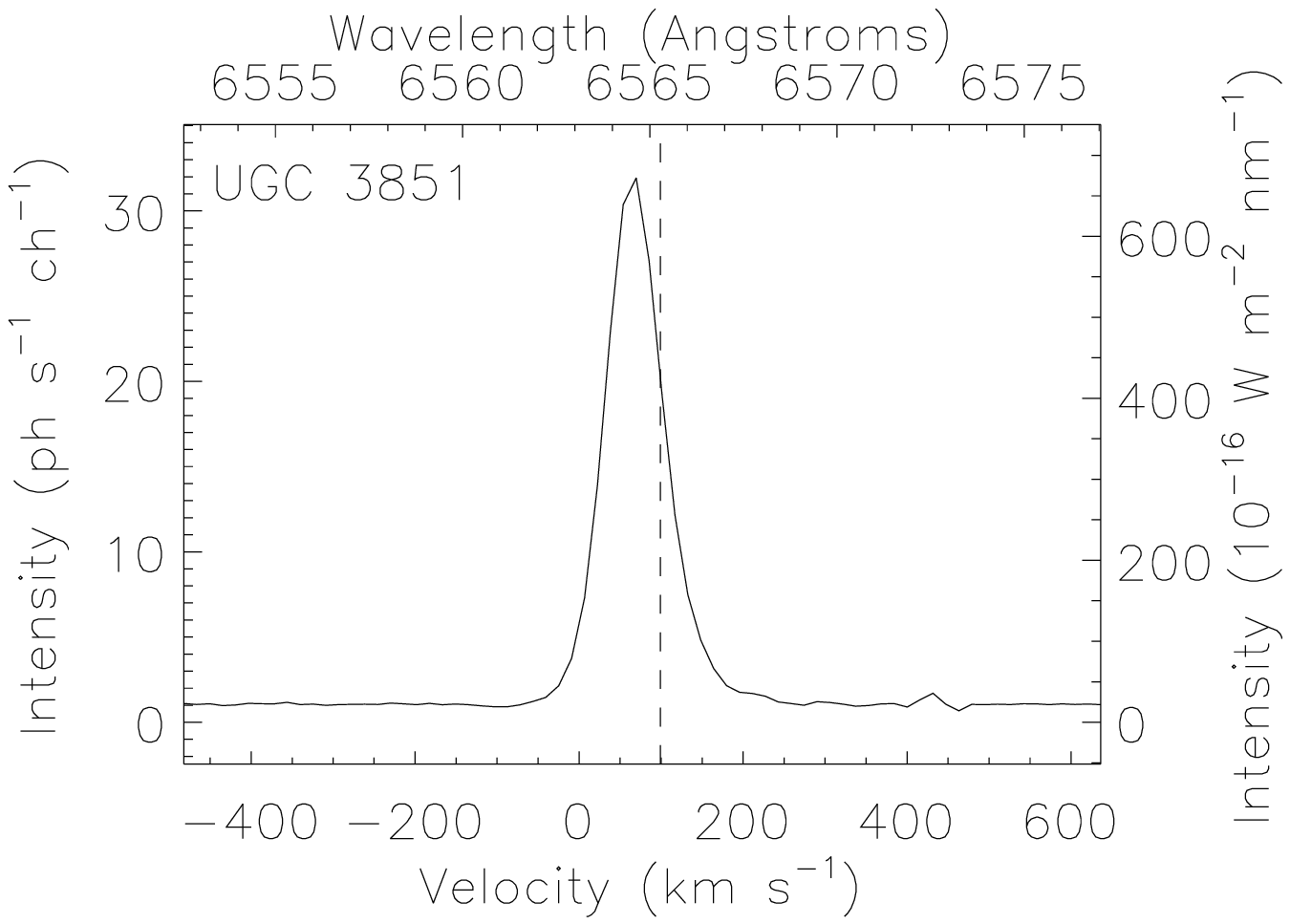}
\end{center}
\end{minipage}
\end{figure}
\clearpage
\begin{figure}
\begin{minipage}{180mm}
\begin{center}
\includegraphics[width=3.5cm]{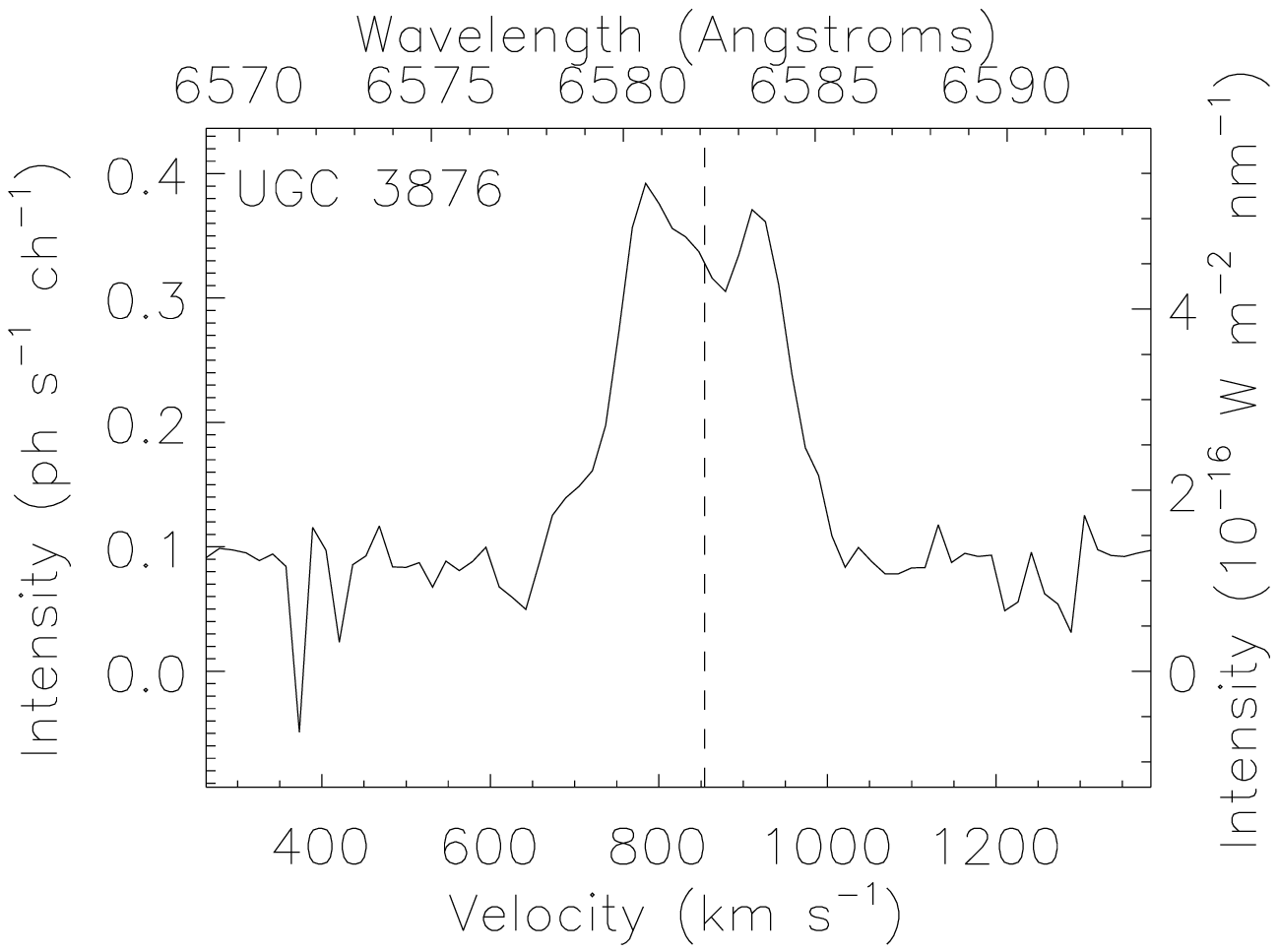}
\includegraphics[width=3.5cm]{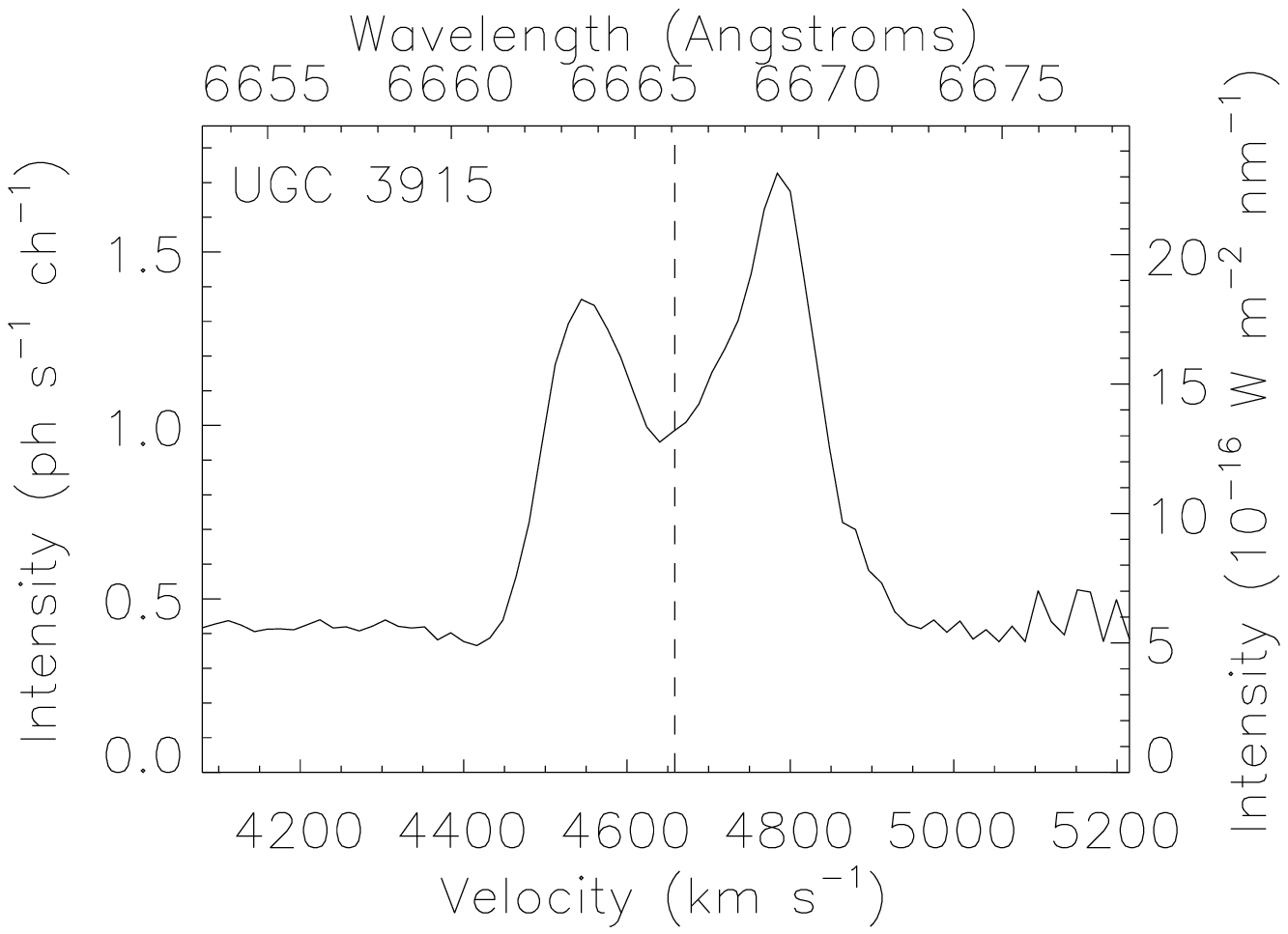}
\includegraphics[width=3.5cm]{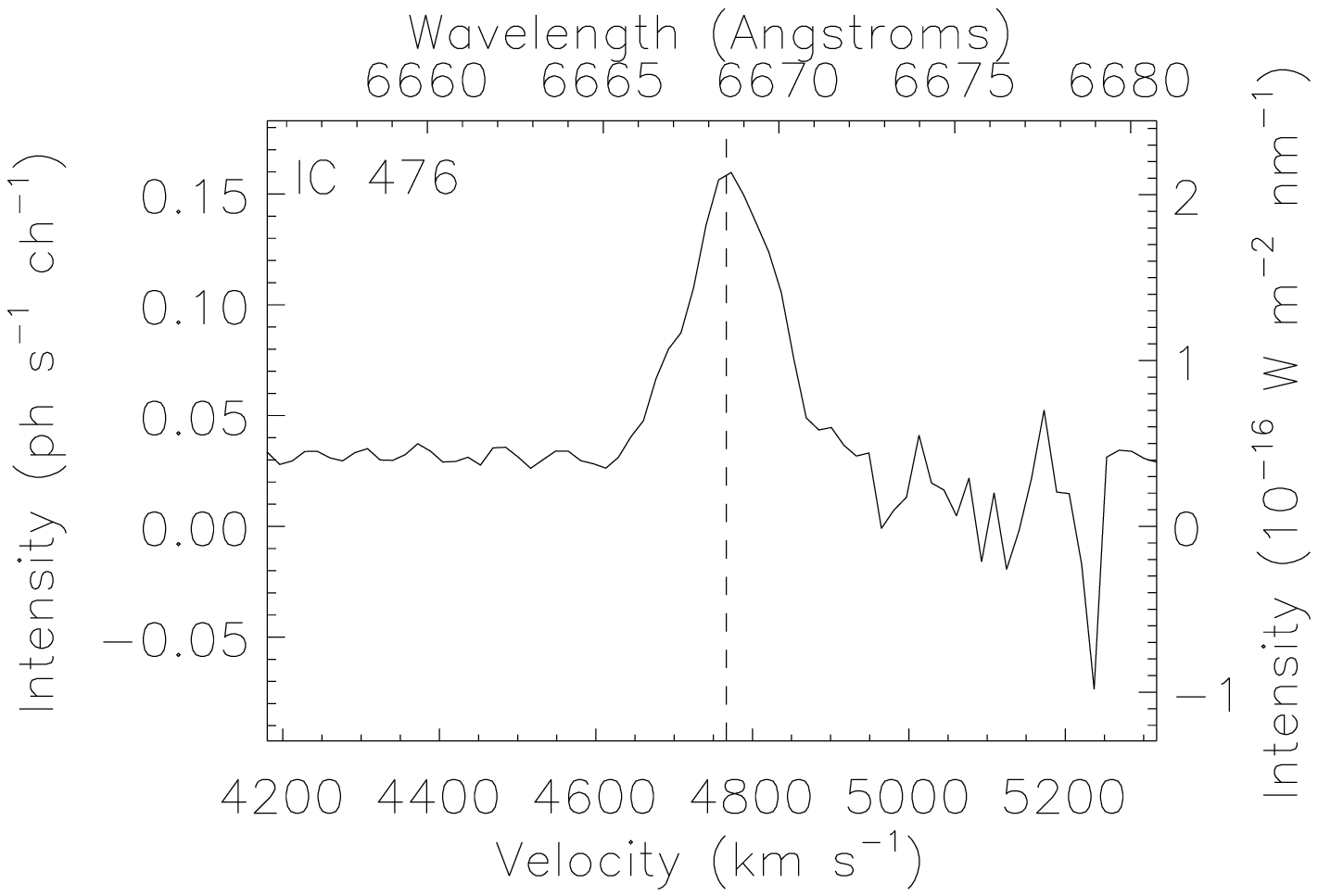}
\includegraphics[width=3.5cm]{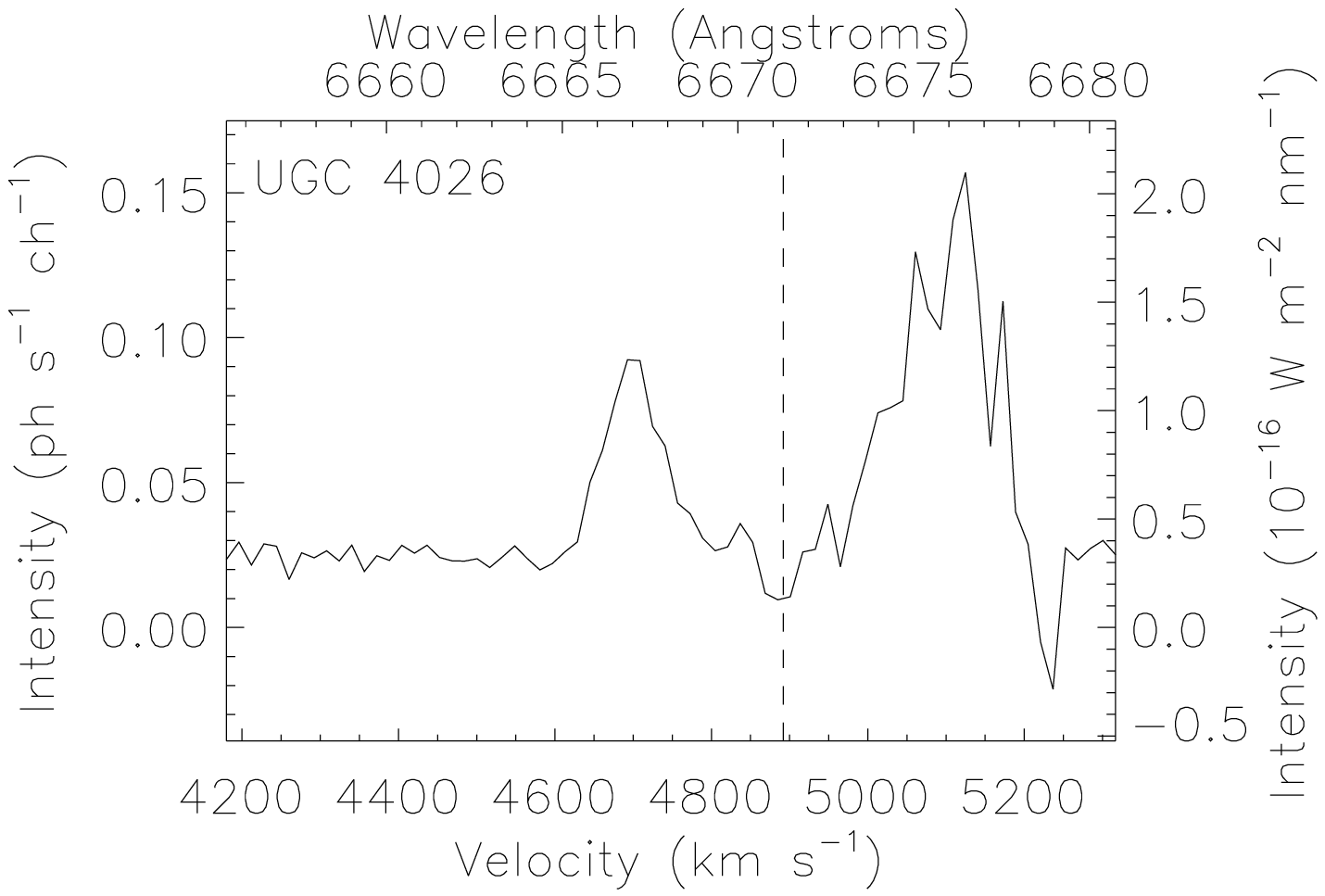}
\includegraphics[width=3.5cm]{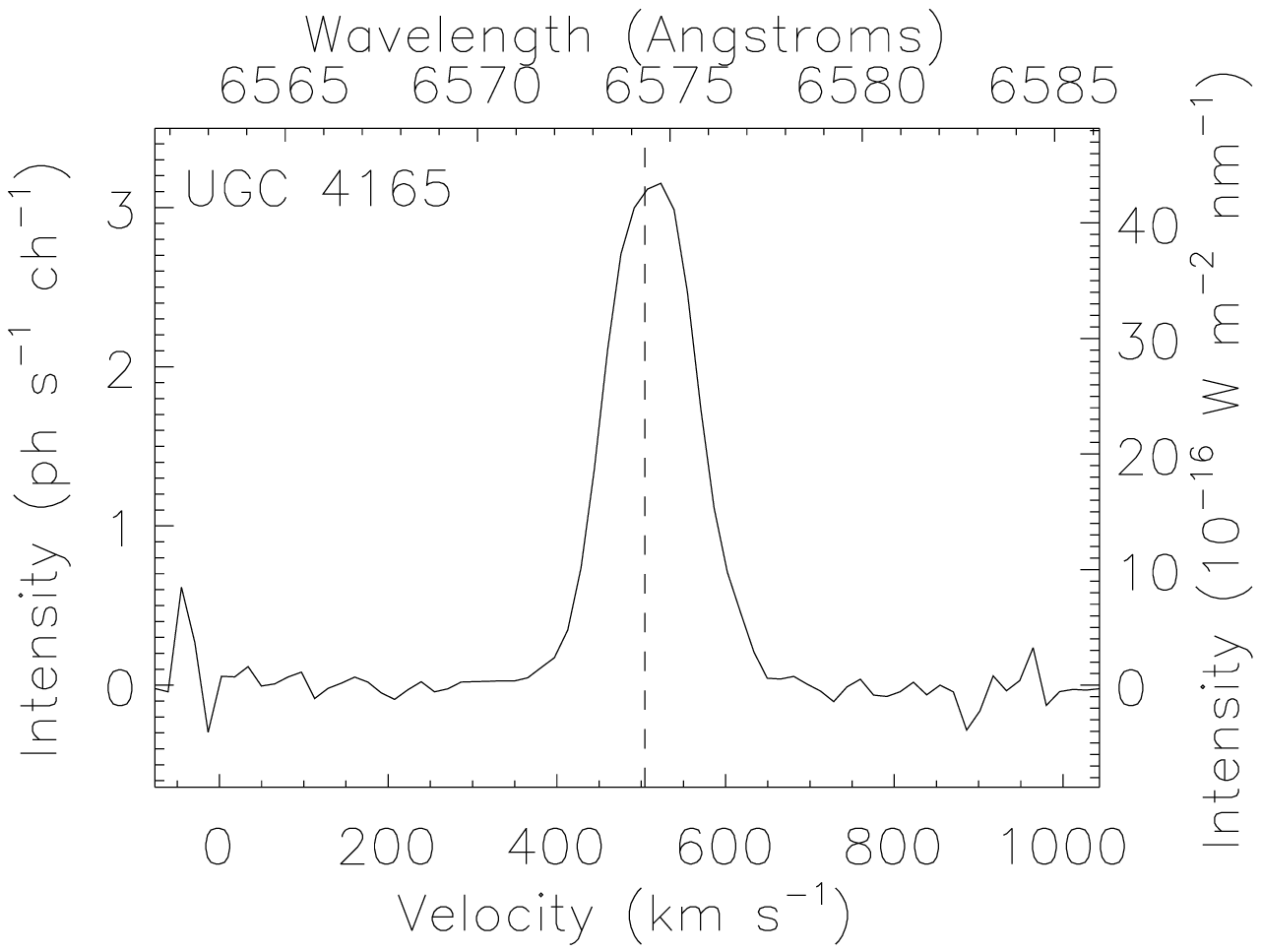}
\includegraphics[width=3.5cm]{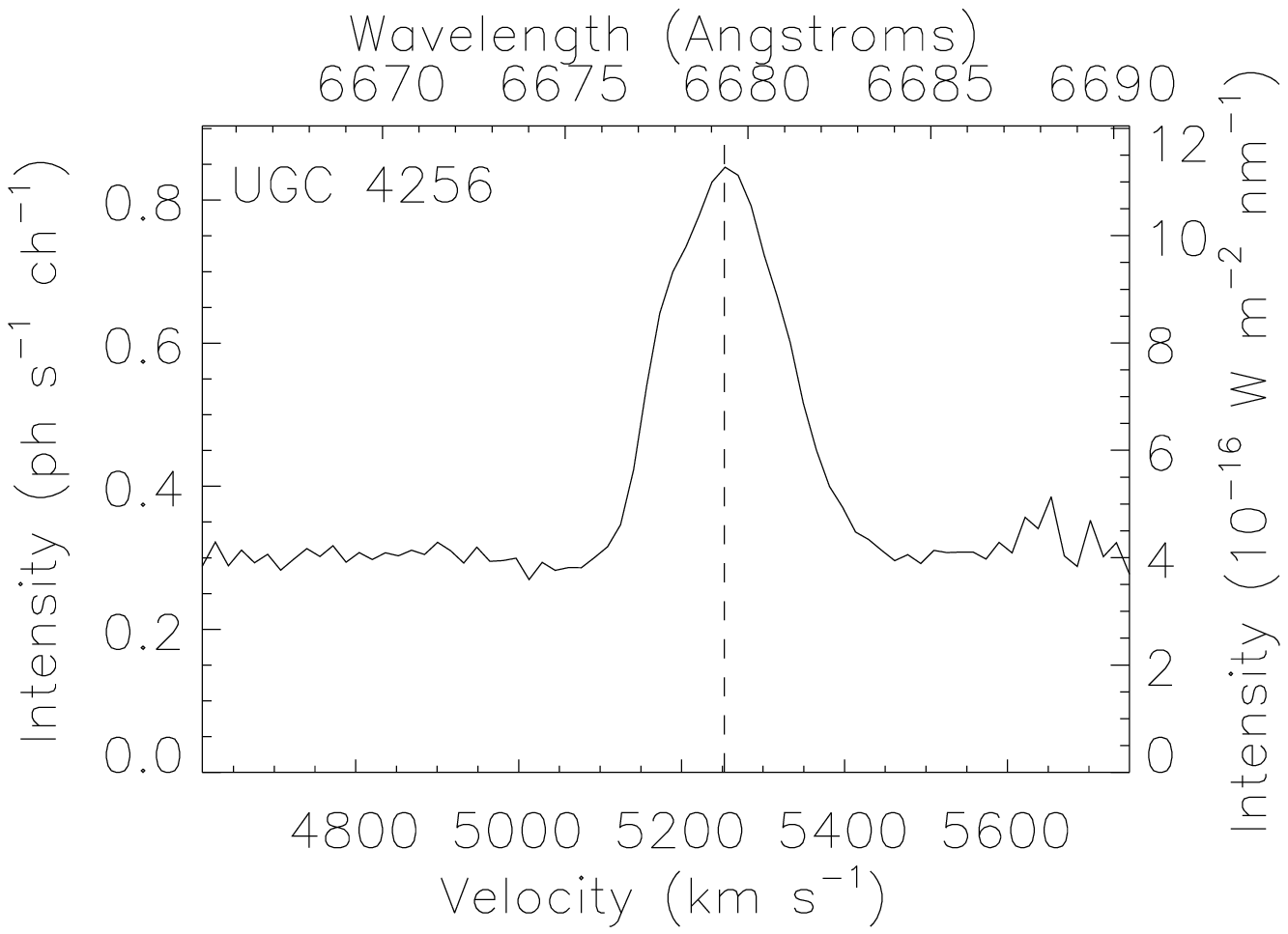}
\includegraphics[width=3.5cm]{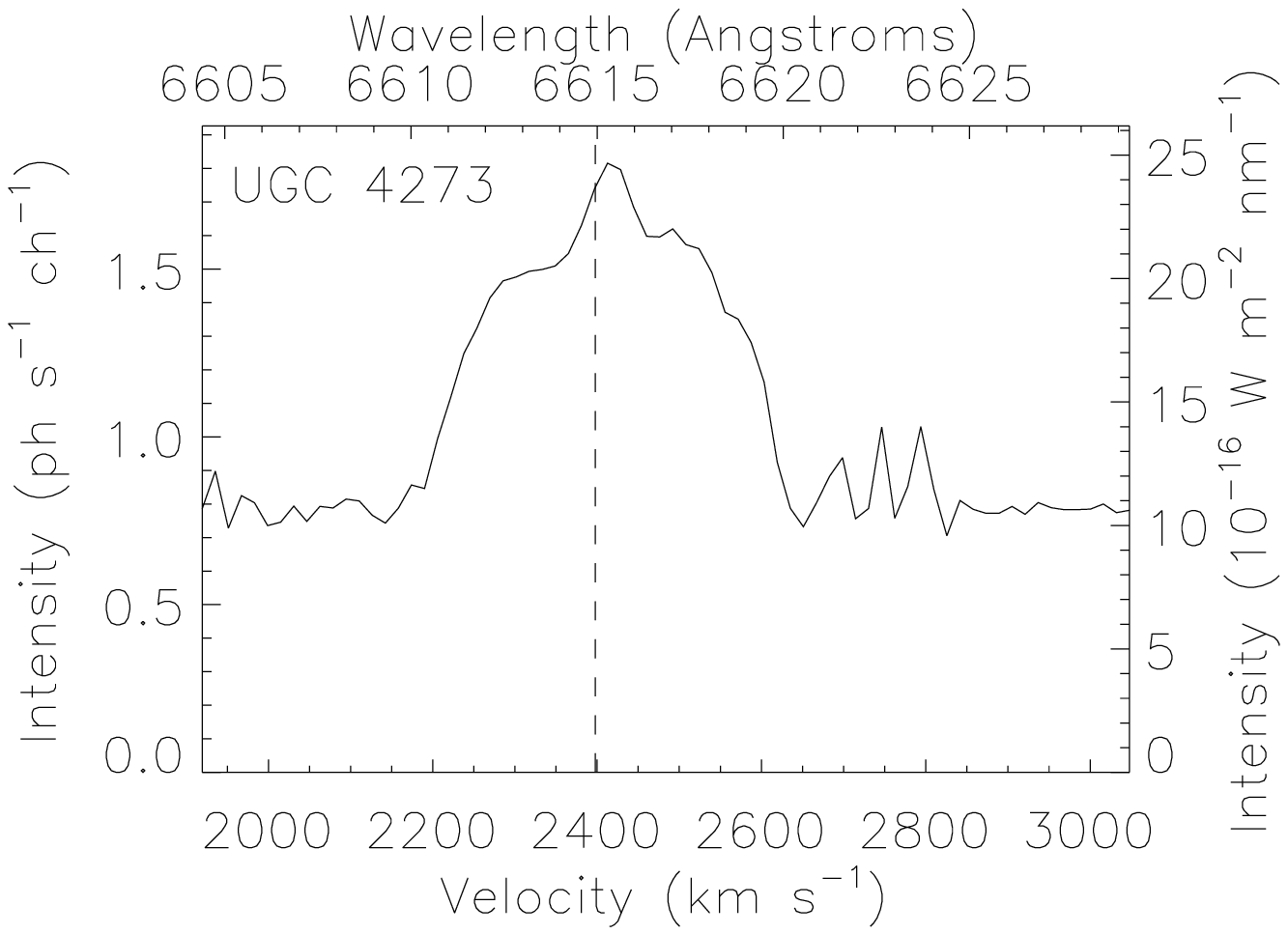}
\includegraphics[width=3.5cm]{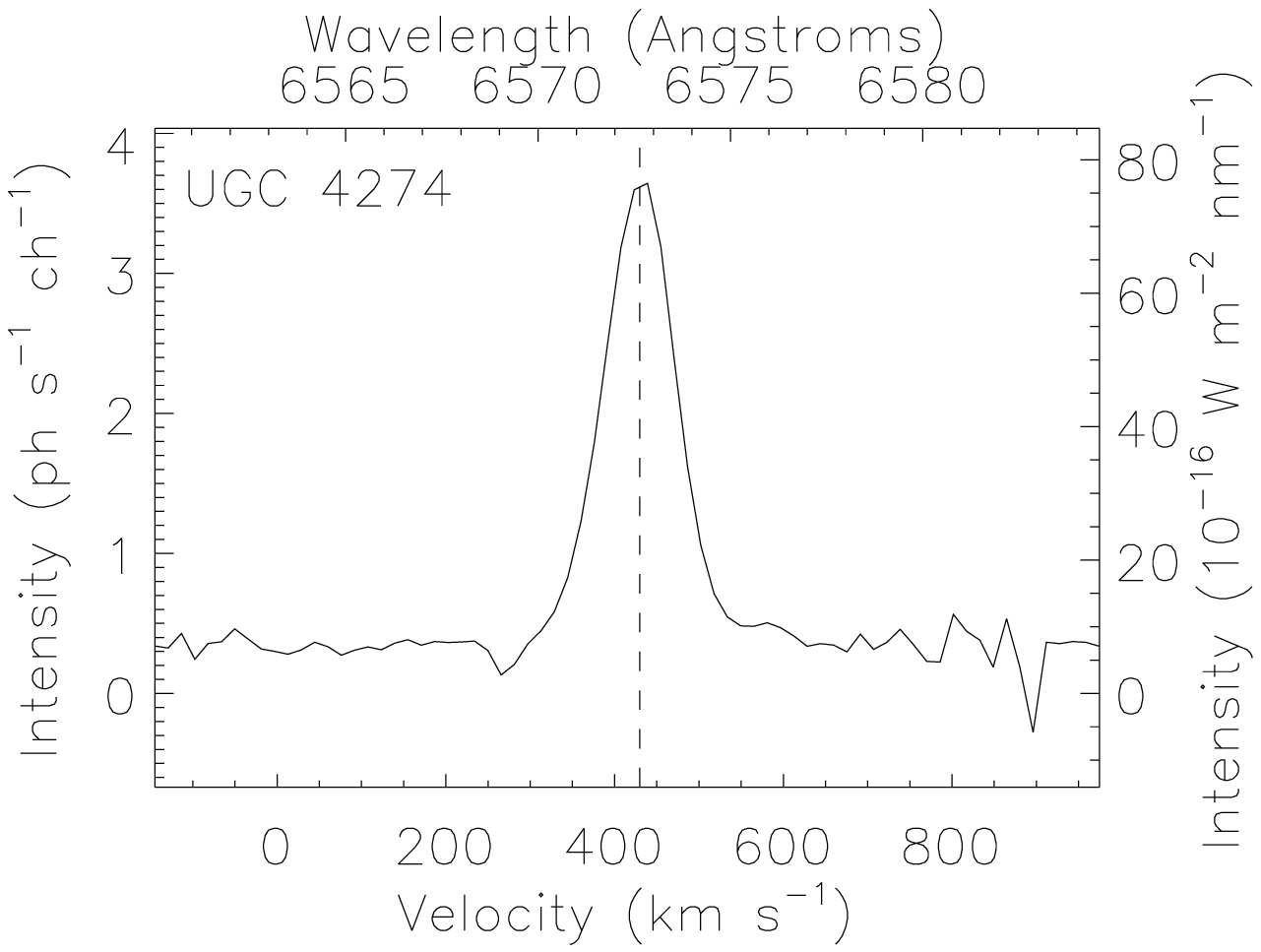}
\includegraphics[width=3.5cm]{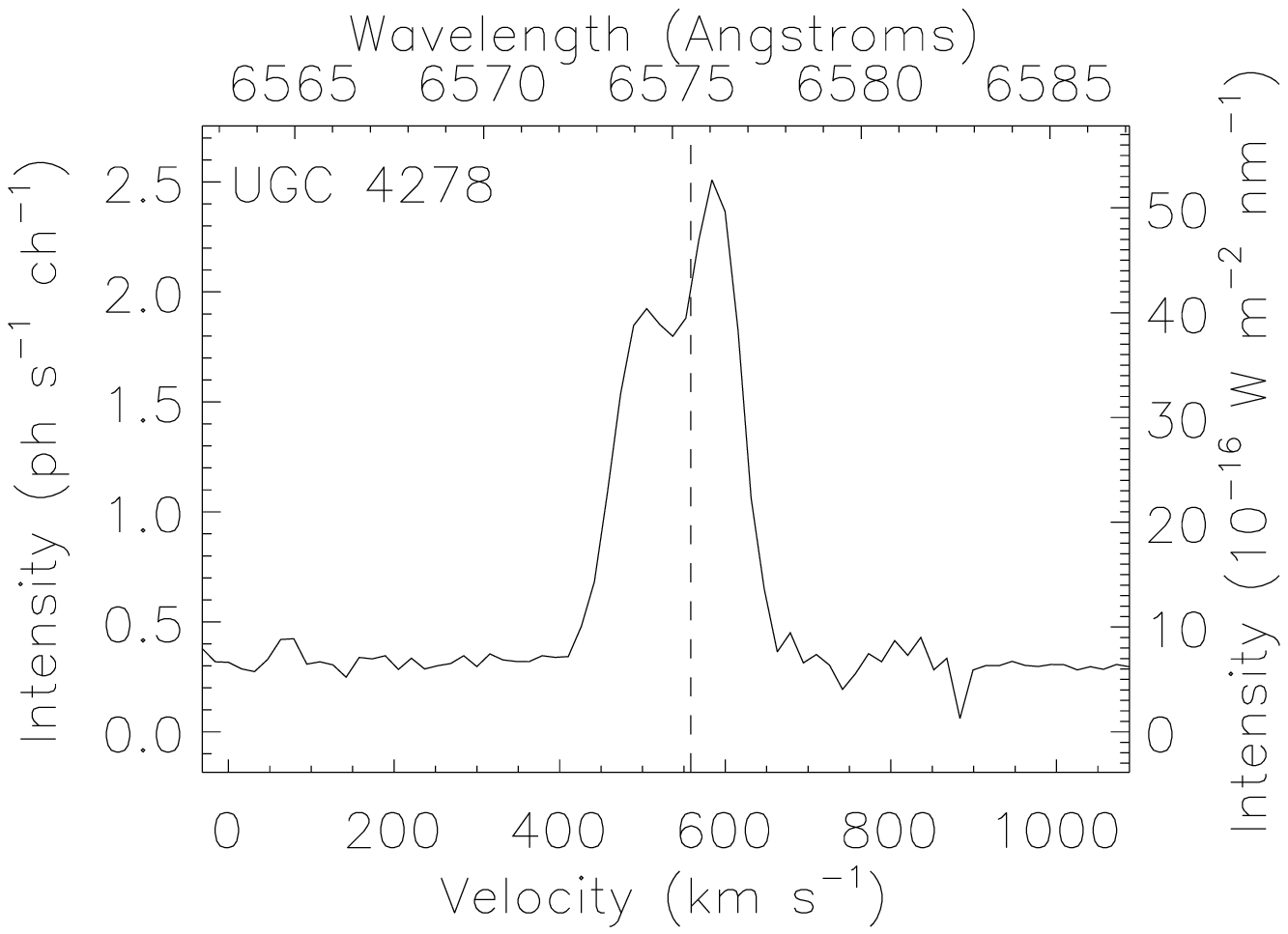}
\includegraphics[width=3.5cm]{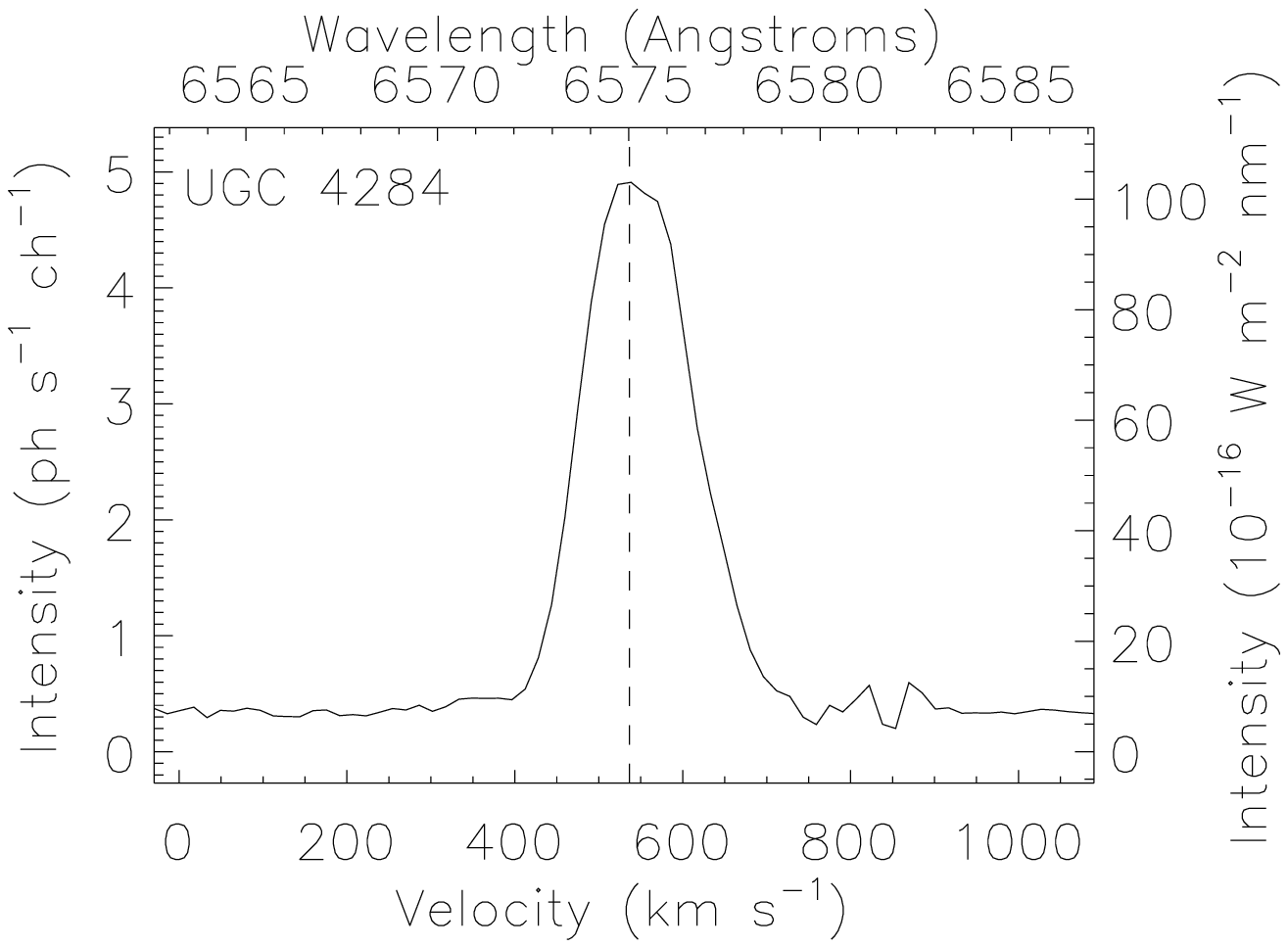}
\includegraphics[width=3.5cm]{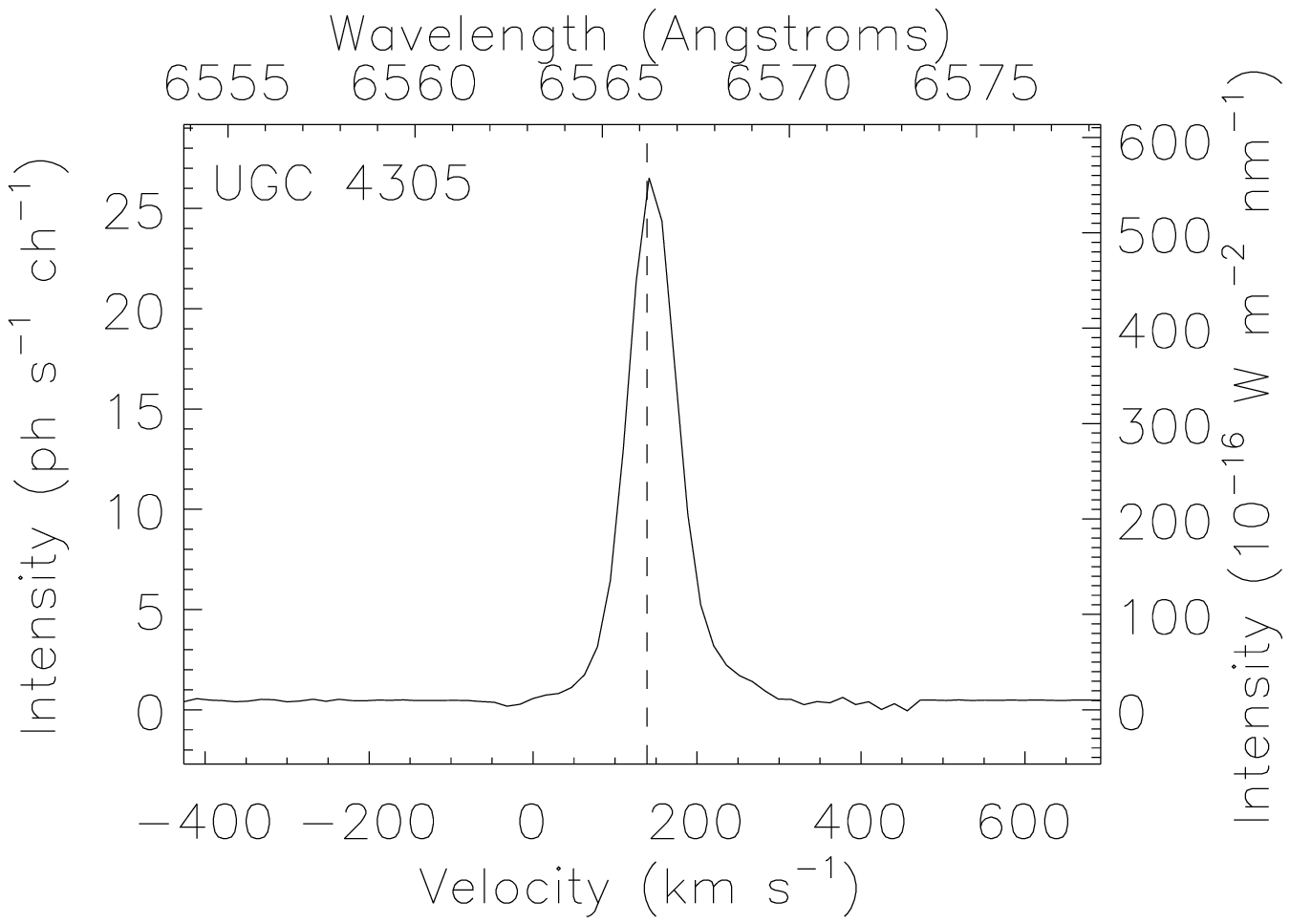}
\includegraphics[width=3.5cm]{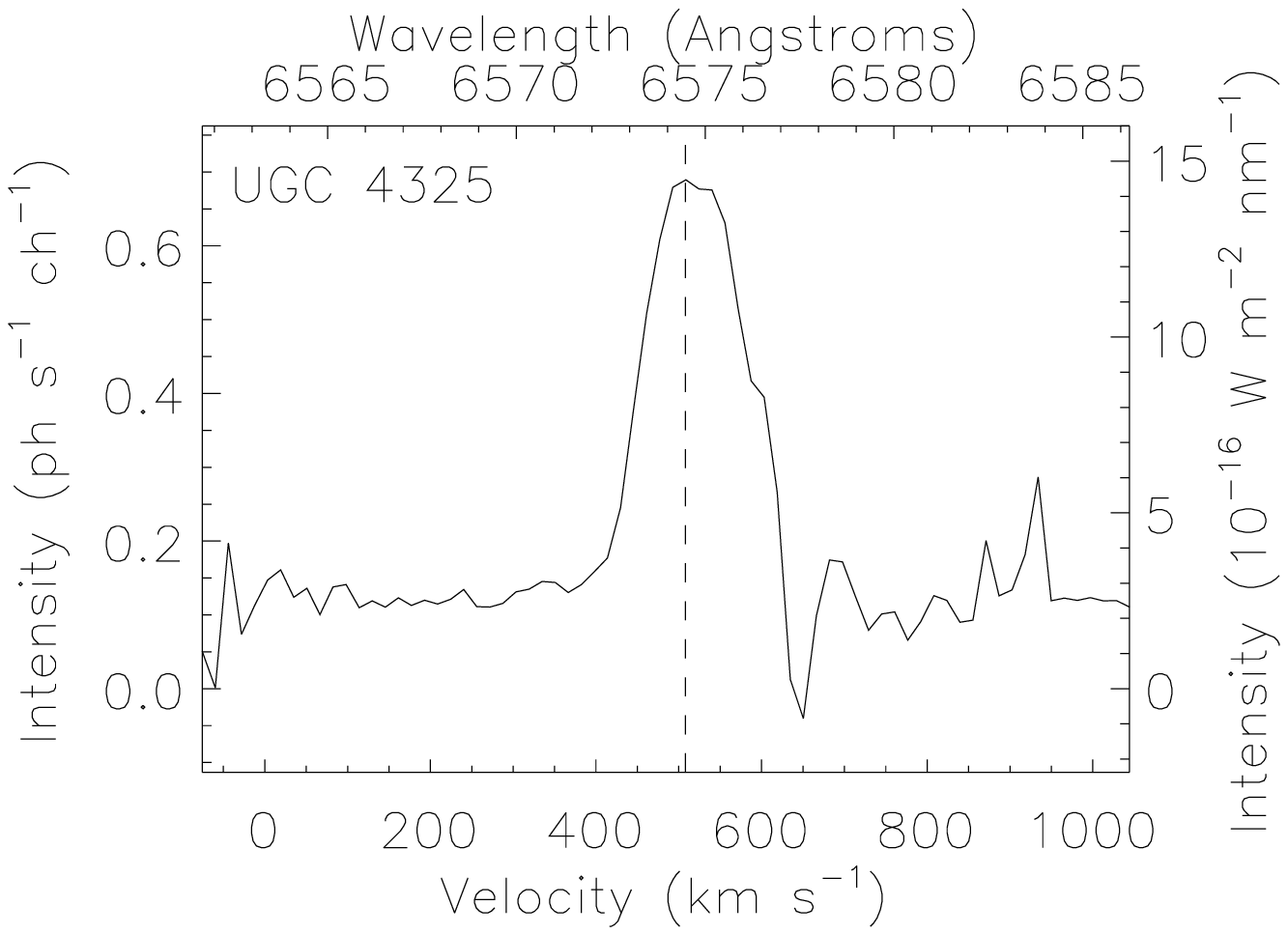}
\includegraphics[width=3.5cm]{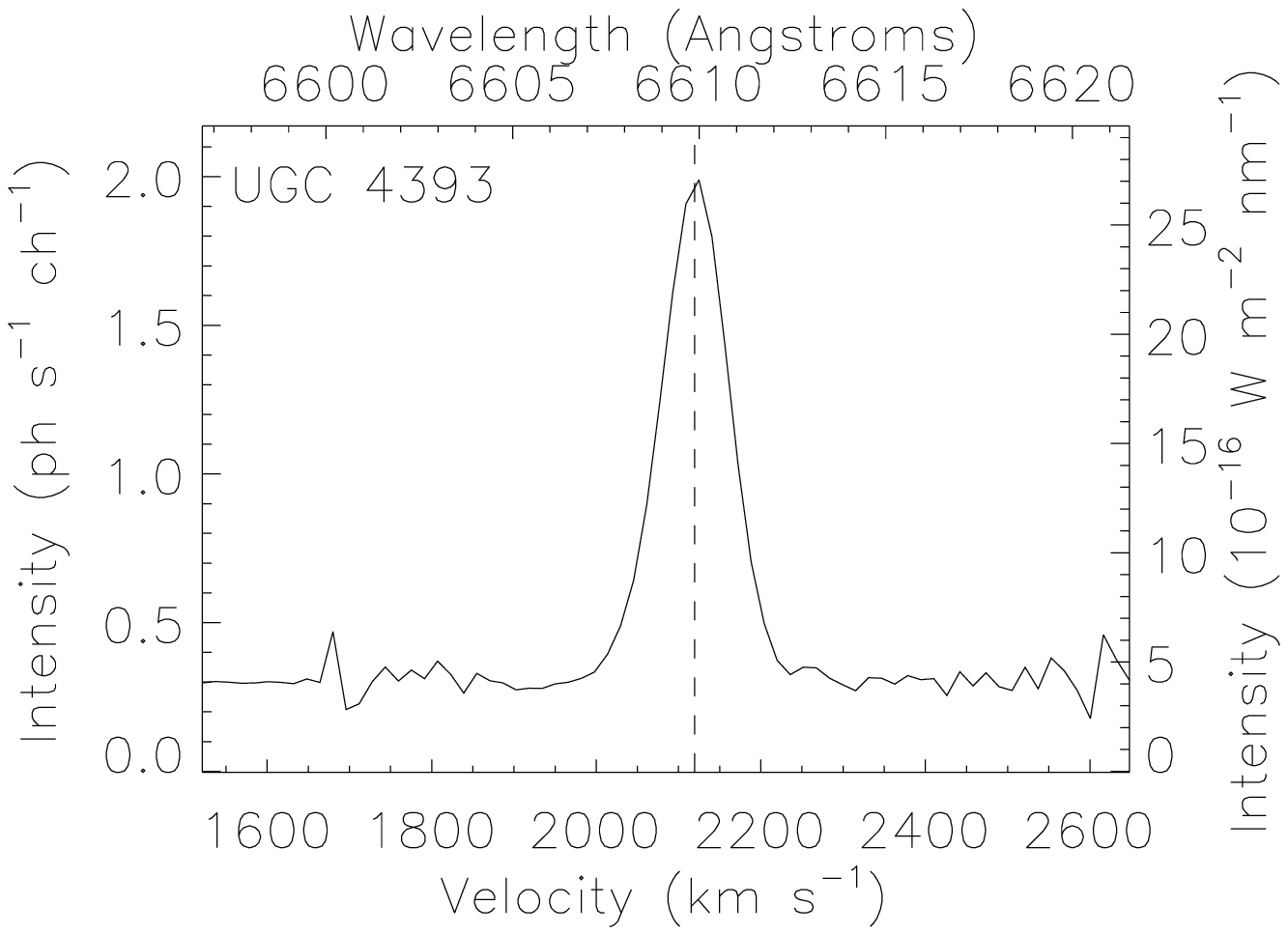}
\includegraphics[width=3.5cm]{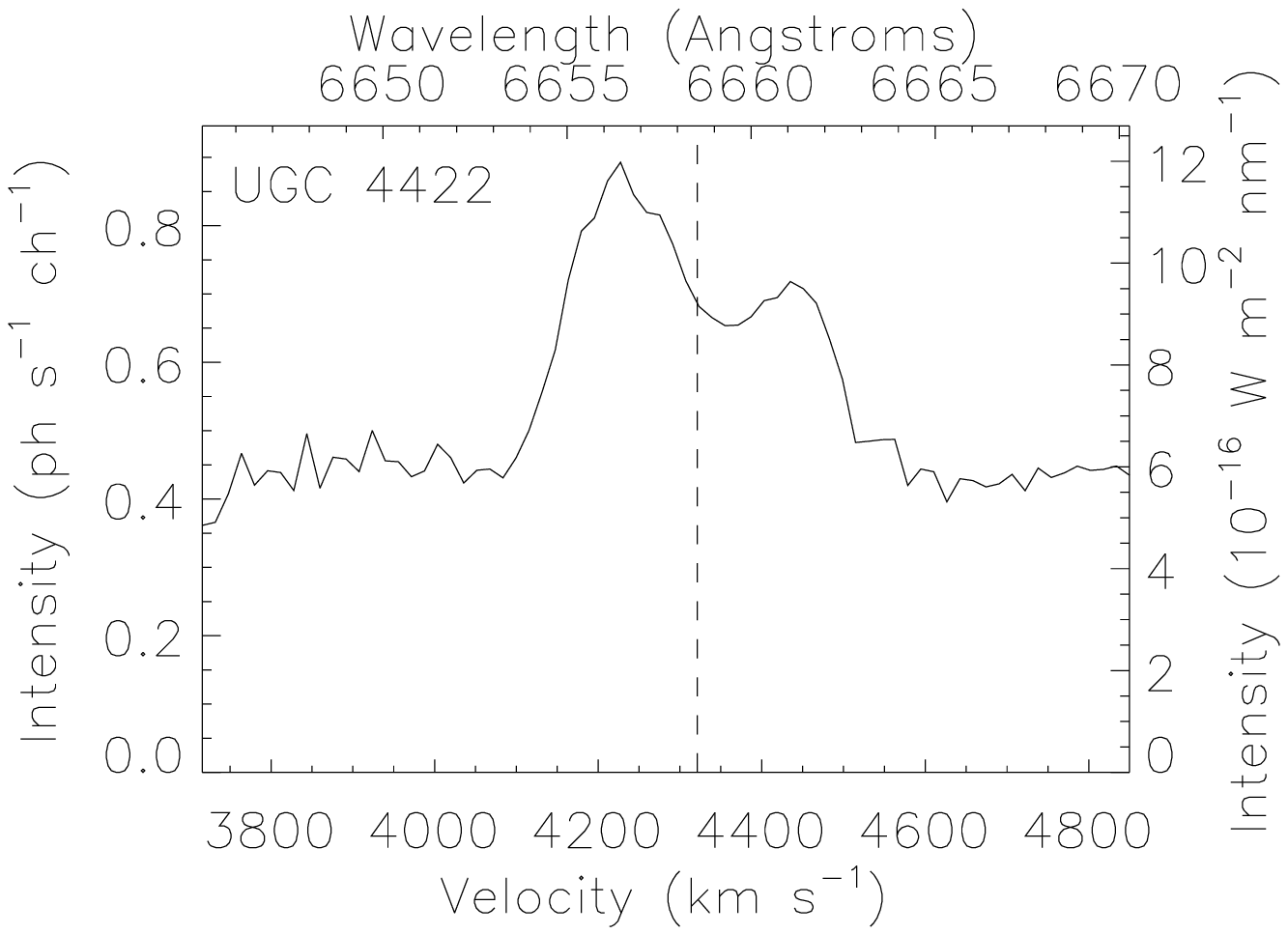}
\includegraphics[width=3.5cm]{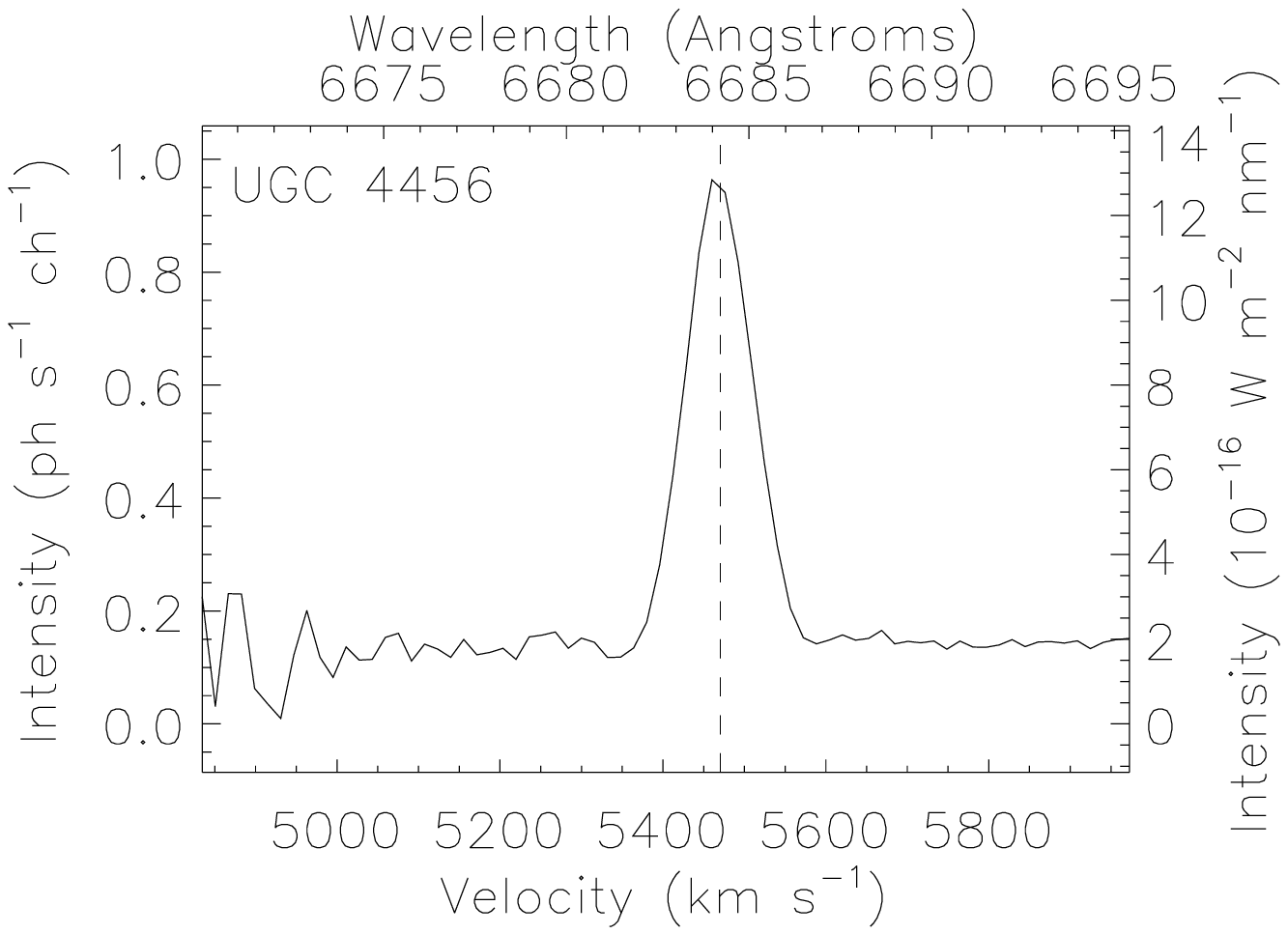}
\includegraphics[width=3.5cm]{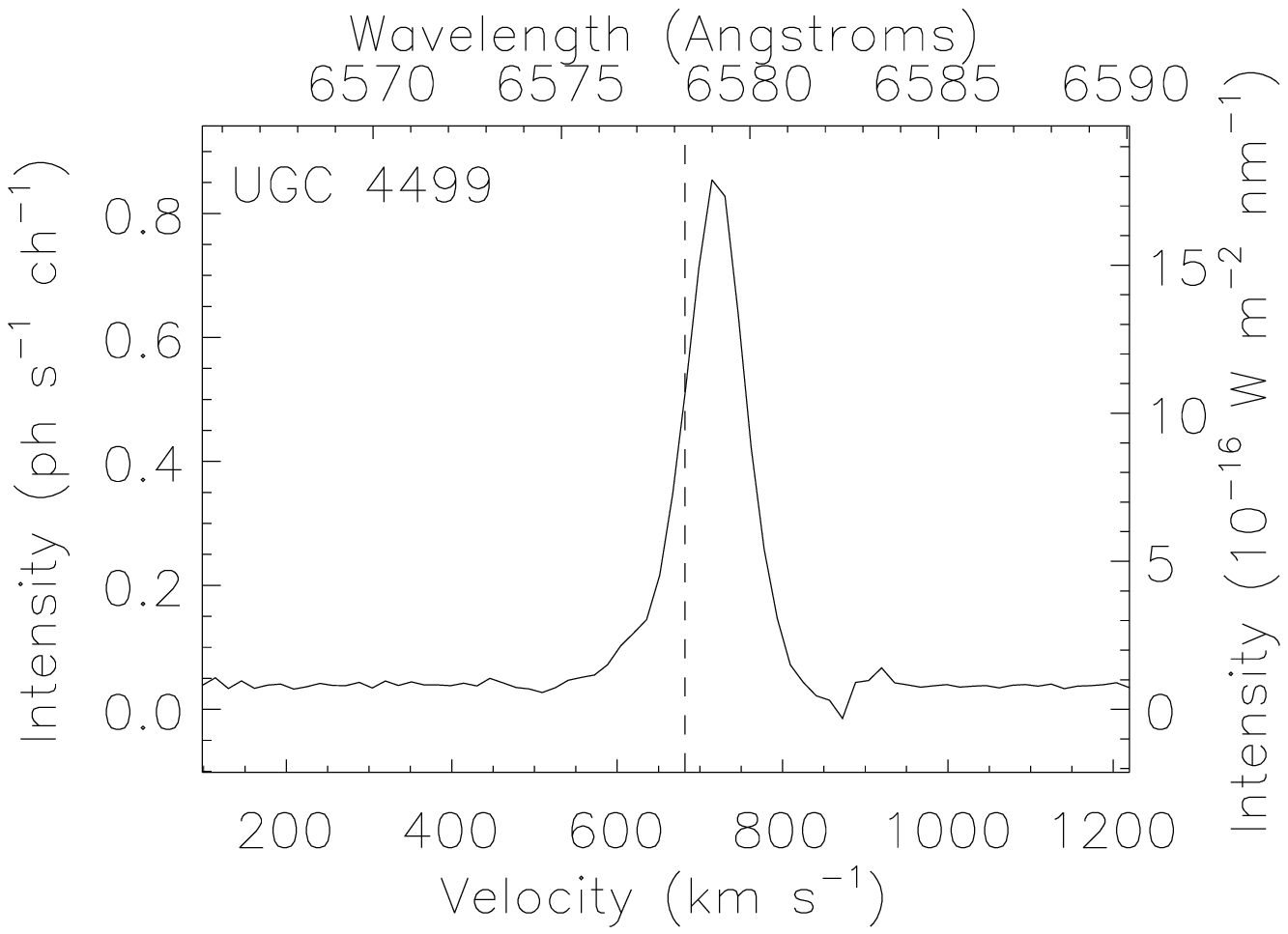}
\includegraphics[width=3.5cm]{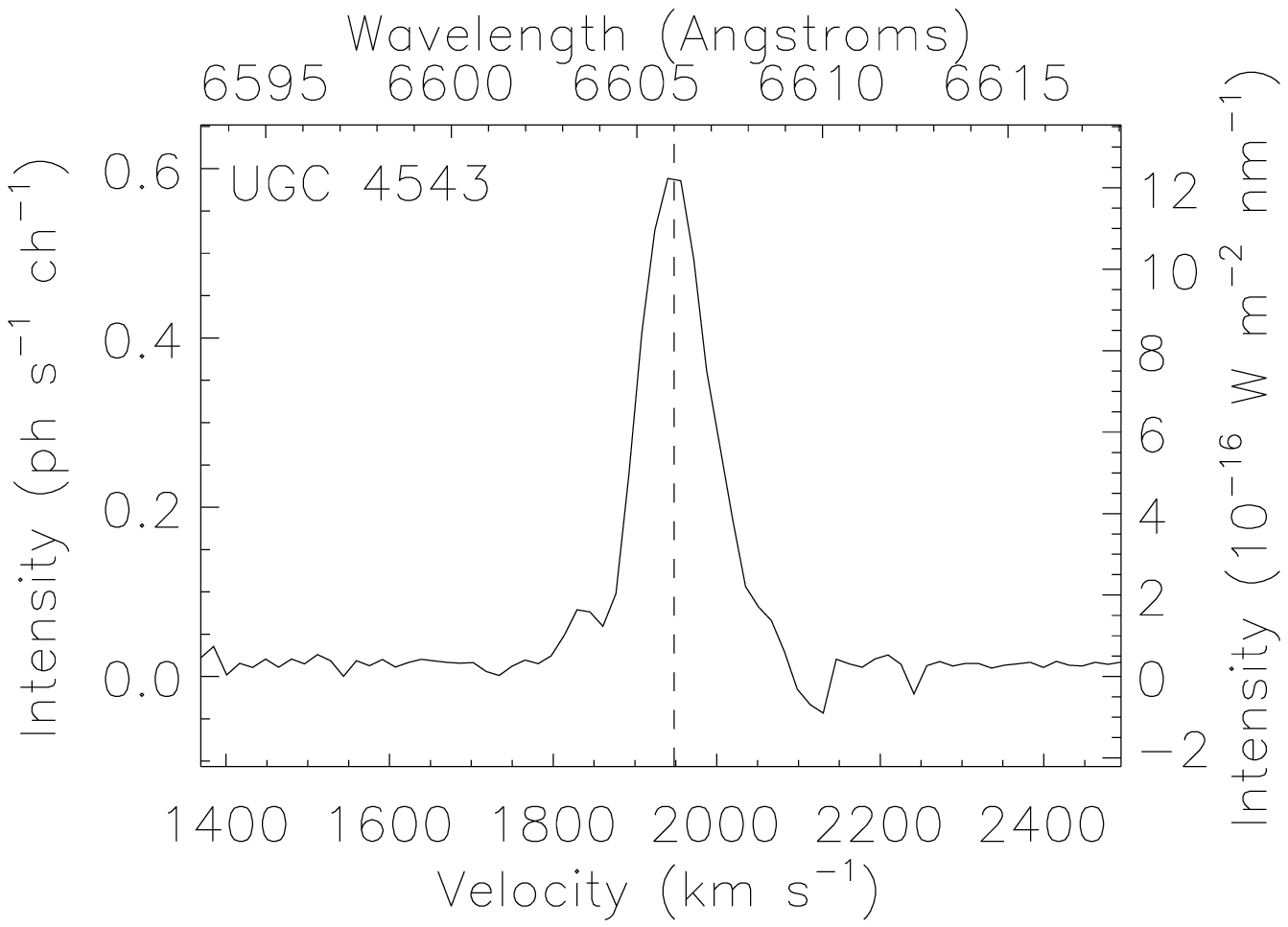}
\includegraphics[width=3.5cm]{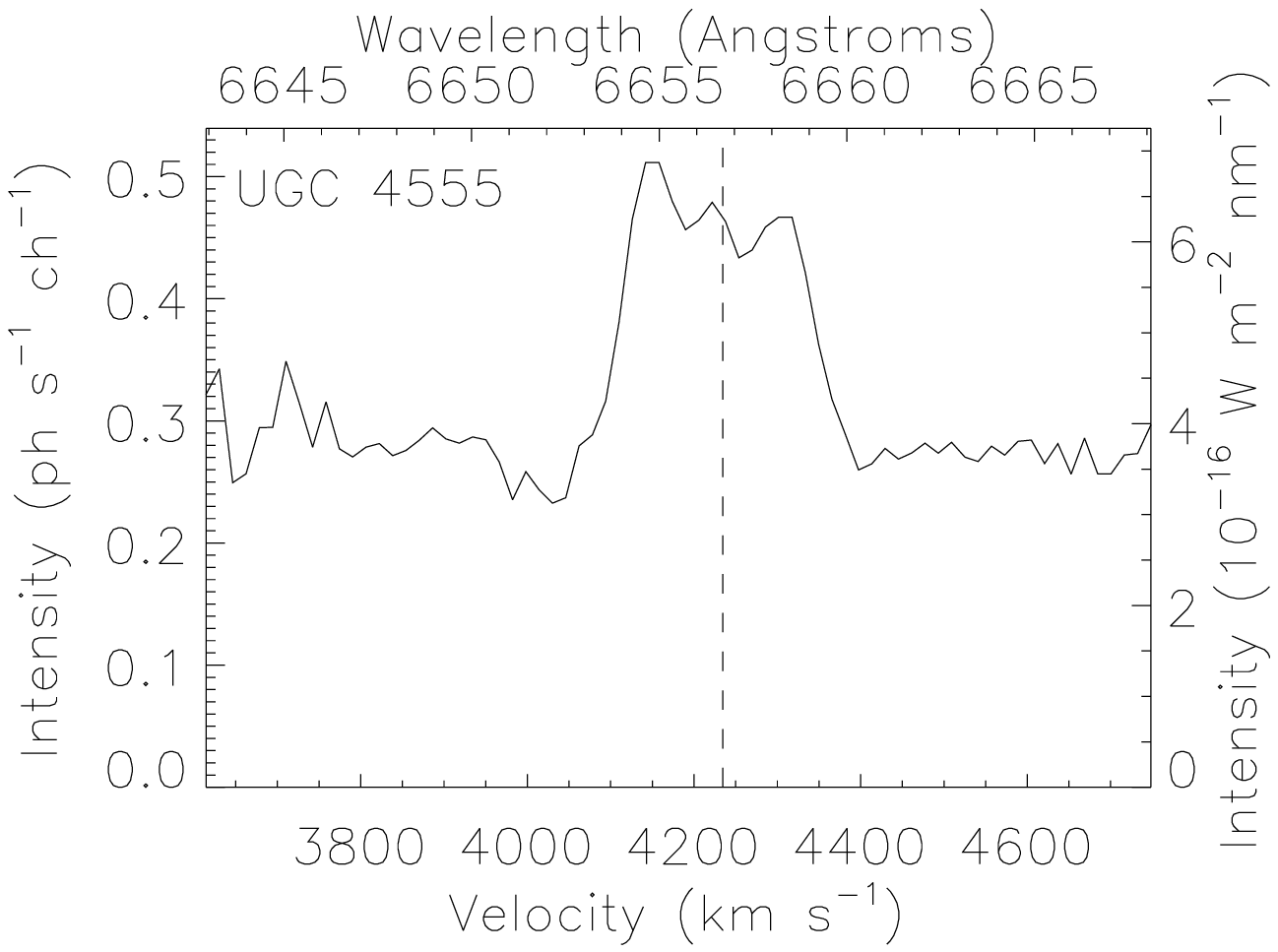}
\includegraphics[width=3.5cm]{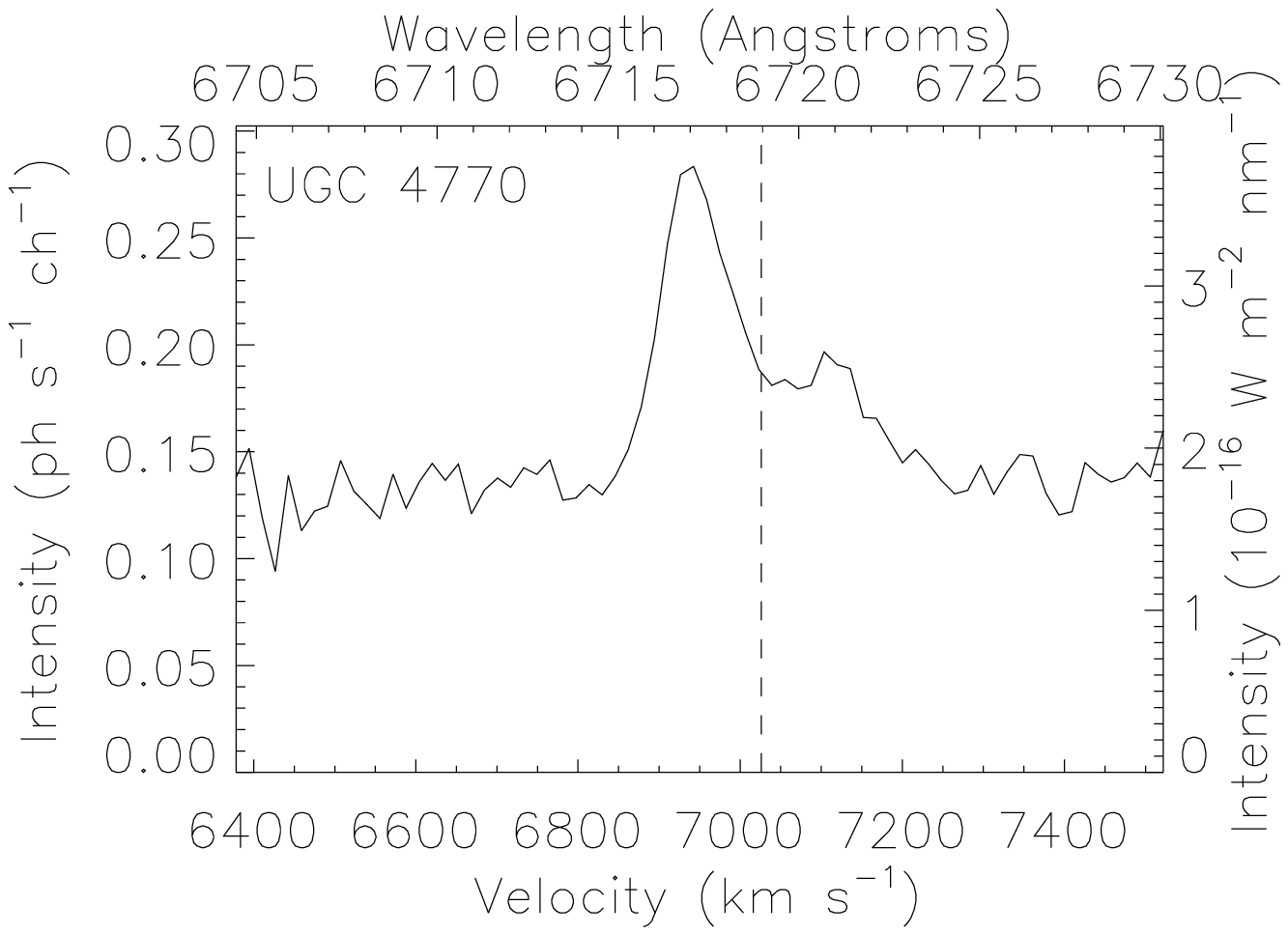}
\includegraphics[width=3.5cm]{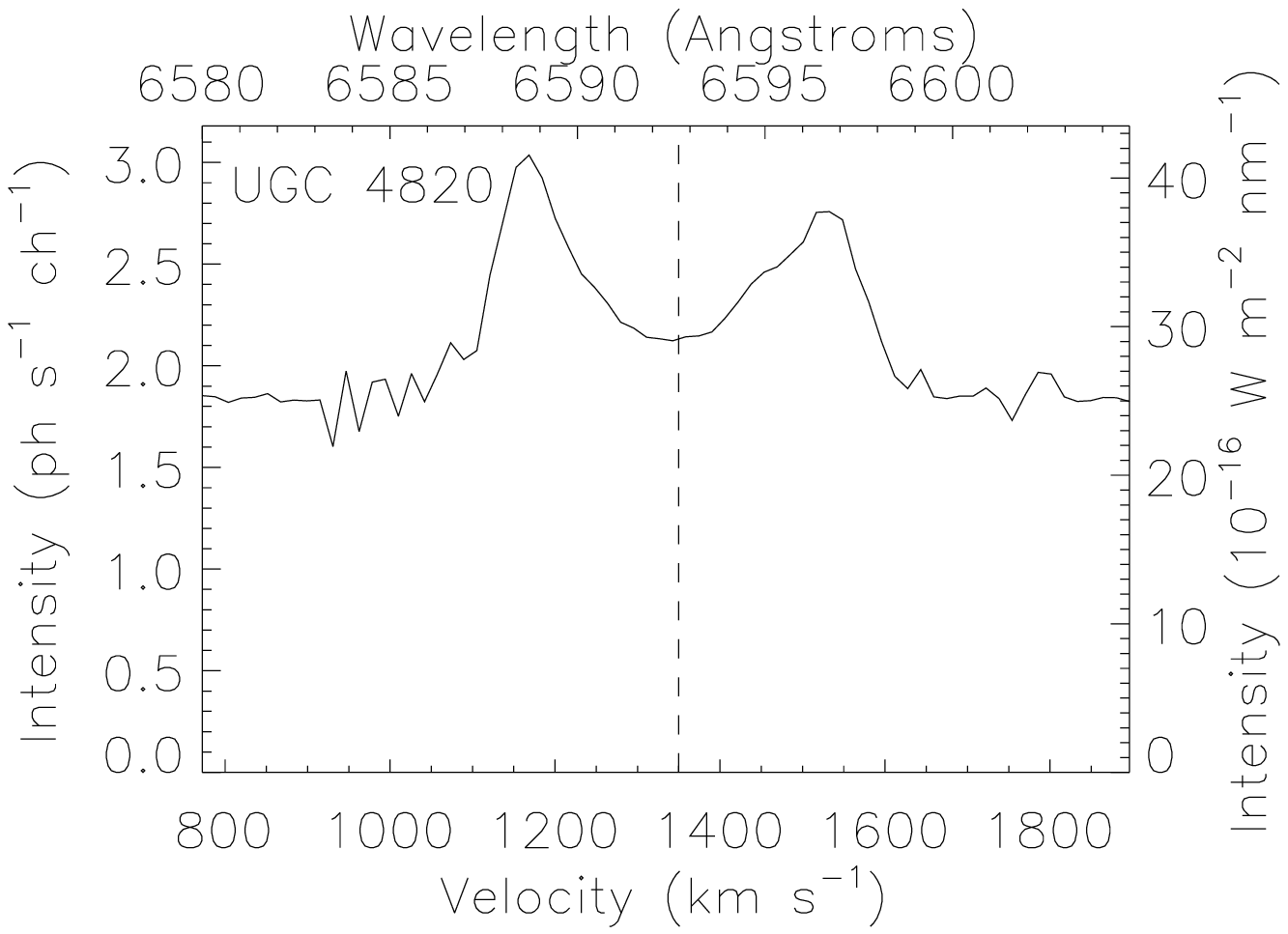}
\includegraphics[width=3.5cm]{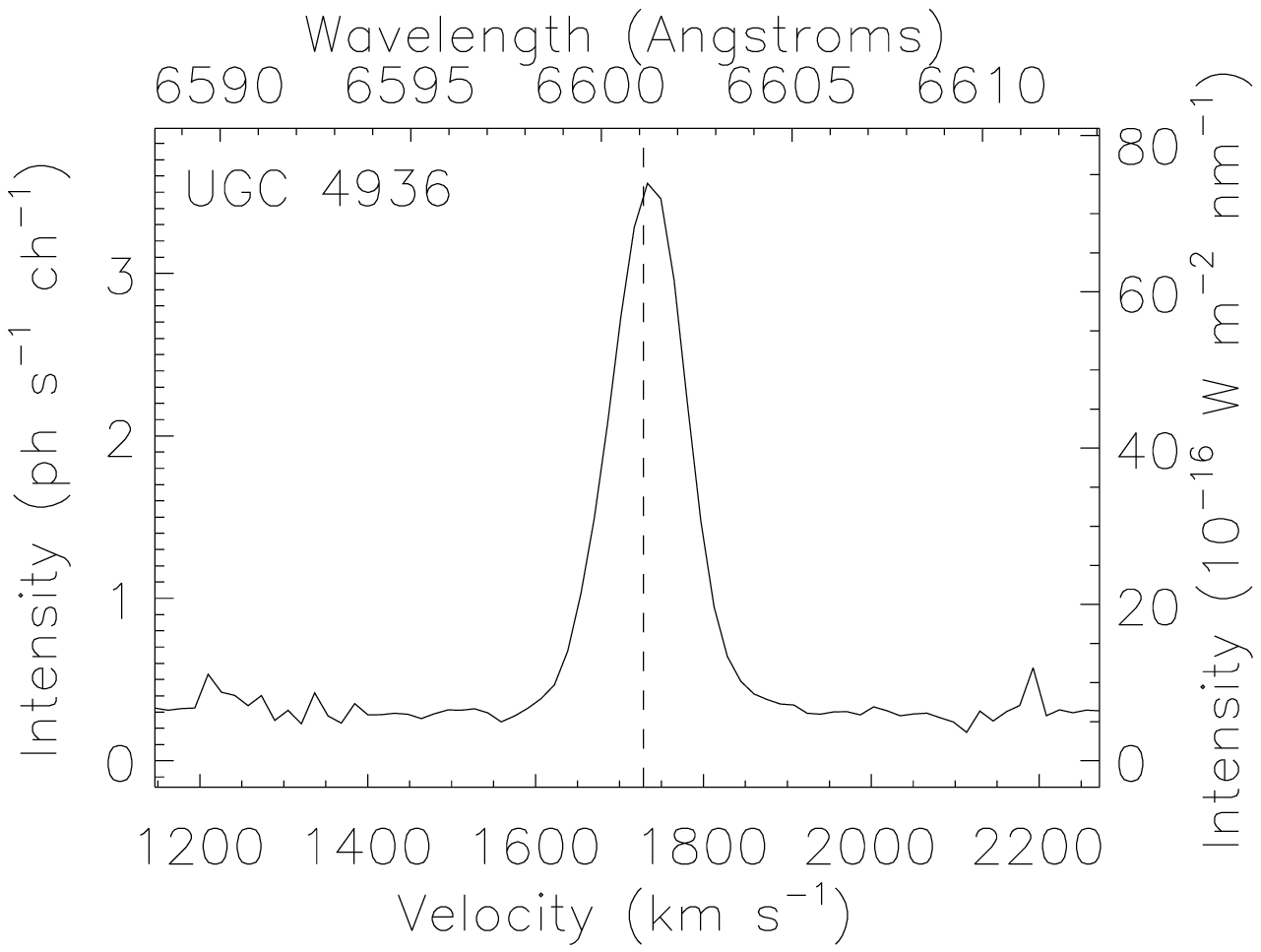}
\includegraphics[width=3.5cm]{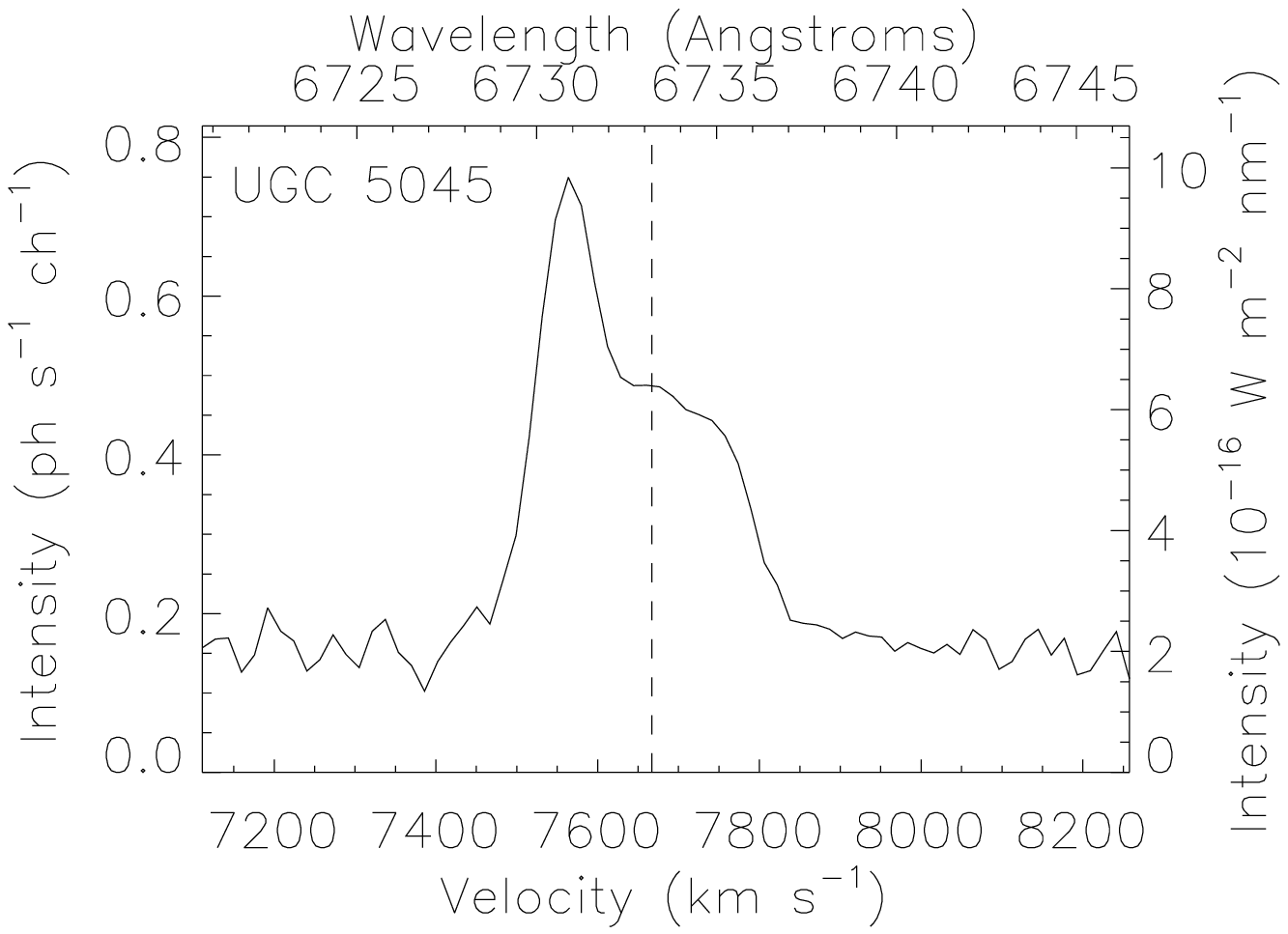}
\includegraphics[width=3.5cm]{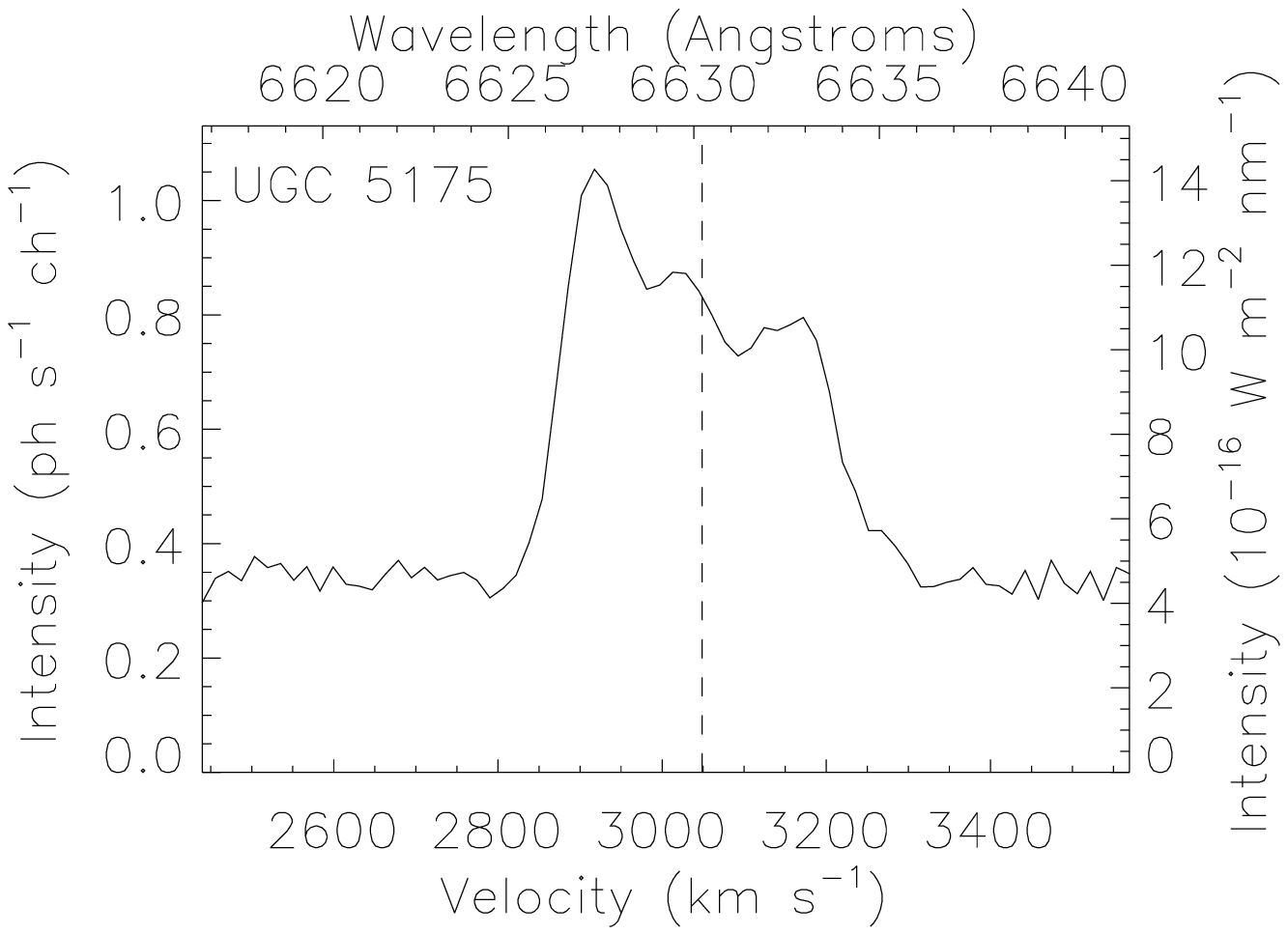}
\includegraphics[width=3.5cm]{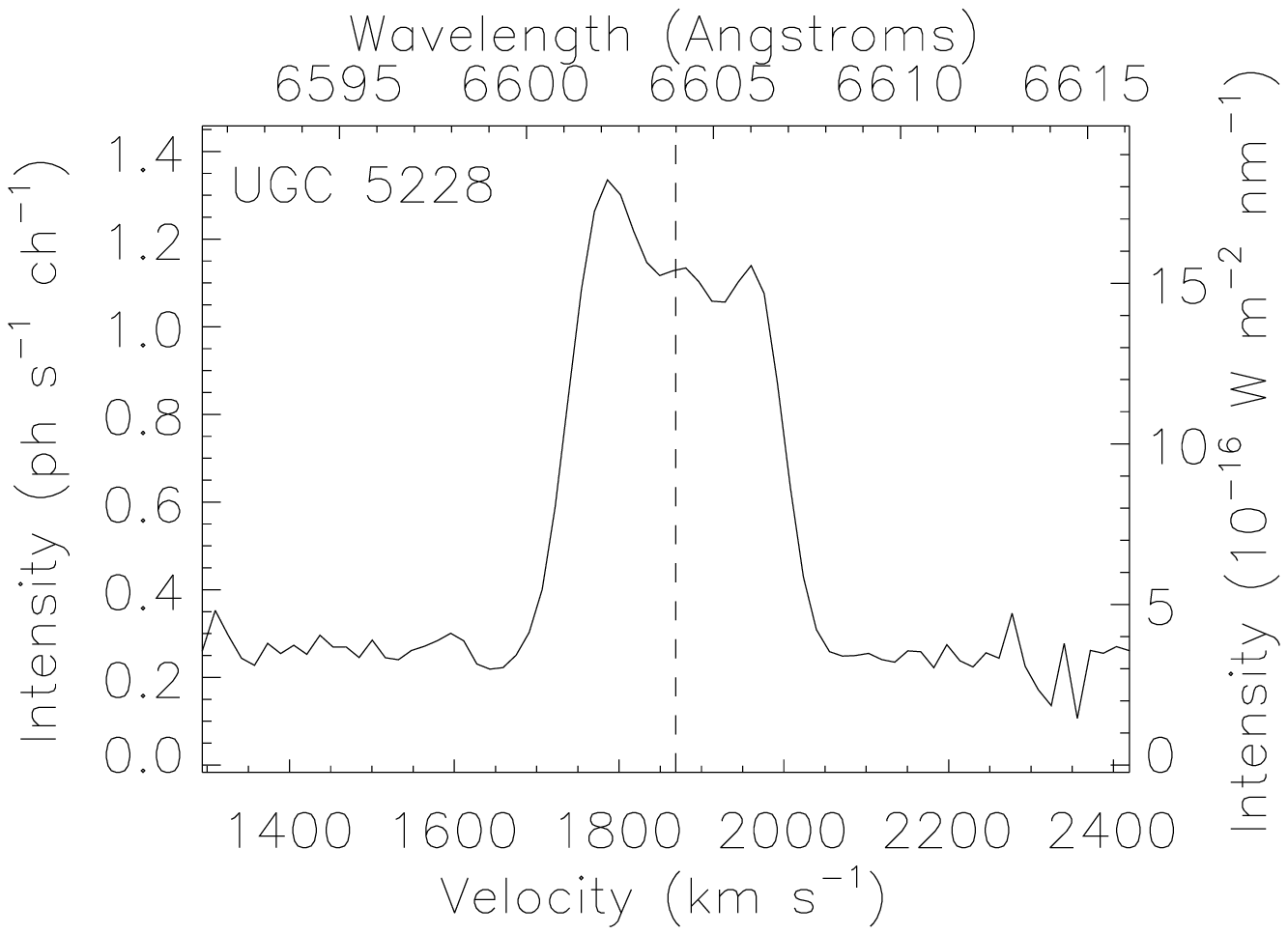}
\includegraphics[width=3.5cm]{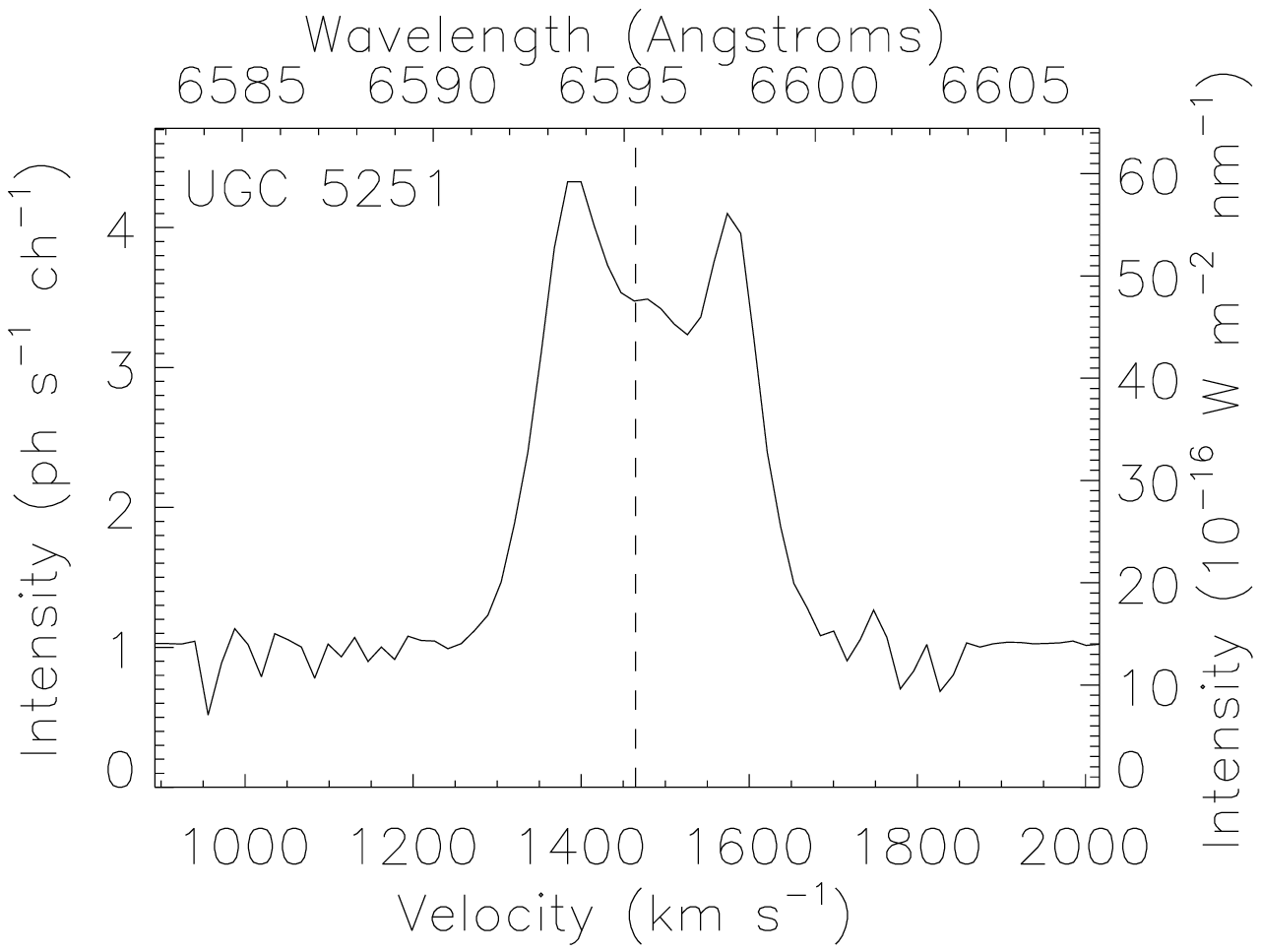}
\includegraphics[width=3.5cm]{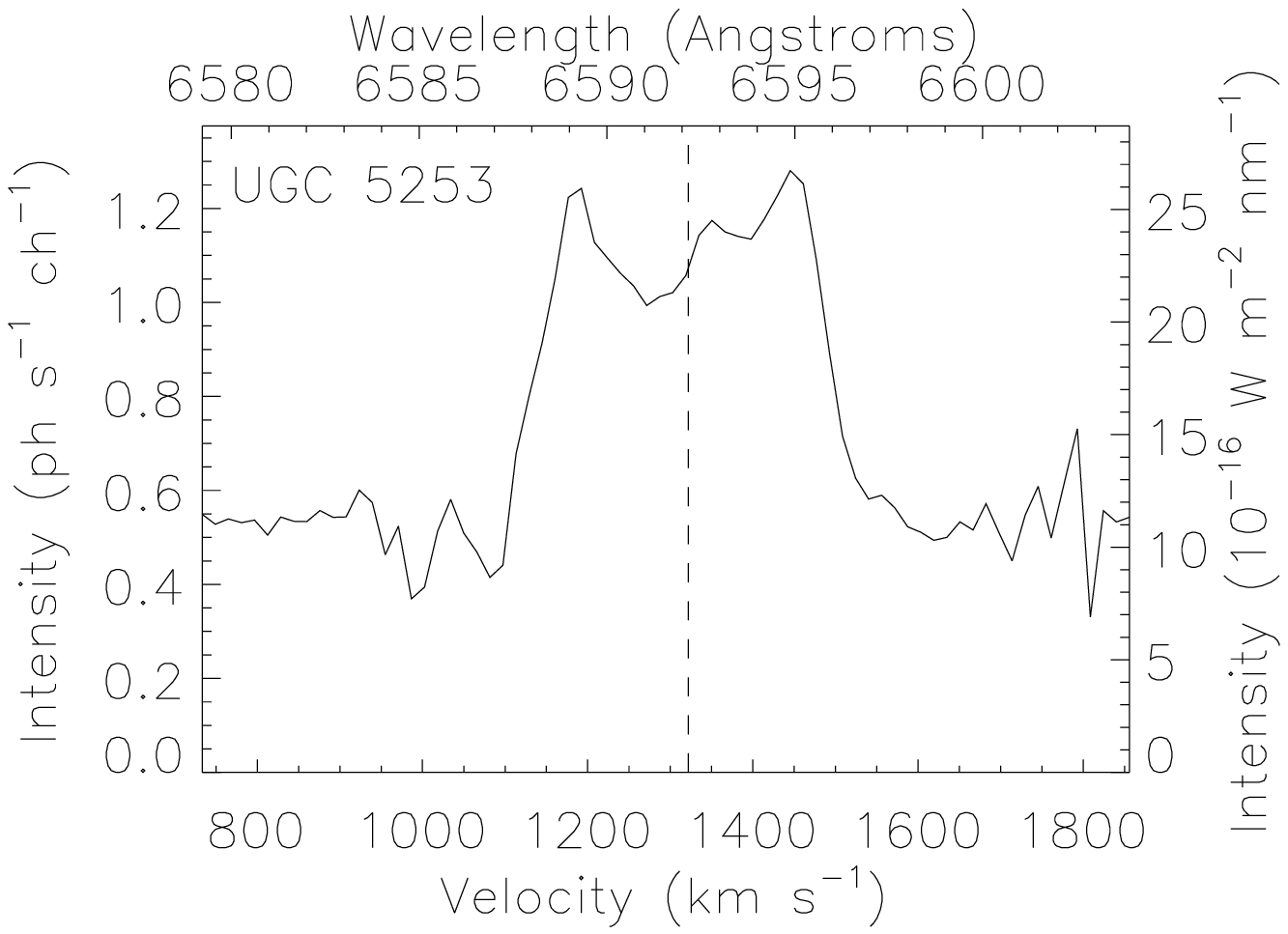}
\includegraphics[width=3.5cm]{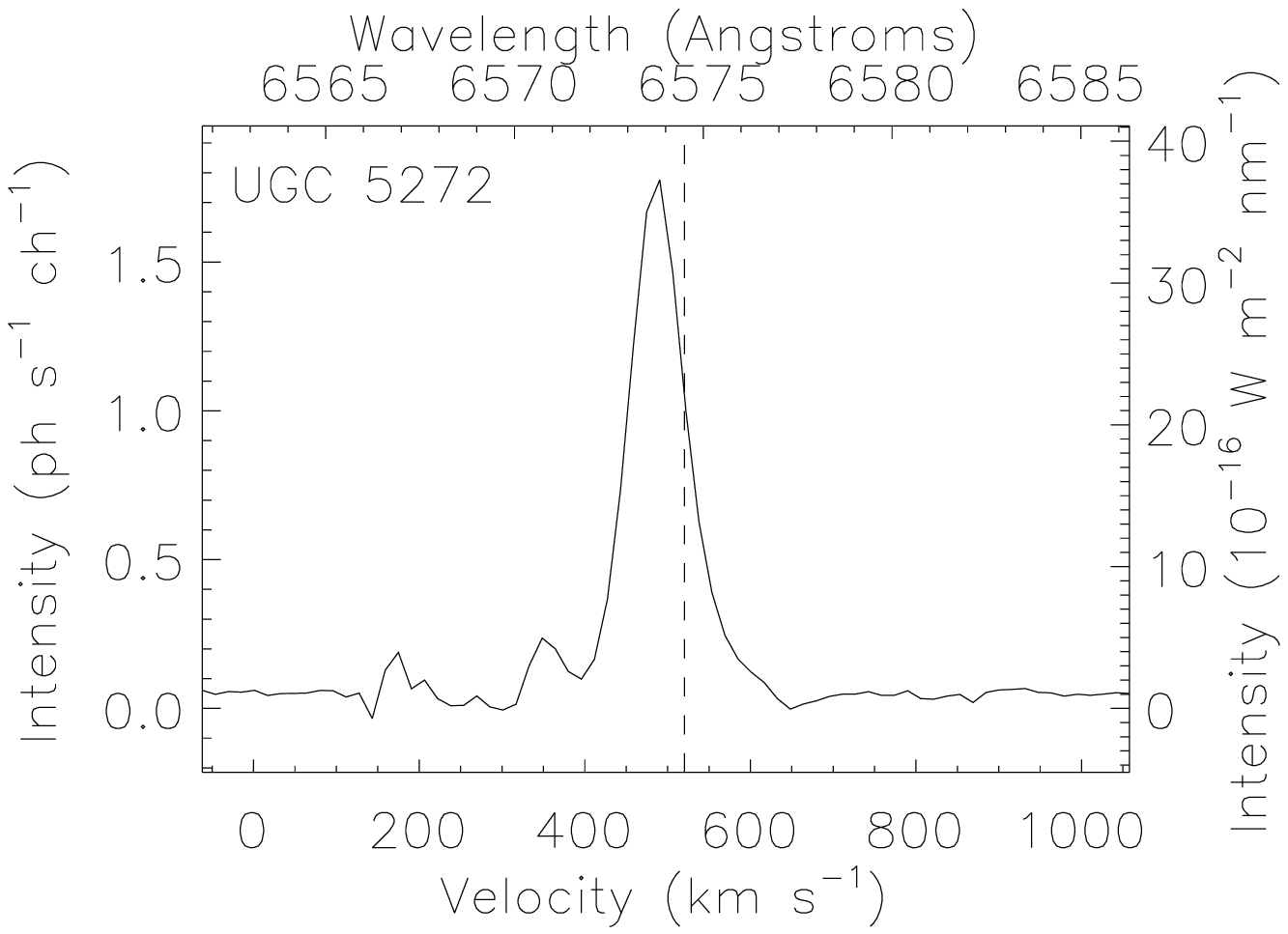}
\includegraphics[width=3.5cm]{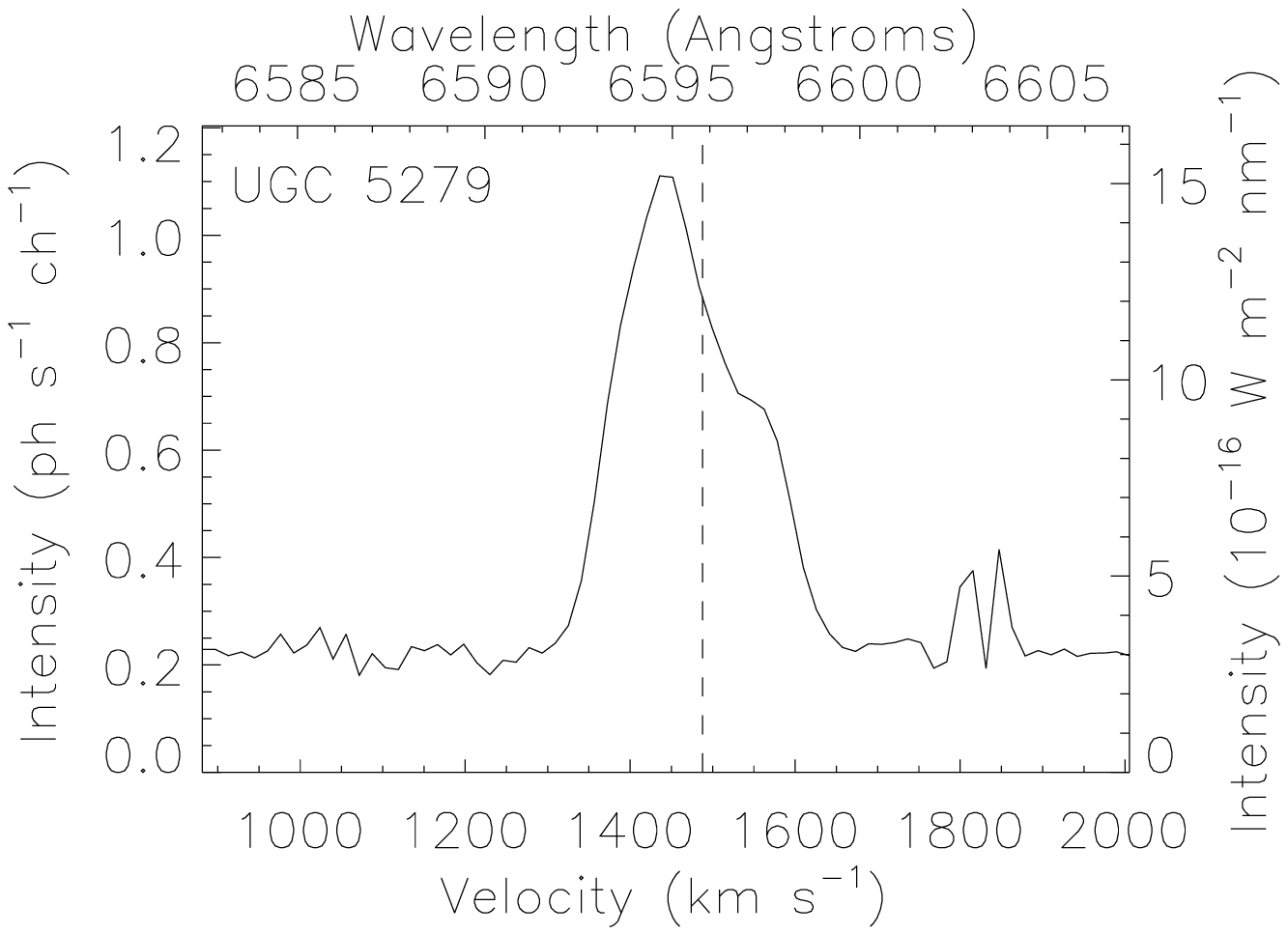}
\includegraphics[width=3.5cm]{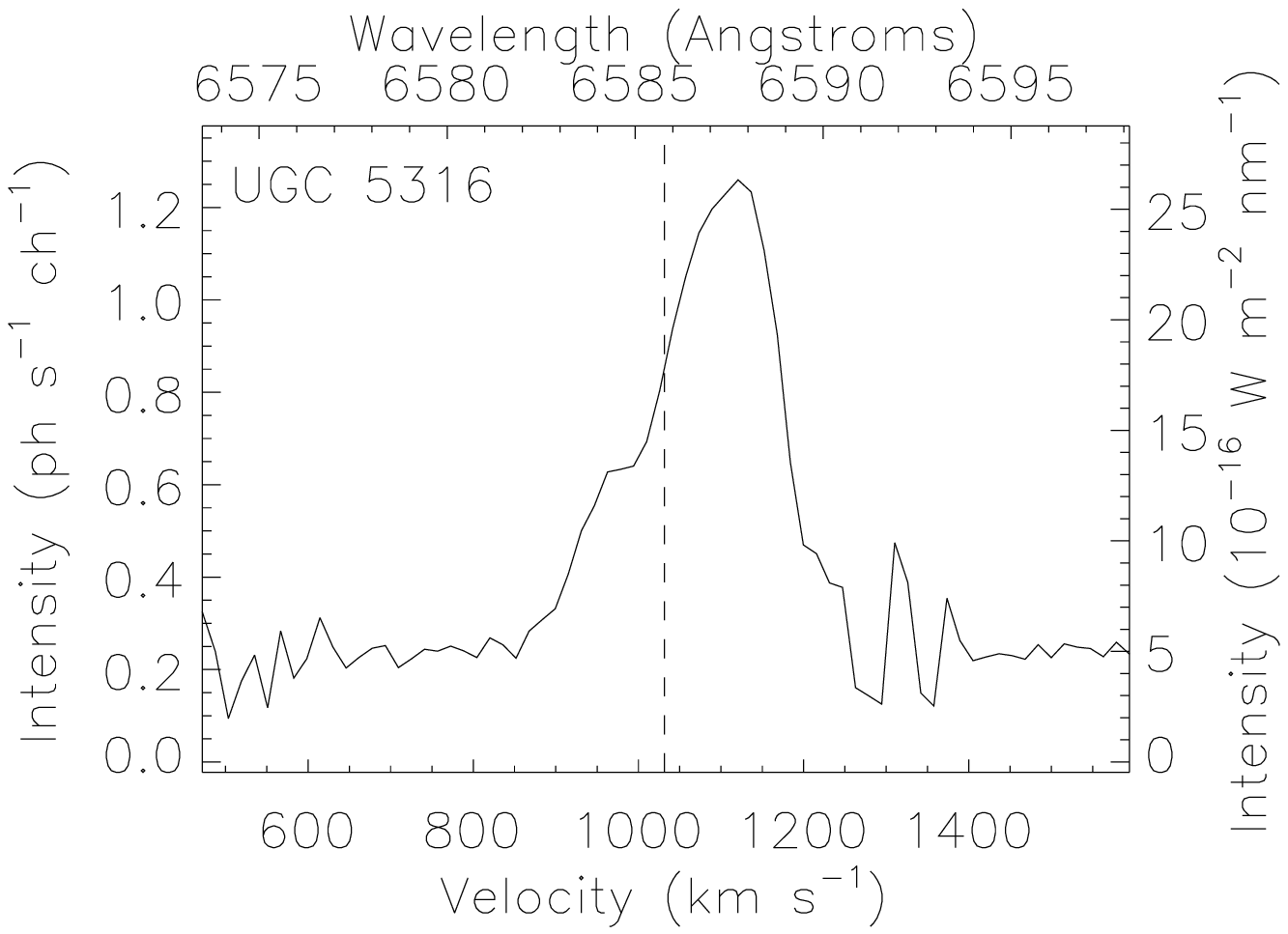}
\includegraphics[width=3.5cm]{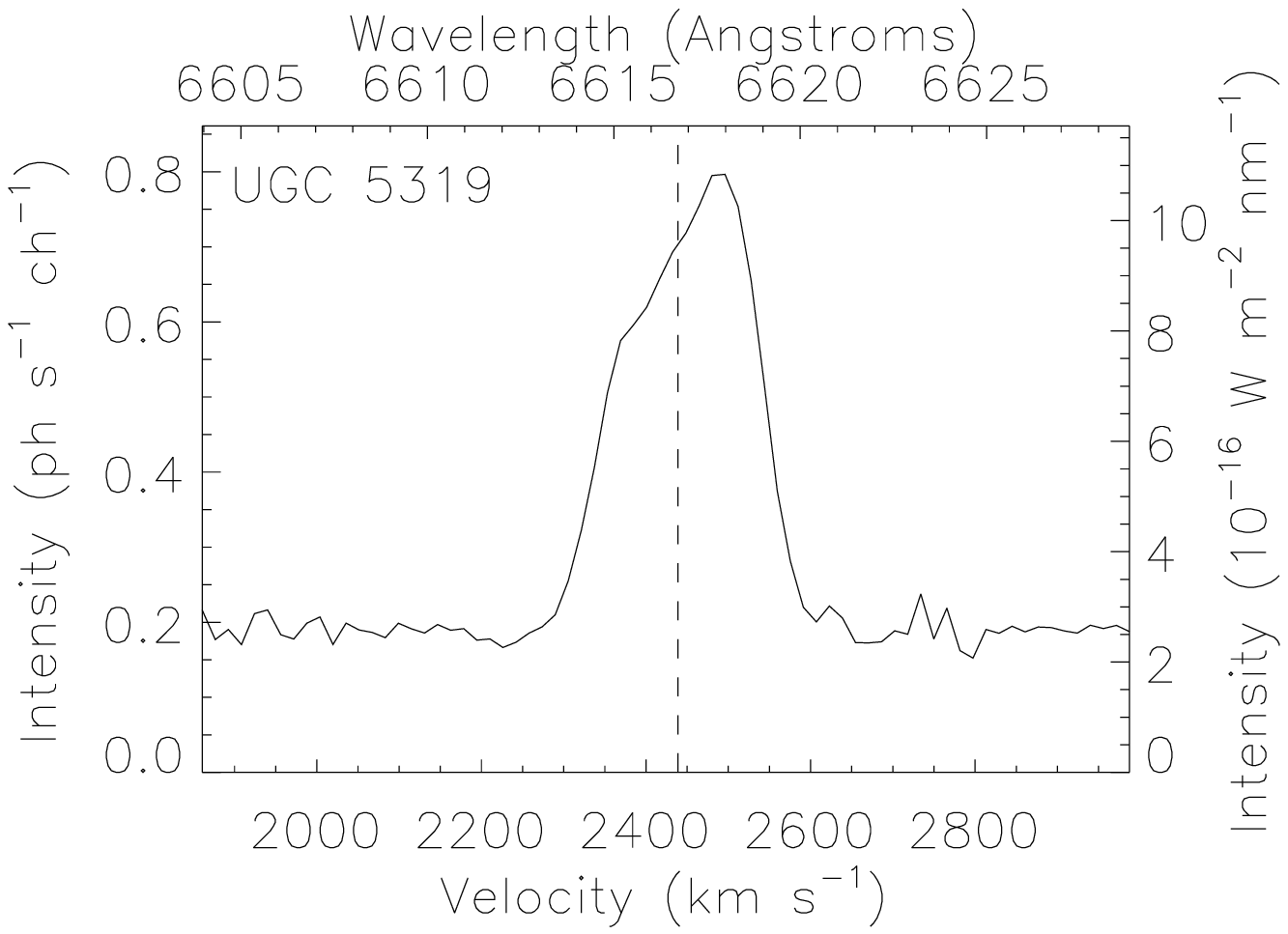}
\includegraphics[width=3.5cm]{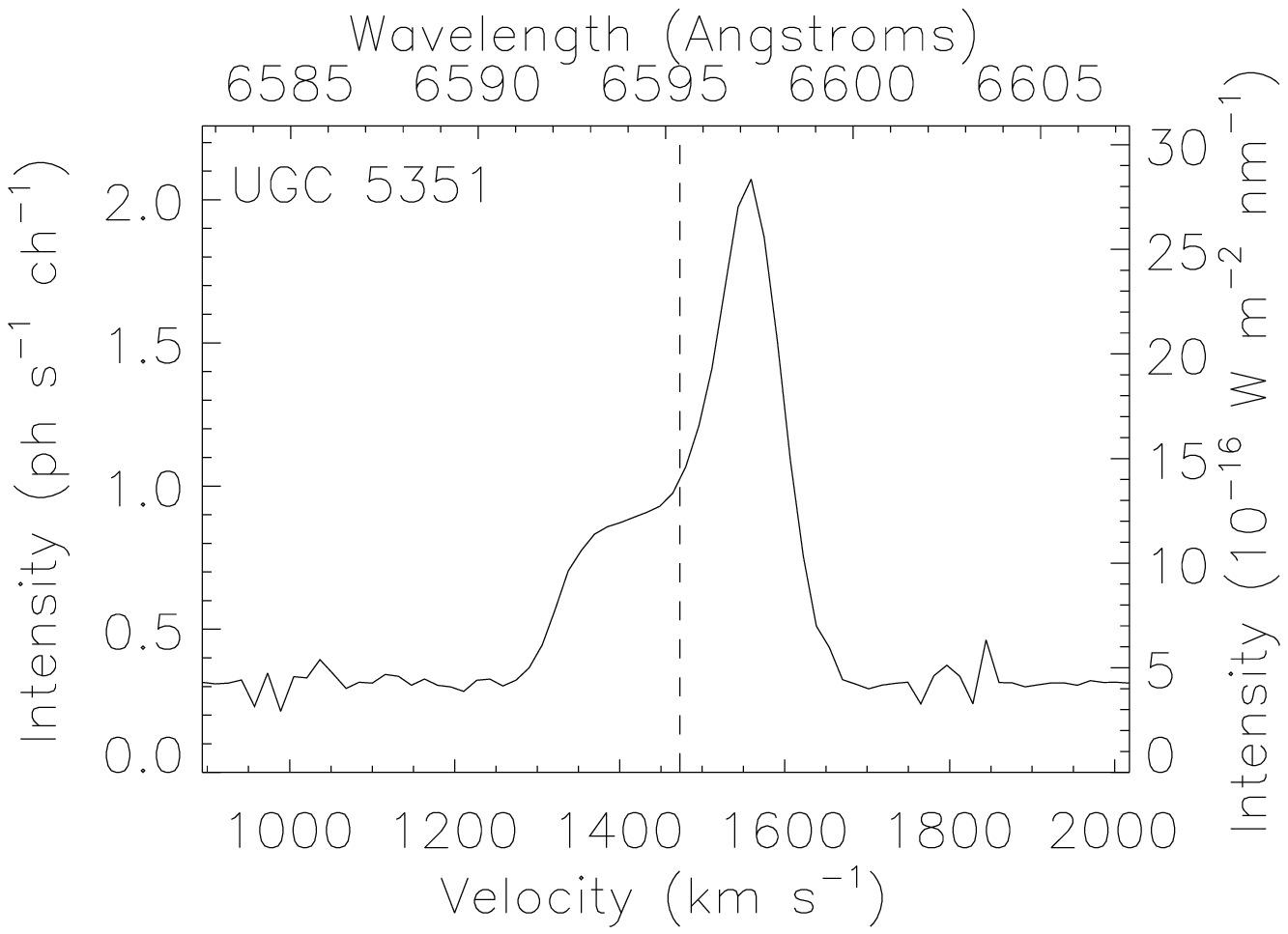}
\includegraphics[width=3.5cm]{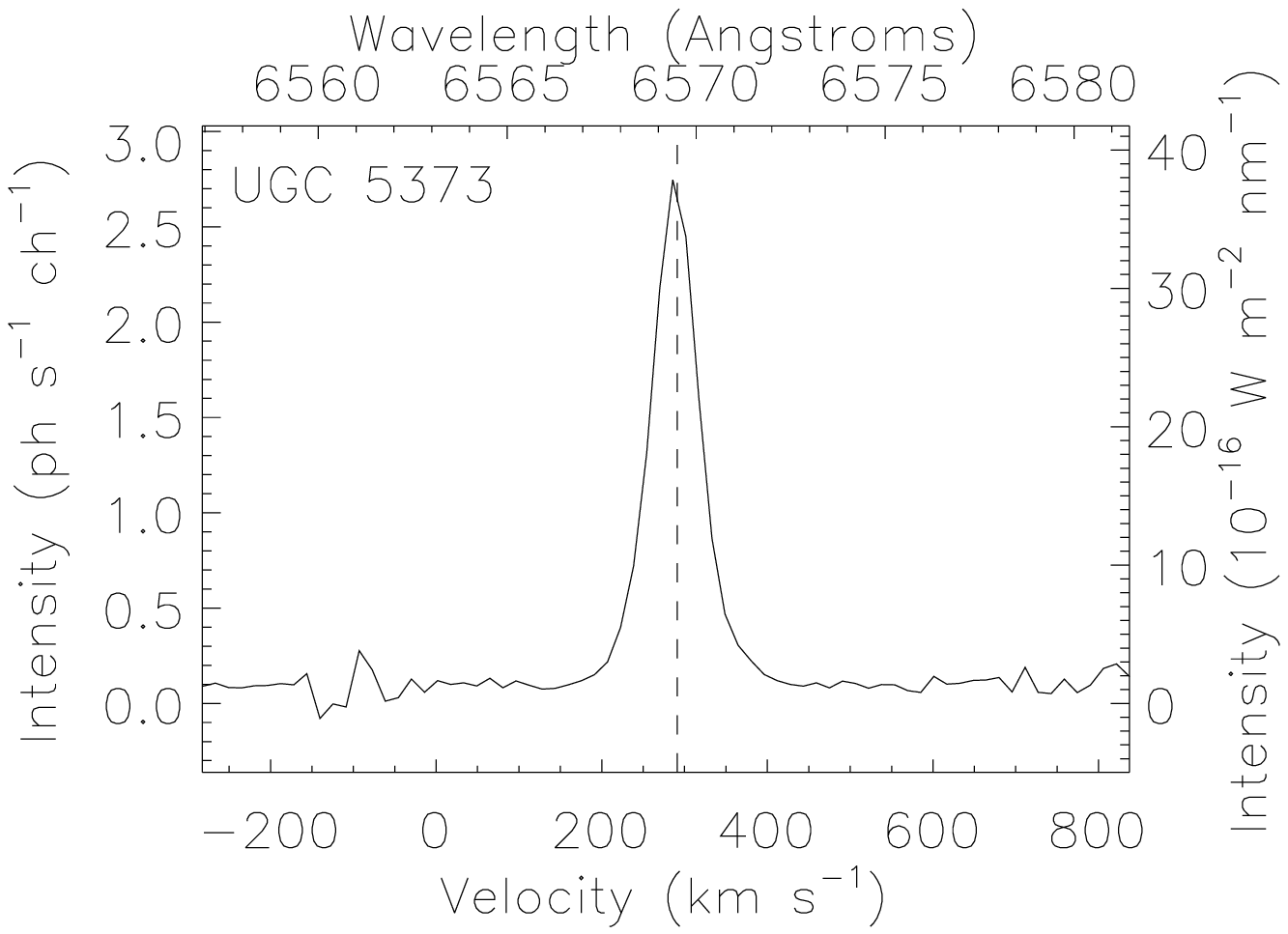}
\includegraphics[width=3.5cm]{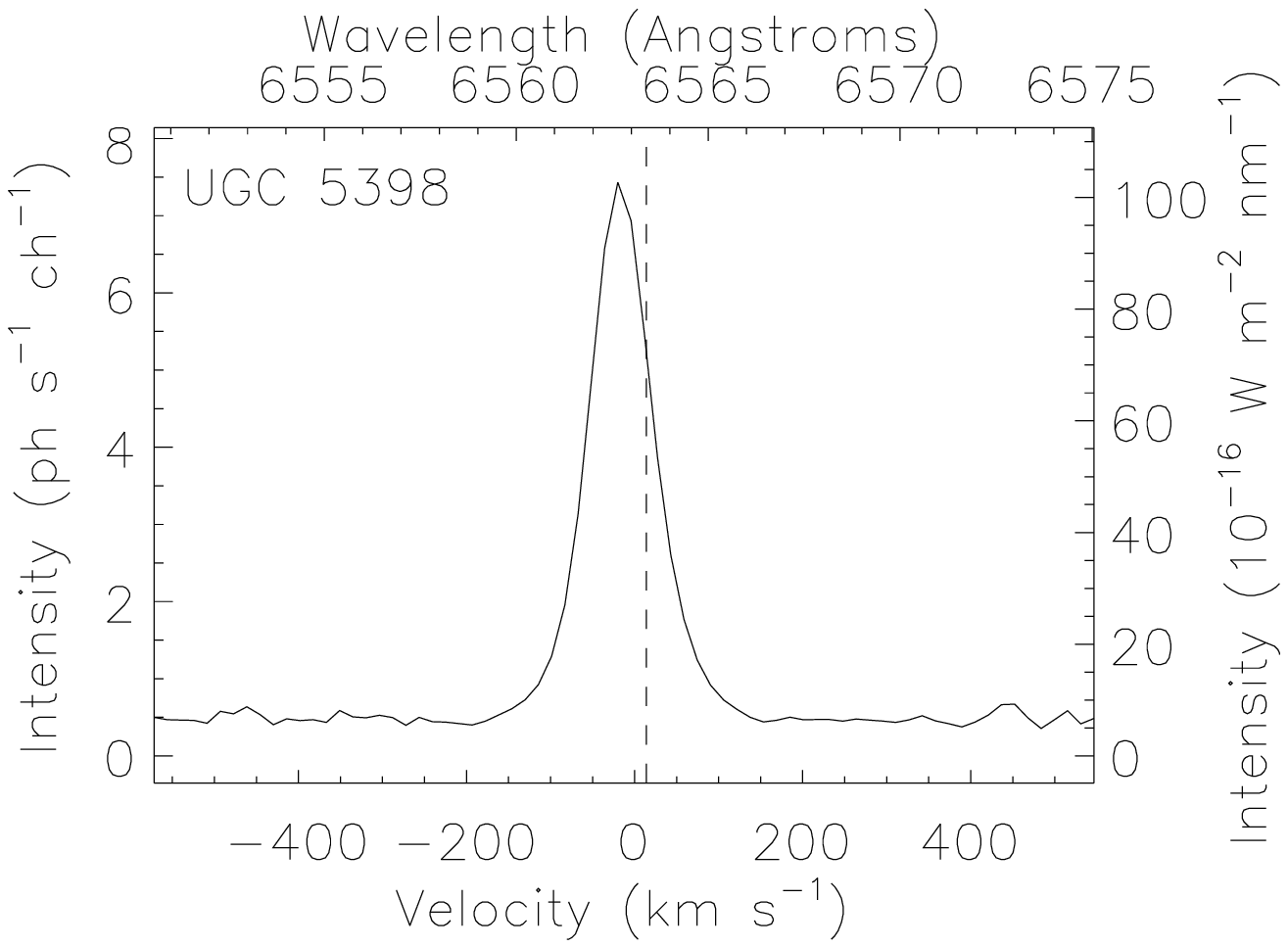}
\includegraphics[width=3.5cm]{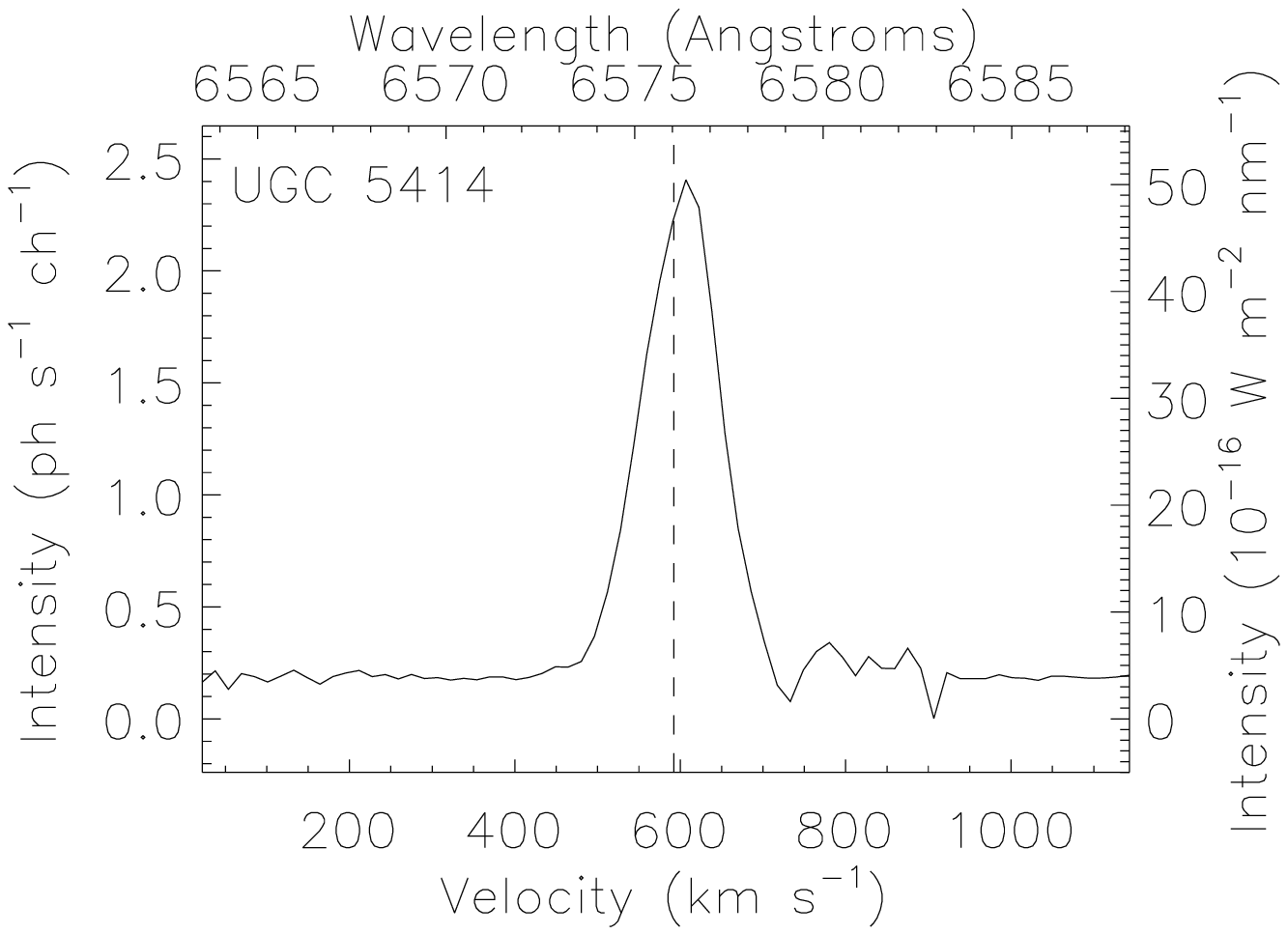}
\includegraphics[width=3.5cm]{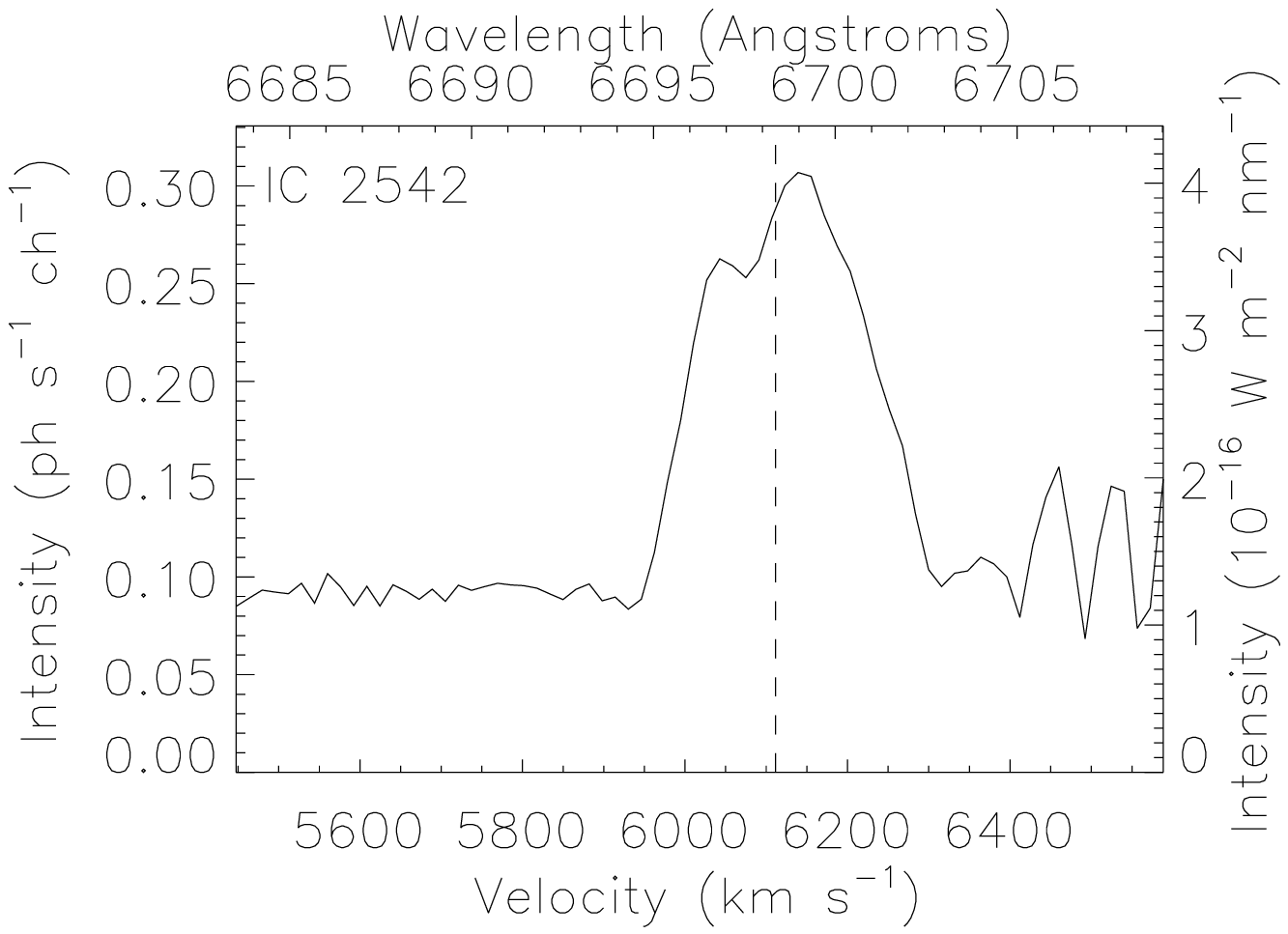}
\includegraphics[width=3.5cm]{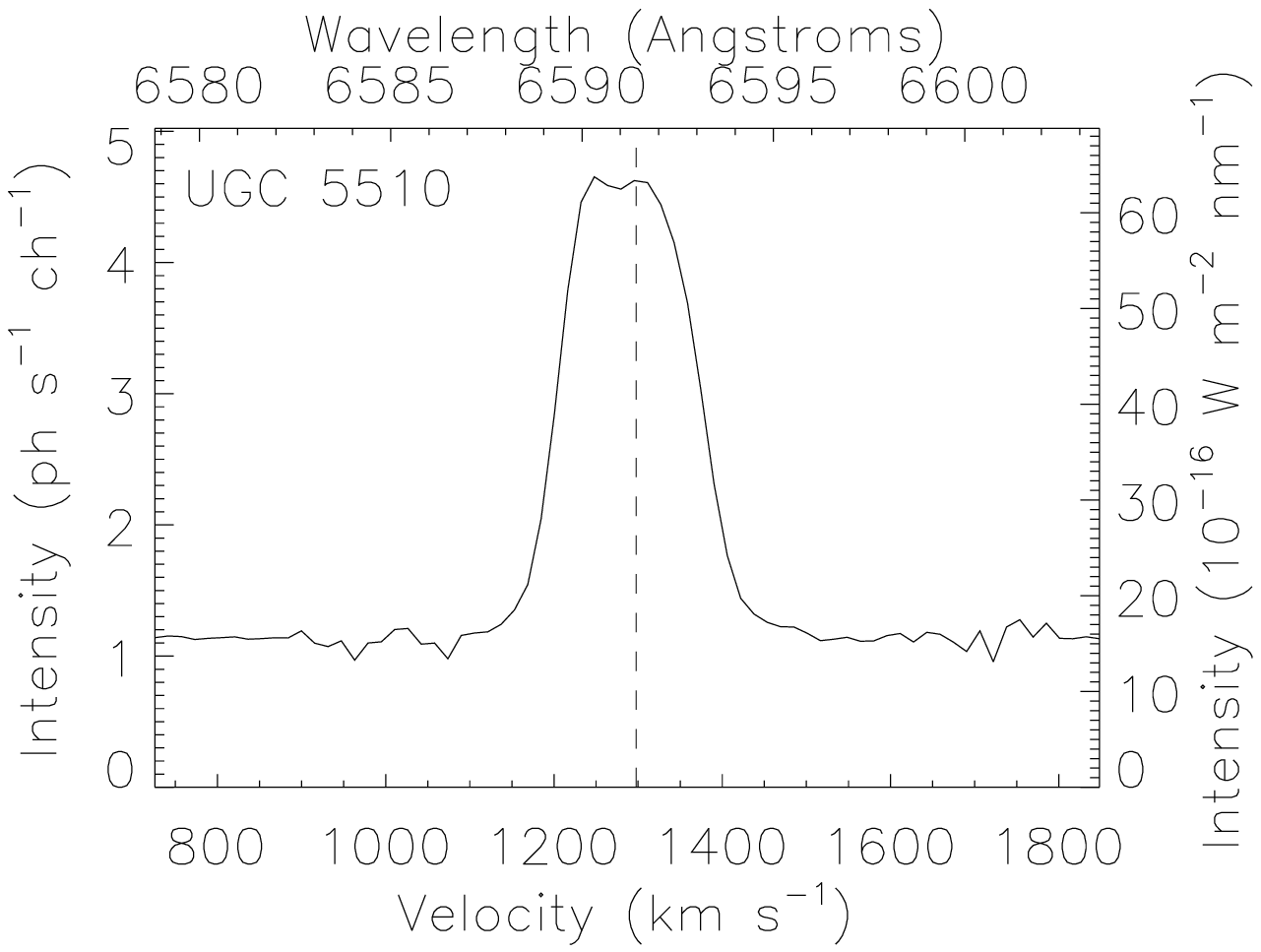}
\includegraphics[width=3.5cm]{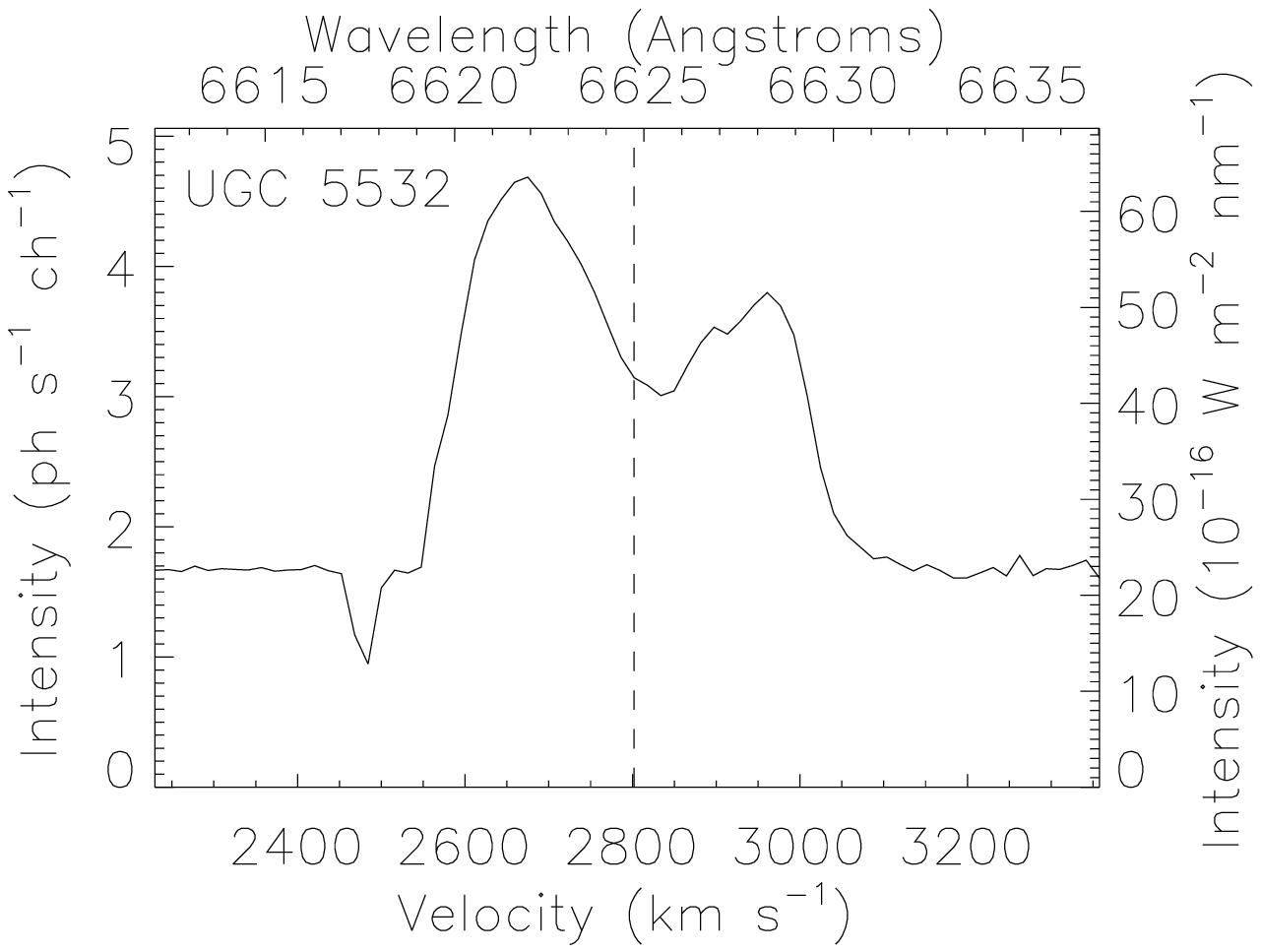}
\includegraphics[width=3.5cm]{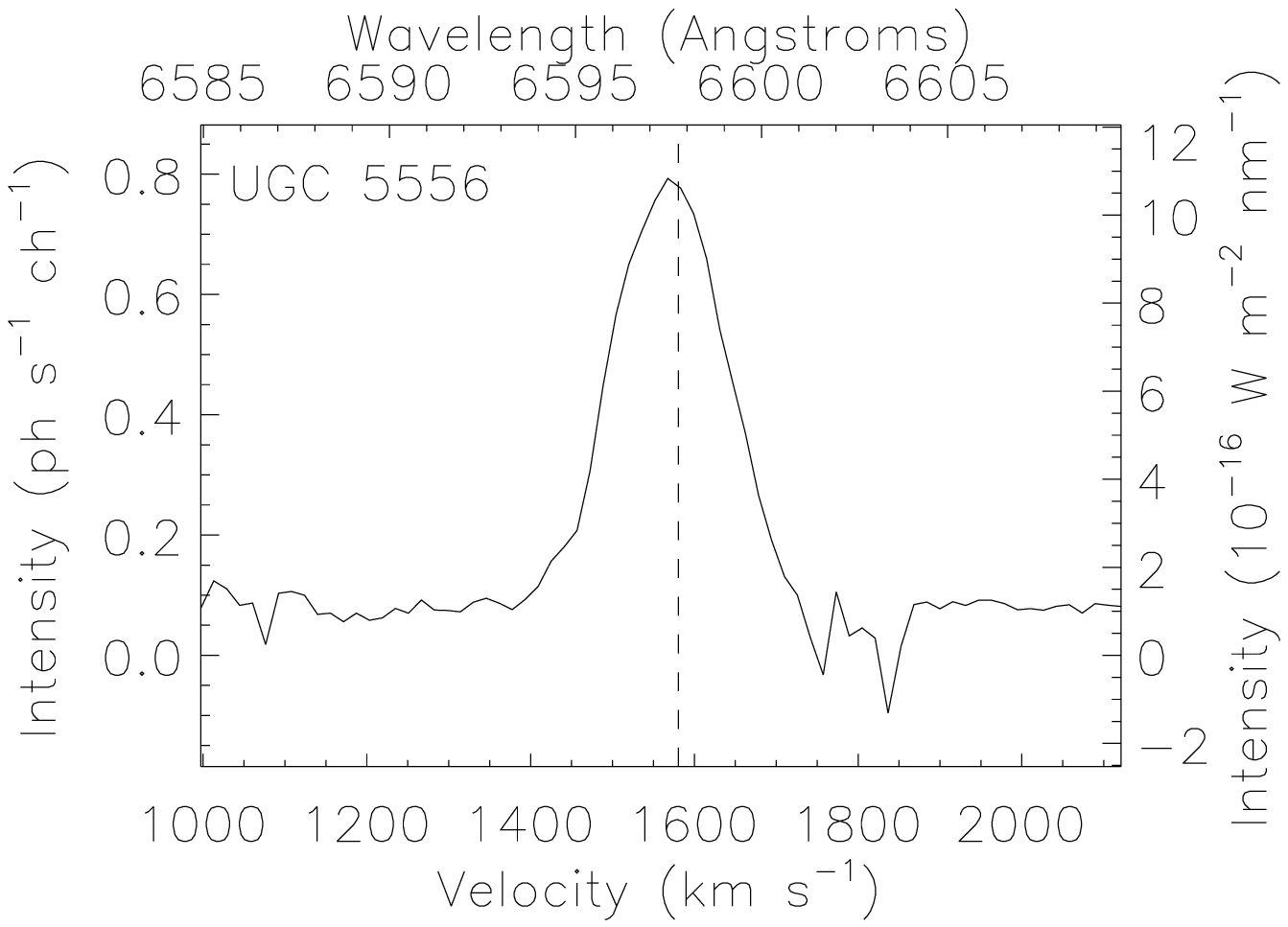}
\includegraphics[width=3.5cm]{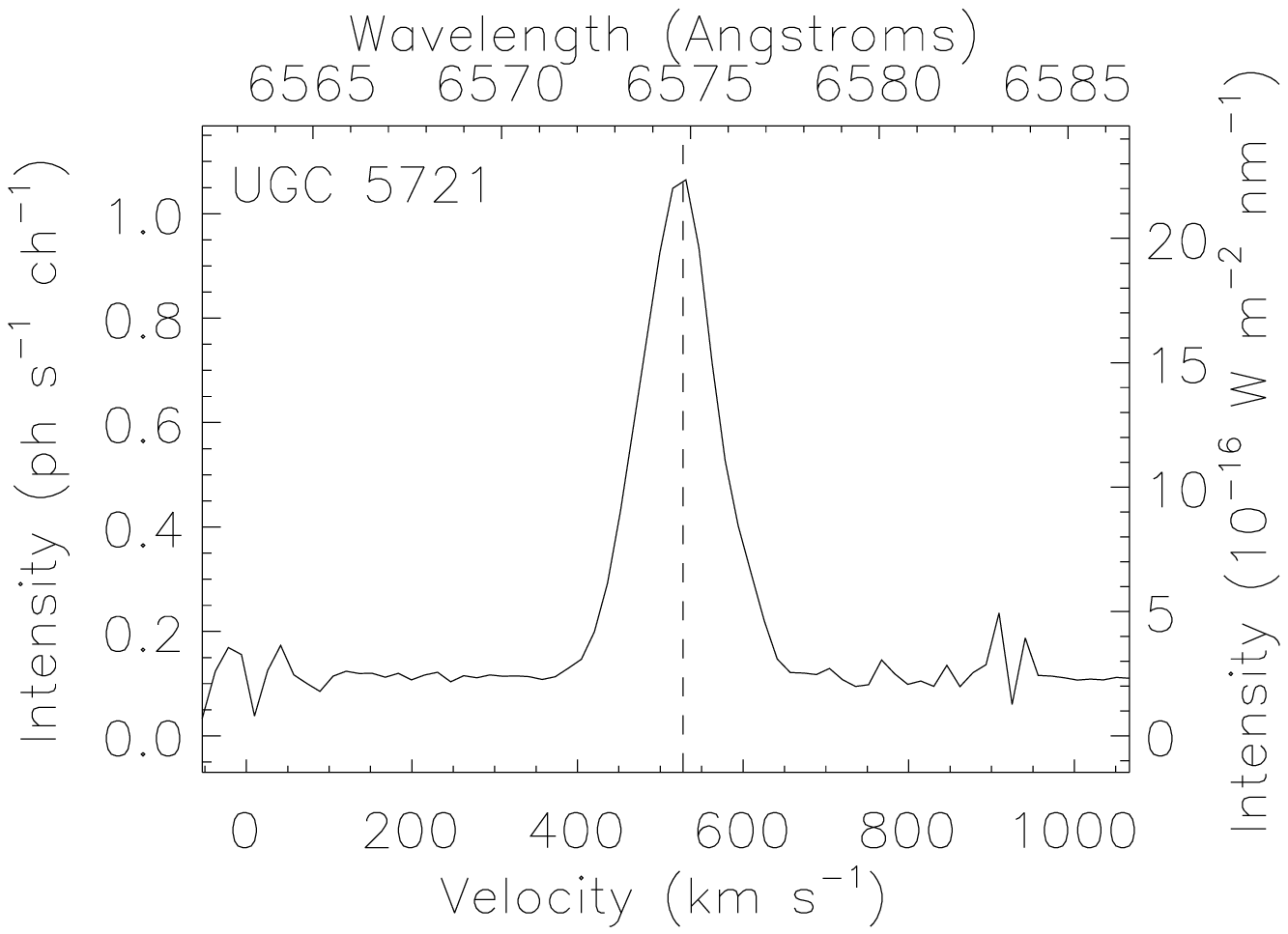}
\includegraphics[width=3.5cm]{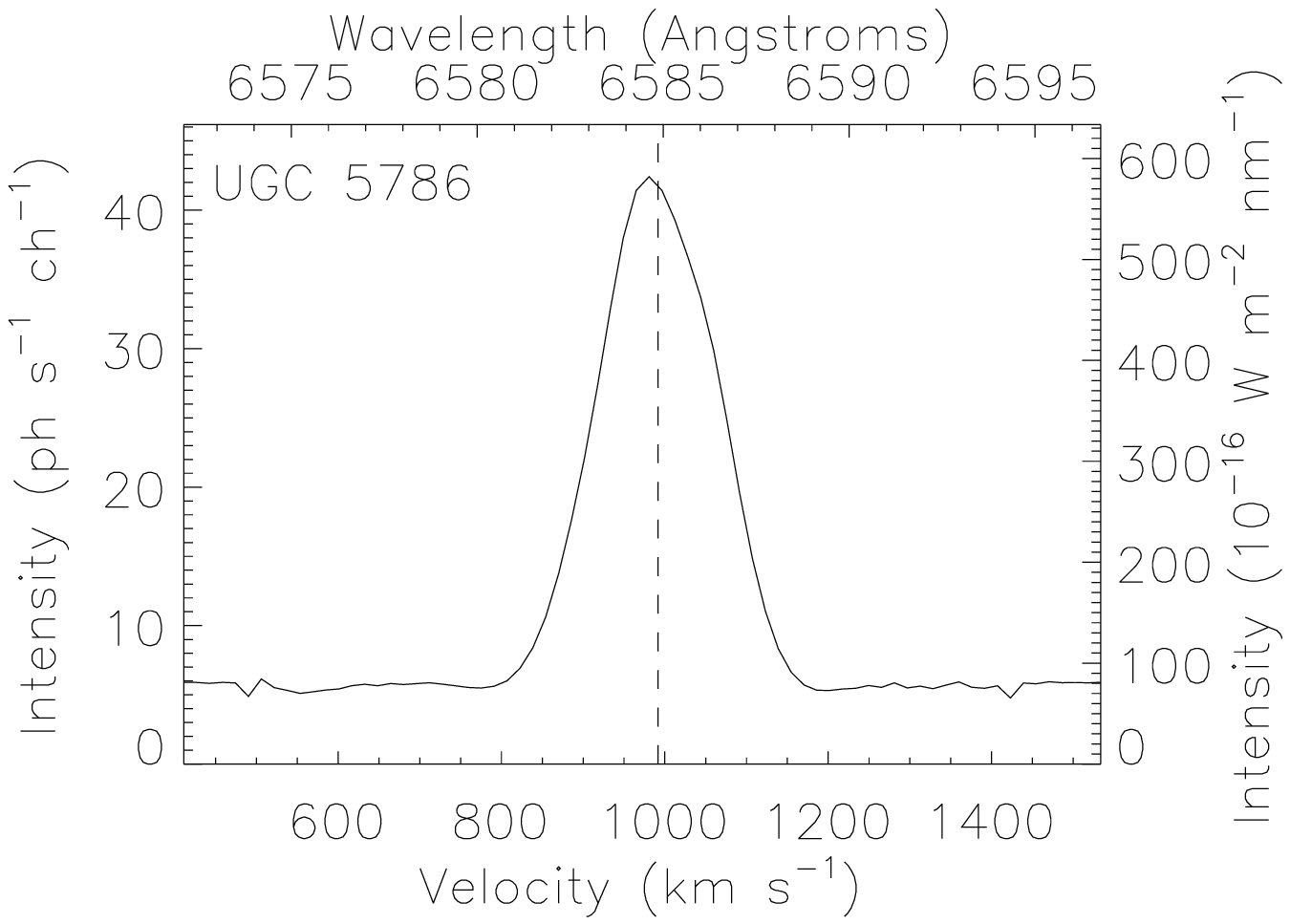}
\includegraphics[width=3.5cm]{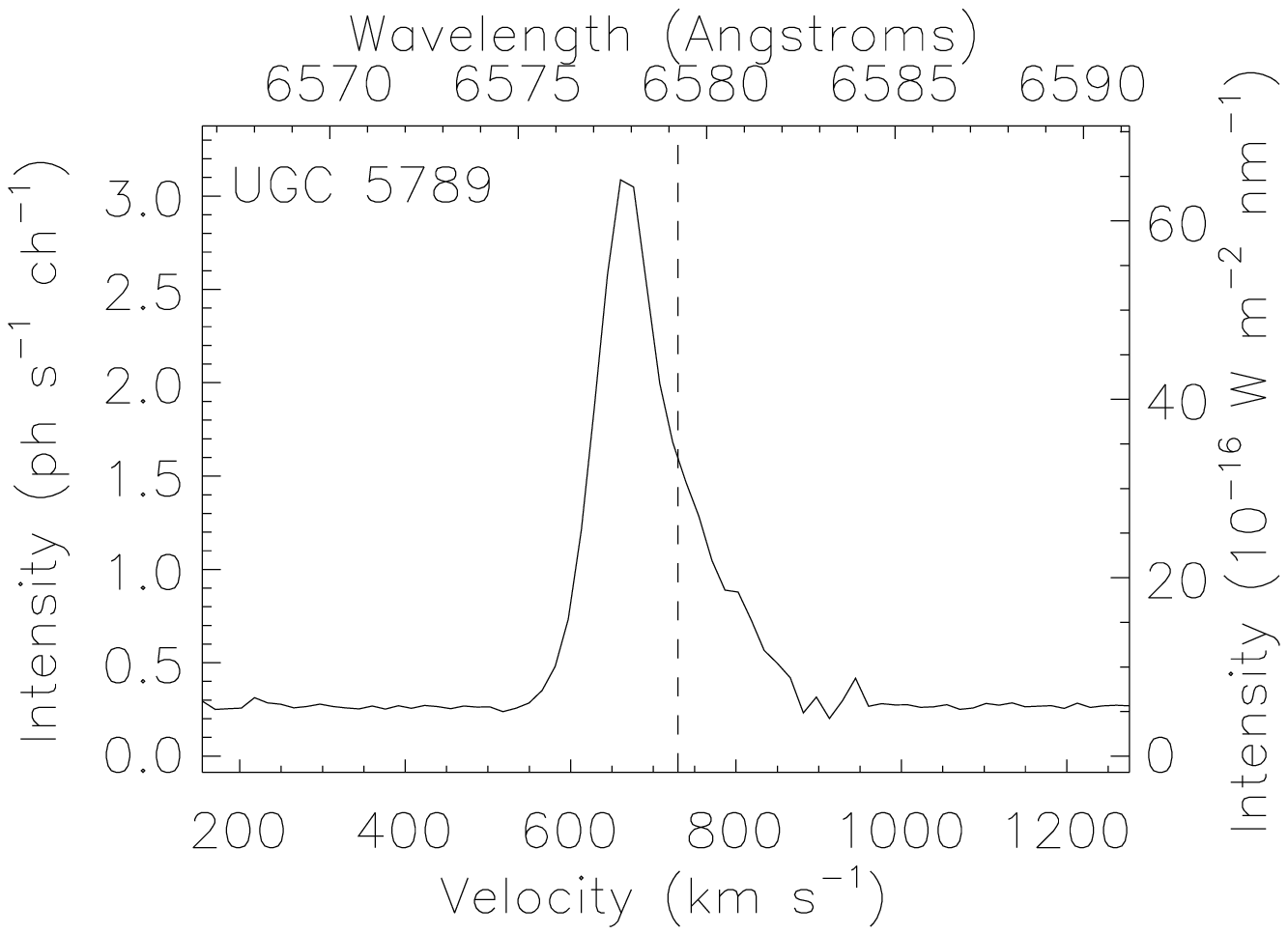}
\includegraphics[width=3.5cm]{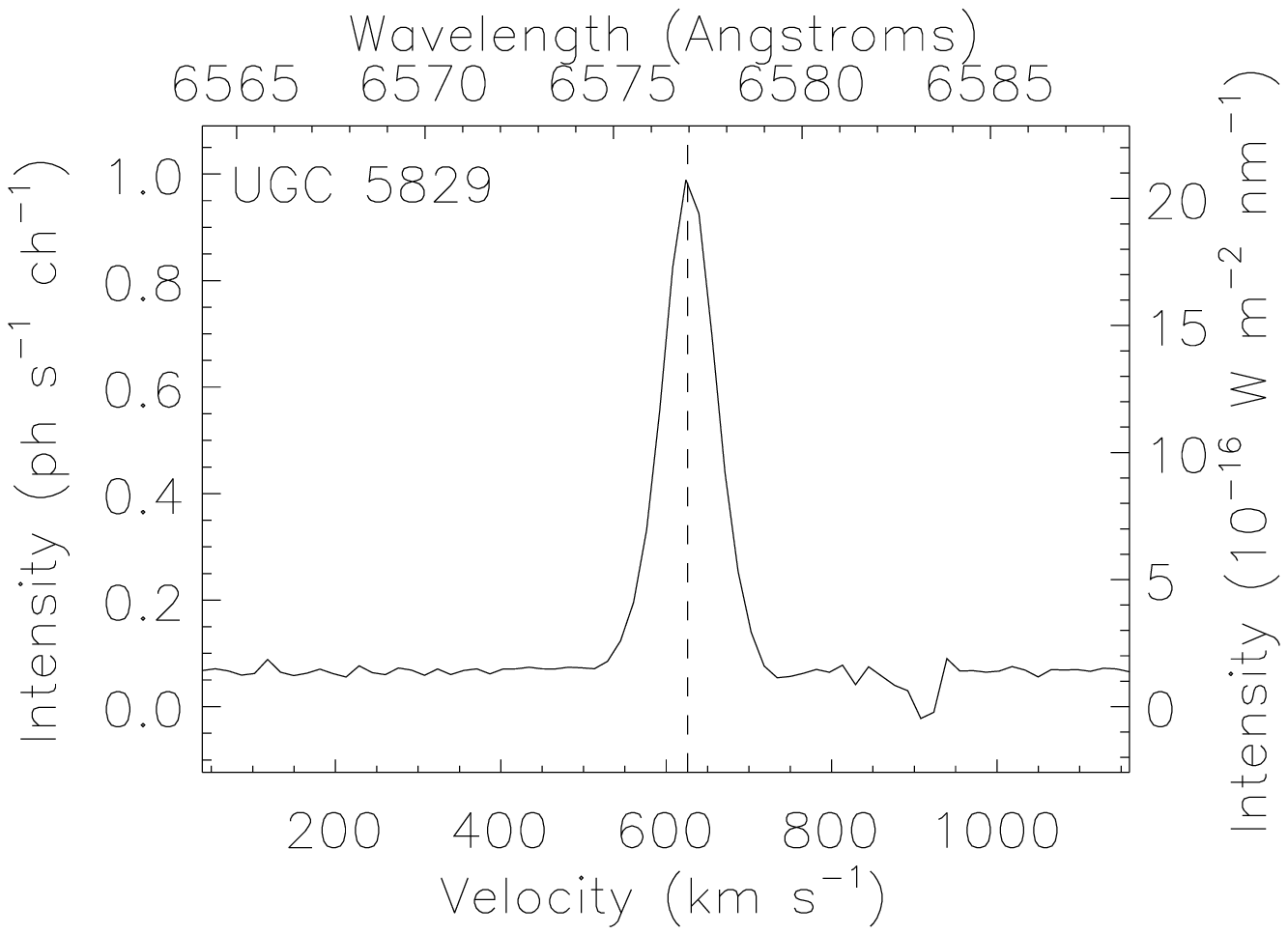}
\includegraphics[width=3.5cm]{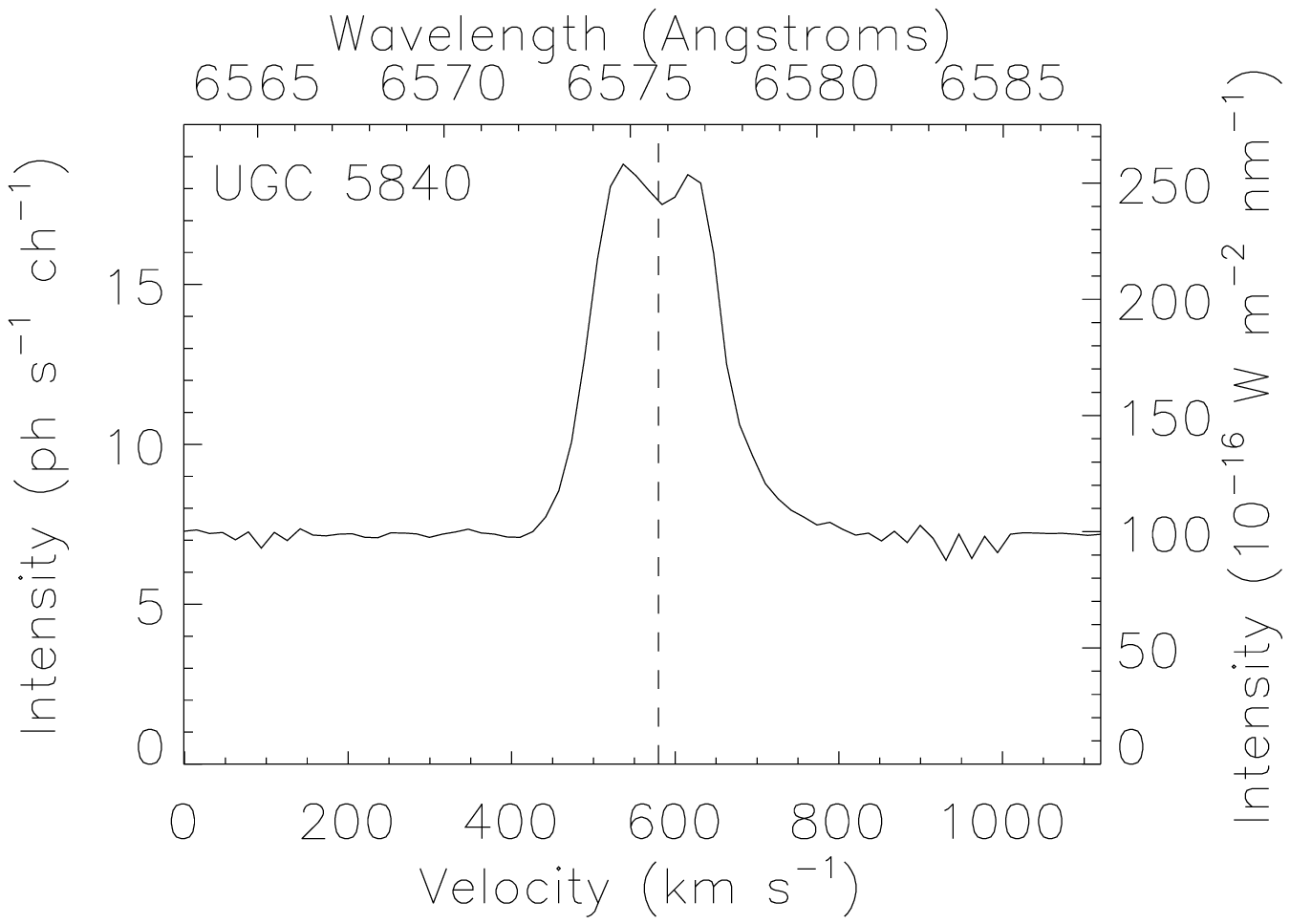}
\includegraphics[width=3.5cm]{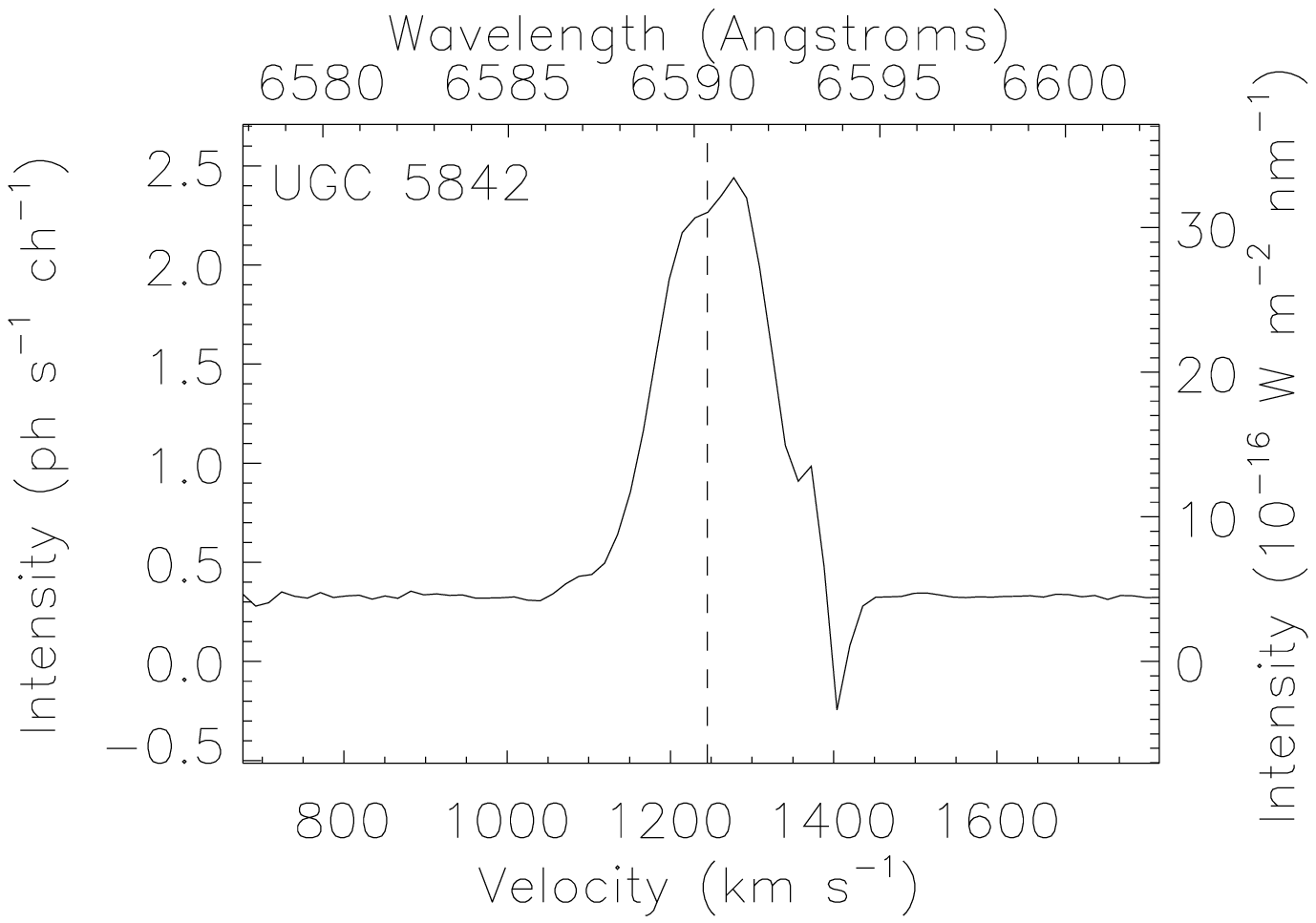}
\includegraphics[width=3.5cm]{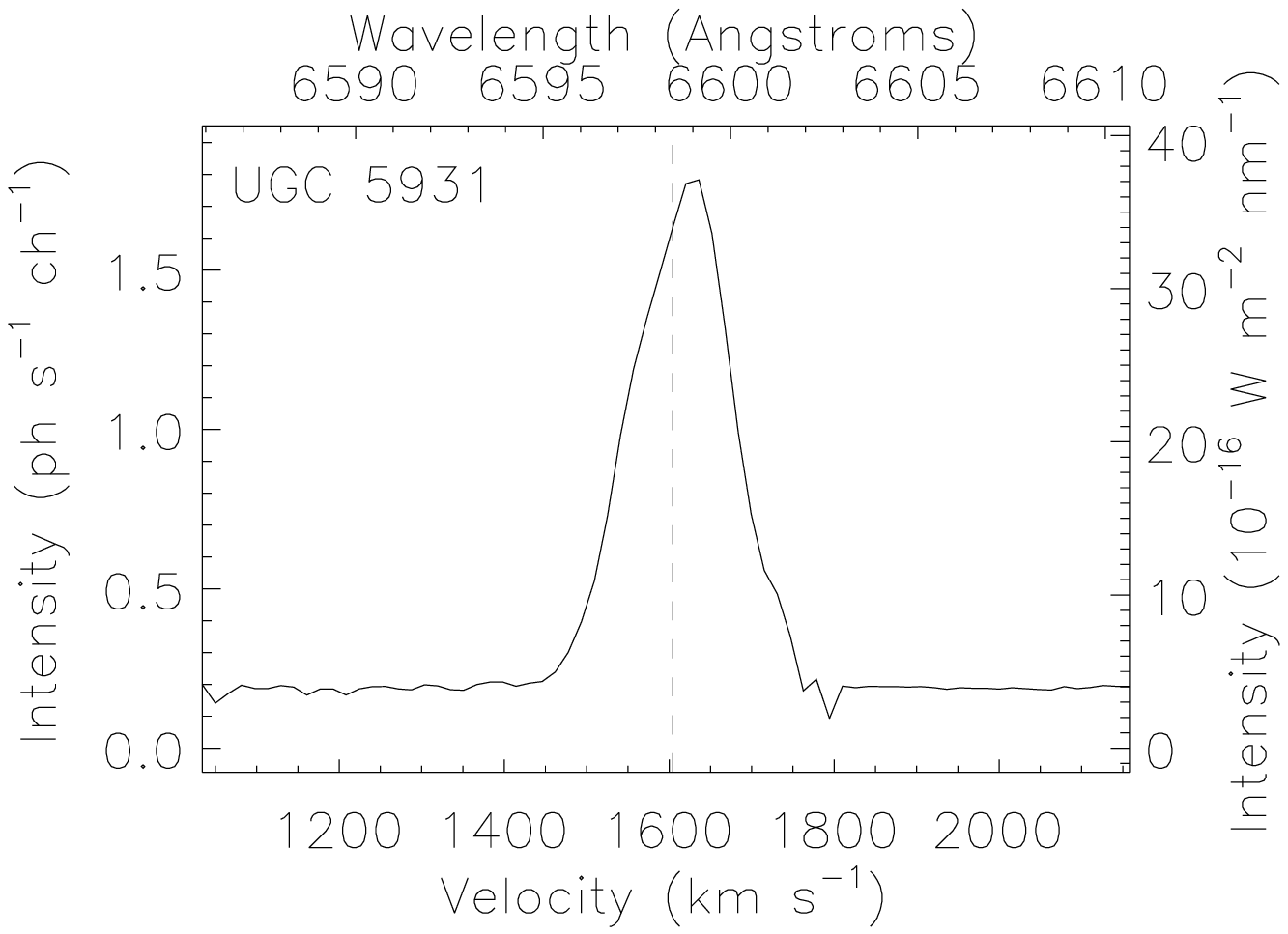}
\includegraphics[width=3.5cm]{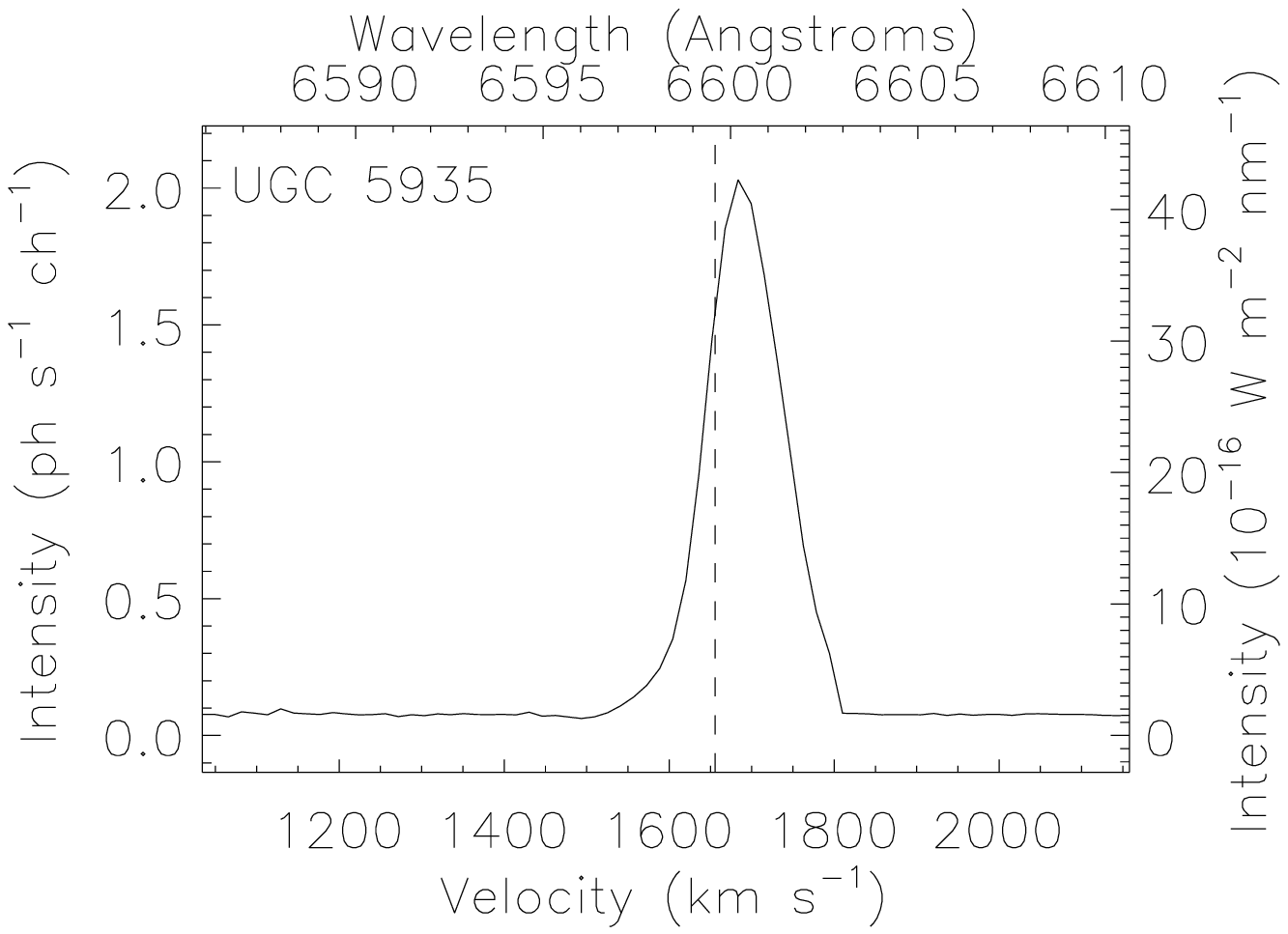}
\includegraphics[width=3.5cm]{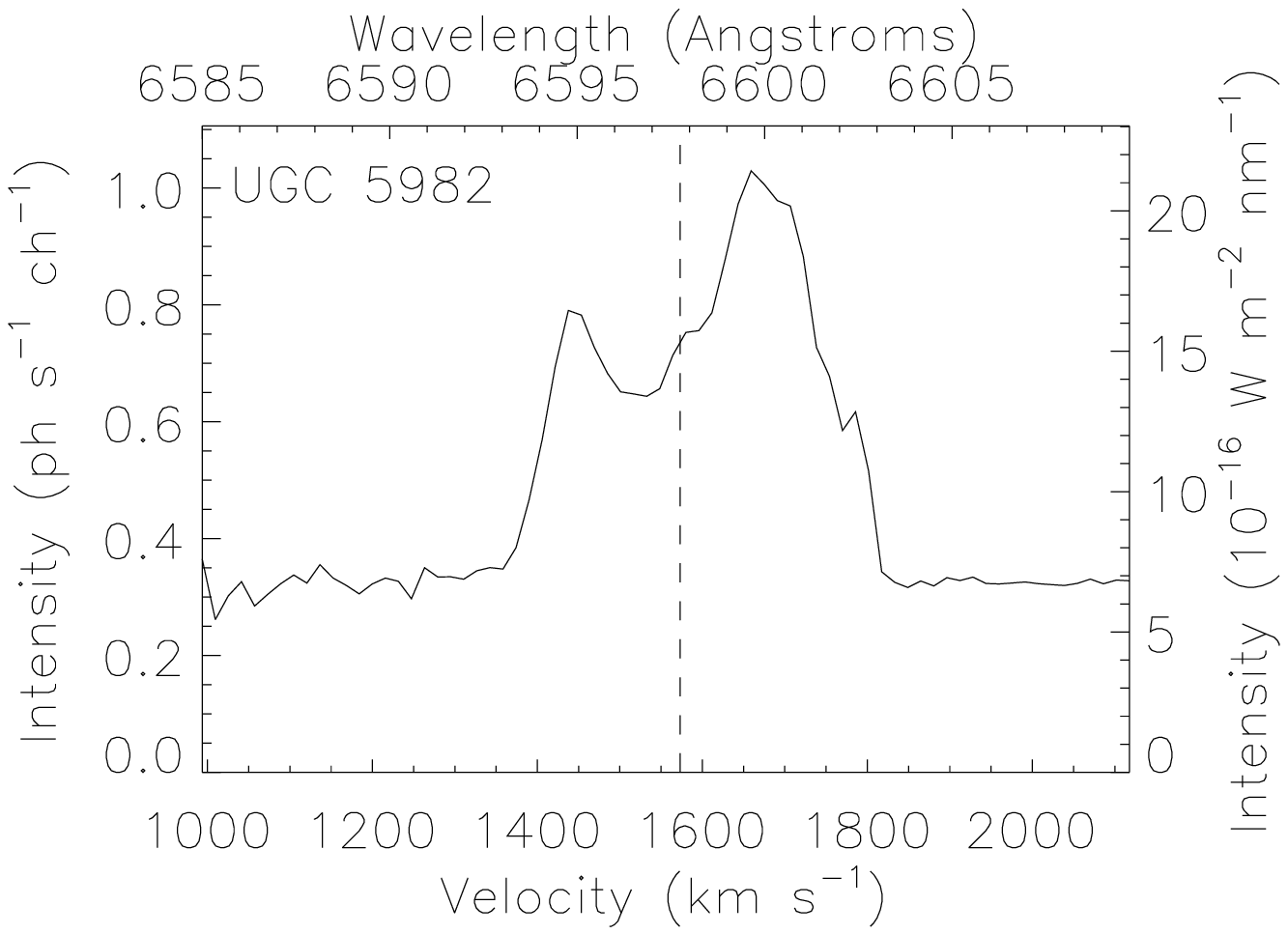}
\includegraphics[width=3.5cm]{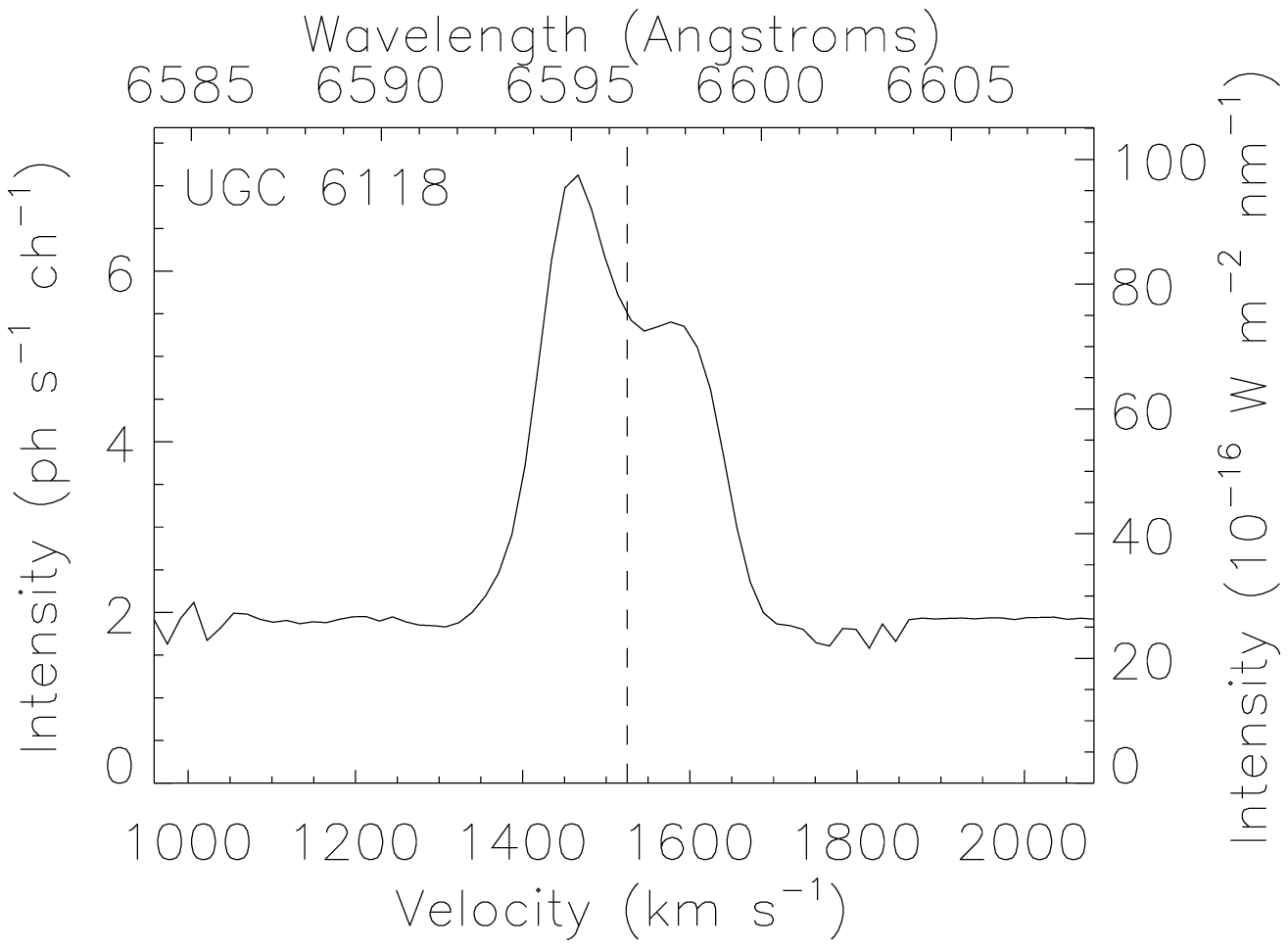}
\includegraphics[width=3.5cm]{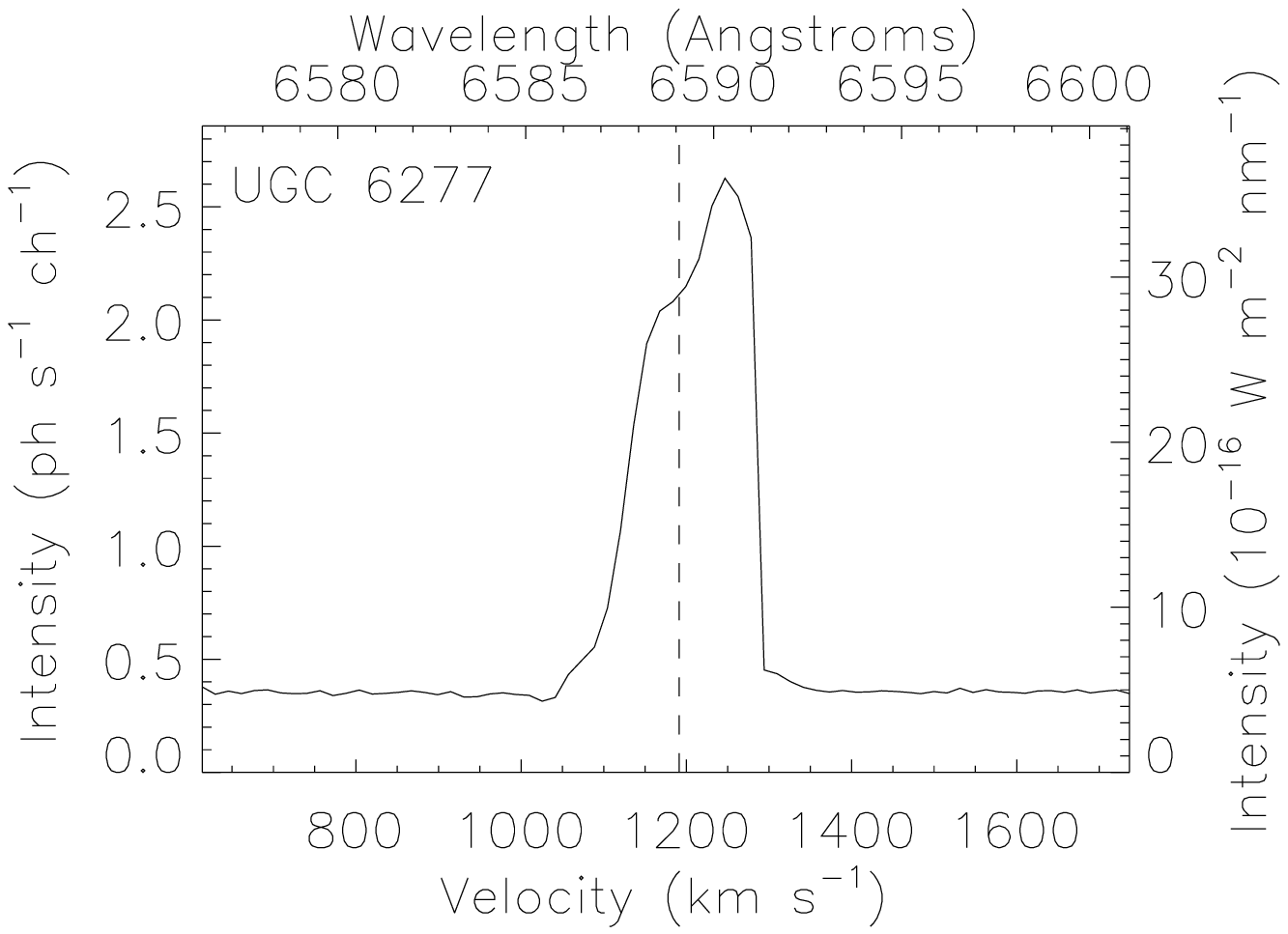}
\includegraphics[width=3.5cm]{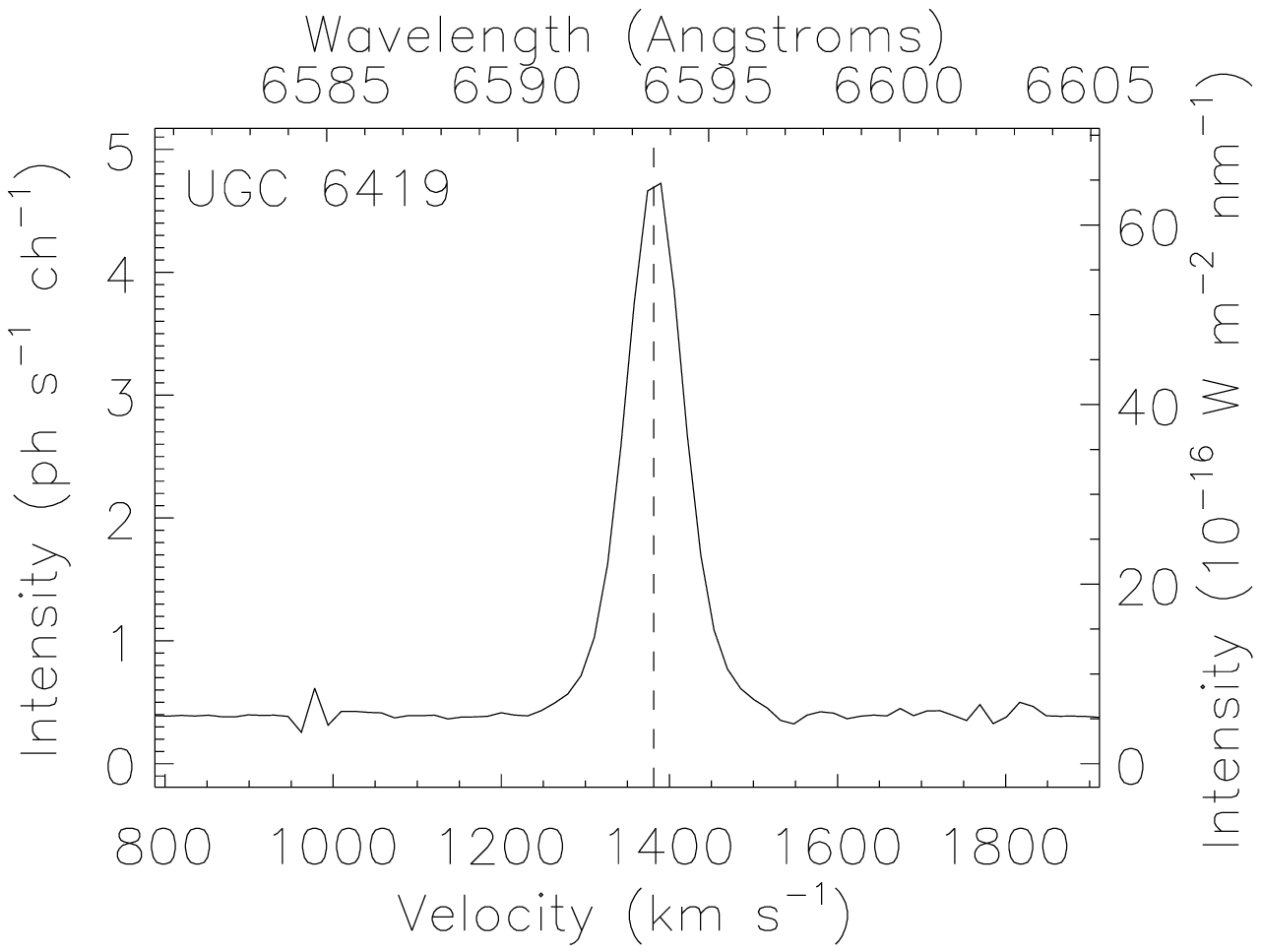}
\end{center}
\end{minipage}
\end{figure}
\clearpage
\begin{figure}
\begin{minipage}{180mm}
\begin{center}
\includegraphics[width=3.5cm]{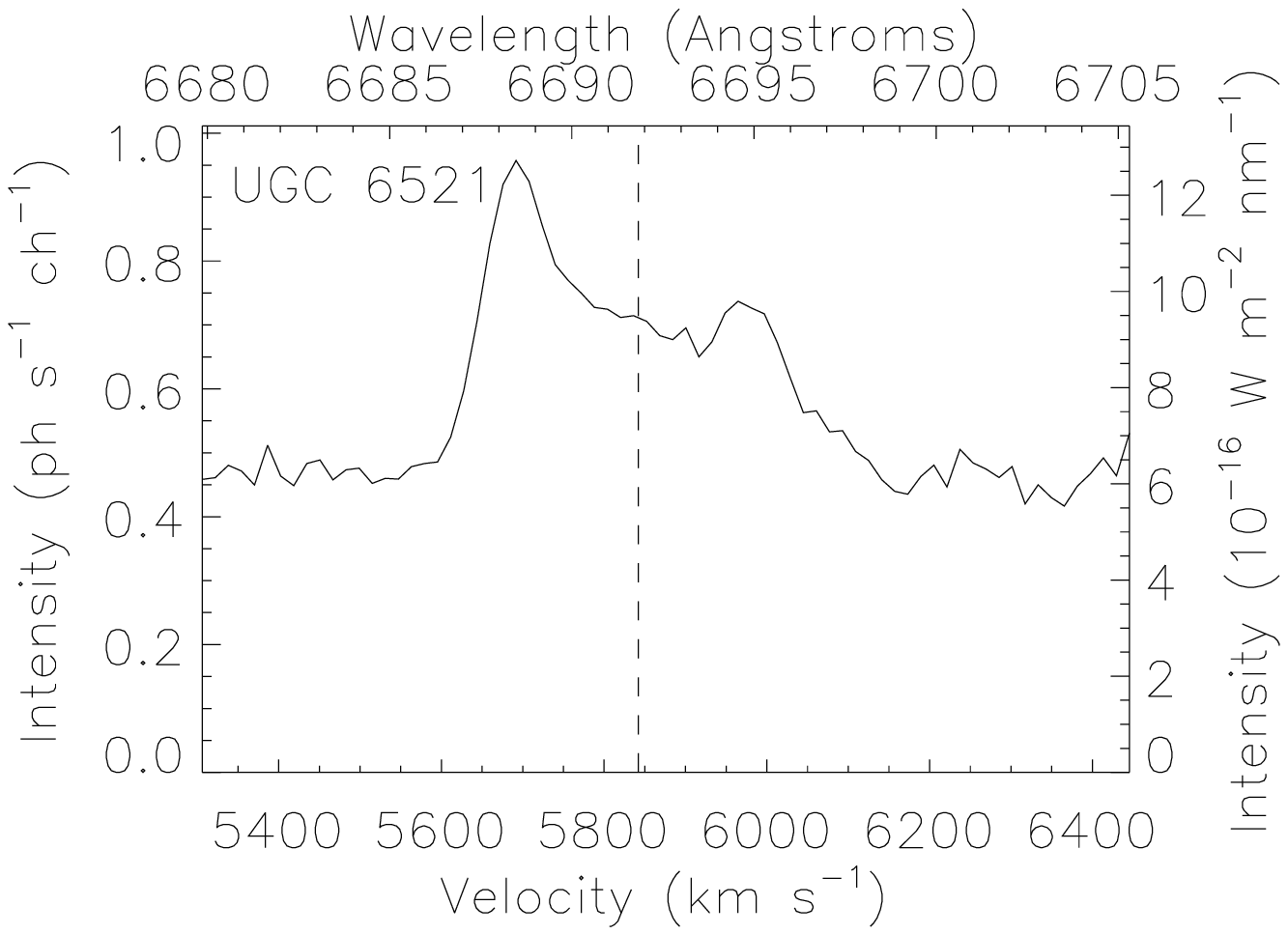}
\includegraphics[width=3.5cm]{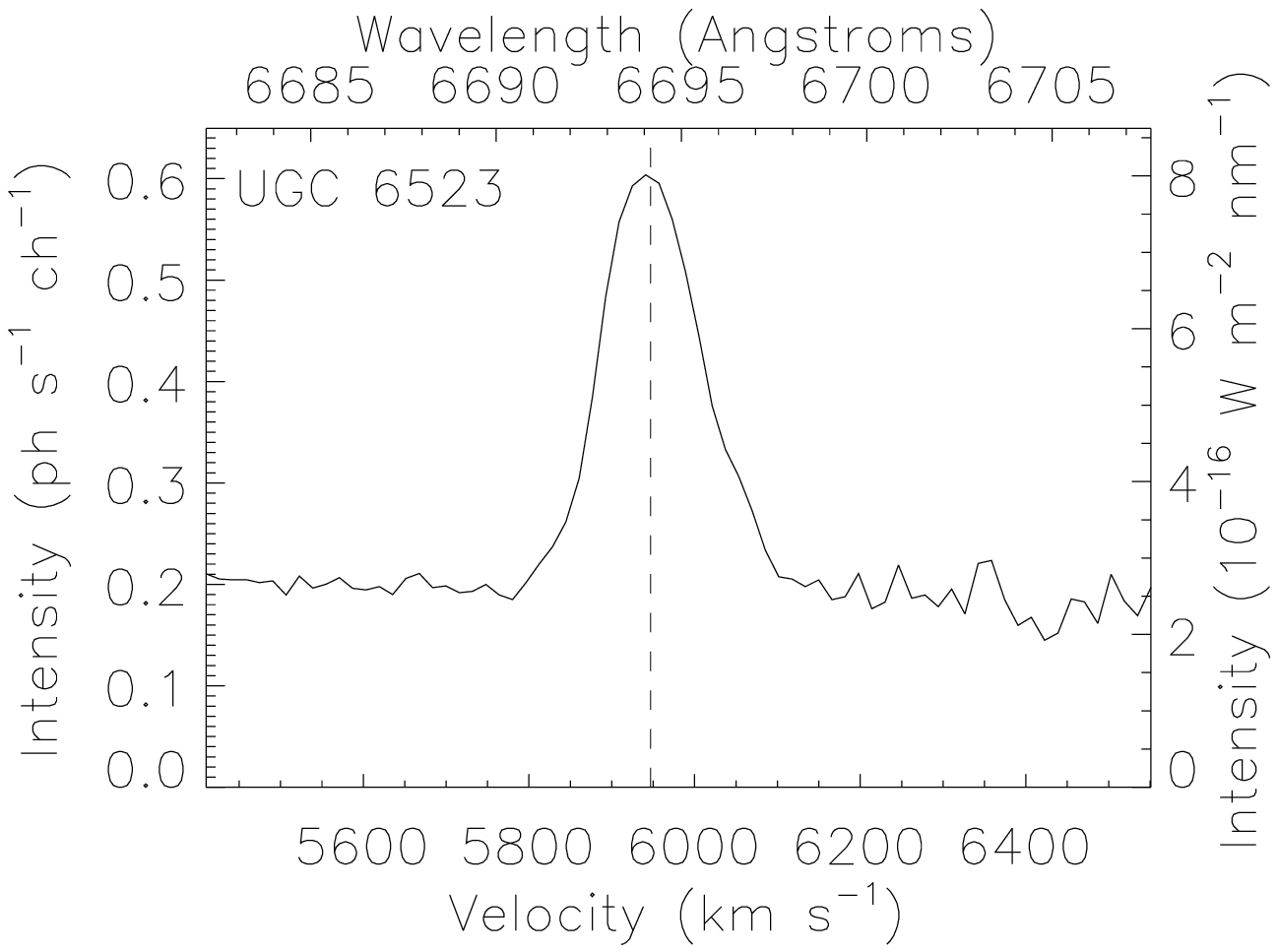}
\includegraphics[width=3.5cm]{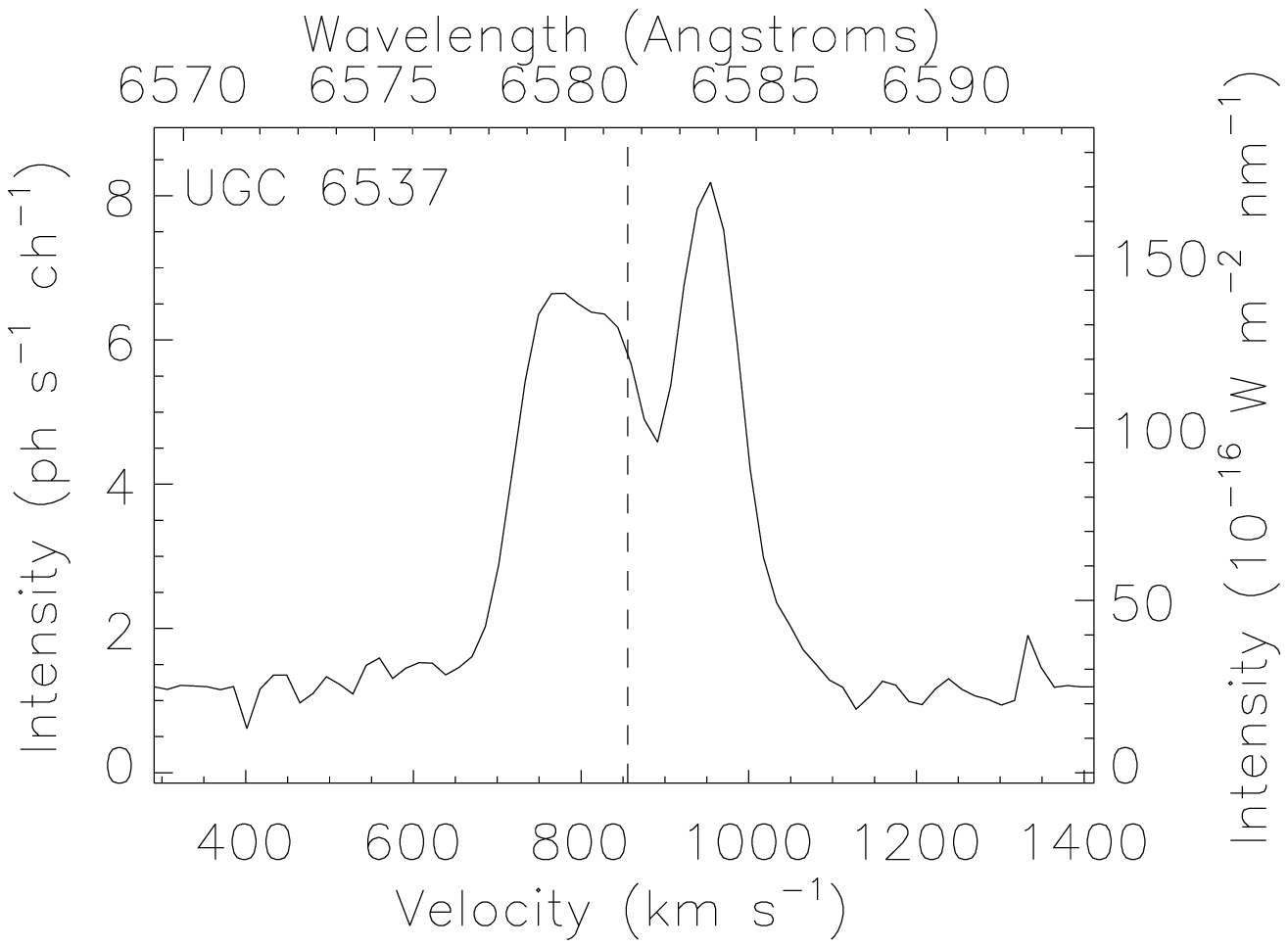}
\includegraphics[width=3.5cm]{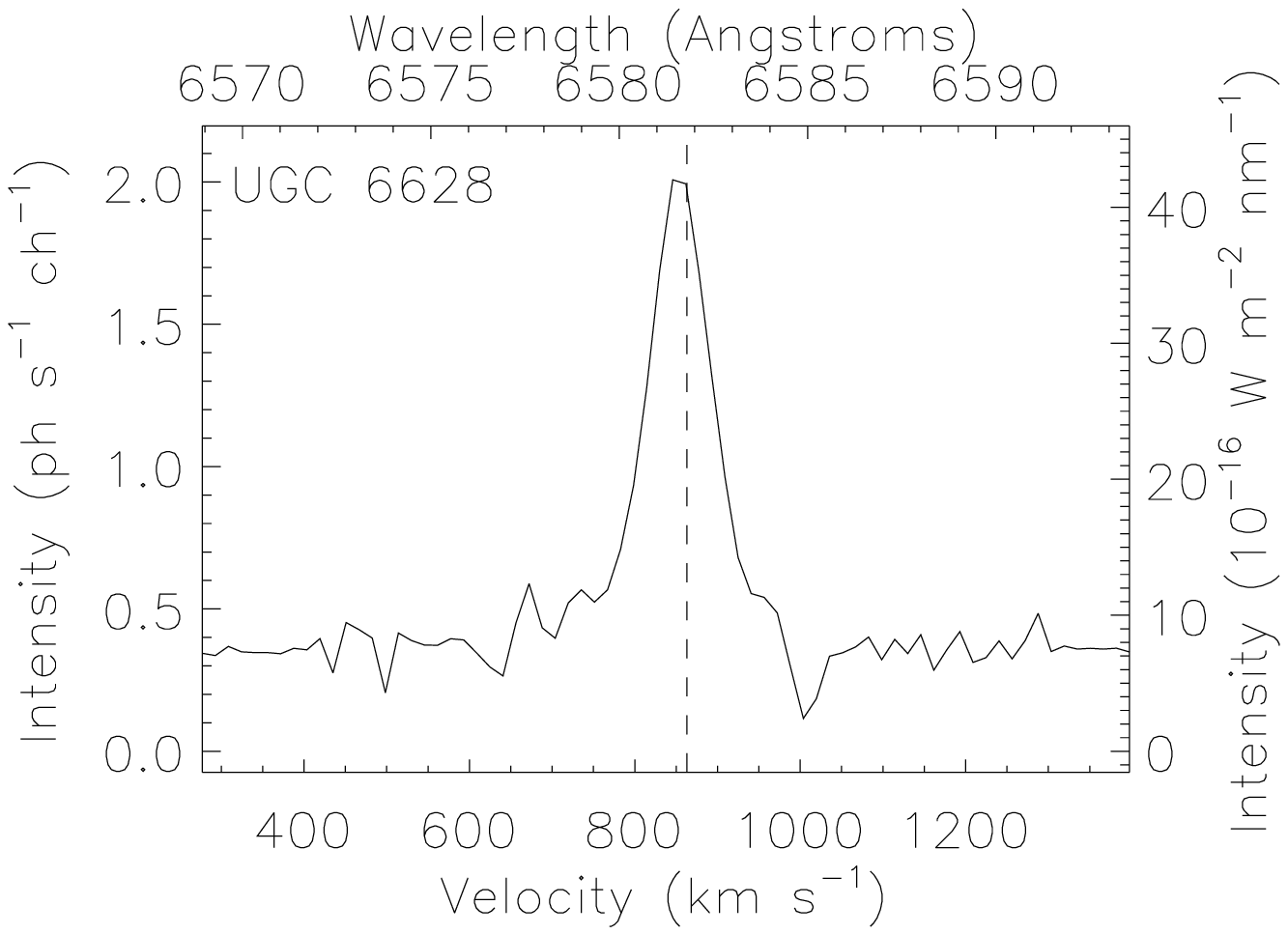}
\includegraphics[width=3.5cm]{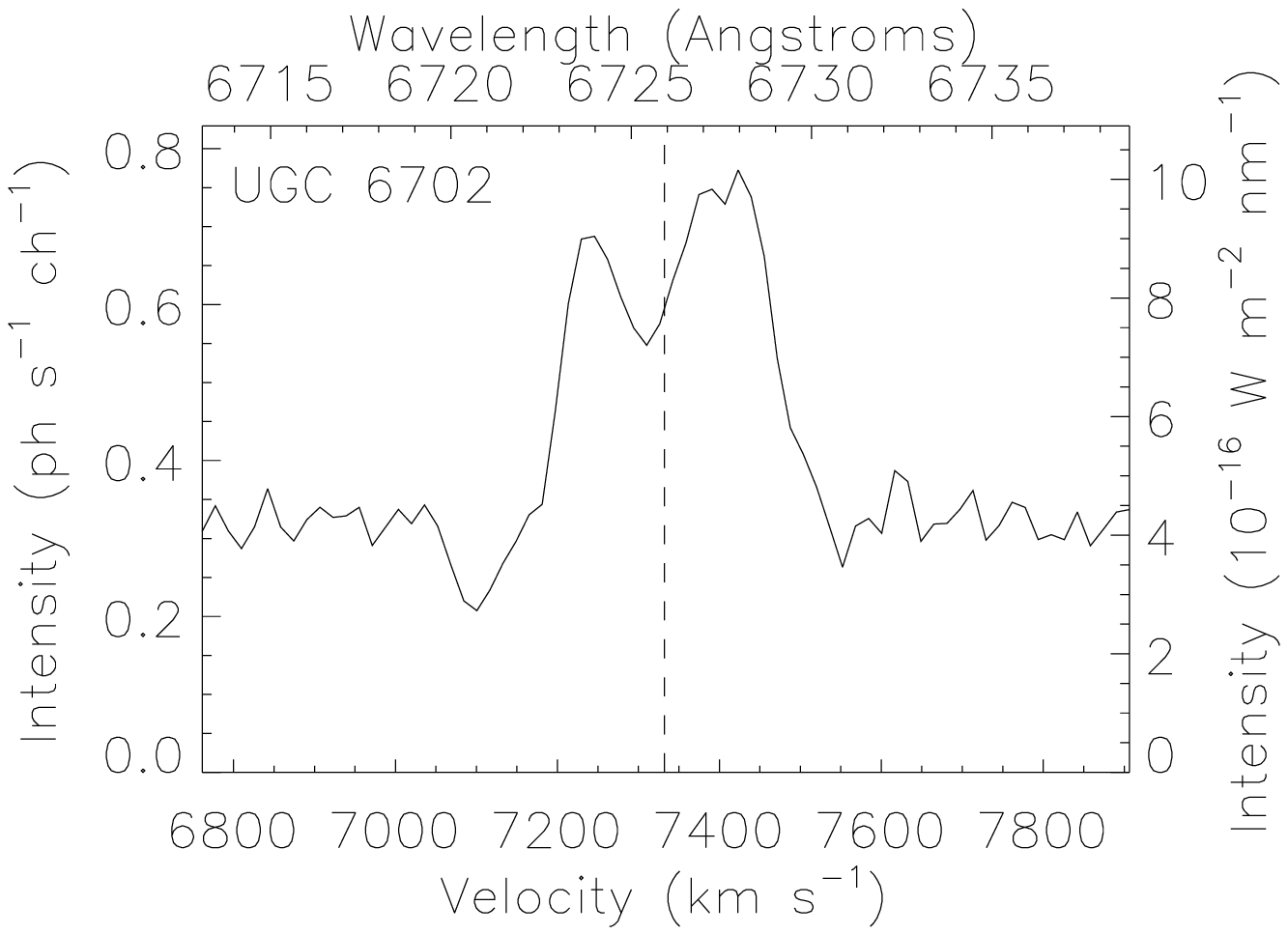}
\includegraphics[width=3.5cm]{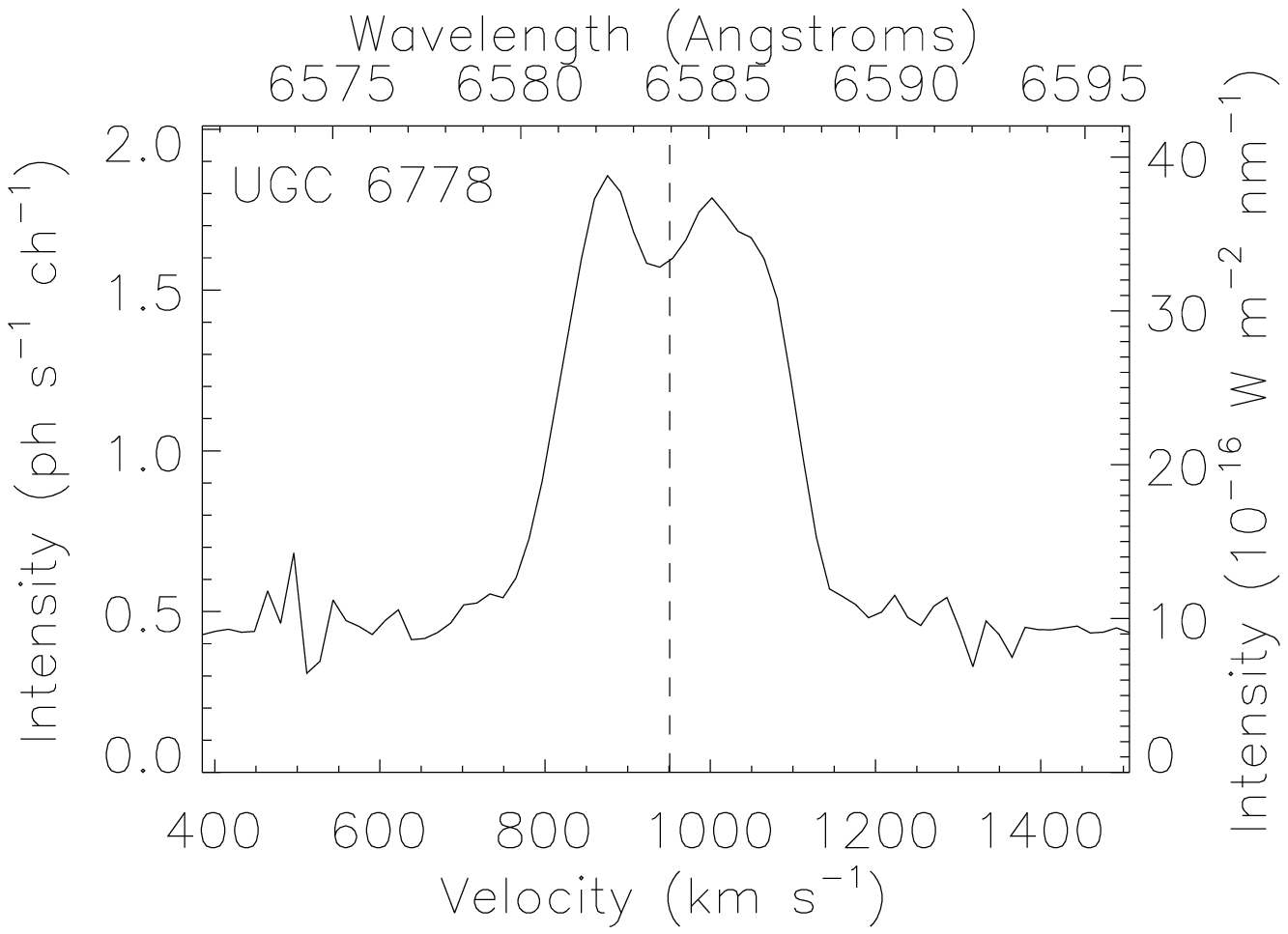}
\includegraphics[width=3.5cm]{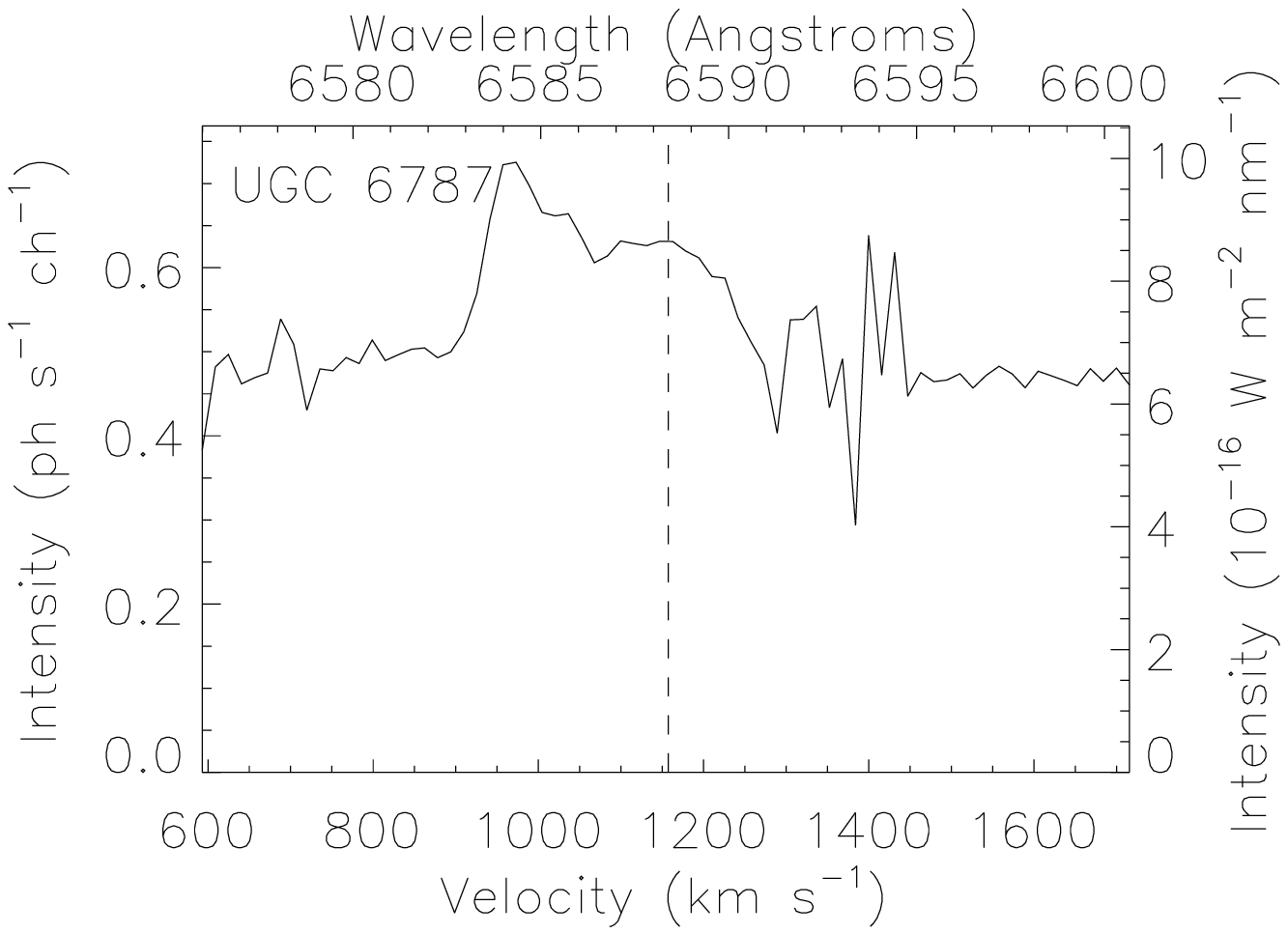}
\includegraphics[width=3.5cm]{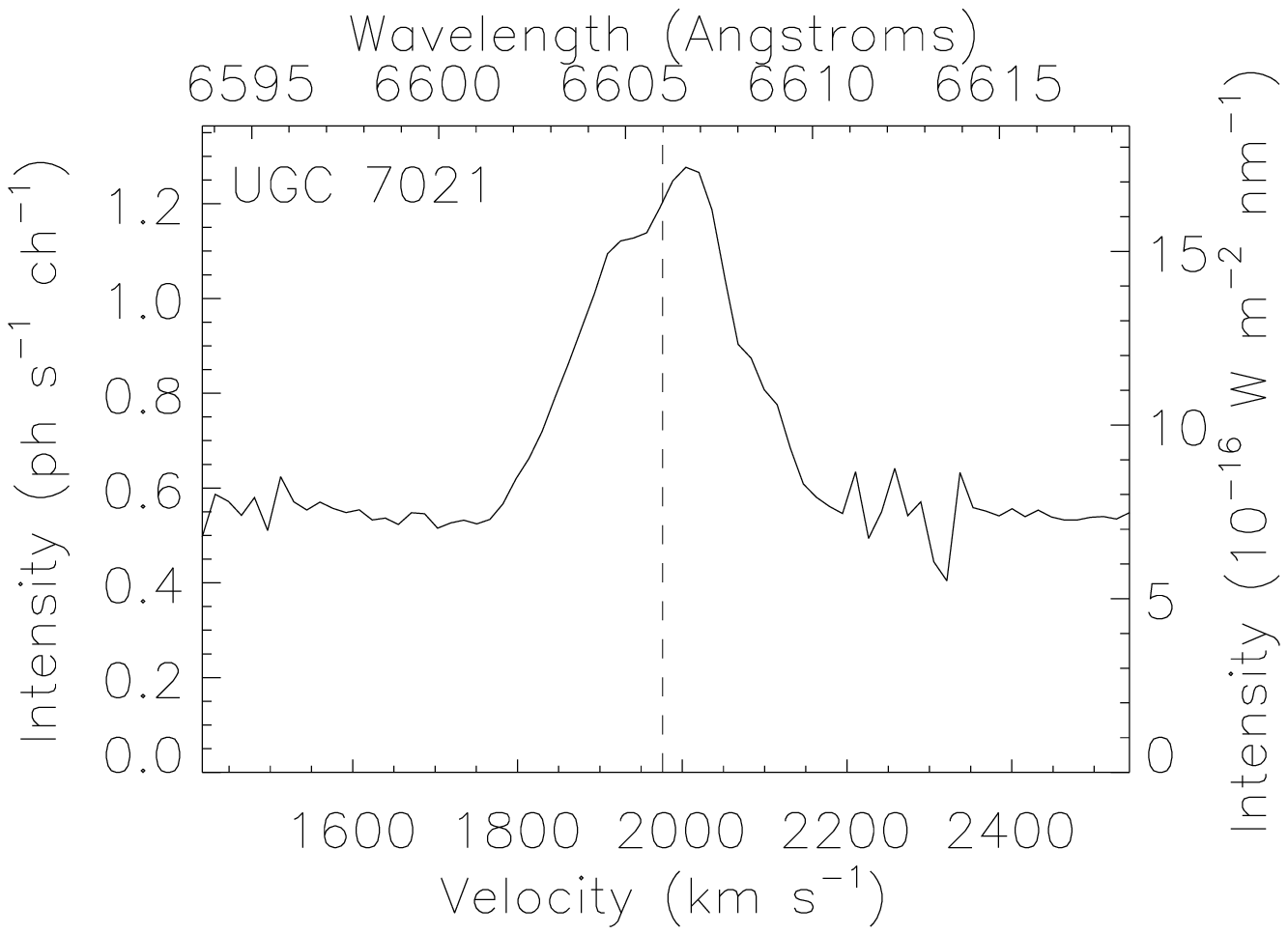}
\includegraphics[width=3.5cm]{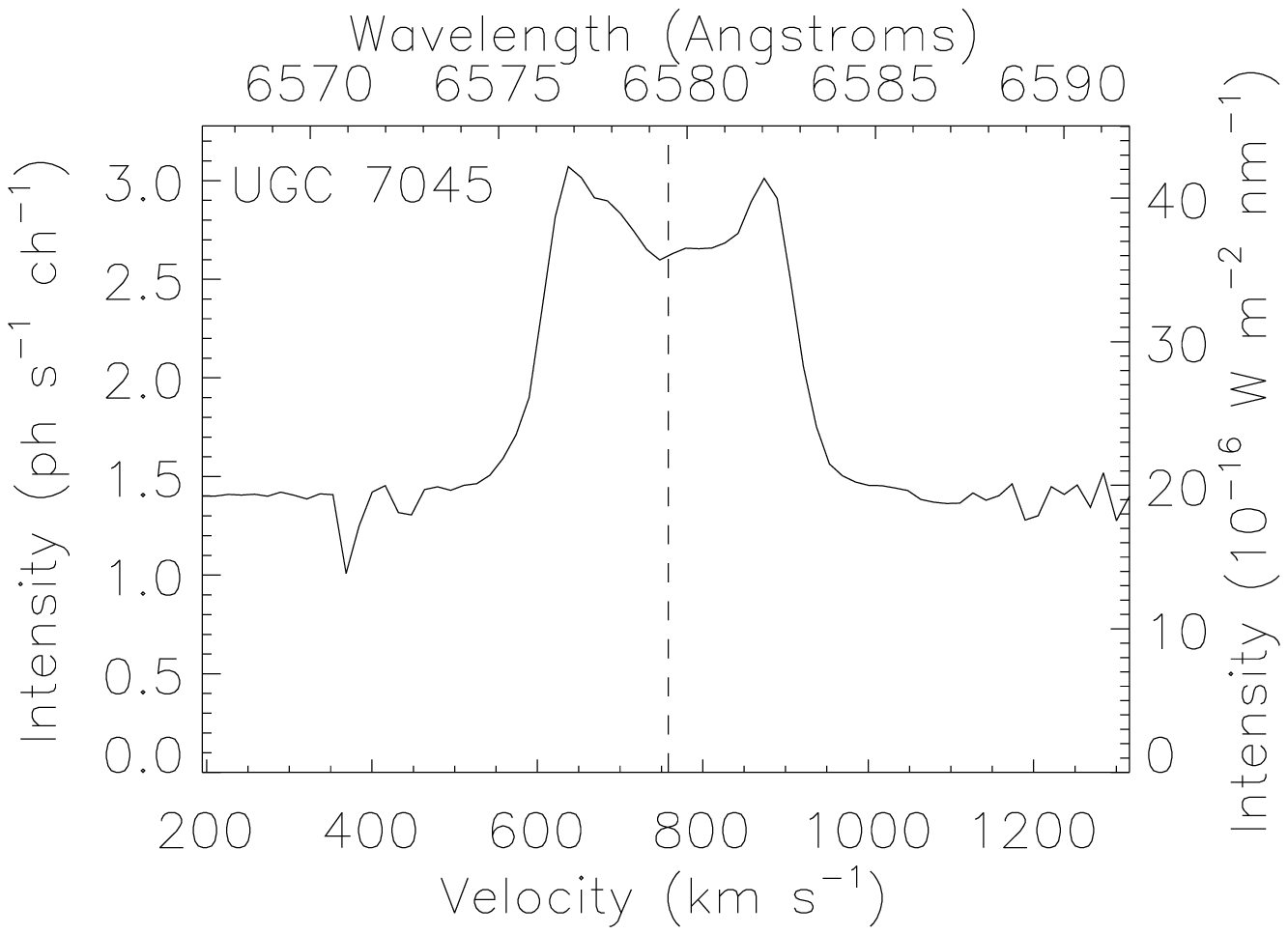}
\includegraphics[width=3.5cm]{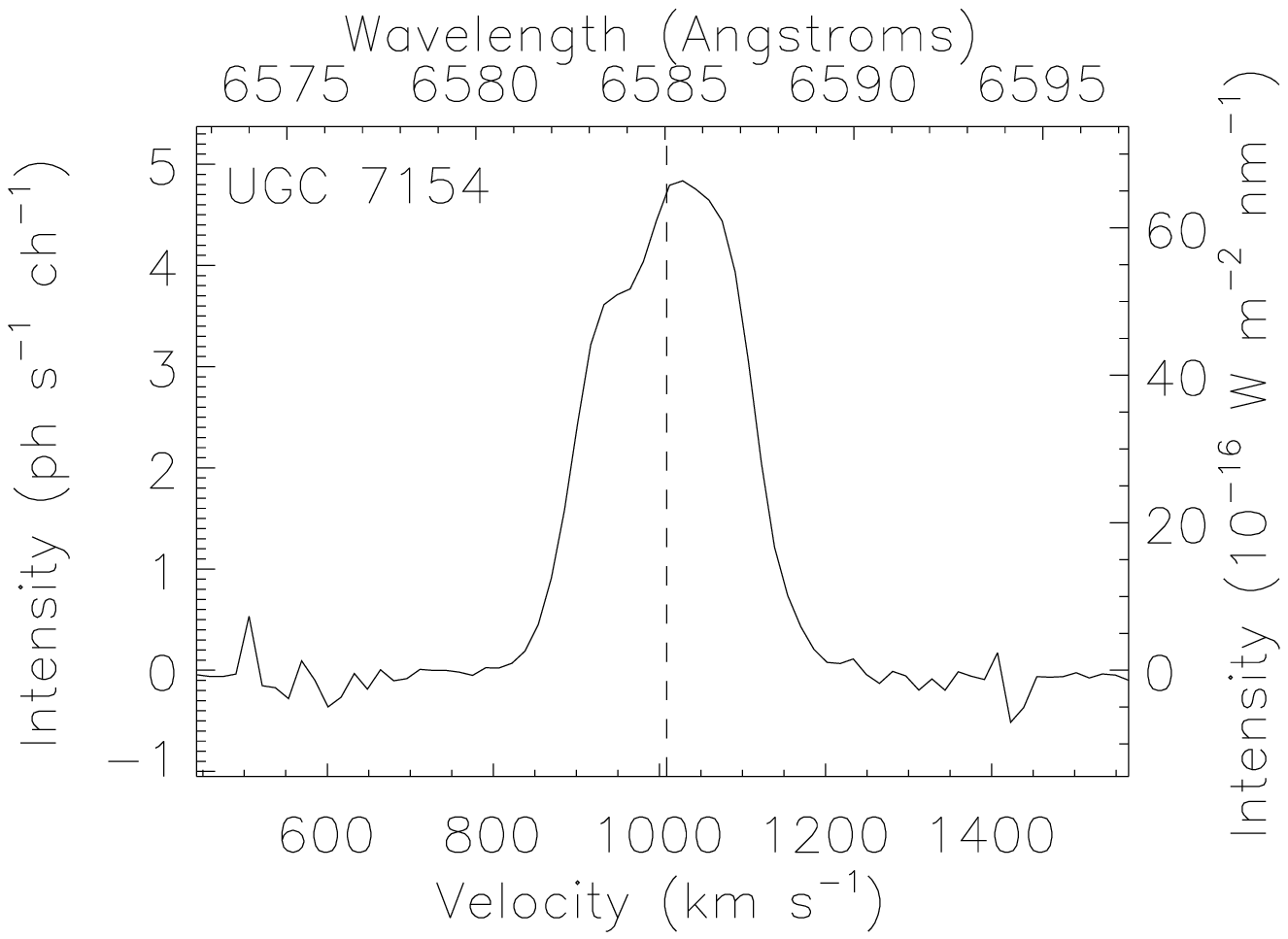}
\includegraphics[width=3.5cm]{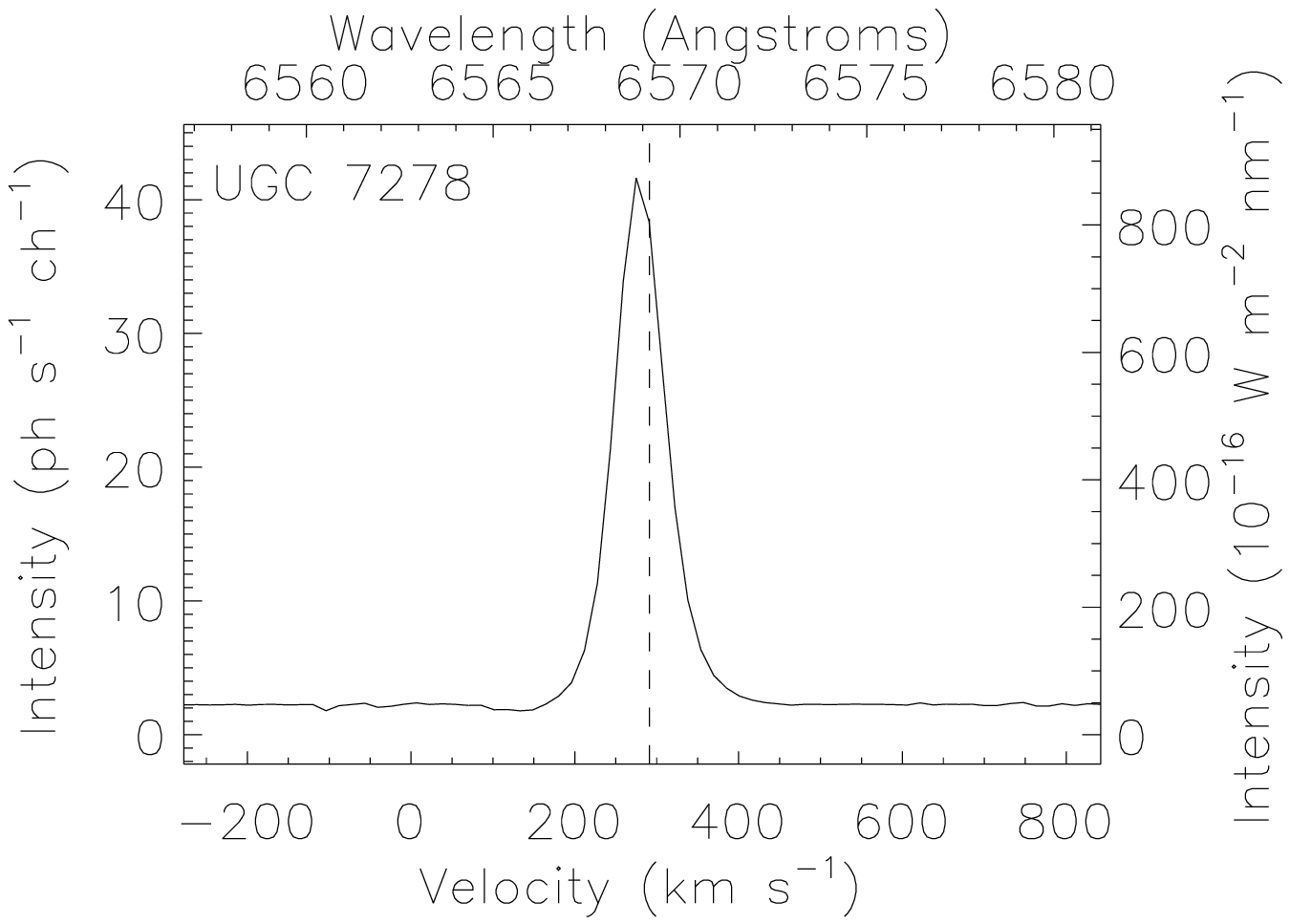}
\includegraphics[width=3.5cm]{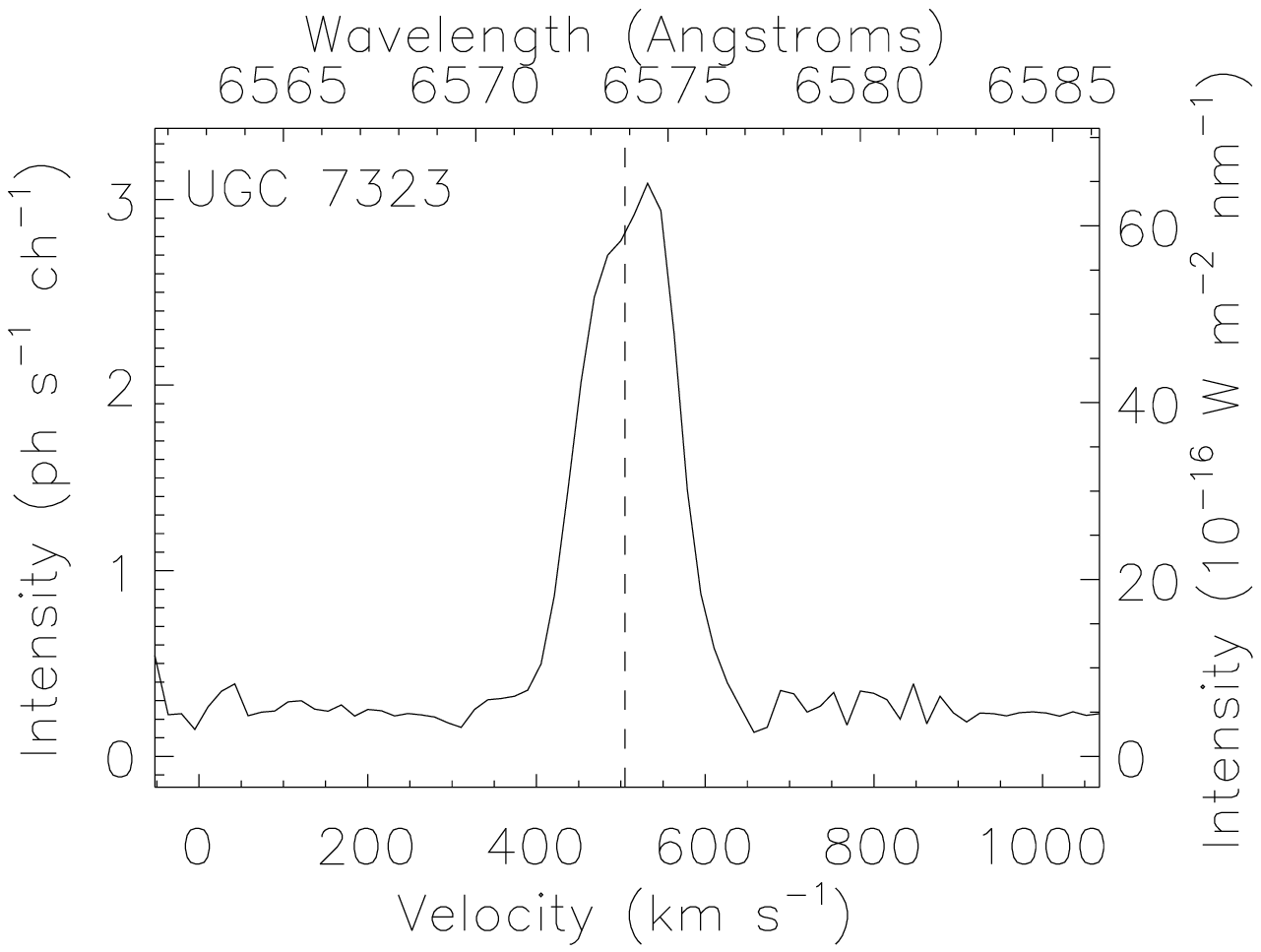}
\includegraphics[width=3.5cm]{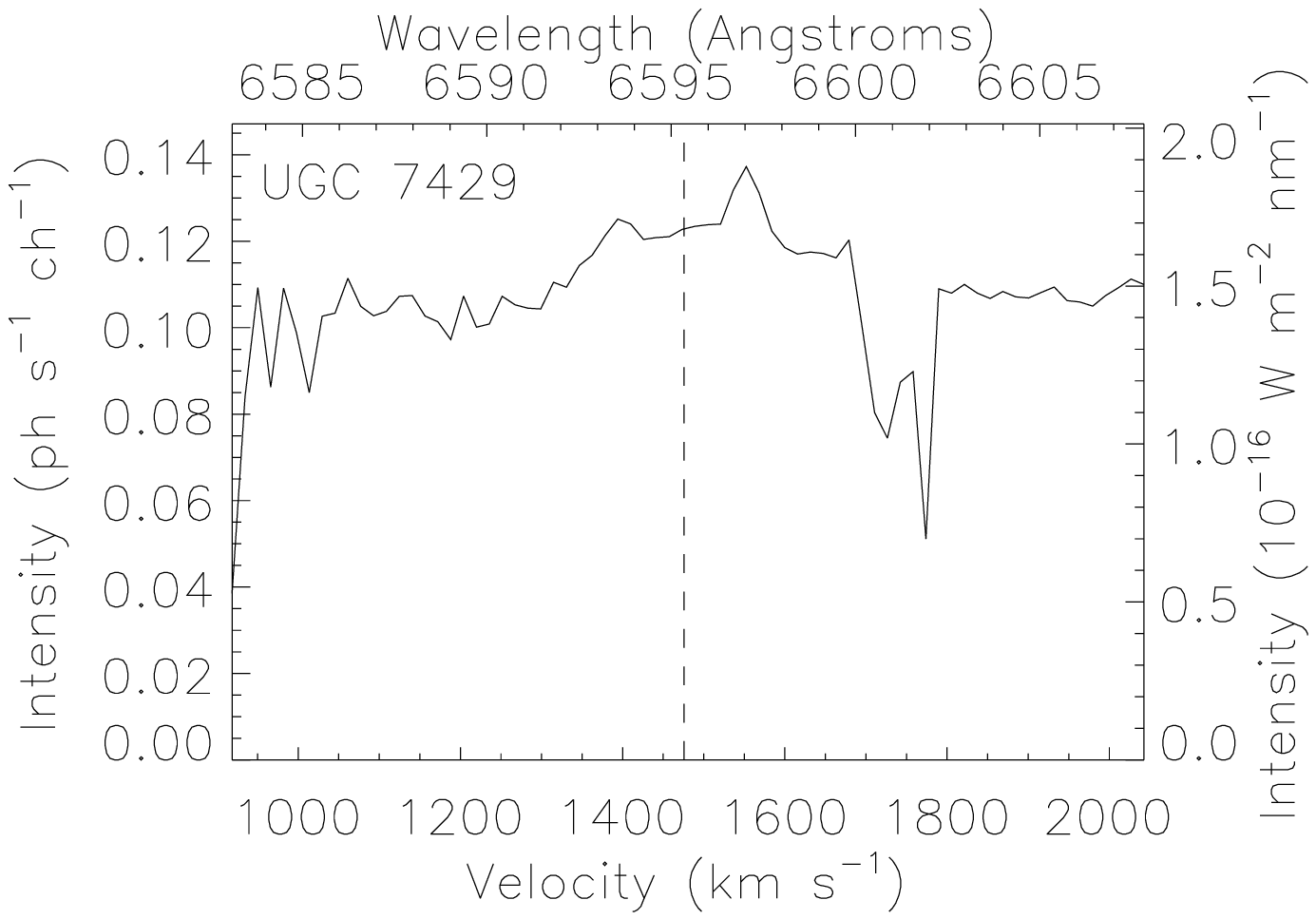}
\includegraphics[width=3.5cm]{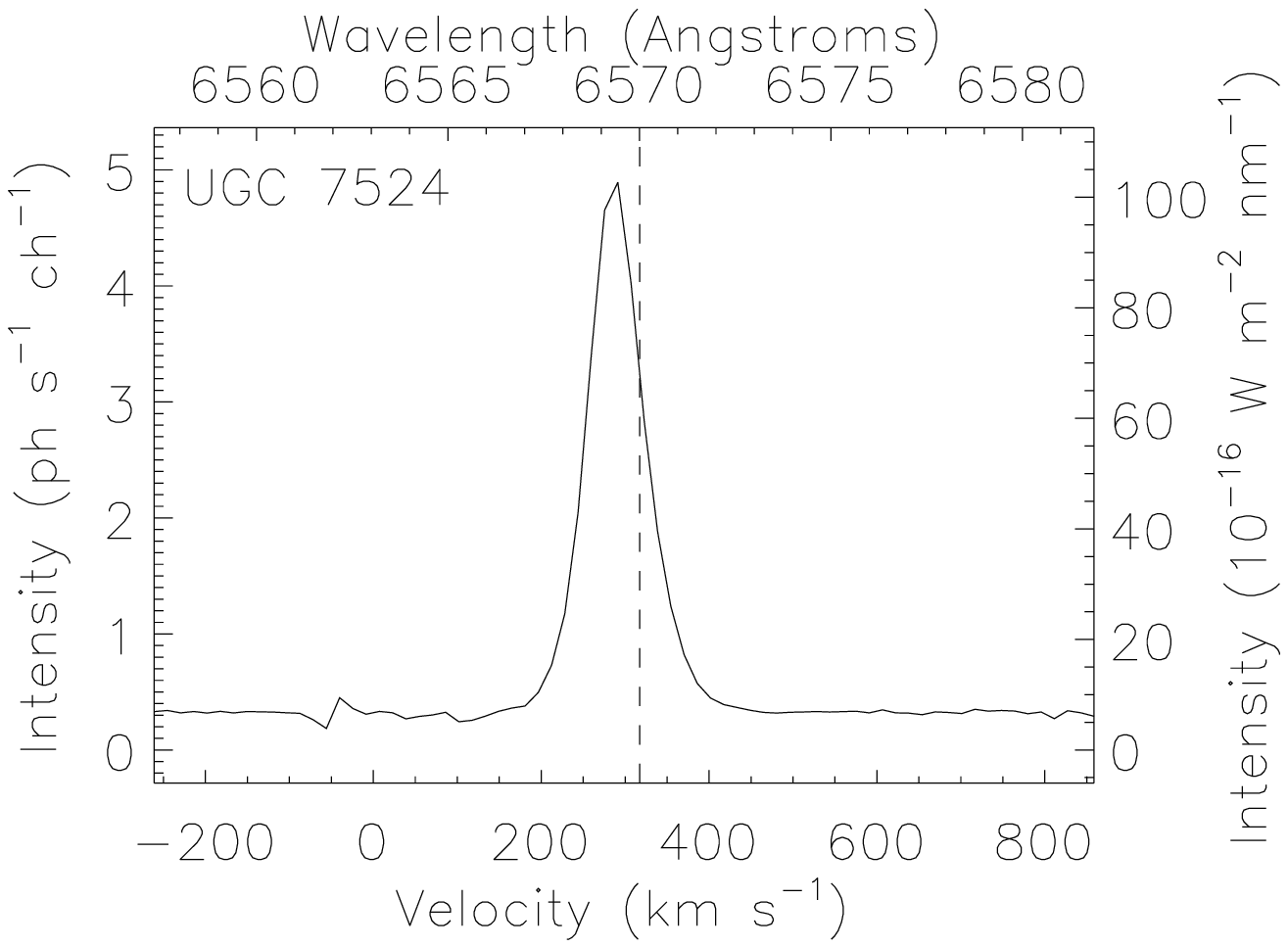}
\includegraphics[width=3.5cm]{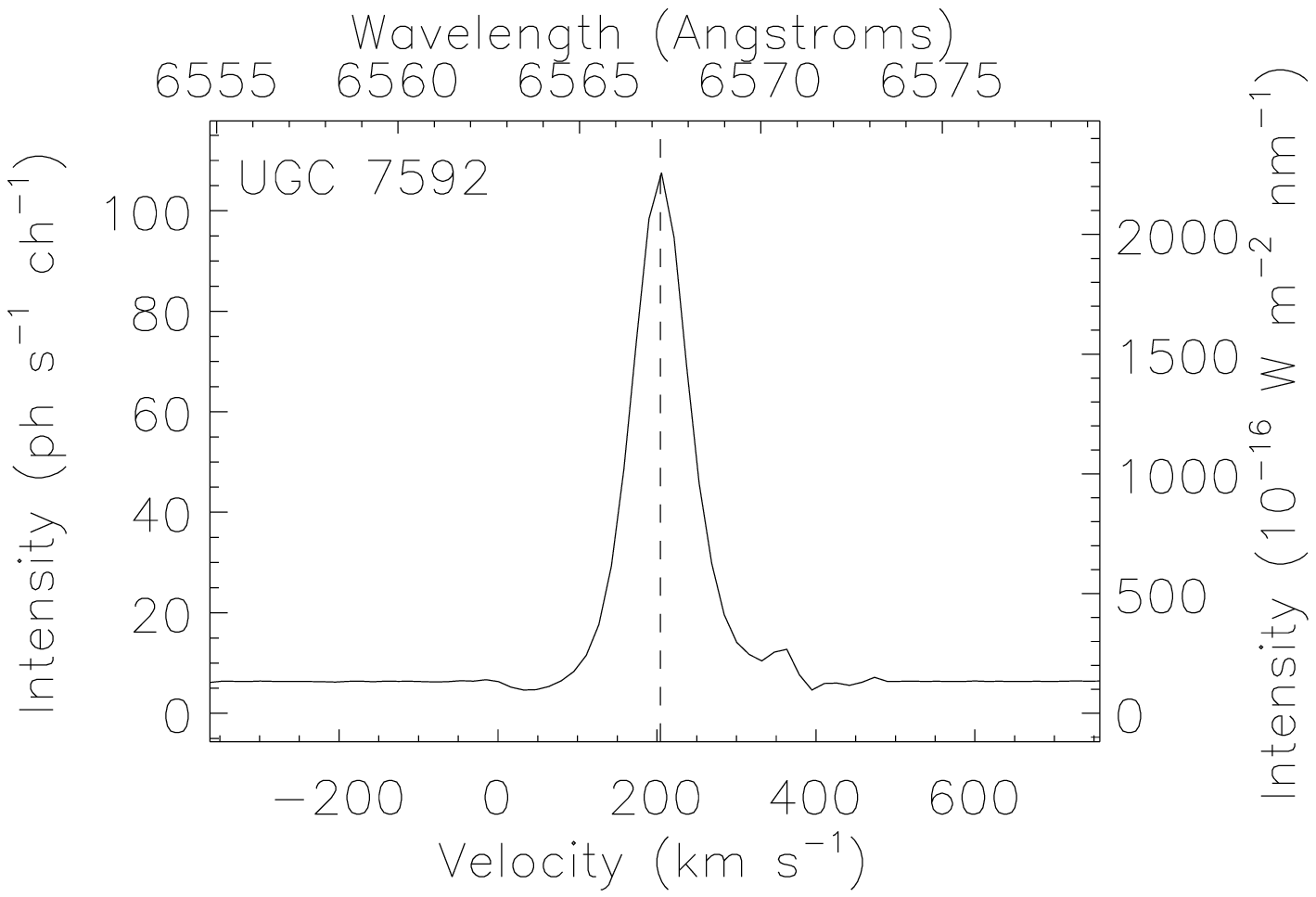}
\includegraphics[width=3.5cm]{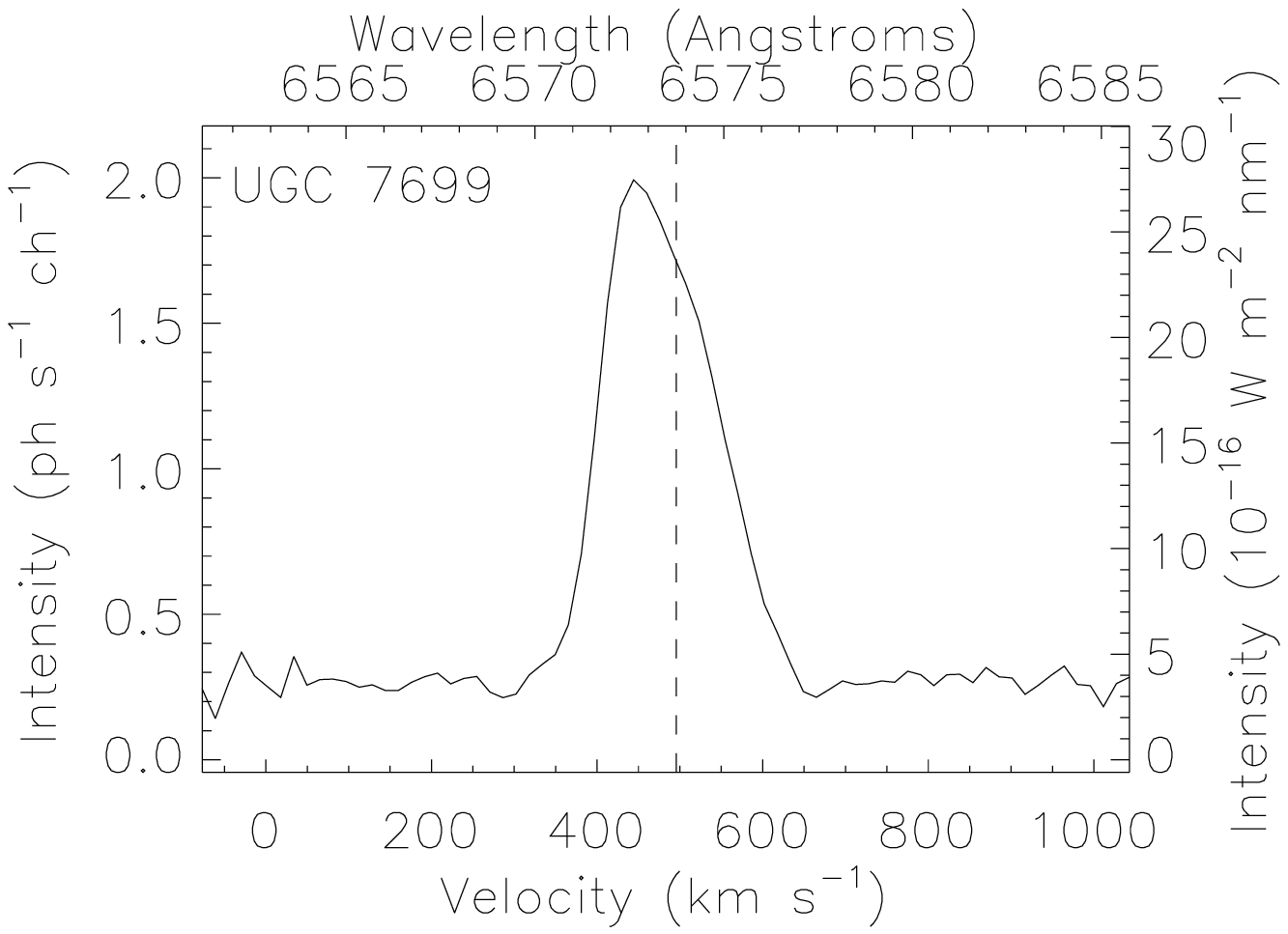}
\includegraphics[width=3.5cm]{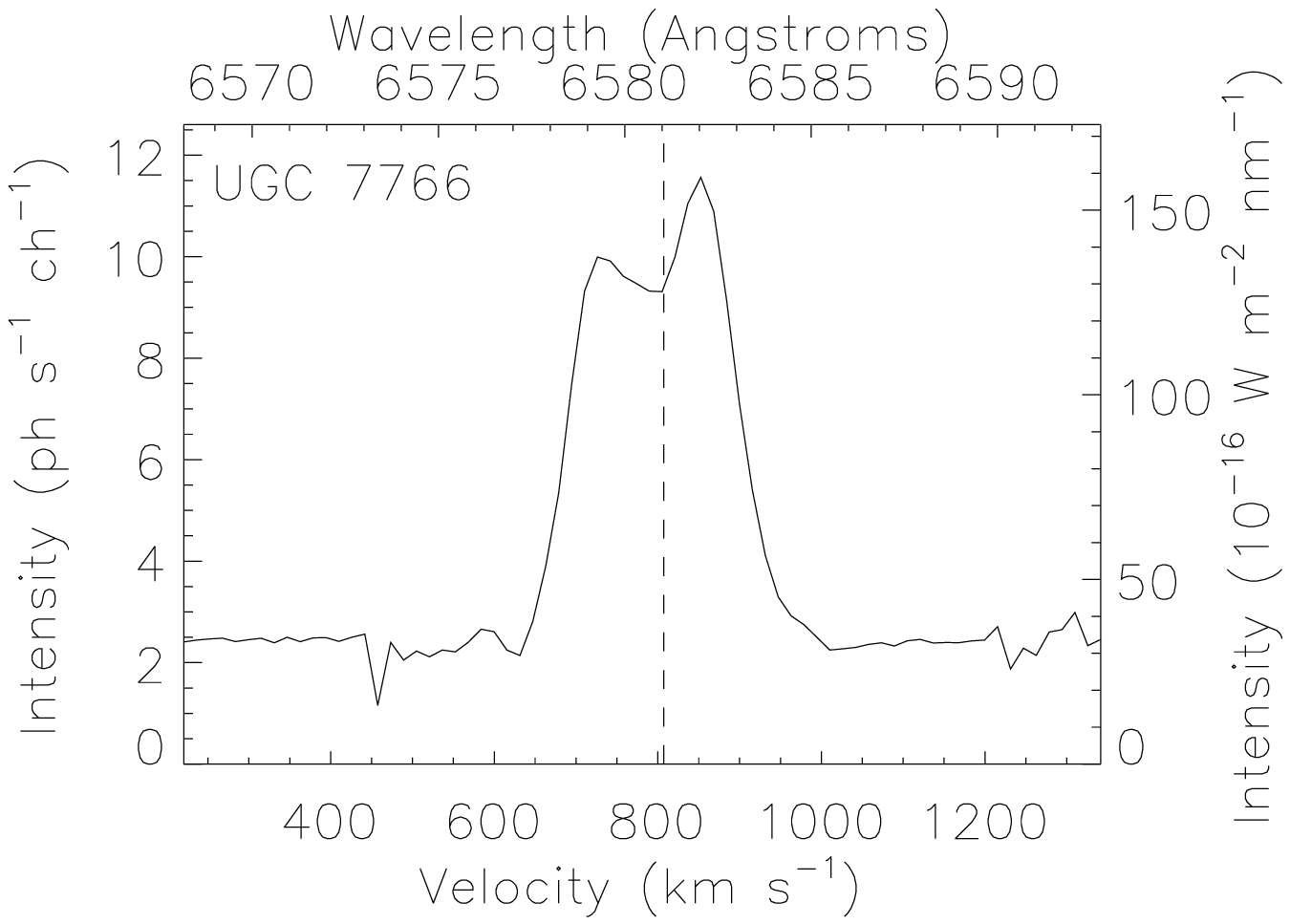}
\includegraphics[width=3.5cm]{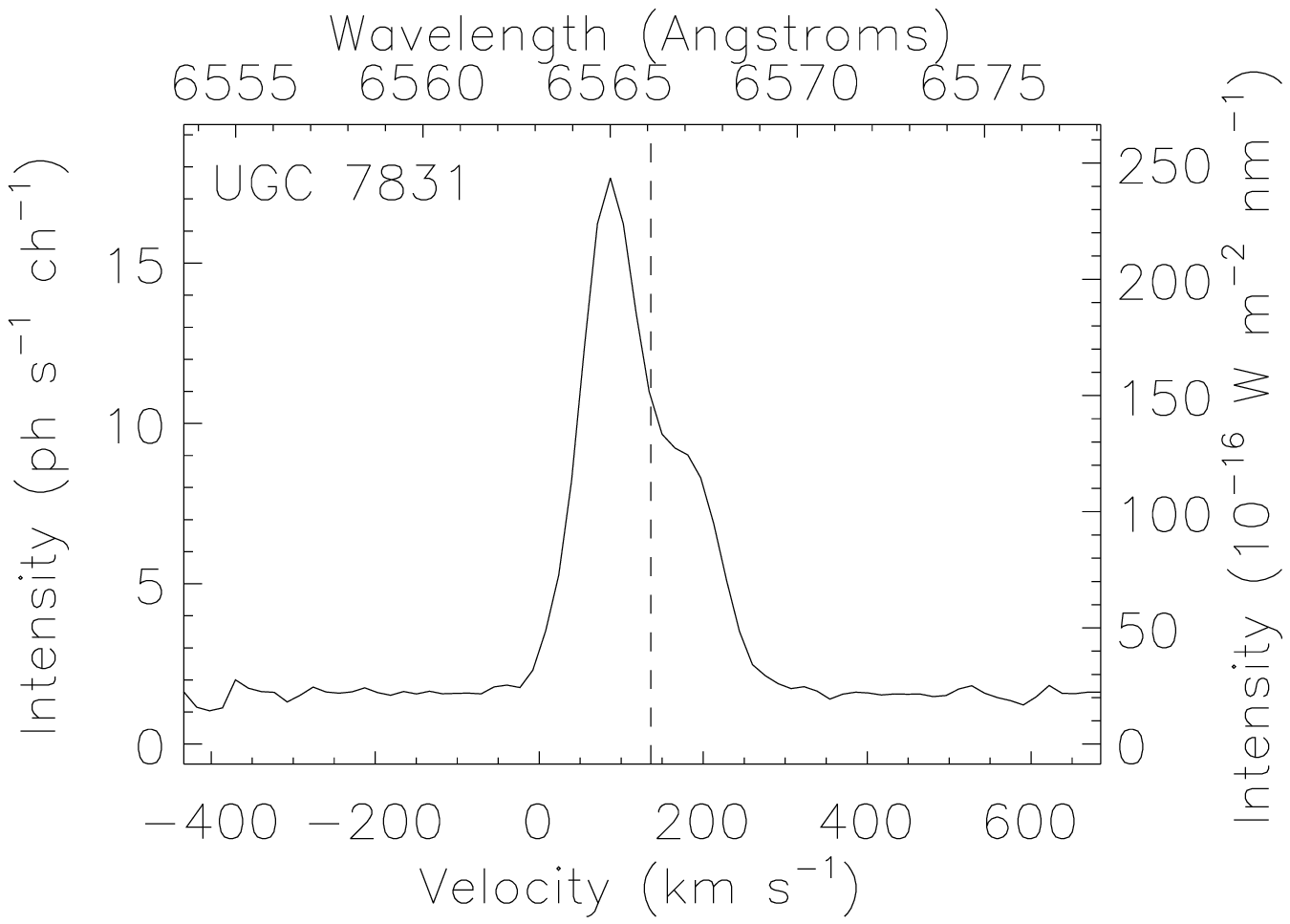}
\includegraphics[width=3.5cm]{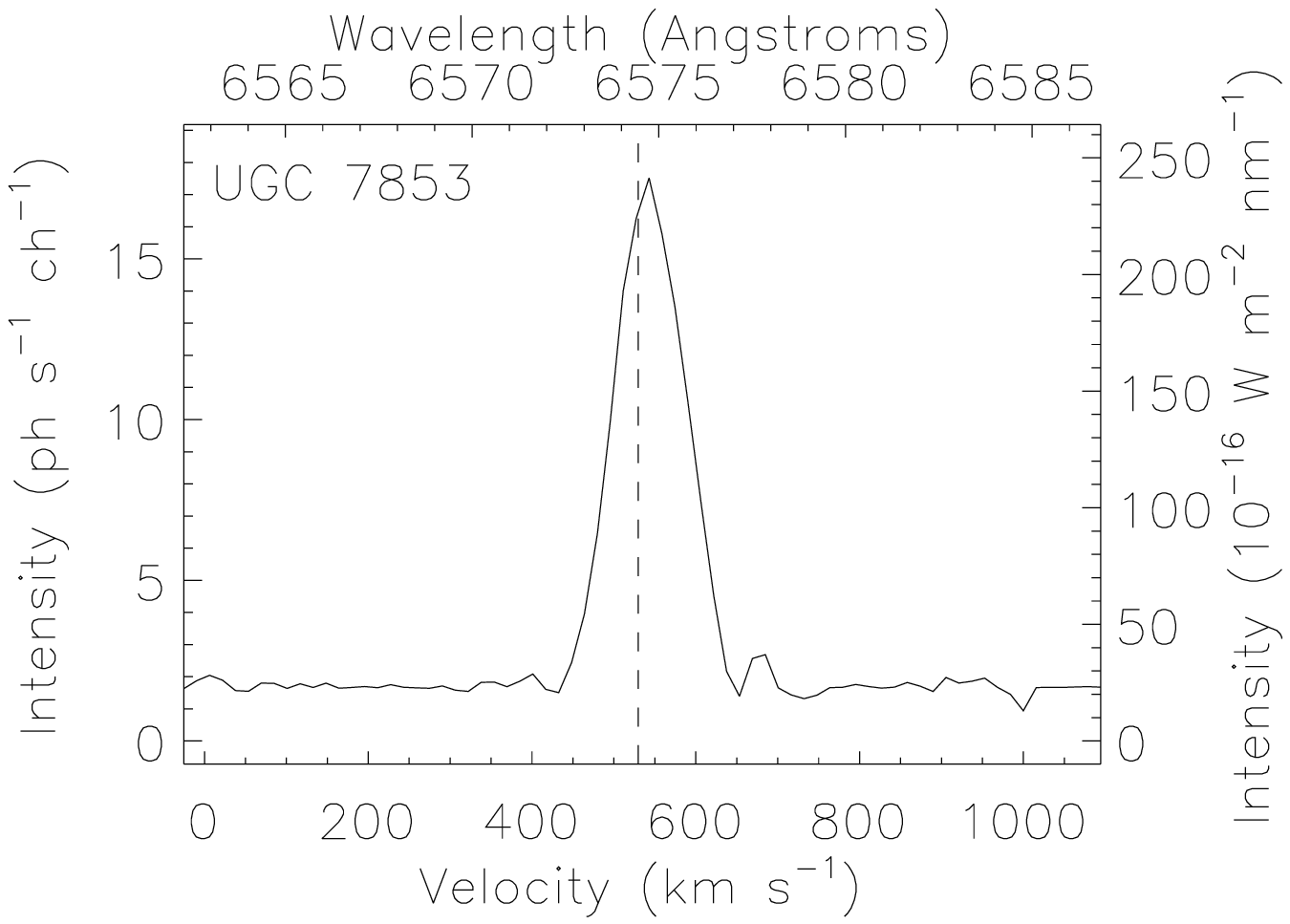}
\includegraphics[width=3.5cm]{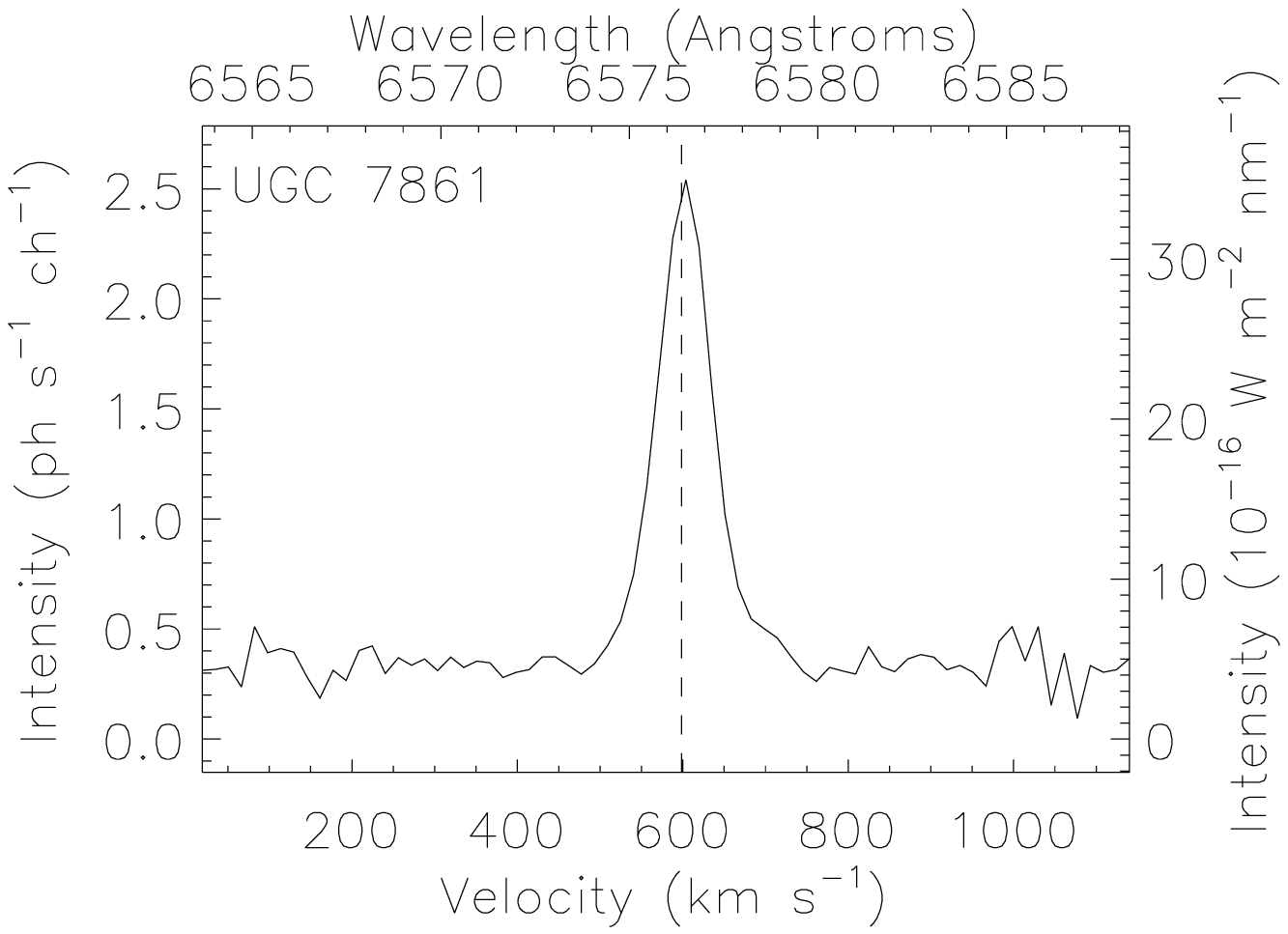}
\includegraphics[width=3.5cm]{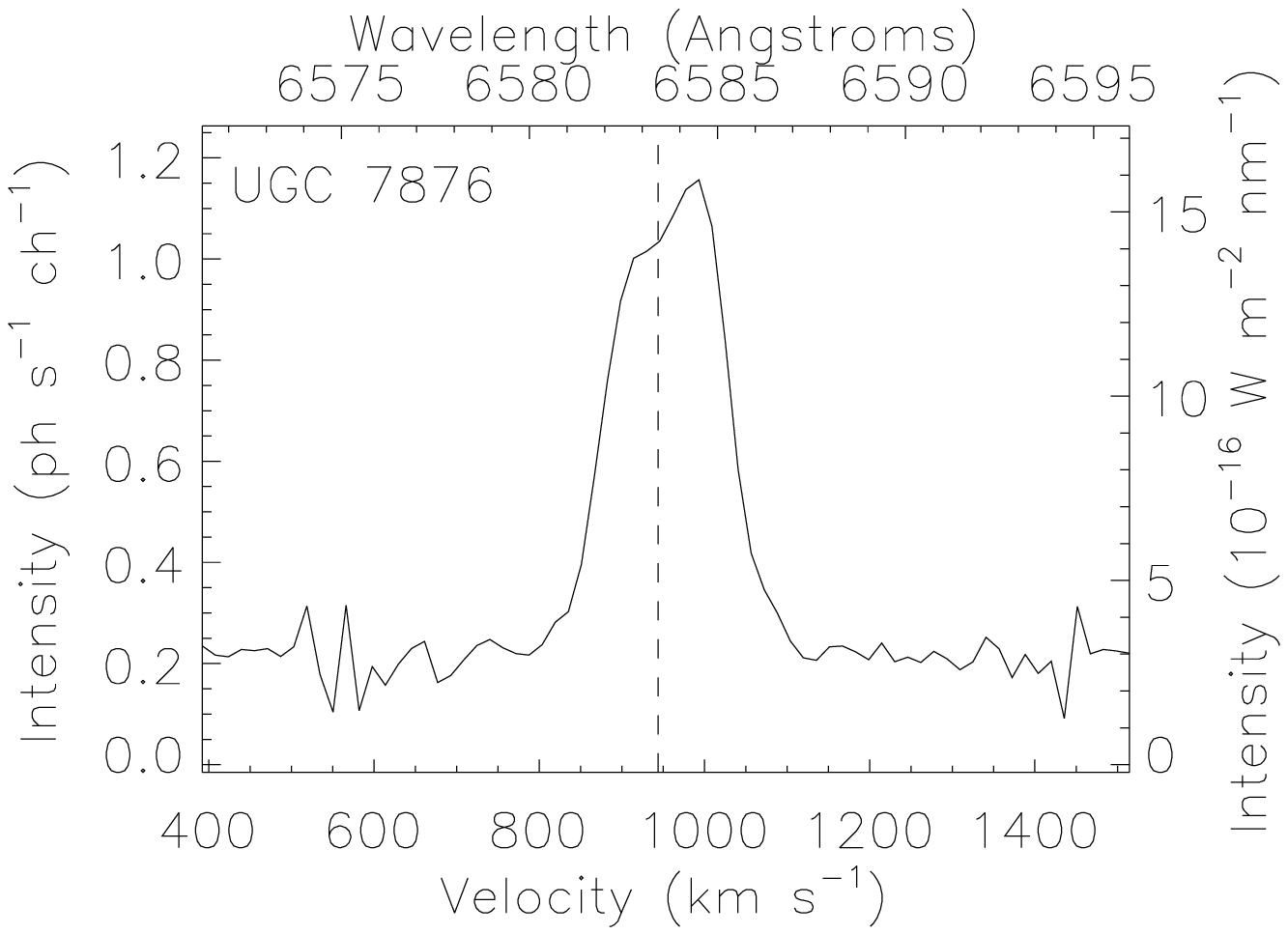}
\includegraphics[width=3.5cm]{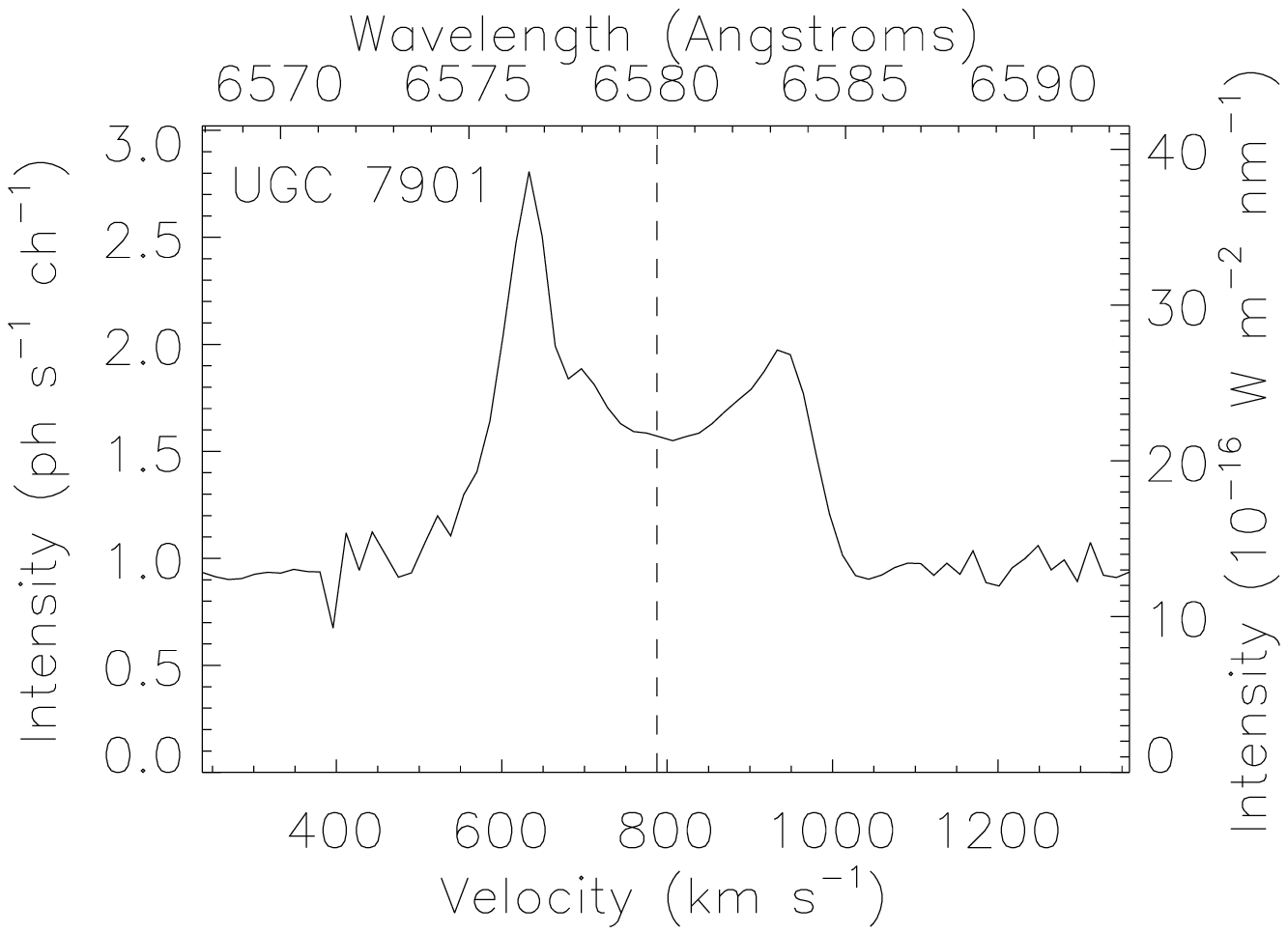}
\includegraphics[width=3.5cm]{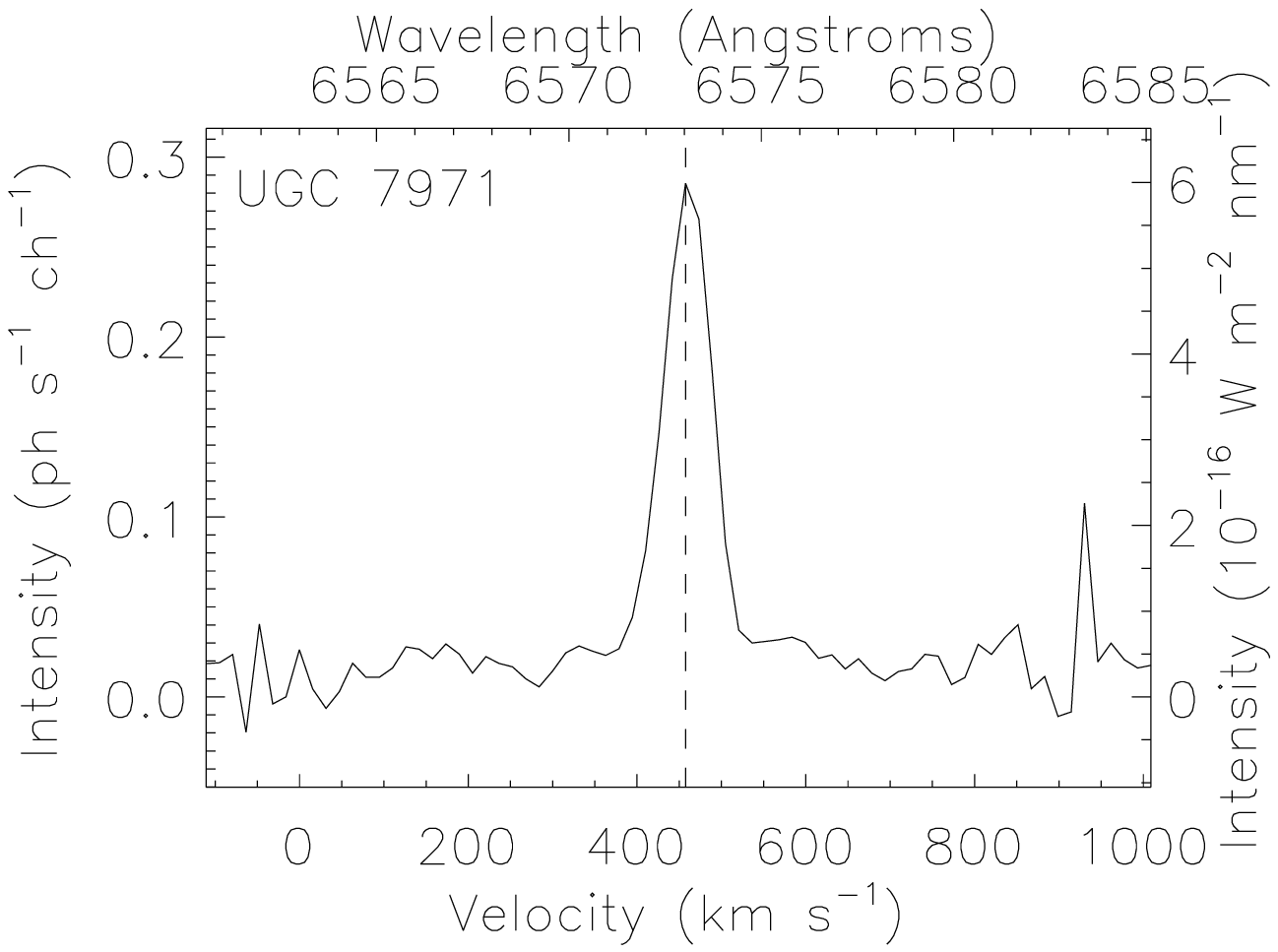}
\includegraphics[width=3.5cm]{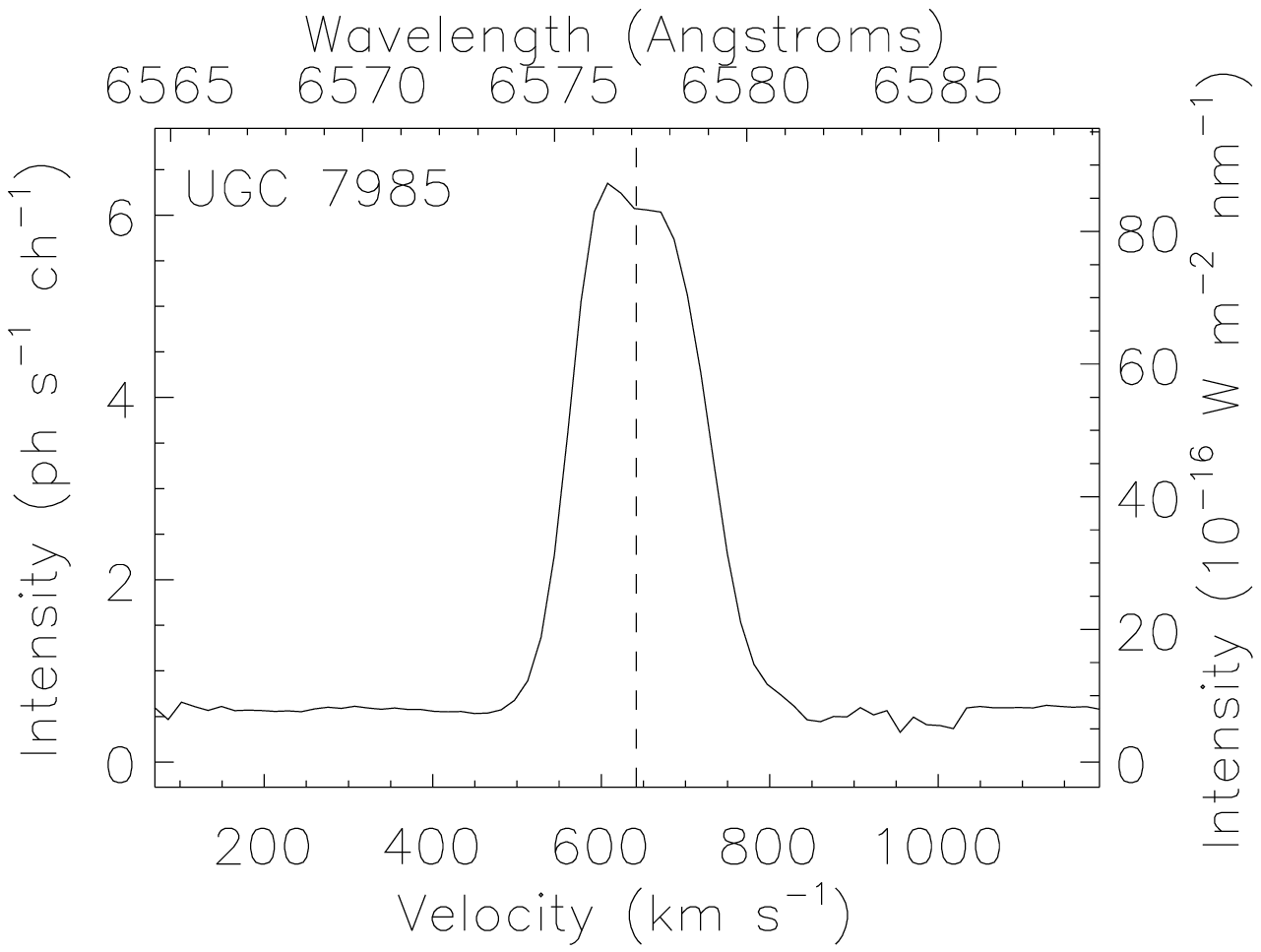}
\includegraphics[width=3.5cm]{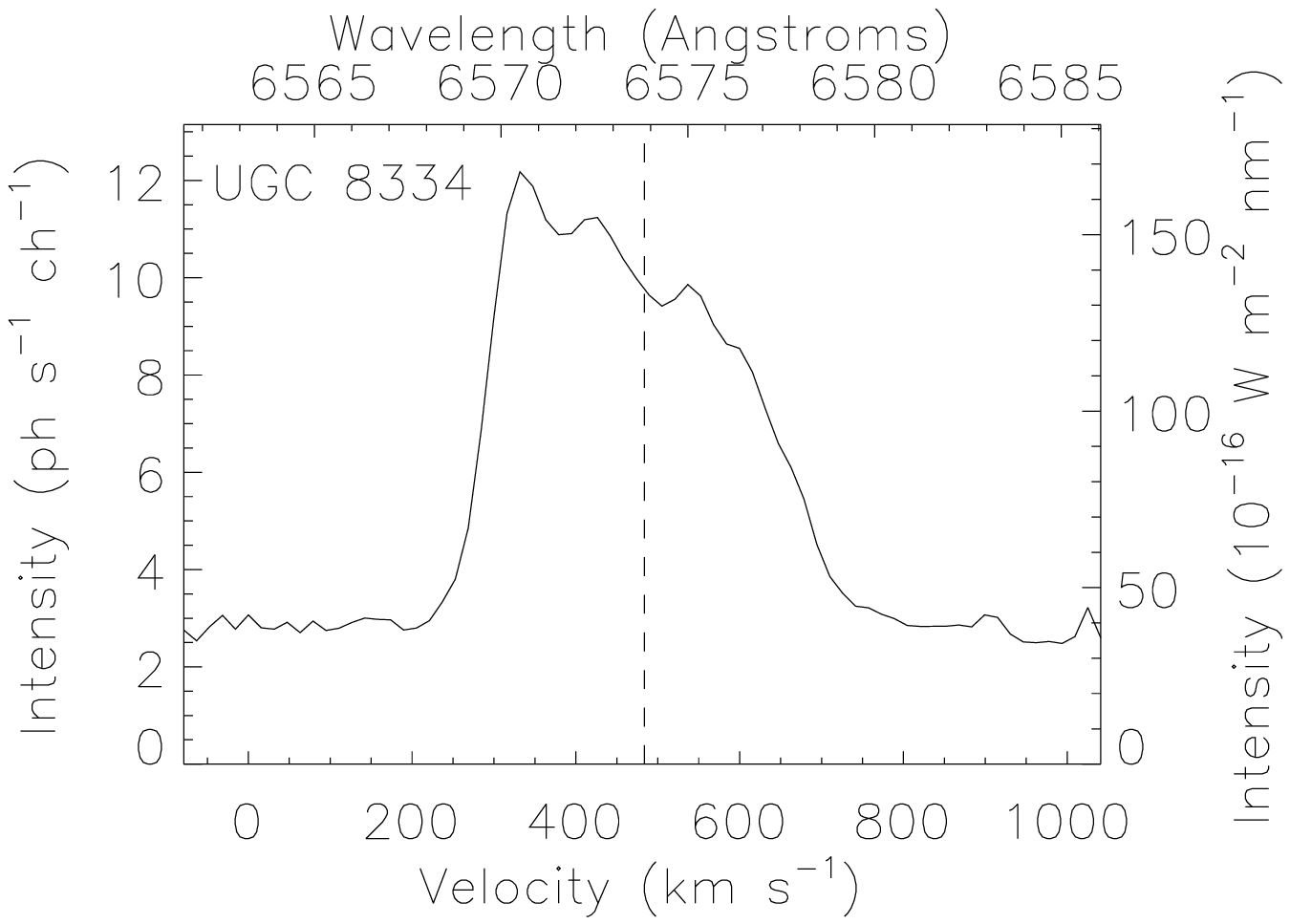}
\includegraphics[width=3.5cm]{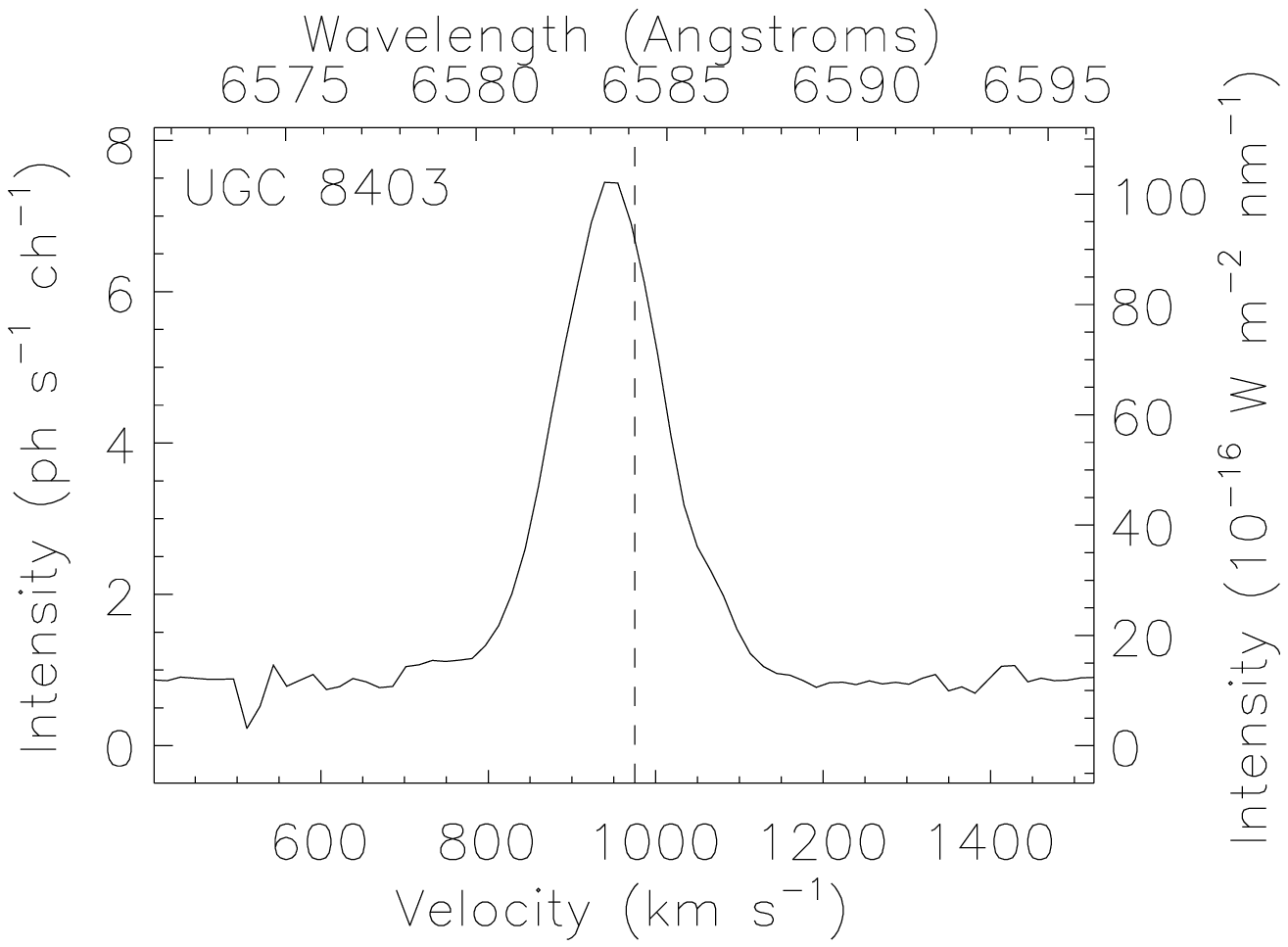}
\includegraphics[width=3.5cm]{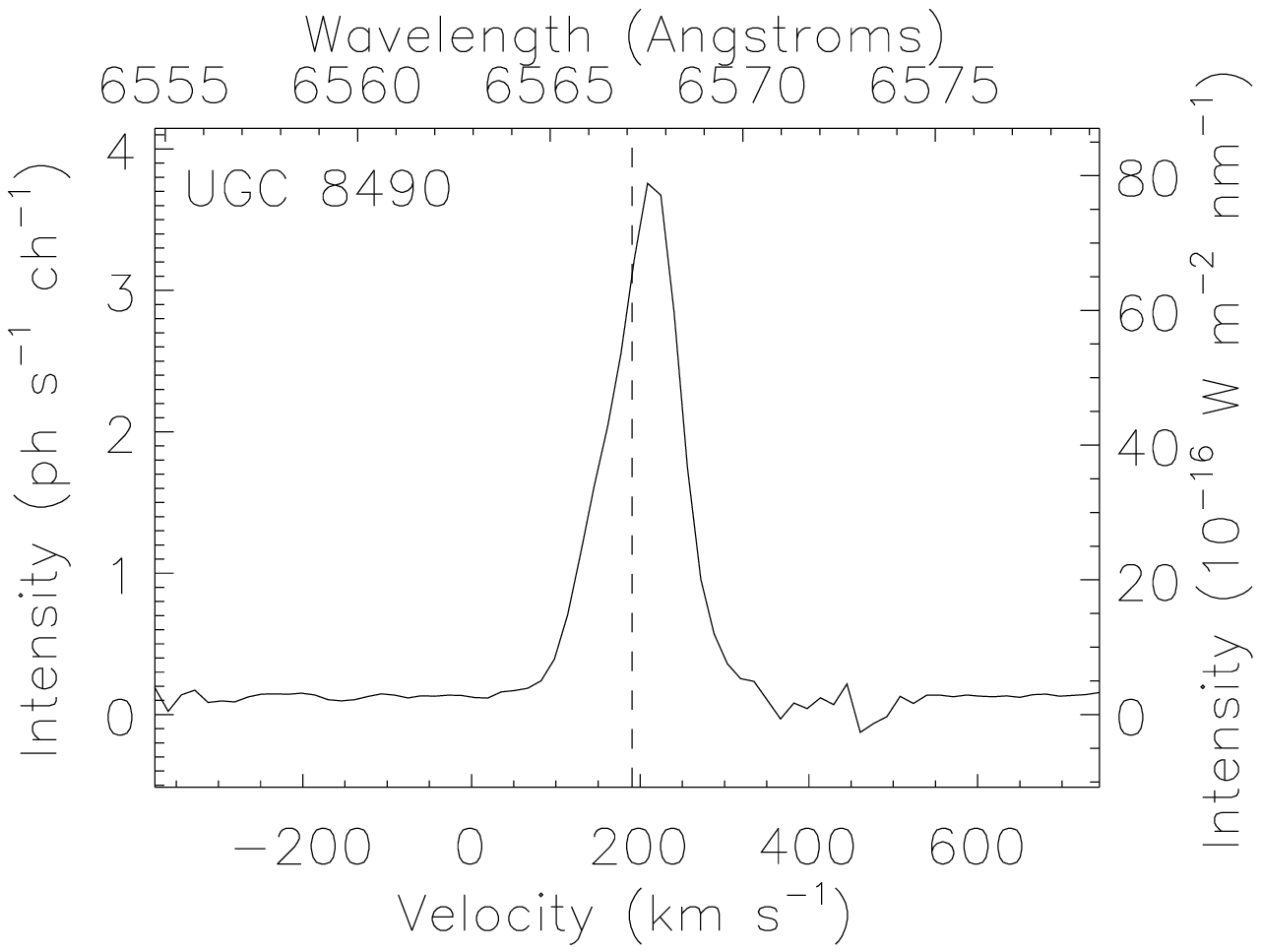}
\includegraphics[width=3.5cm]{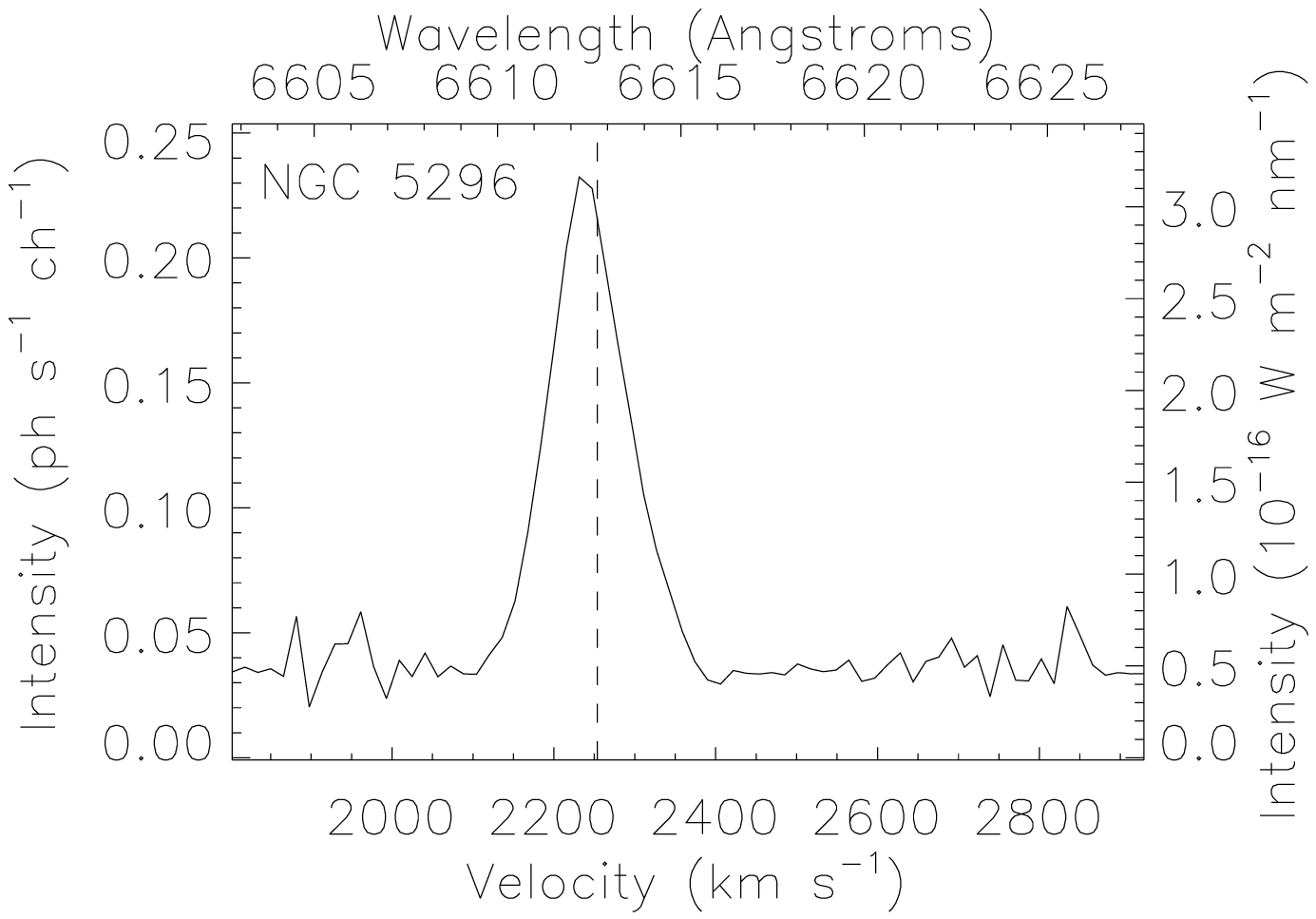}
\includegraphics[width=3.5cm]{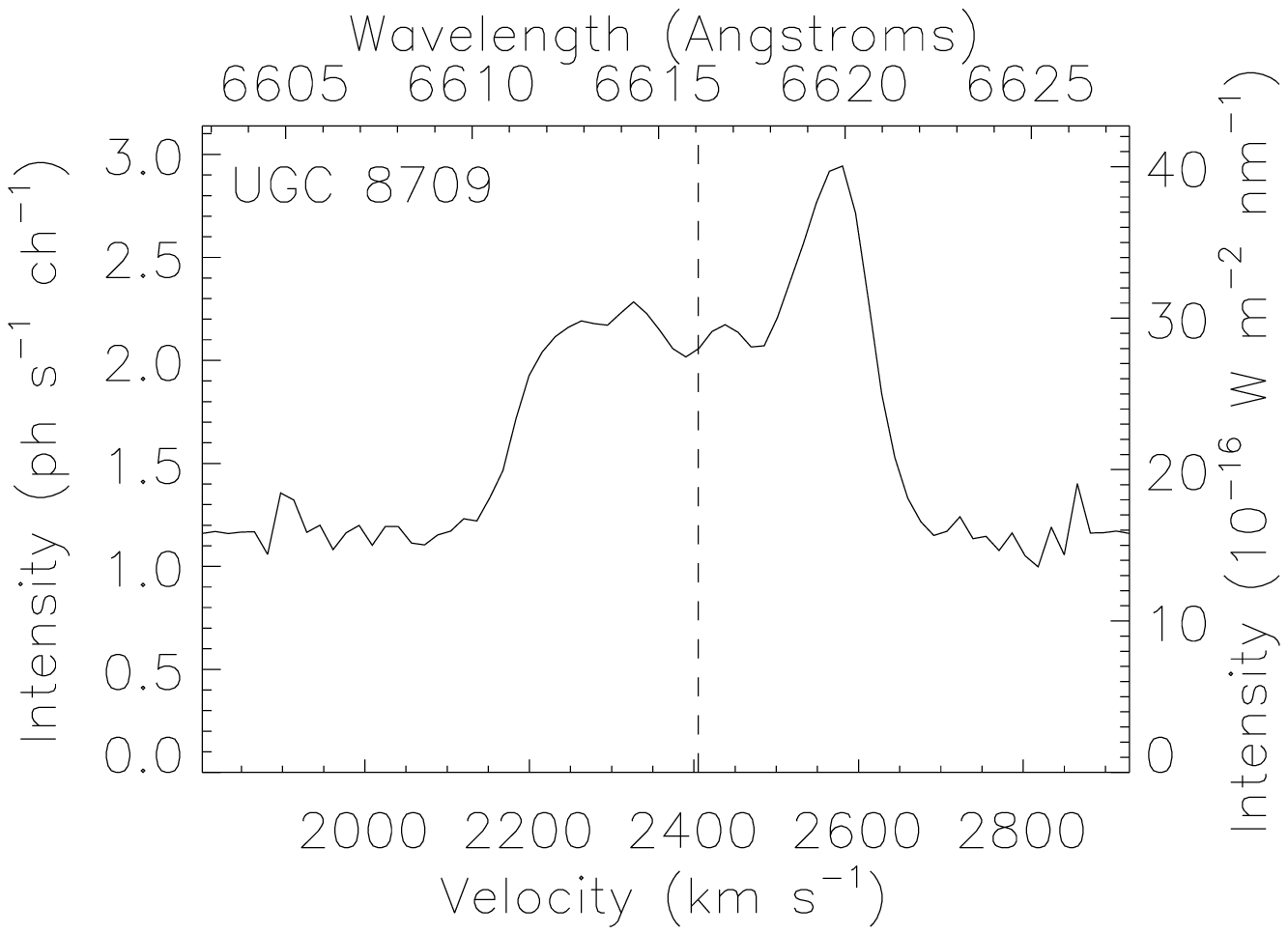}
\includegraphics[width=3.5cm]{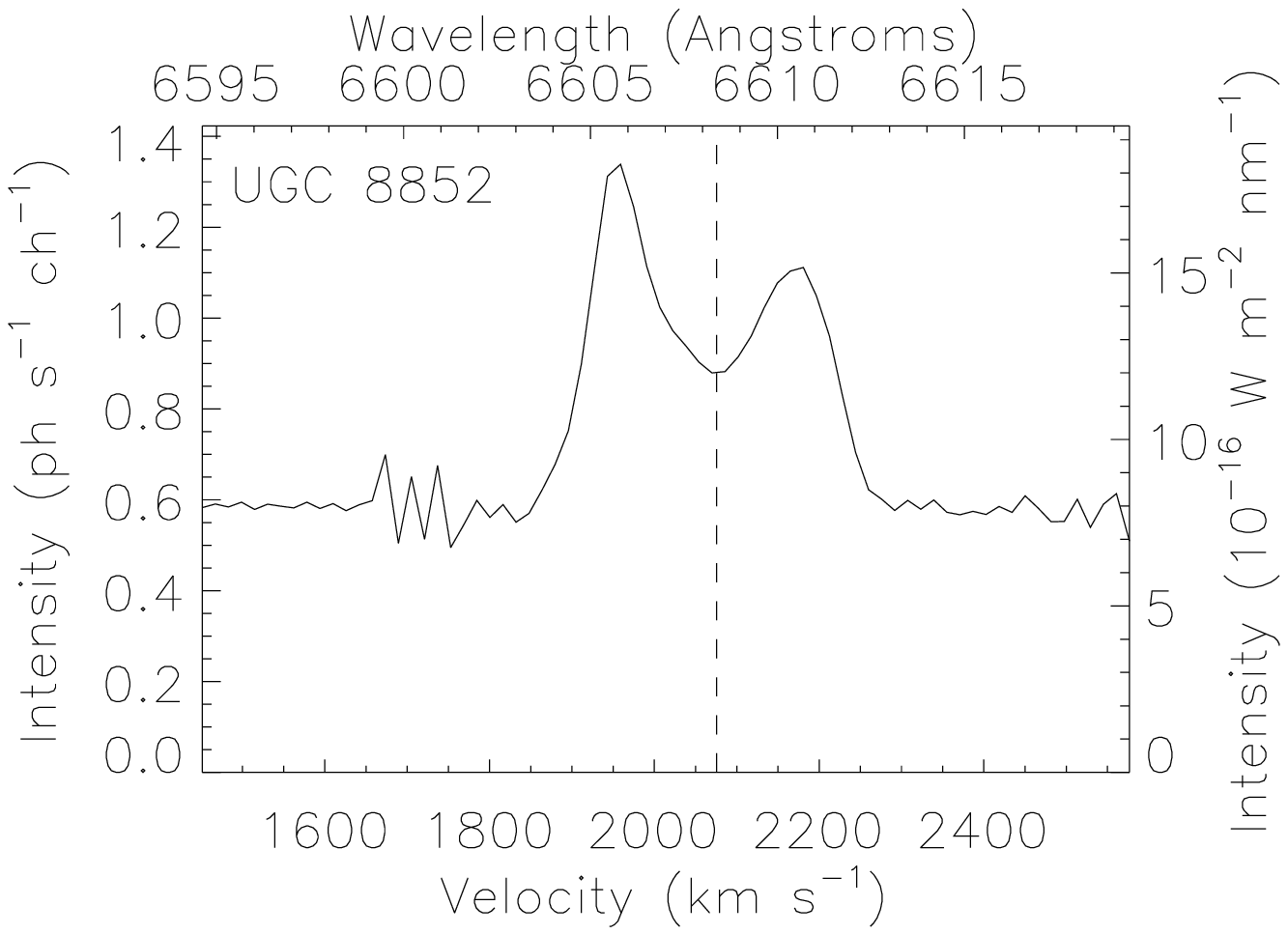}
\includegraphics[width=3.5cm]{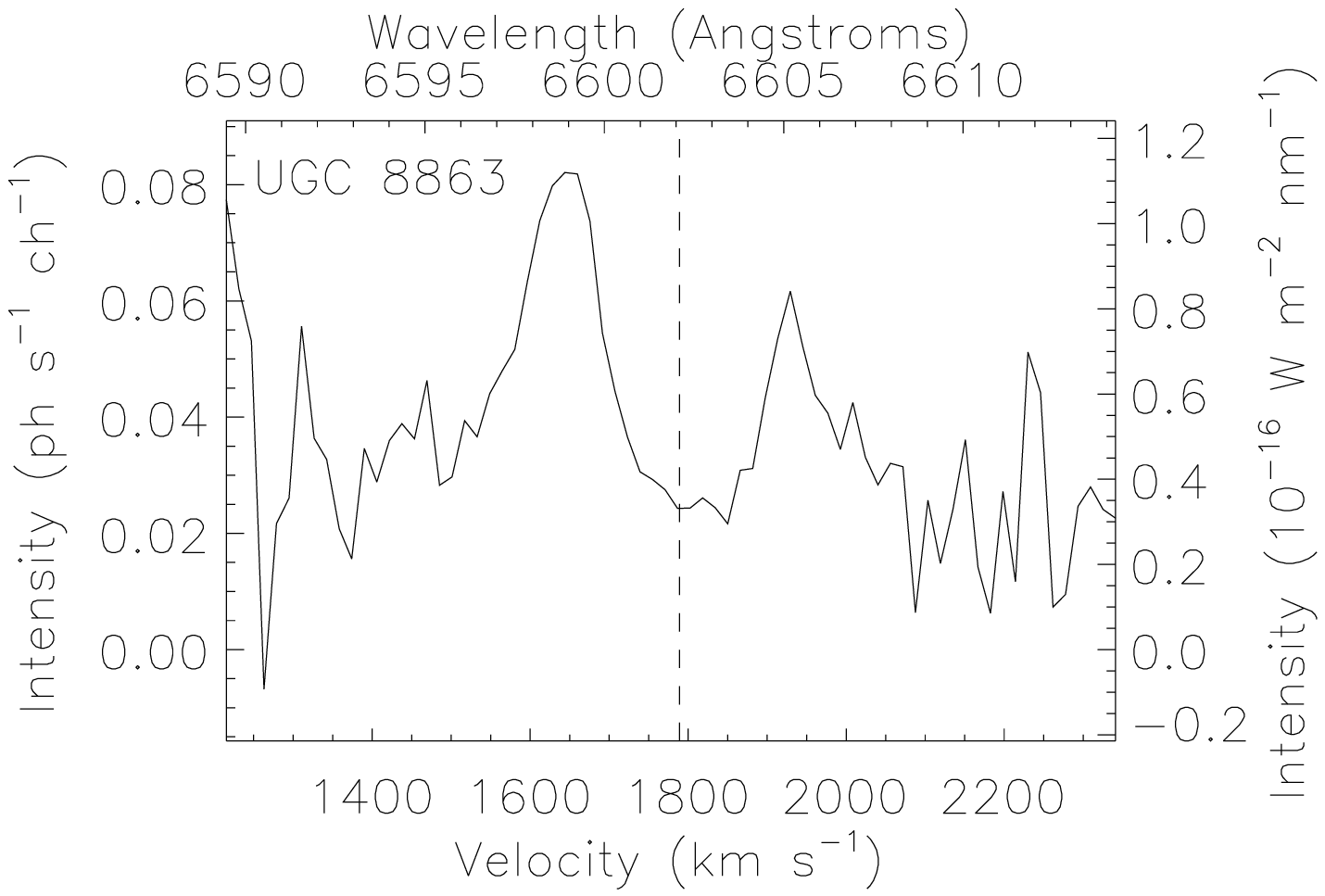}
\includegraphics[width=3.5cm]{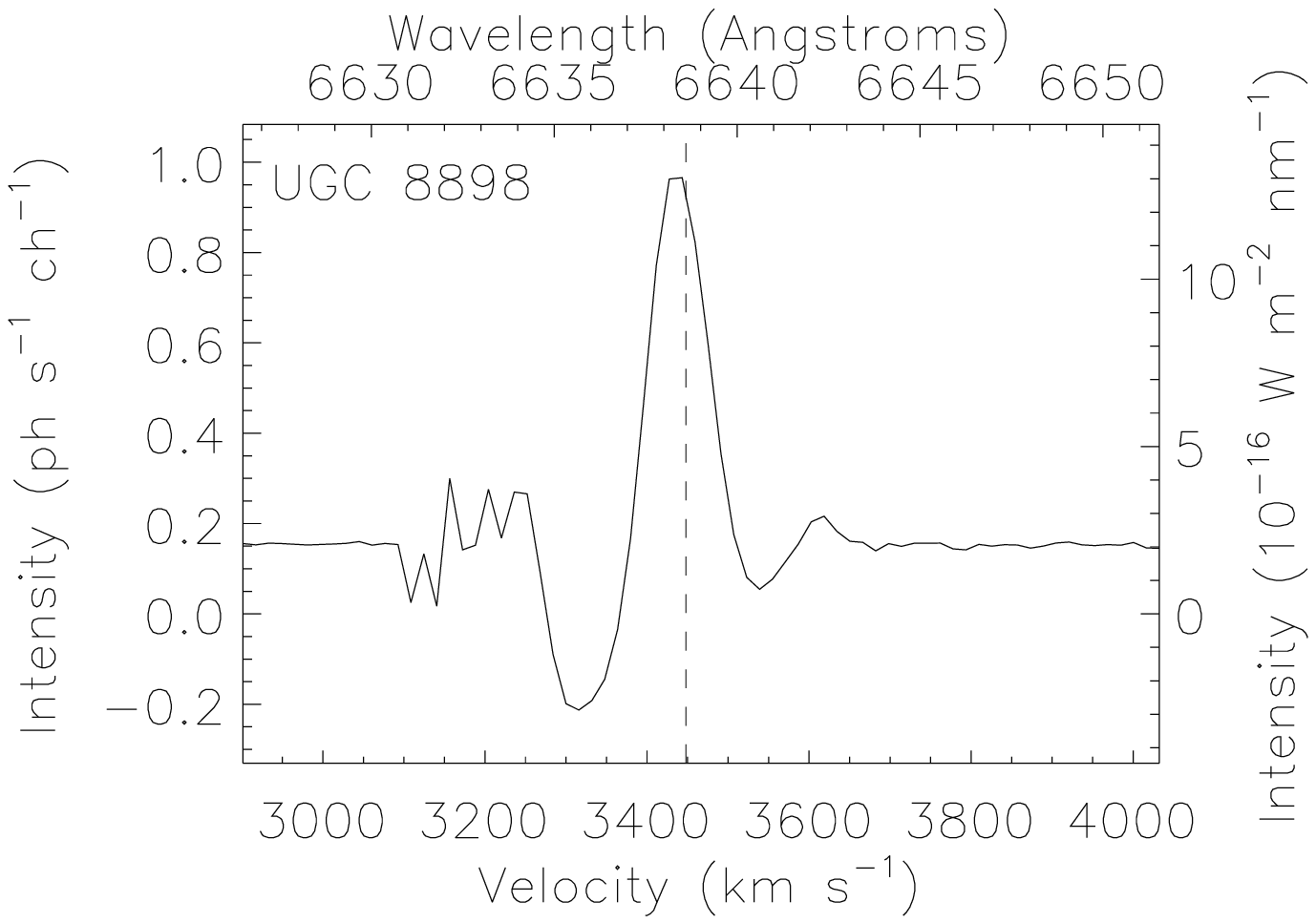}
\includegraphics[width=3.5cm]{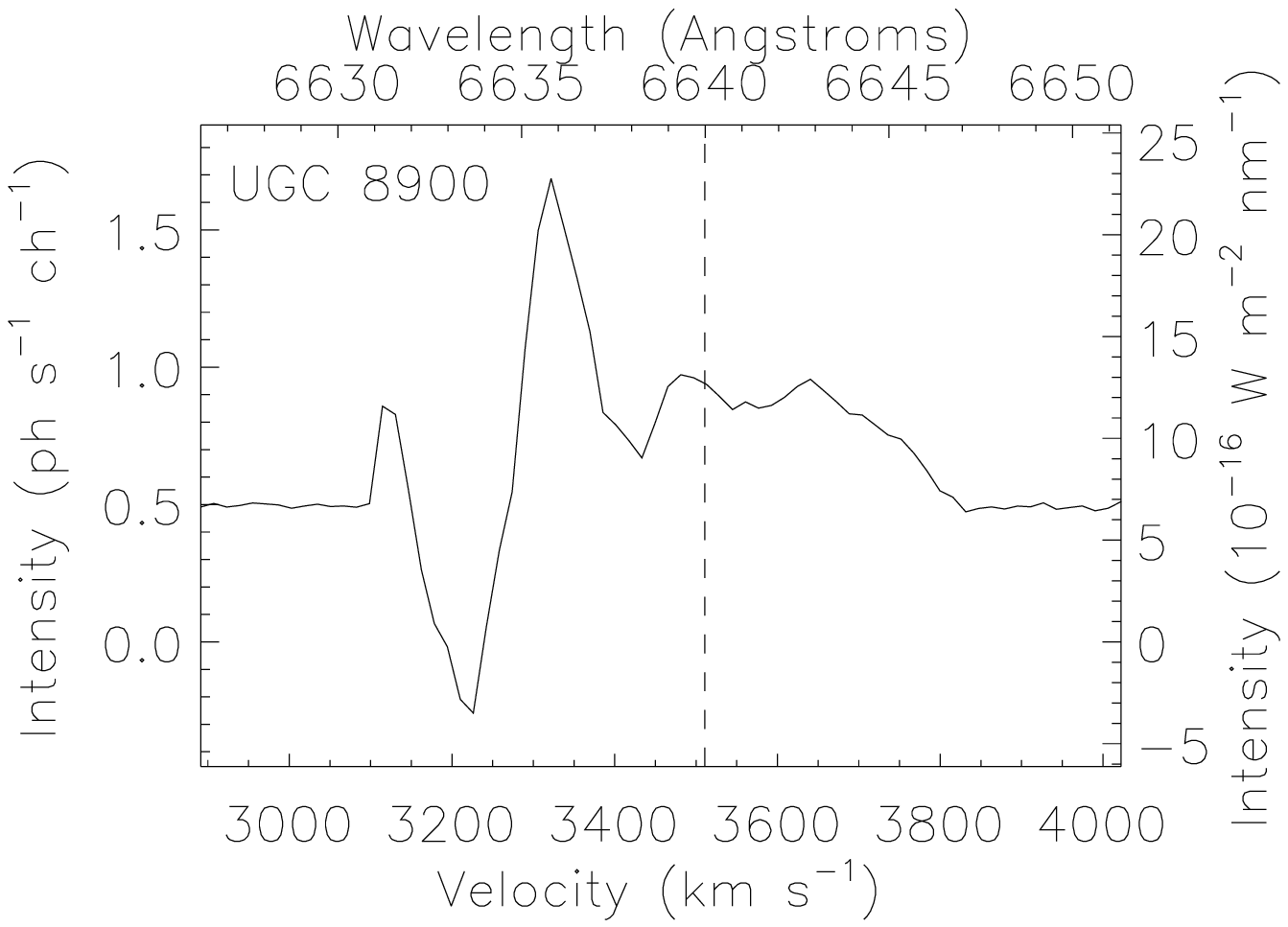}
\includegraphics[width=3.5cm]{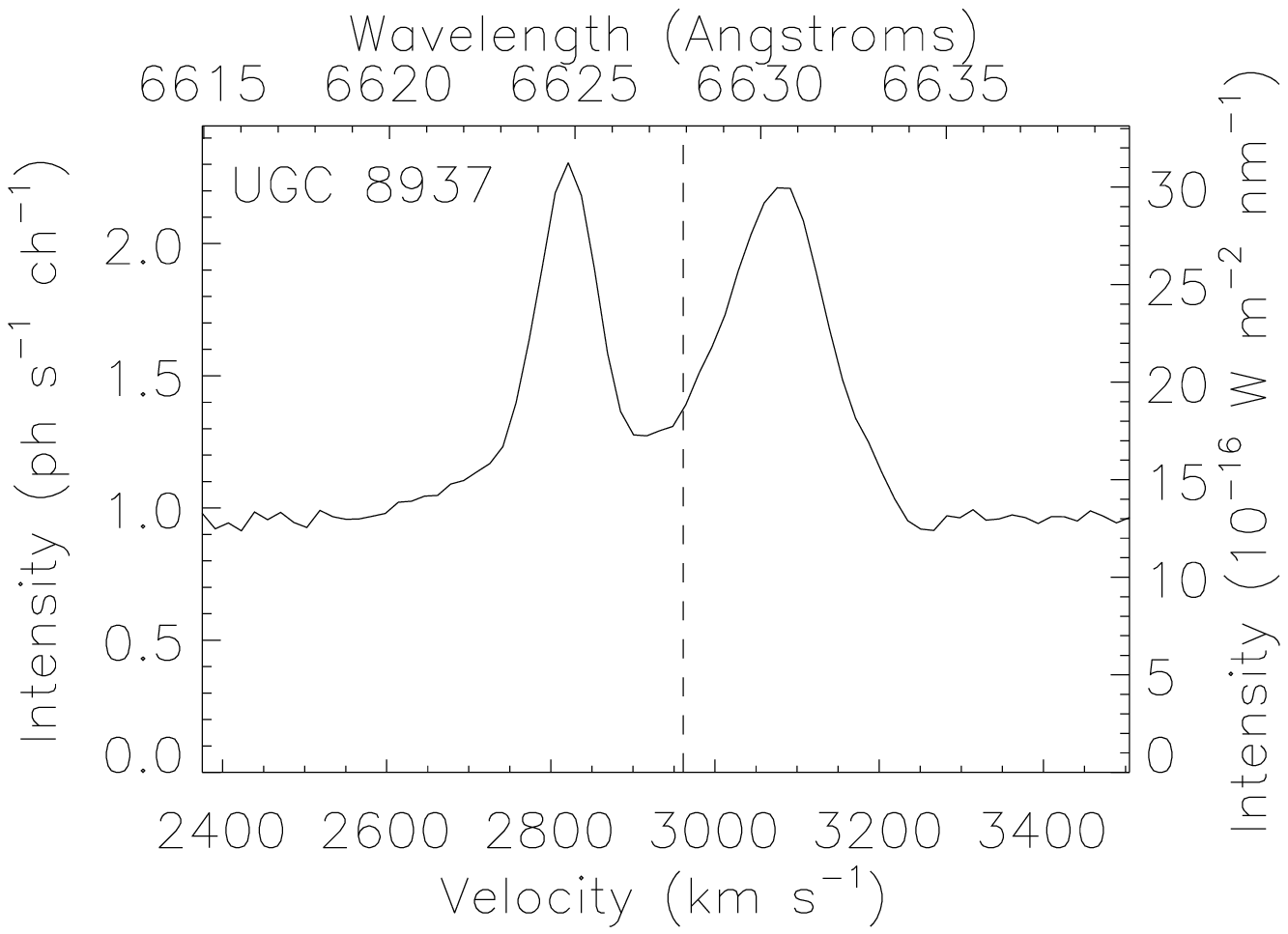}
\includegraphics[width=3.5cm]{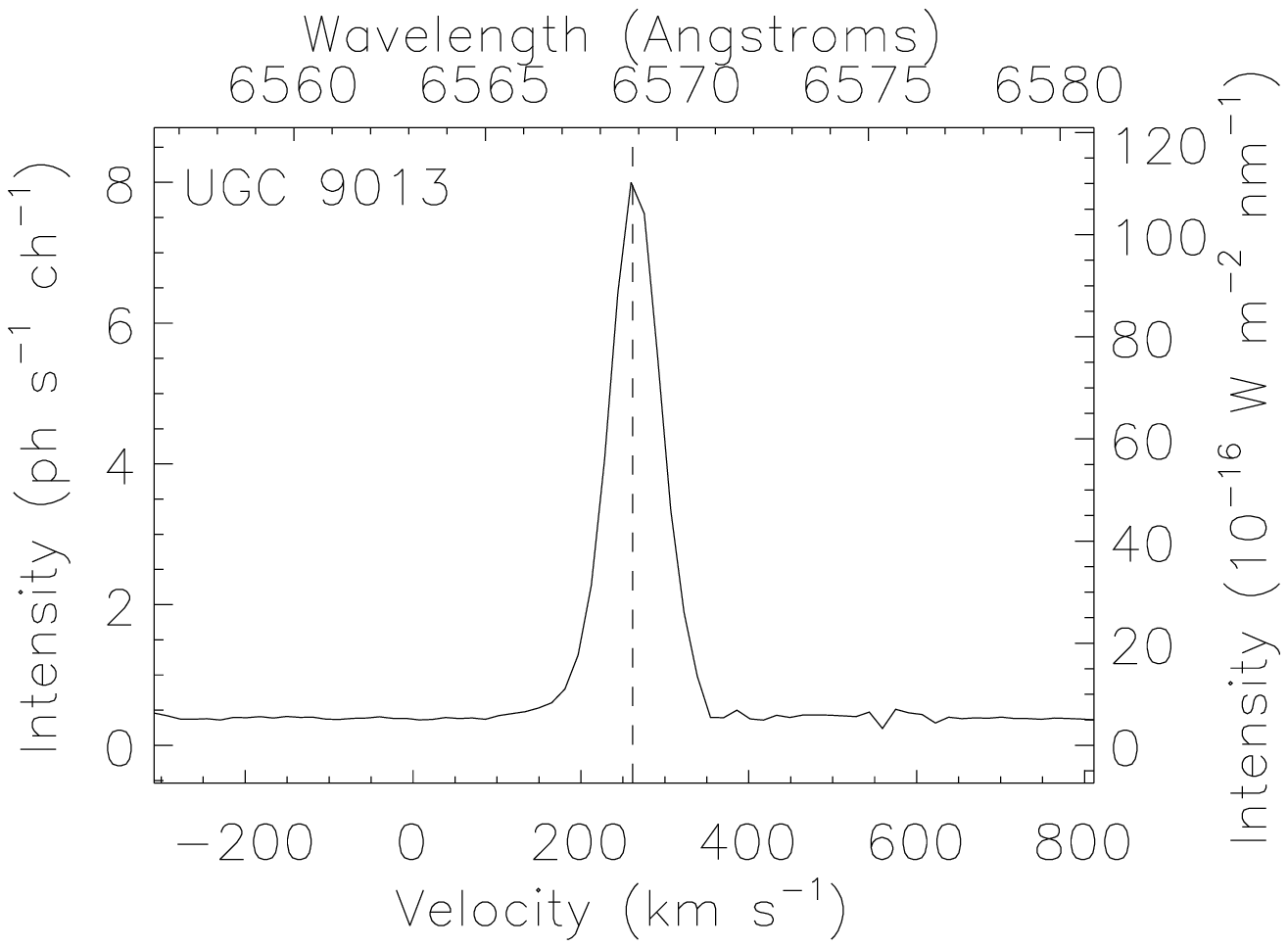}
\includegraphics[width=3.5cm]{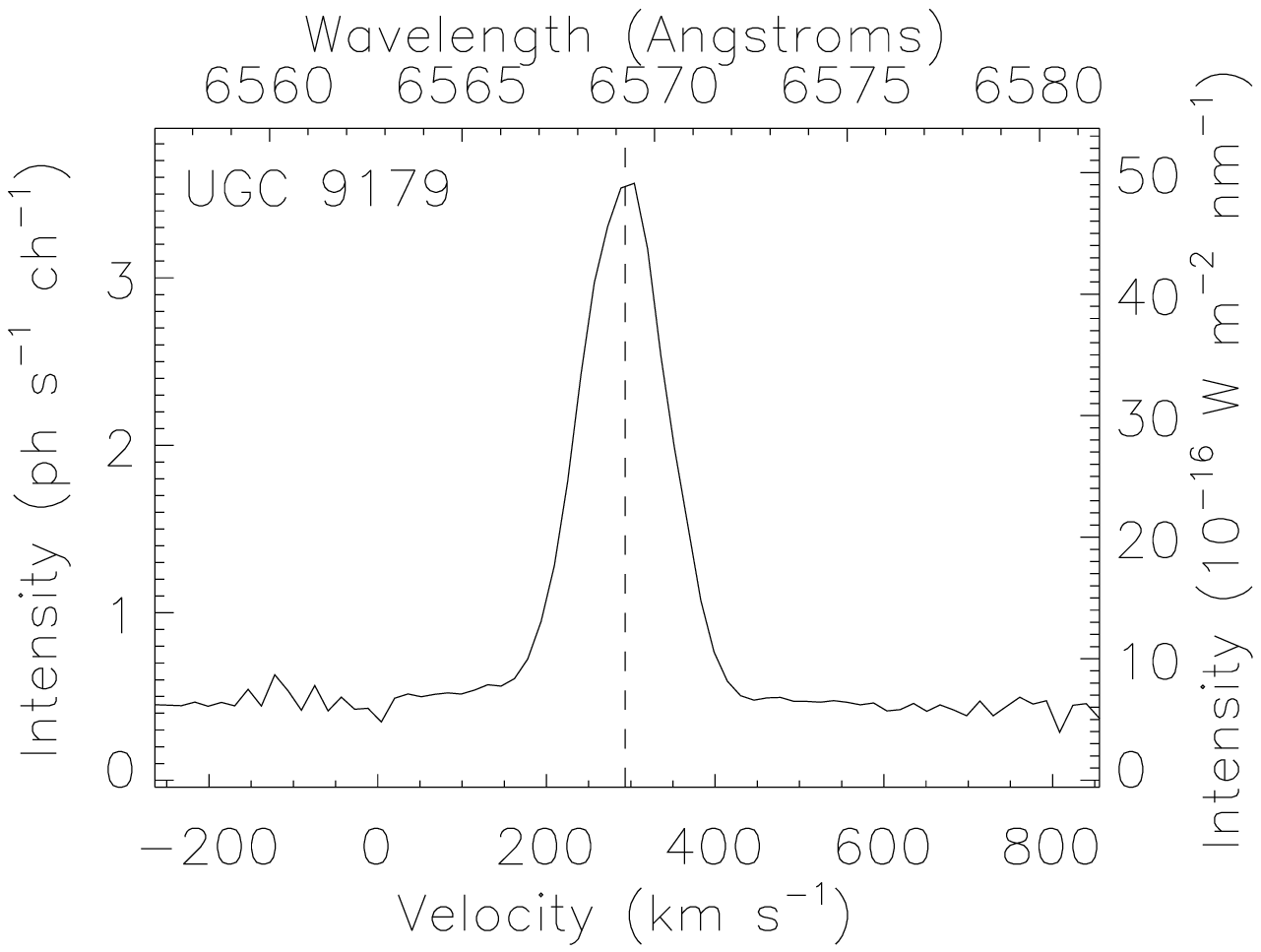}
\includegraphics[width=3.5cm]{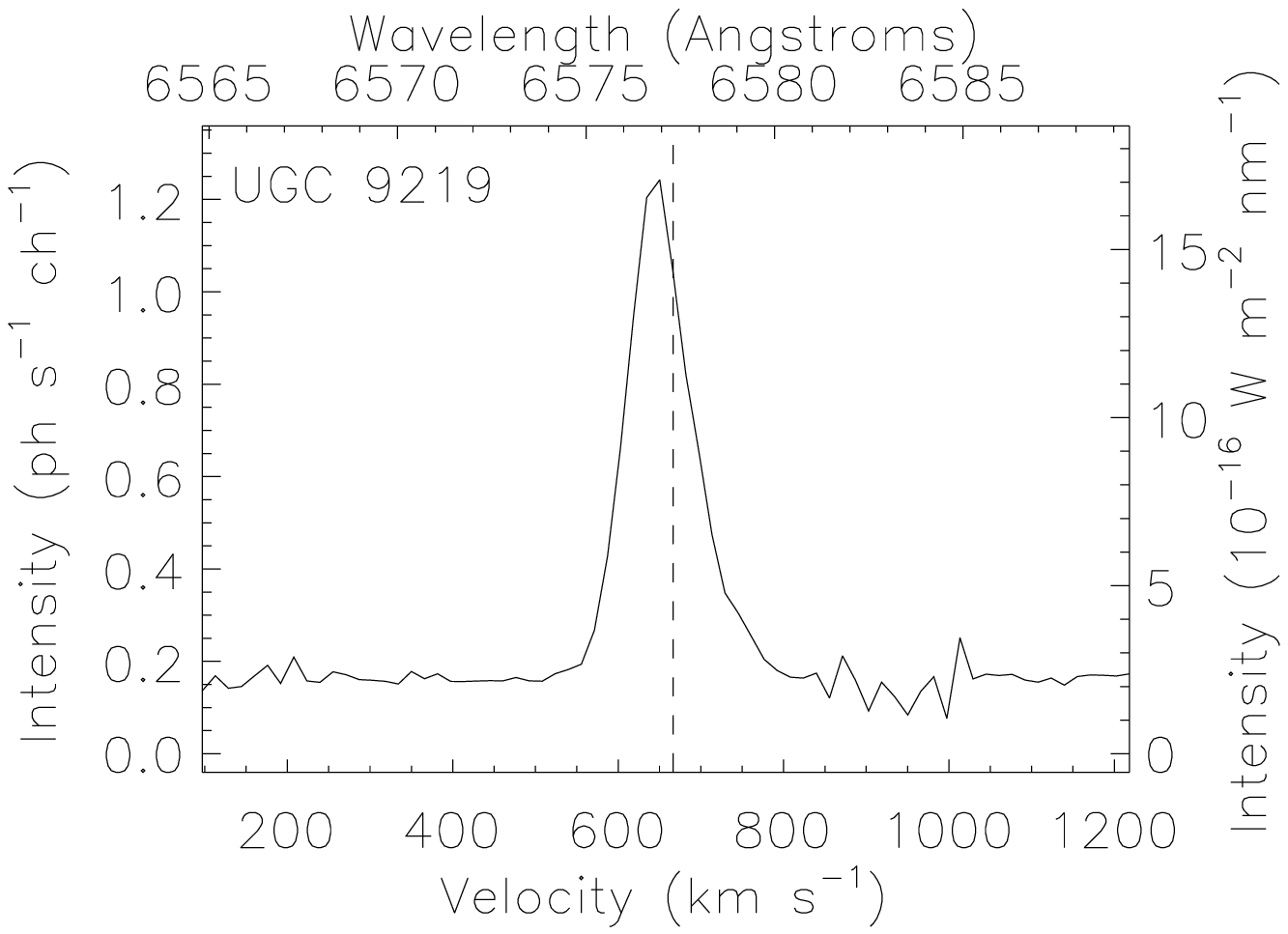}
\includegraphics[width=3.5cm]{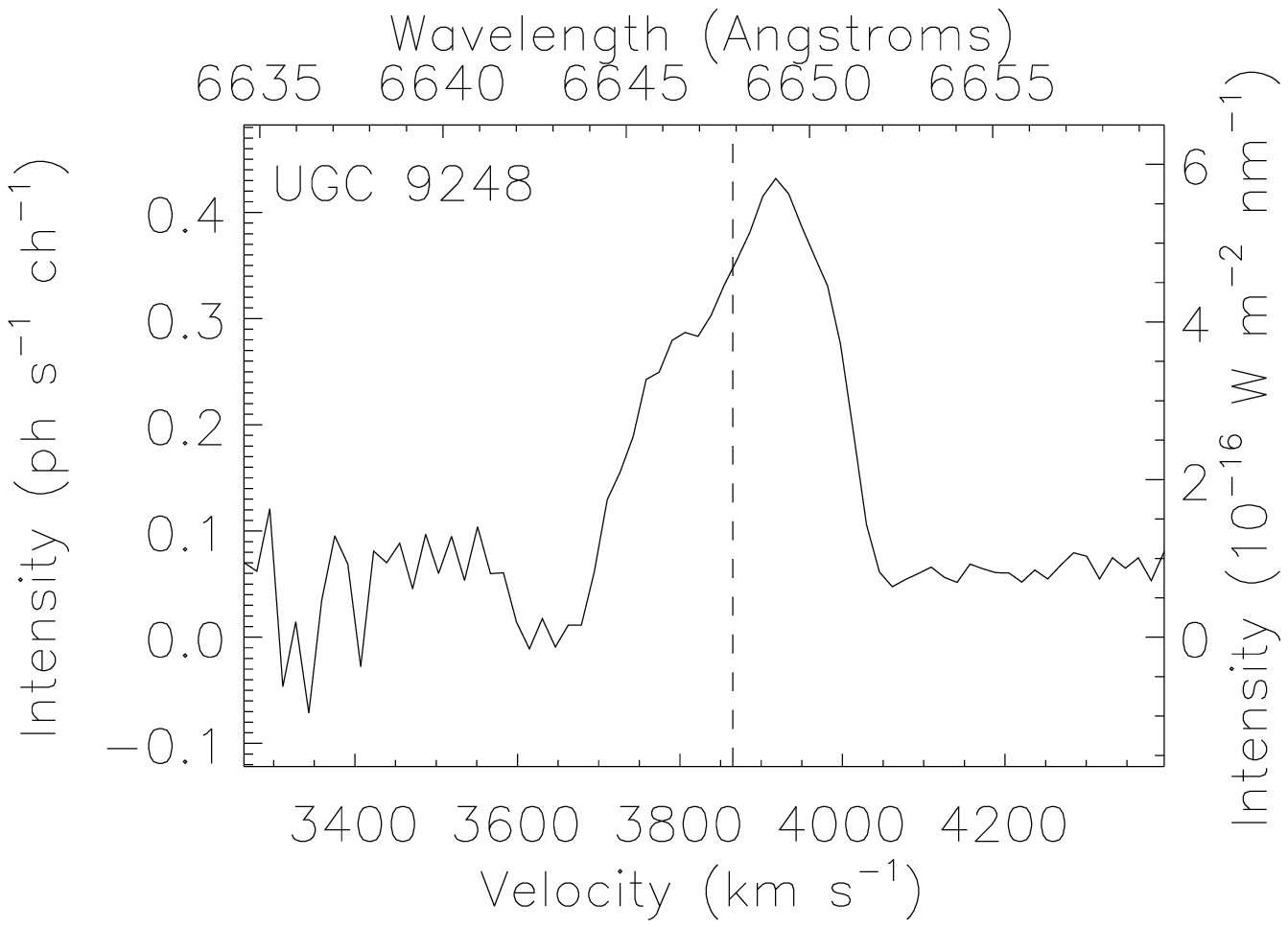}
\includegraphics[width=3.5cm]{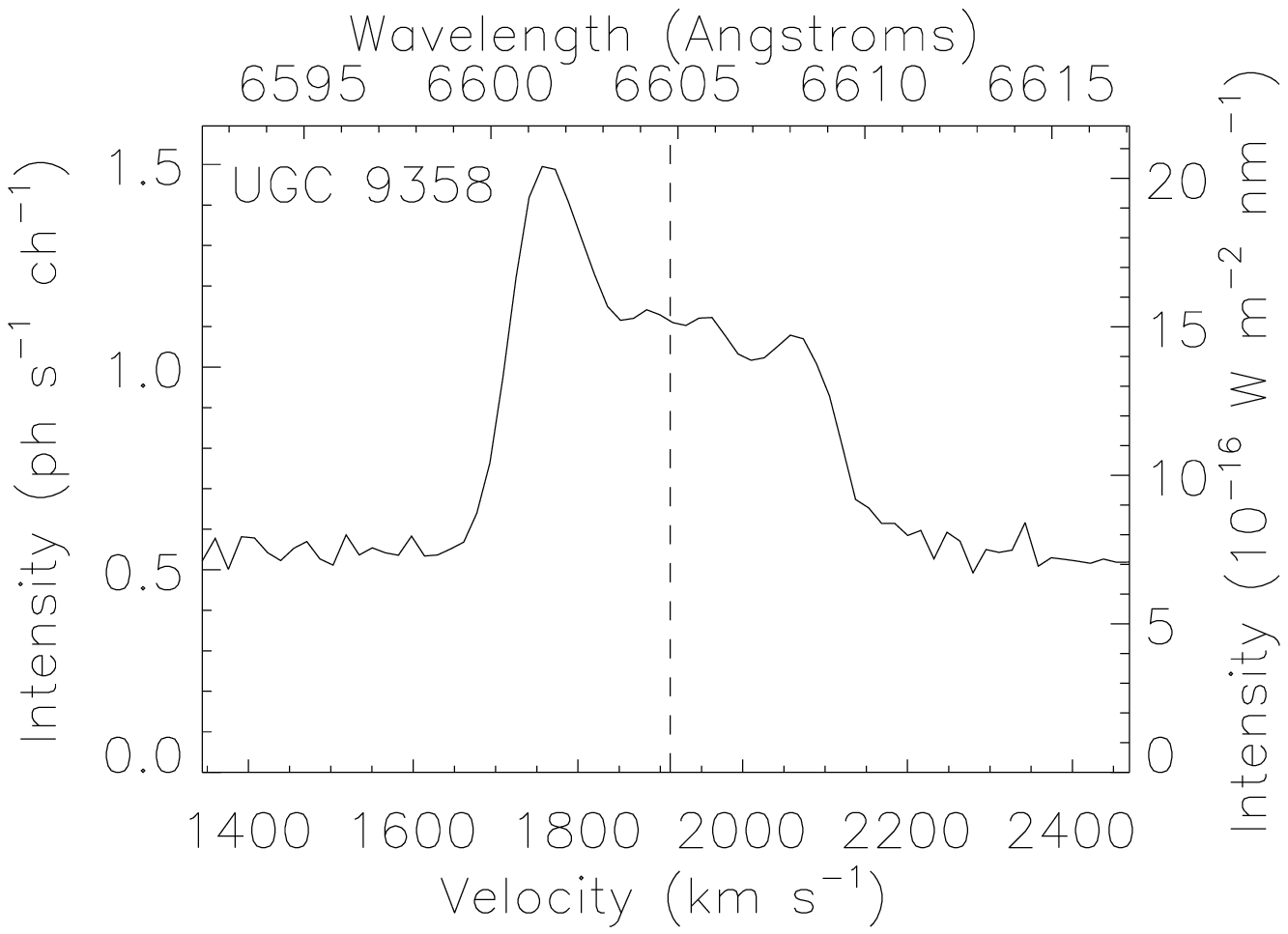}
\includegraphics[width=3.5cm]{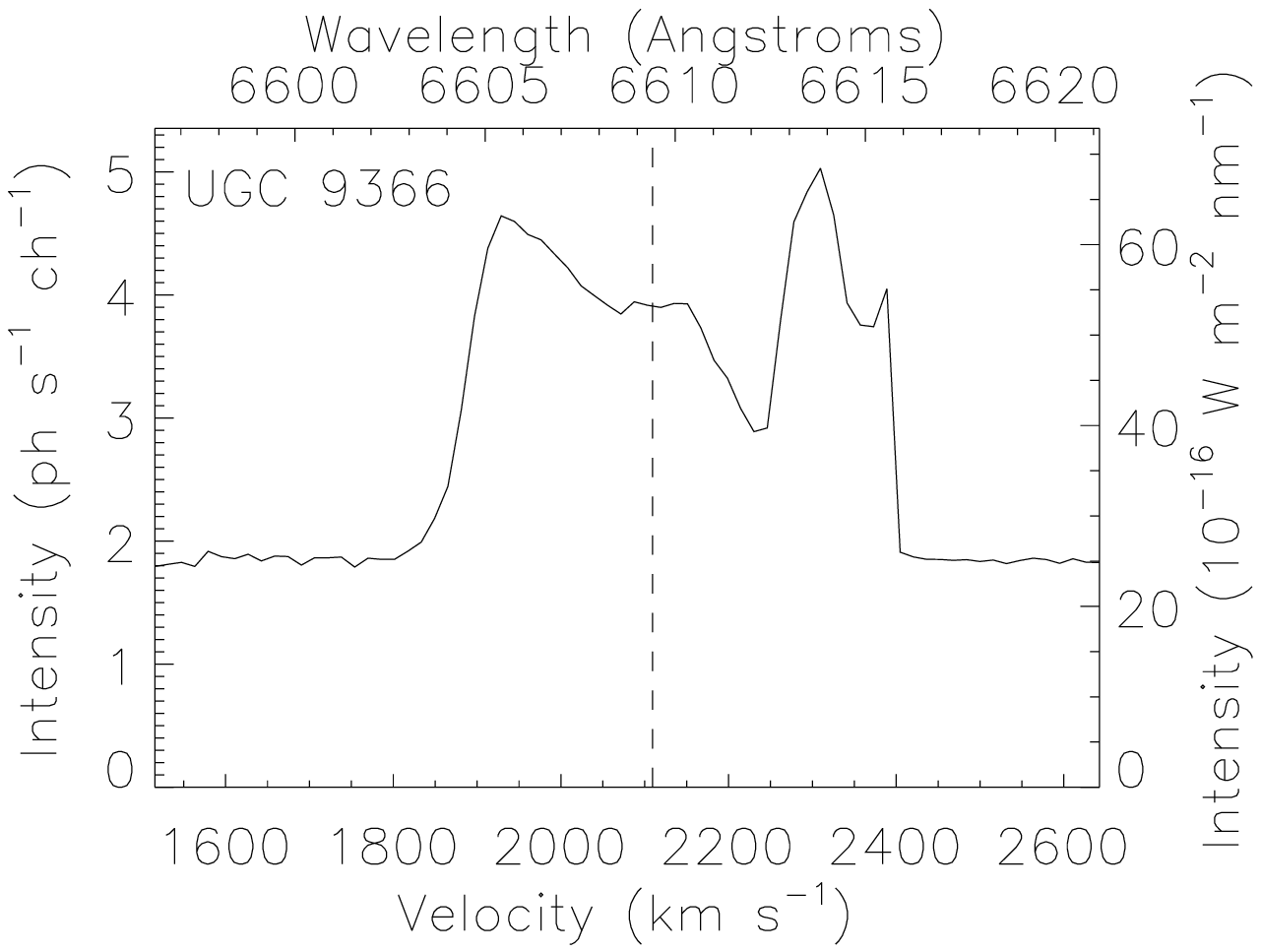}
\includegraphics[width=3.5cm]{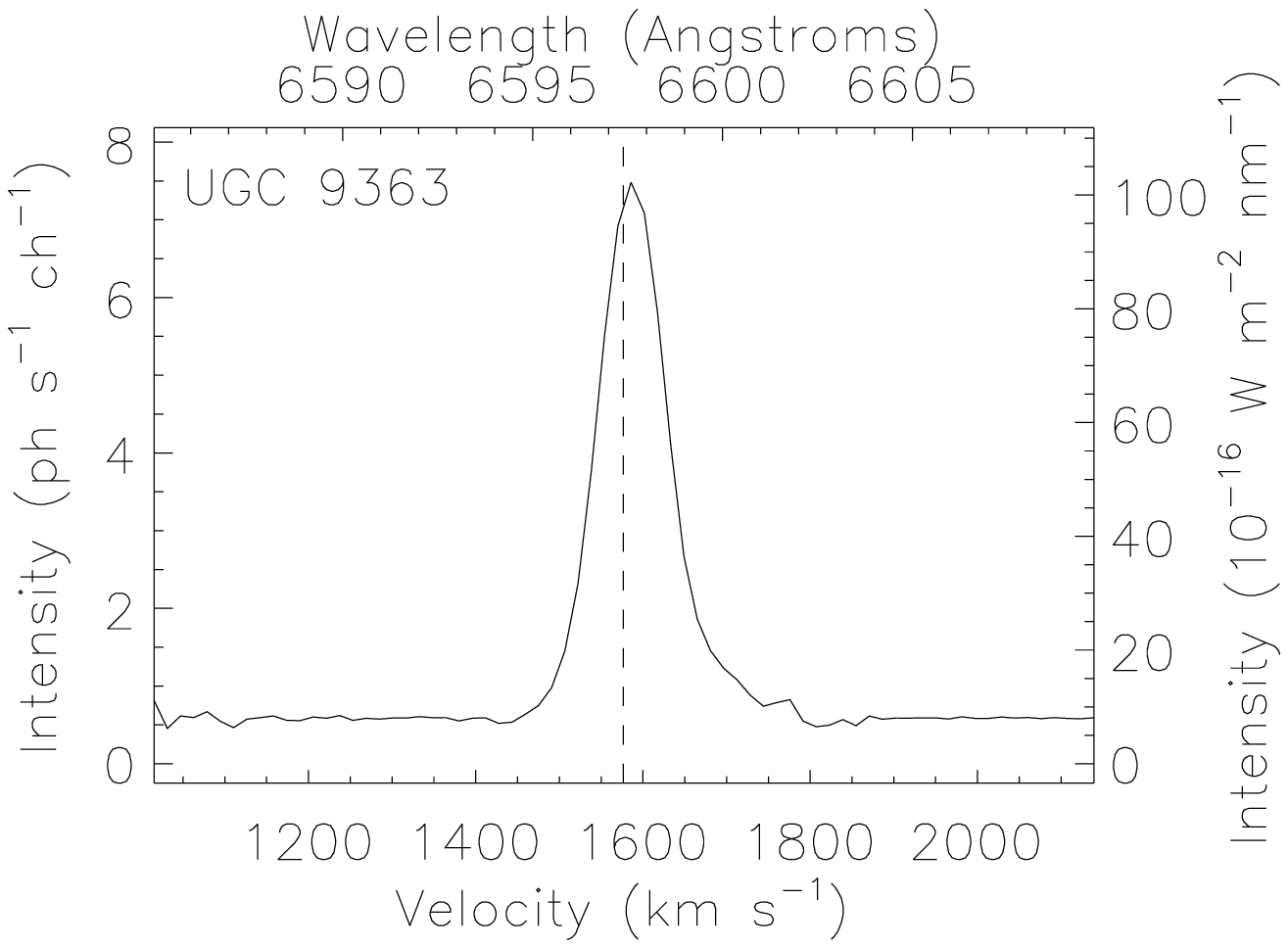}
\includegraphics[width=3.5cm]{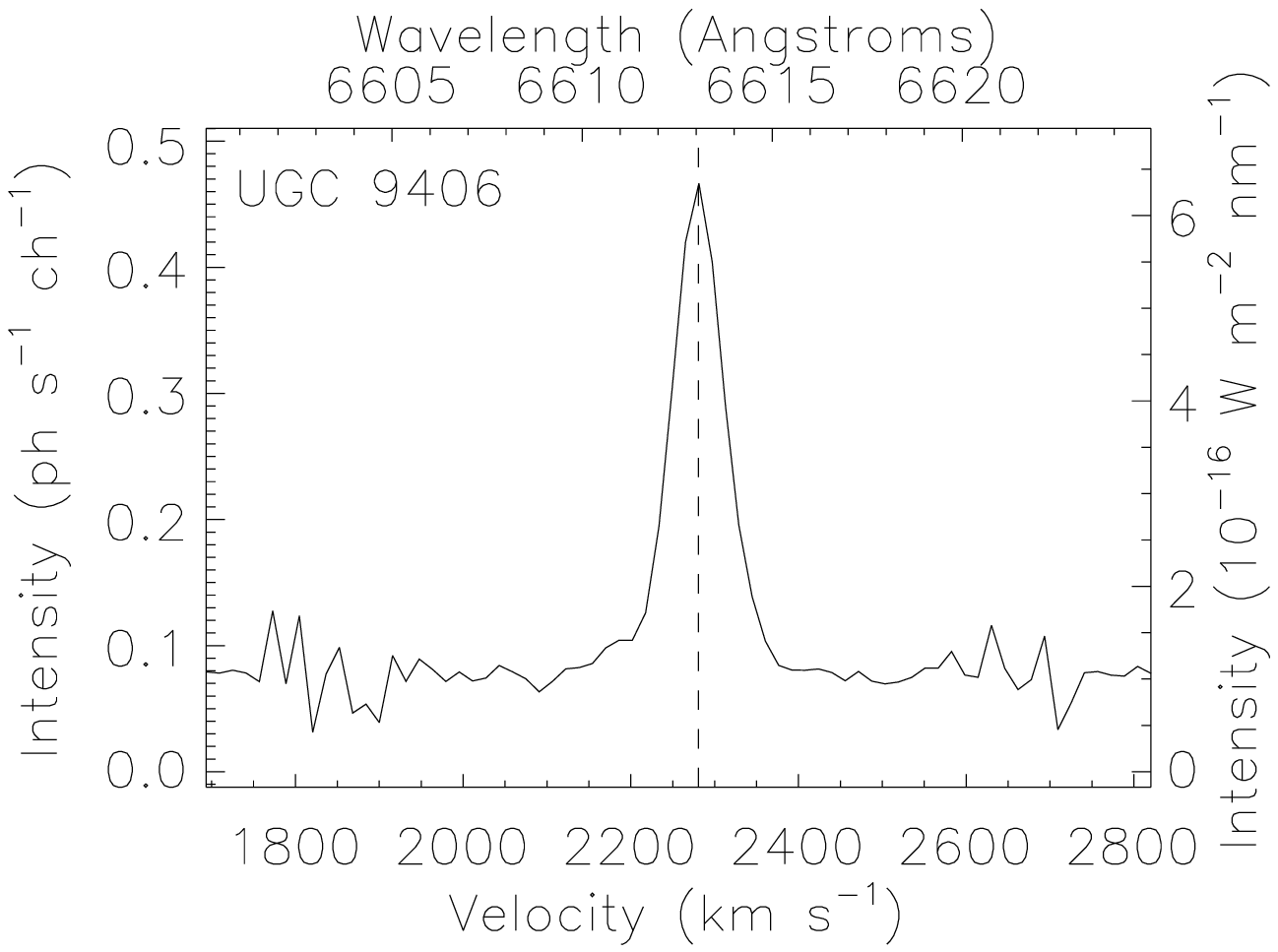}
\includegraphics[width=3.5cm]{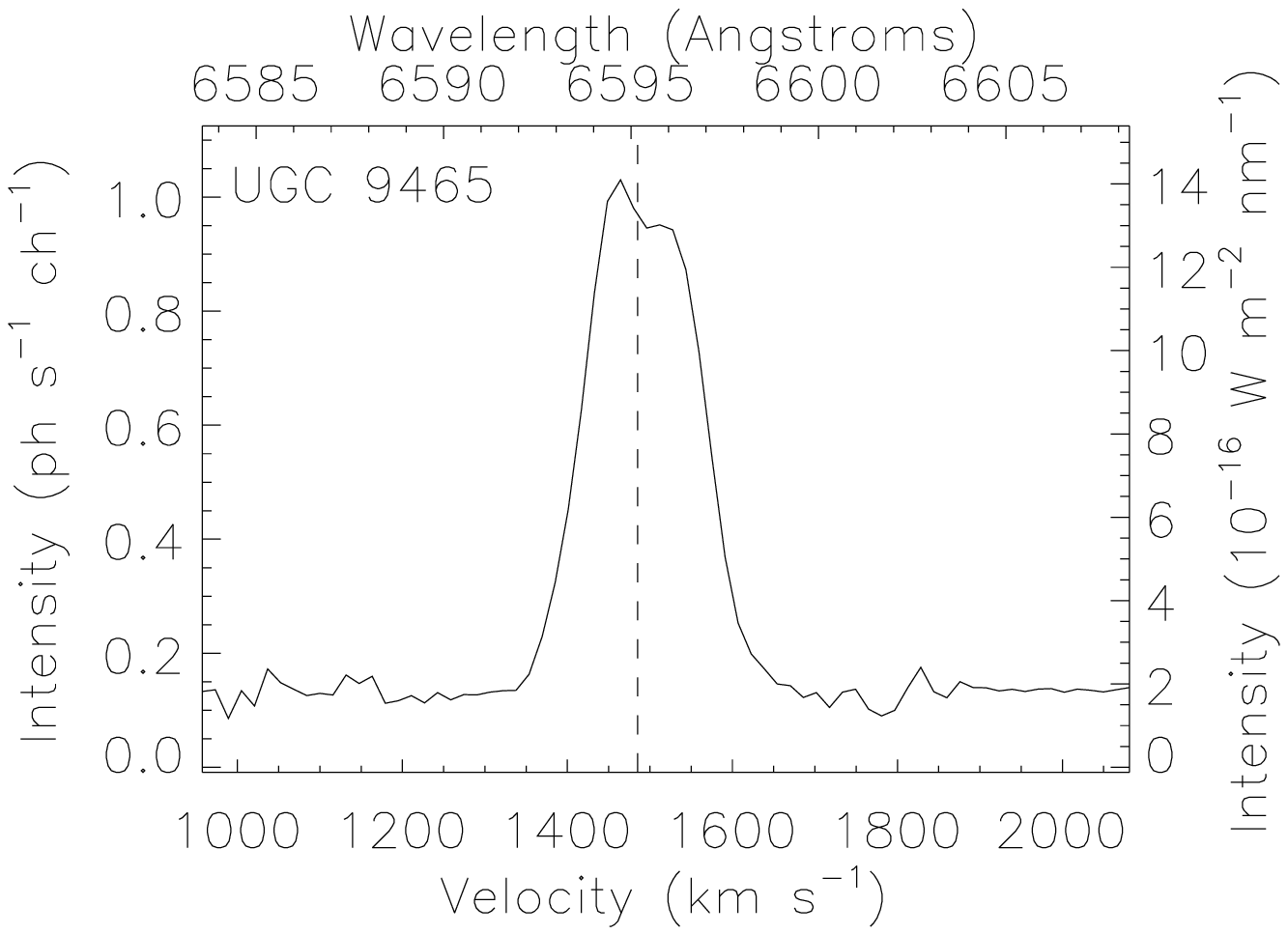}
\includegraphics[width=3.5cm]{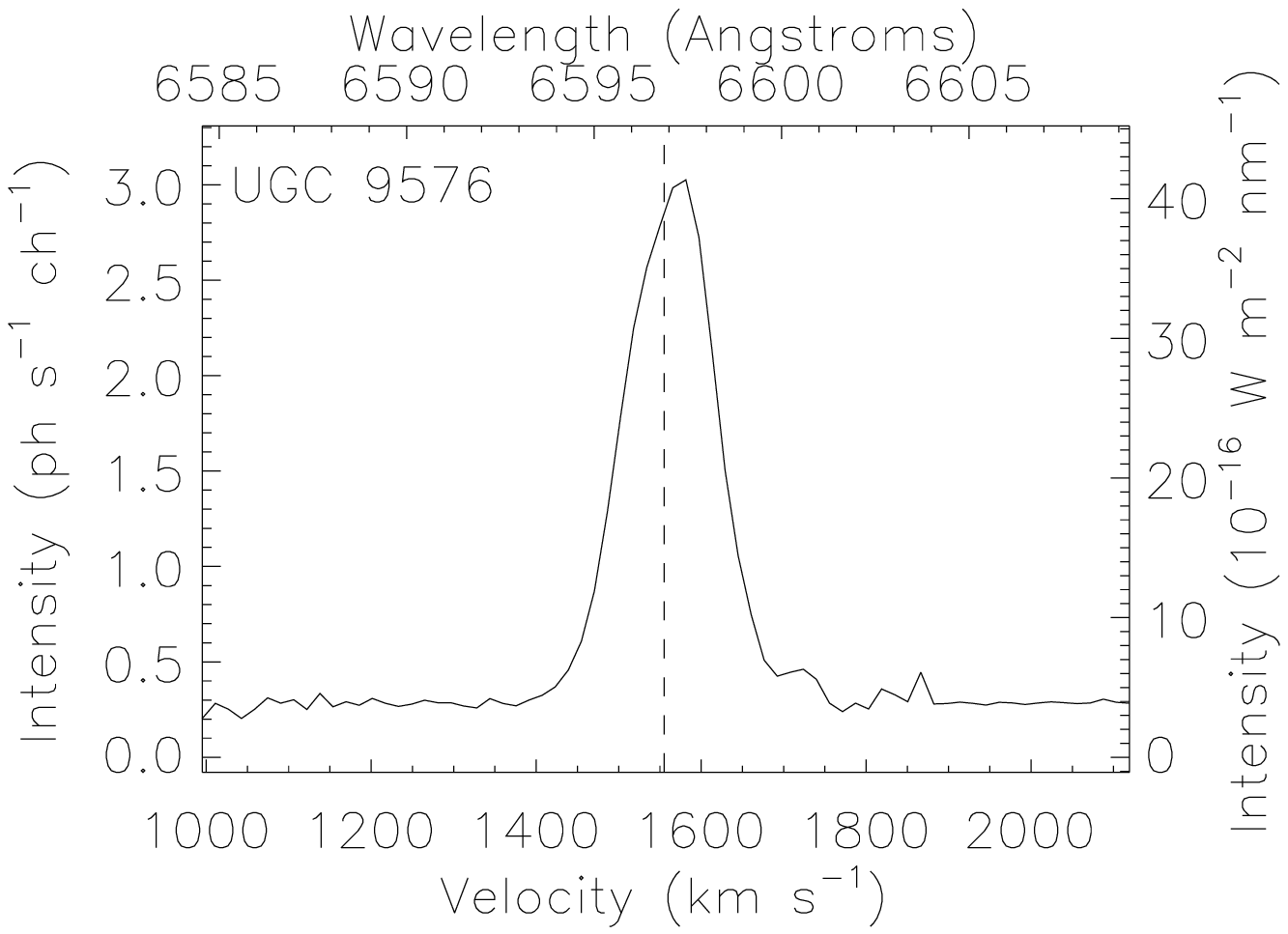}
\includegraphics[width=3.5cm]{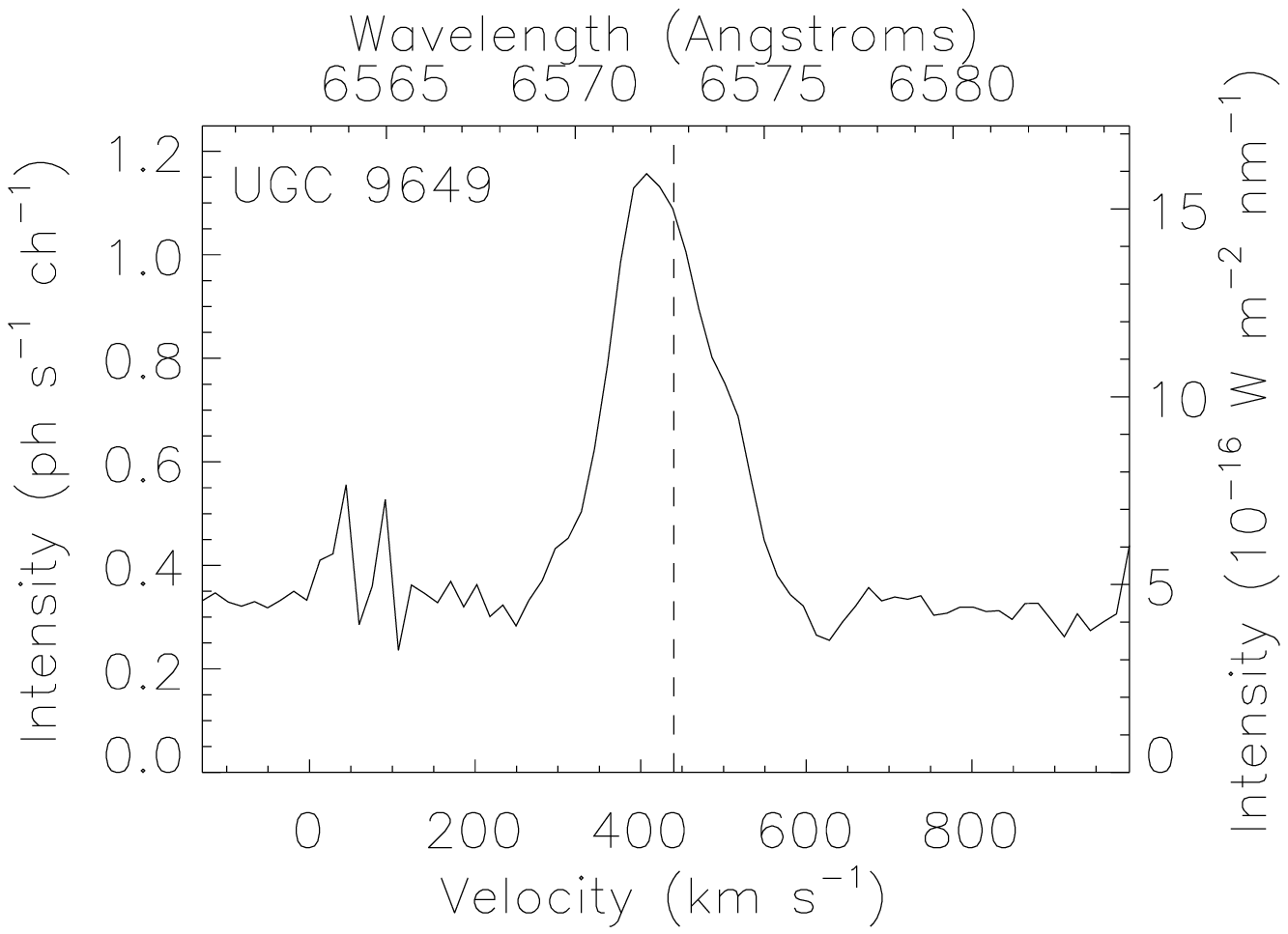}
\includegraphics[width=3.5cm]{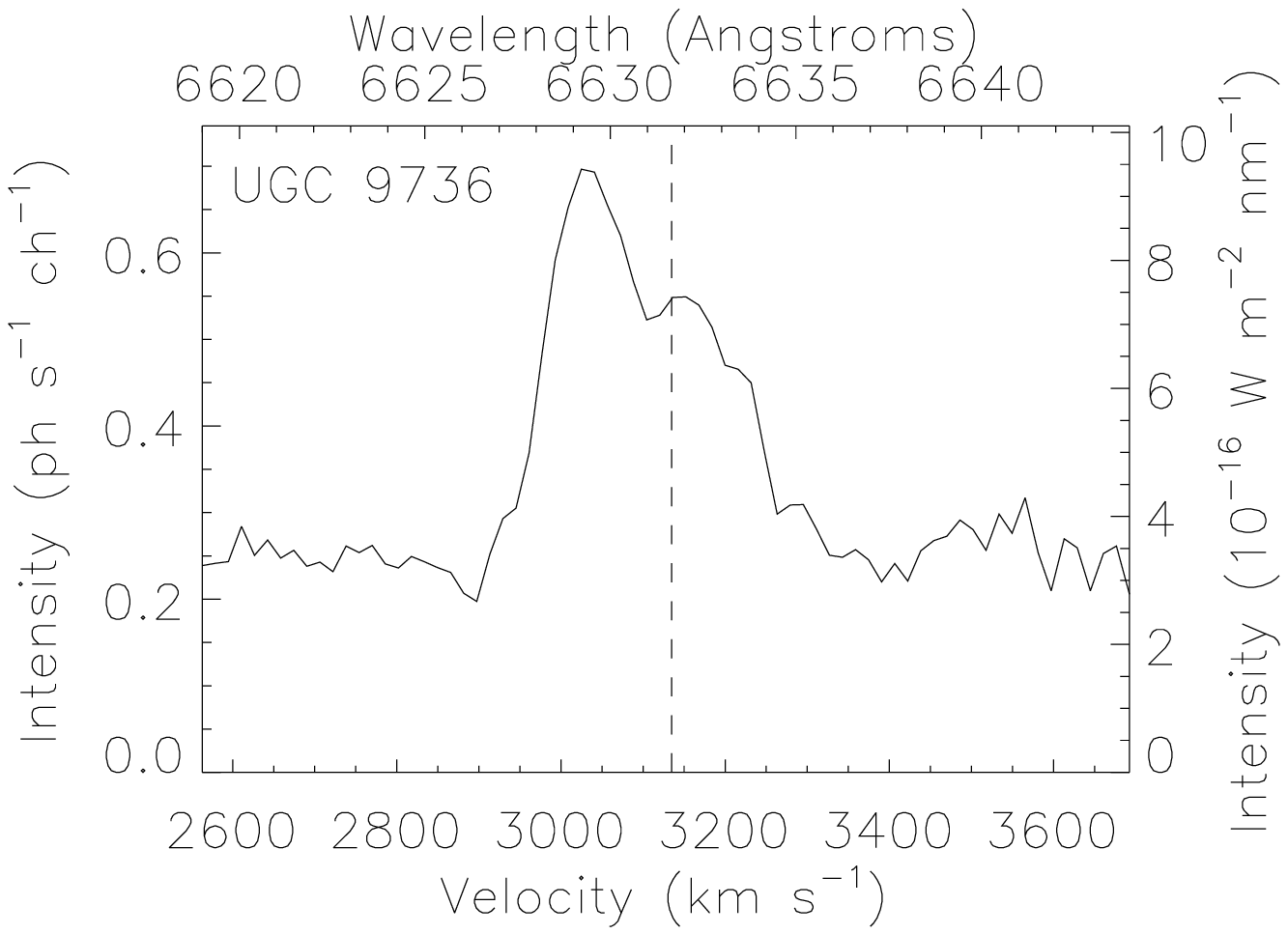}
\includegraphics[width=3.5cm]{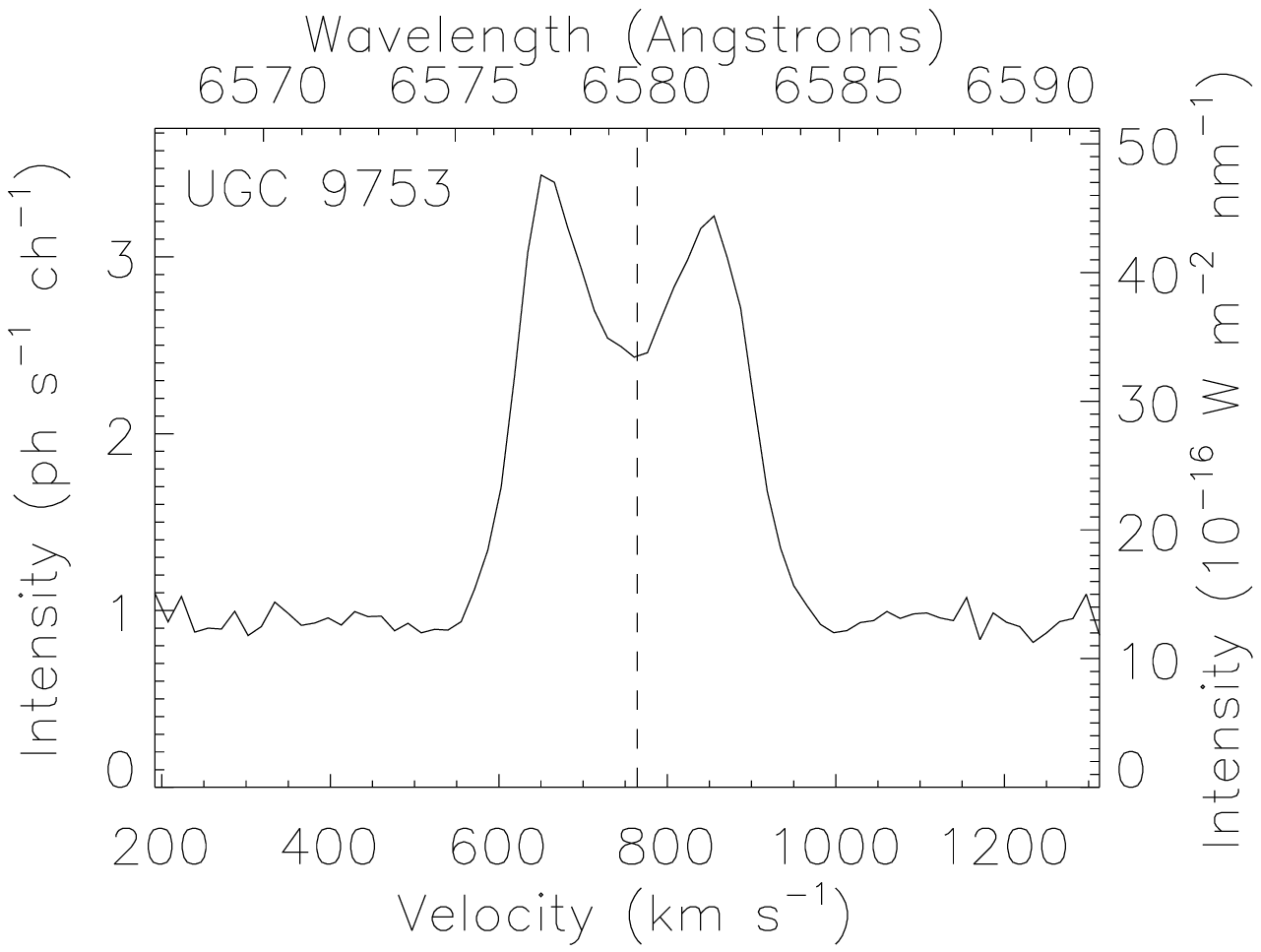}
\includegraphics[width=3.5cm]{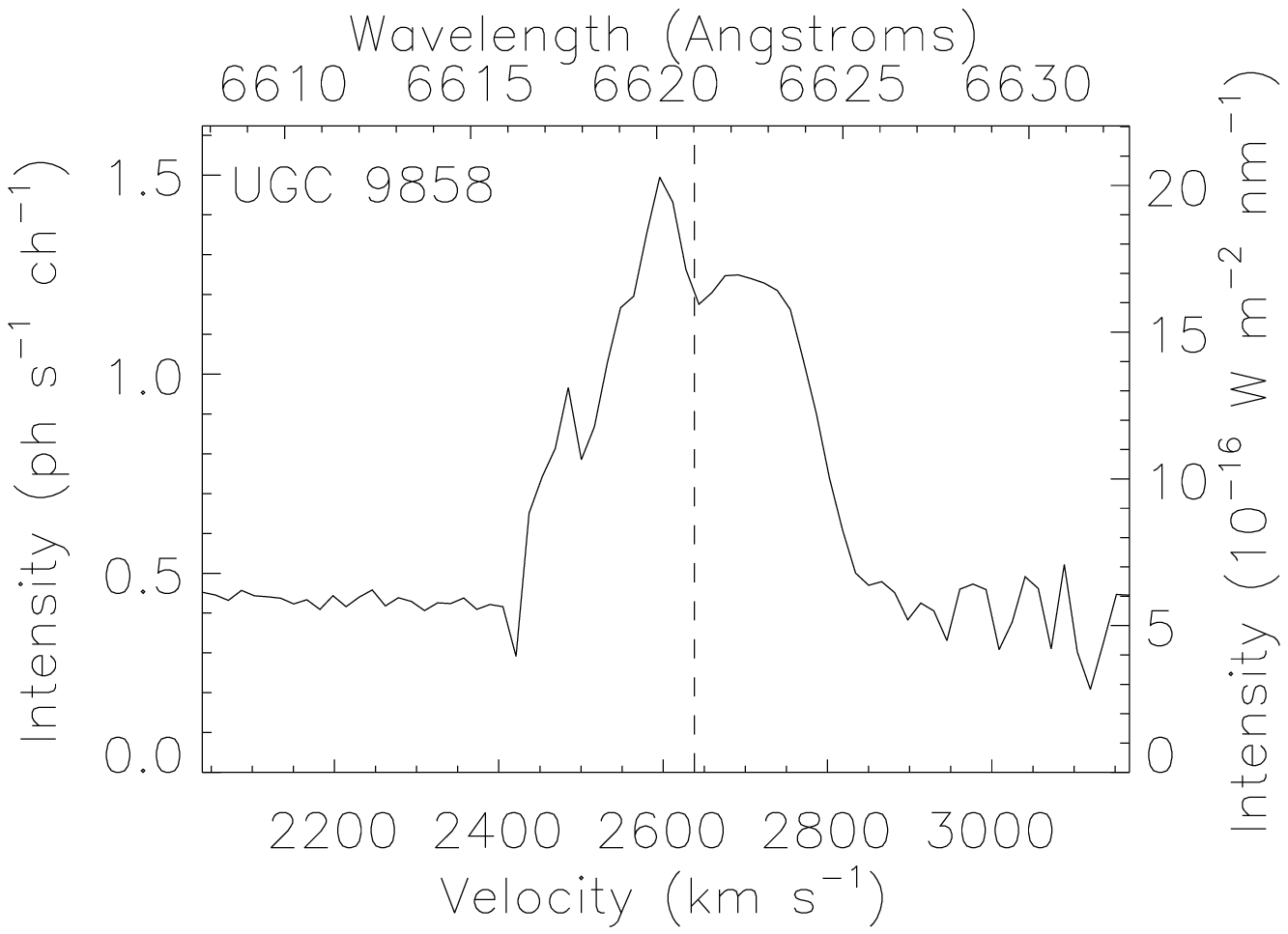}
\includegraphics[width=3.5cm]{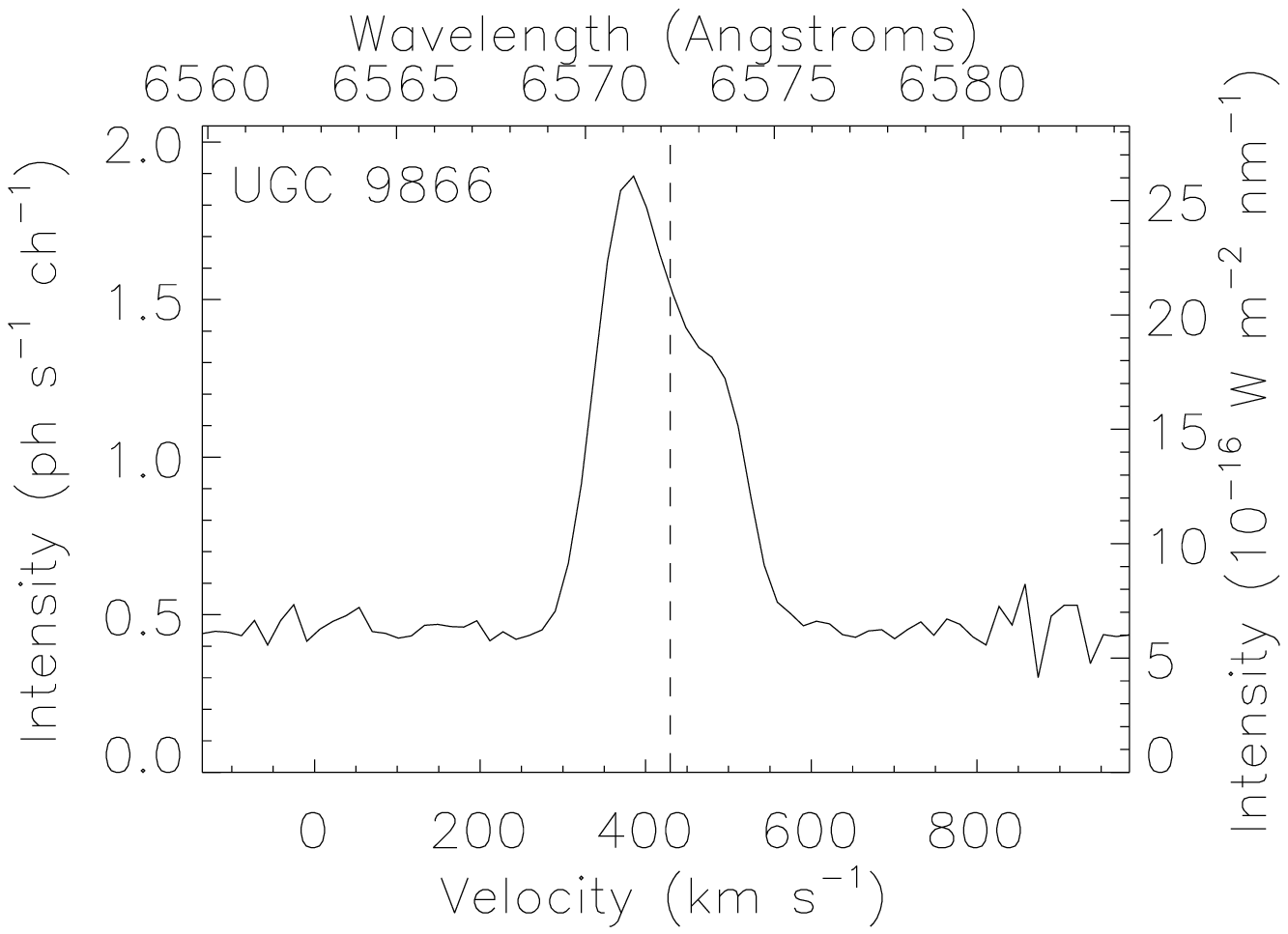}
\includegraphics[width=3.5cm]{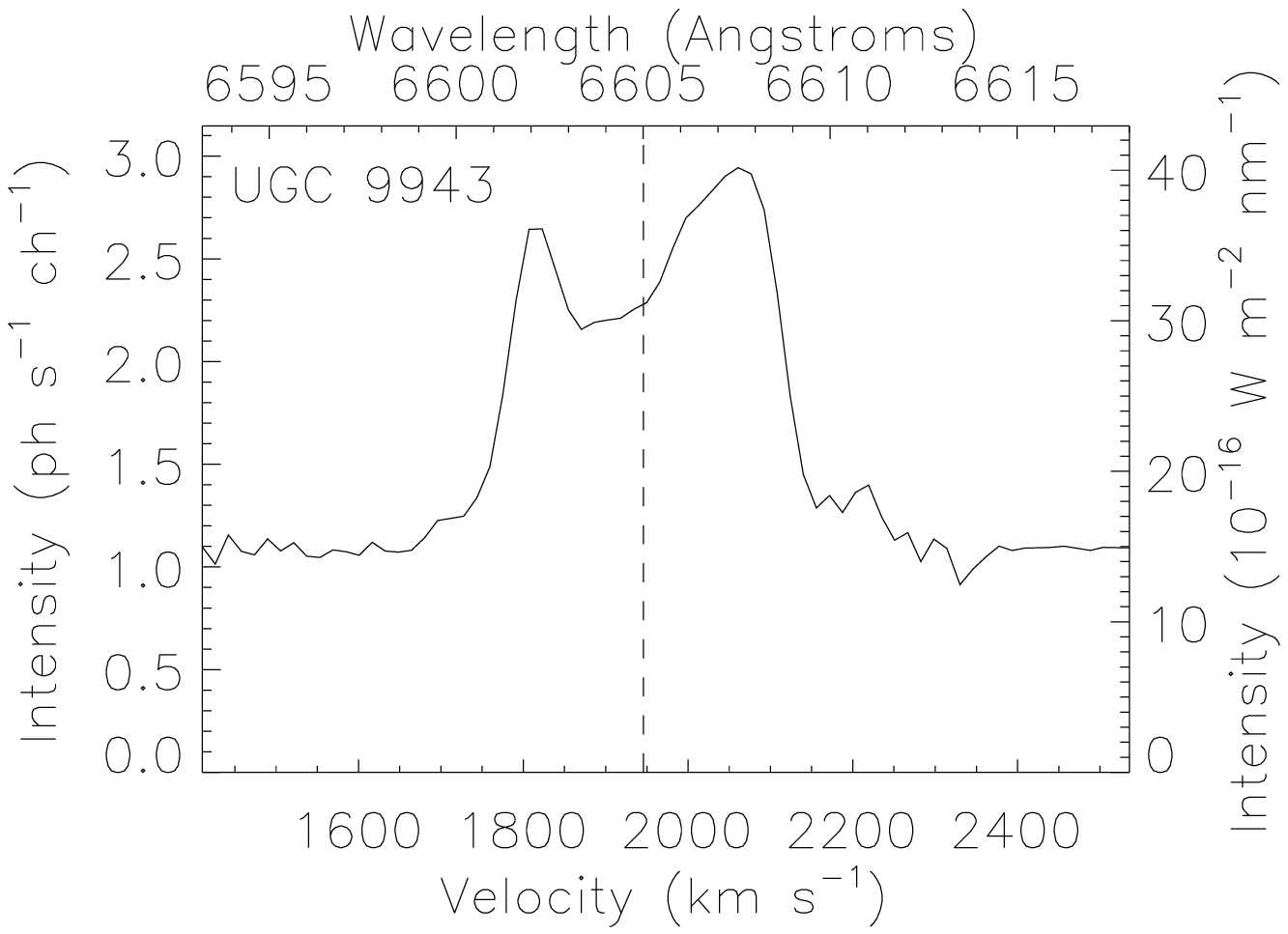}
\end{center}
\end{minipage}
\end{figure}
\clearpage
\begin{figure}
\begin{minipage}{180mm}
\begin{center}
\includegraphics[width=3.5cm]{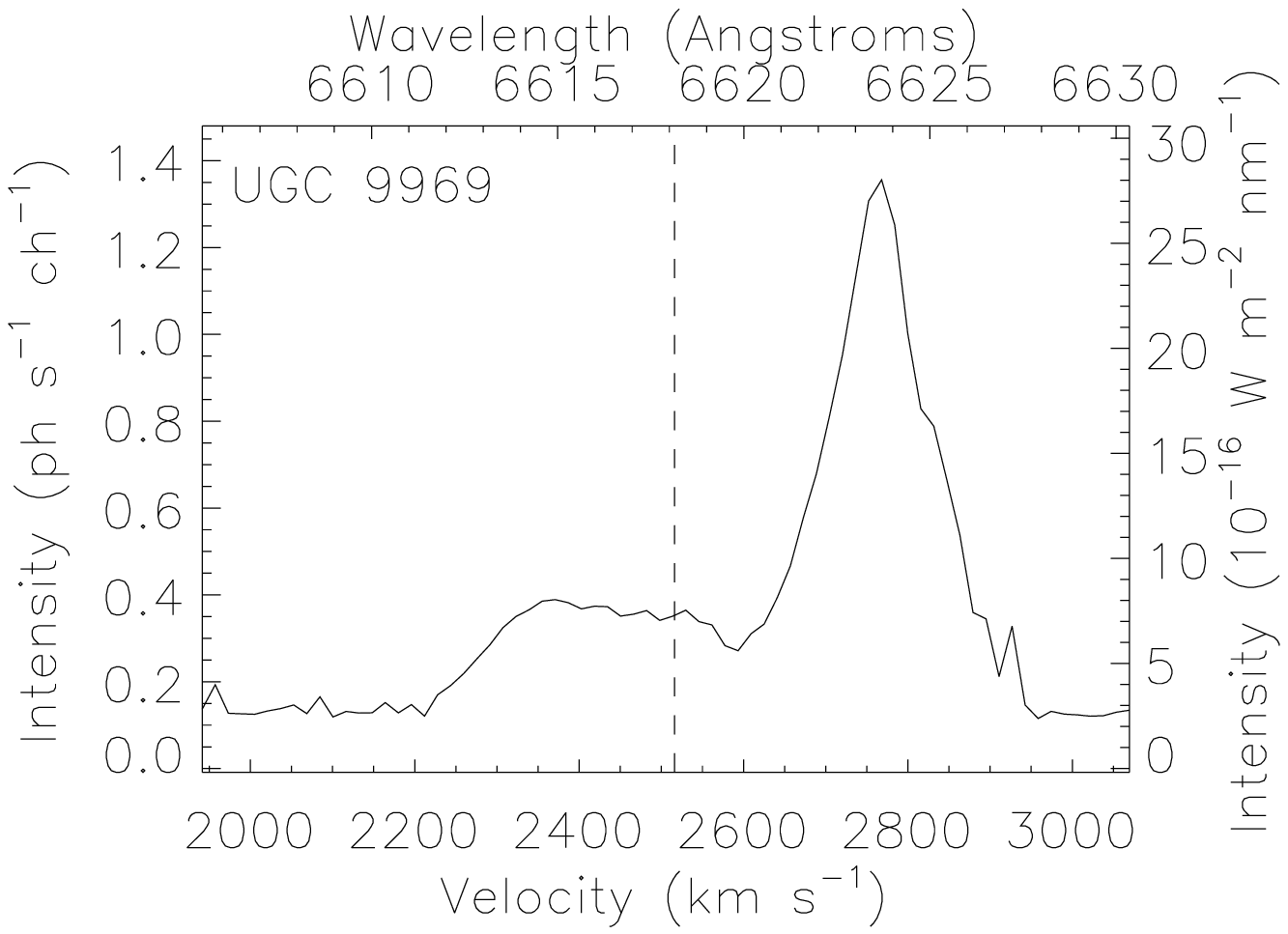}
\includegraphics[width=3.5cm]{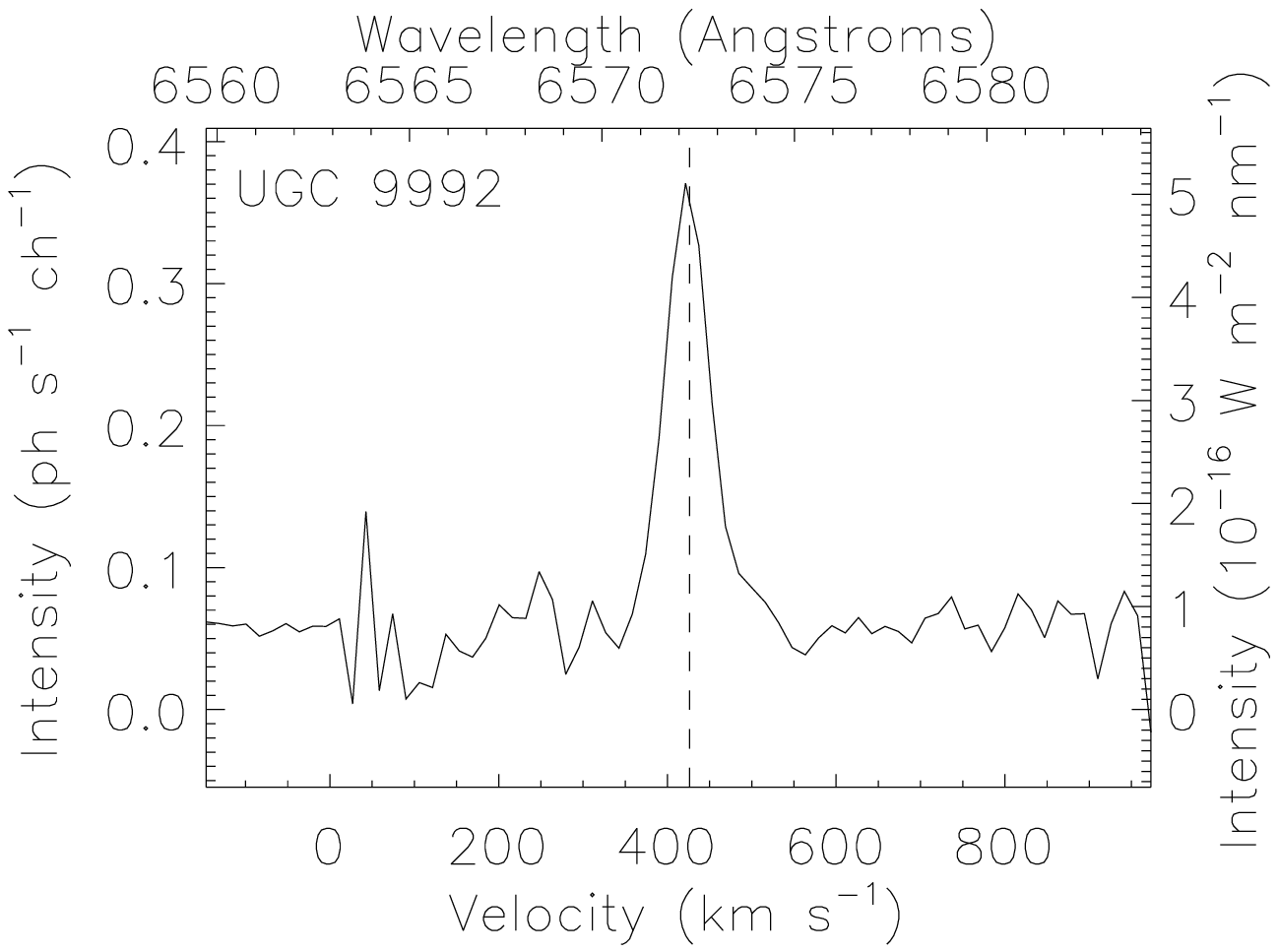}
\includegraphics[width=3.5cm]{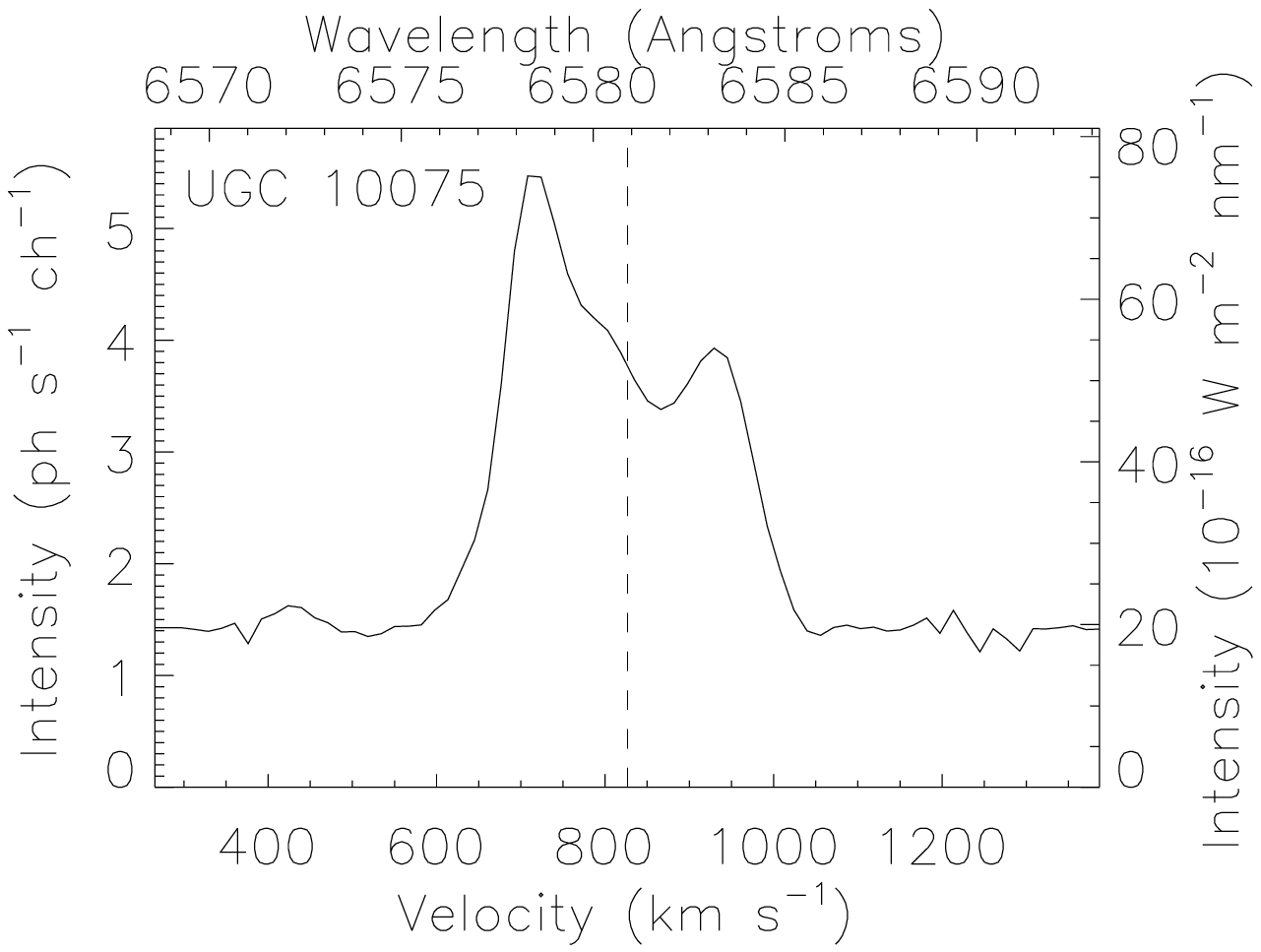}
\includegraphics[width=3.5cm]{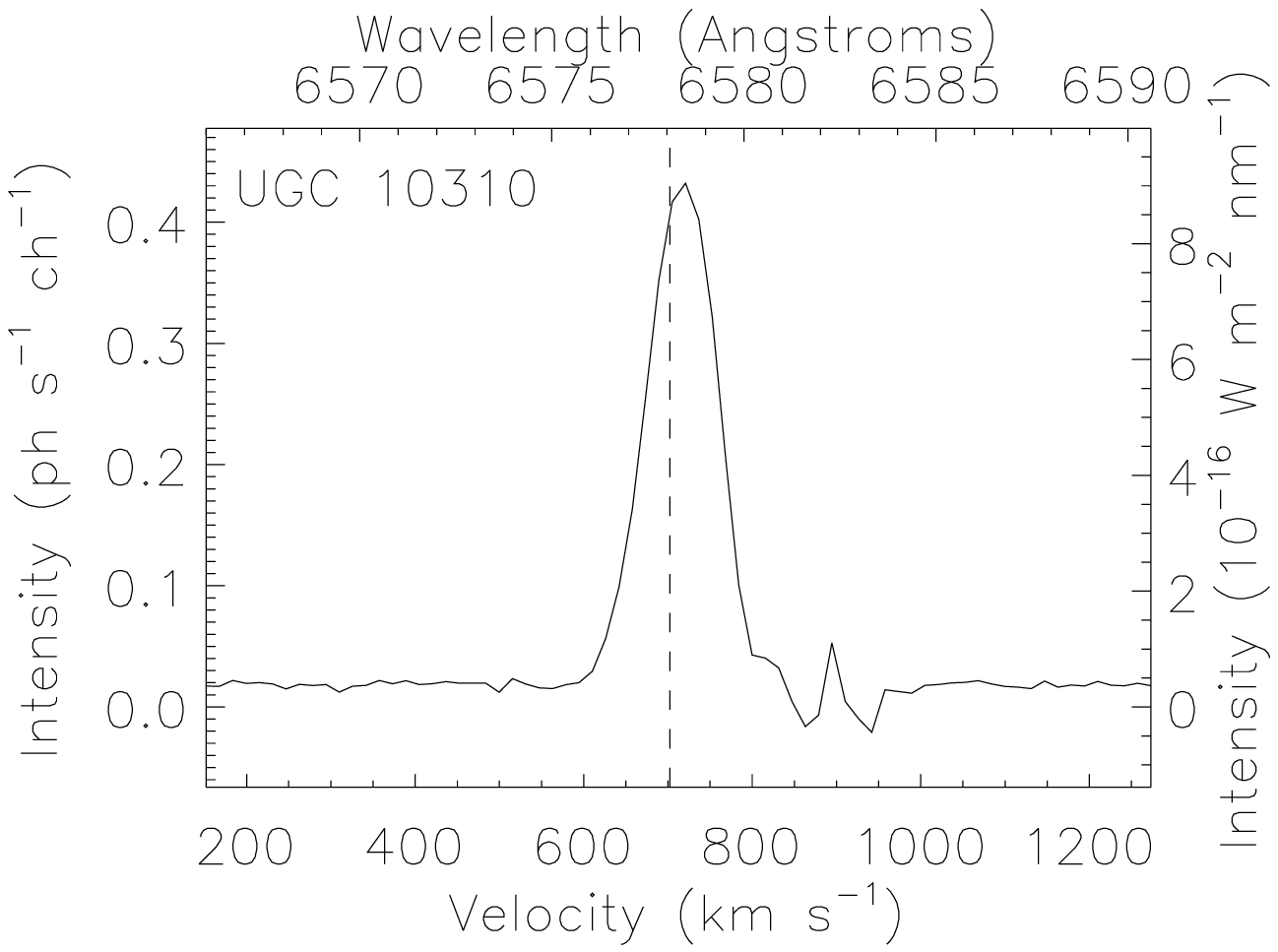}
\includegraphics[width=3.5cm]{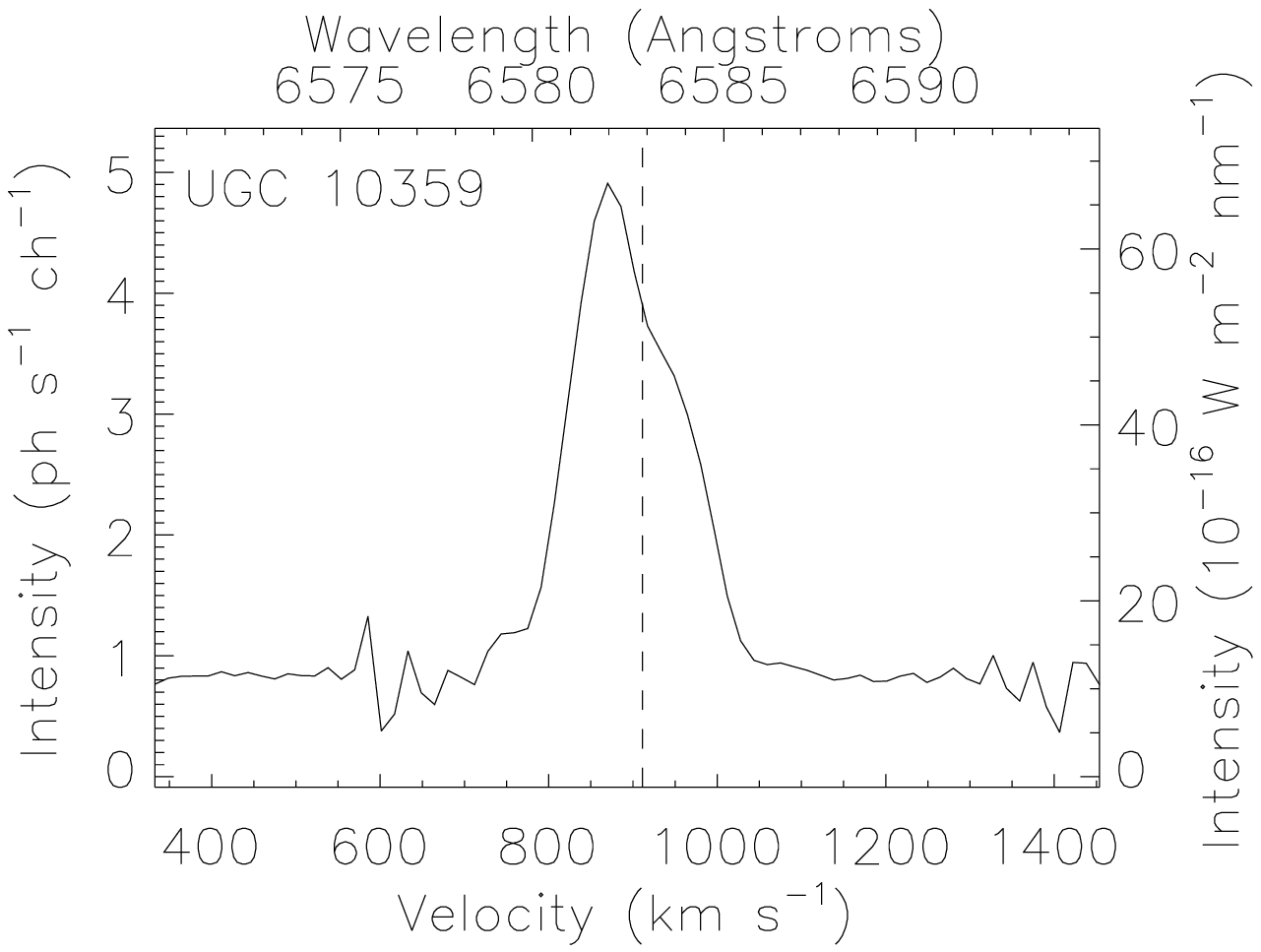}
\includegraphics[width=3.5cm]{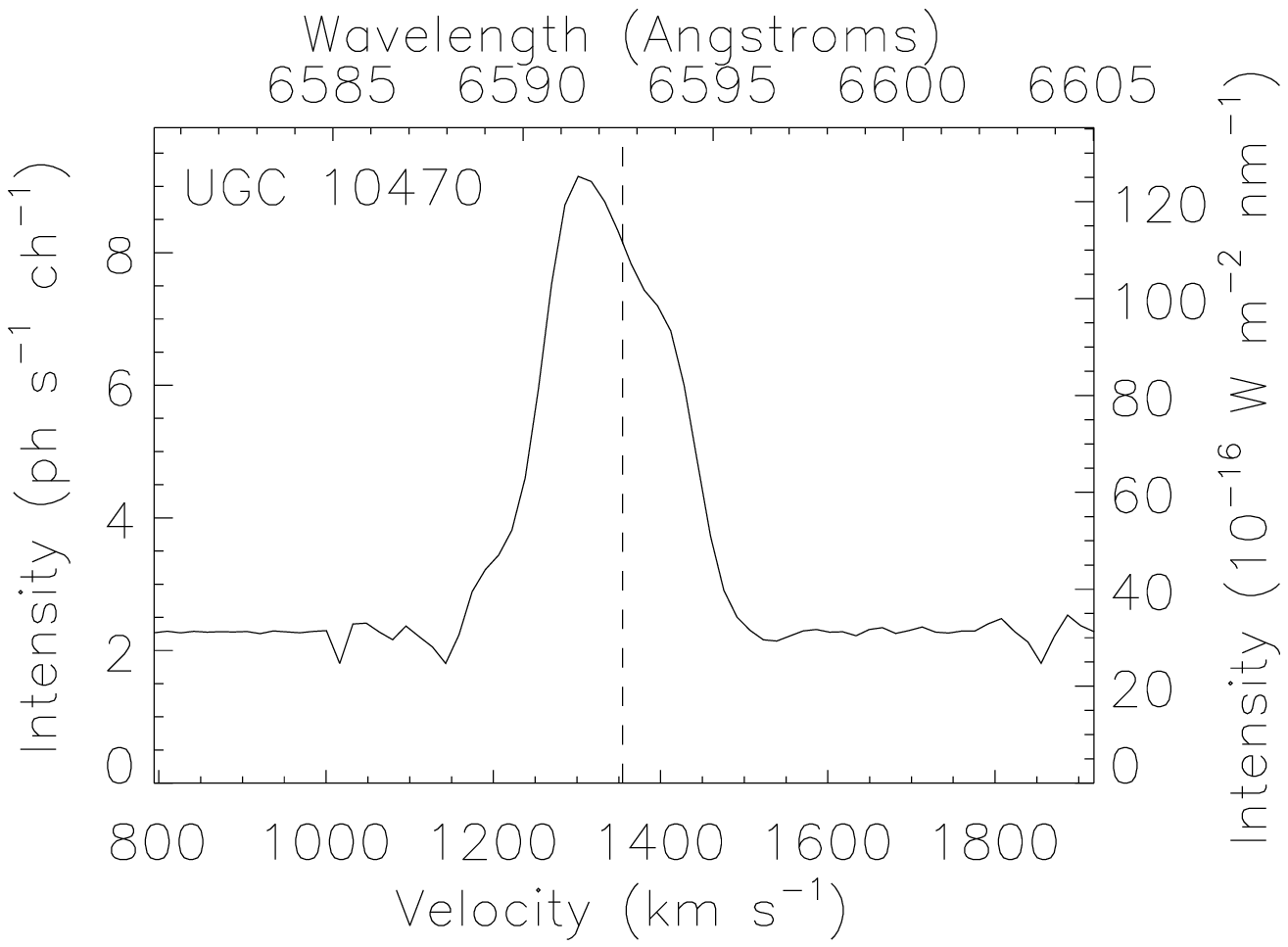}
\includegraphics[width=3.5cm]{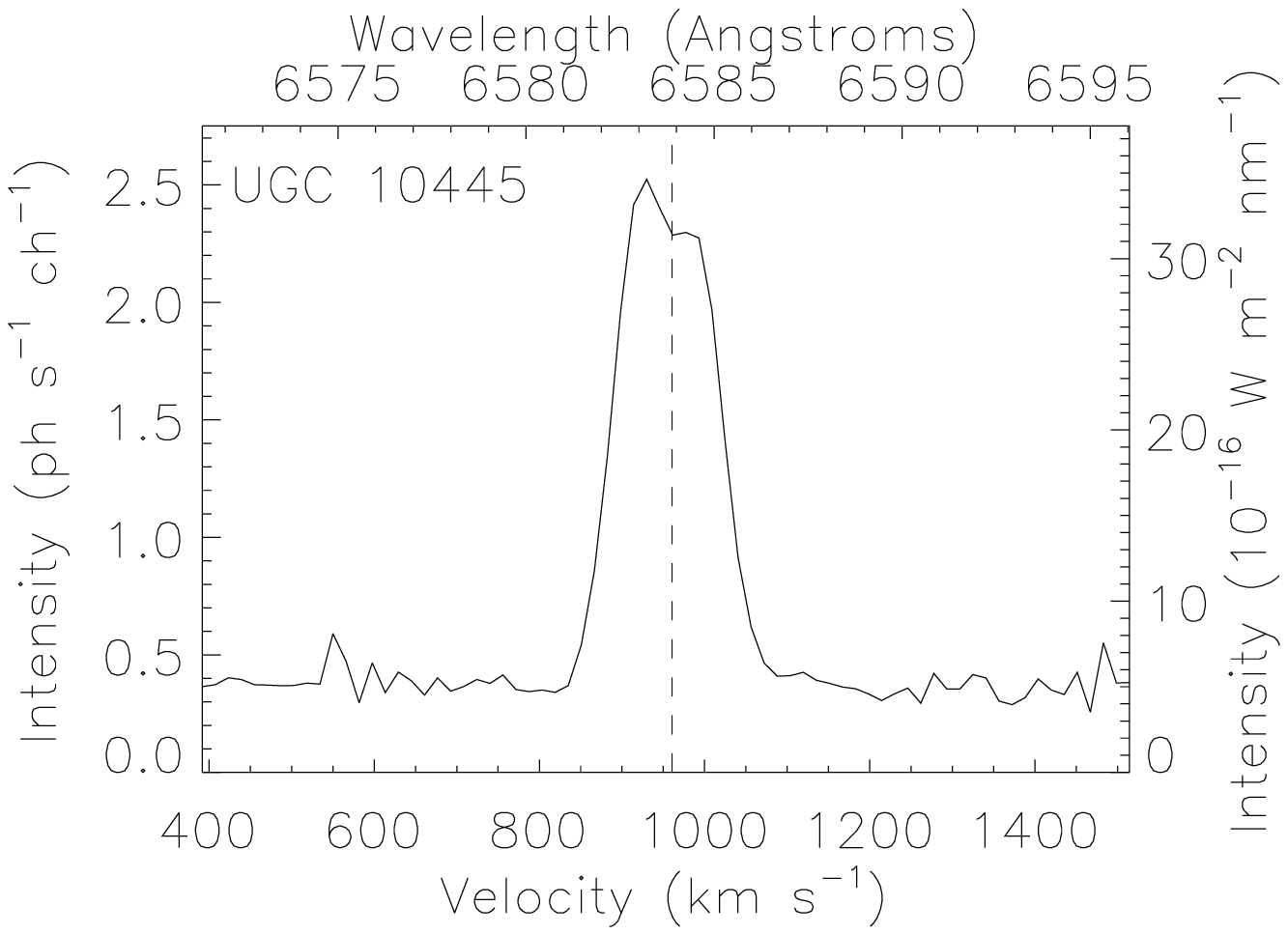}
\includegraphics[width=3.5cm]{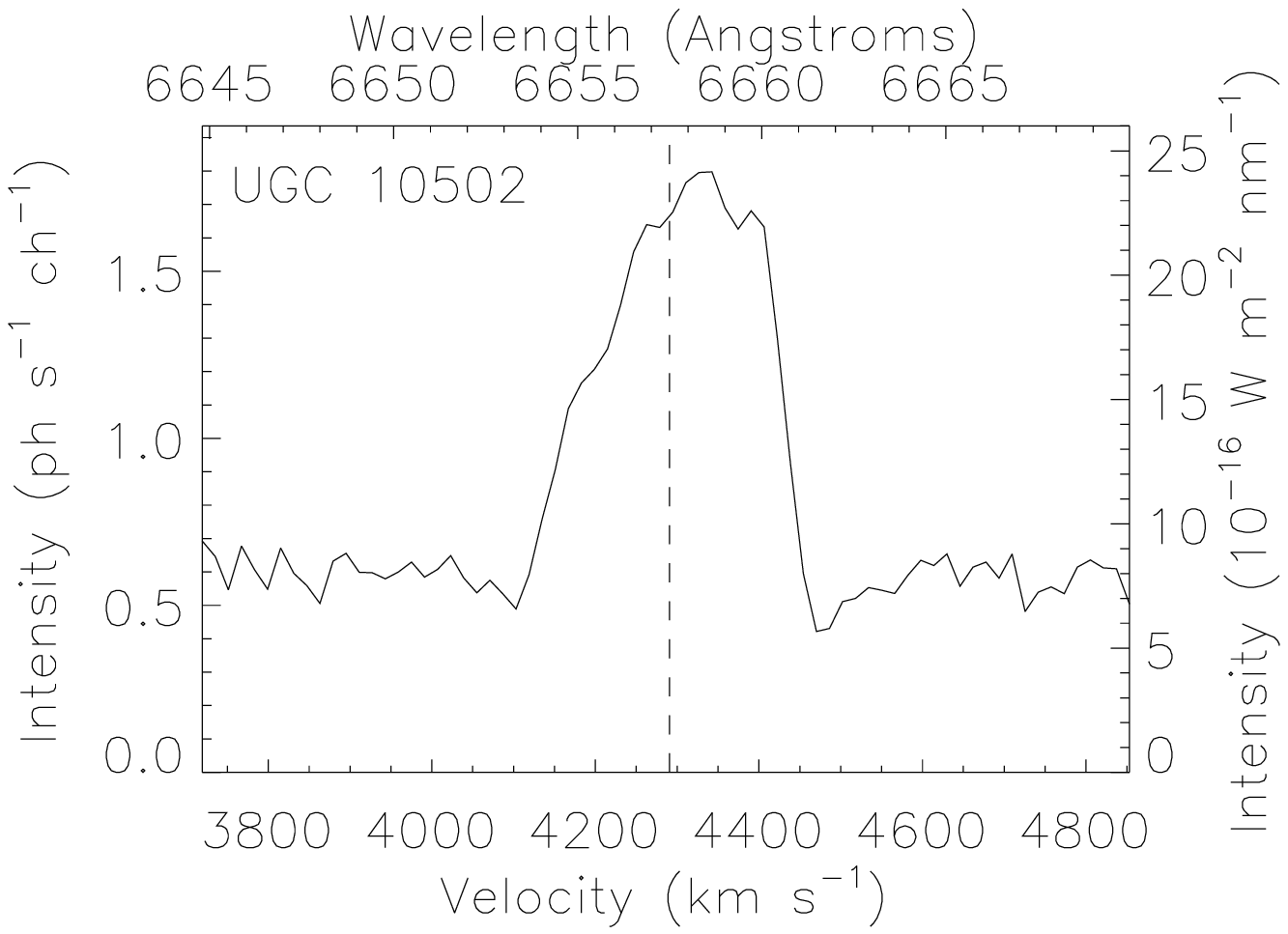}
\includegraphics[width=3.5cm]{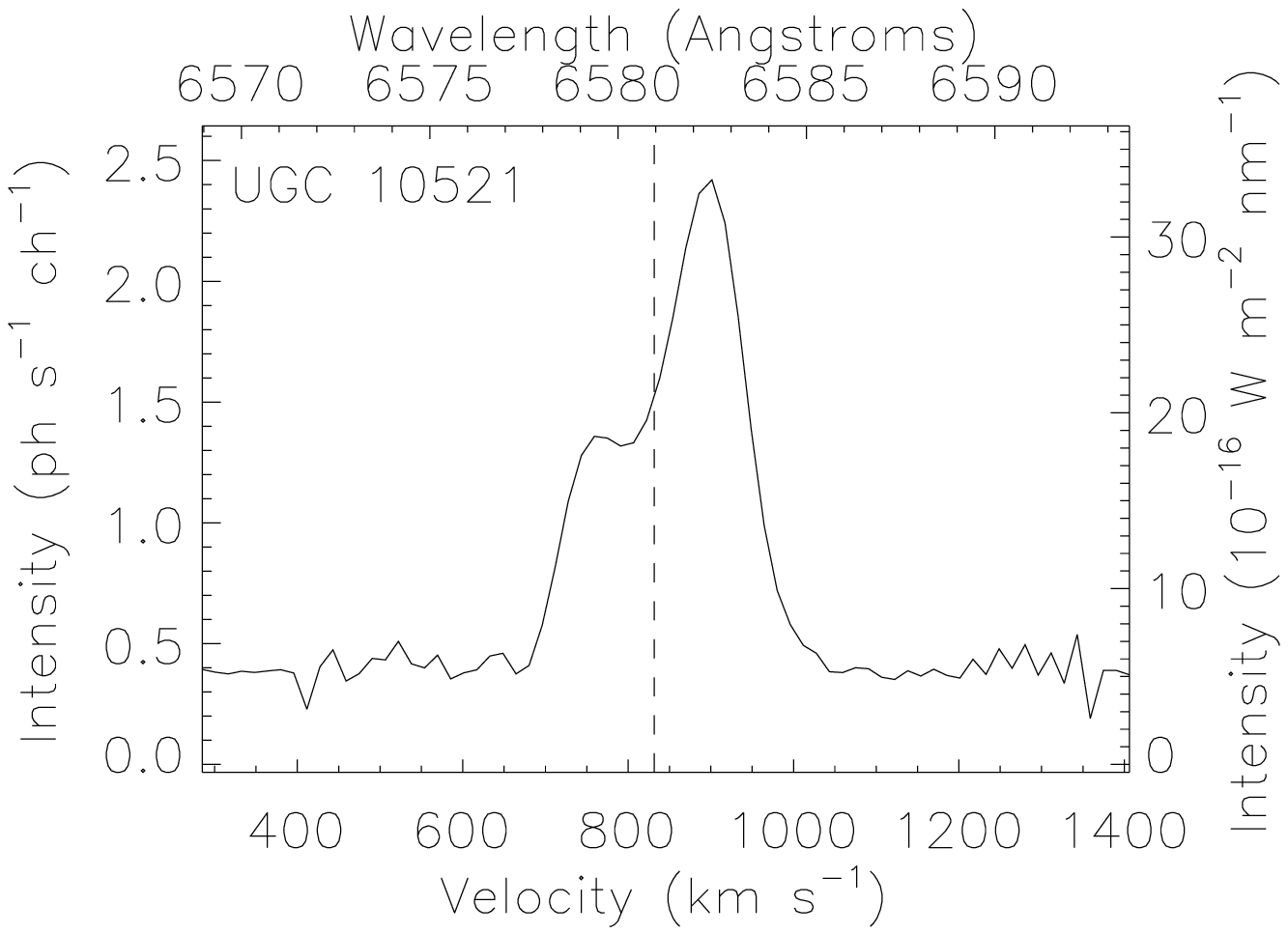}
\includegraphics[width=3.5cm]{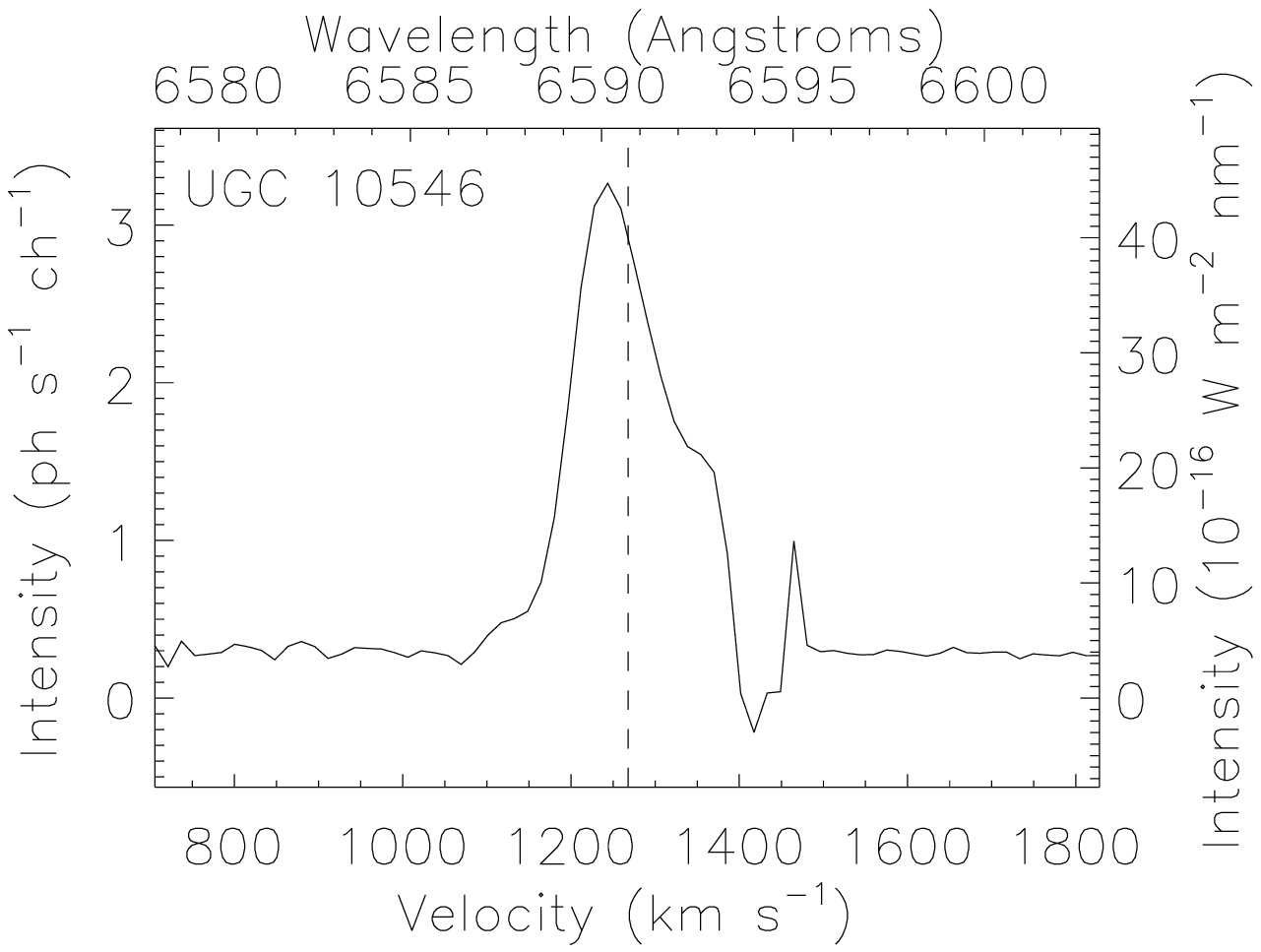}
\includegraphics[width=3.5cm]{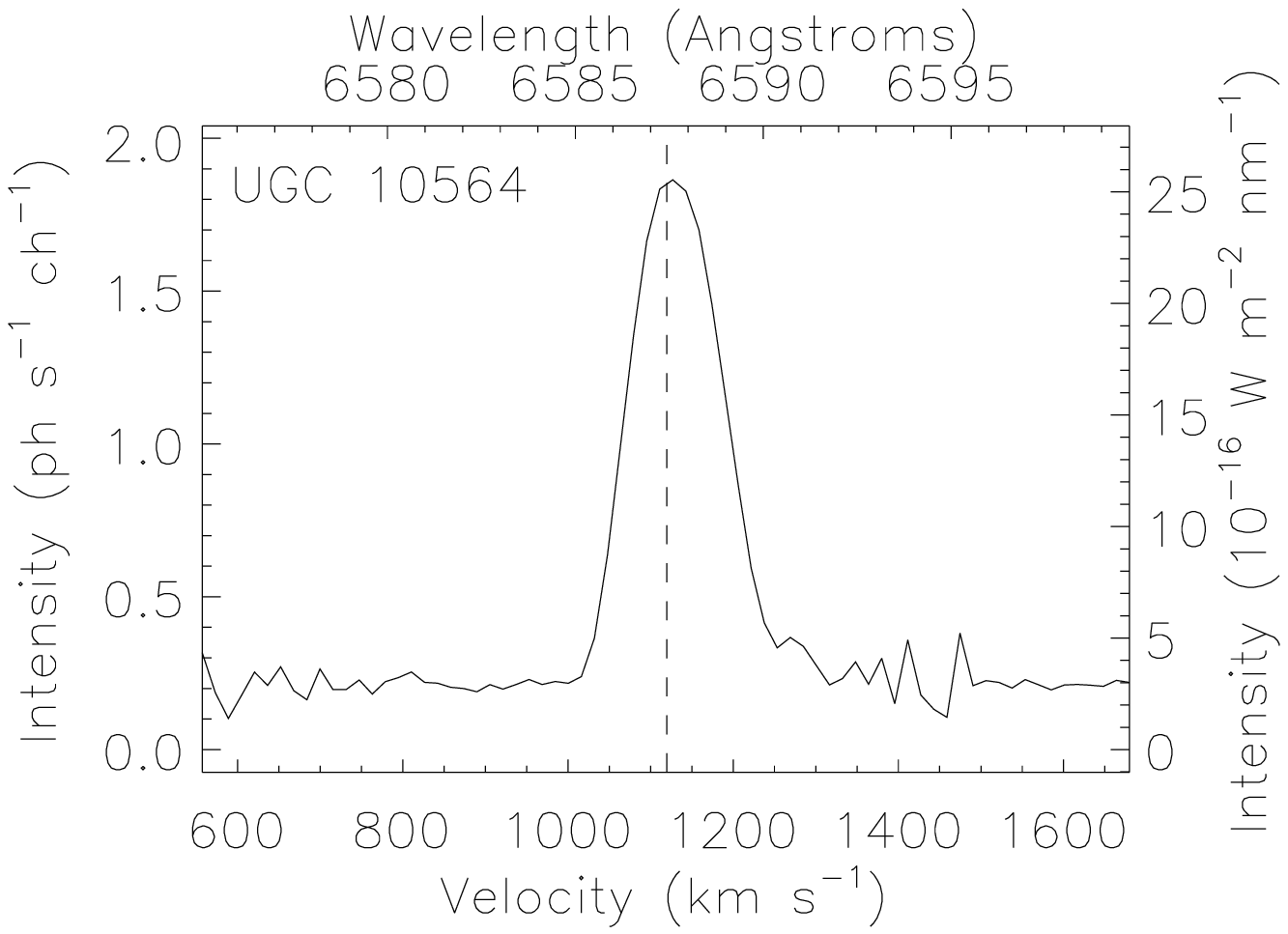}
\includegraphics[width=3.5cm]{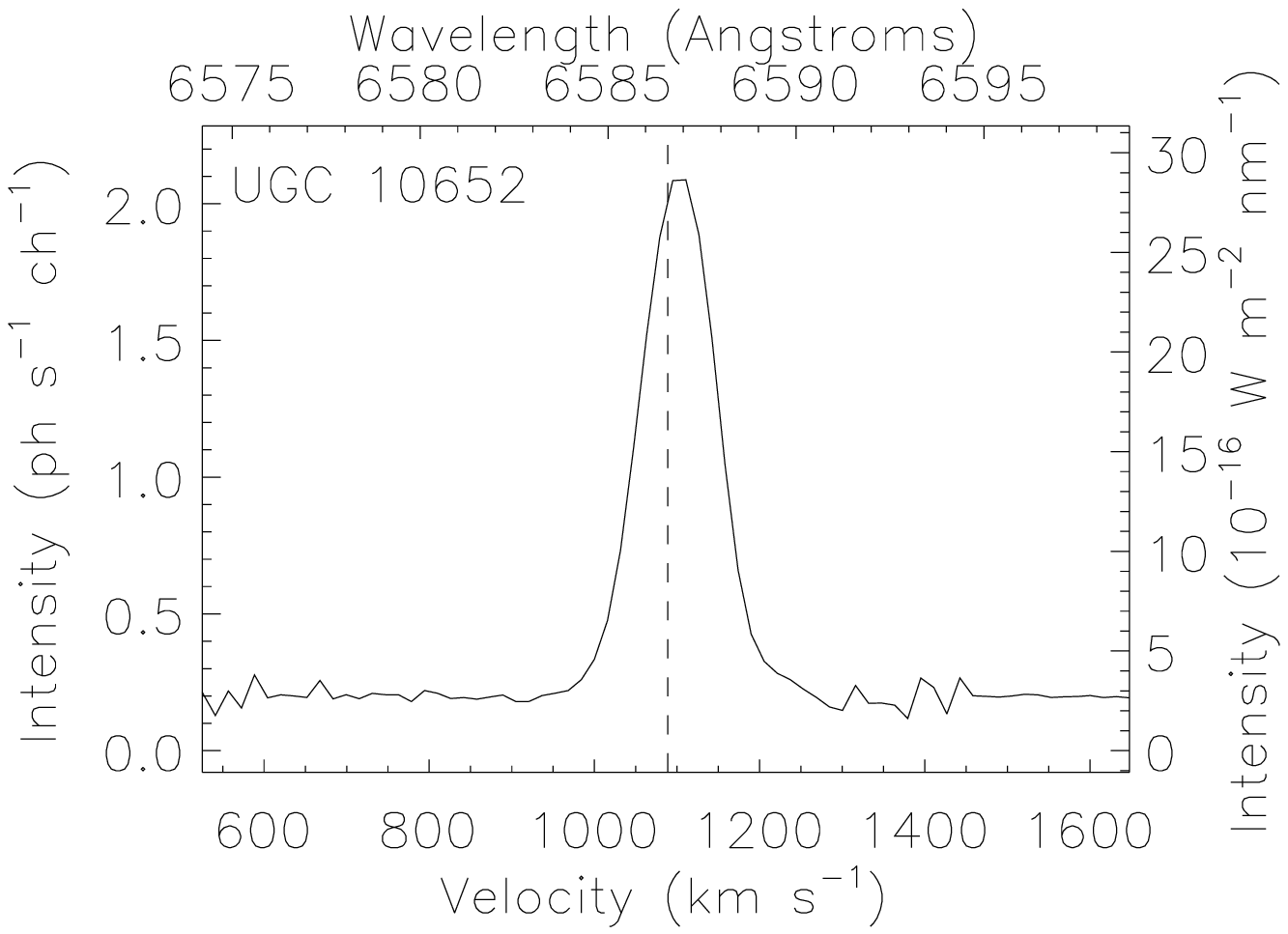}
\includegraphics[width=3.5cm]{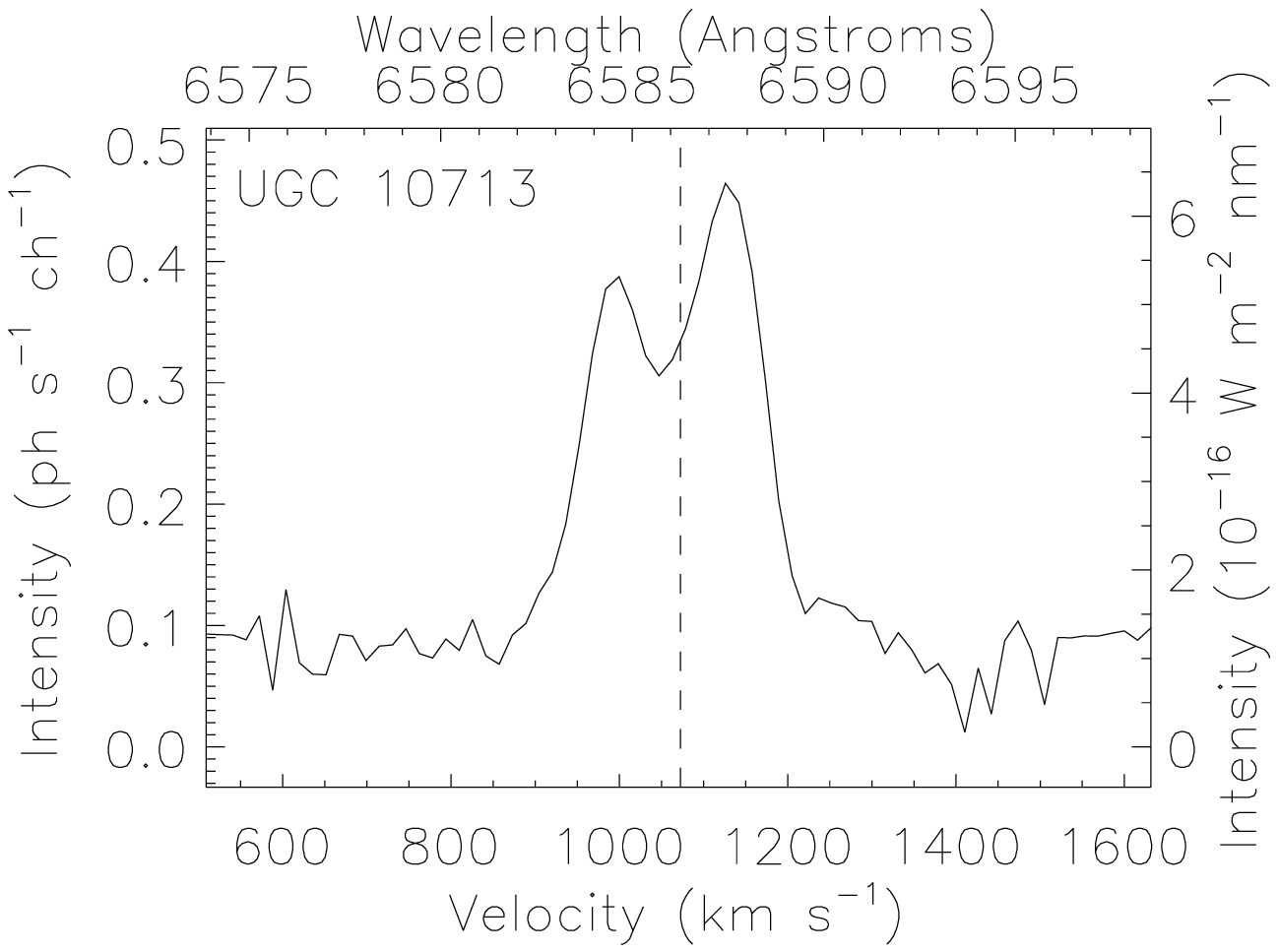}
\includegraphics[width=3.5cm]{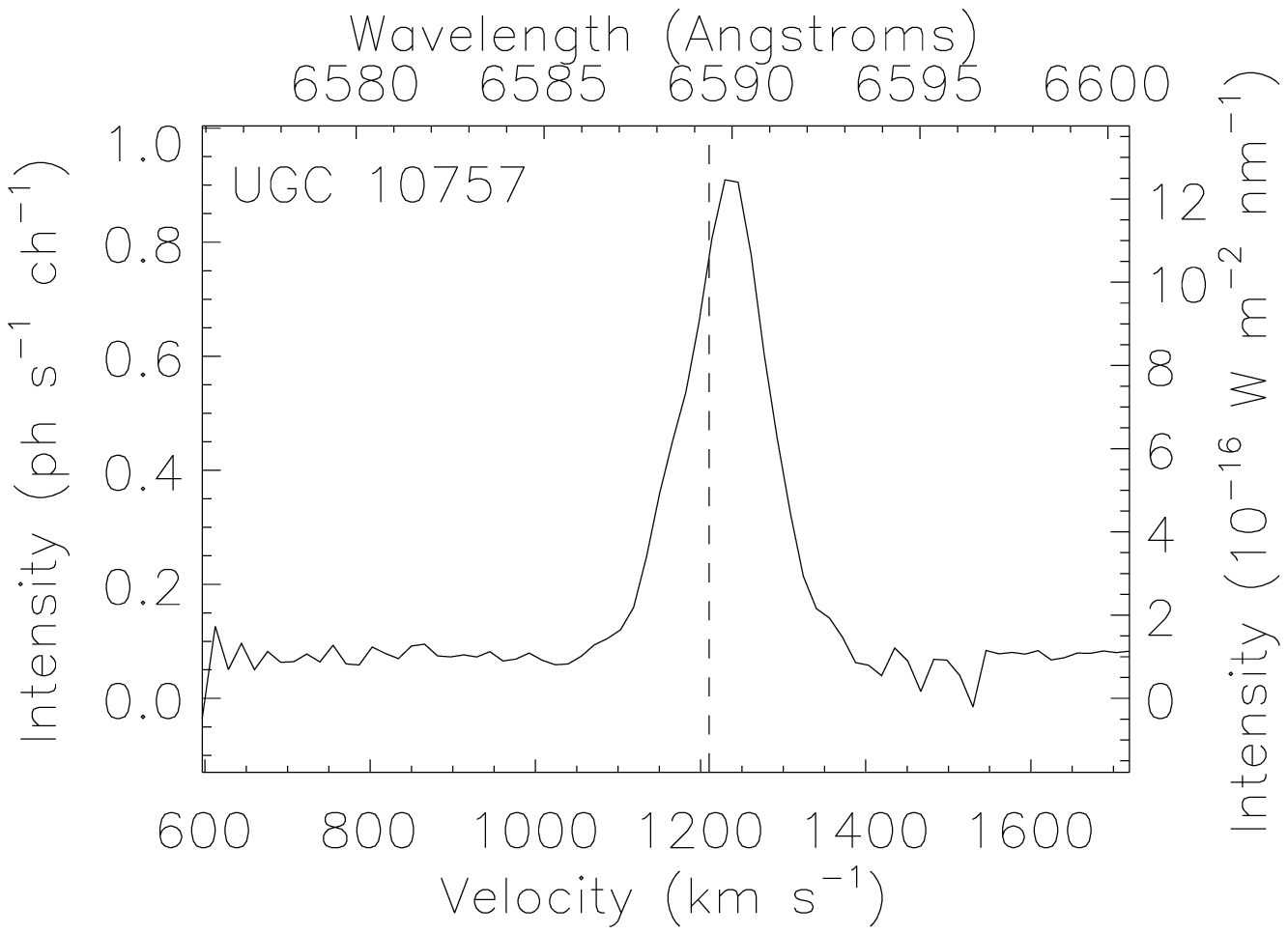}
\includegraphics[width=3.5cm]{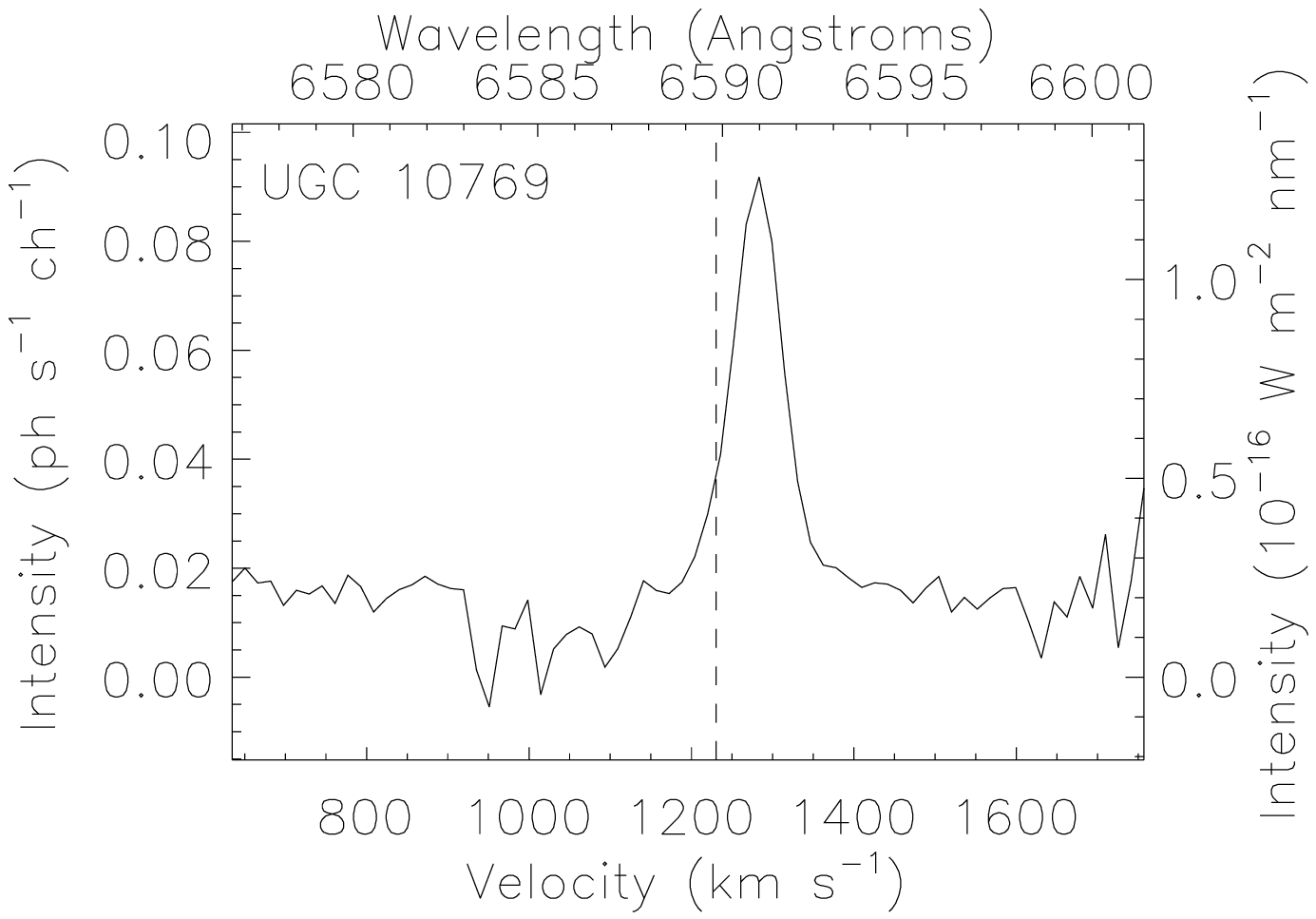}
\includegraphics[width=3.5cm]{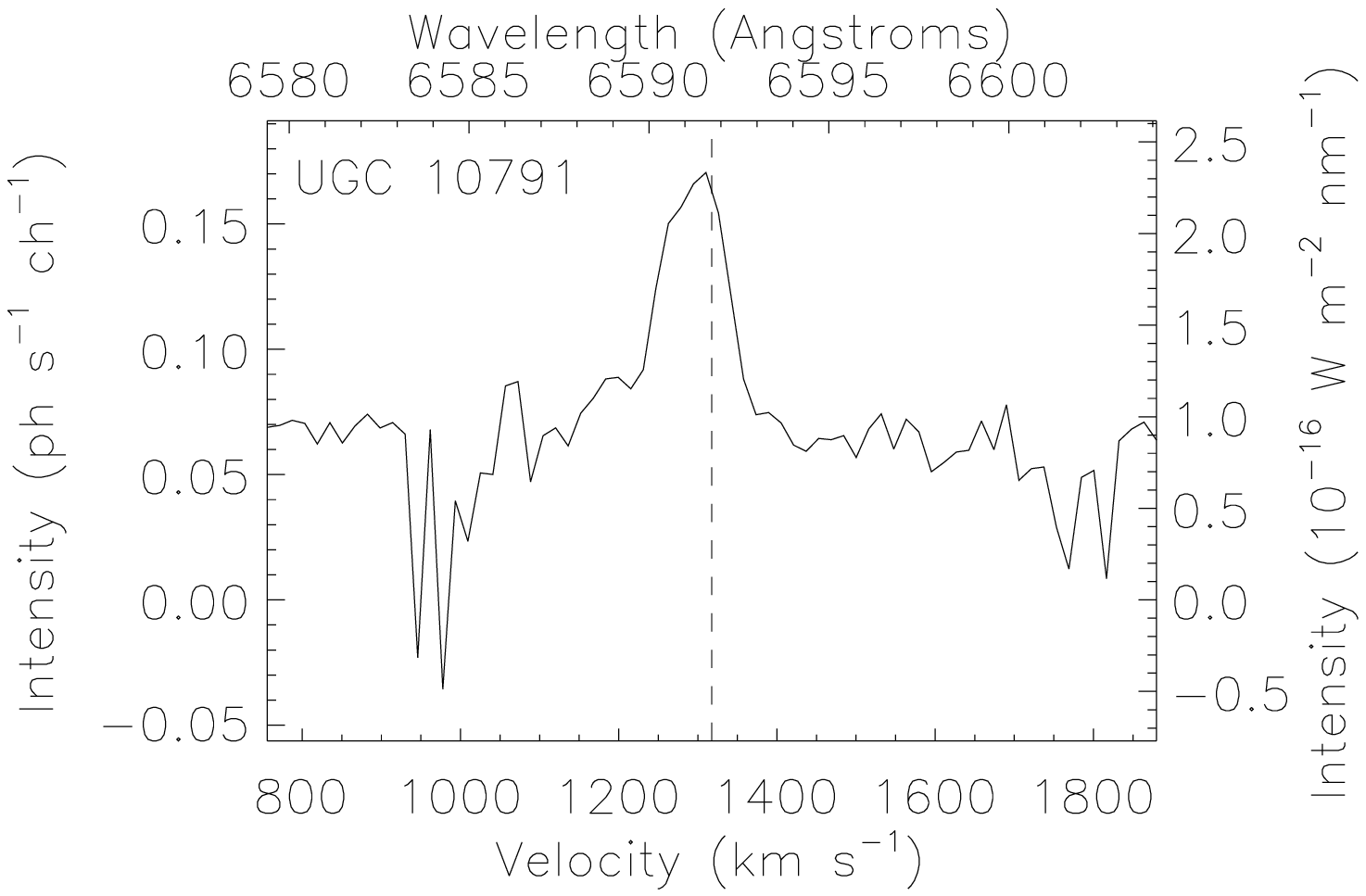}
\includegraphics[width=3.5cm]{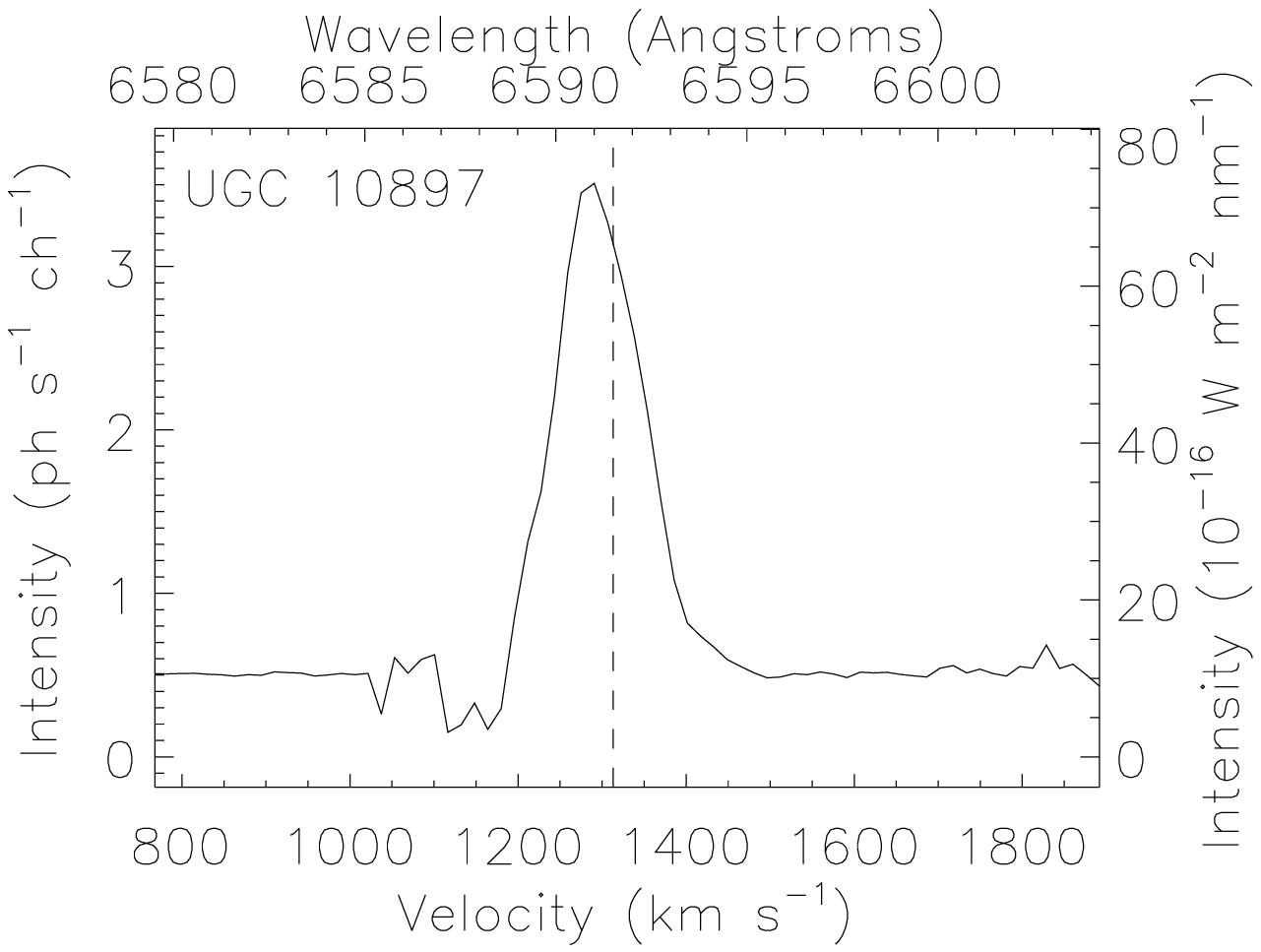}
\includegraphics[width=3.5cm]{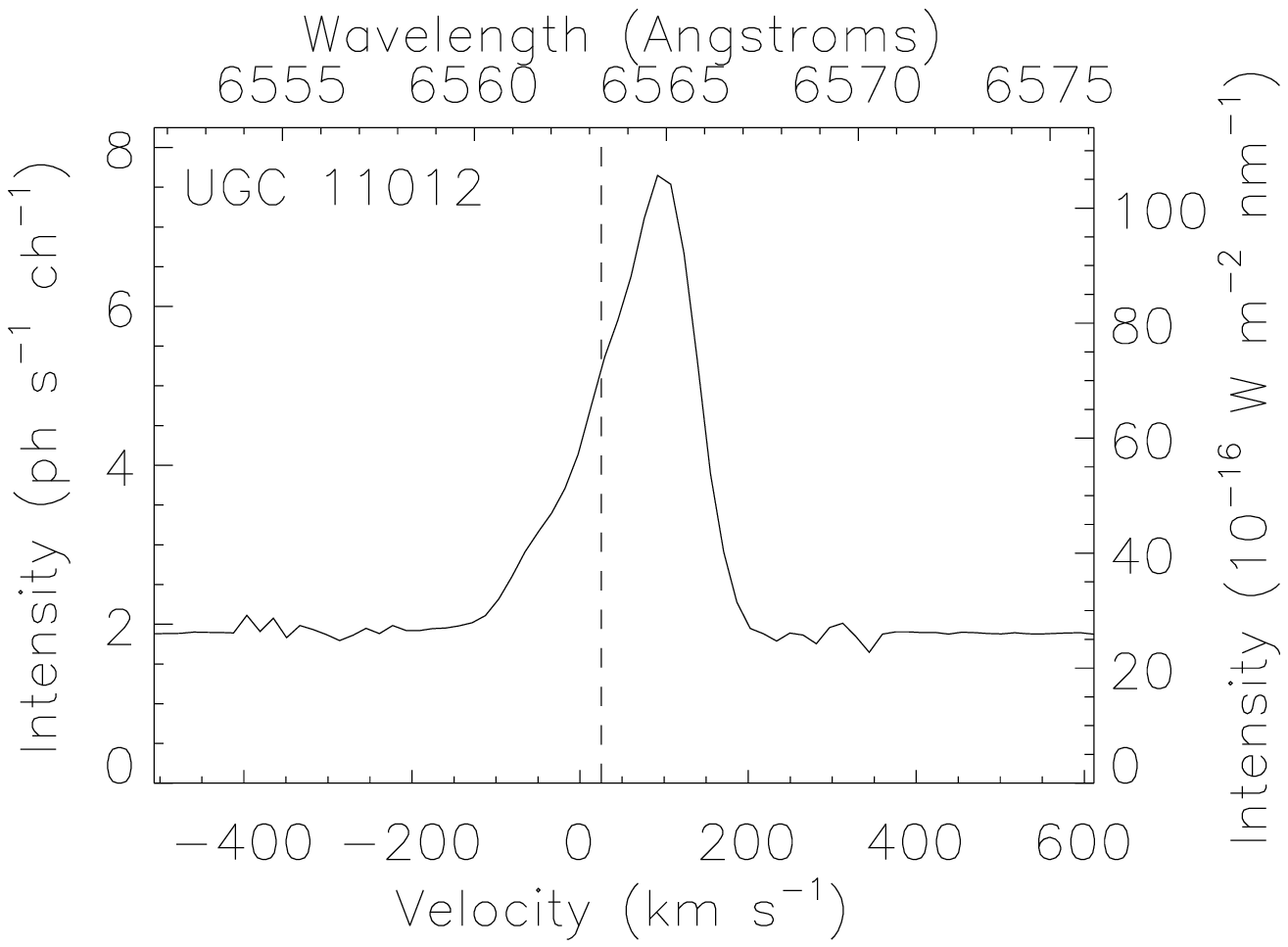}
\includegraphics[width=3.5cm]{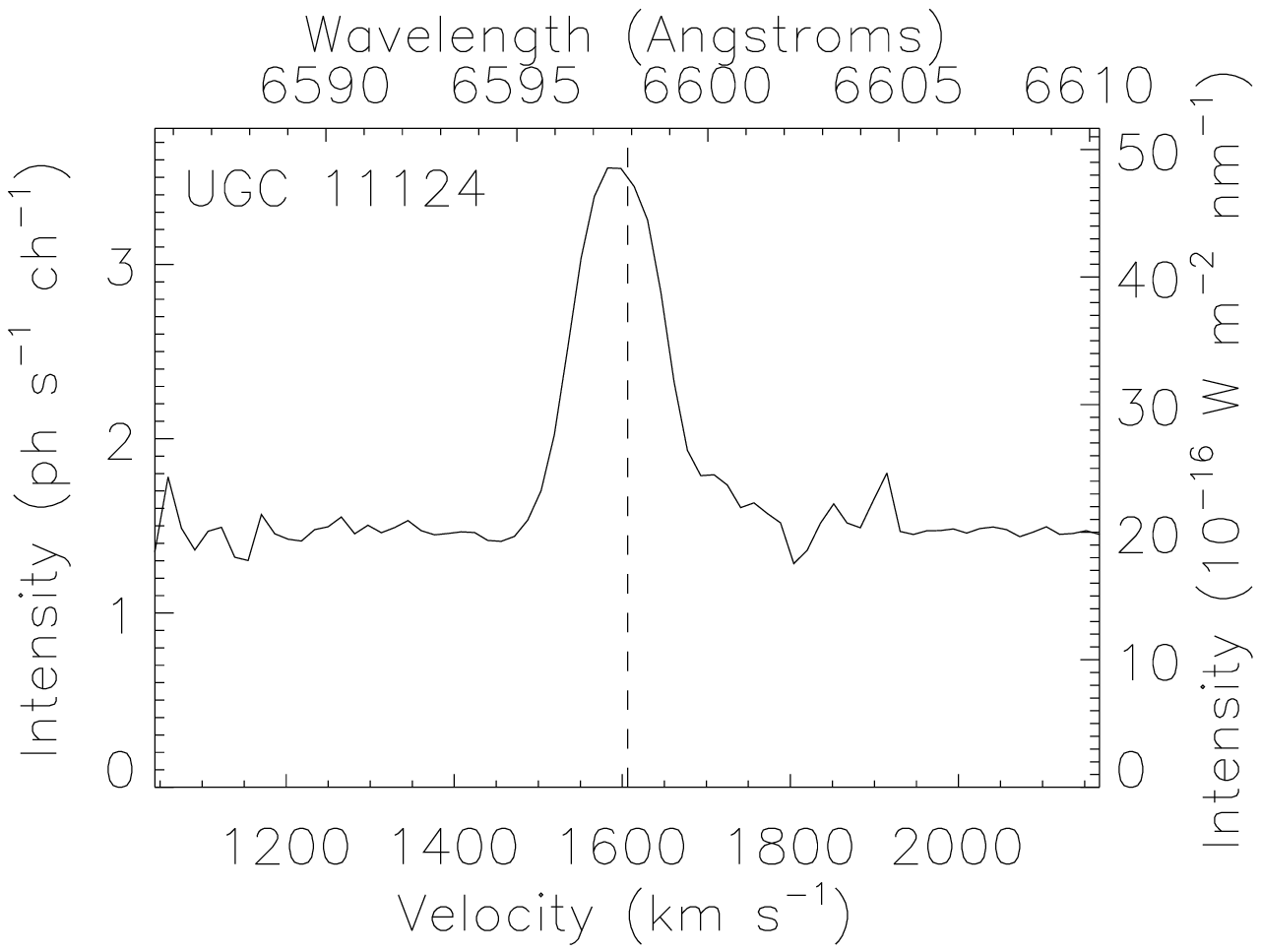}
\includegraphics[width=3.5cm]{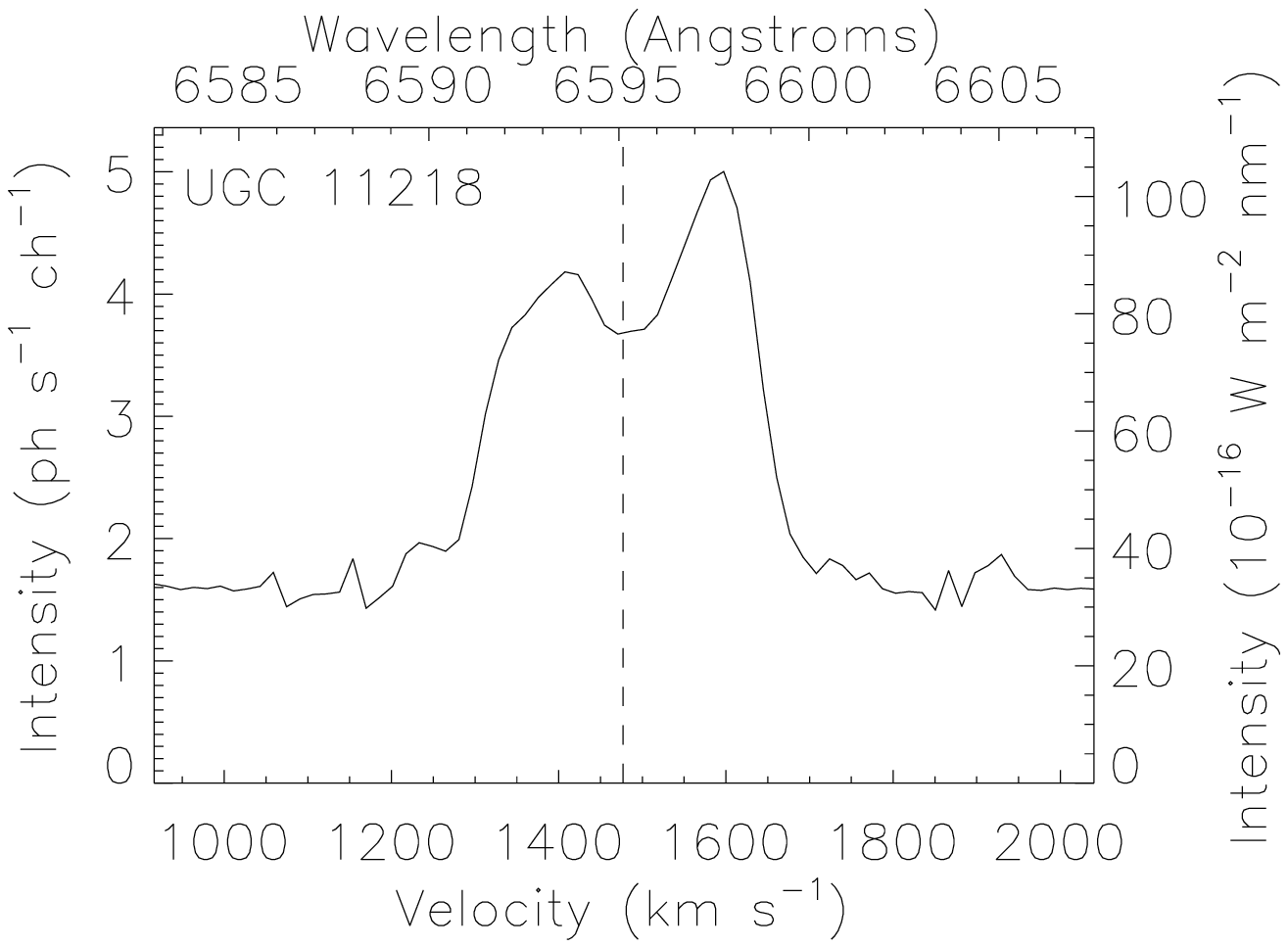}
\includegraphics[width=3.5cm]{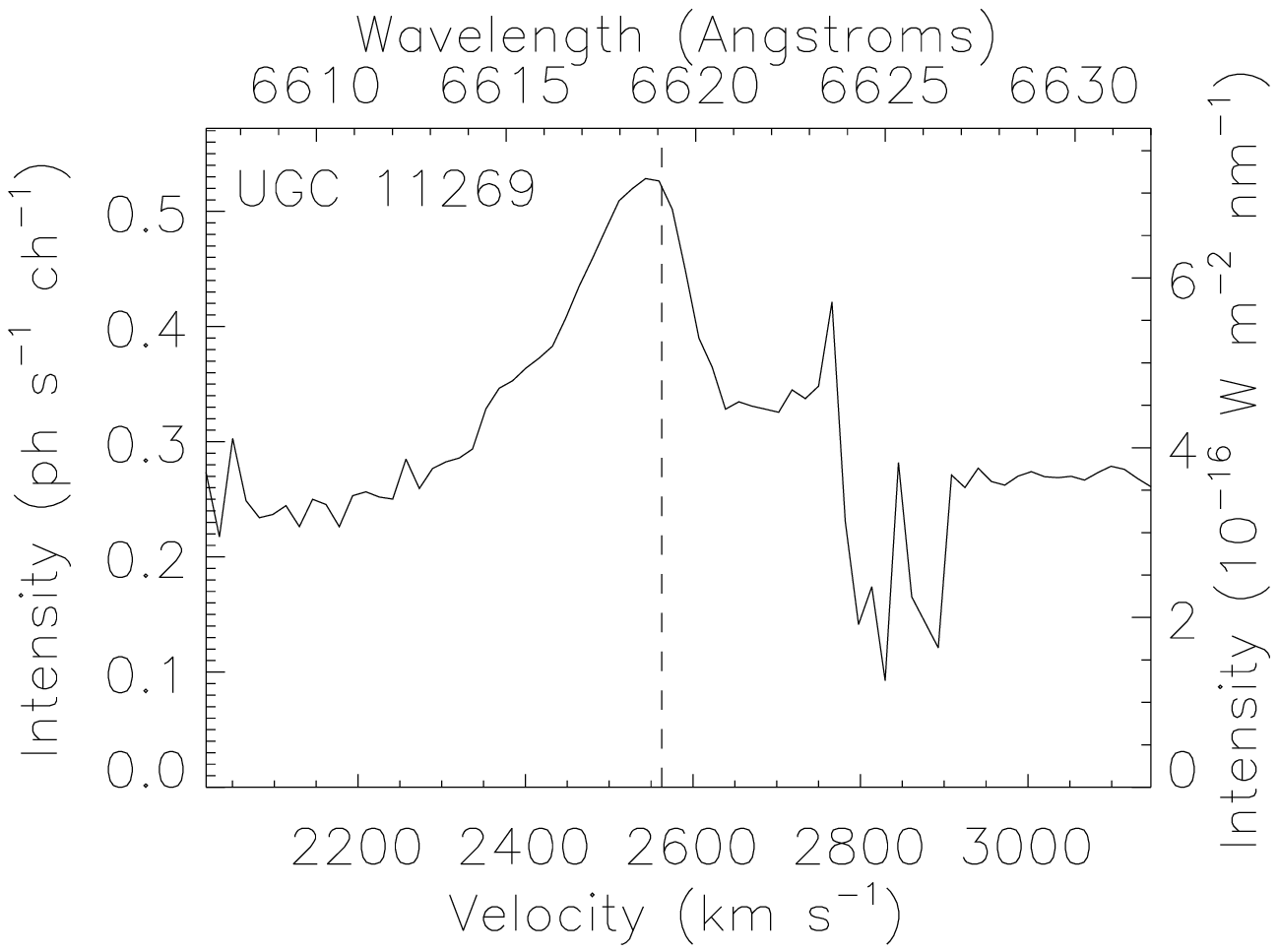}
\includegraphics[width=3.5cm]{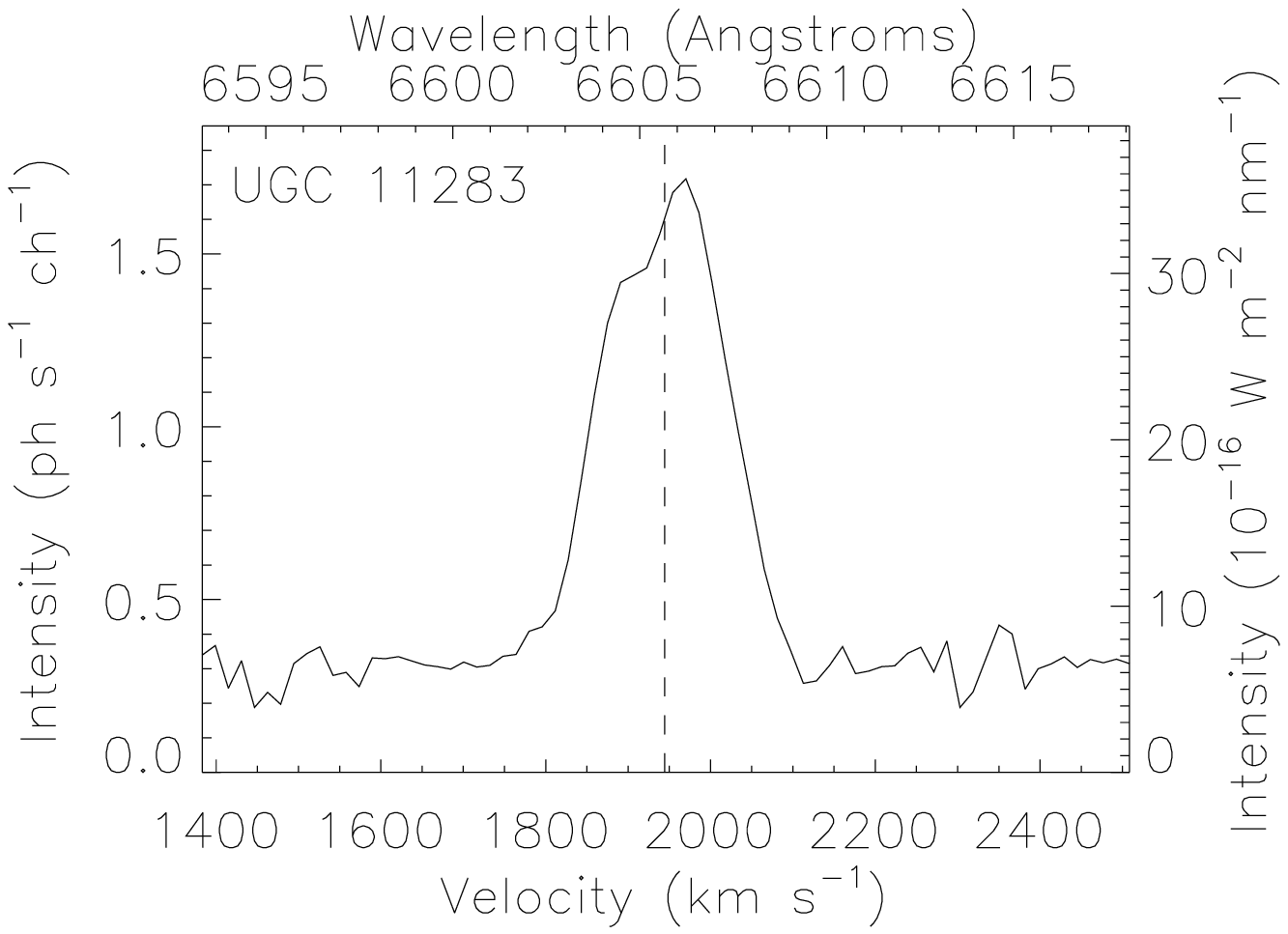}
\includegraphics[width=3.5cm]{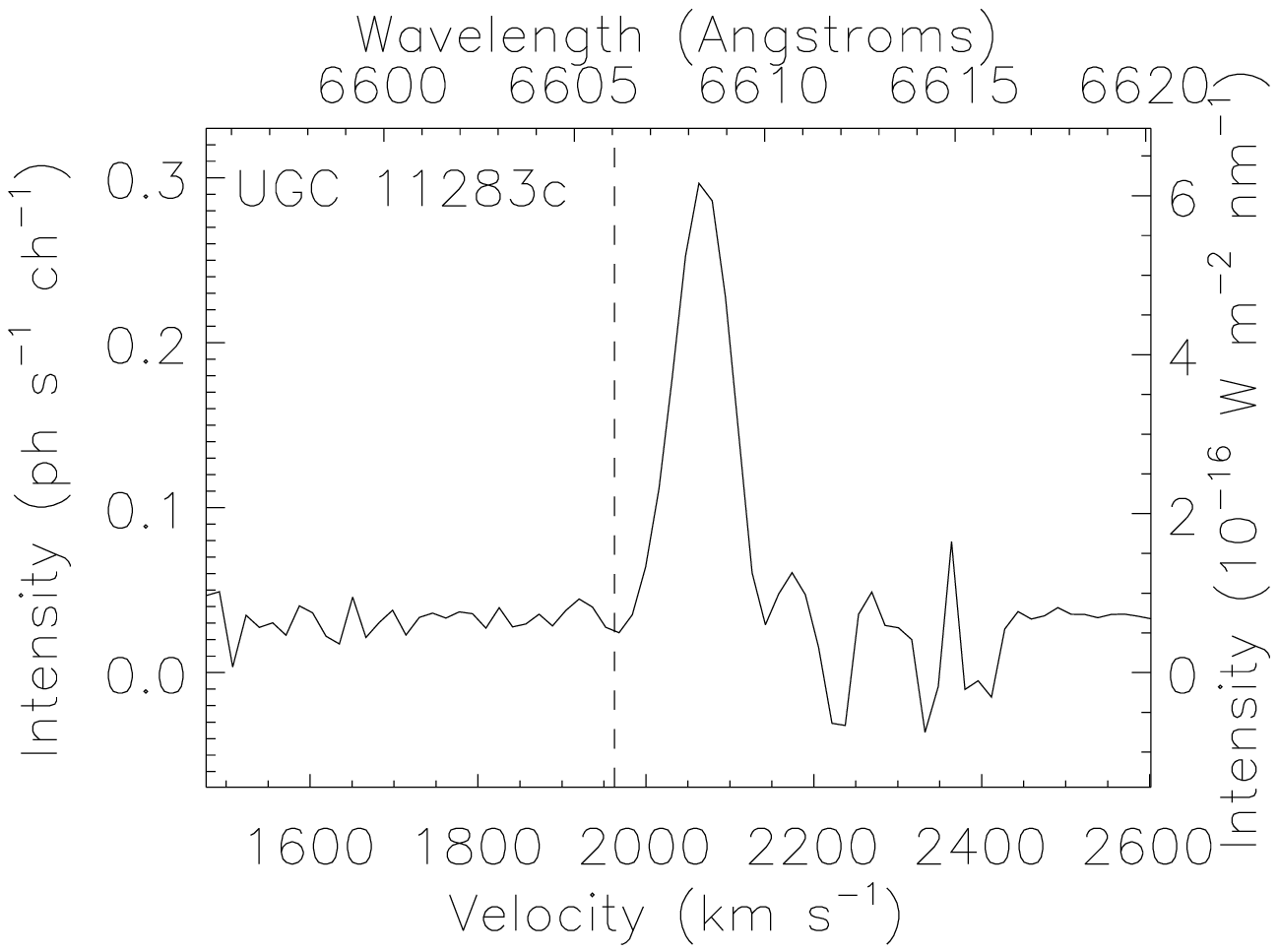}
\includegraphics[width=3.5cm]{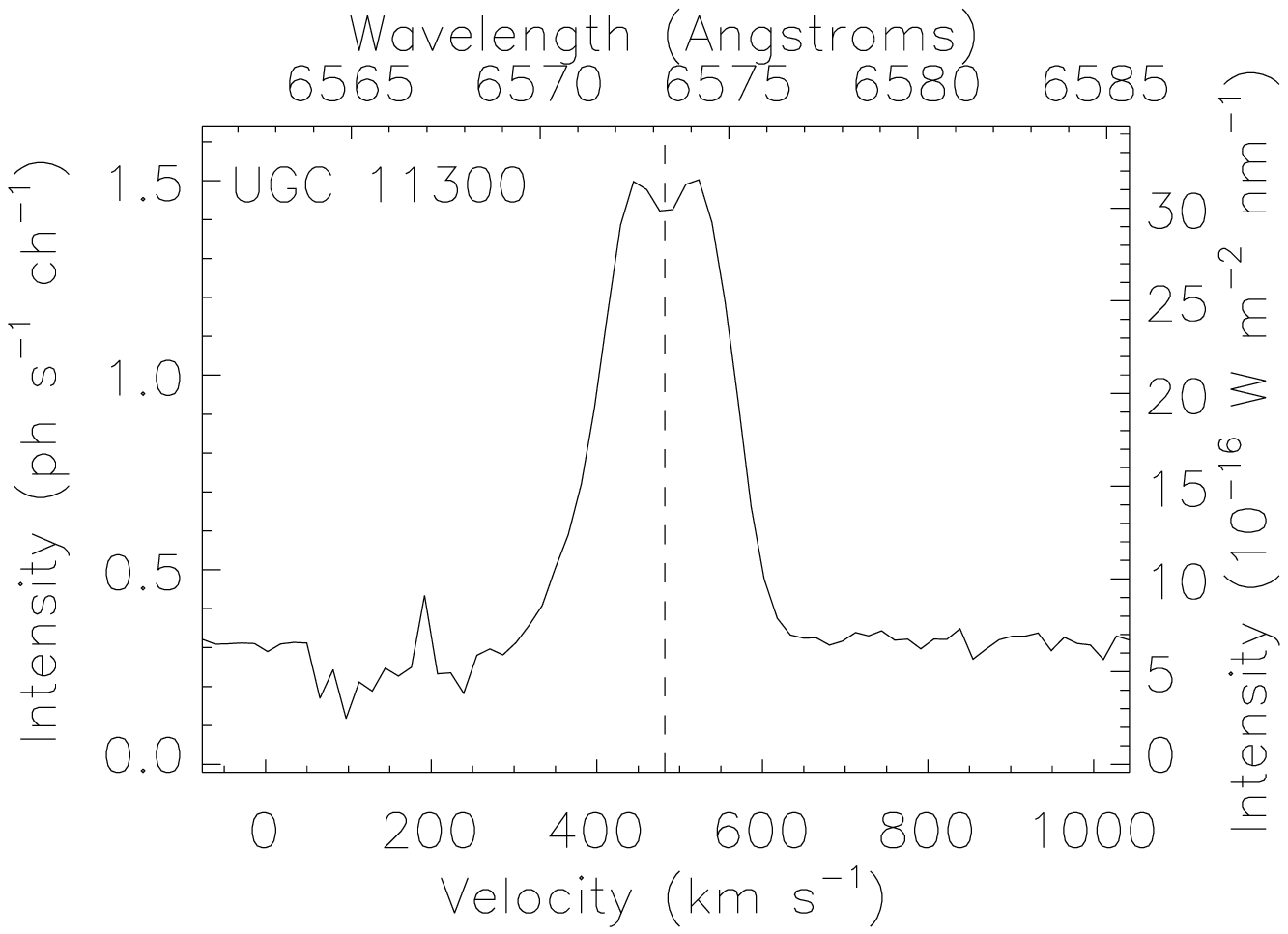}
\includegraphics[width=3.5cm]{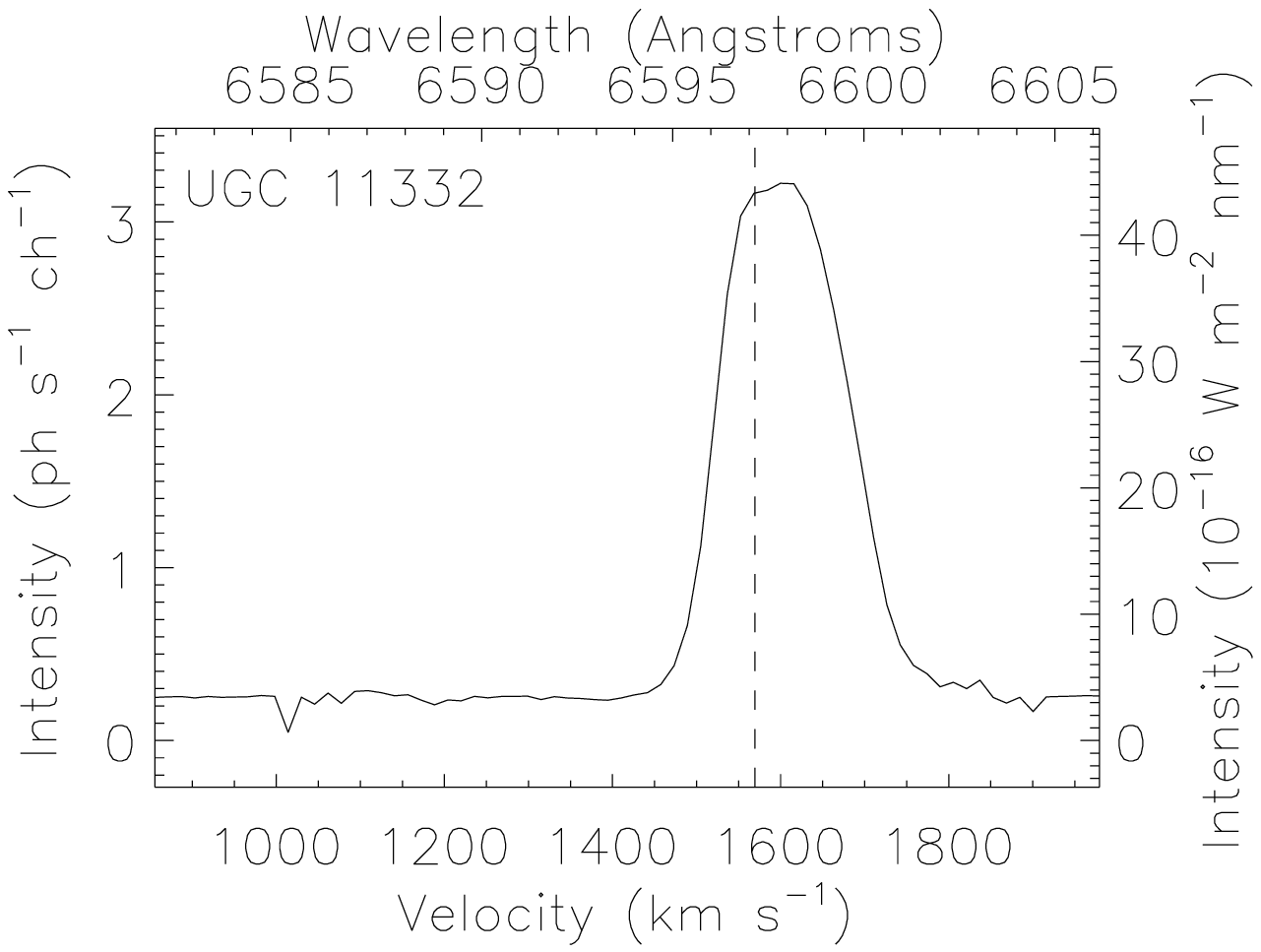}
\includegraphics[width=3.5cm]{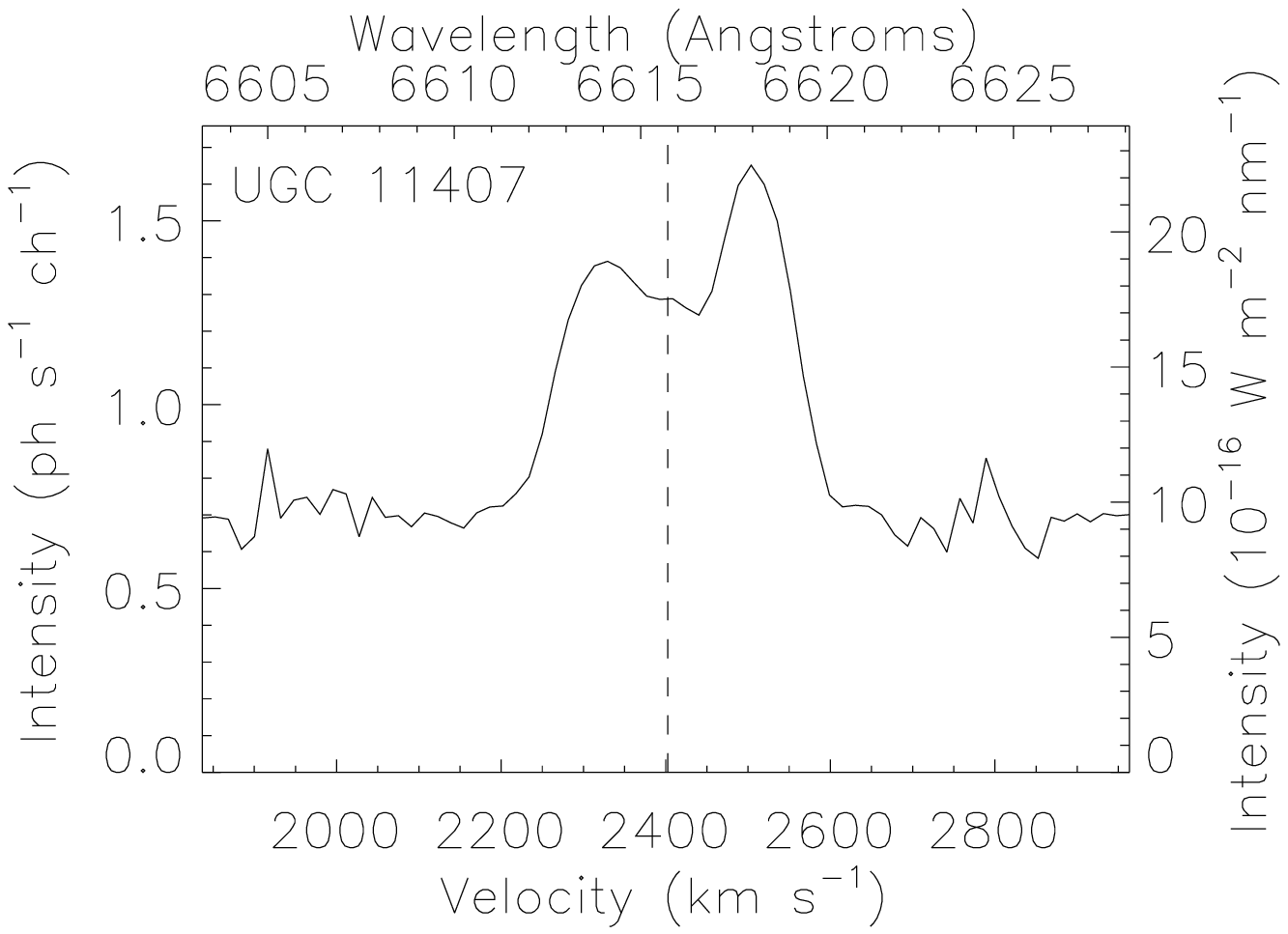}
\includegraphics[width=3.5cm]{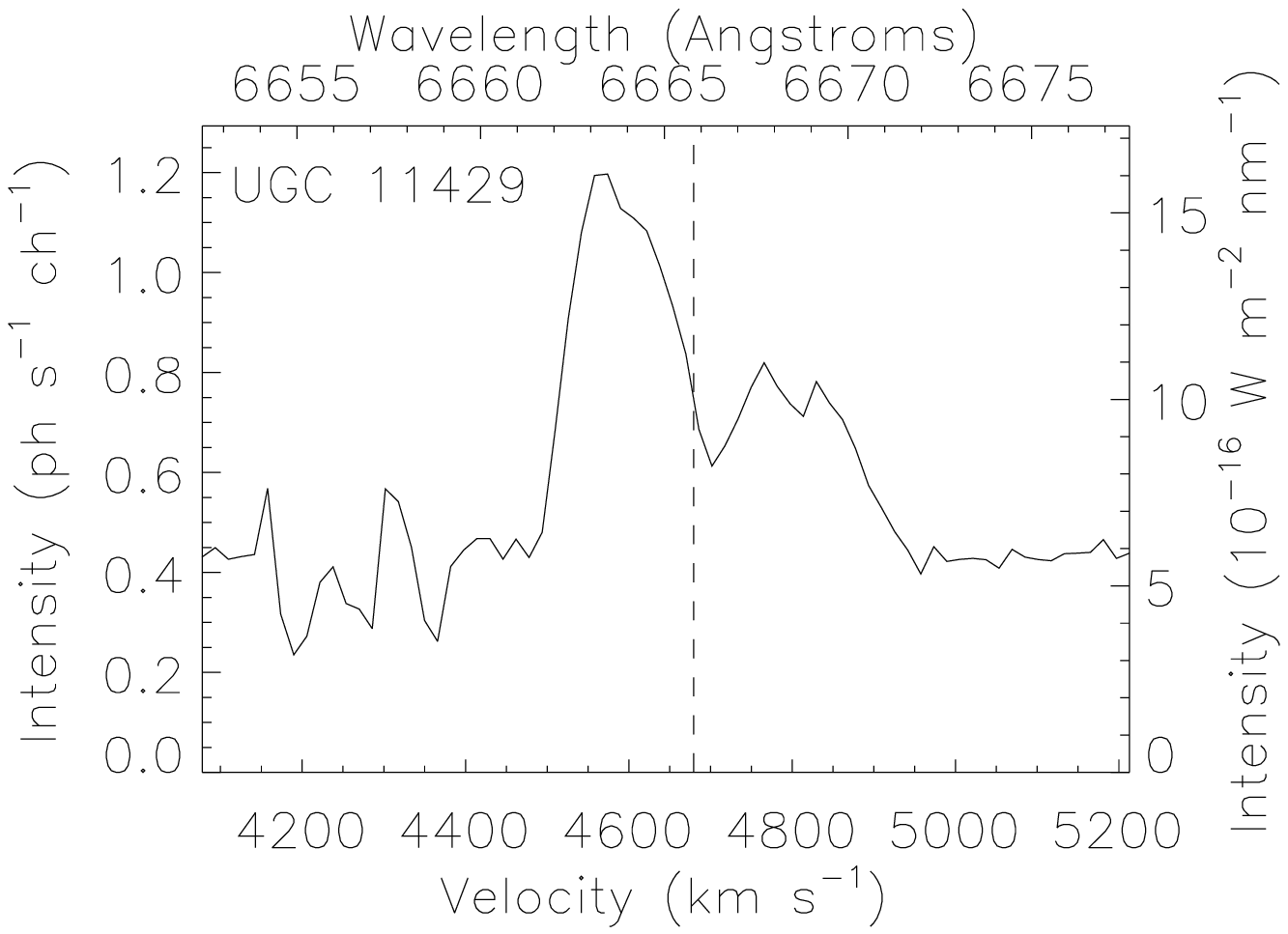}
\includegraphics[width=3.5cm]{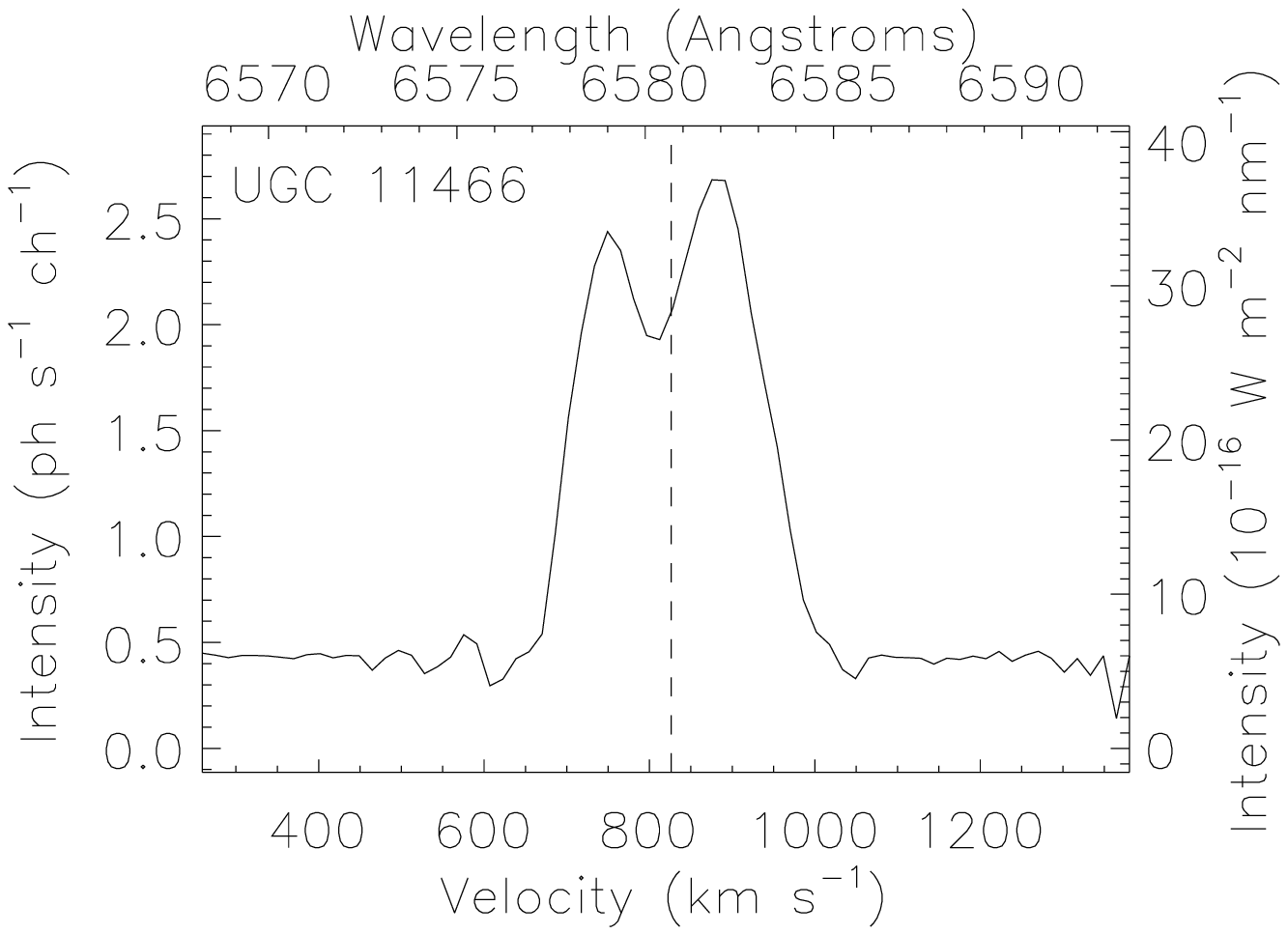}
\includegraphics[width=3.5cm]{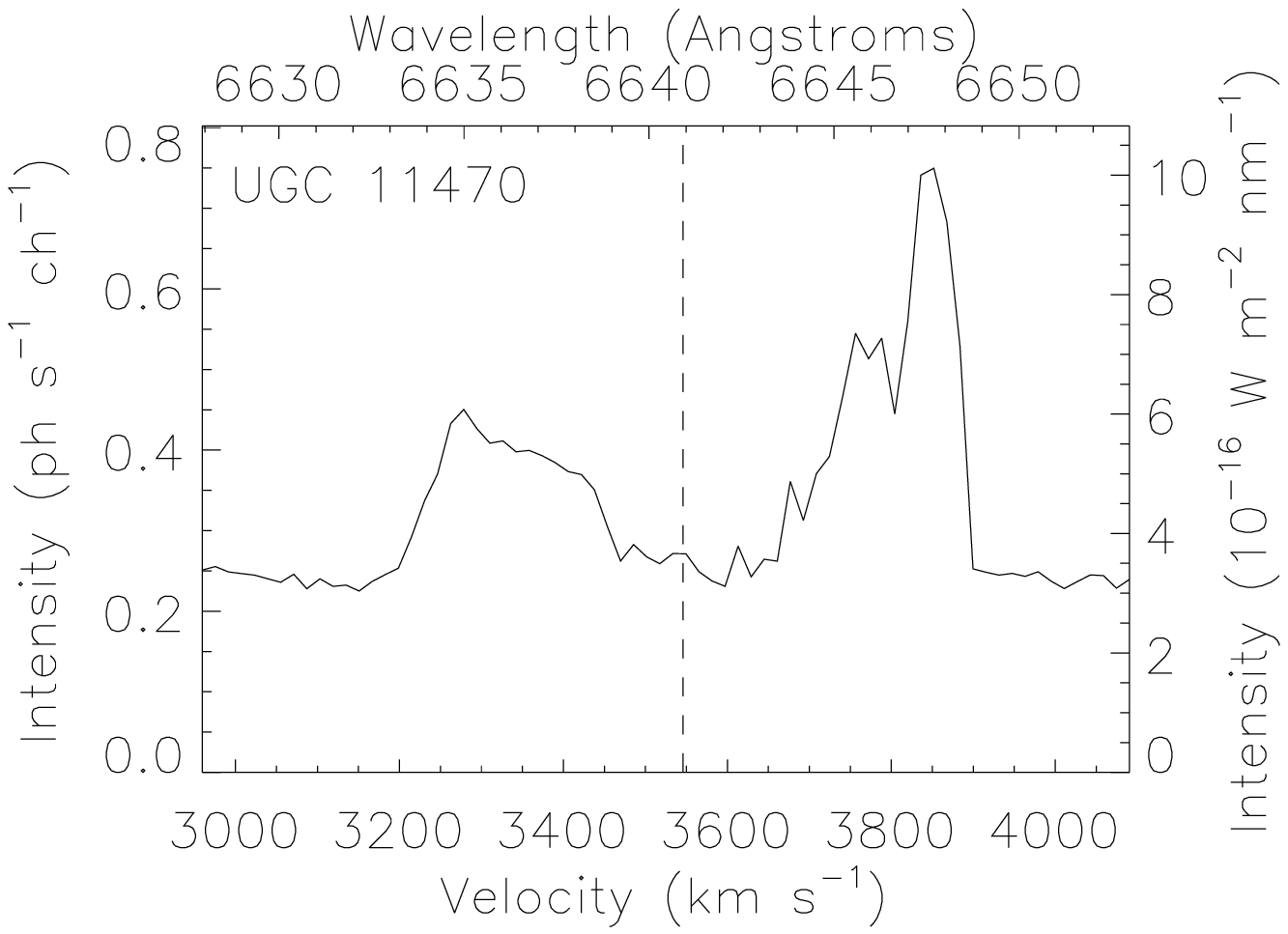}
\includegraphics[width=3.5cm]{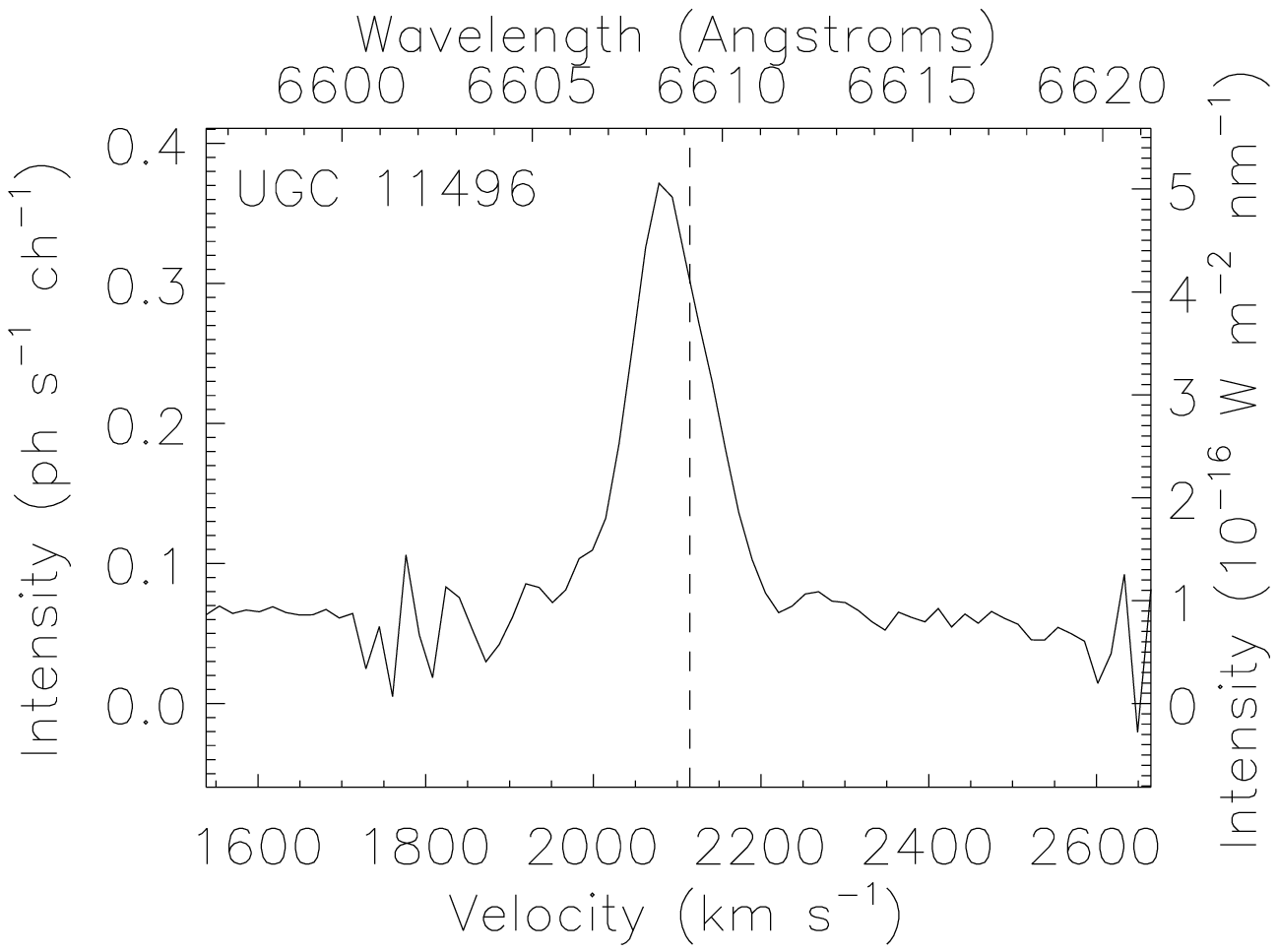}
\includegraphics[width=3.5cm]{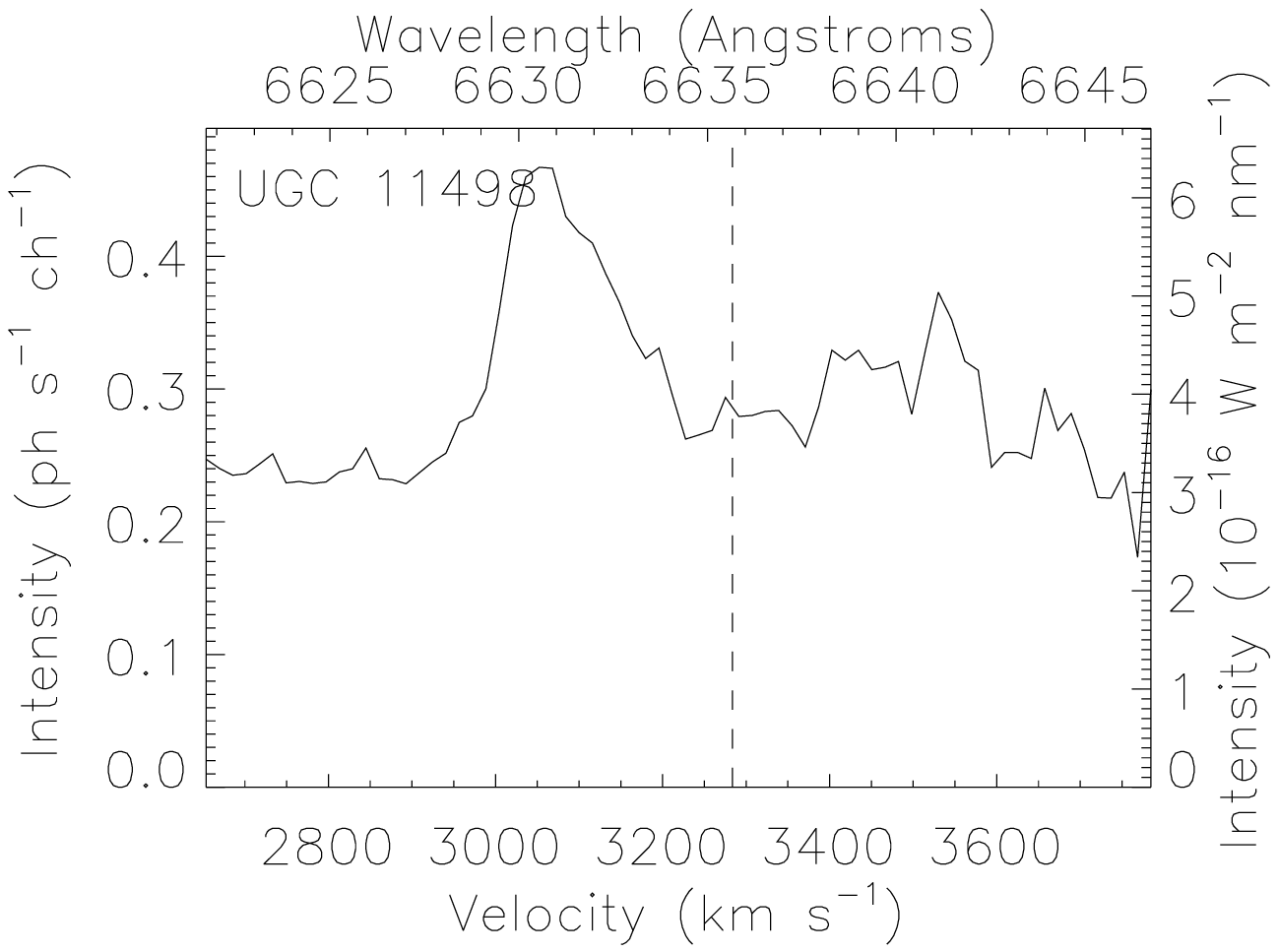}
\includegraphics[width=3.5cm]{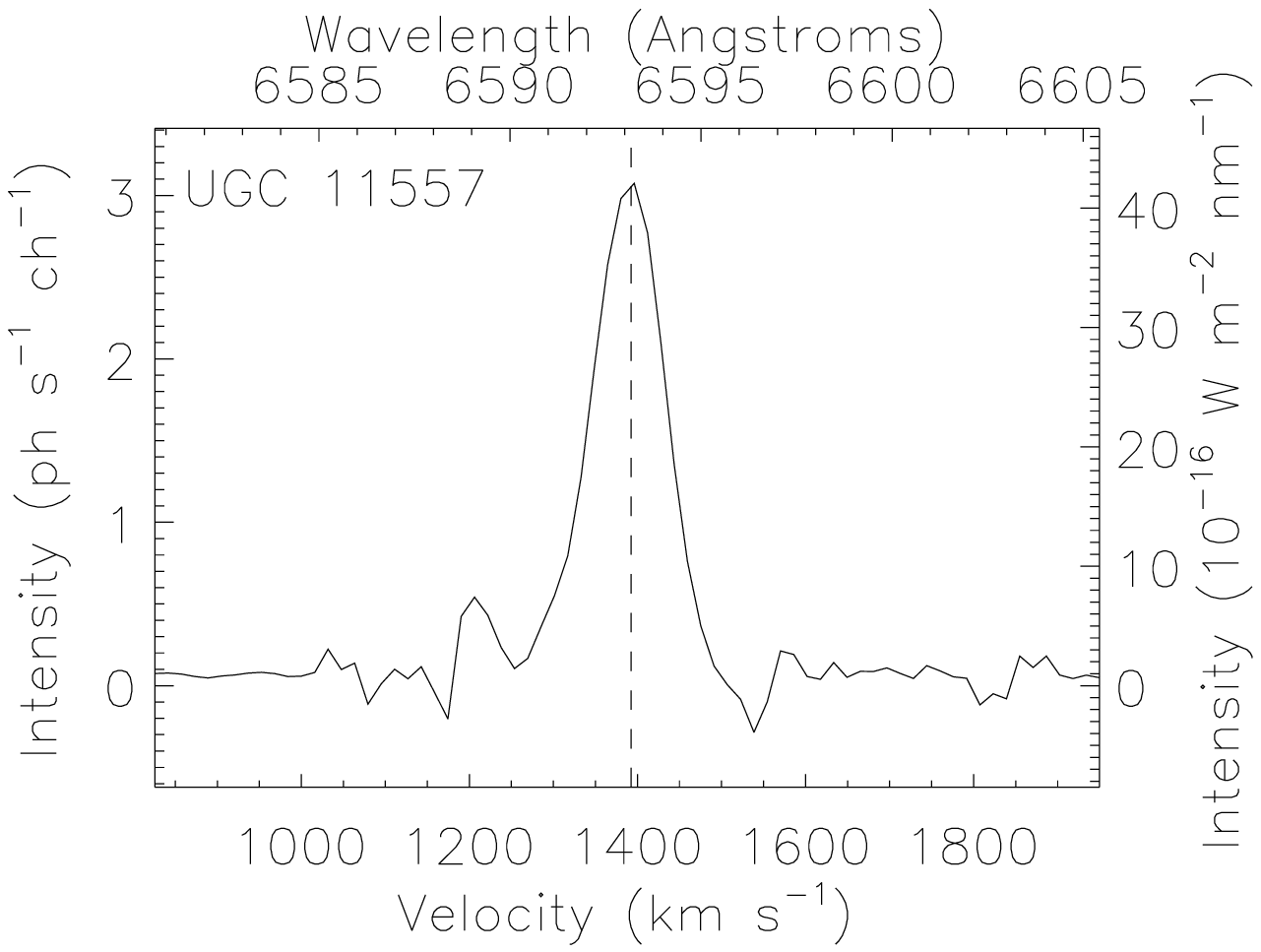}
\includegraphics[width=3.5cm]{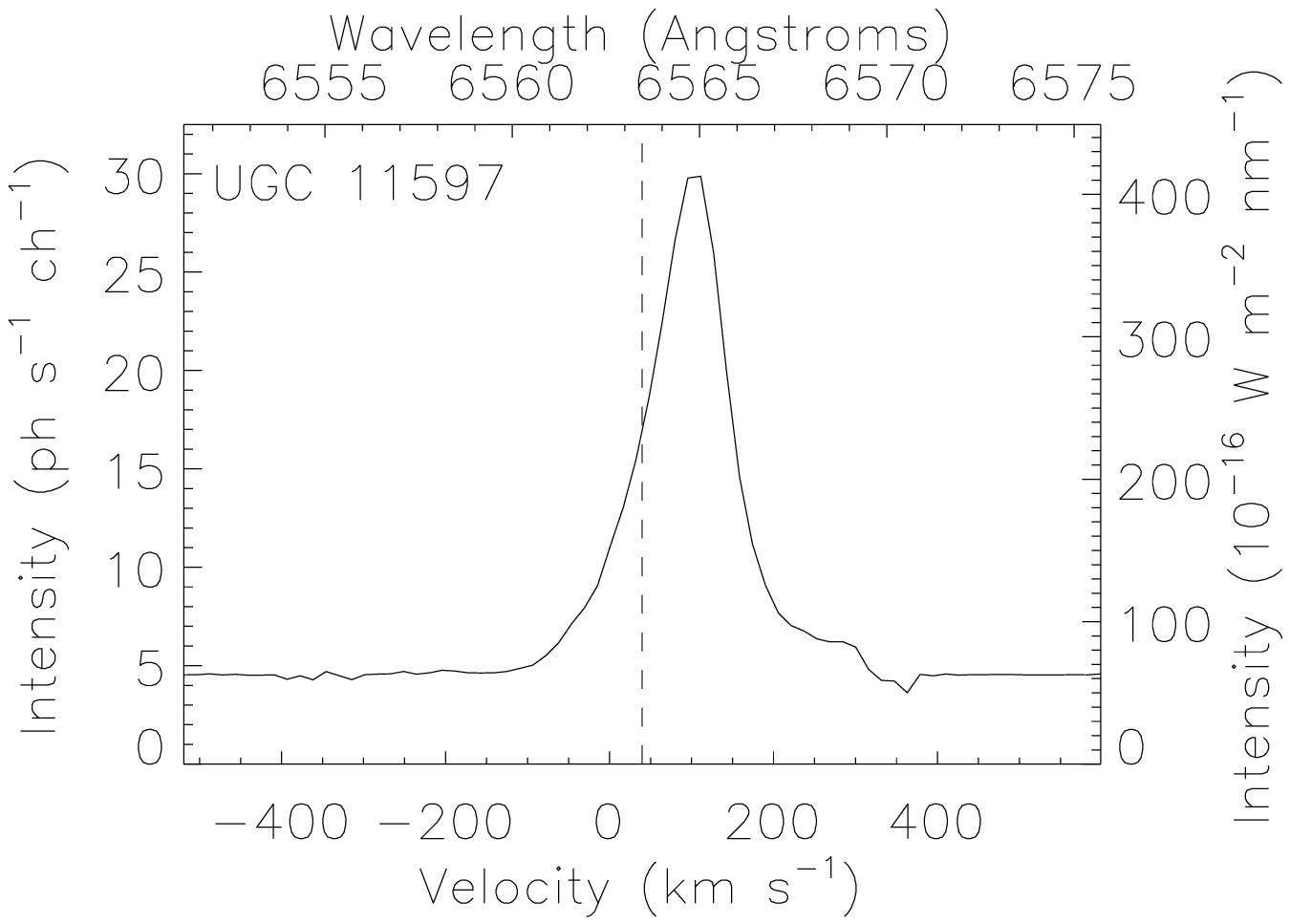}
\includegraphics[width=3.5cm]{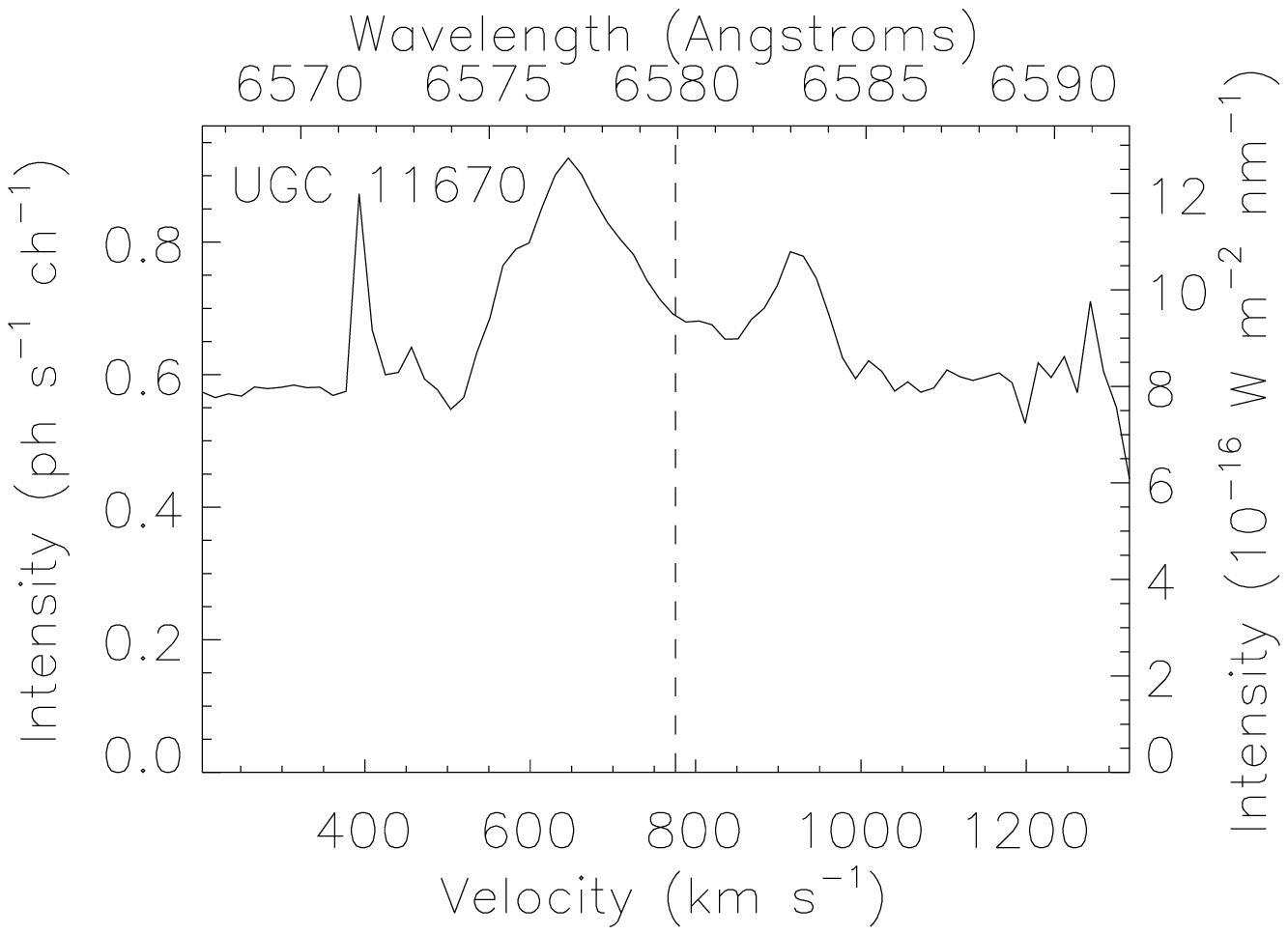}
\includegraphics[width=3.5cm]{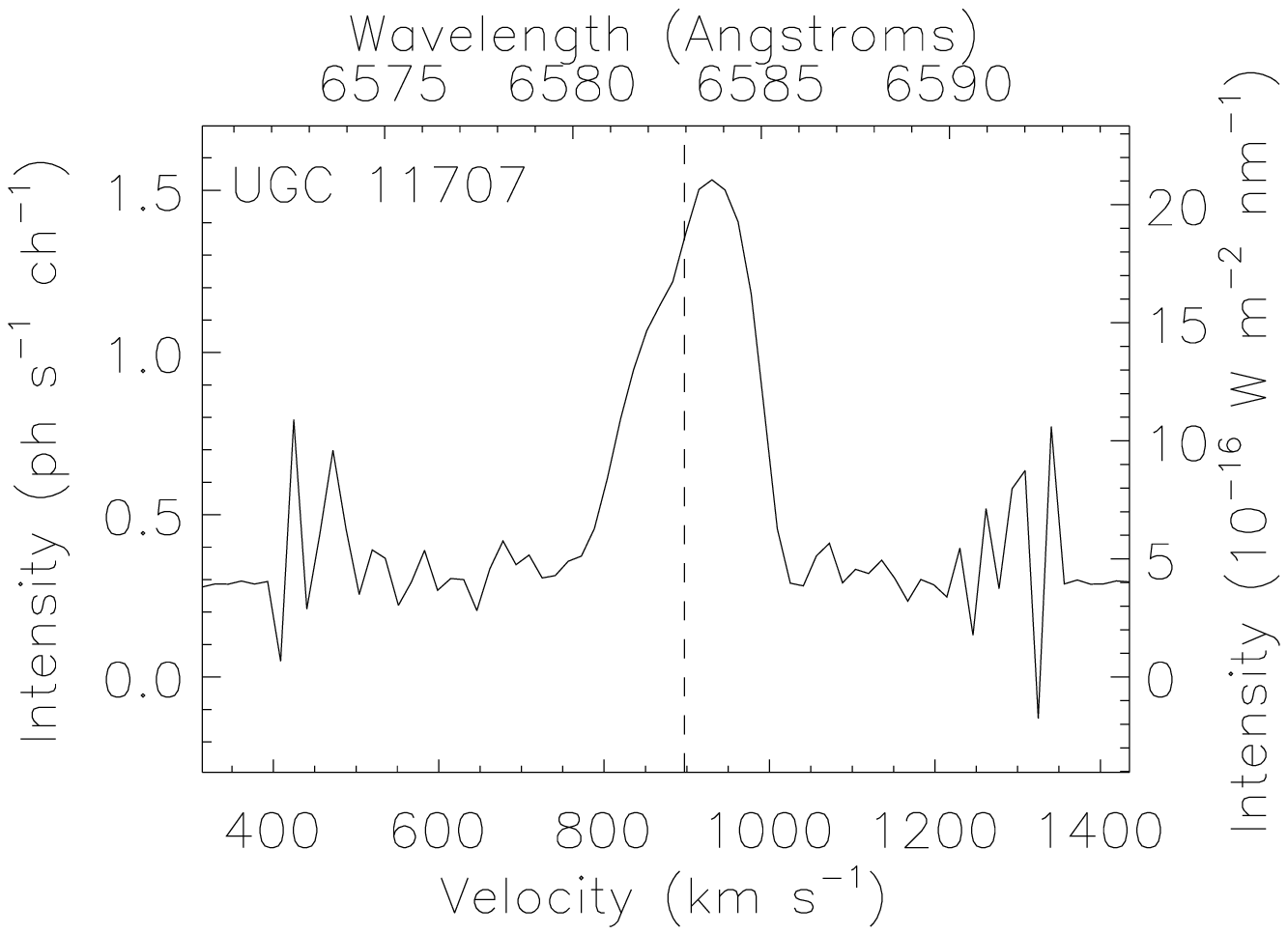}
\includegraphics[width=3.5cm]{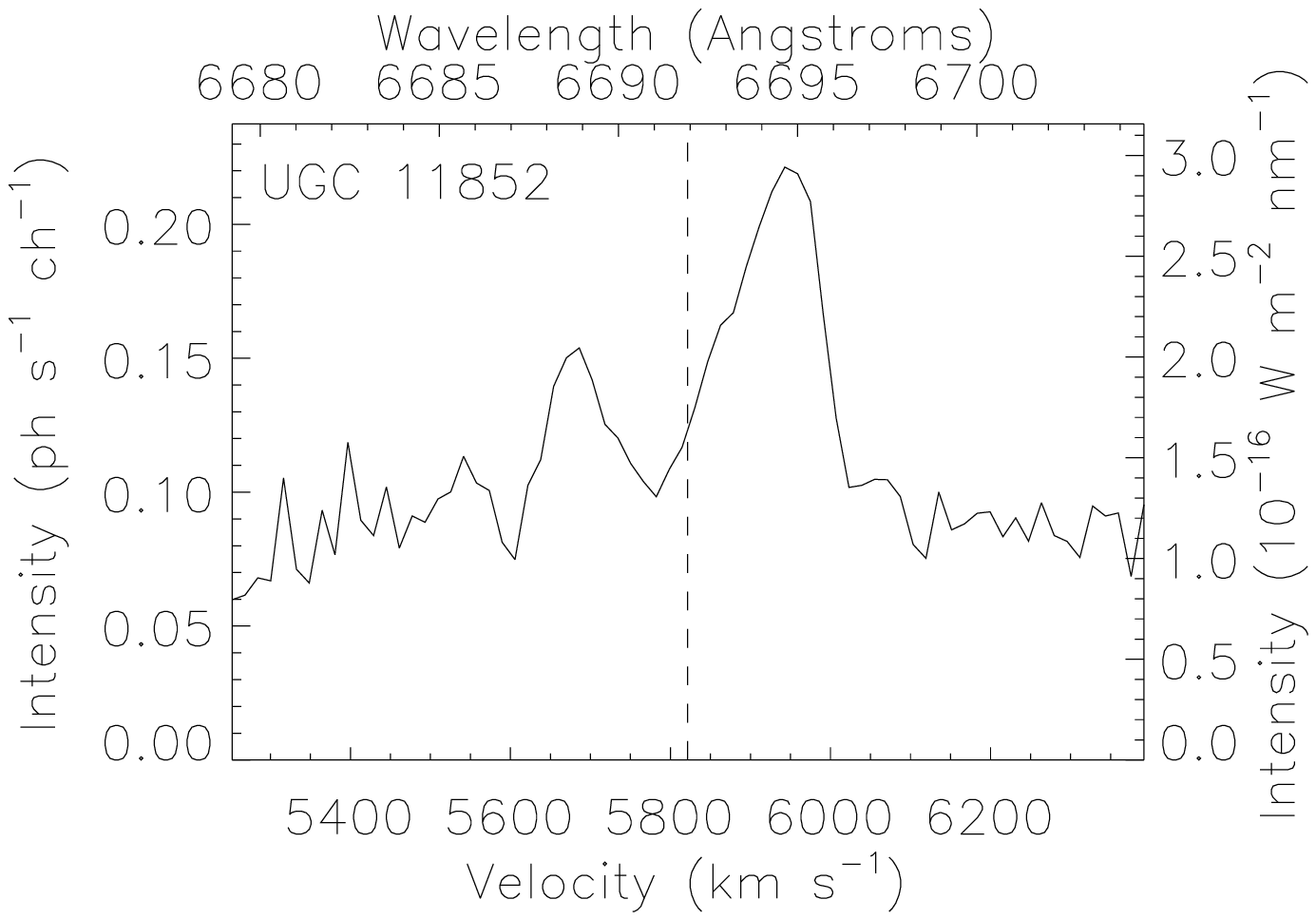}
\includegraphics[width=3.5cm]{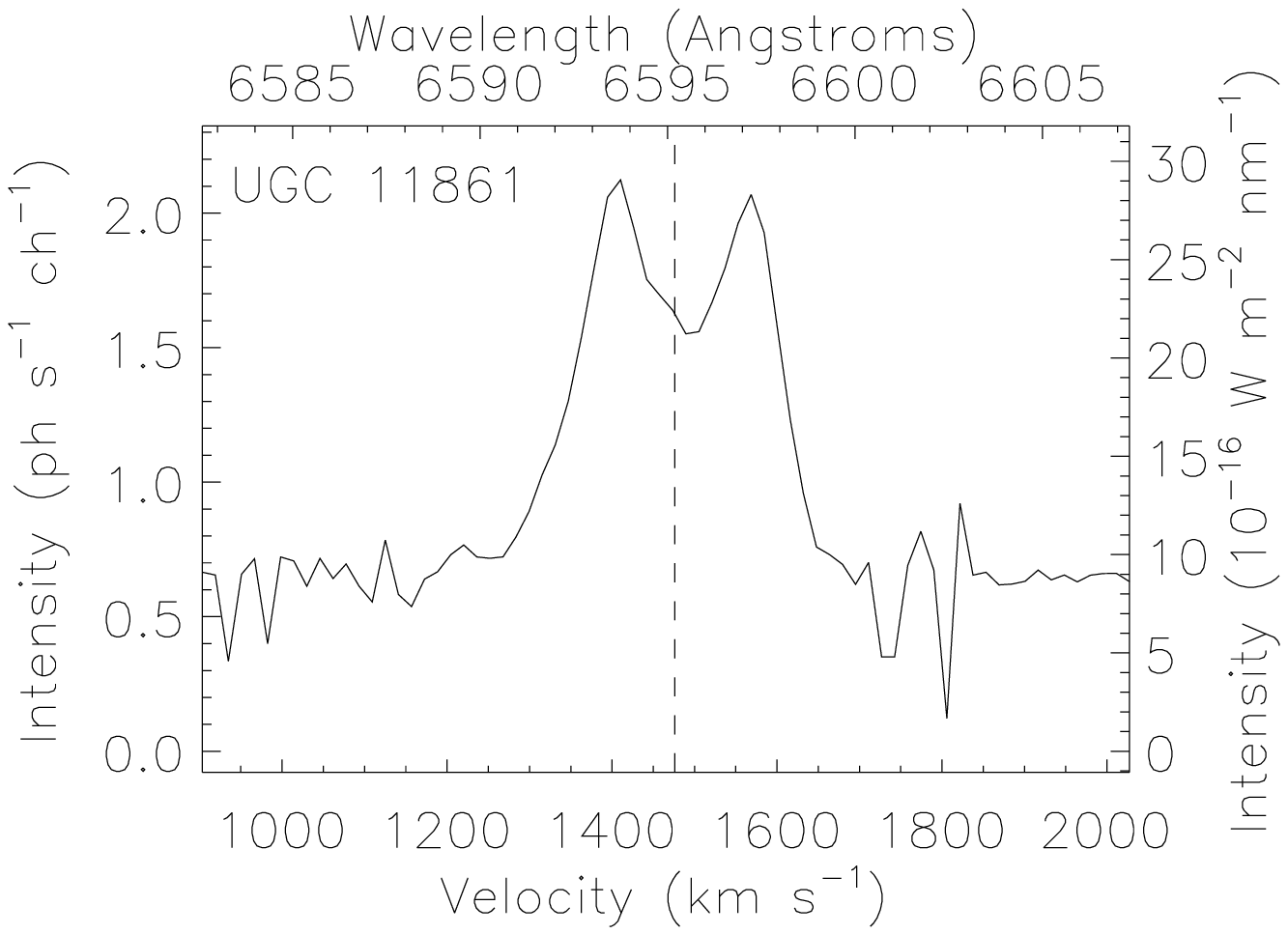}
\includegraphics[width=3.5cm]{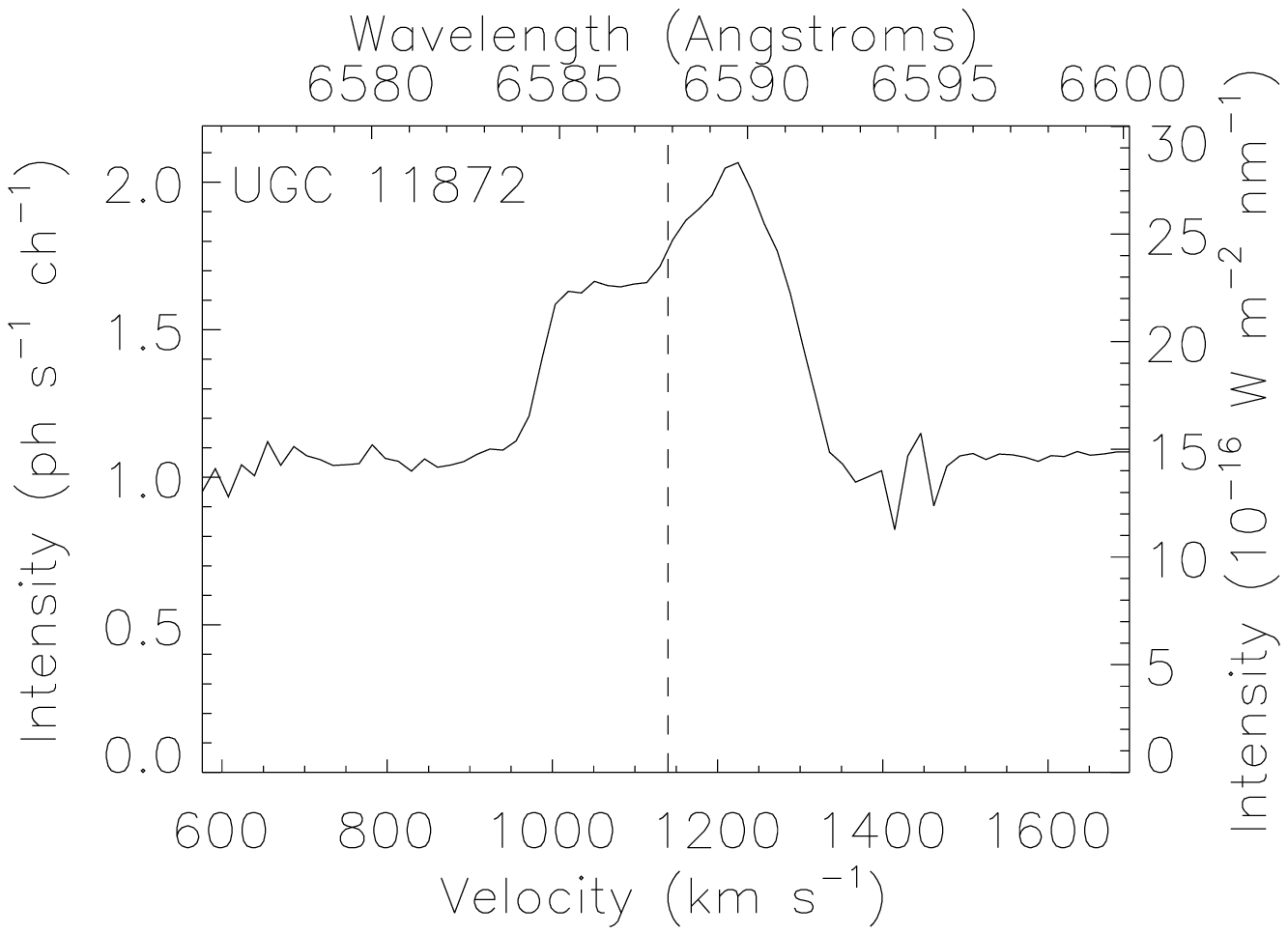}
\includegraphics[width=3.5cm]{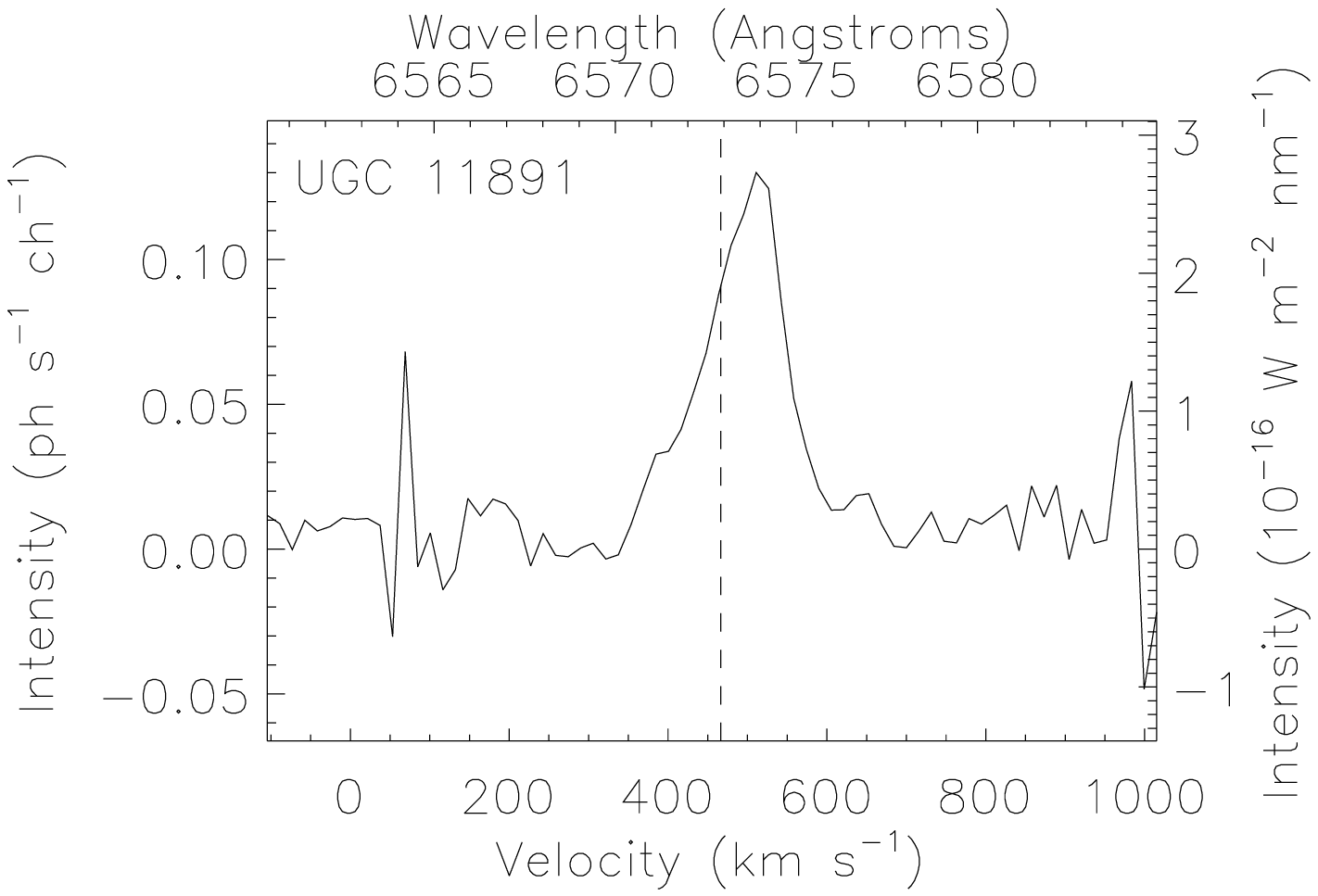}
\includegraphics[width=3.5cm]{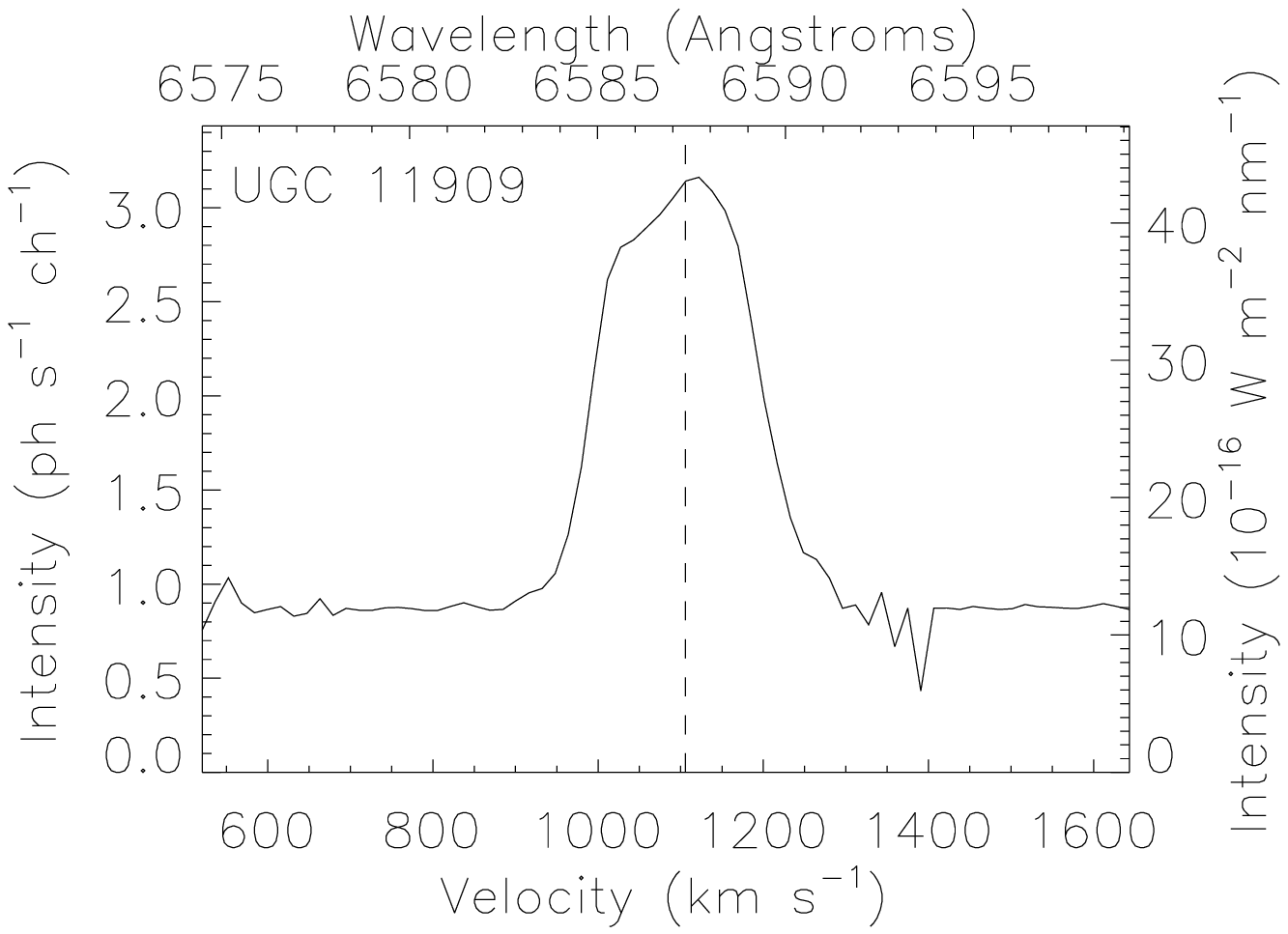}
\includegraphics[width=3.5cm]{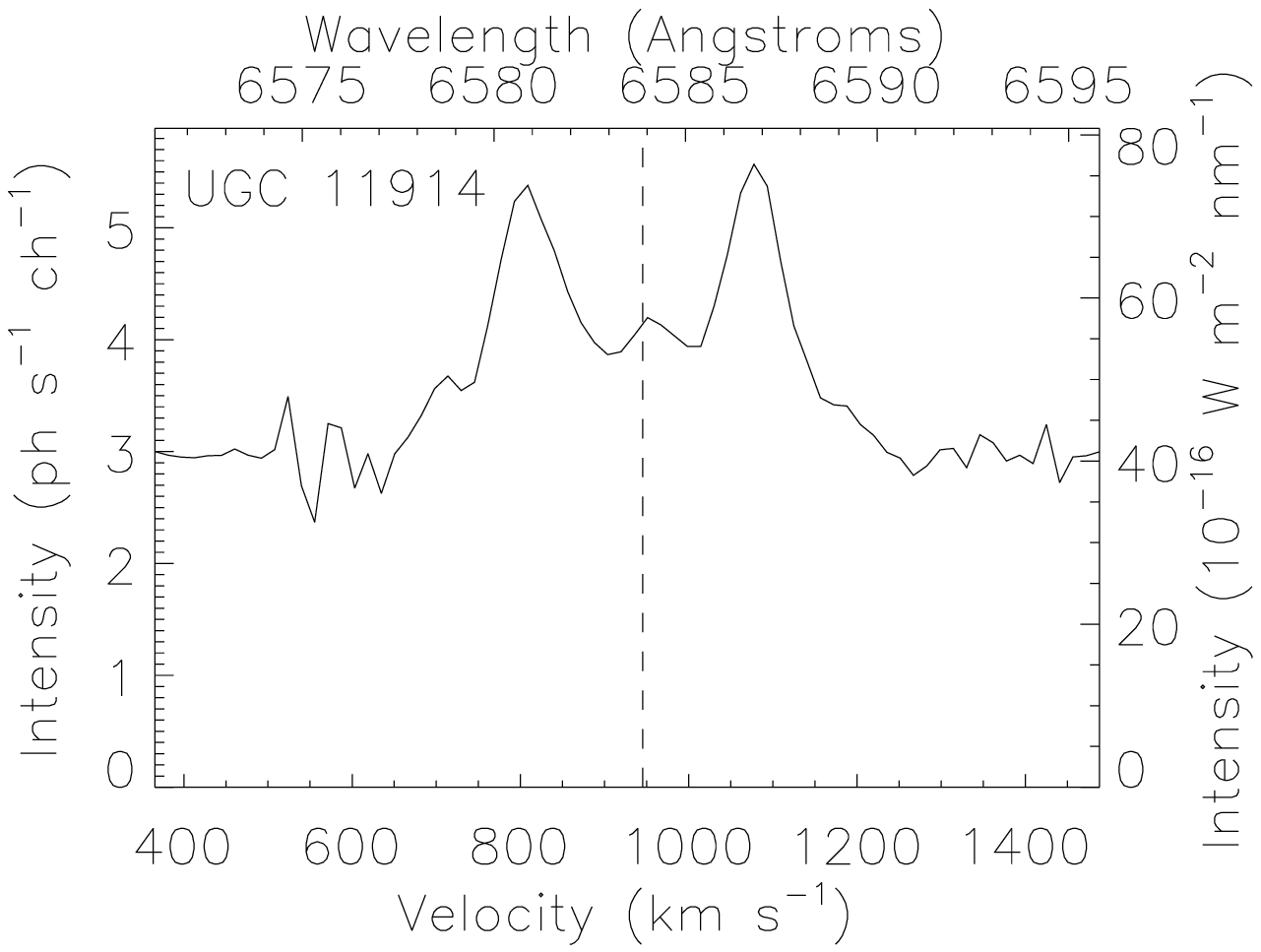}
\includegraphics[width=3.5cm]{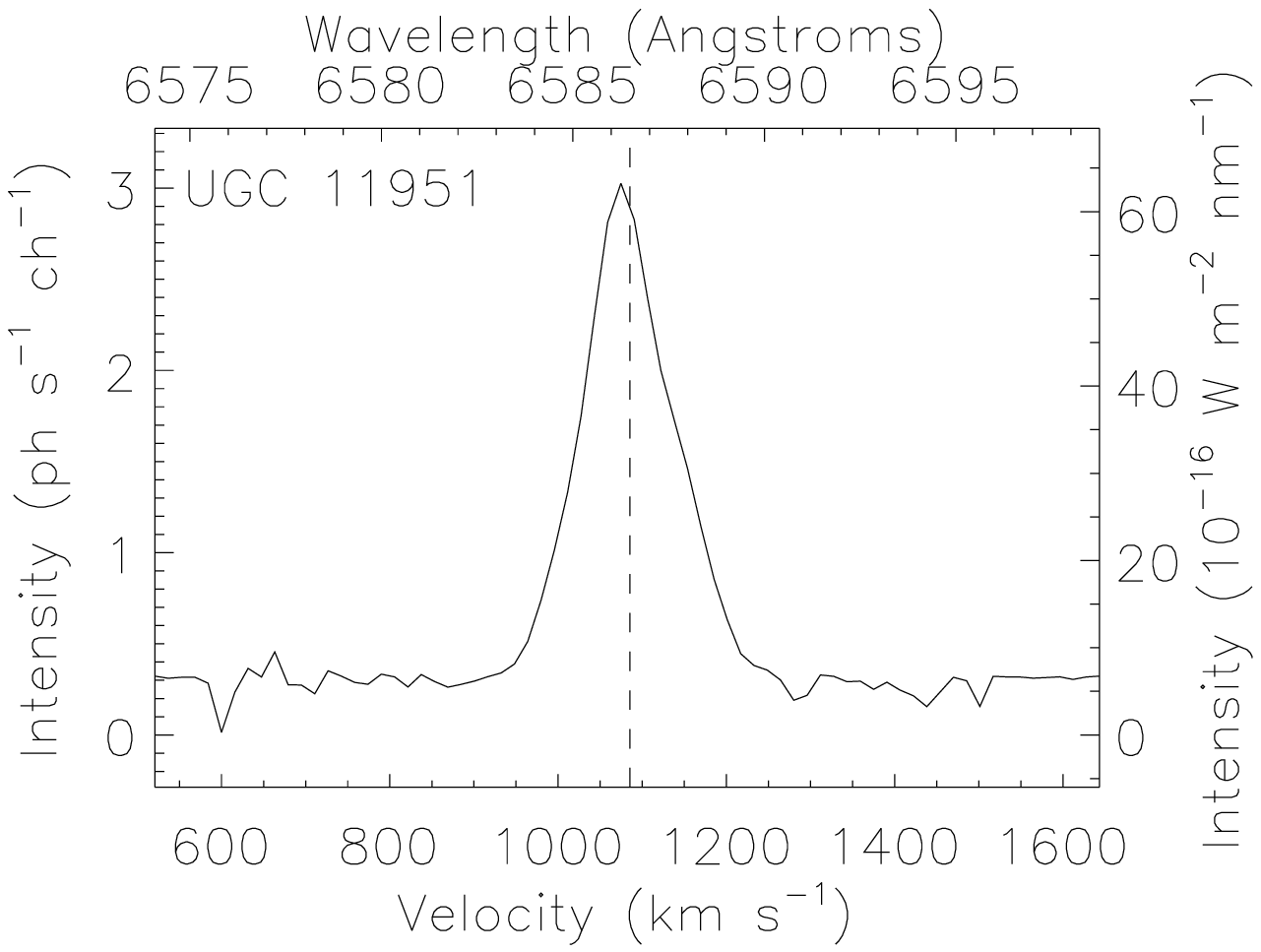}
\includegraphics[width=3.5cm]{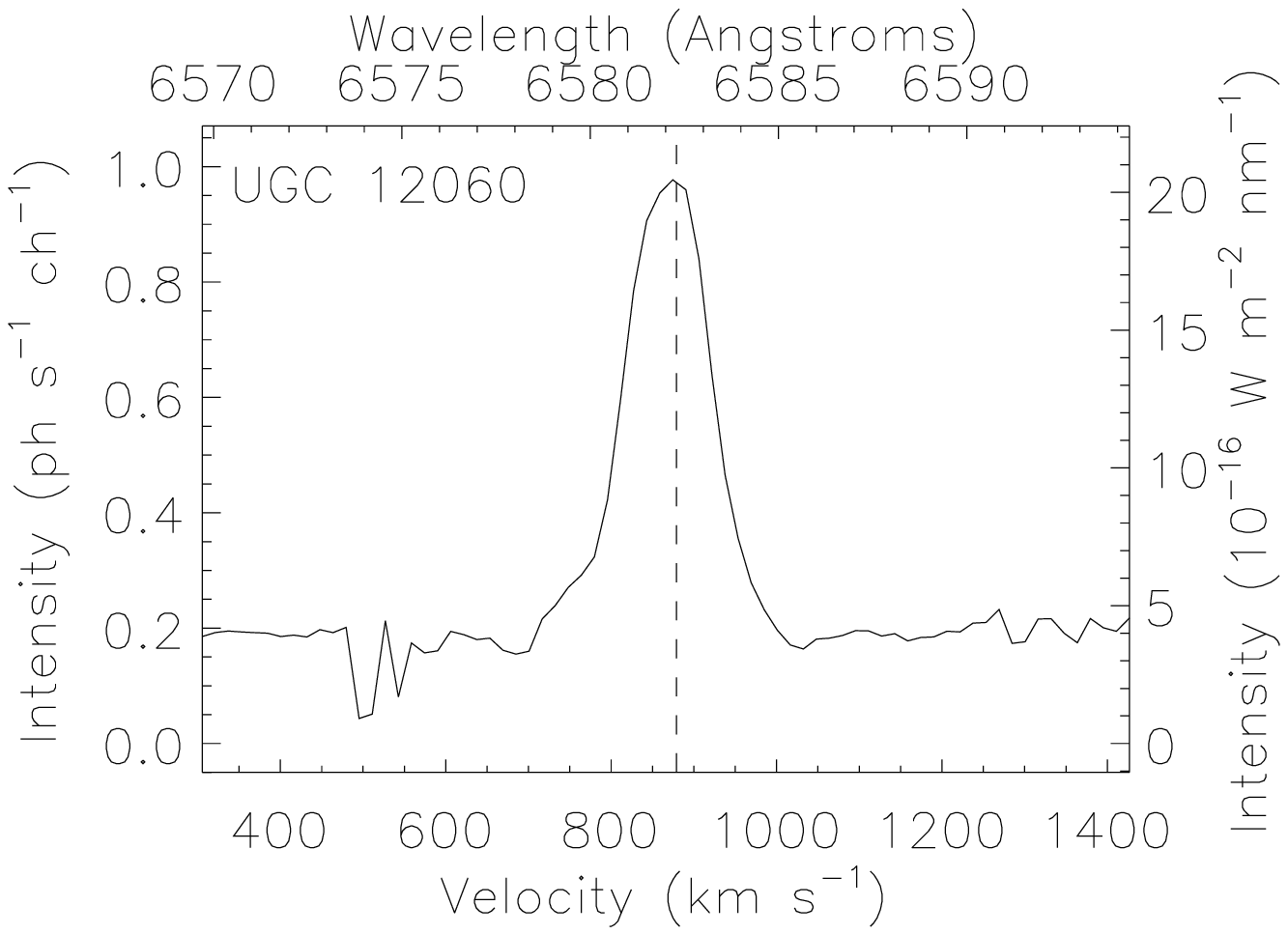}
\includegraphics[width=3.5cm]{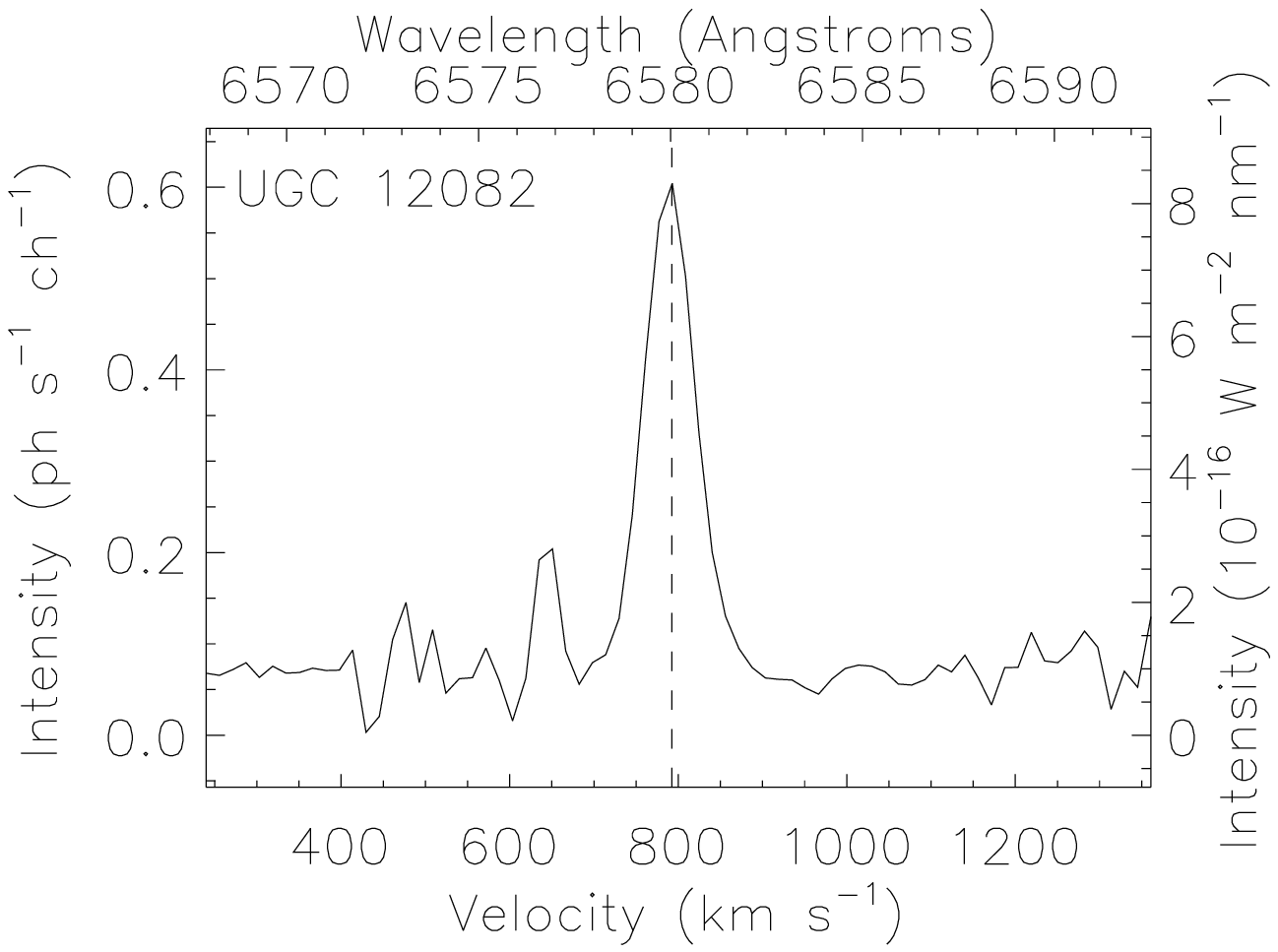}
\includegraphics[width=3.5cm]{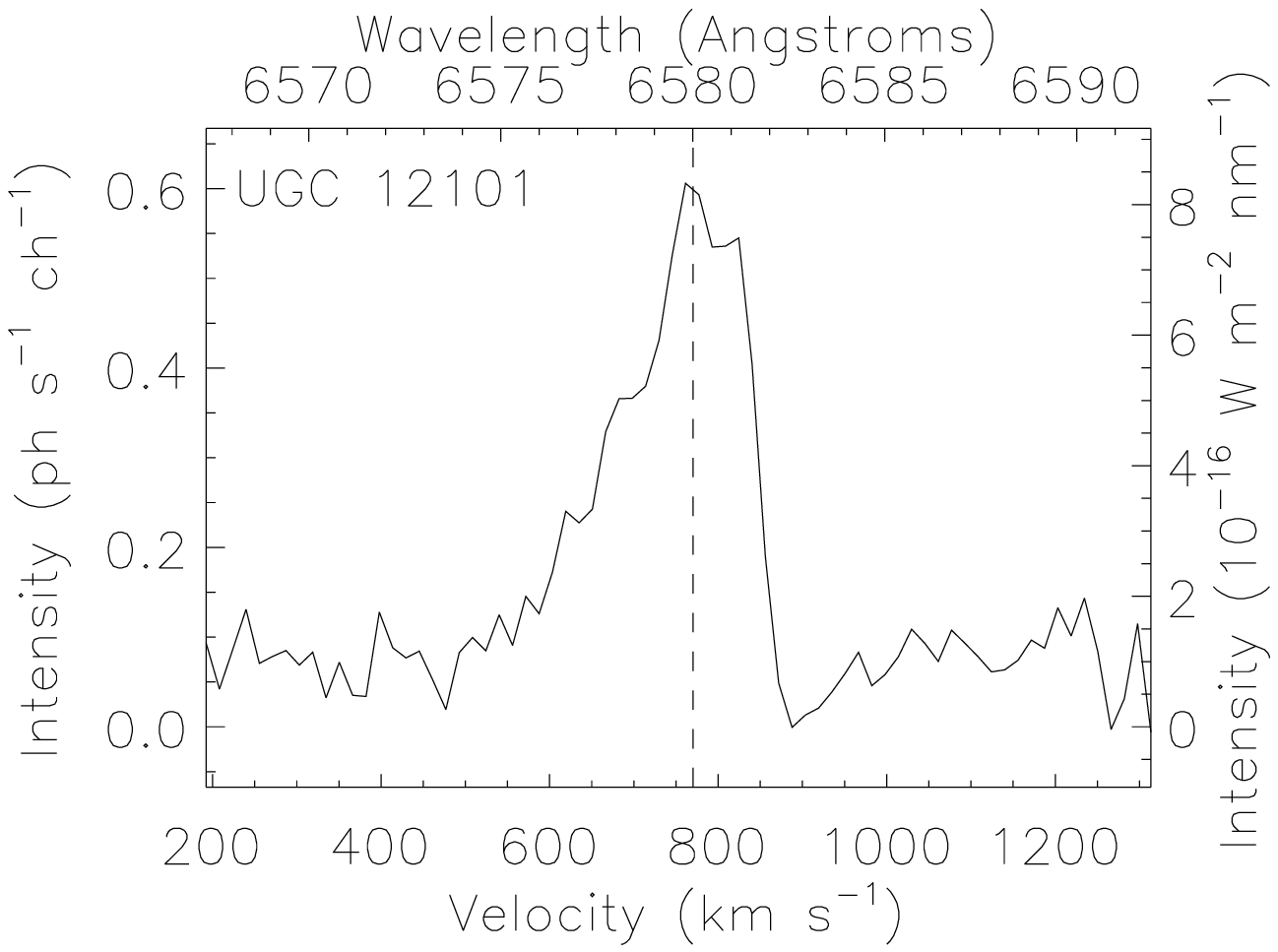}
\includegraphics[width=3.5cm]{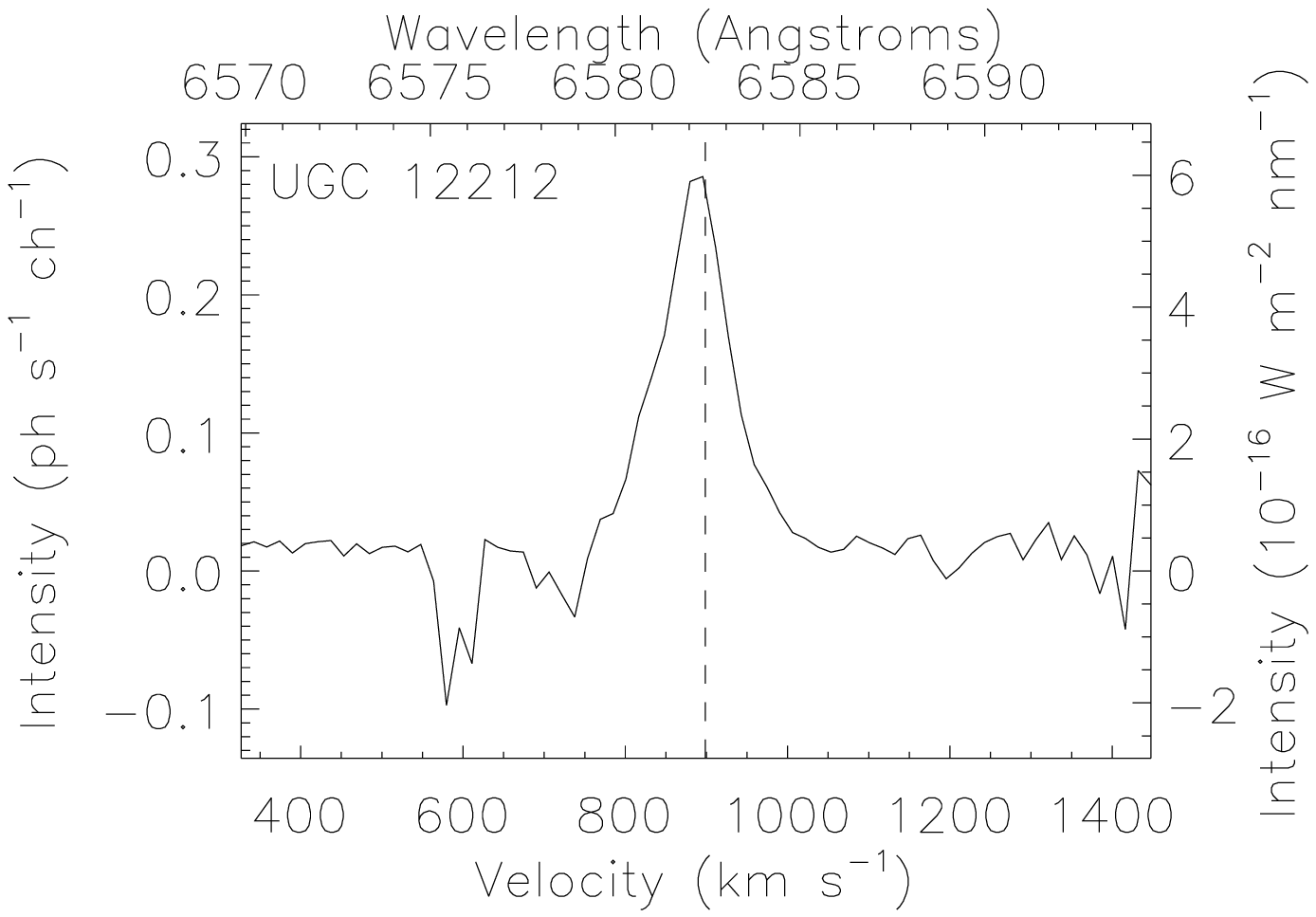}
\includegraphics[width=3.5cm]{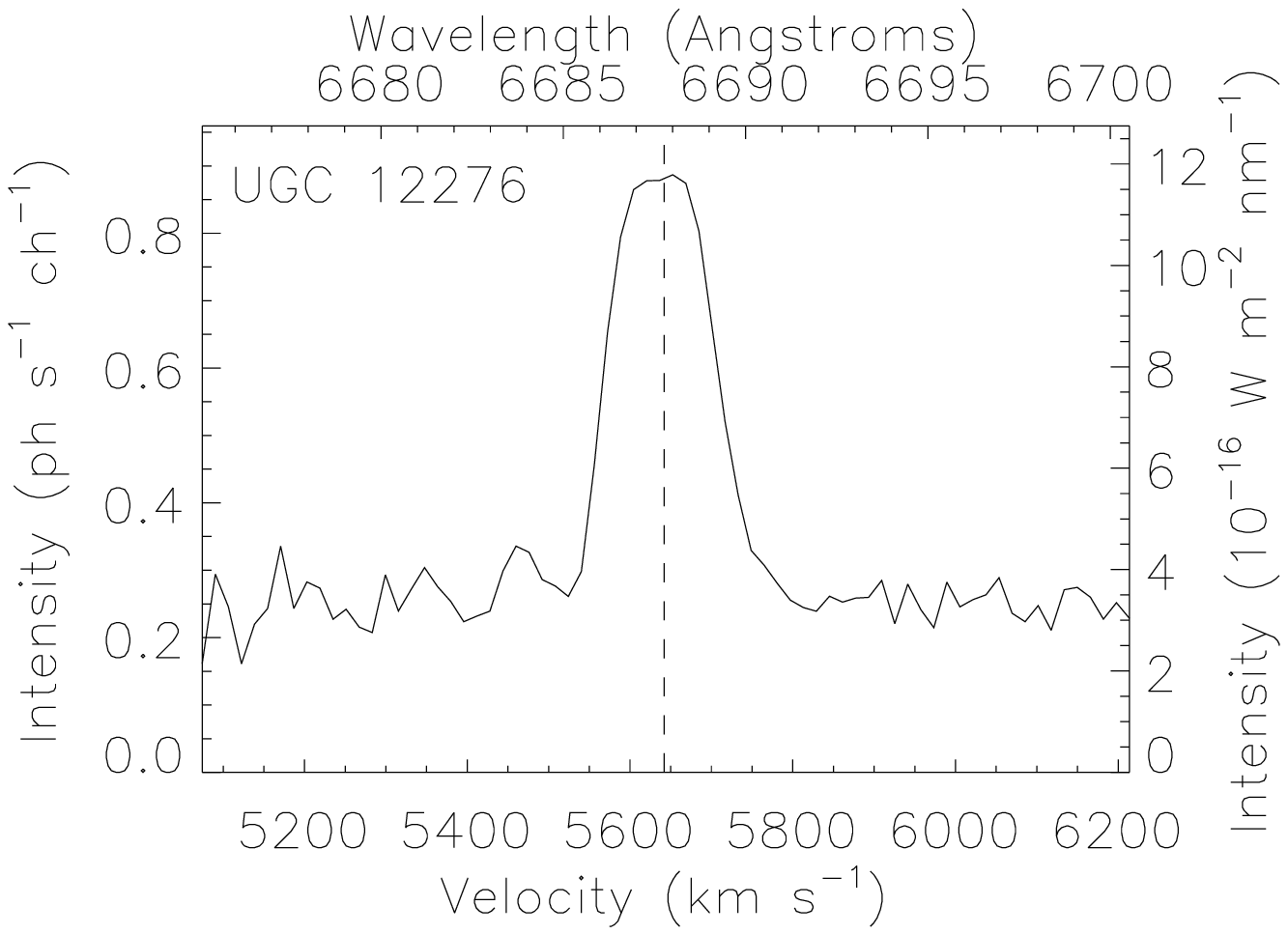}
\includegraphics[width=3.5cm]{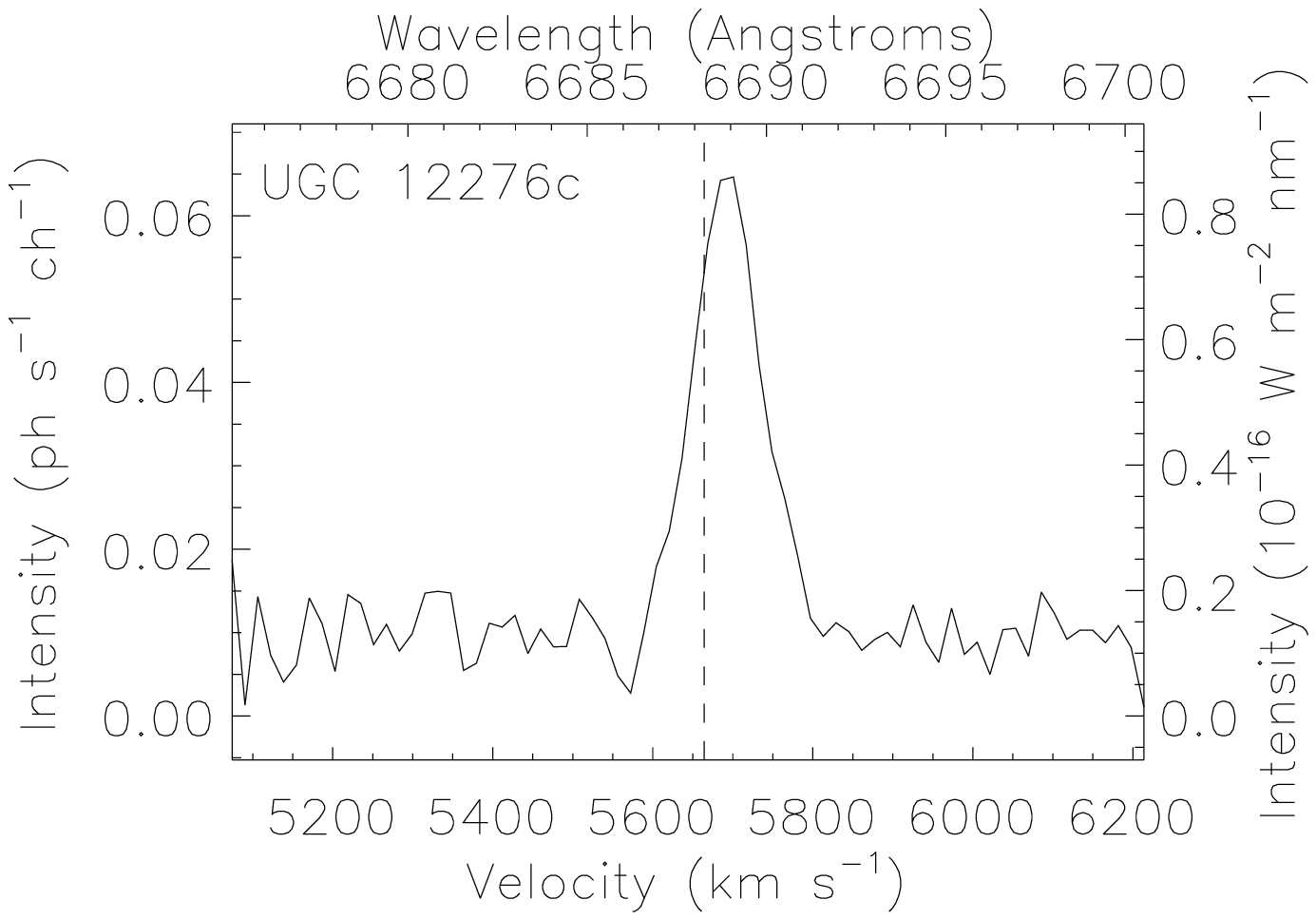}
\includegraphics[width=3.5cm]{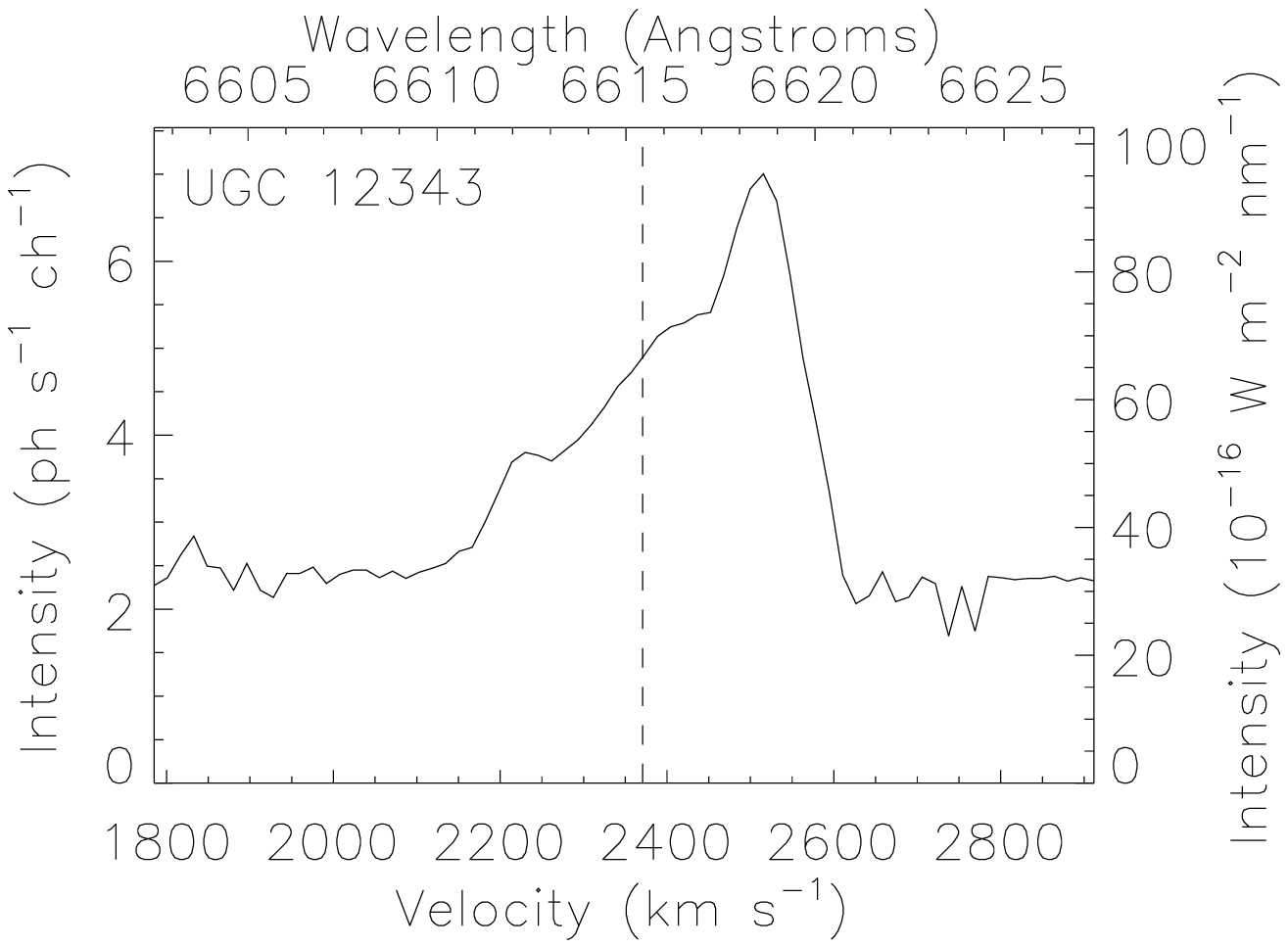}
\includegraphics[width=3.5cm]{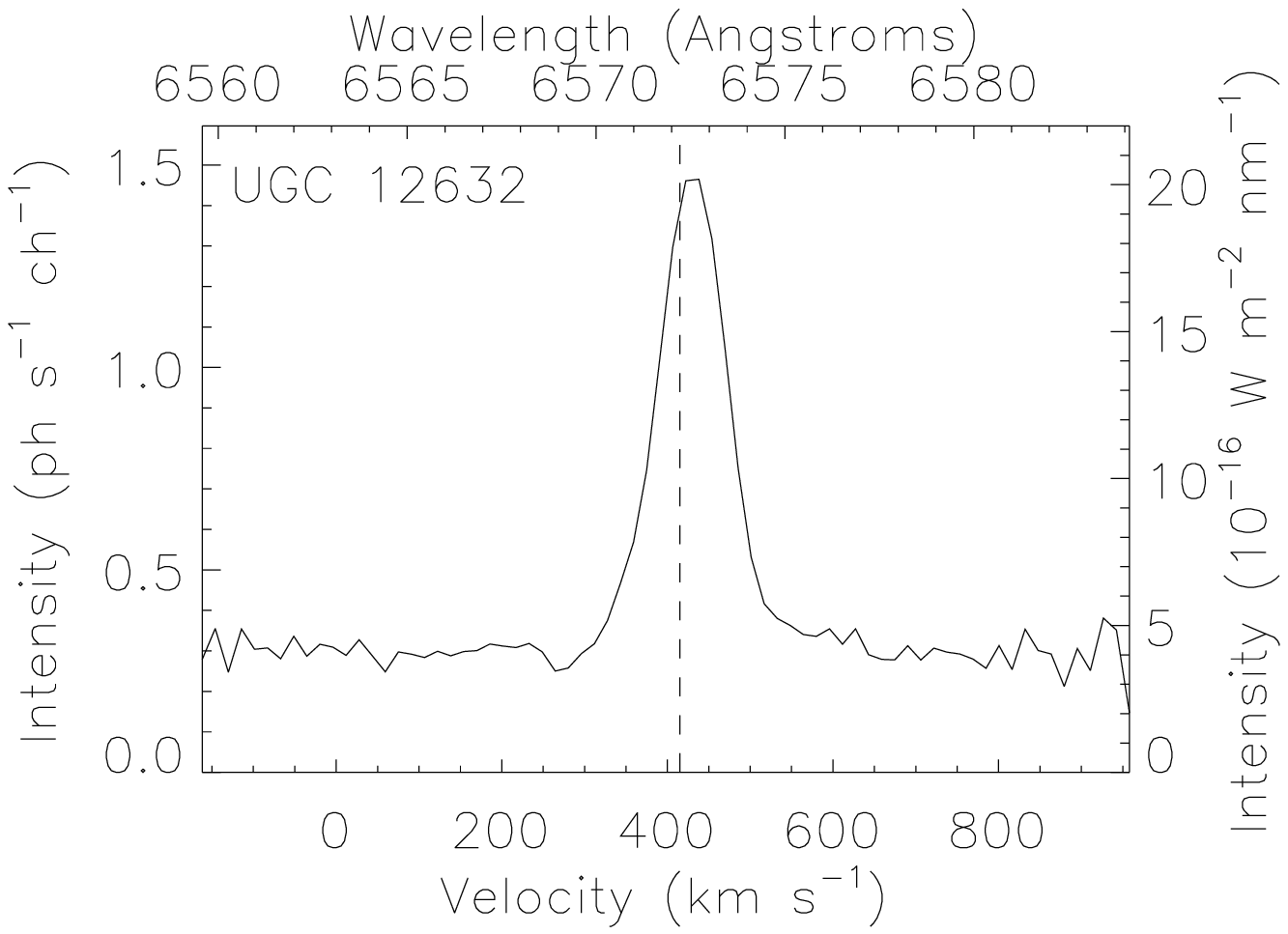}
\end{center}
\end{minipage}
\end{figure}
\clearpage
\begin{figure}
\begin{minipage}{180mm}
\begin{center}
\includegraphics[width=3.5cm]{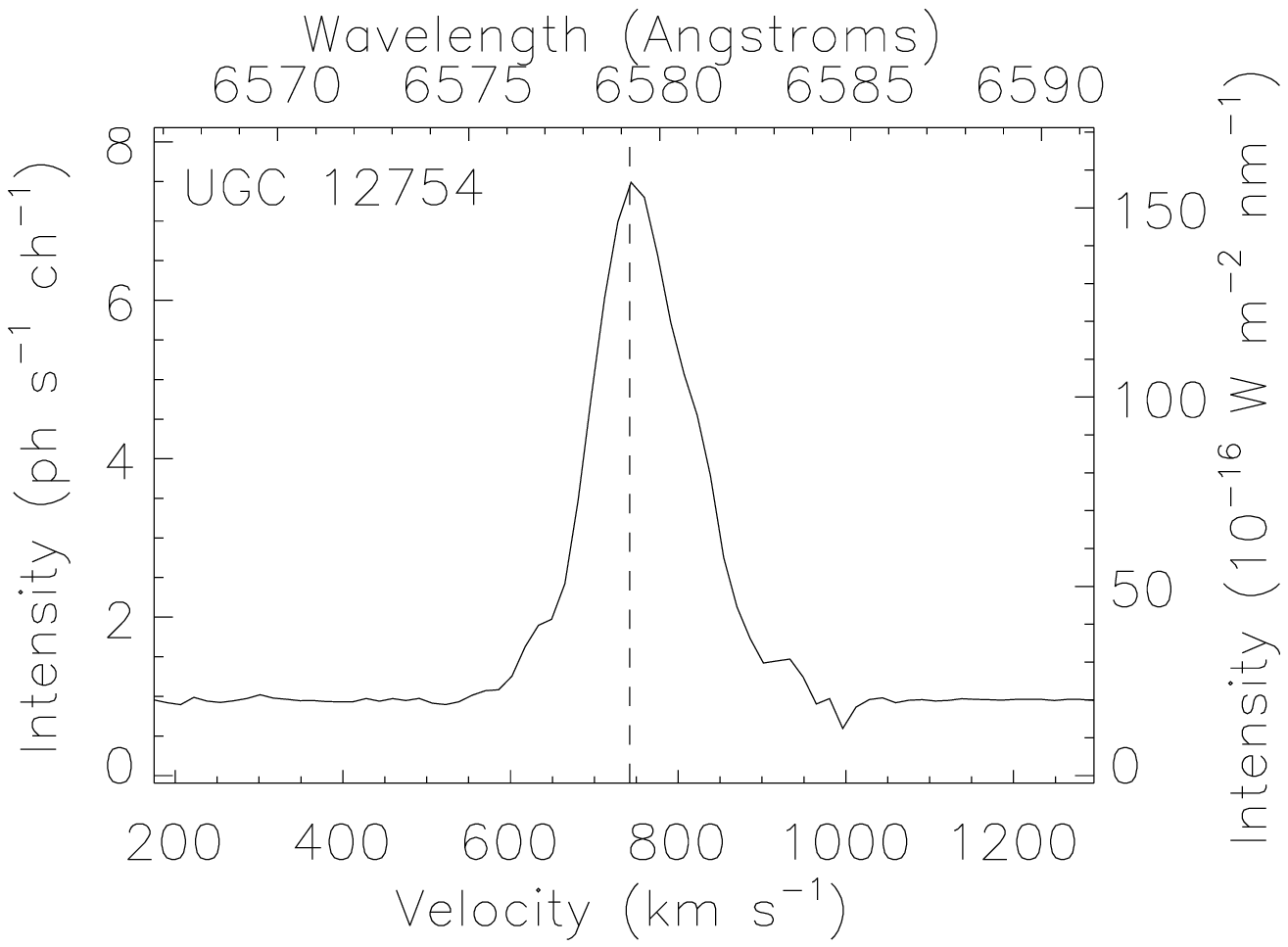}
\caption{Integrated \Ha~profiles. The profiles have been displayed over three times the spectral range ($\sim$25$\AA$, top label or $\sim$1100\kms, bottom label). The instrumental intensity in photo-electron per second and per channel is given on the left Y-axis. The calibrated intensity is displayed on the right Y-axis. The dashed vertical line indicates the systemic velocity provided by our kinematical models (see Table \ref{tablemod}).}
\label{plot_profiles}
\end{center}
\end{minipage}
\end{figure}
\clearpage

\clearpage
\section{Individual maps and position-velocity diagrams}
\label{maps}
\begin{figure*}
\begin{minipage}{180mm}
\begin{center}
   \includegraphics[width=17cm]{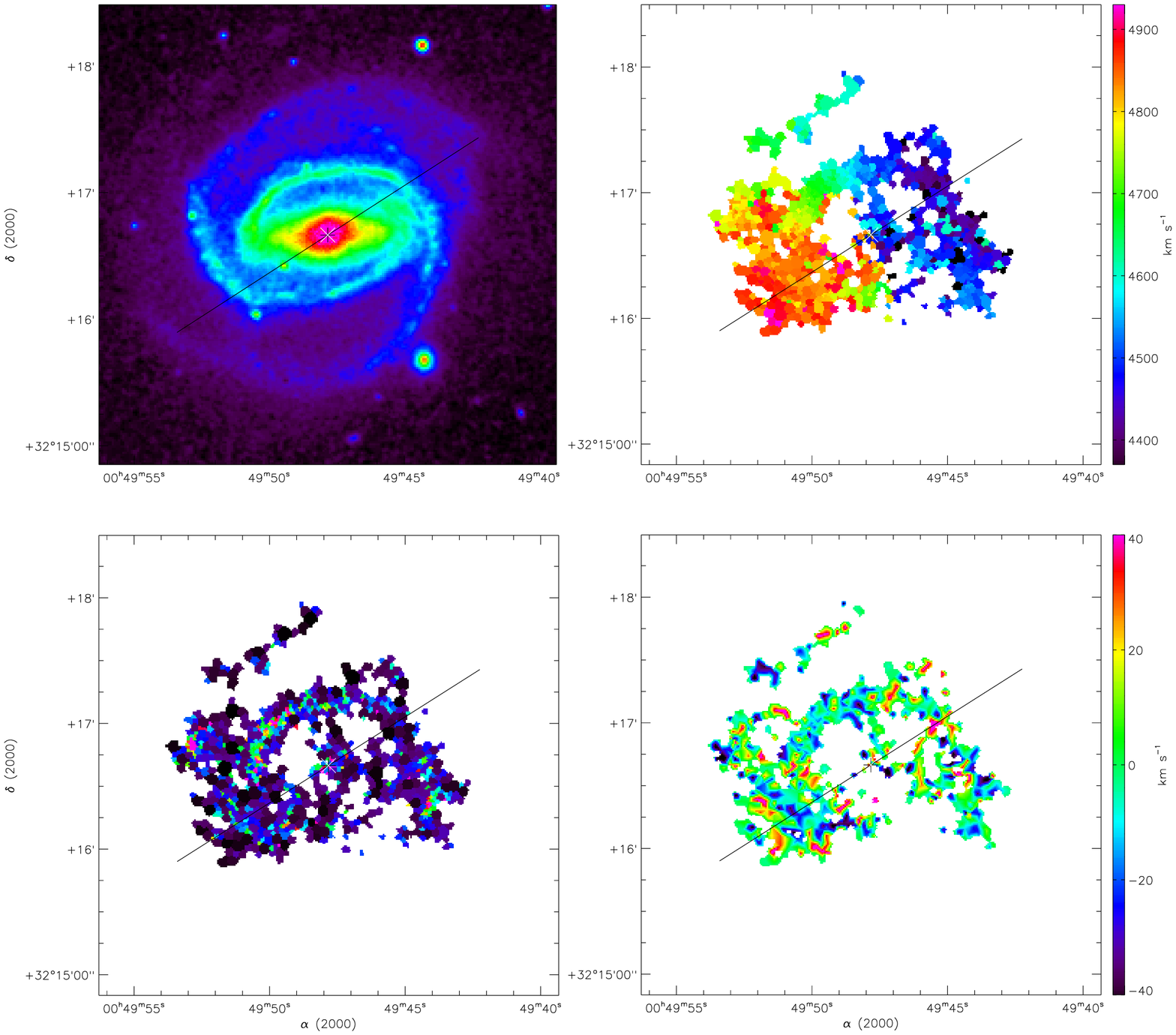}
   \includegraphics[width=19cm]{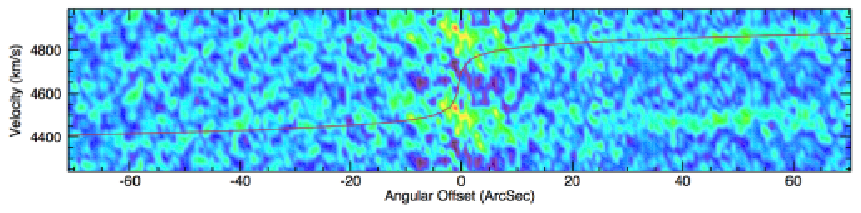}
\end{center}
\caption{UGC 508. \textbf{Top left}: XDSS Blue Band image.
\textbf{Top right}: \ha~\VF. \textbf{Middle
left}: \ha~monochromatic image.
\textbf{Middle right}: \ha~residual \VF.
The white \& black cross is the kinematical center.
The black line is the major axis, its length represents the $D_{25}$.
\textbf{Bottom}: Position-velocity diagram along the major axis (full width of 7 pixels), arbitrary flux units.
The red line plots the \RC~computed from the model \VF~along the major axis (full width of 7 pixels).
} \label{ugc508}
\end{minipage}
\end{figure*}
\clearpage

\section{Rotation curves}
\label{rc}
\begin{figure*}
\begin{minipage}{180mm}
\begin{center}
   \includegraphics[width=8cm]{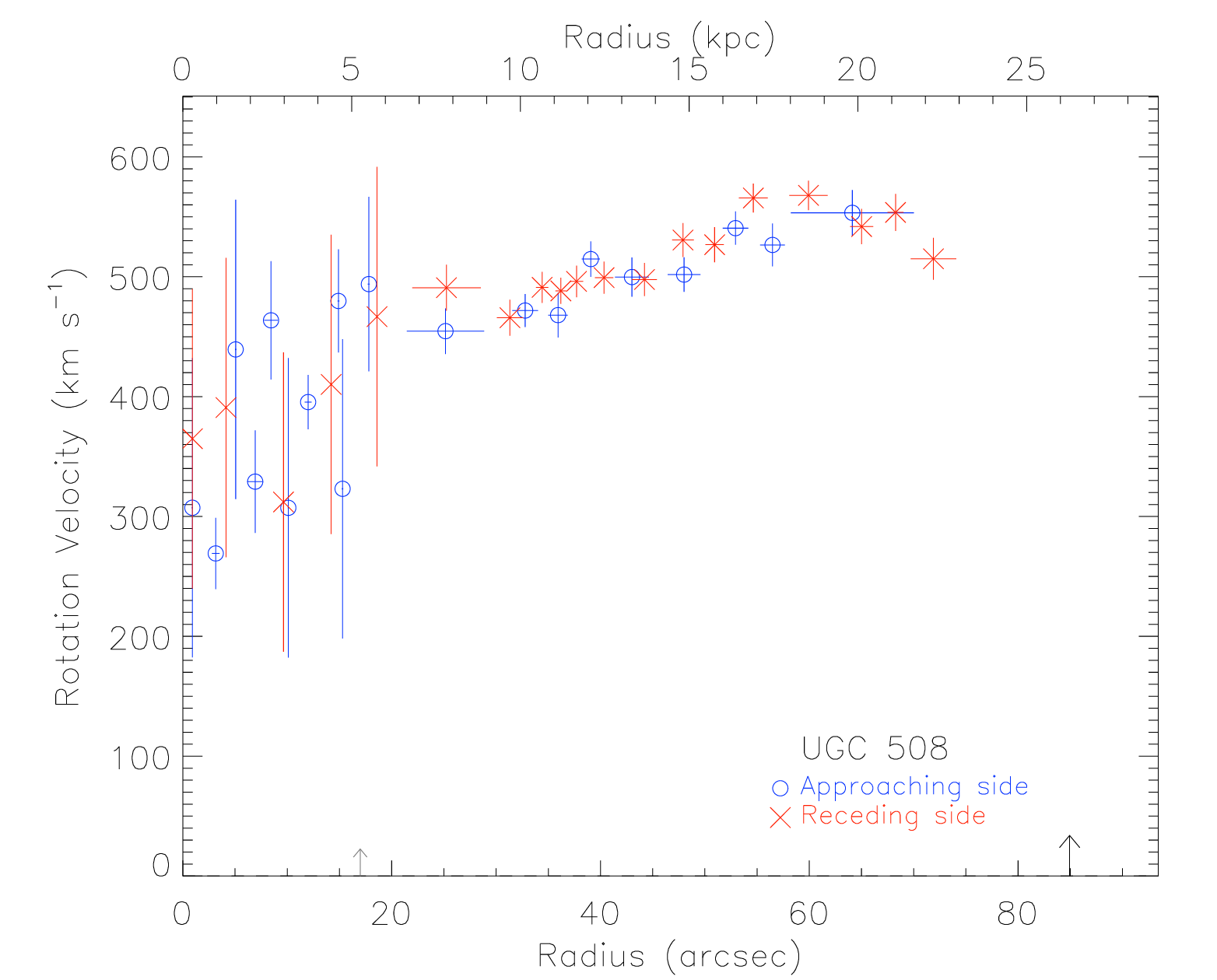}
   \includegraphics[width=8cm]{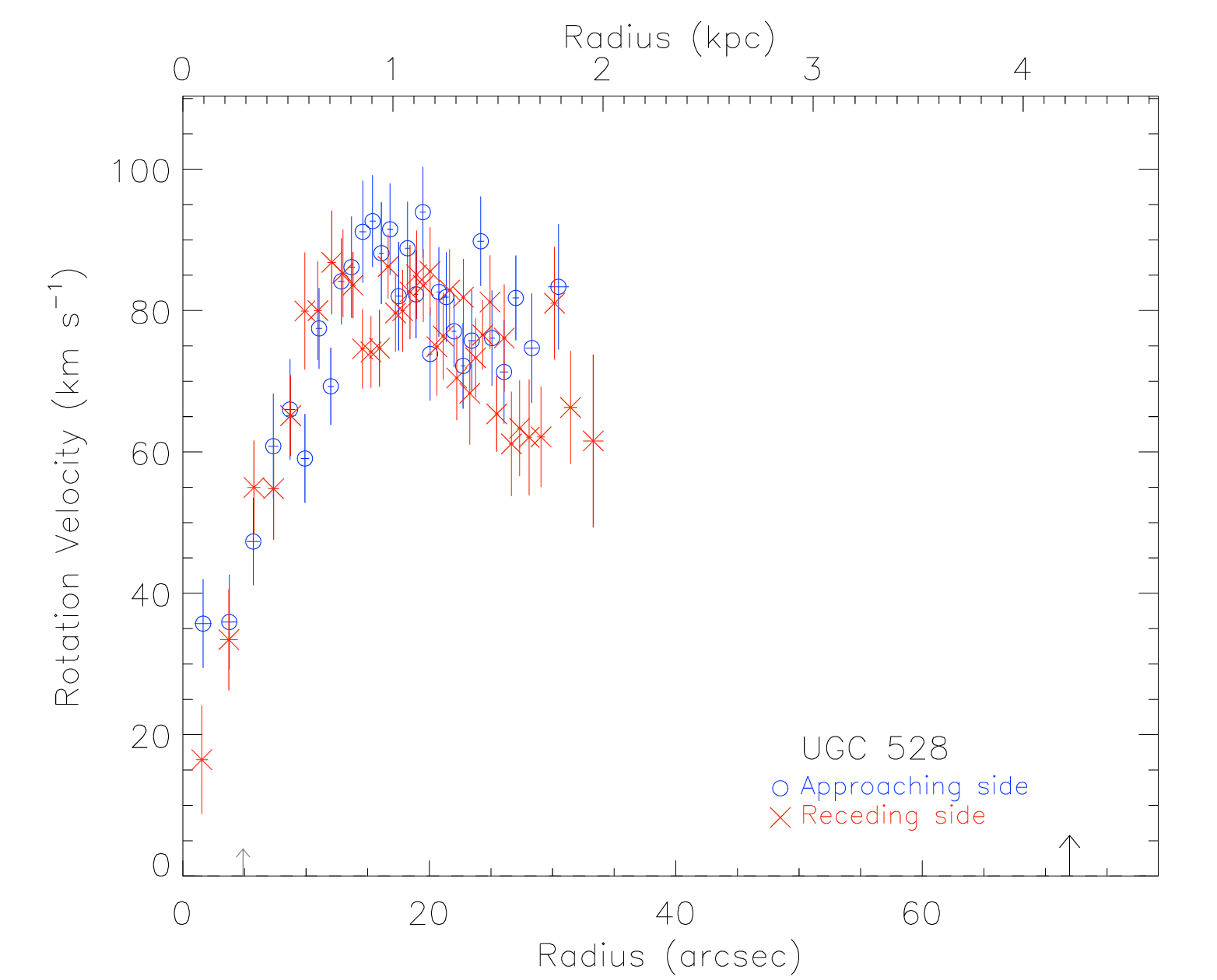}
   \includegraphics[width=8cm]{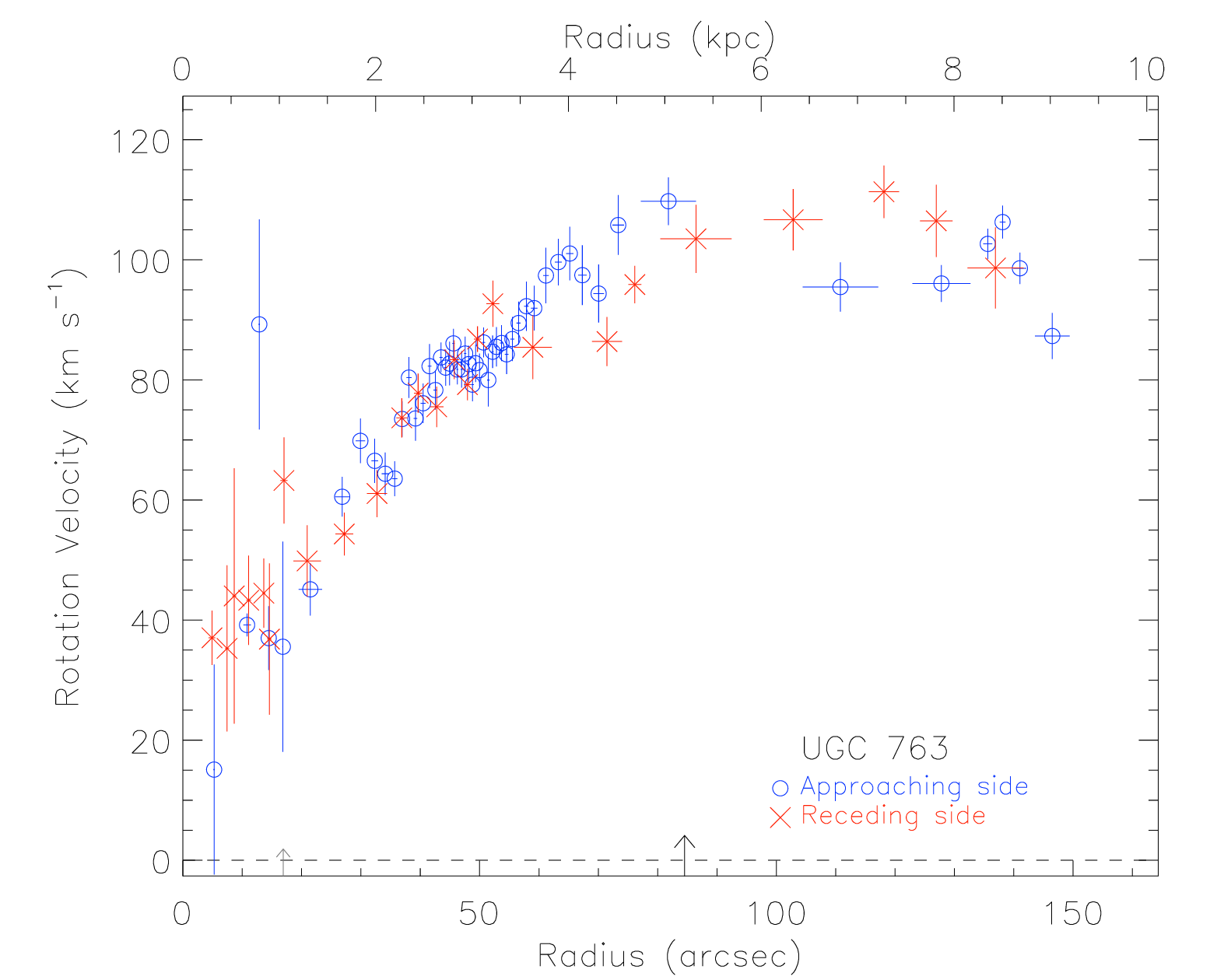}
   \includegraphics[width=8cm]{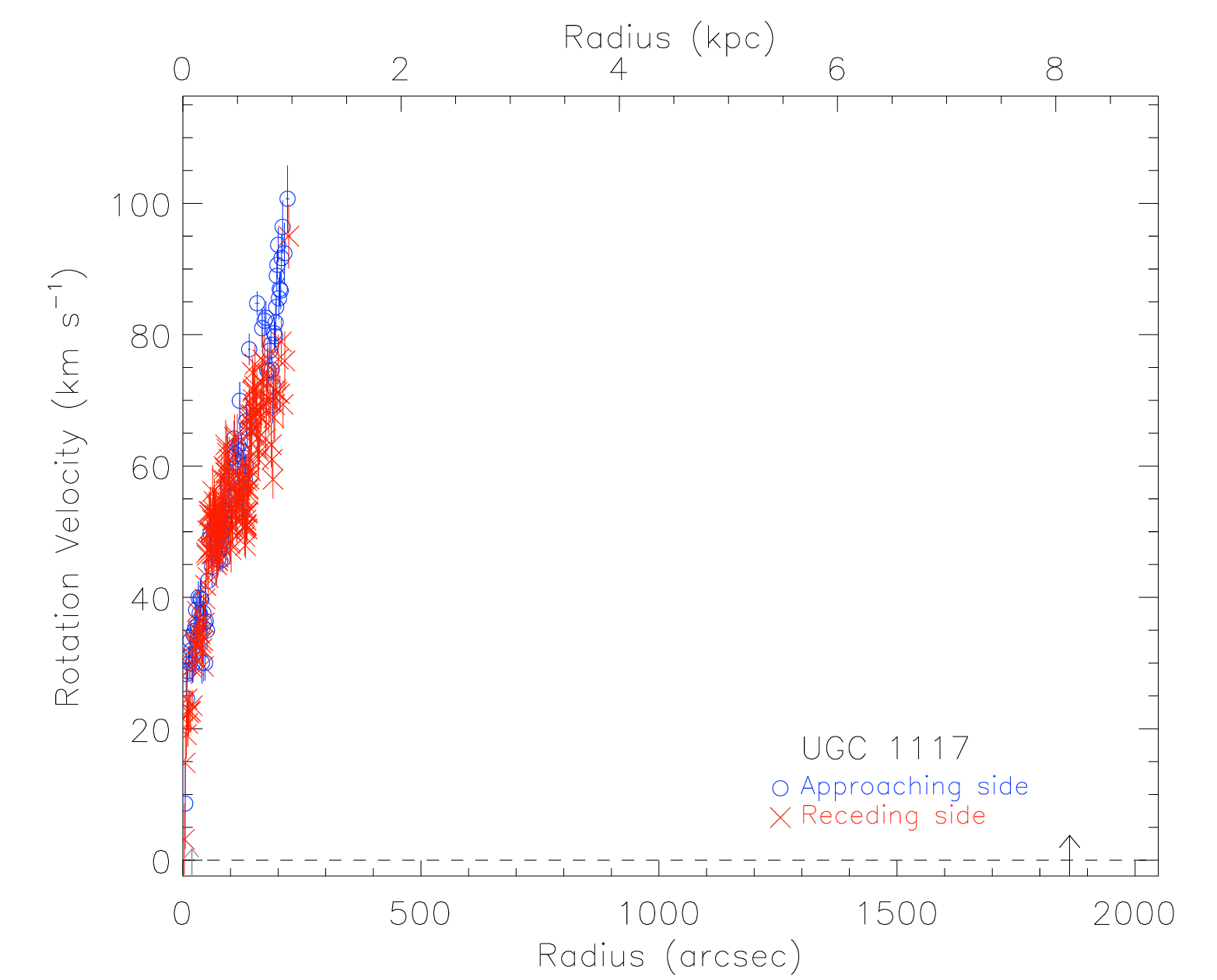}
   \includegraphics[width=8cm]{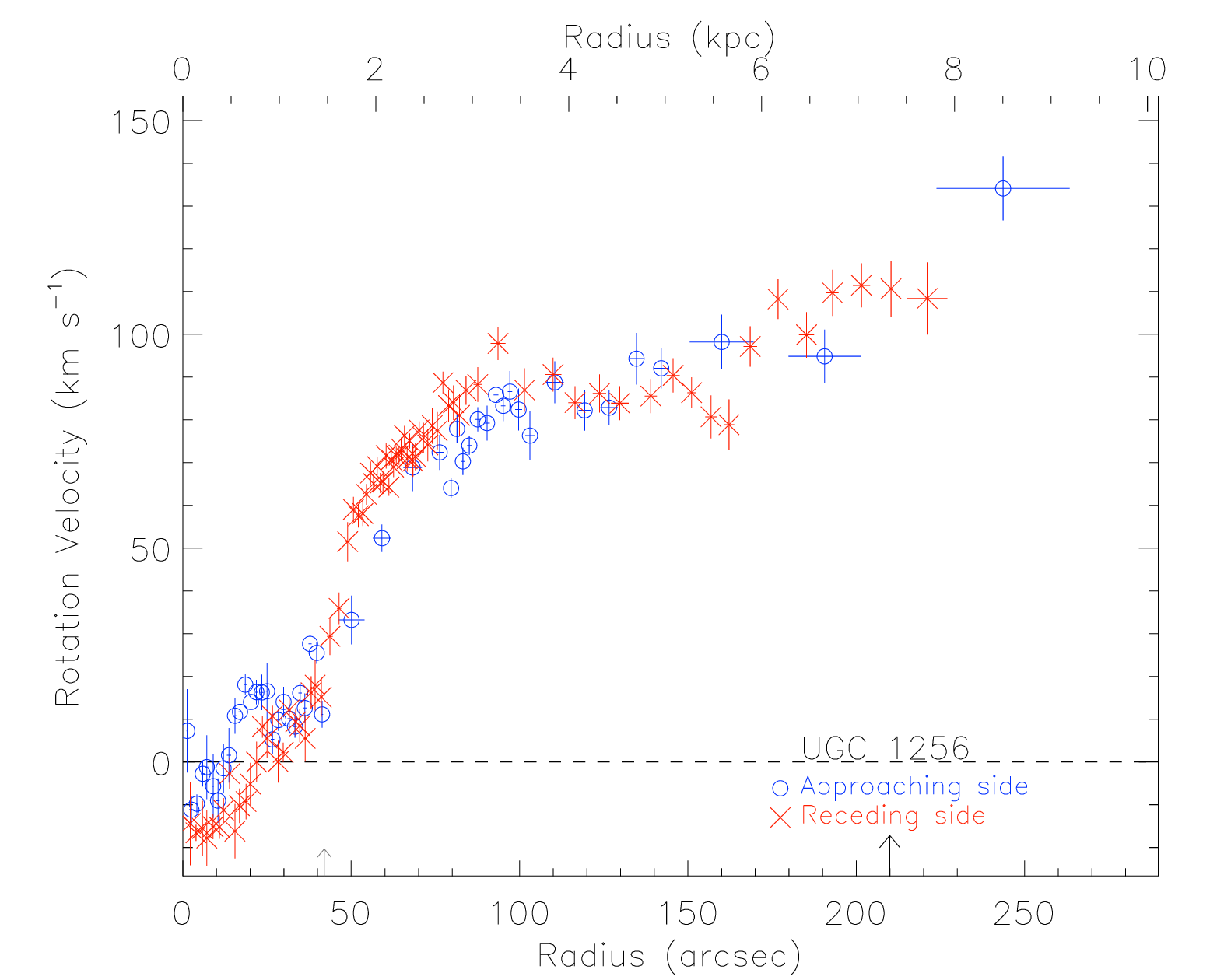}
   \includegraphics[width=8cm]{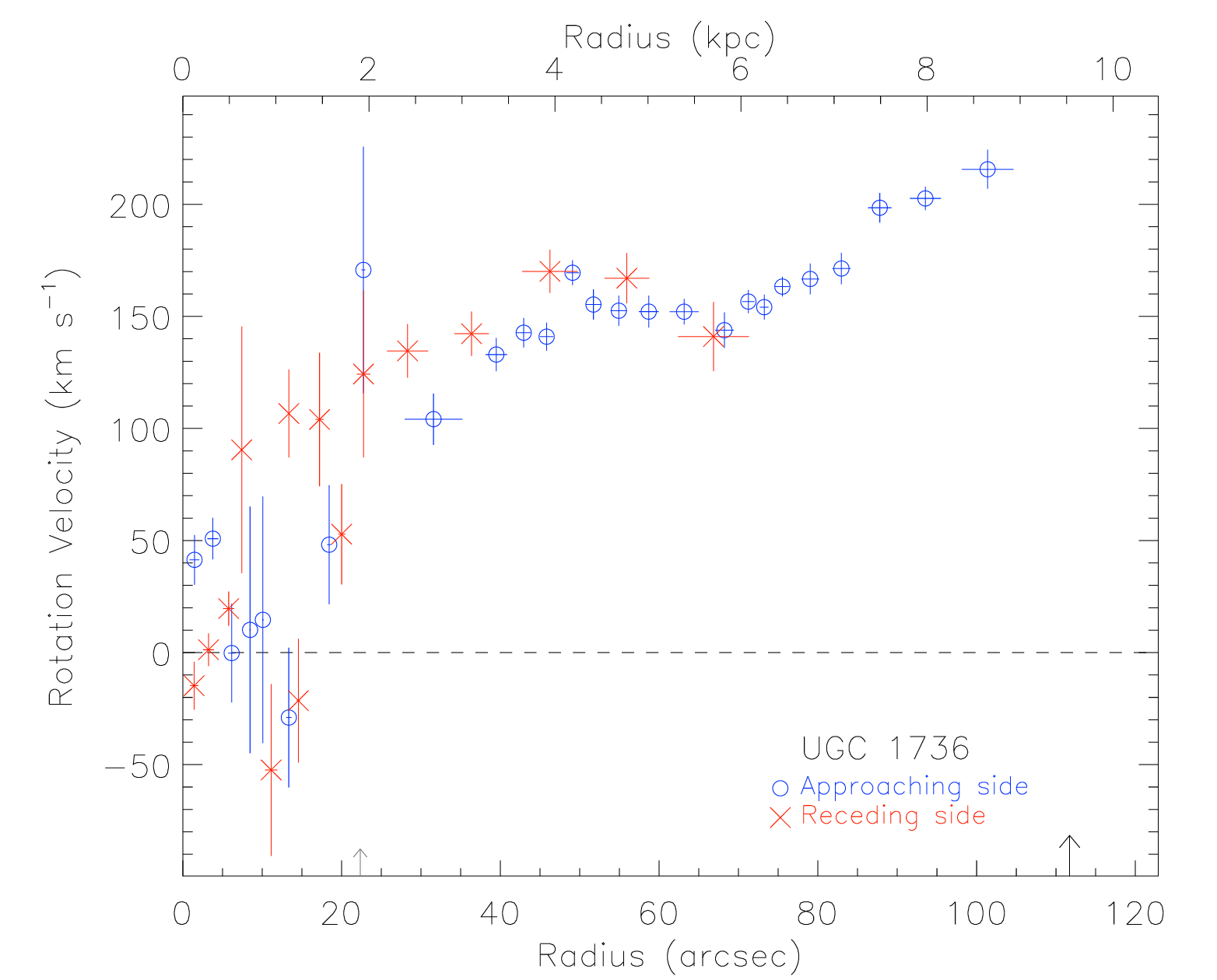}
\end{center}
\caption{From top left to bottom right: \ha~\RC~of UGC 508, UGC 528, UGC 763, UGC 1117, UGC 1256, and UGC 1736.
}
\end{minipage}
\end{figure*}
\clearpage
\begin{figure*}
\begin{minipage}{180mm}
\begin{center}
   \includegraphics[width=8cm]{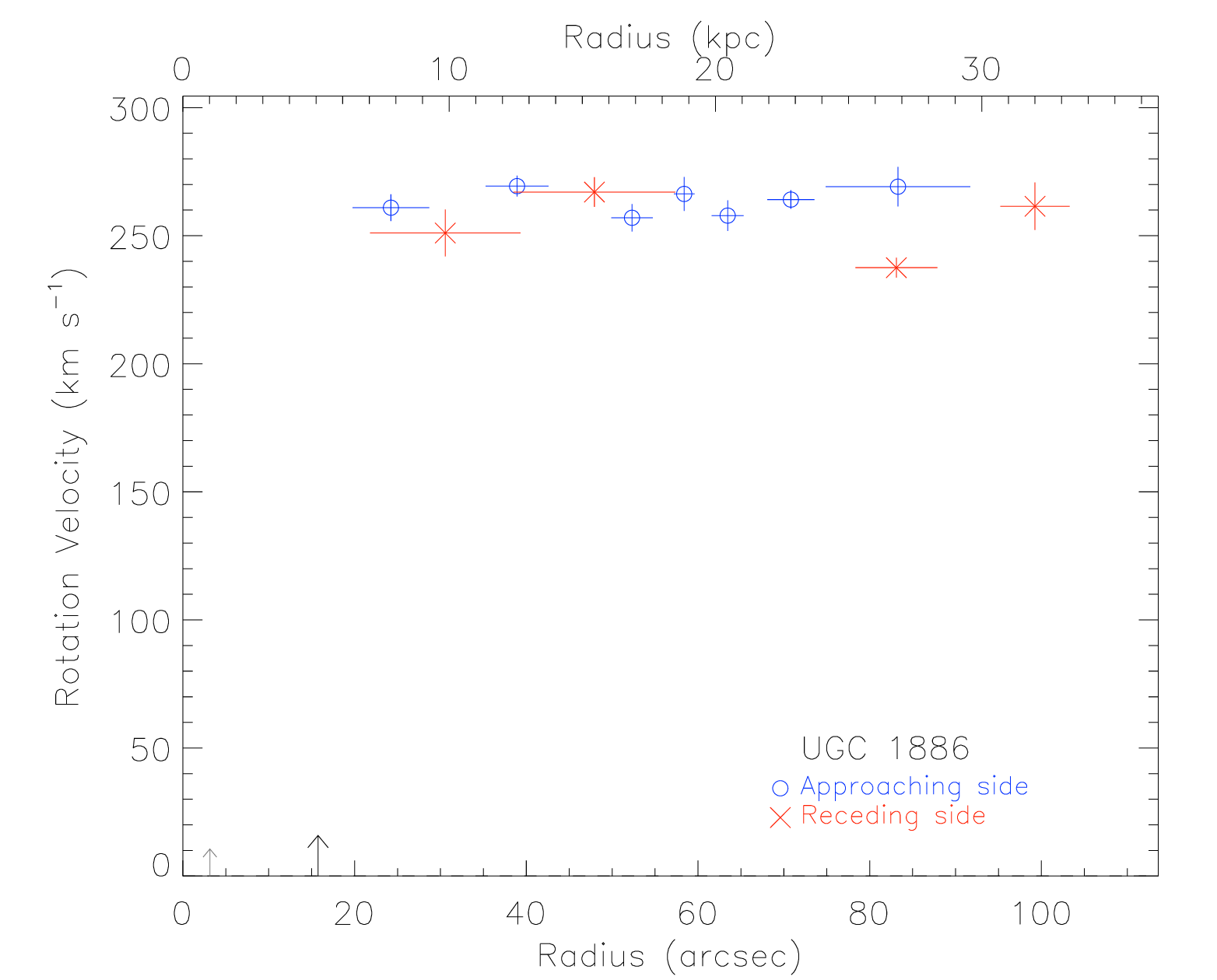}
   \includegraphics[width=8cm]{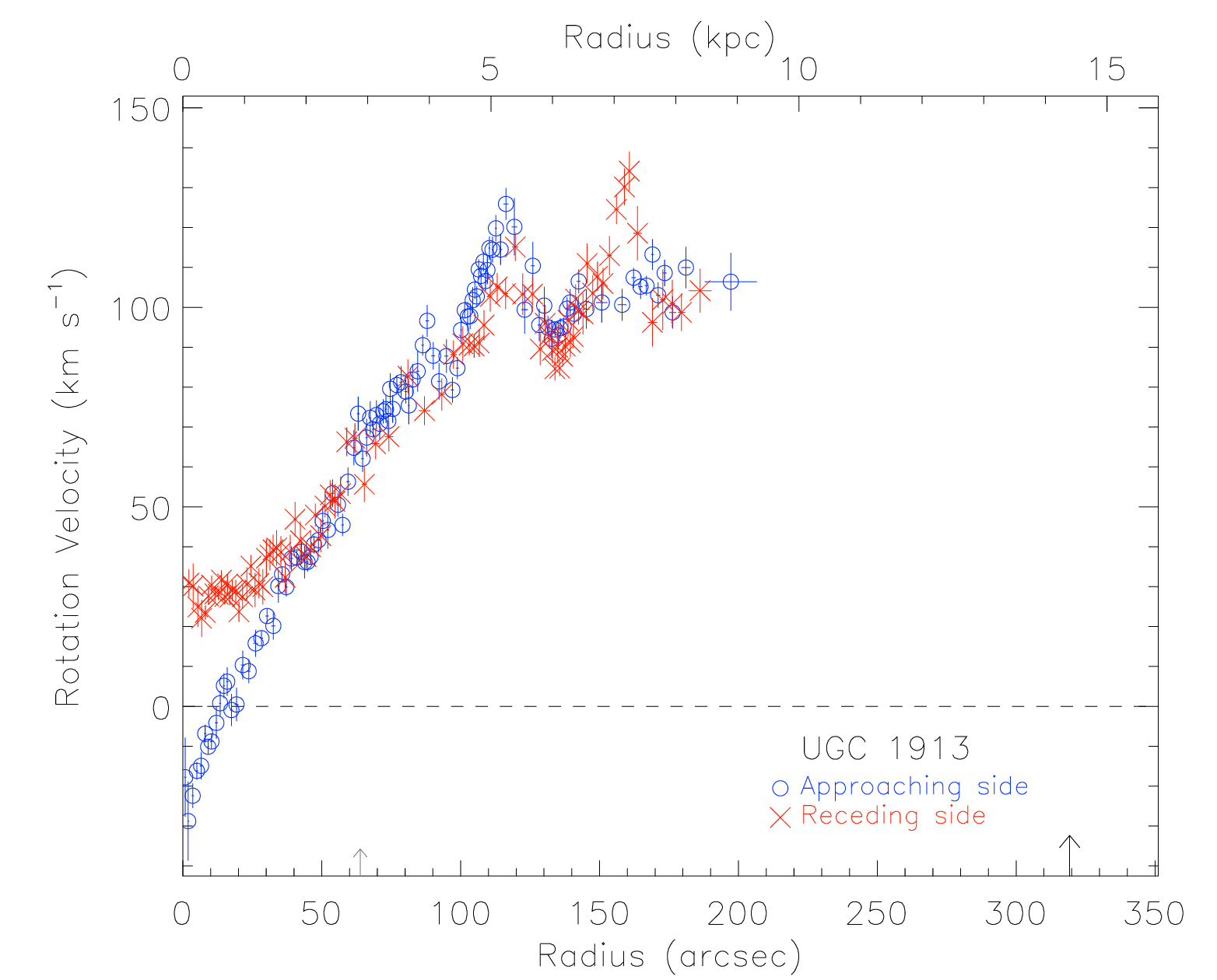}
   \includegraphics[width=8cm]{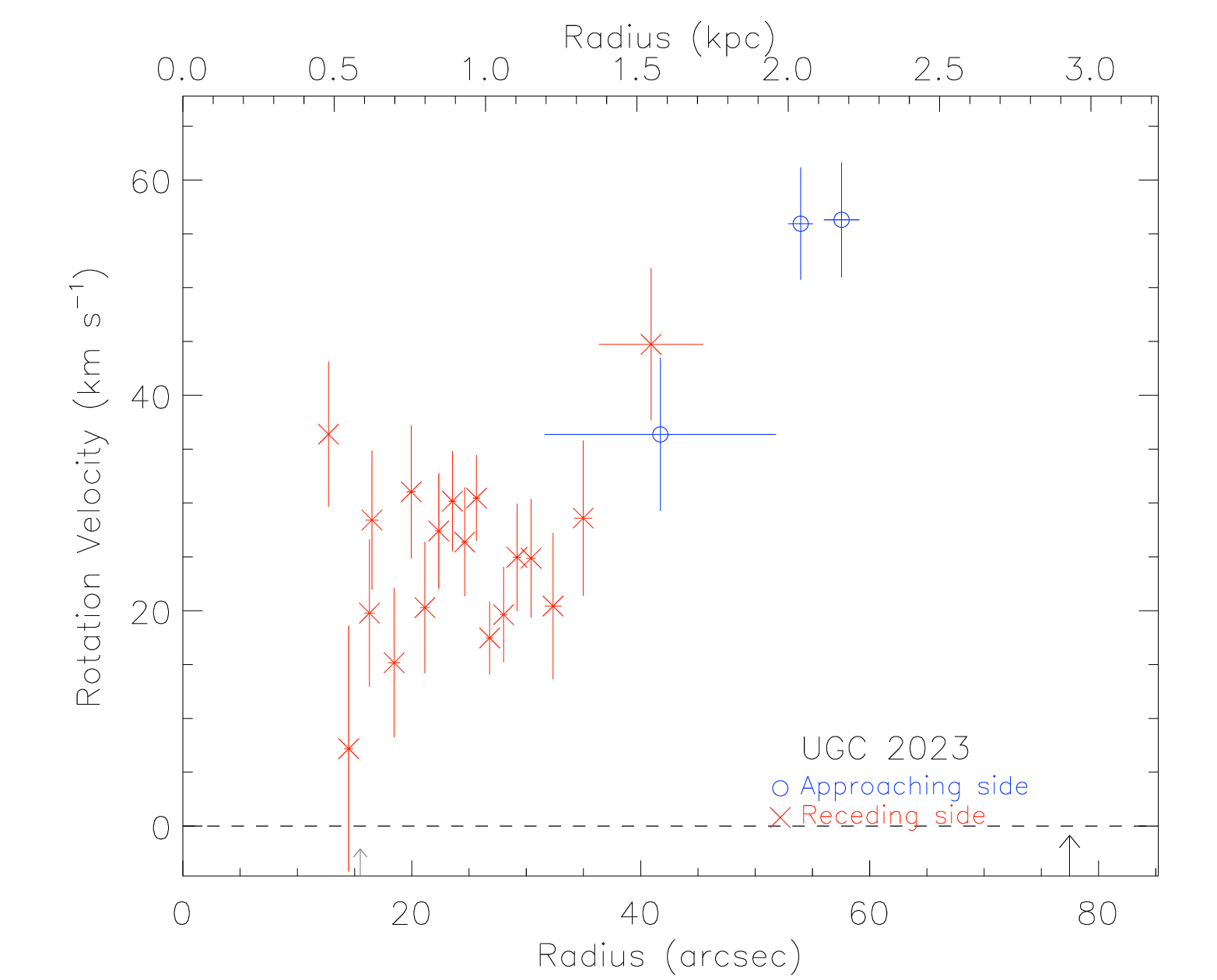}
   \includegraphics[width=8cm]{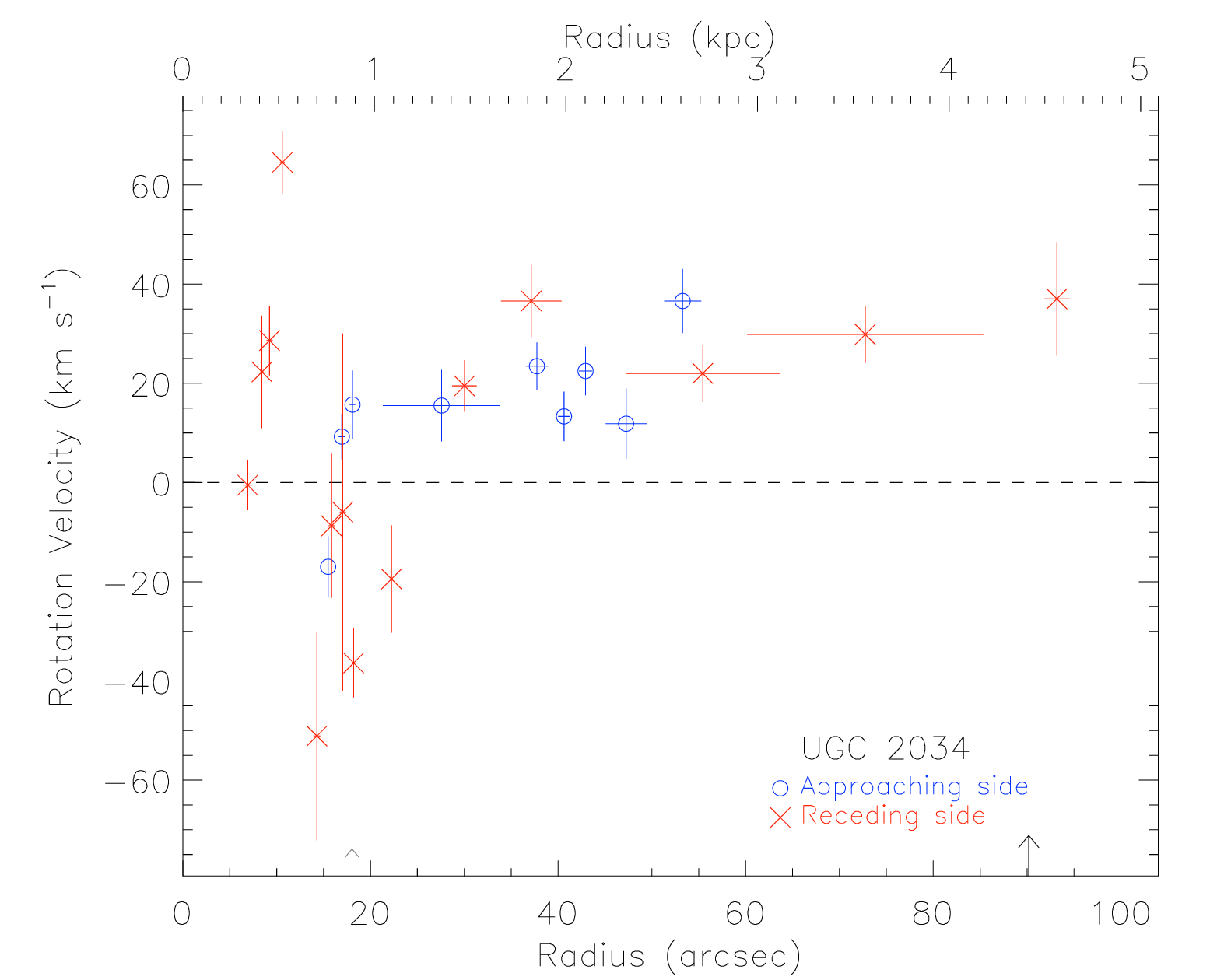}
   \includegraphics[width=8cm]{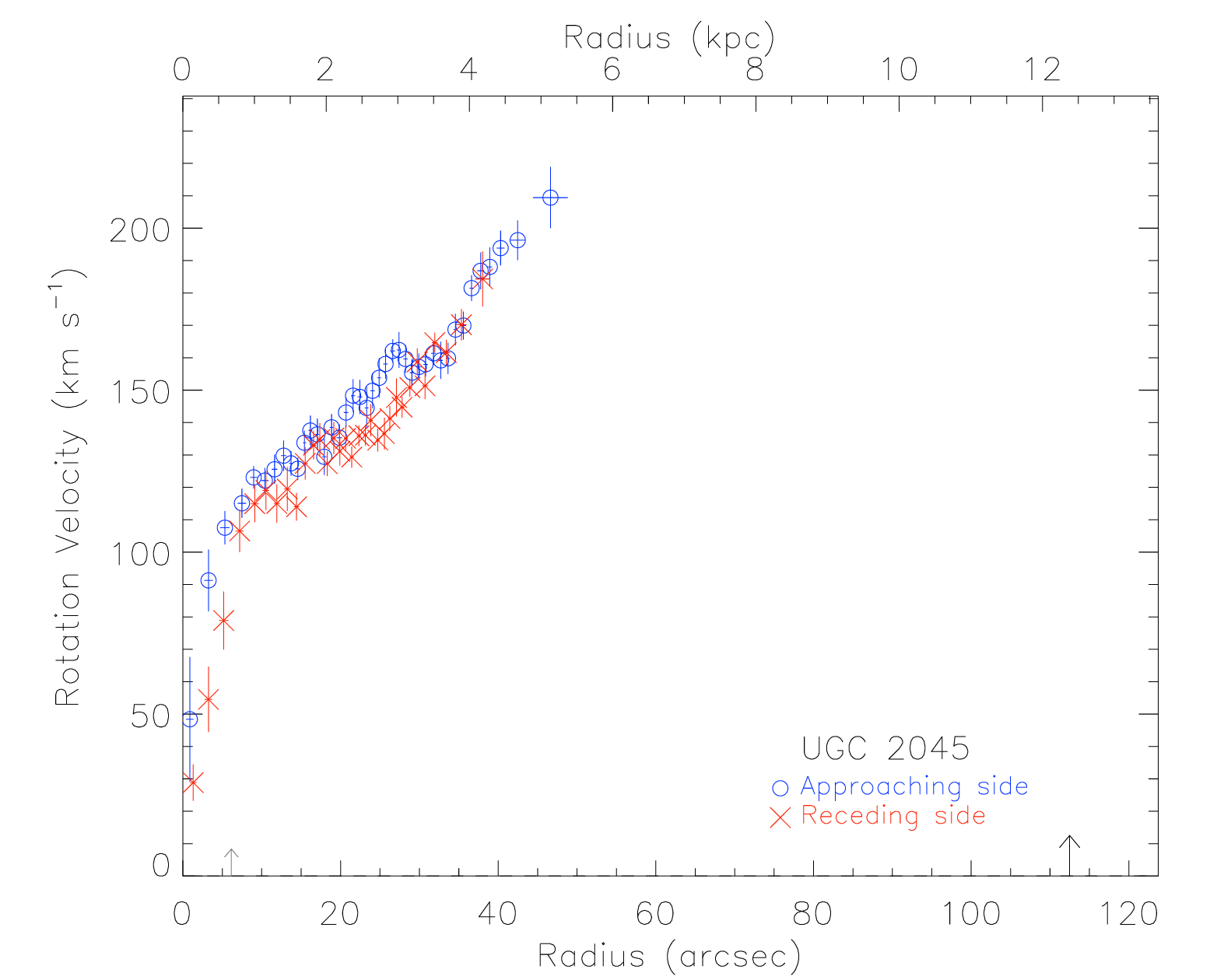}
   \includegraphics[width=8cm]{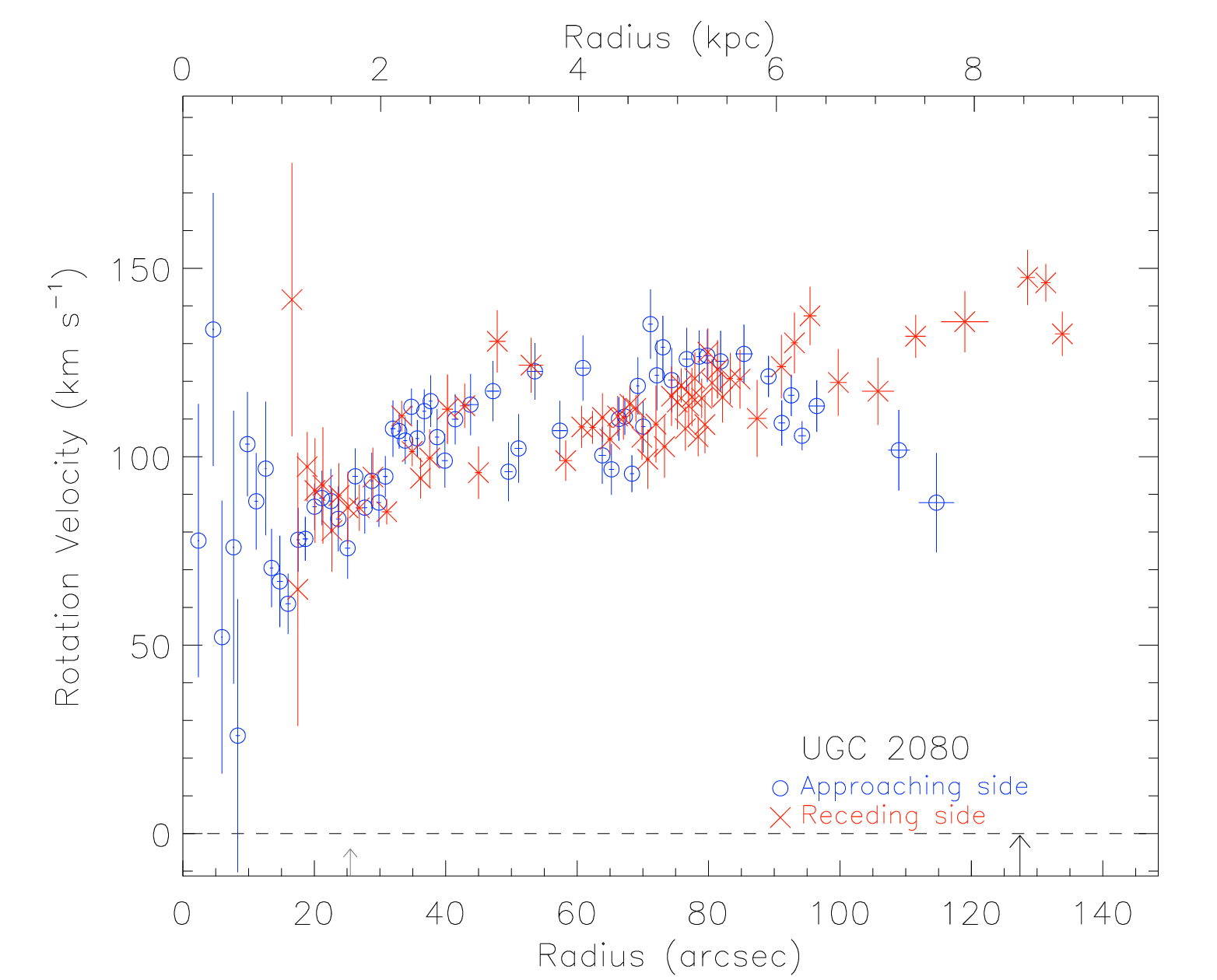}
\end{center}
\caption{From top left to bottom right: \ha~\RC~of UGC 1886, UGC 1913, UGC 2023, UGC 2034, UGC 2045, and UGC 2080.
}
\end{minipage}
\end{figure*}
\clearpage
\begin{figure*}
\begin{minipage}{180mm}
\begin{center}
   \includegraphics[width=8cm]{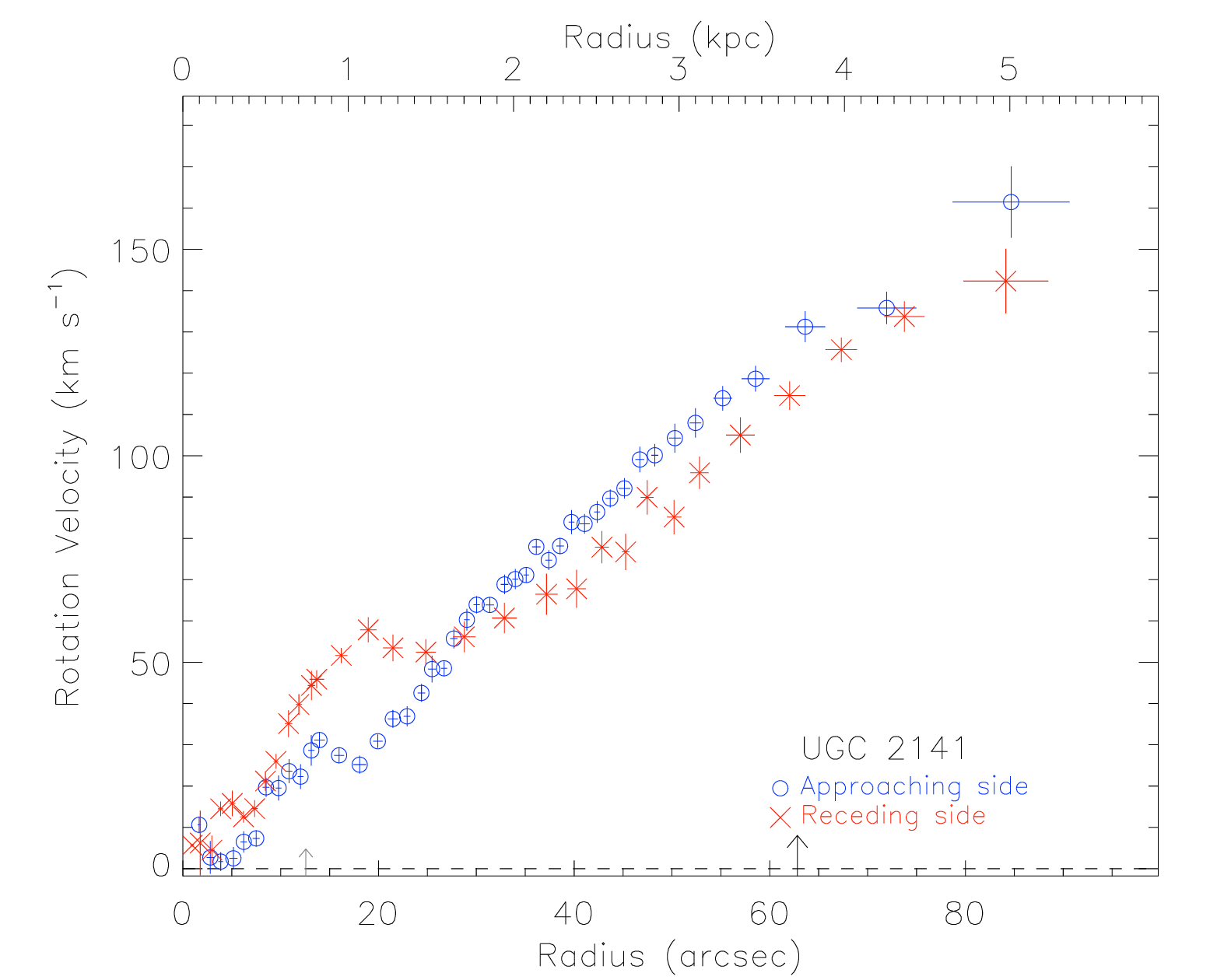}
   \includegraphics[width=8cm]{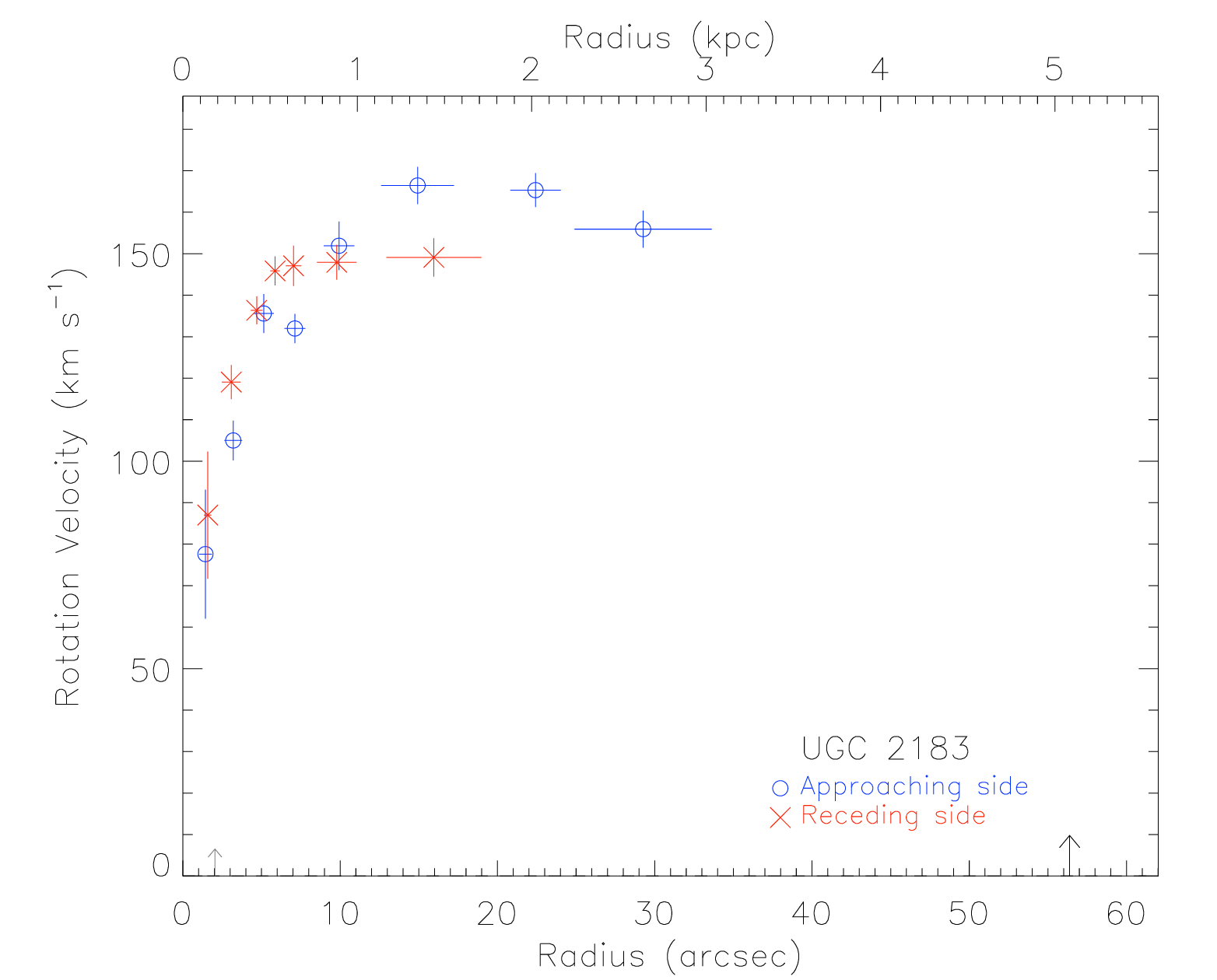}
   \includegraphics[width=8cm]{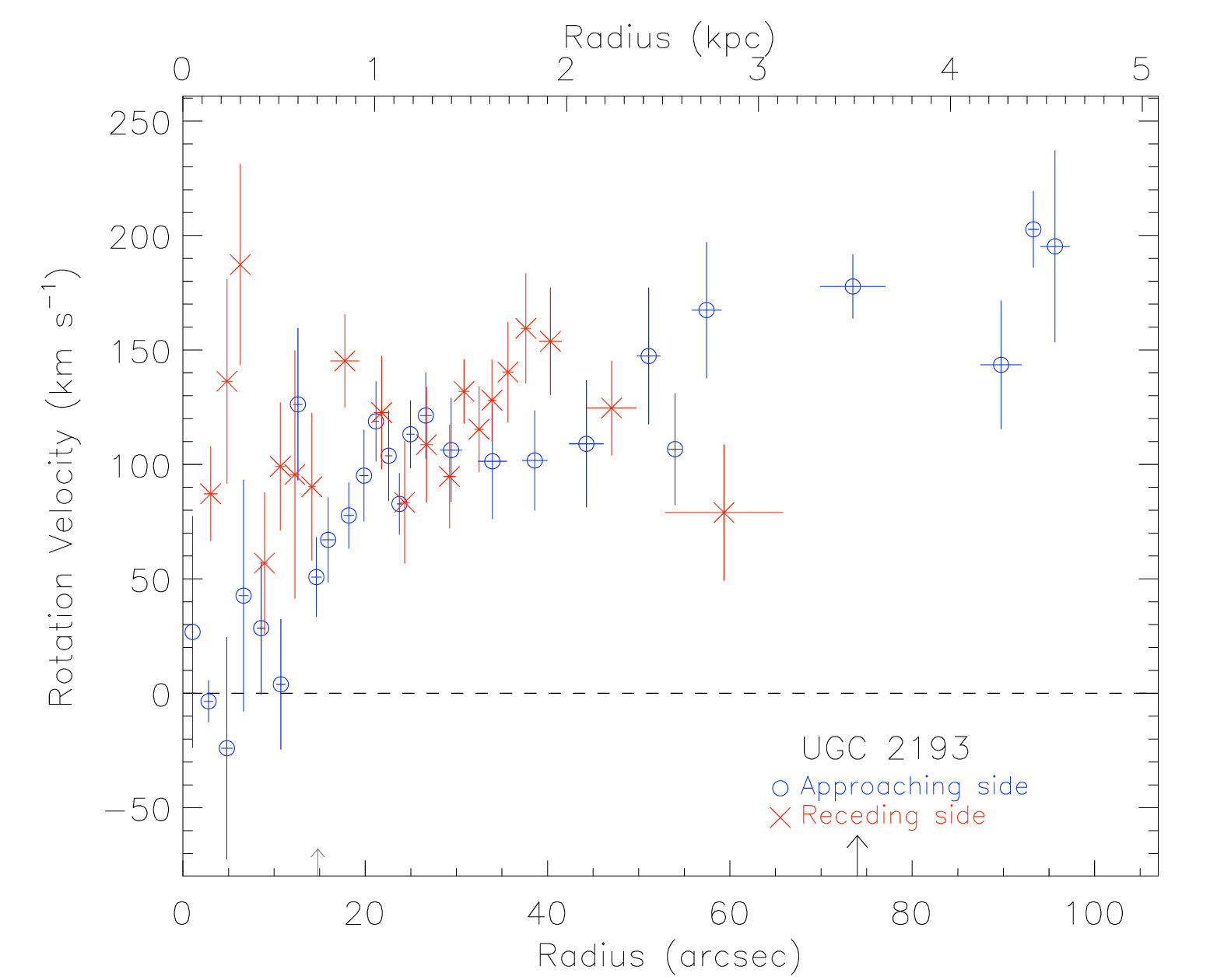}
   \includegraphics[width=8cm]{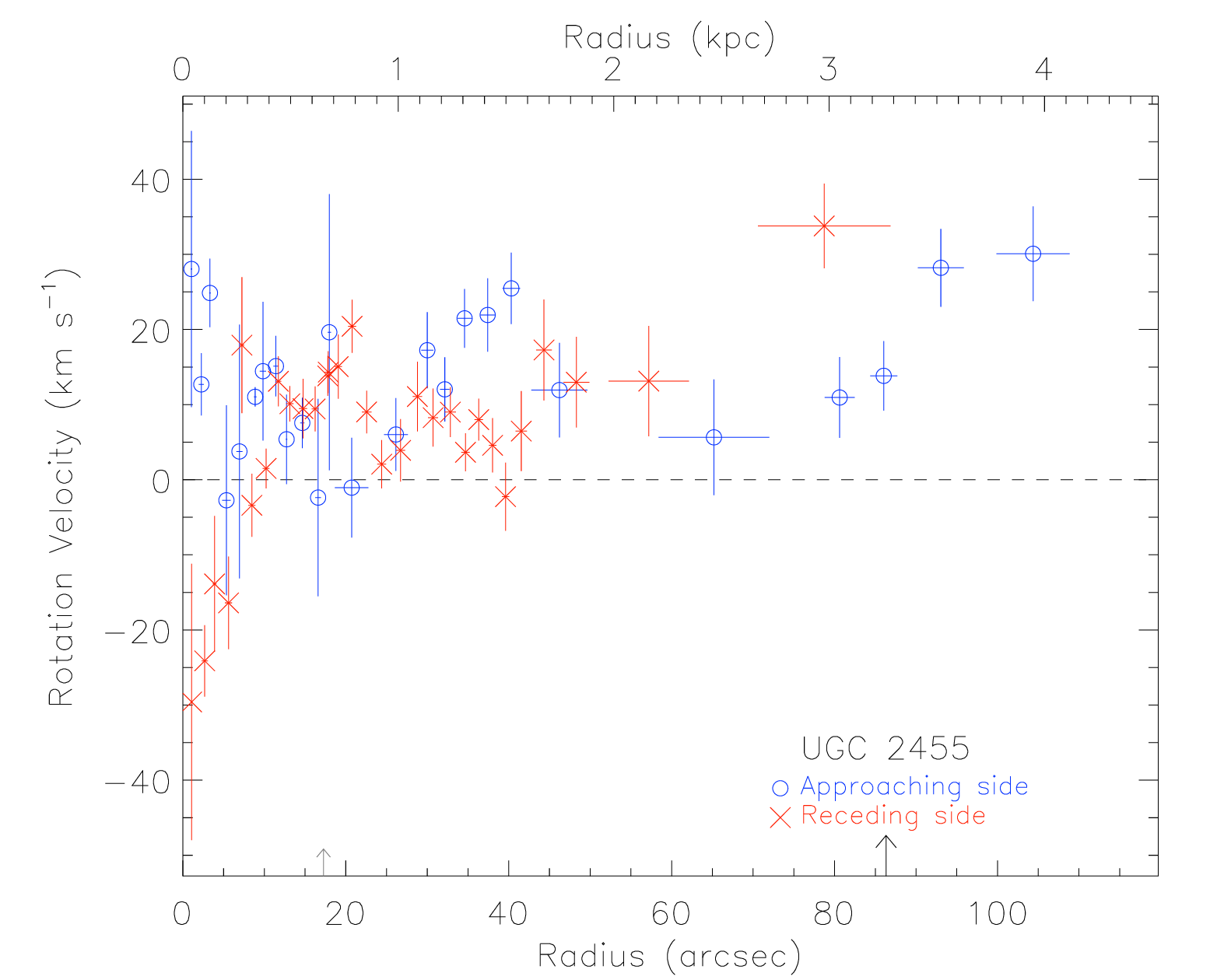}
   \includegraphics[width=8cm]{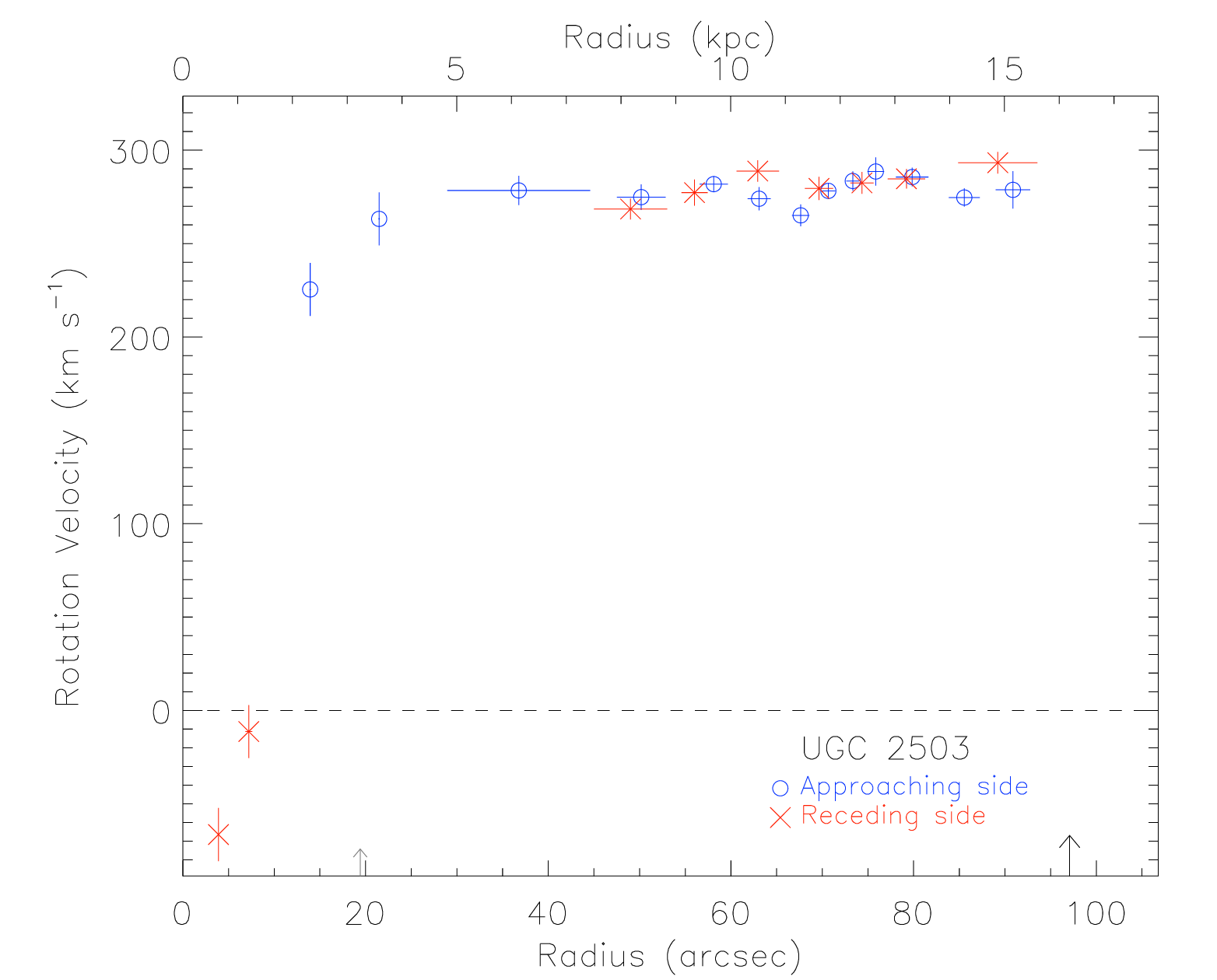}
   \includegraphics[width=8cm]{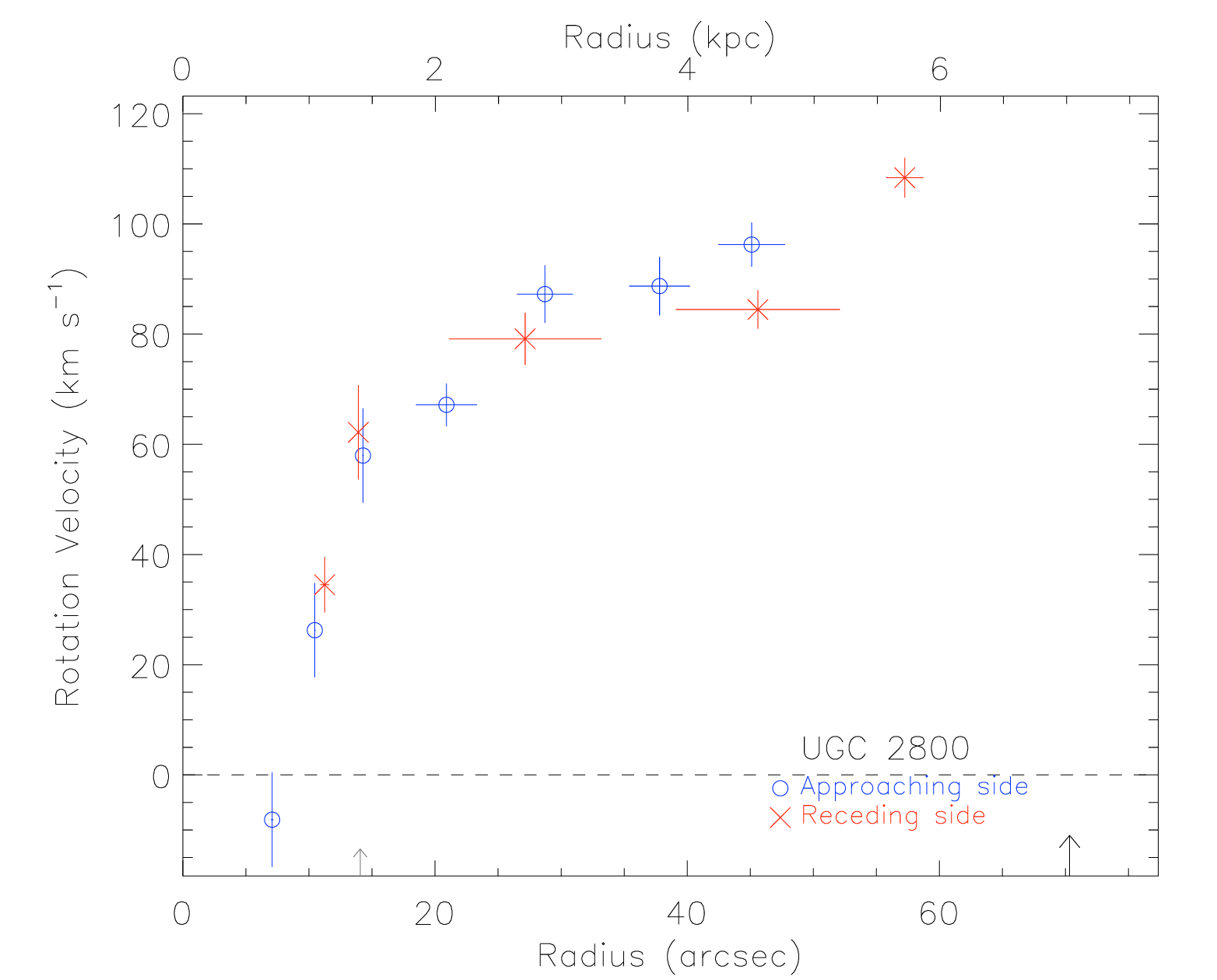}
\end{center}
\caption{From top left to bottom right: \ha~\RC~of UGC 2141, UGC 2183, UGC 2193, UGC 2455, UGC 2503, and UGC 2800.
}
\end{minipage}
\end{figure*}
\clearpage
\begin{figure*}
\begin{minipage}{180mm}
\begin{center}
   \includegraphics[width=8cm]{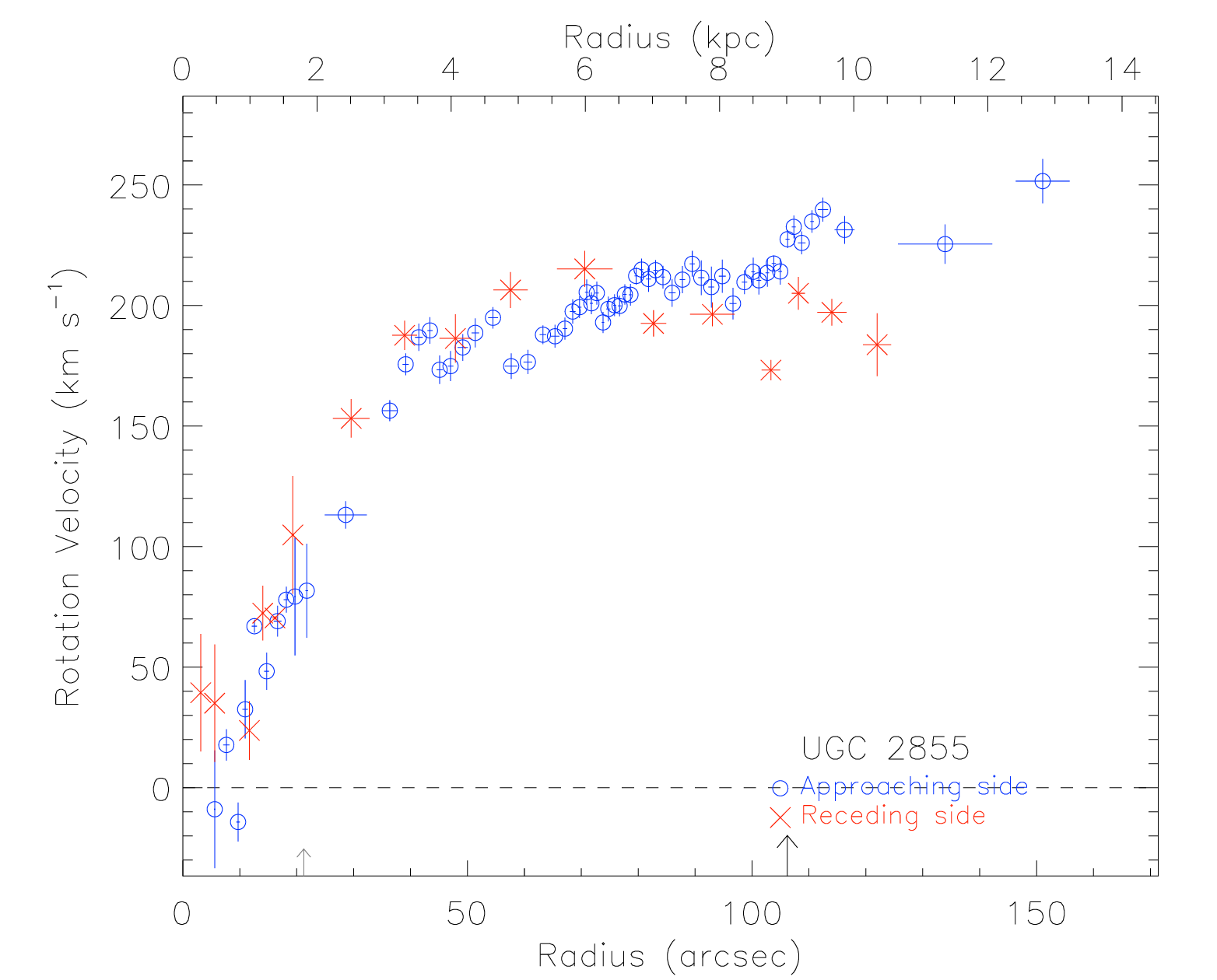}
   \includegraphics[width=8cm]{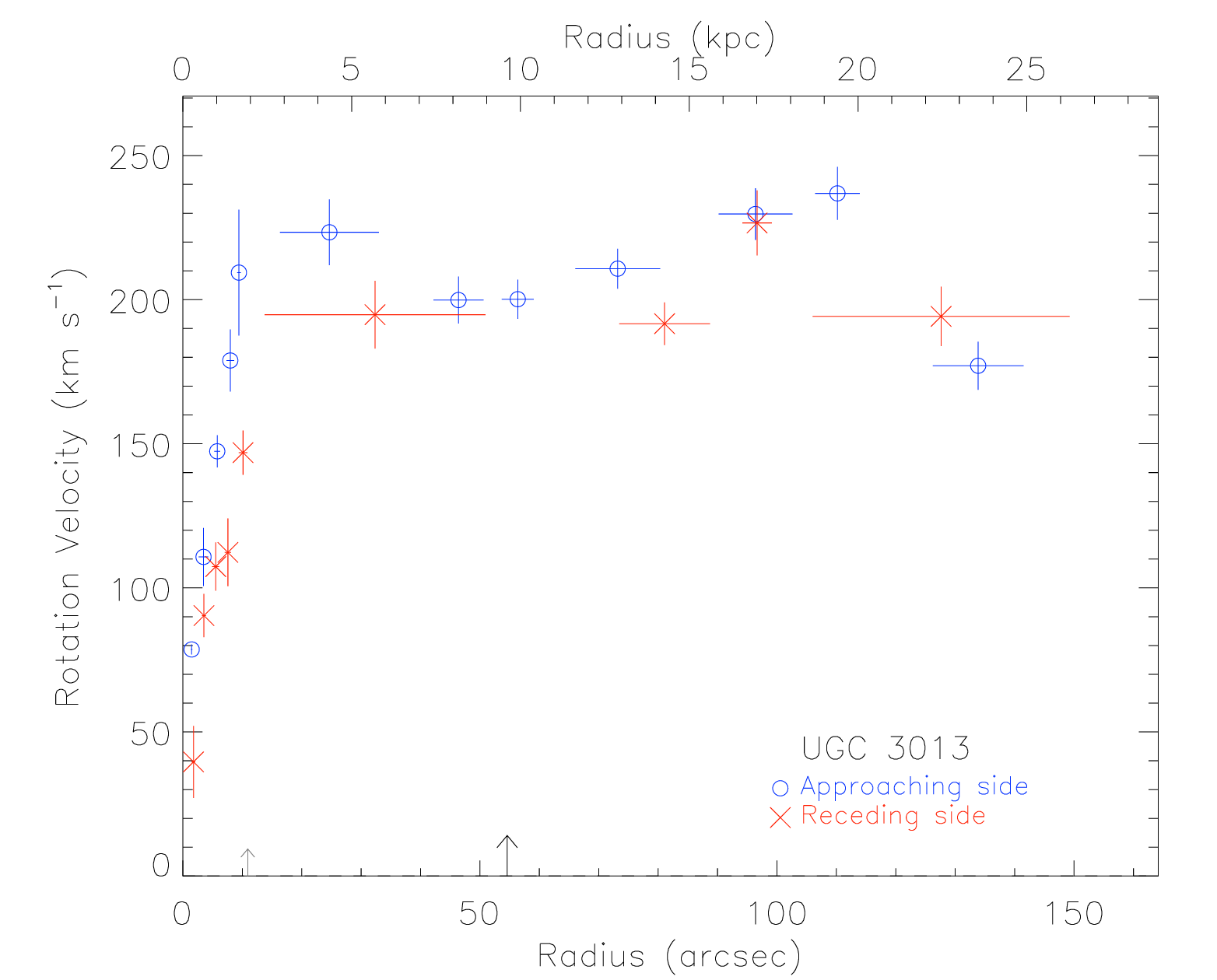}
   \includegraphics[width=8cm]{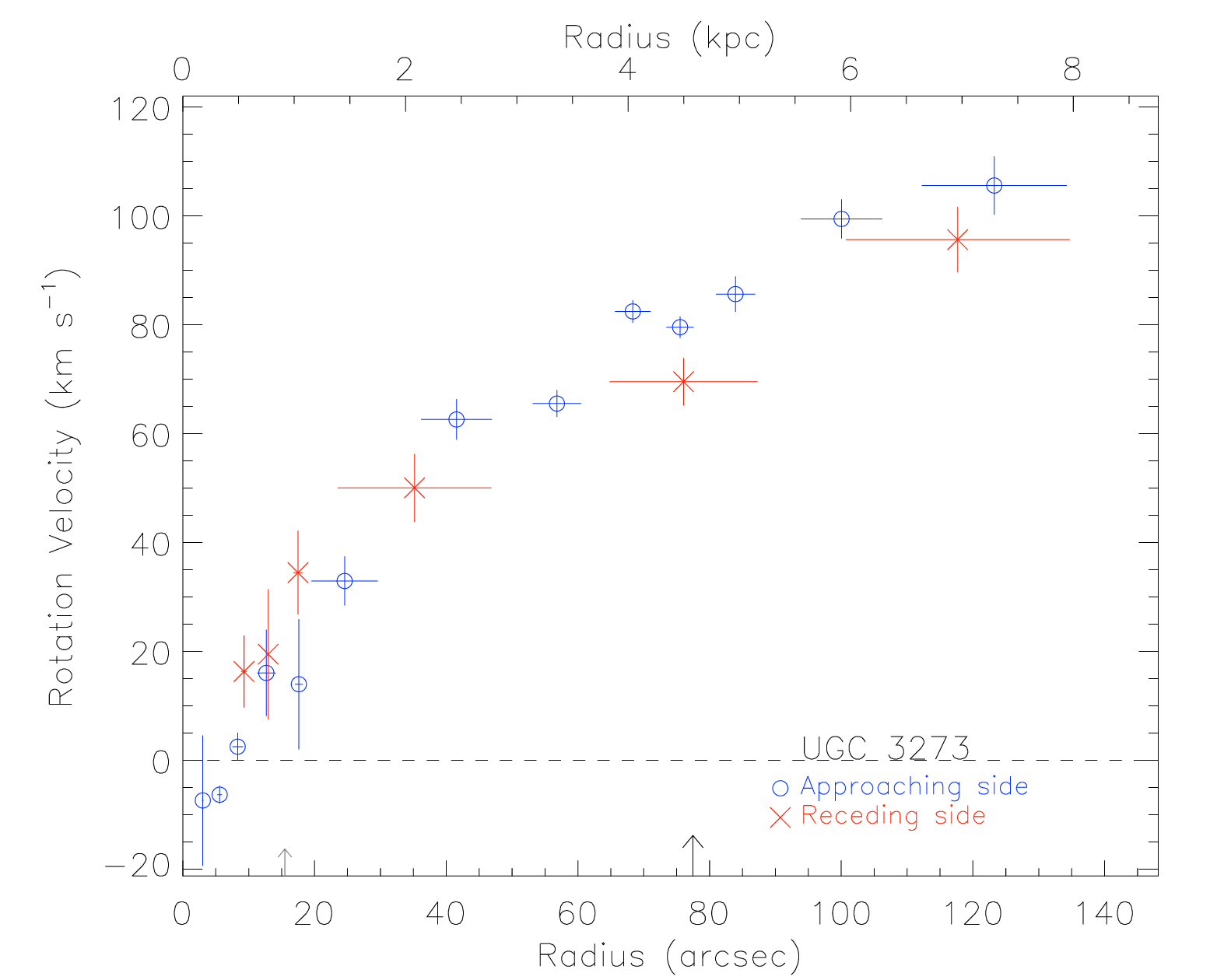}
   \includegraphics[width=8cm]{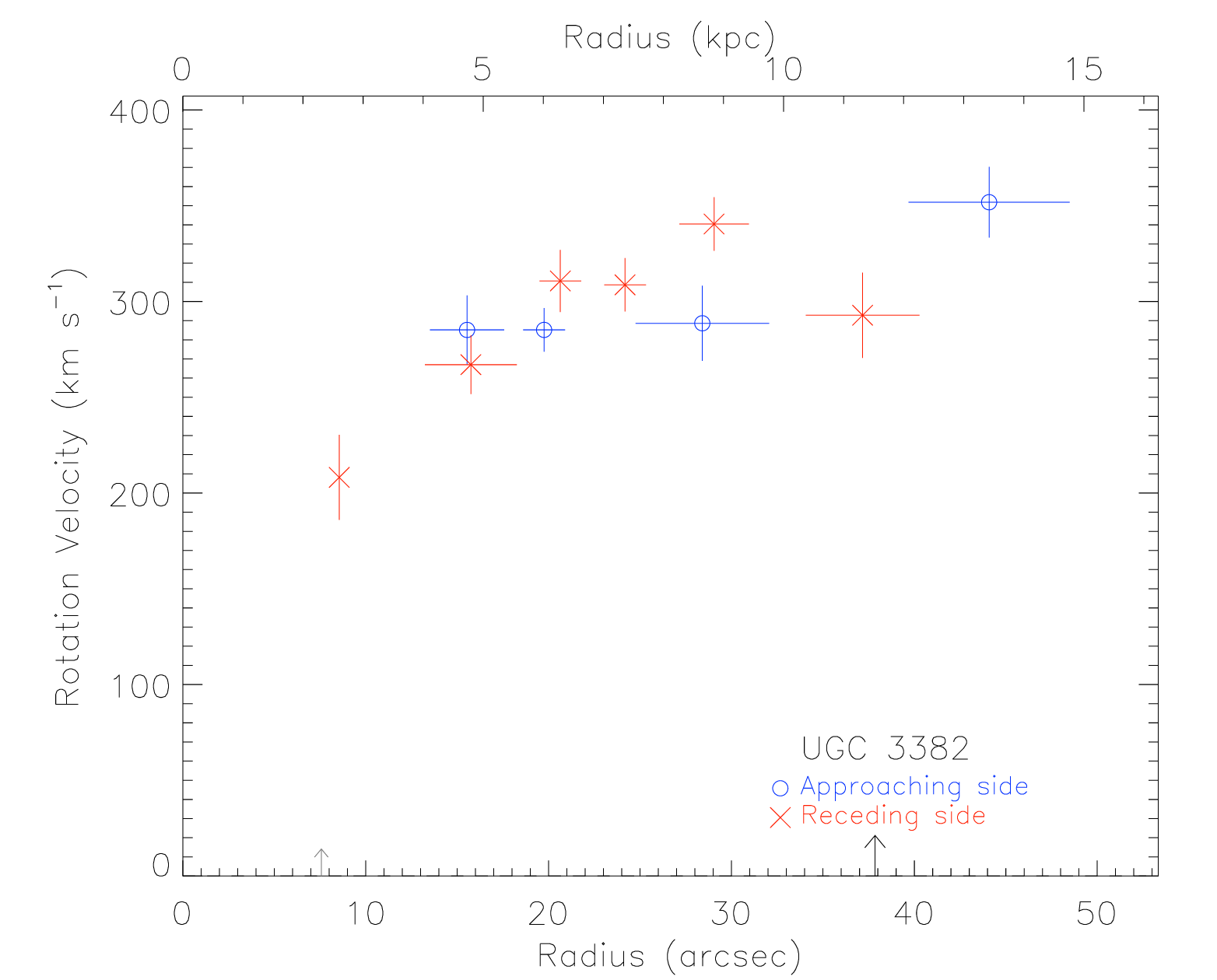}
   \includegraphics[width=8cm]{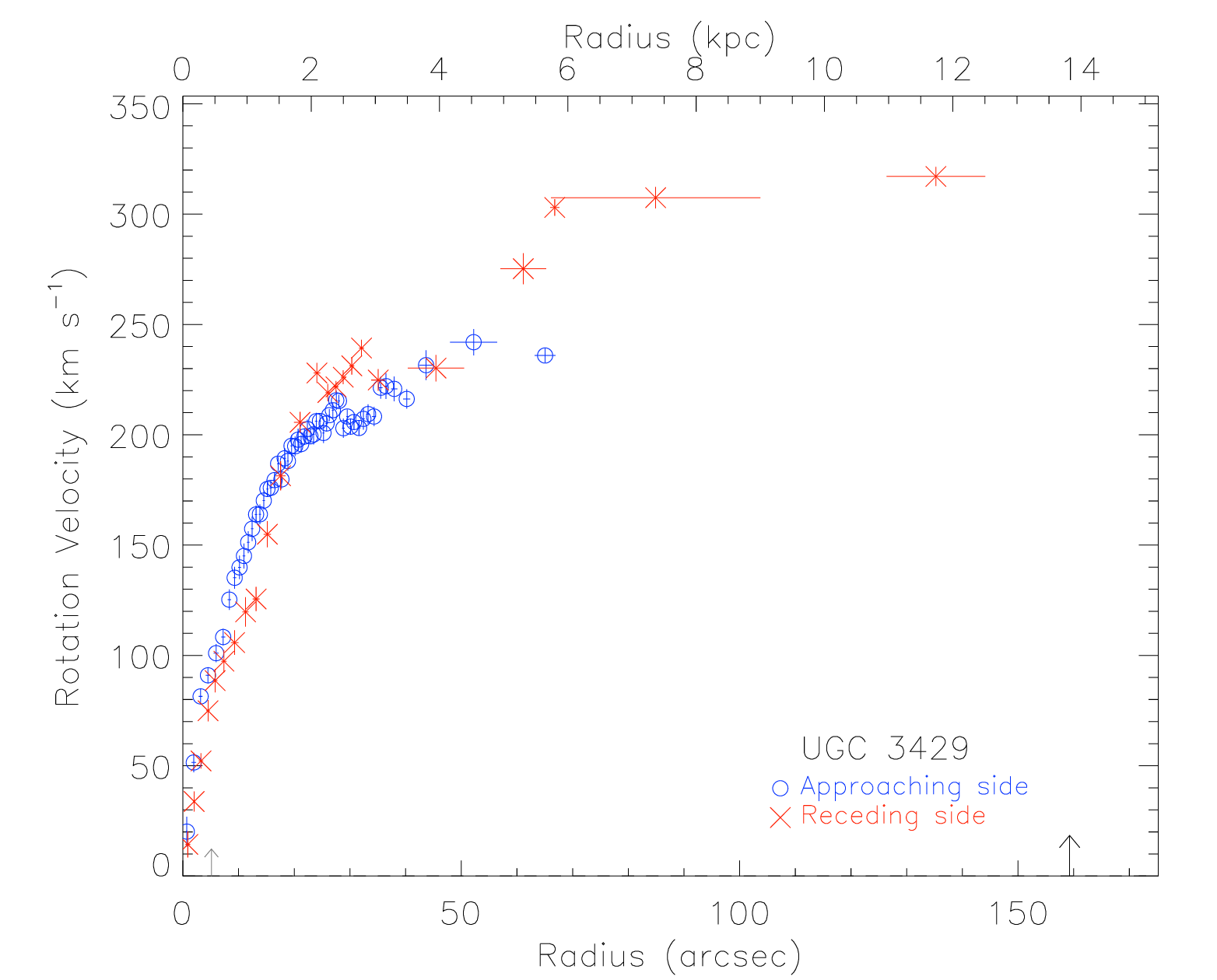}
   \includegraphics[width=8cm]{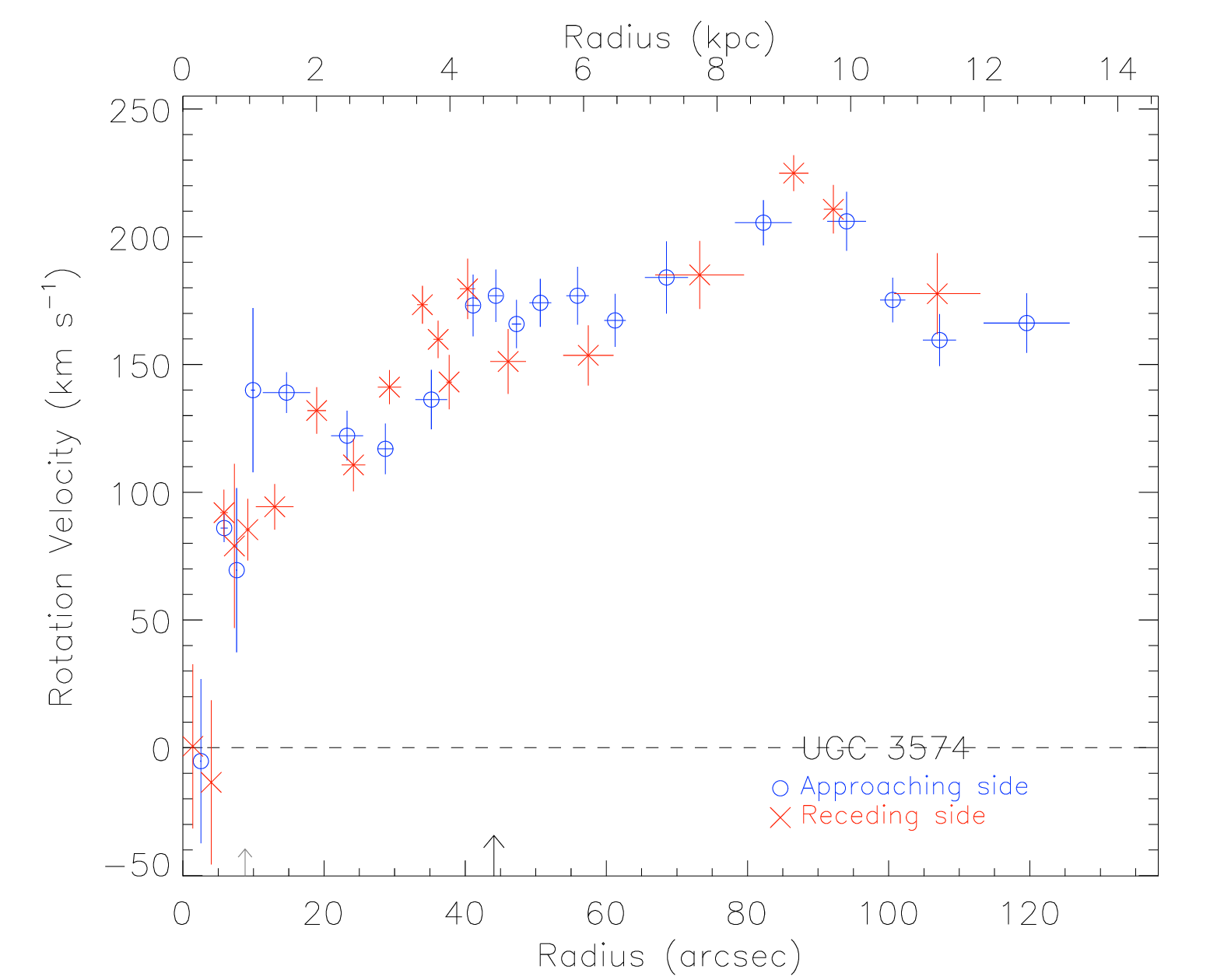}
\end{center}
\caption{From top left to bottom right: \ha~\RC~of UGC 2855, UGC 3013, UGC 3273, UGC 3382, UGC 3429, and UGC 3574.
}
\end{minipage}
\end{figure*}
\clearpage
\begin{figure*}
\begin{minipage}{180mm}
\begin{center}
   \includegraphics[width=8cm]{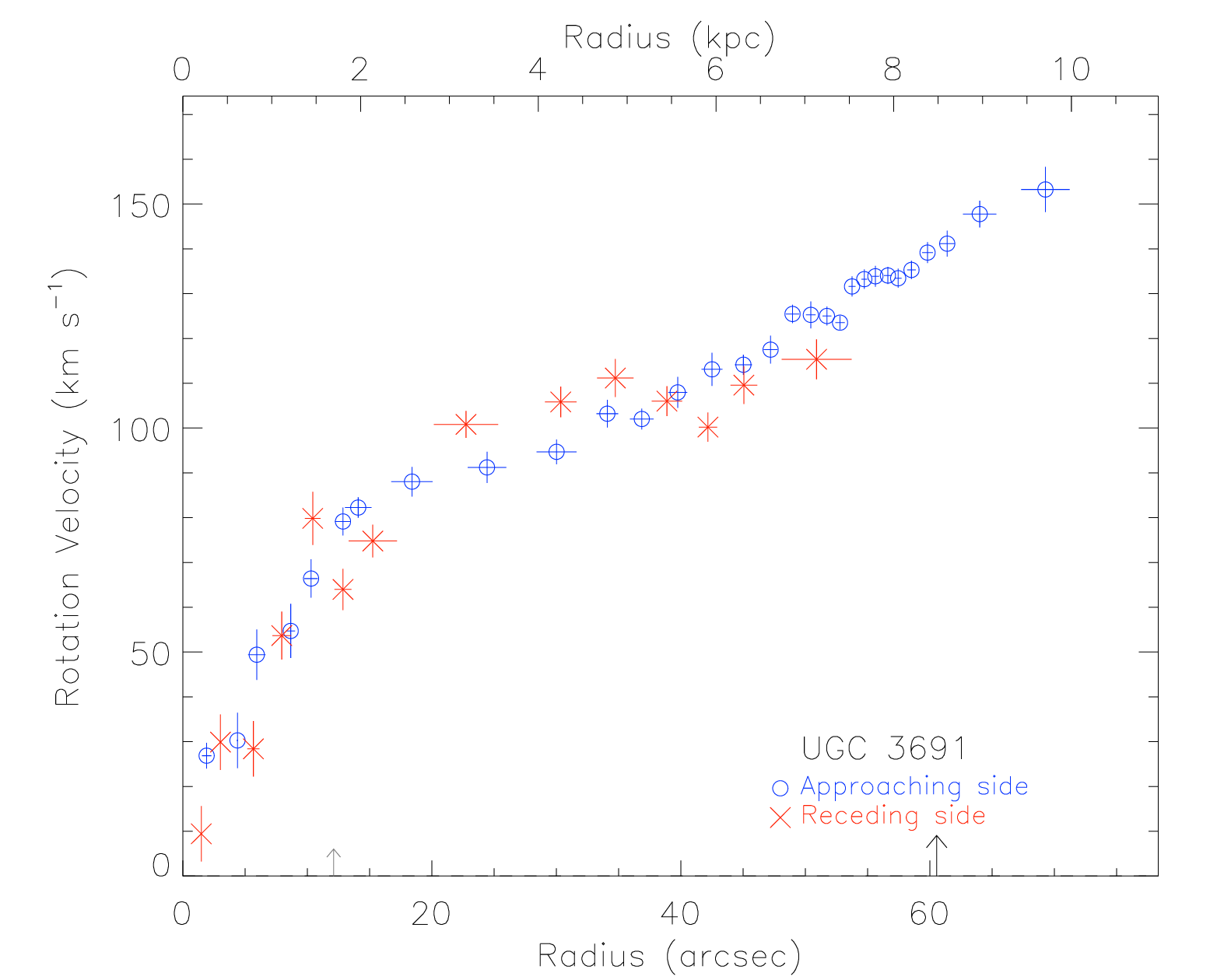}
   \includegraphics[width=8cm]{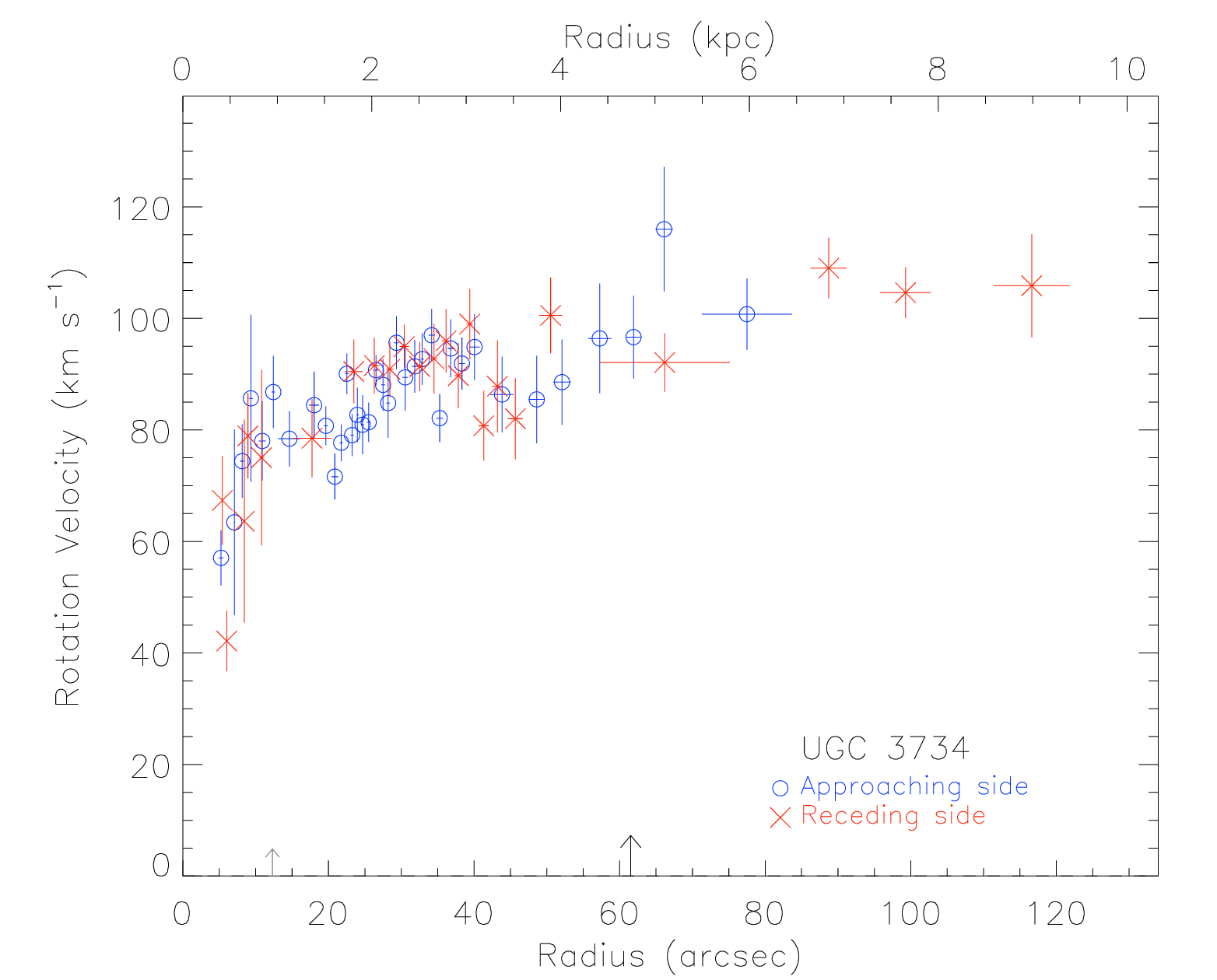}
   \includegraphics[width=8cm]{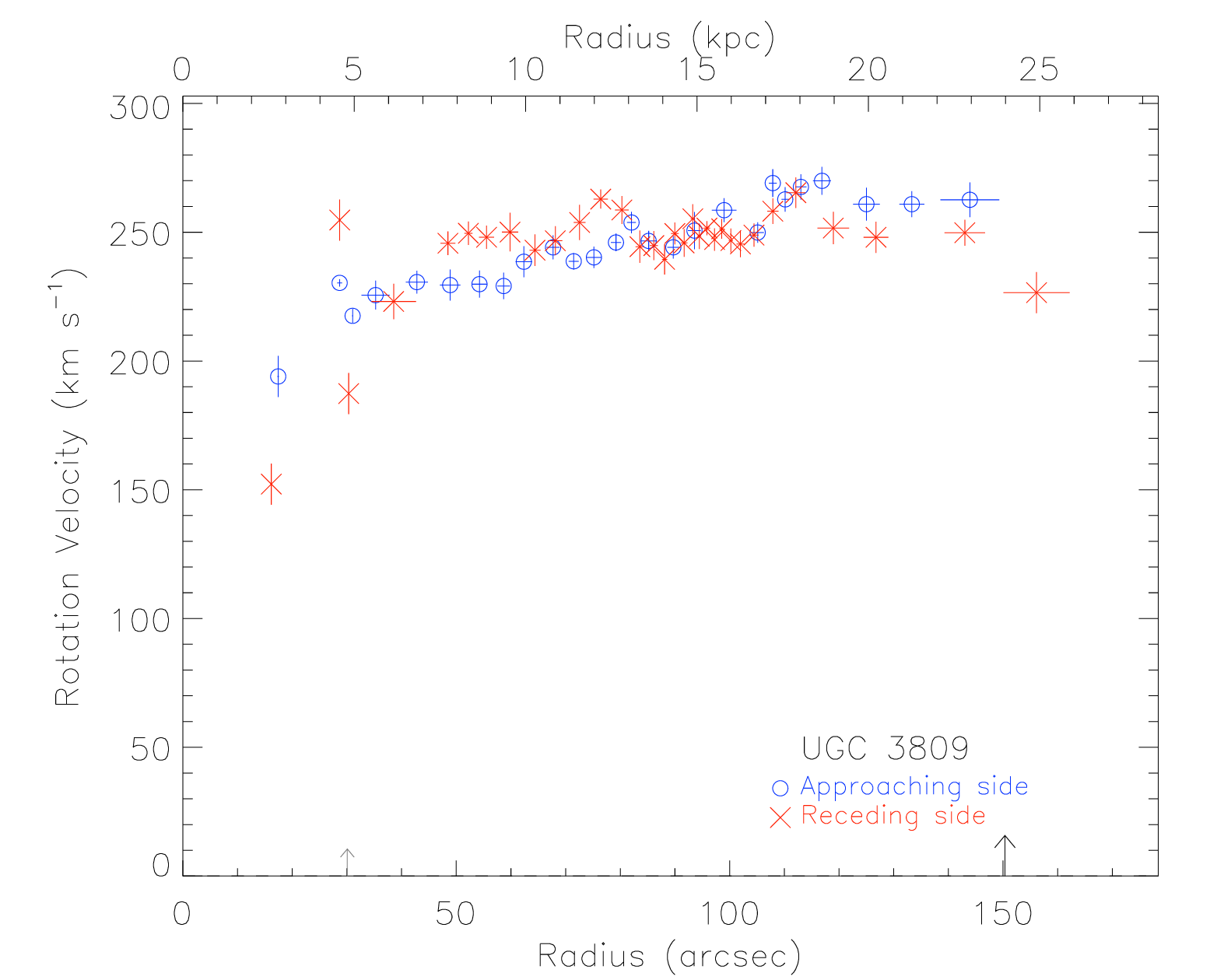}
   \includegraphics[width=8cm]{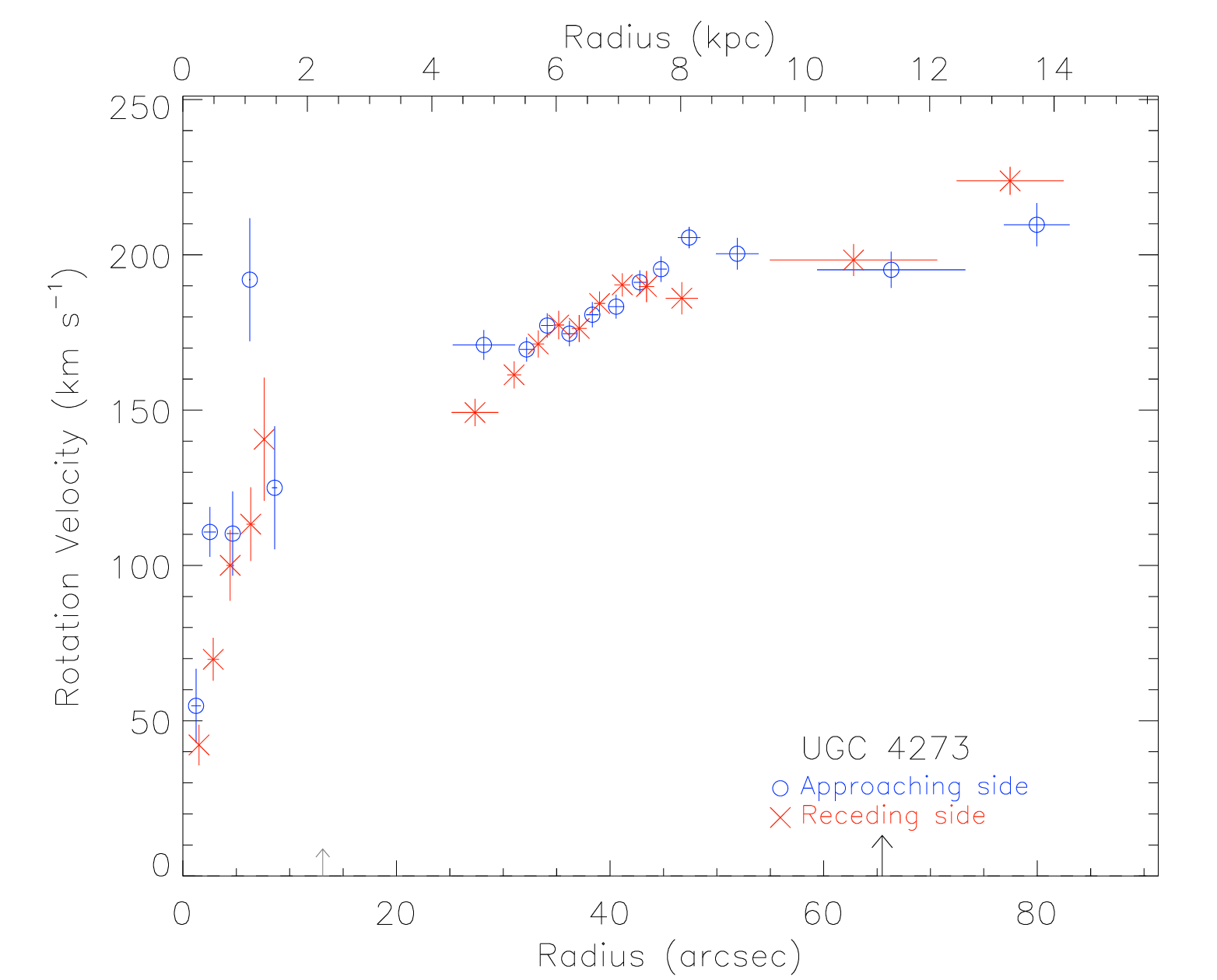}
   \includegraphics[width=8cm]{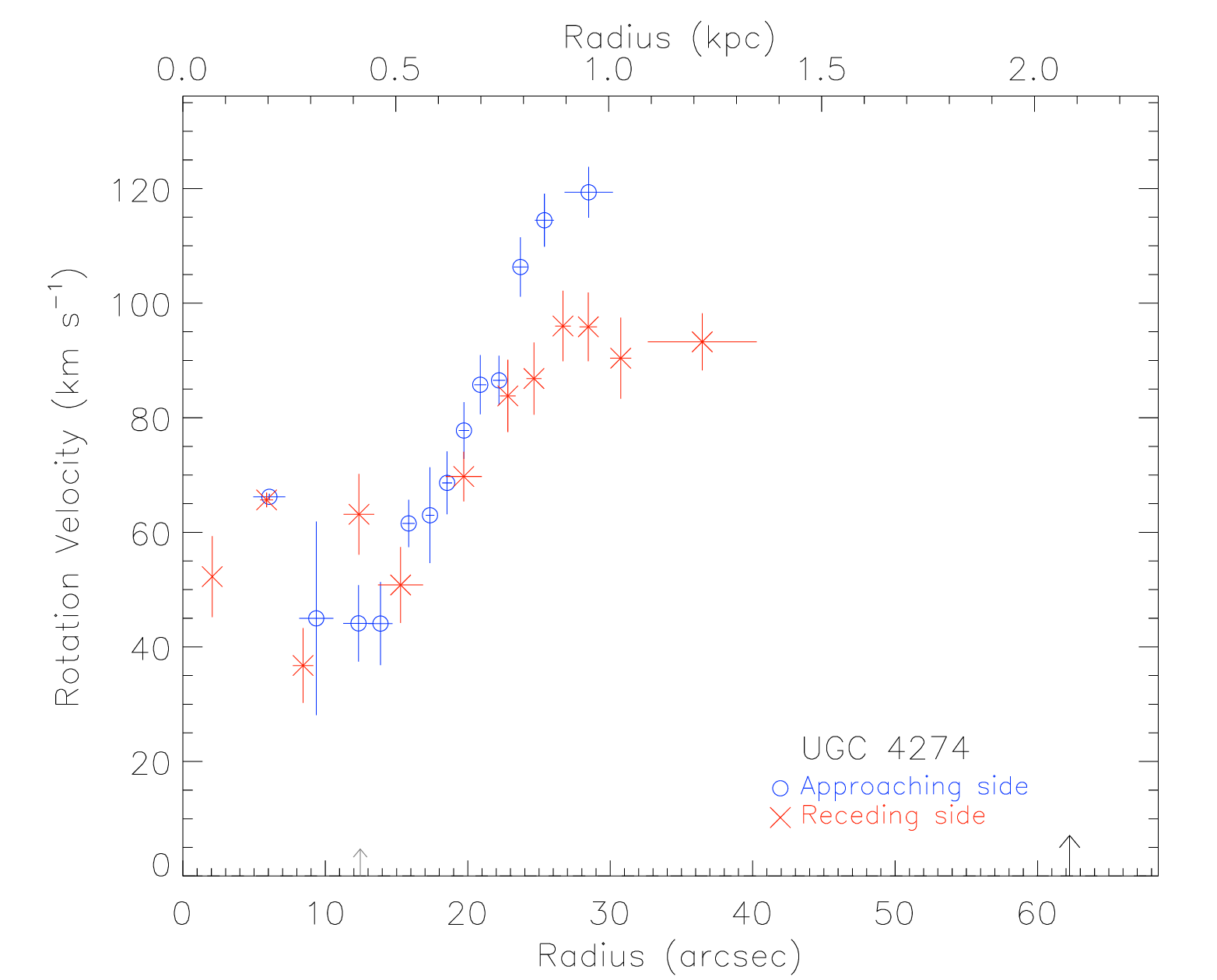}
   \includegraphics[width=8cm]{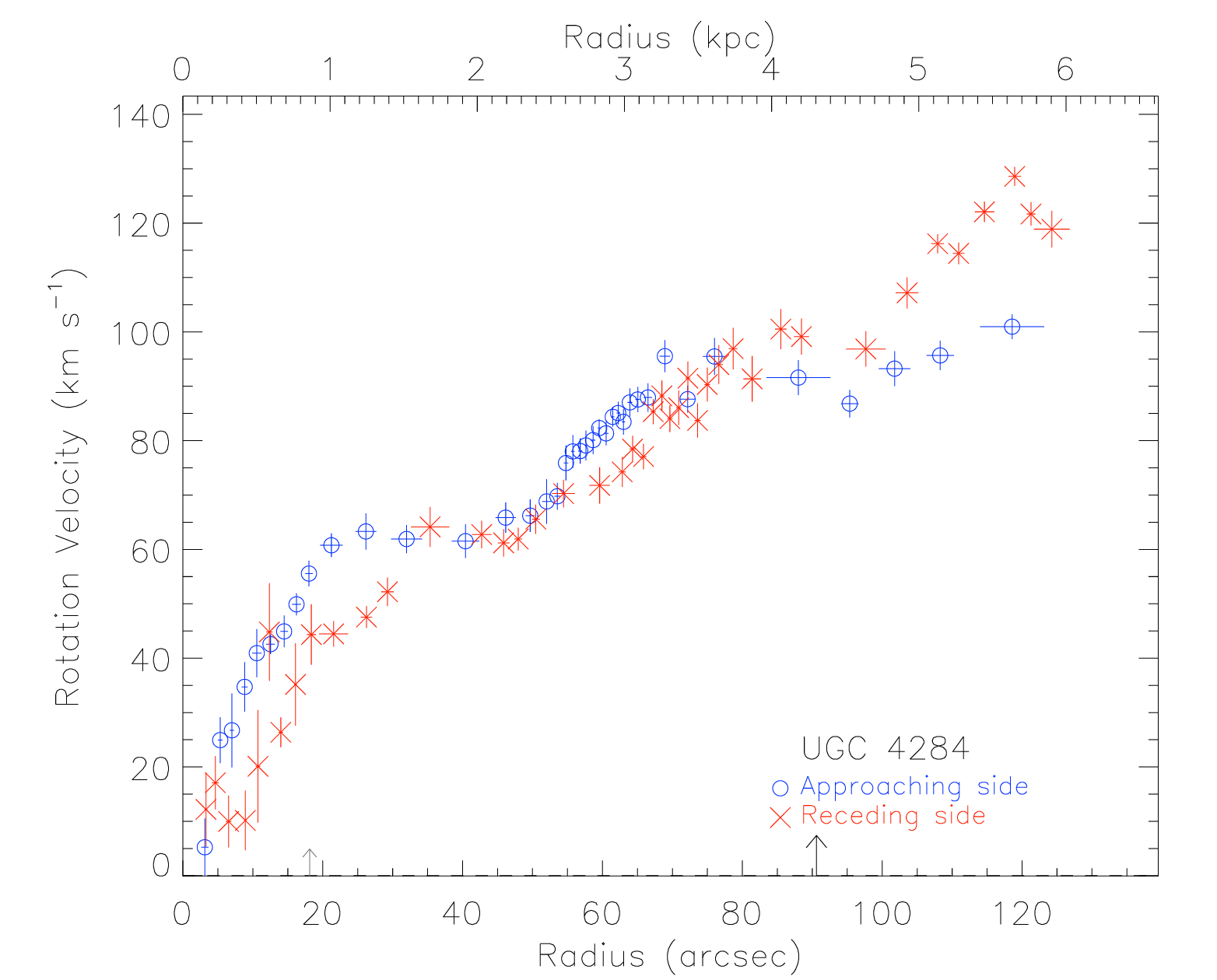}
\end{center}
\caption{From top left to bottom right: \ha~\RC~of UGC 3691, UGC 3734, UGC 3809, UGC 4273, UGC 4274, and UGC 4284.
}
\end{minipage}
\end{figure*}
\clearpage
\begin{figure*}
\begin{minipage}{180mm}
\begin{center}
   \includegraphics[width=8cm]{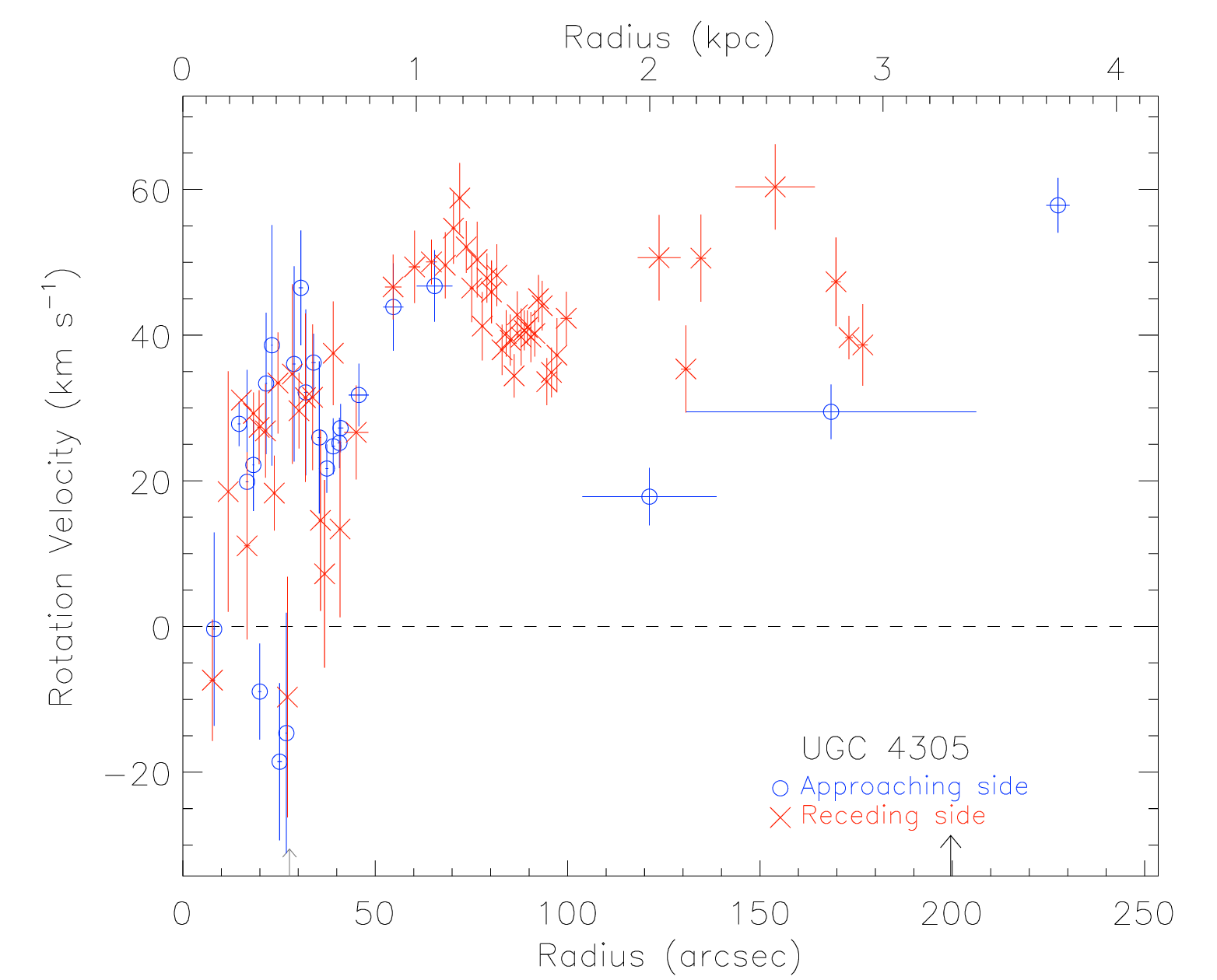}
   \includegraphics[width=8cm]{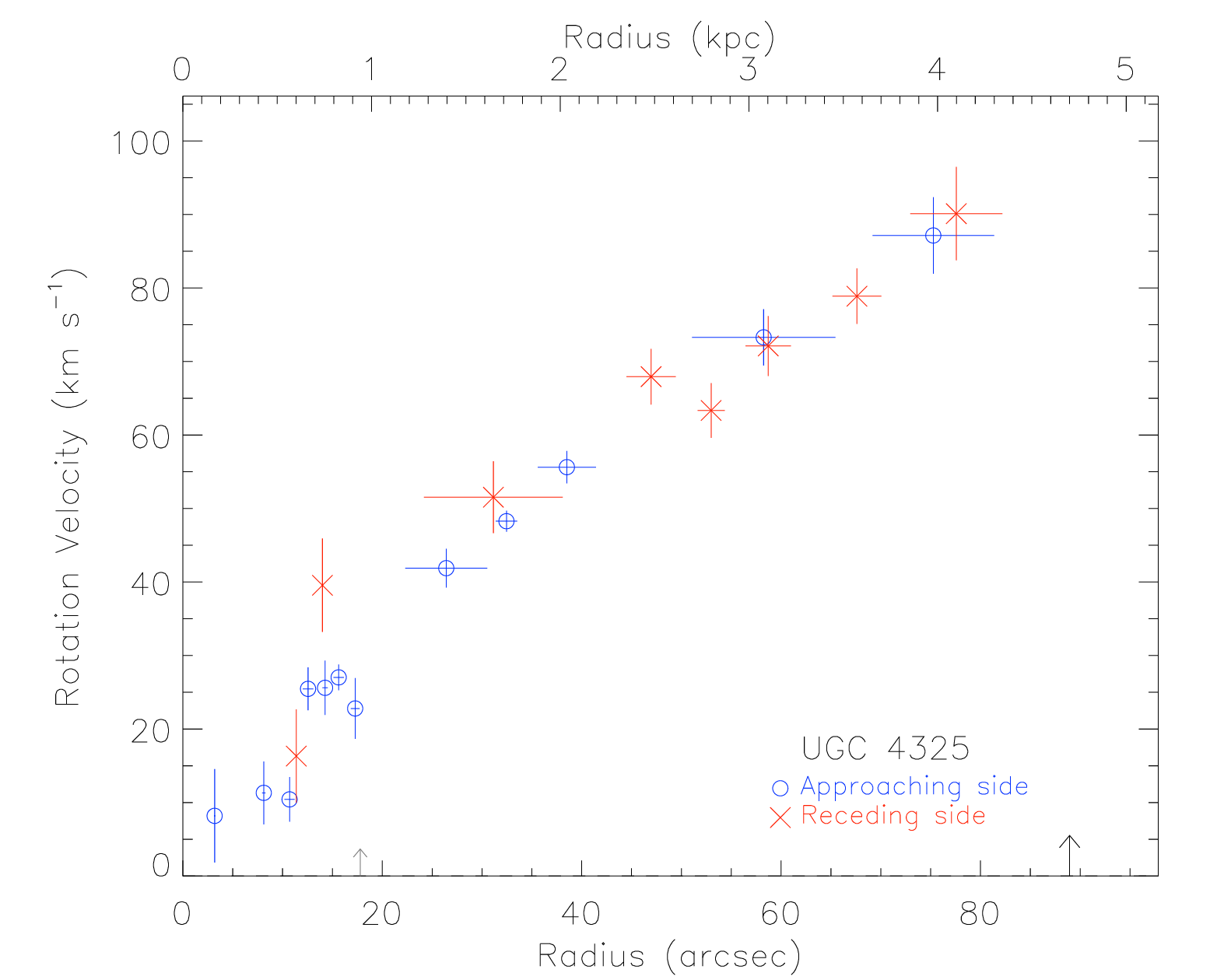}
   \includegraphics[width=8cm]{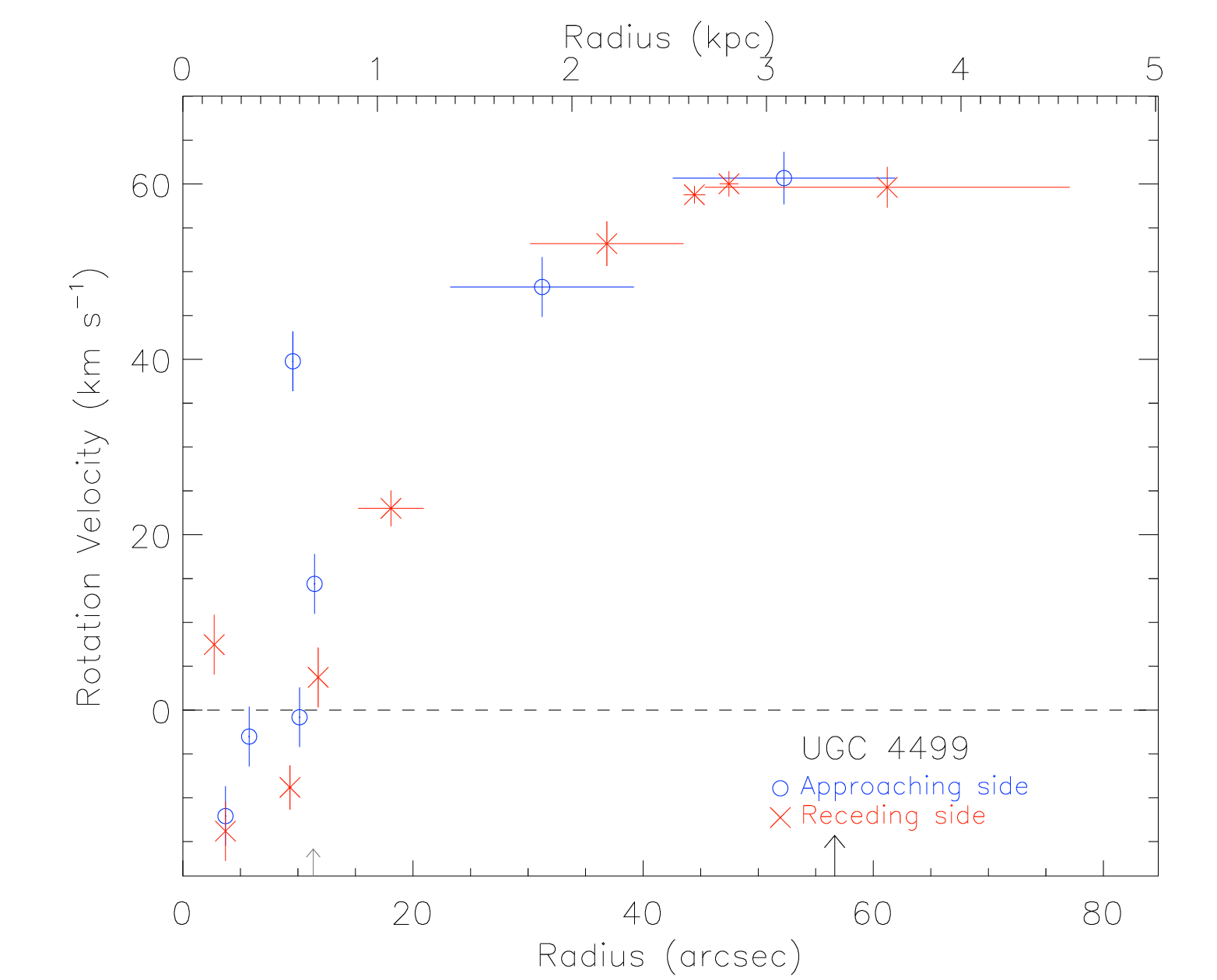}
   \includegraphics[width=8cm]{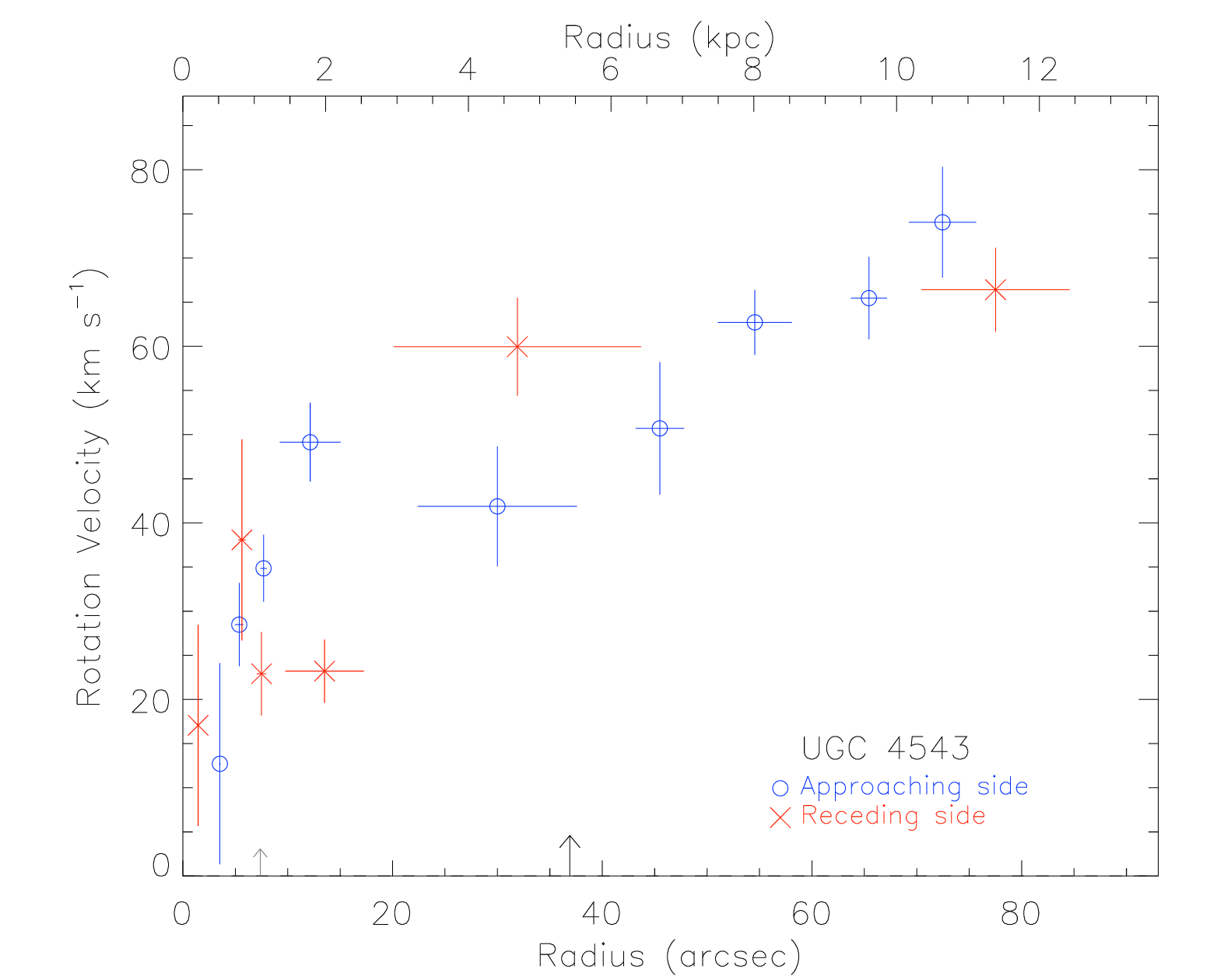}
   \includegraphics[width=8cm]{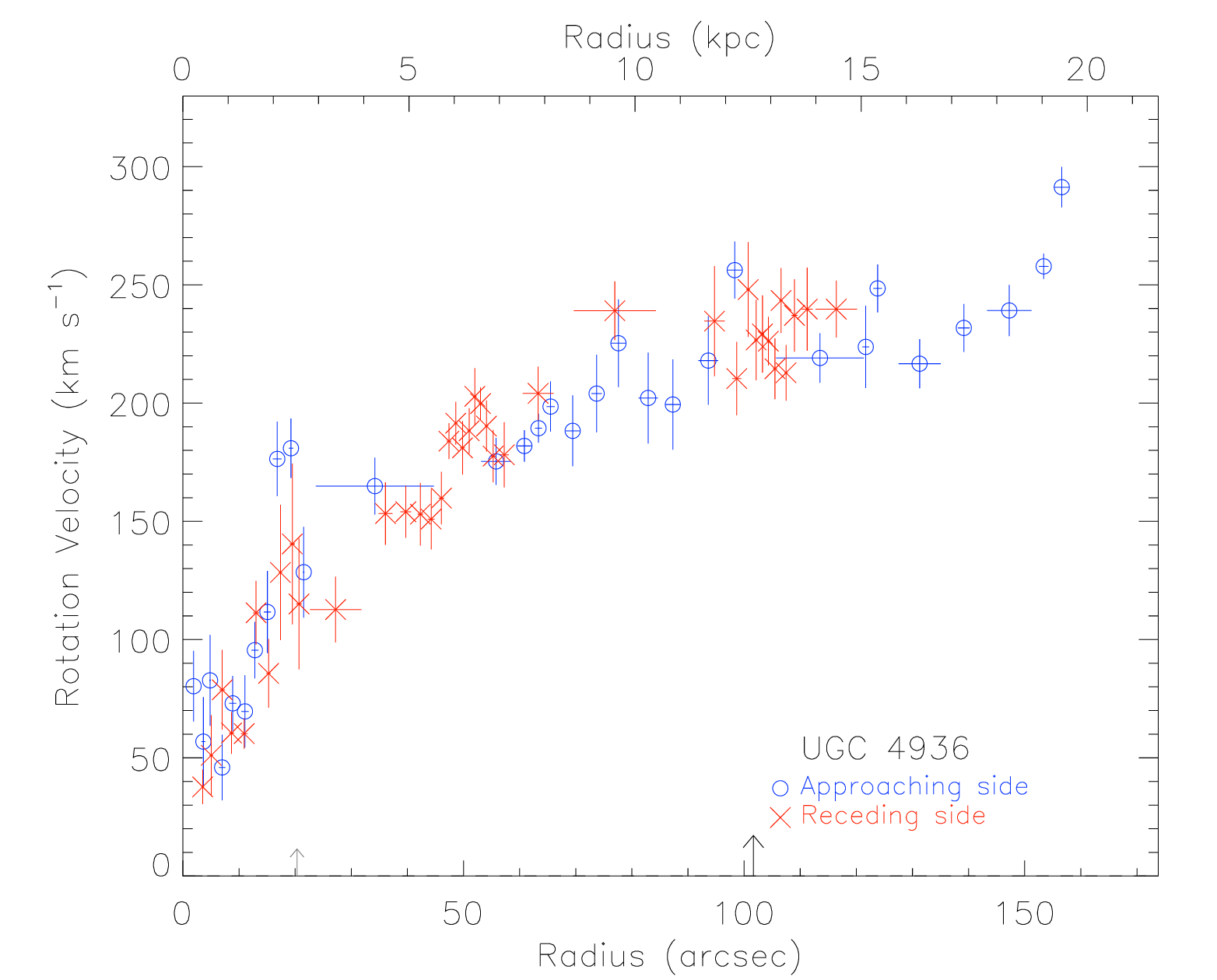}
   \includegraphics[width=8cm]{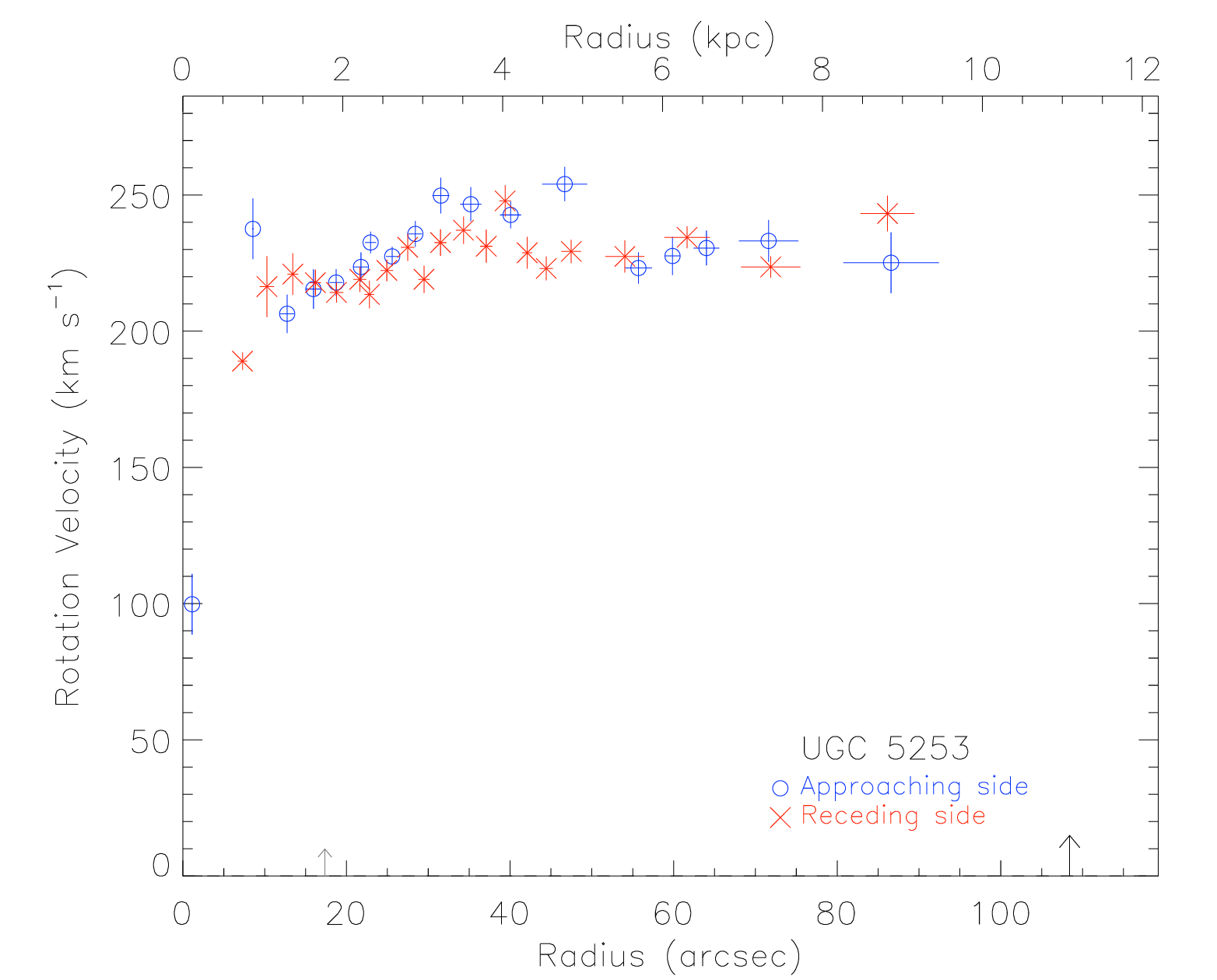}
\end{center}
\caption{From top left to bottom right: \ha~\RC~of UGC 4305, UGC 4325, UGC 4499, UGC 4543, UGC 4936, and UGC 5253.
}
\end{minipage}
\end{figure*}
\clearpage
\begin{figure*}
\begin{minipage}{180mm}
\begin{center}
   \includegraphics[width=8cm]{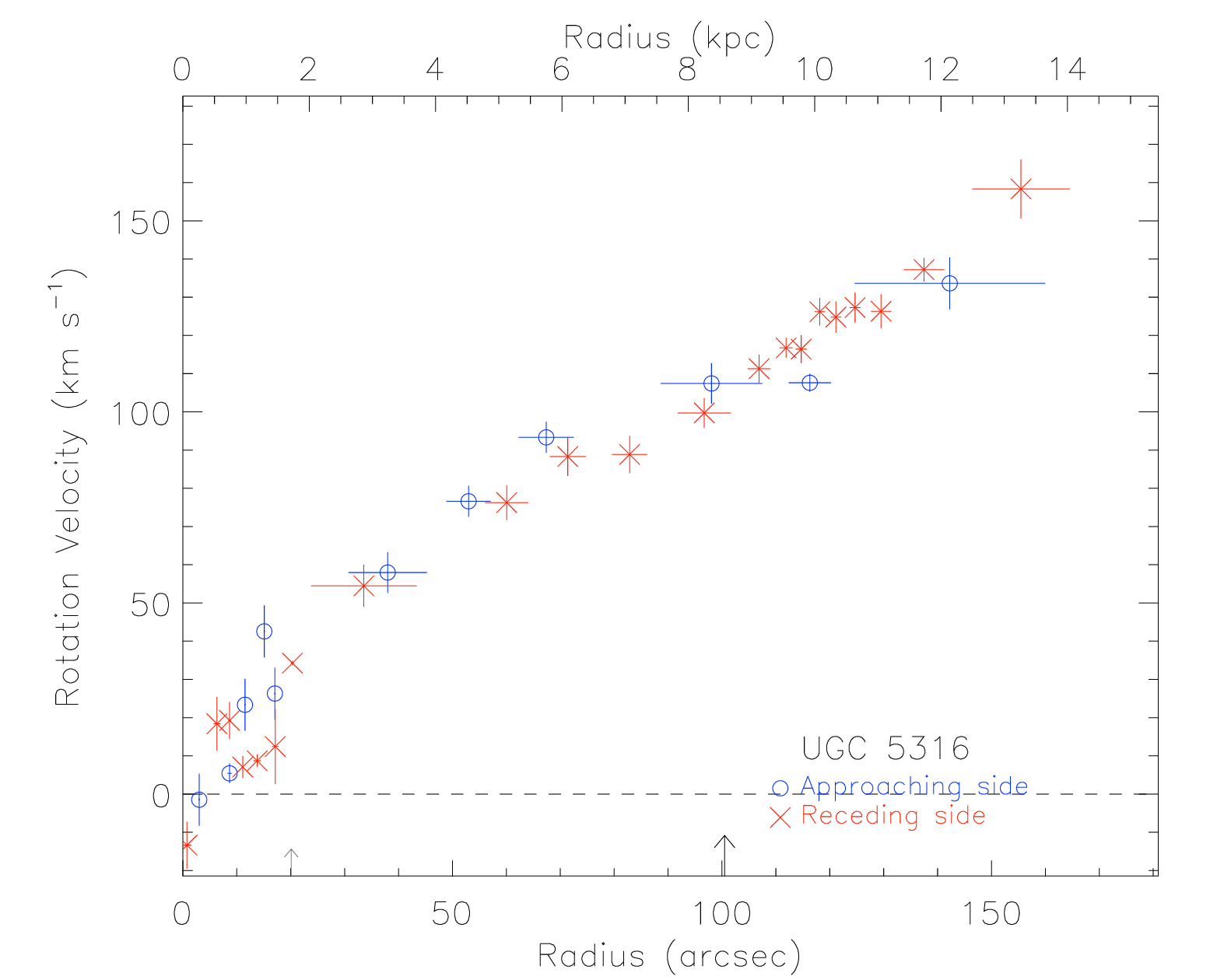}
   \includegraphics[width=8cm]{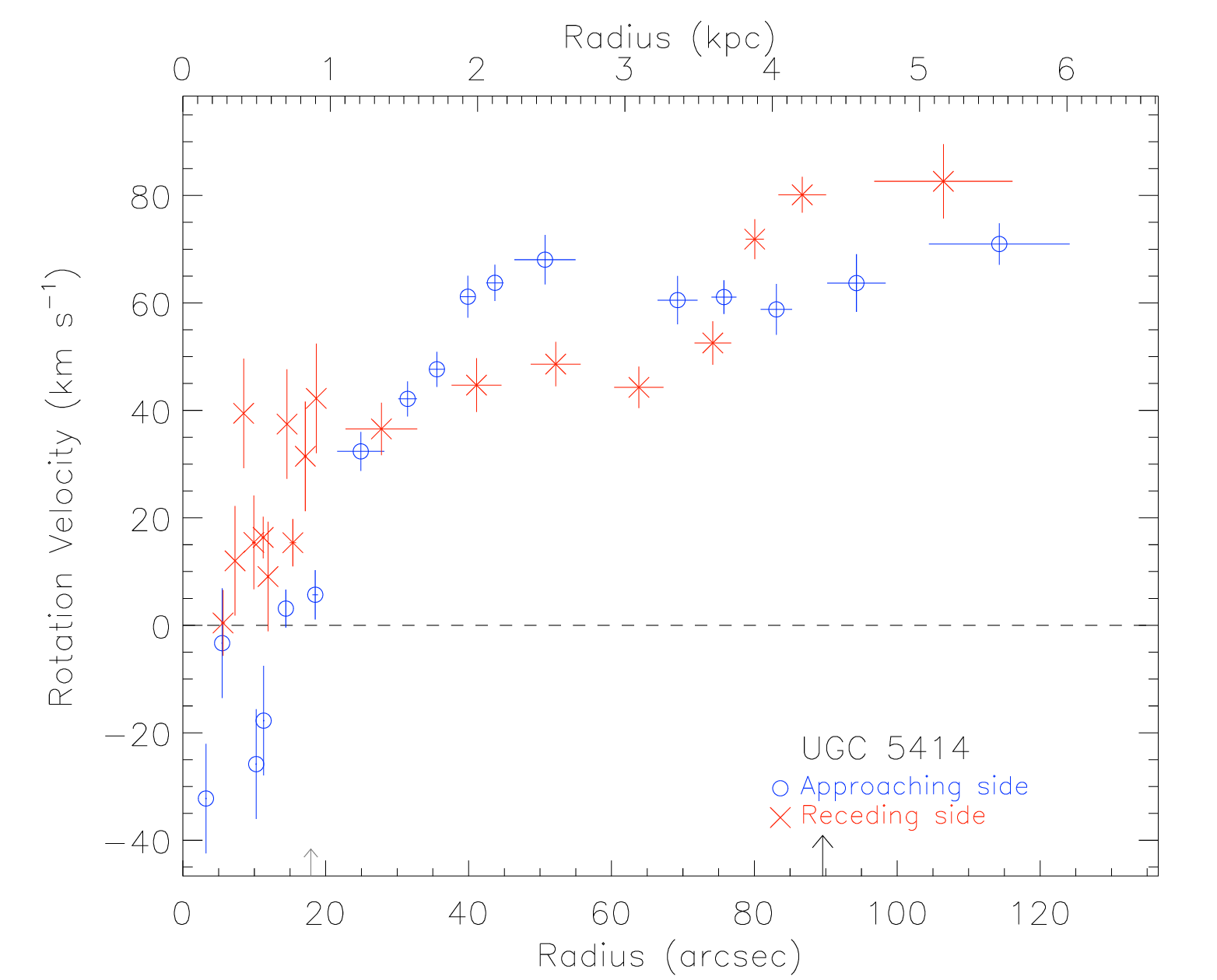}
   \includegraphics[width=8cm]{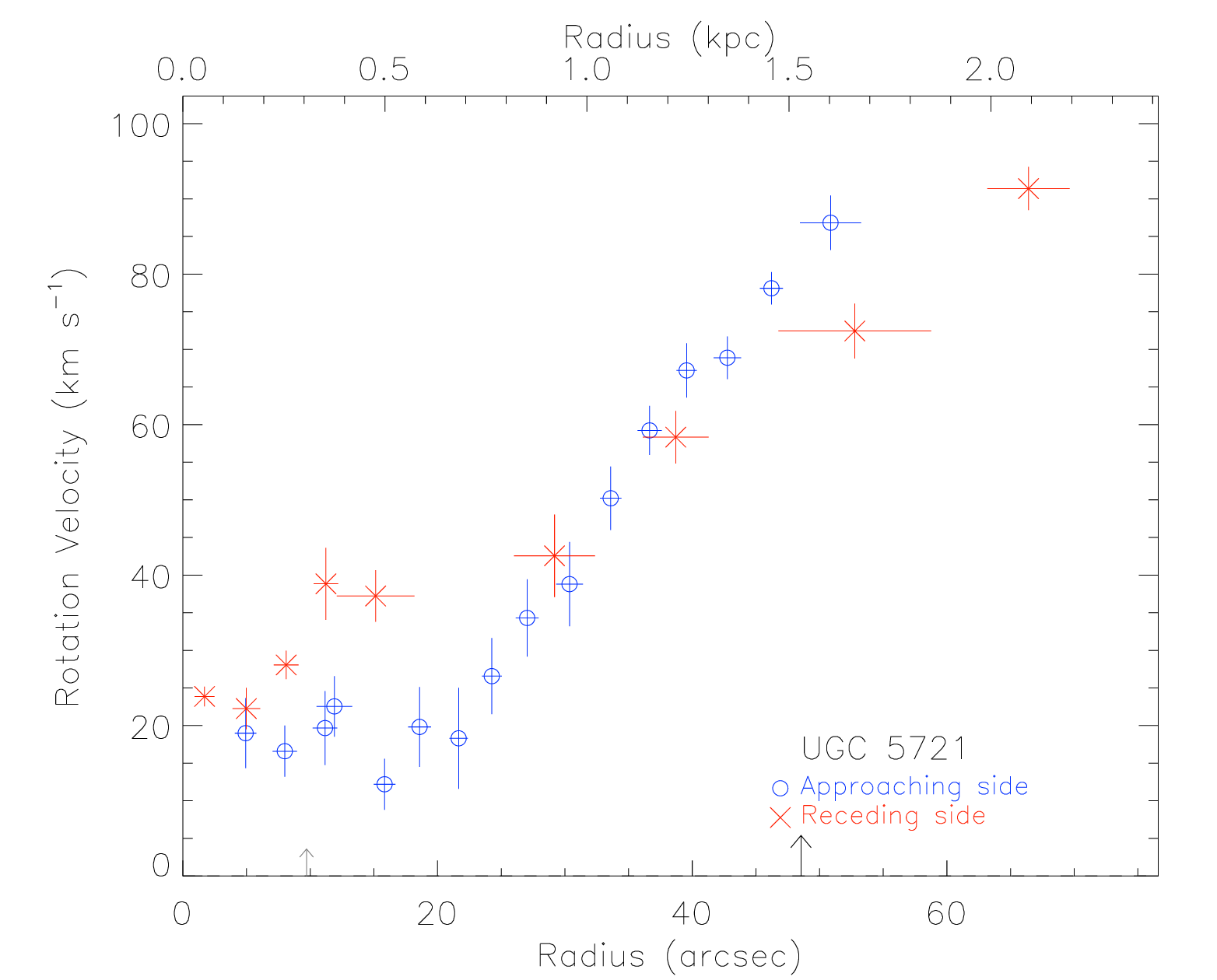}
   \includegraphics[width=8cm]{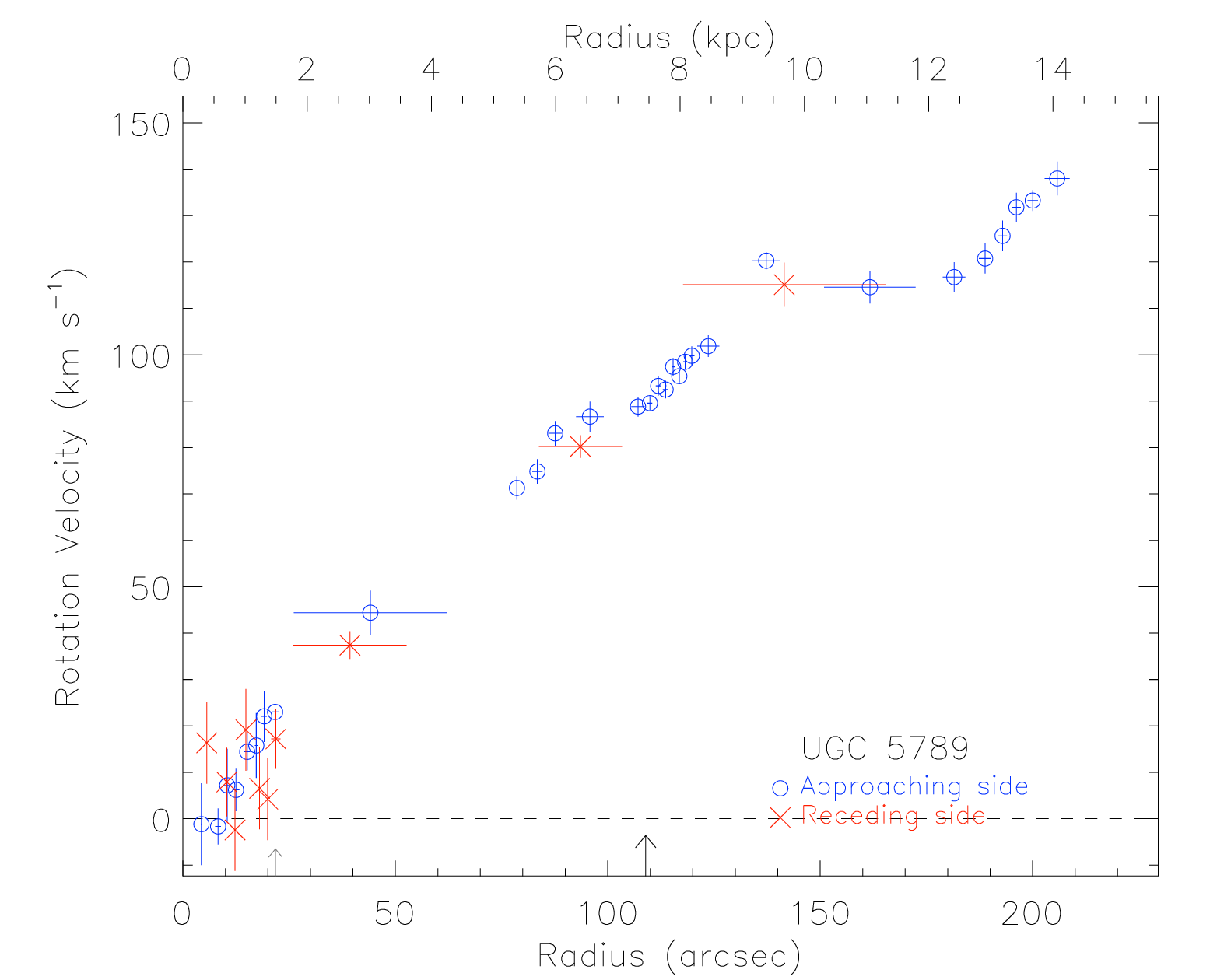}
   \includegraphics[width=8cm]{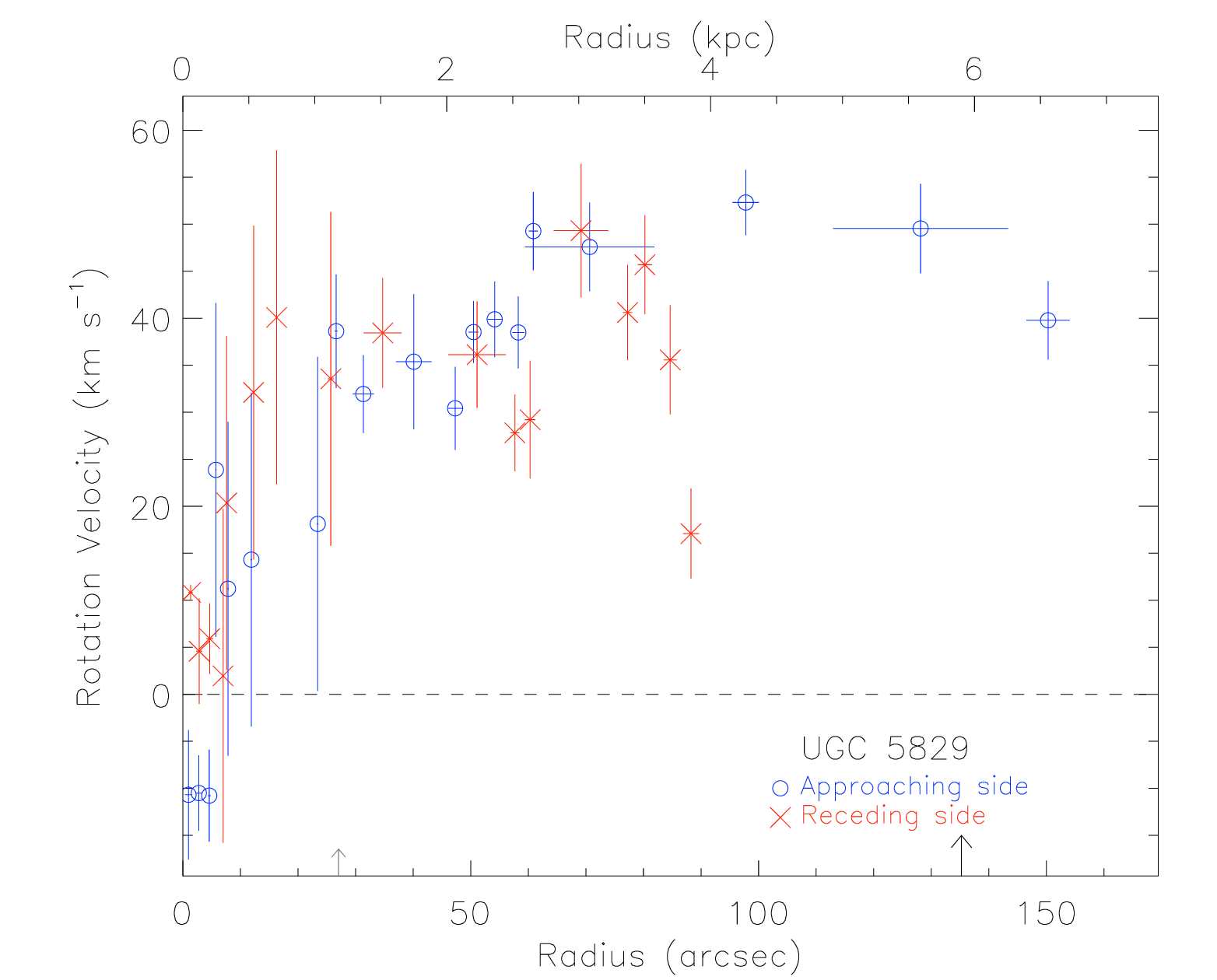}
   \includegraphics[width=8cm]{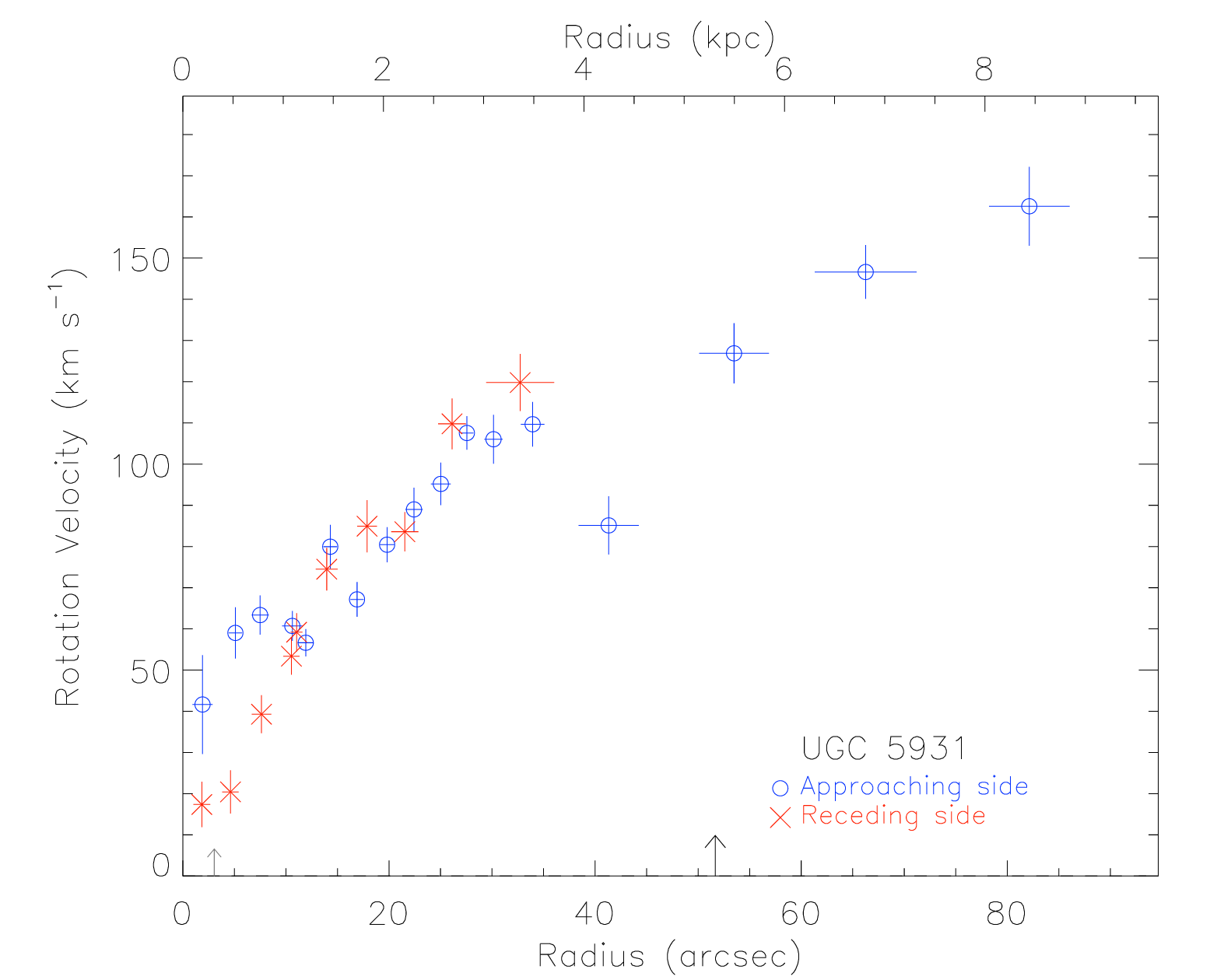}
\end{center}
\caption{From top left to bottom right: \ha~\RC~of UGC 5316, UGC 5414, UGC 5721, UGC 5789, UGC 5829, and UGC 5931.
}
\end{minipage}
\end{figure*}
\clearpage
\begin{figure*}
\begin{minipage}{180mm}
\begin{center}
   \includegraphics[width=8cm]{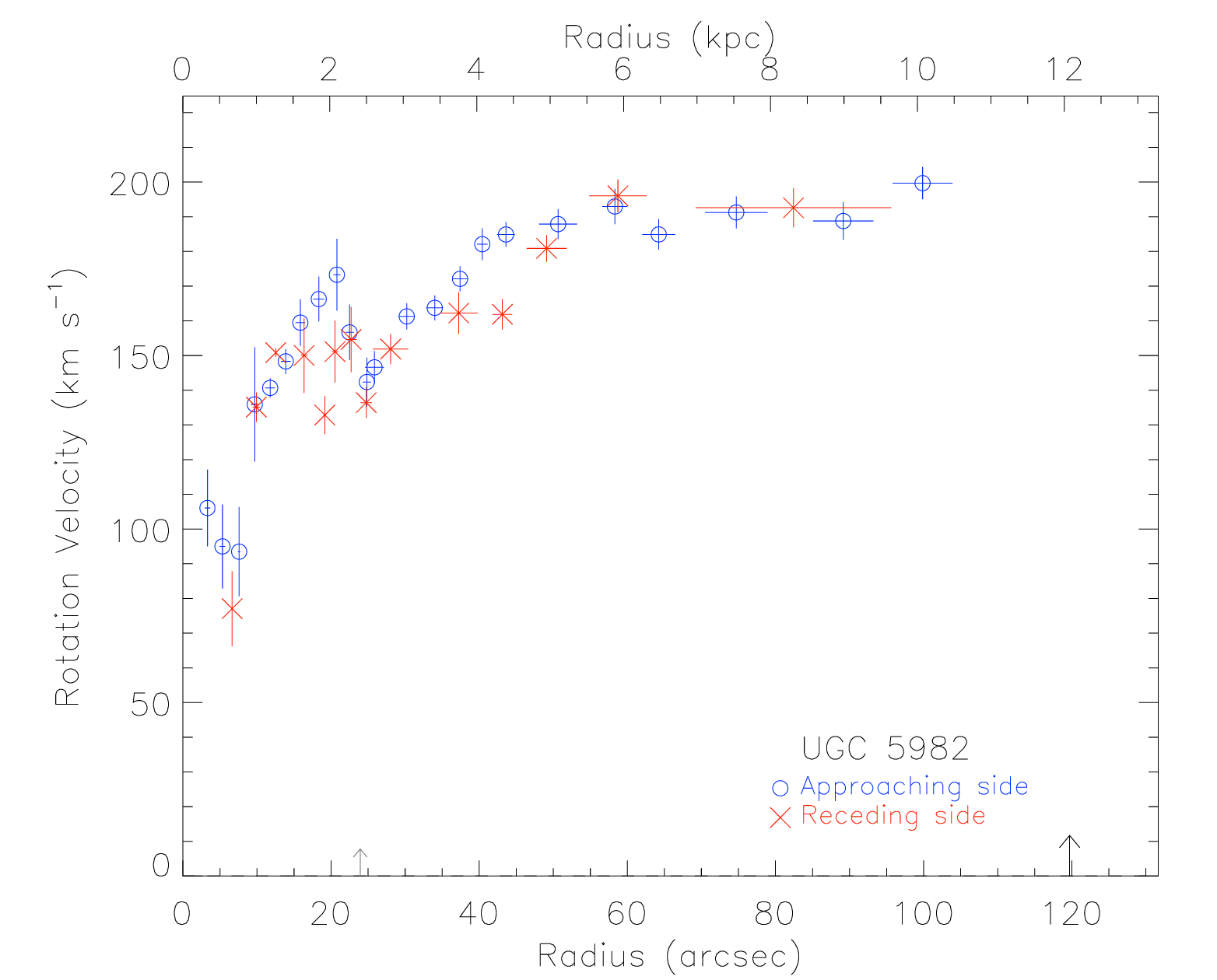}
   \includegraphics[width=8cm]{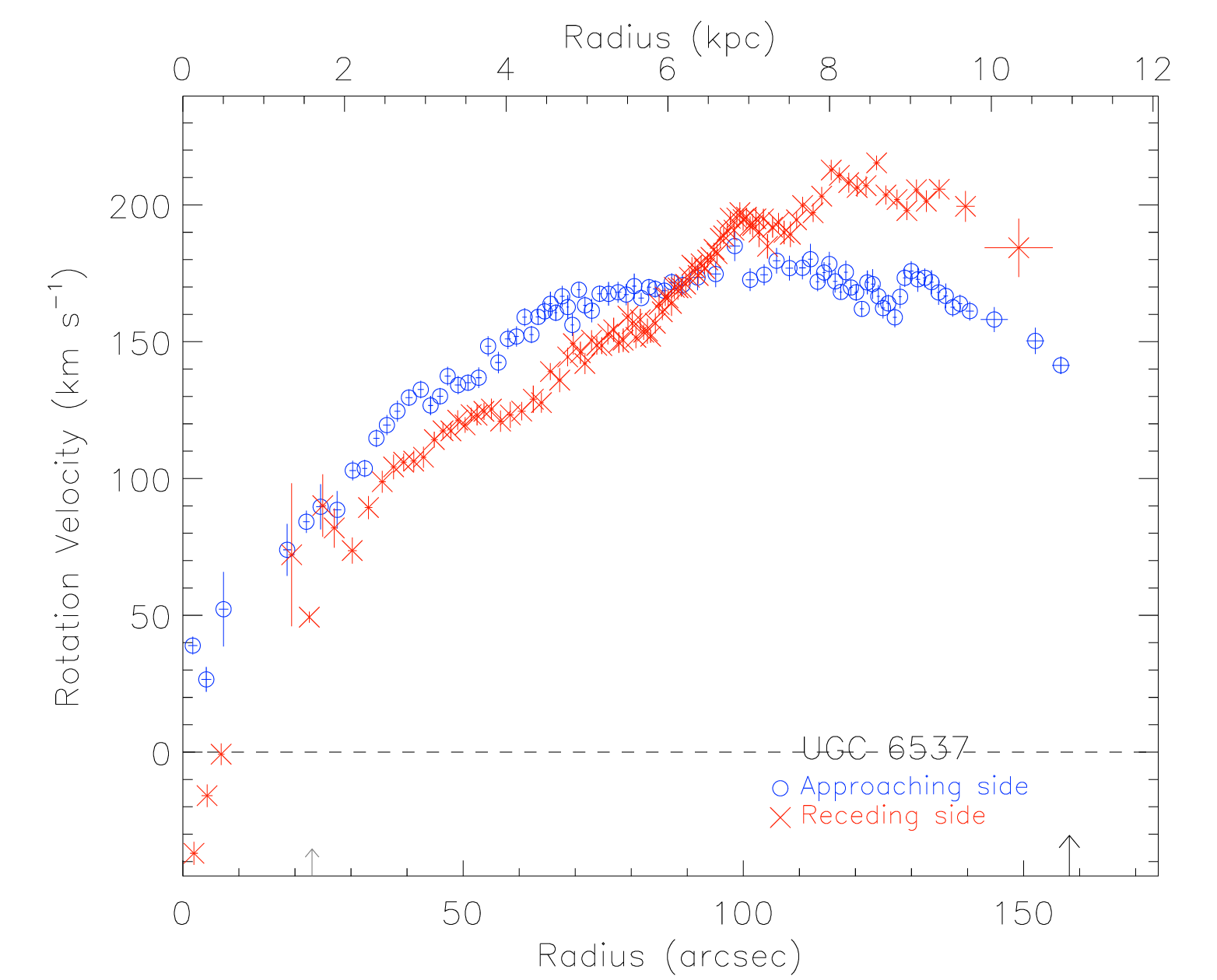}
   \includegraphics[width=8cm]{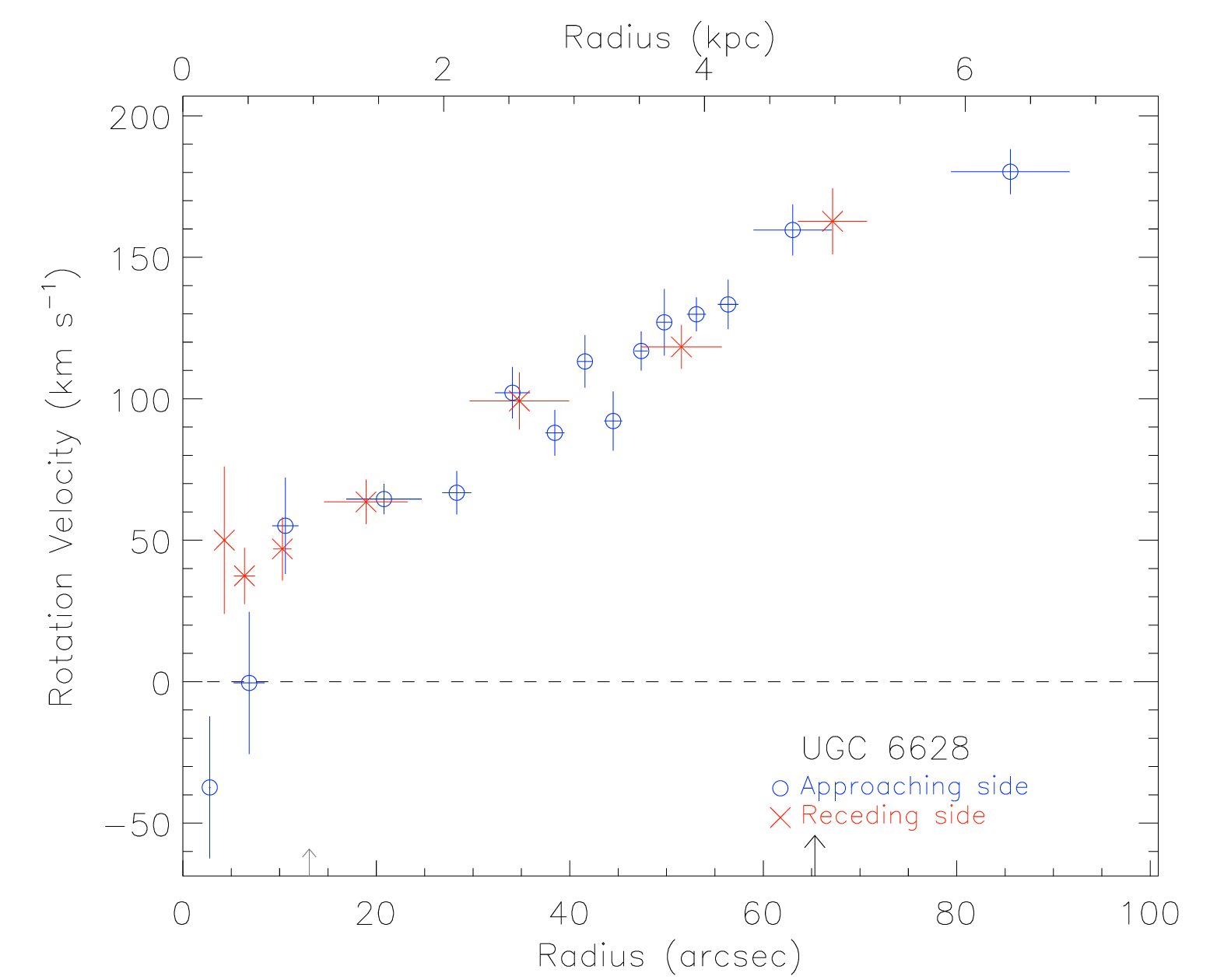}
   \includegraphics[width=8cm]{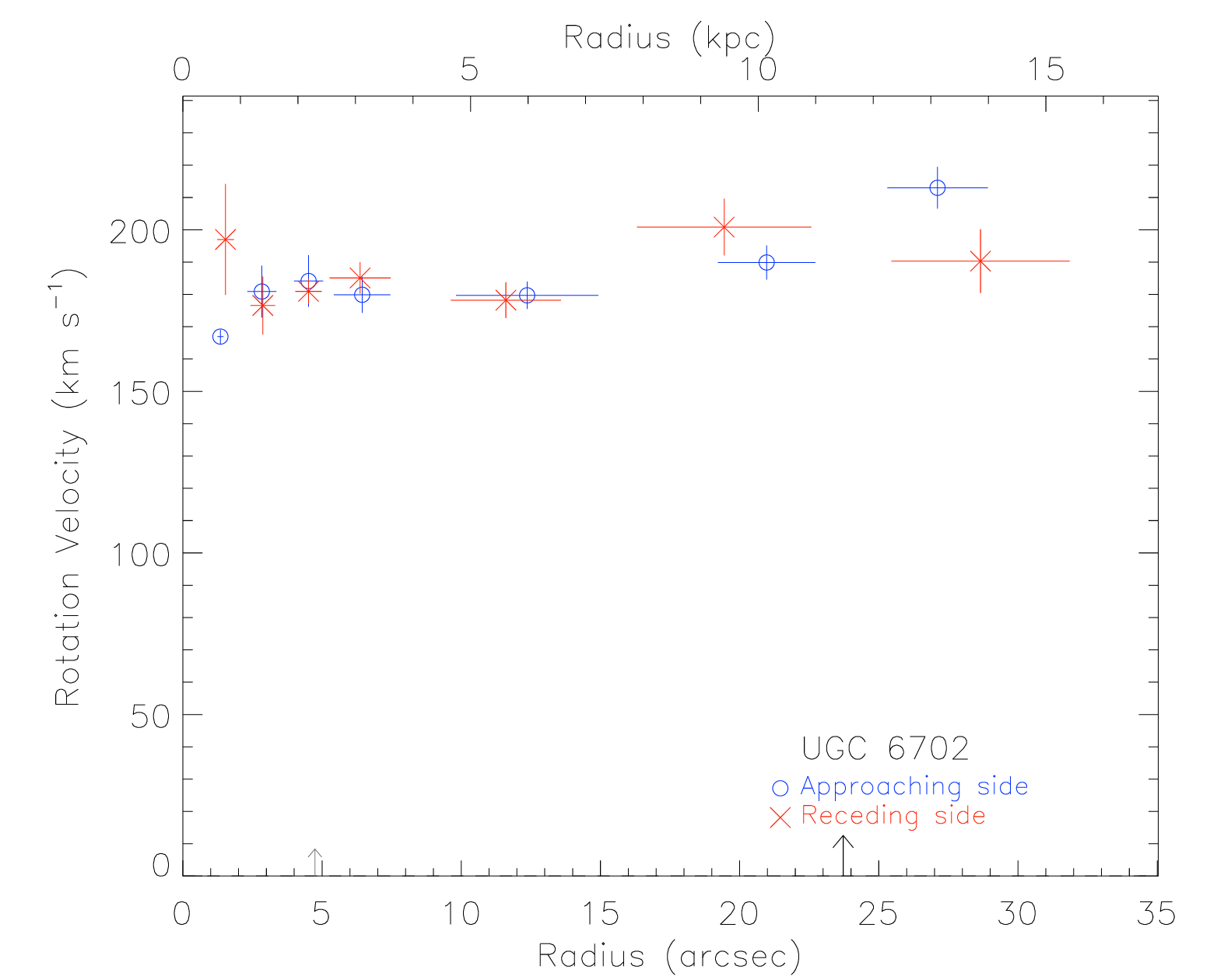}
   \includegraphics[width=8cm]{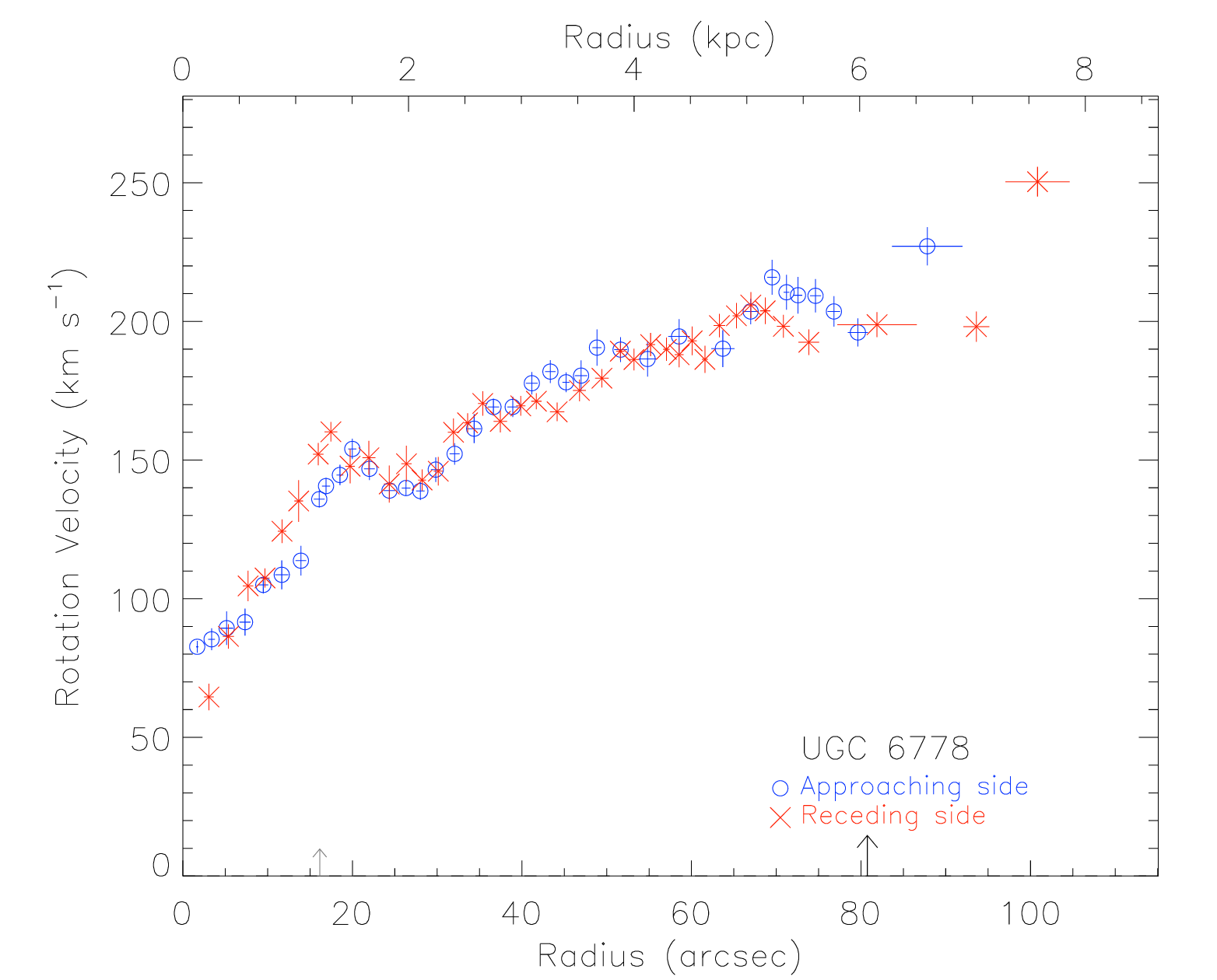}
   \includegraphics[width=8cm]{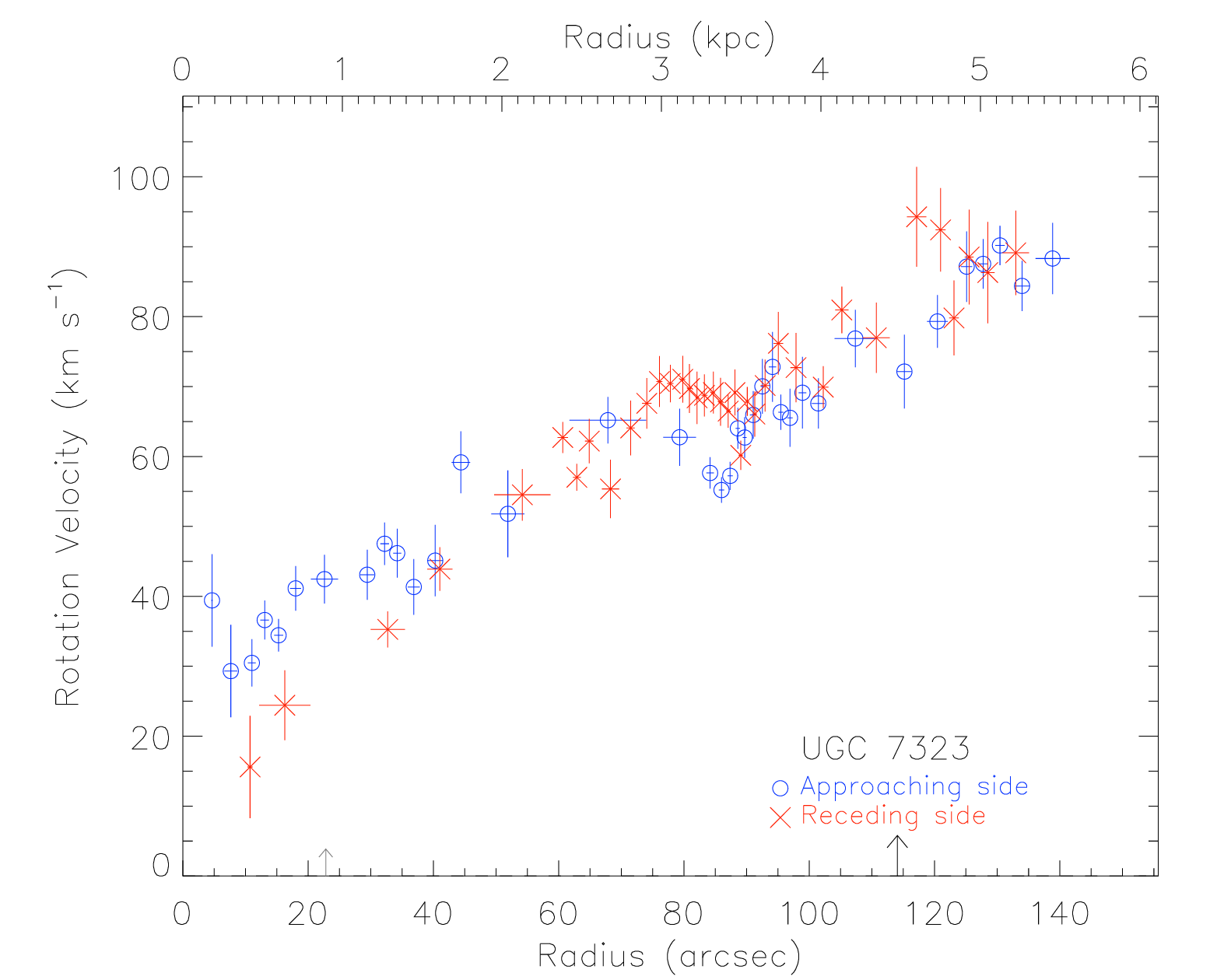}
\end{center}
\caption{From top left to bottom right: \ha~\RC~of UGC 5982, UGC 6537, UGC 6628, UGC 6702, UGC 6778, and UGC 7323.
}
\end{minipage}
\end{figure*}
\clearpage
\begin{figure*}
\begin{minipage}{180mm}
\begin{center}
   \includegraphics[width=8cm]{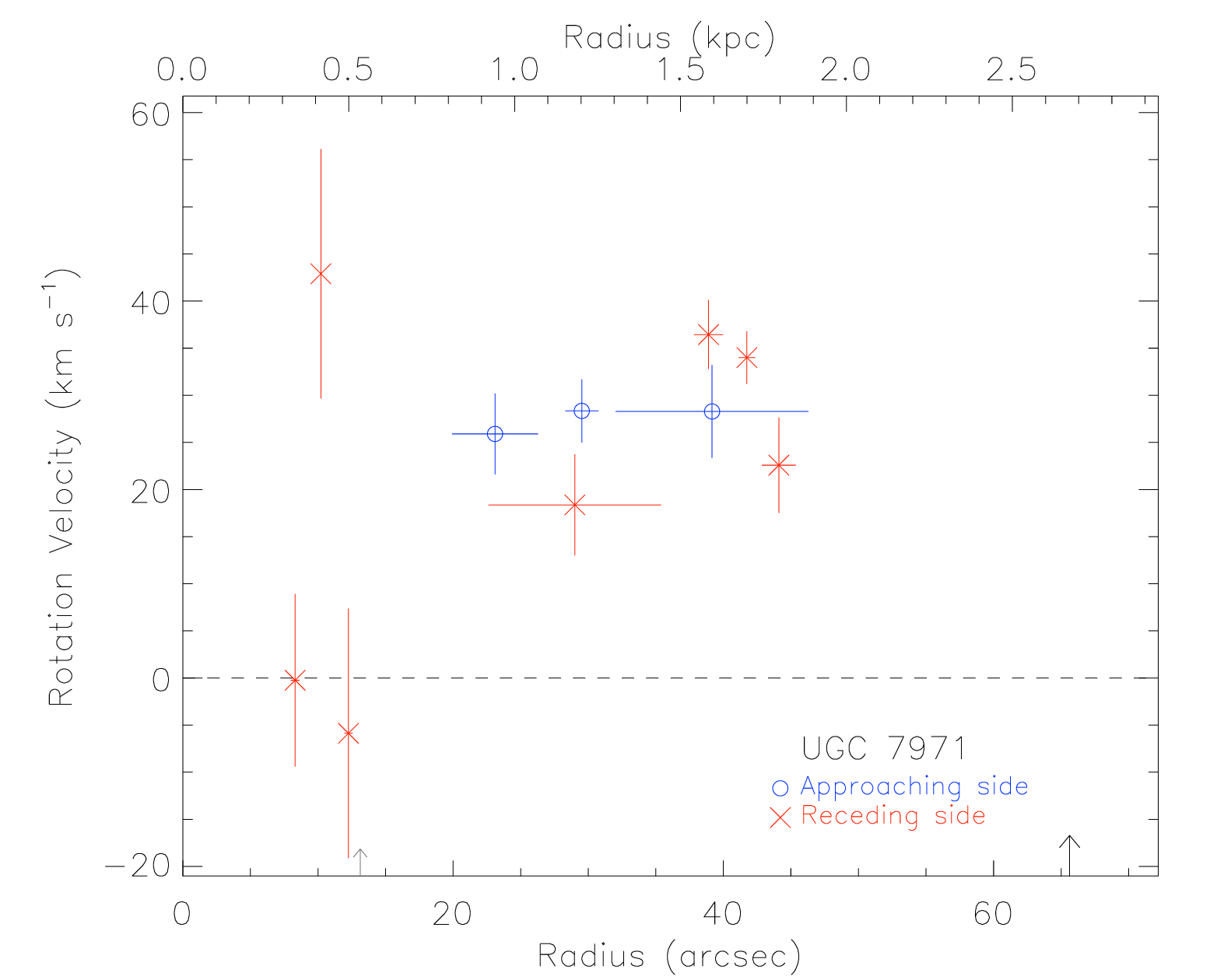}
   \includegraphics[width=8cm]{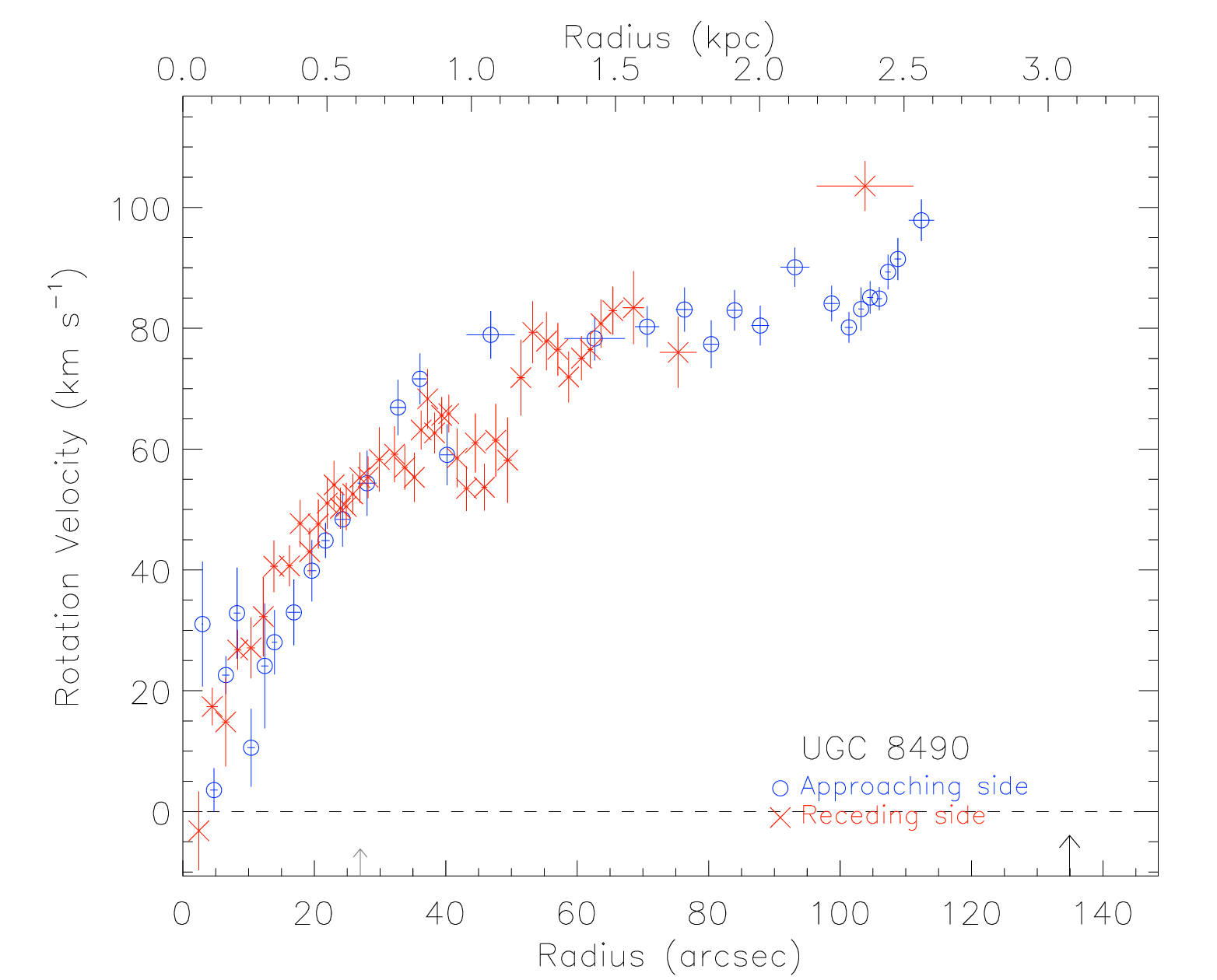}
   \includegraphics[width=8cm]{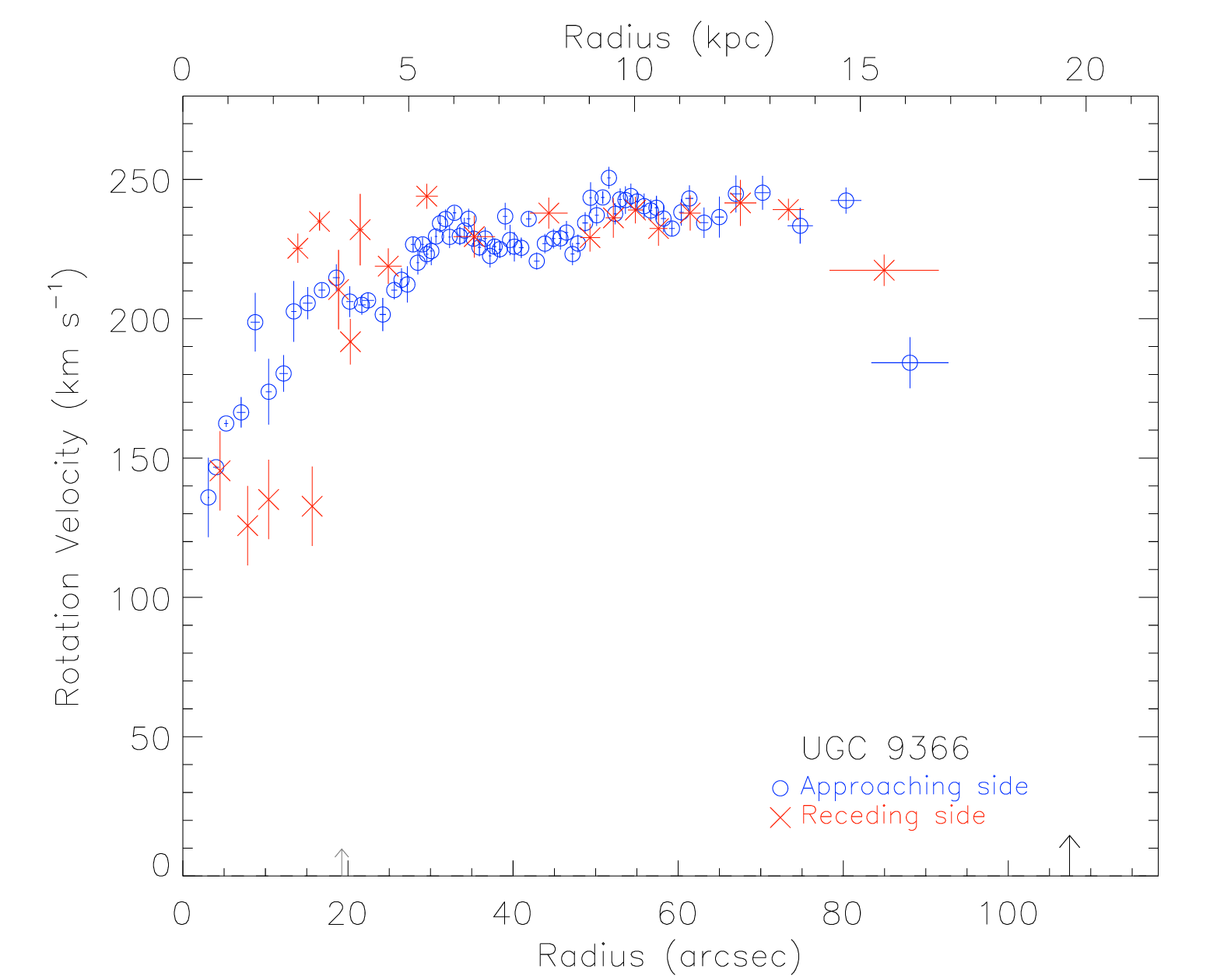}
   \includegraphics[width=8cm]{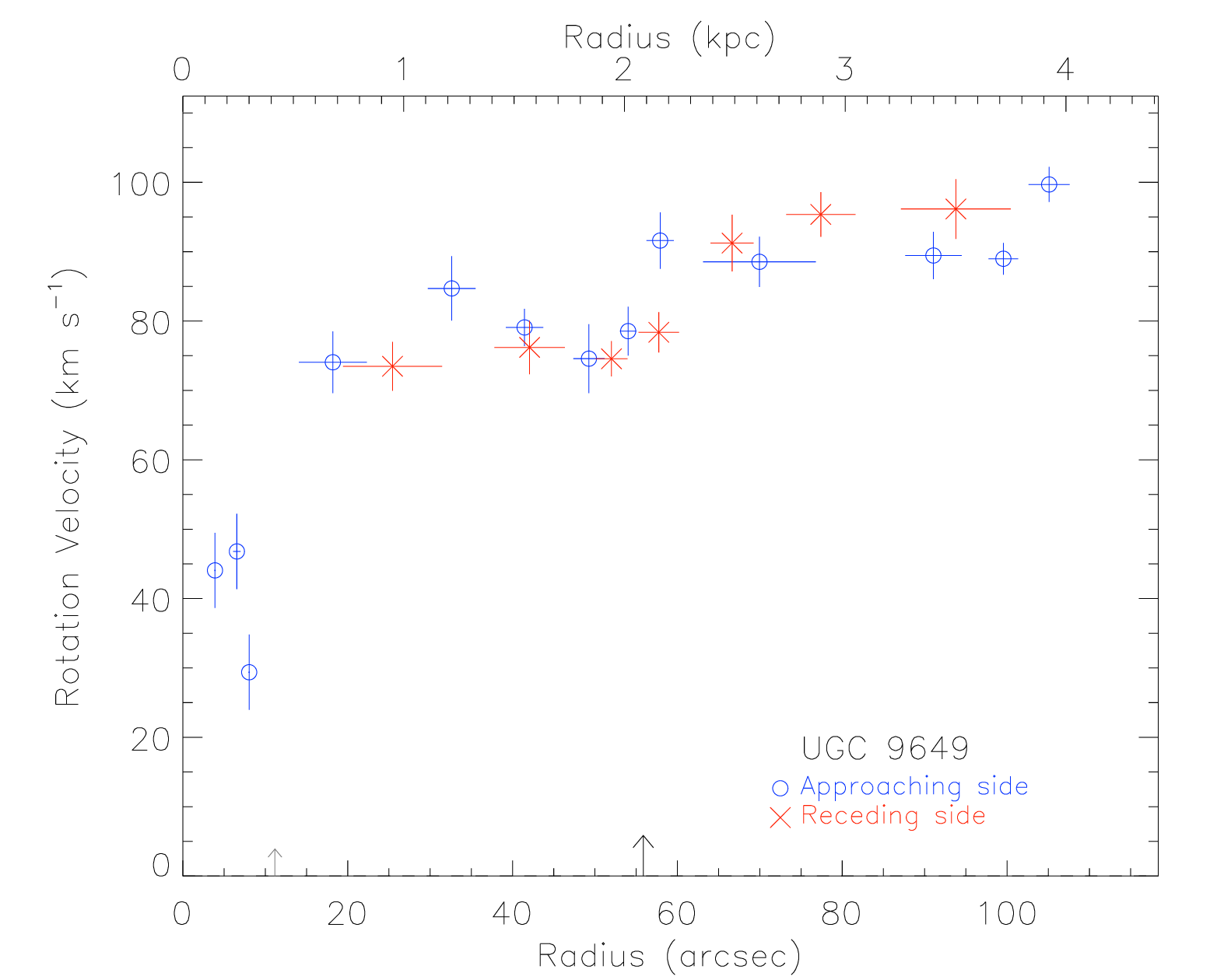}
   \includegraphics[width=8cm]{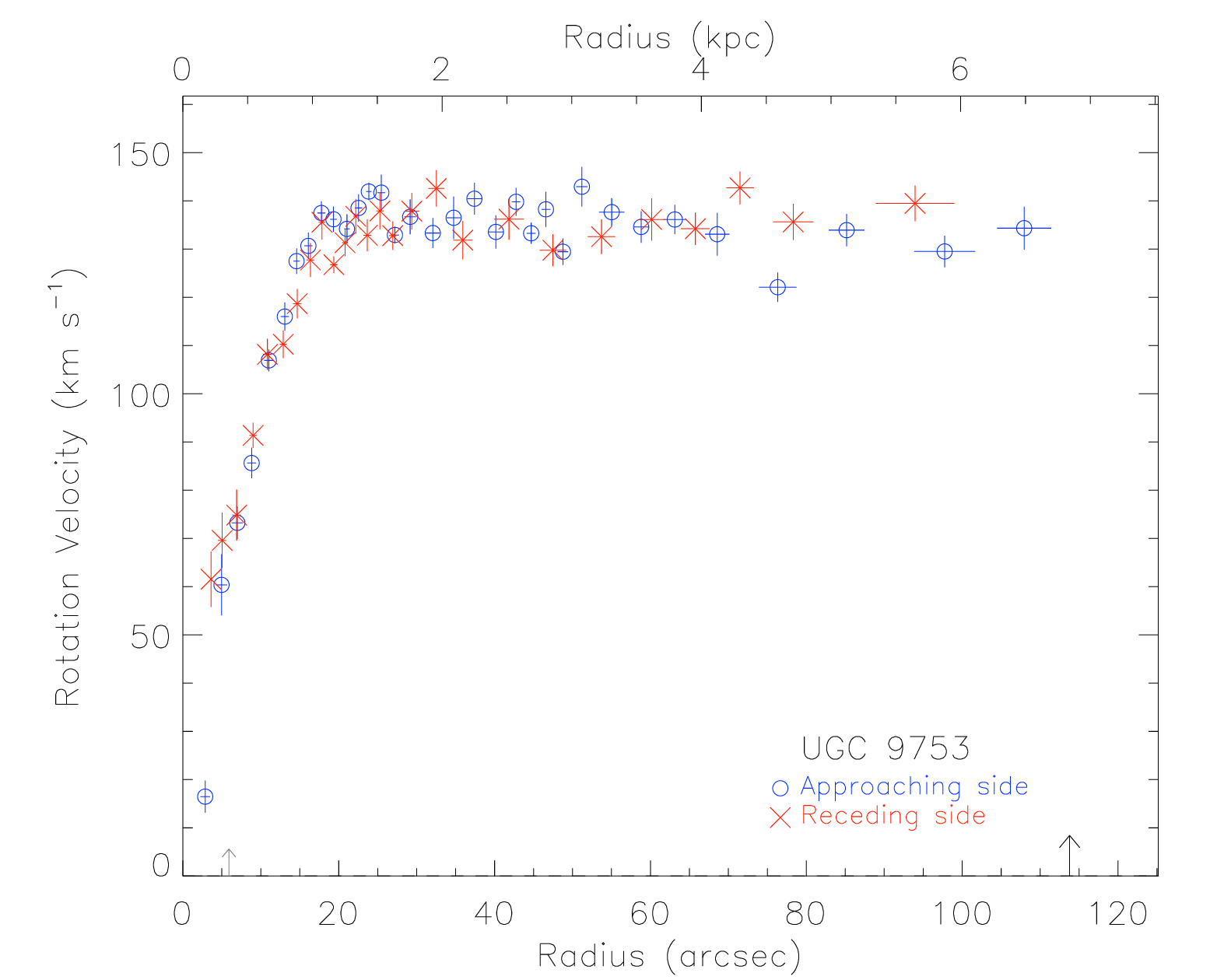}
   \includegraphics[width=8cm]{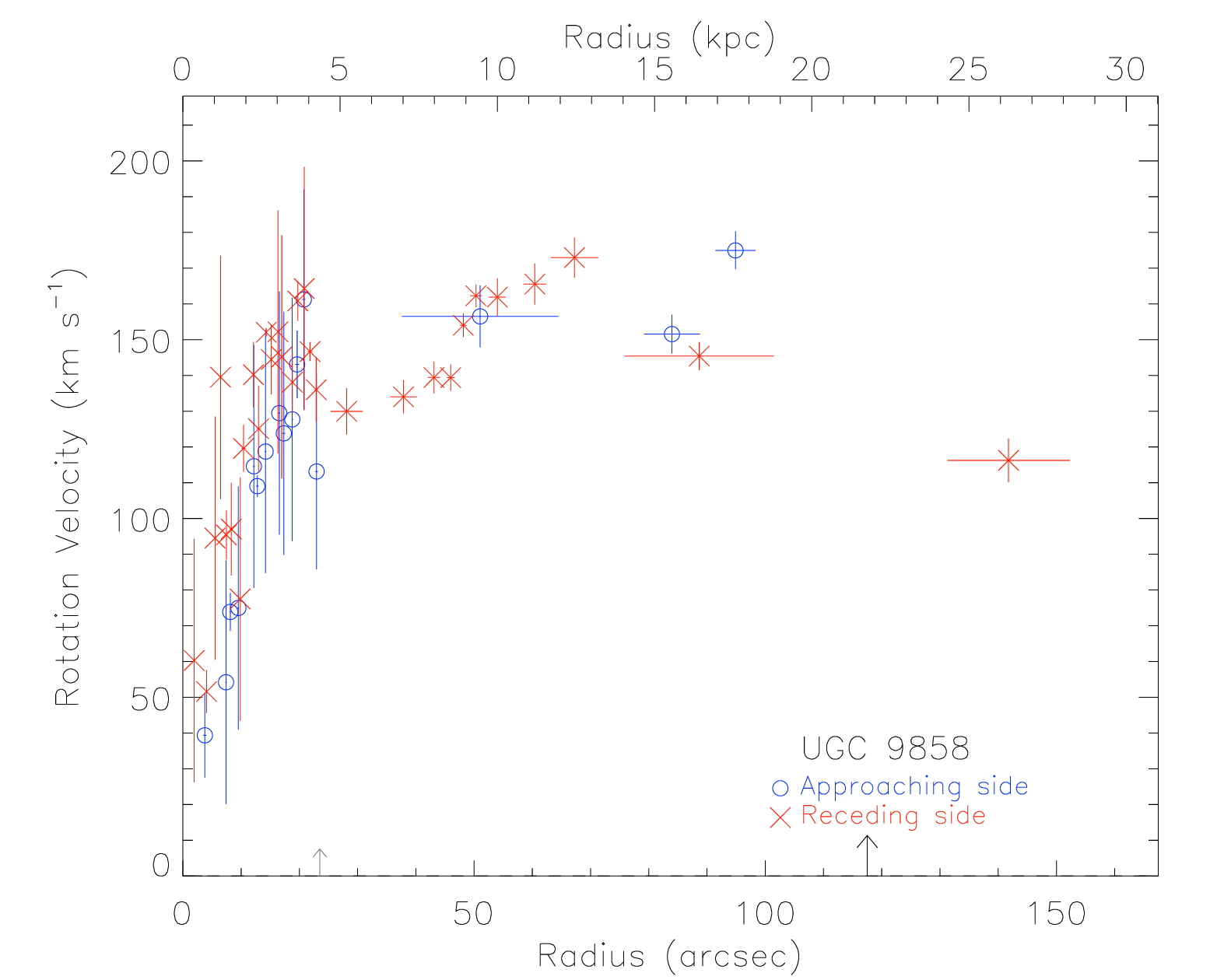}
\end{center}
\caption{From top left to bottom right: \ha~\RC~of UGC 7971, UGC 8490, UGC 9366, UGC 9649, UGC 9753, and UGC 9858.
}
\end{minipage}
\end{figure*}
\clearpage
\begin{figure*}
\begin{minipage}{180mm}
\begin{center}
   \includegraphics[width=8cm]{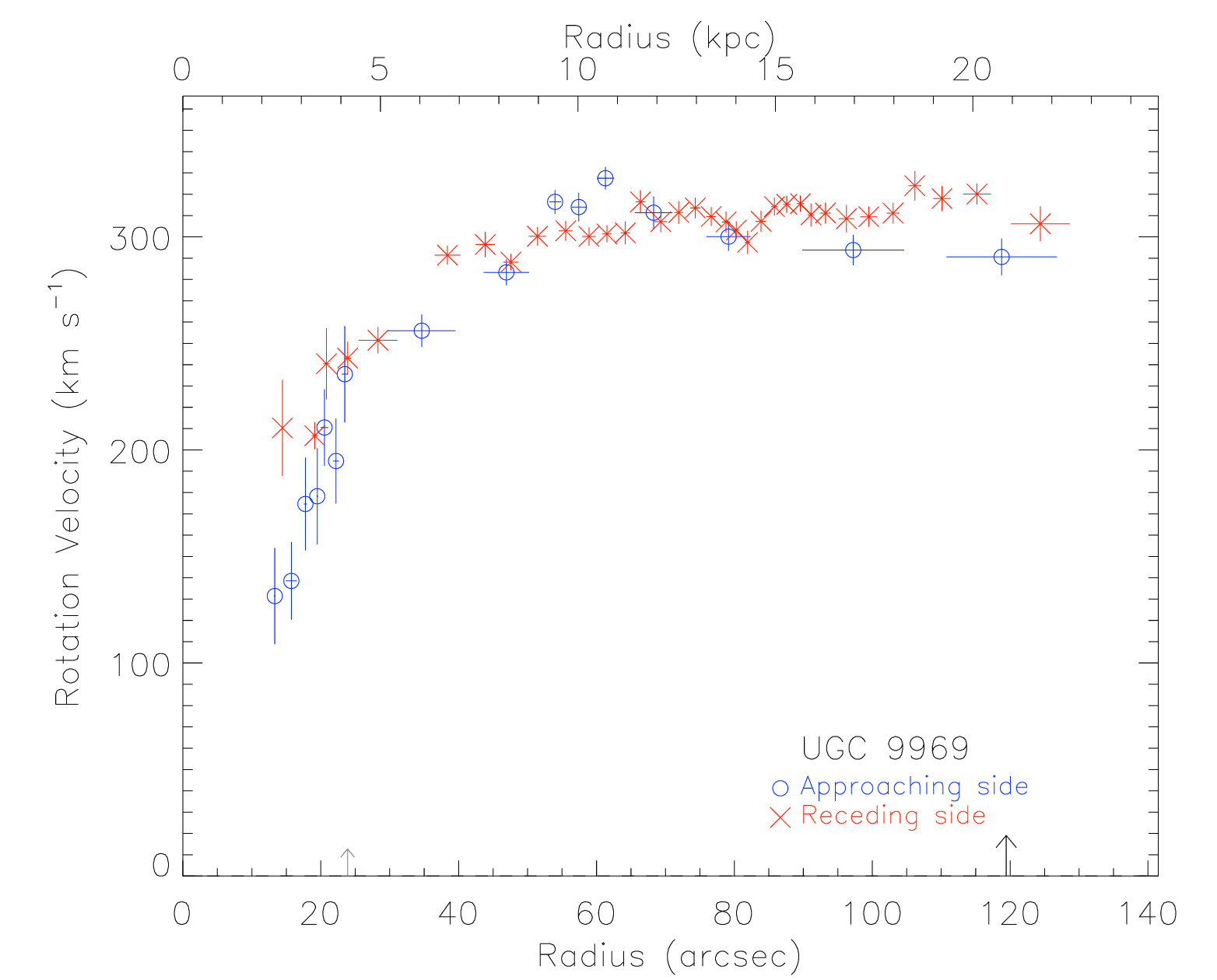}
   \includegraphics[width=8cm]{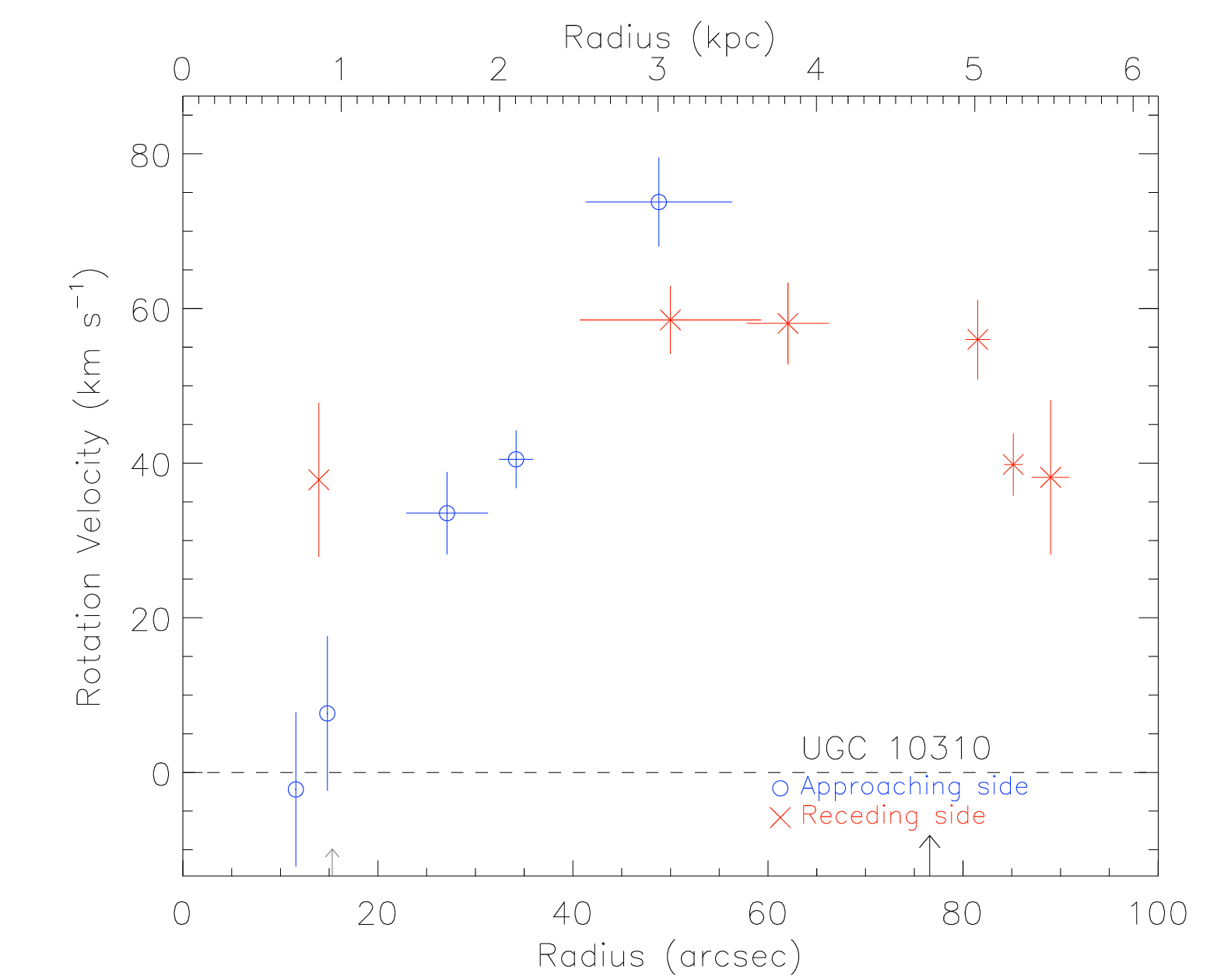}
   \includegraphics[width=8cm]{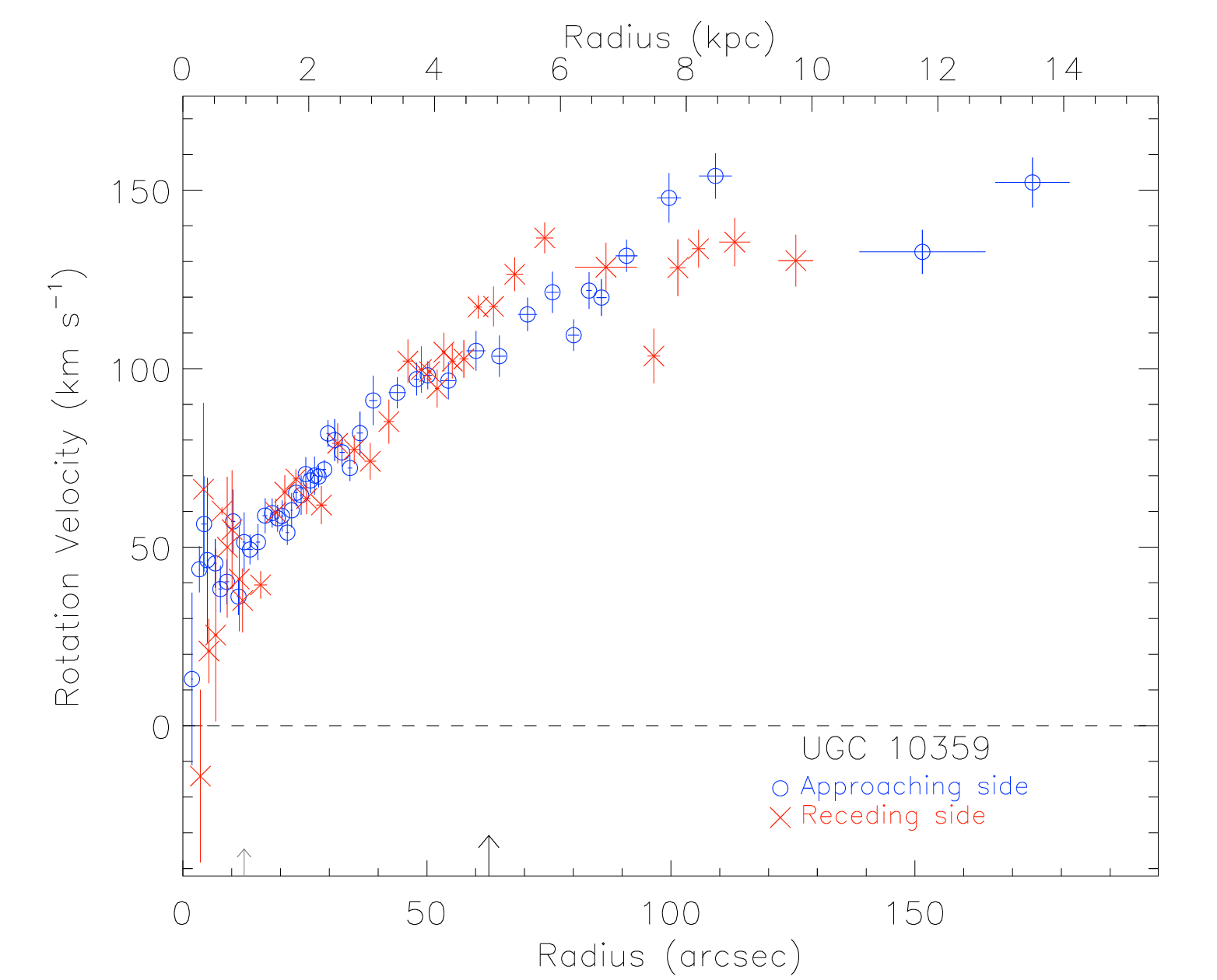}
   \includegraphics[width=8cm]{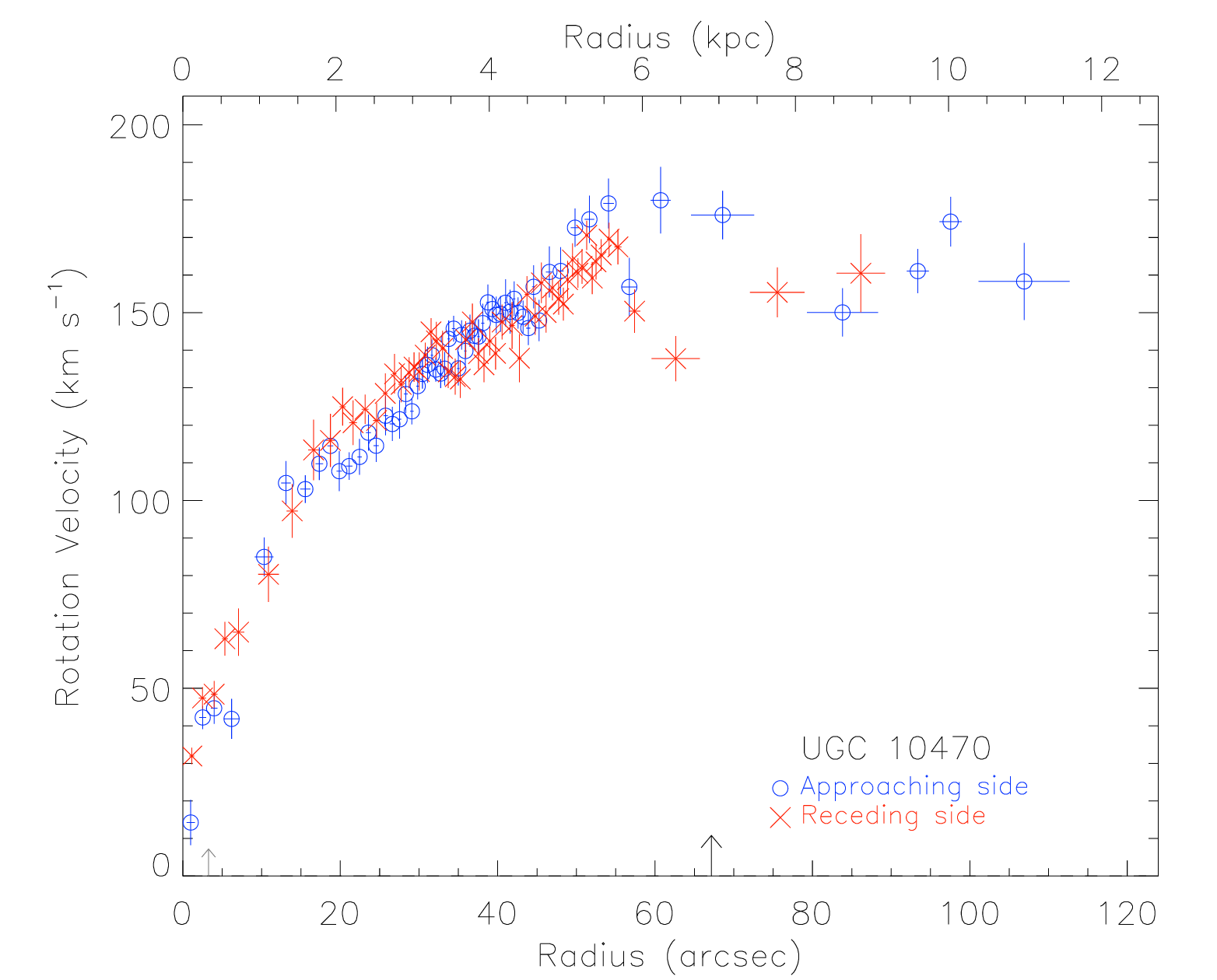}
   \includegraphics[width=8cm]{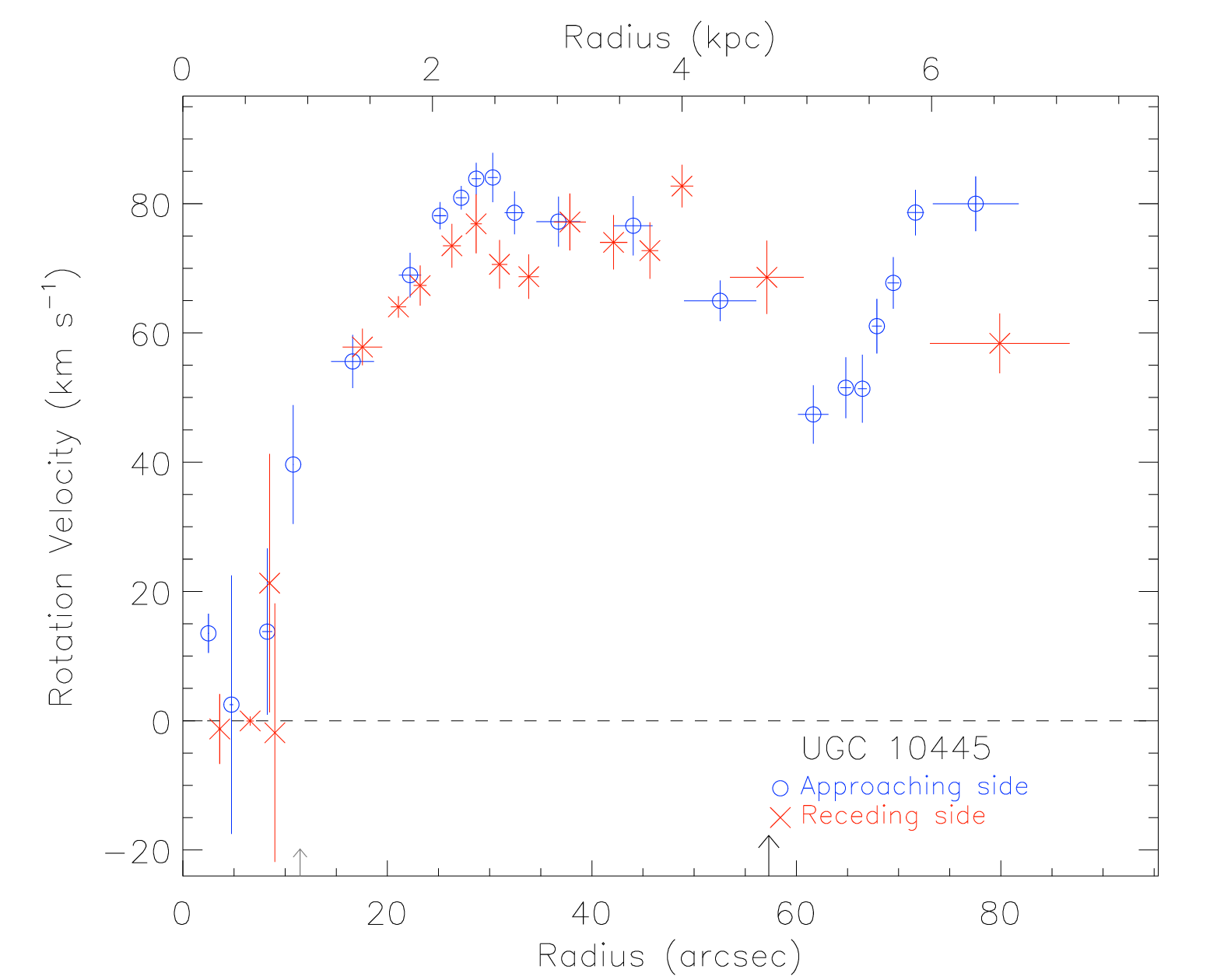}
   \includegraphics[width=8cm]{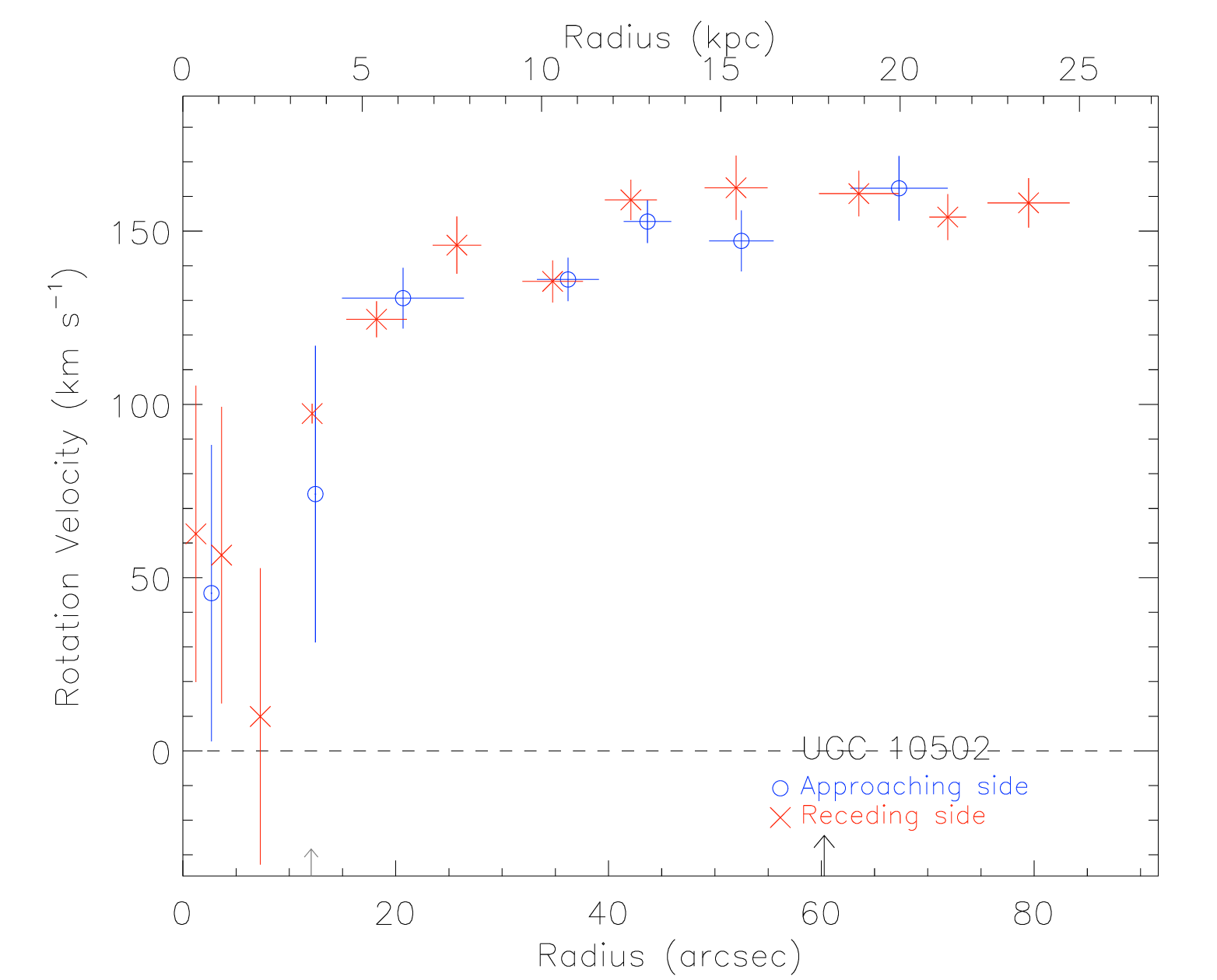}
\end{center}
\caption{From top left to bottom right: \ha~\RC~of UGC 9969, UGC 10310, UGC 10359, UGC 10470, UGC 10445, and UGC 10502.
}
\end{minipage}
\end{figure*}
\clearpage
\begin{figure*}
\begin{minipage}{180mm}
\begin{center}
   \includegraphics[width=8cm]{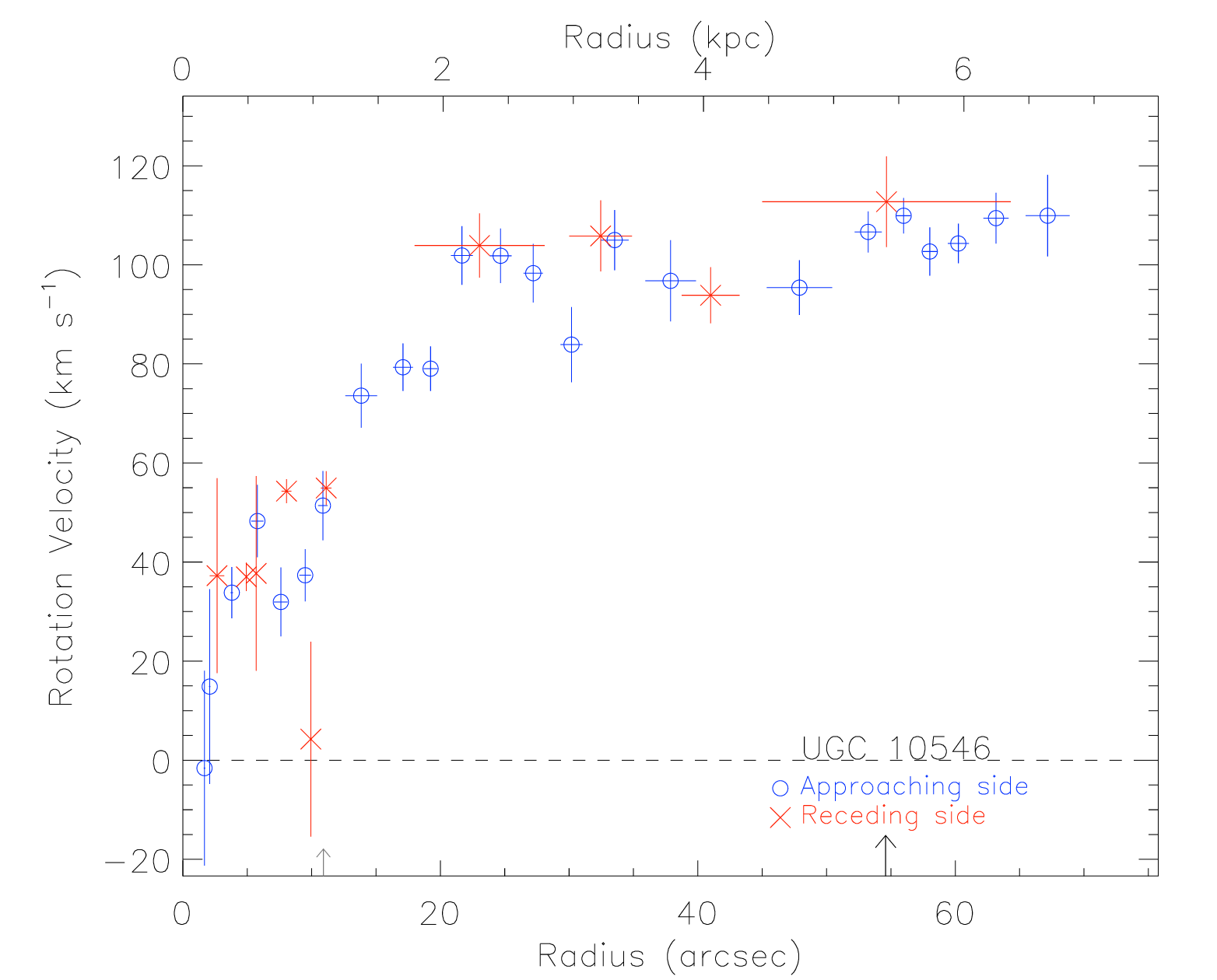}
   \includegraphics[width=8cm]{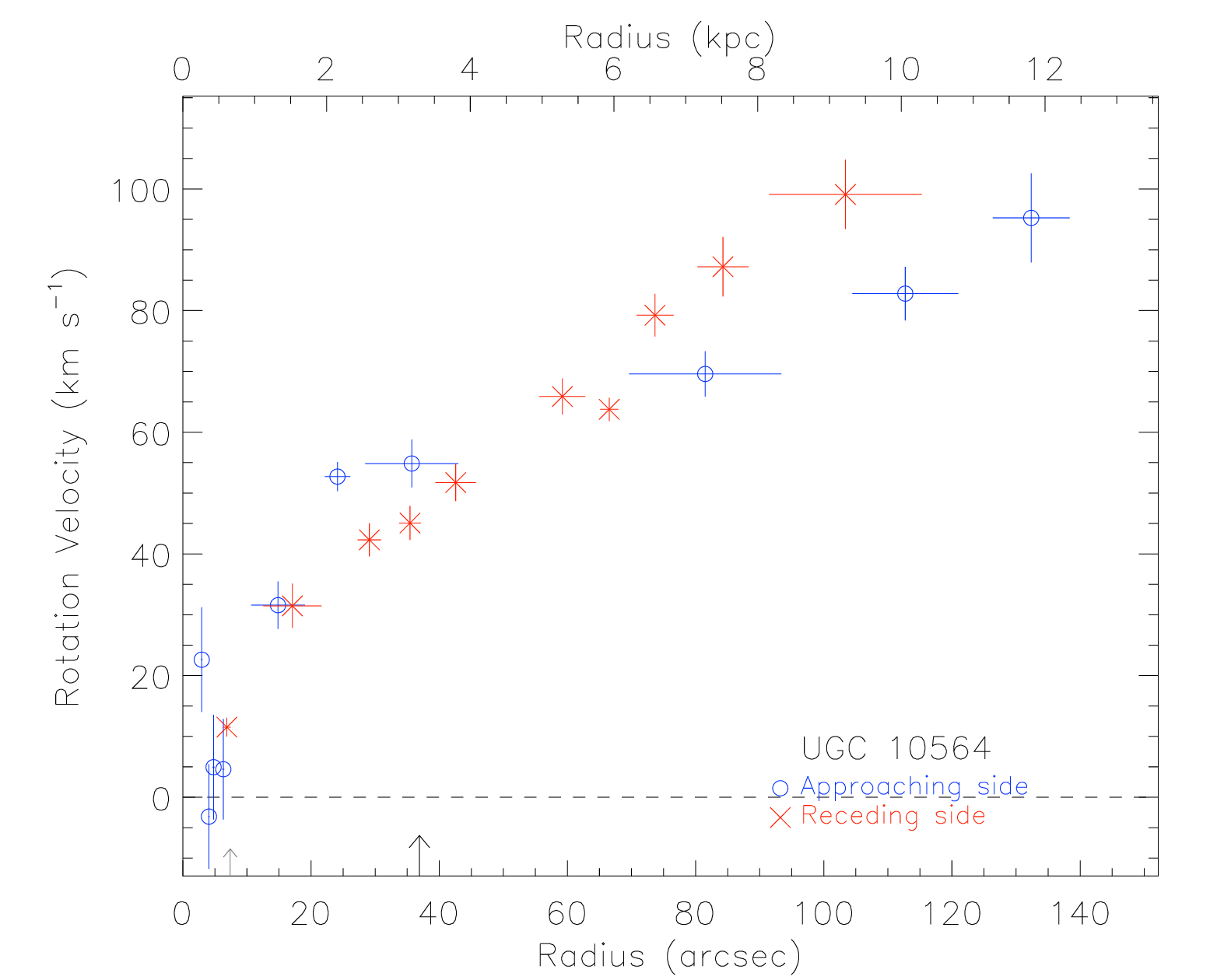}
   \includegraphics[width=8cm]{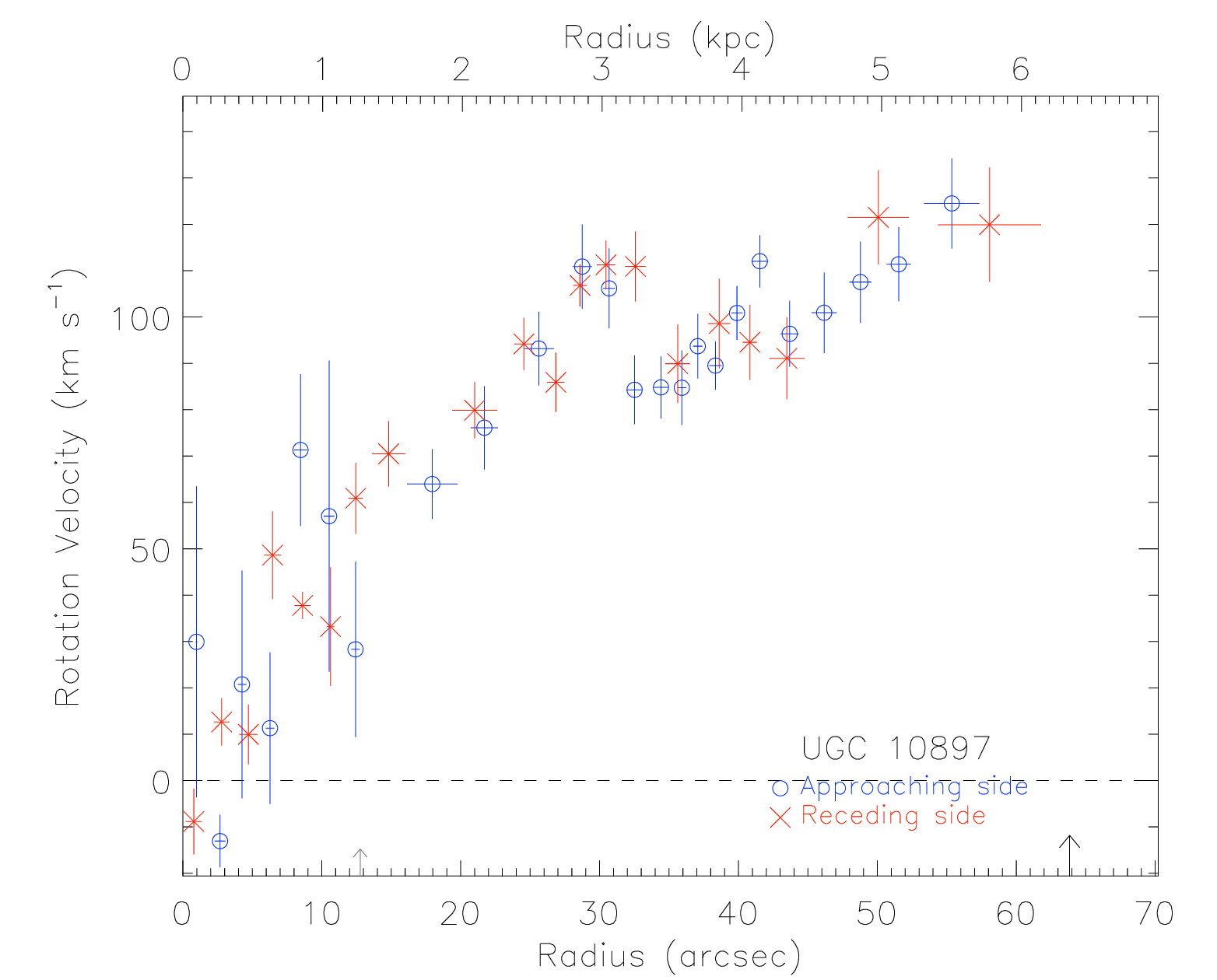}
   \includegraphics[width=8cm]{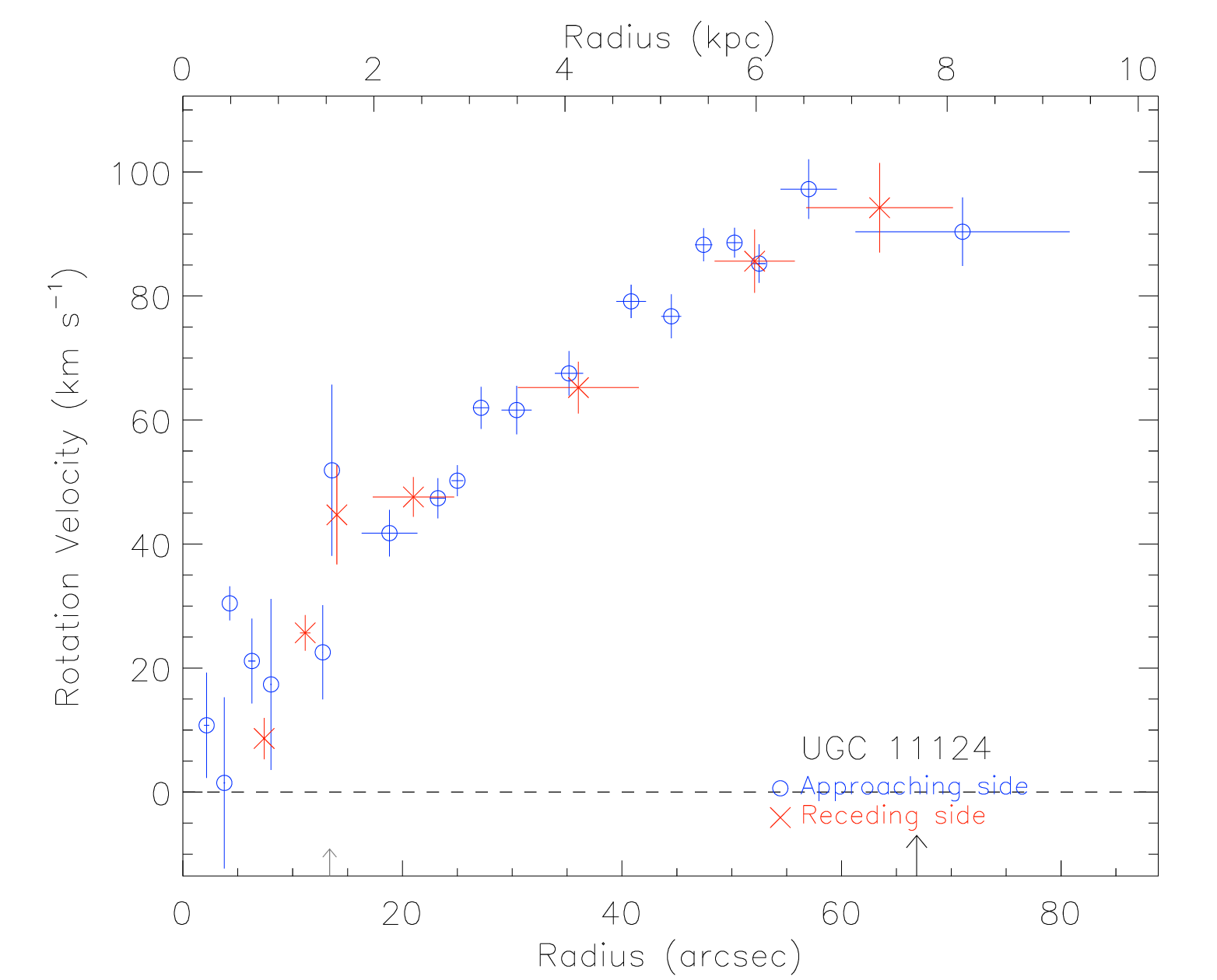}
   \includegraphics[width=8cm]{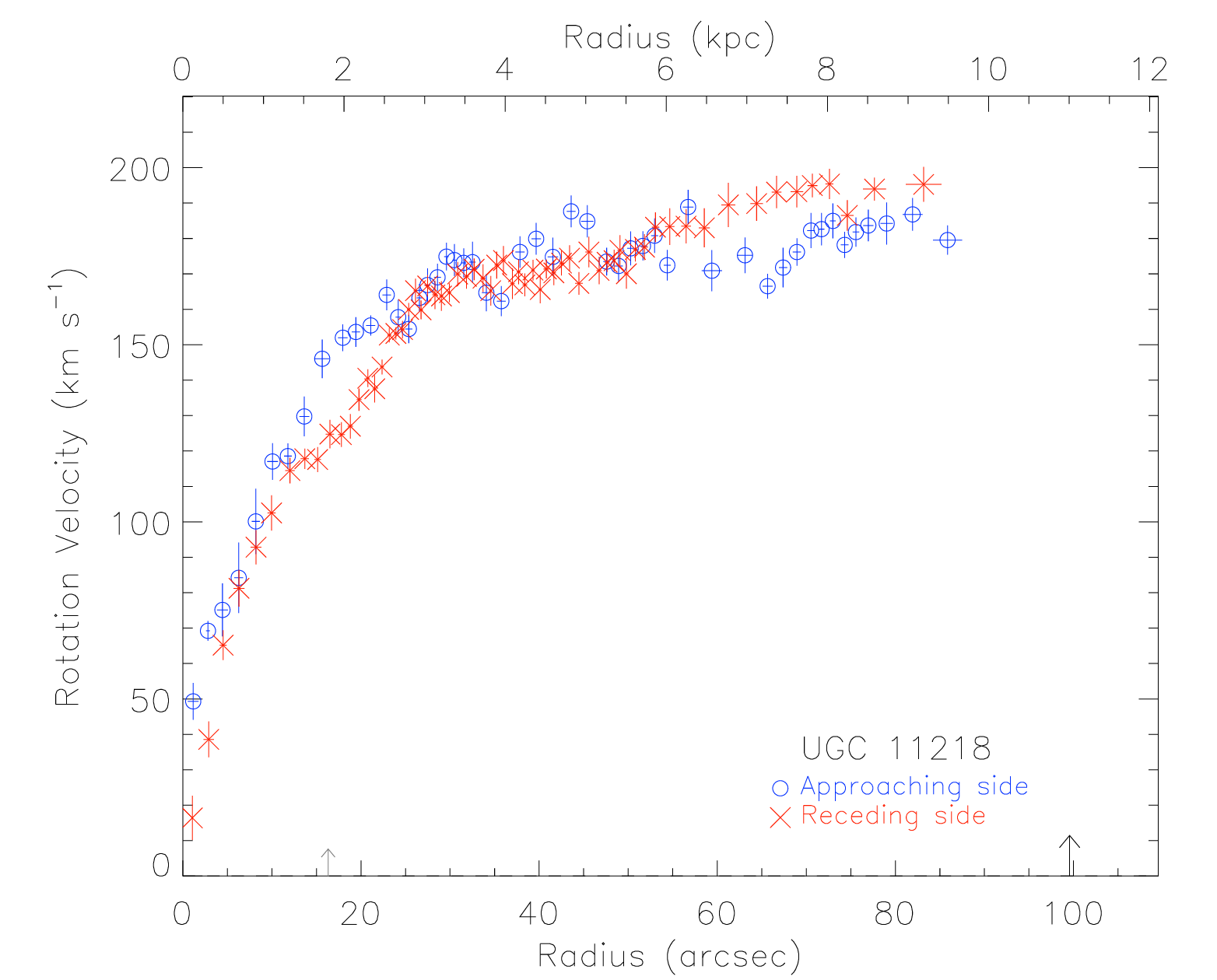}
   \includegraphics[width=8cm]{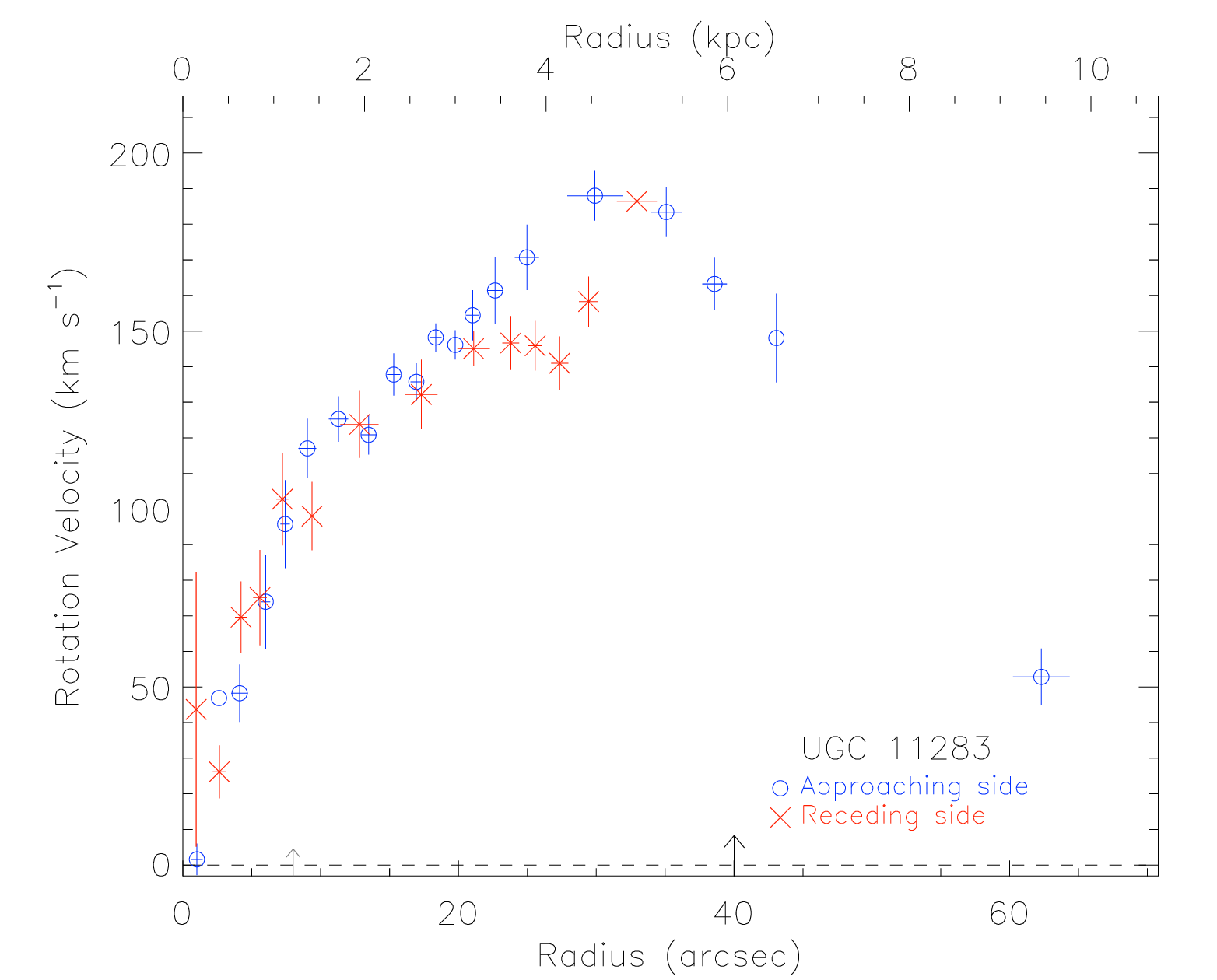}
\end{center}
\caption{From top left to bottom right: \ha~\RC~of UGC 10546, UGC 10564, UGC 10897, UGC 11124, UGC 11218, and UGC 11283.
}
\end{minipage}
\end{figure*}
\clearpage
\begin{figure*}
\begin{minipage}{180mm}
\begin{center}
   \includegraphics[width=8cm]{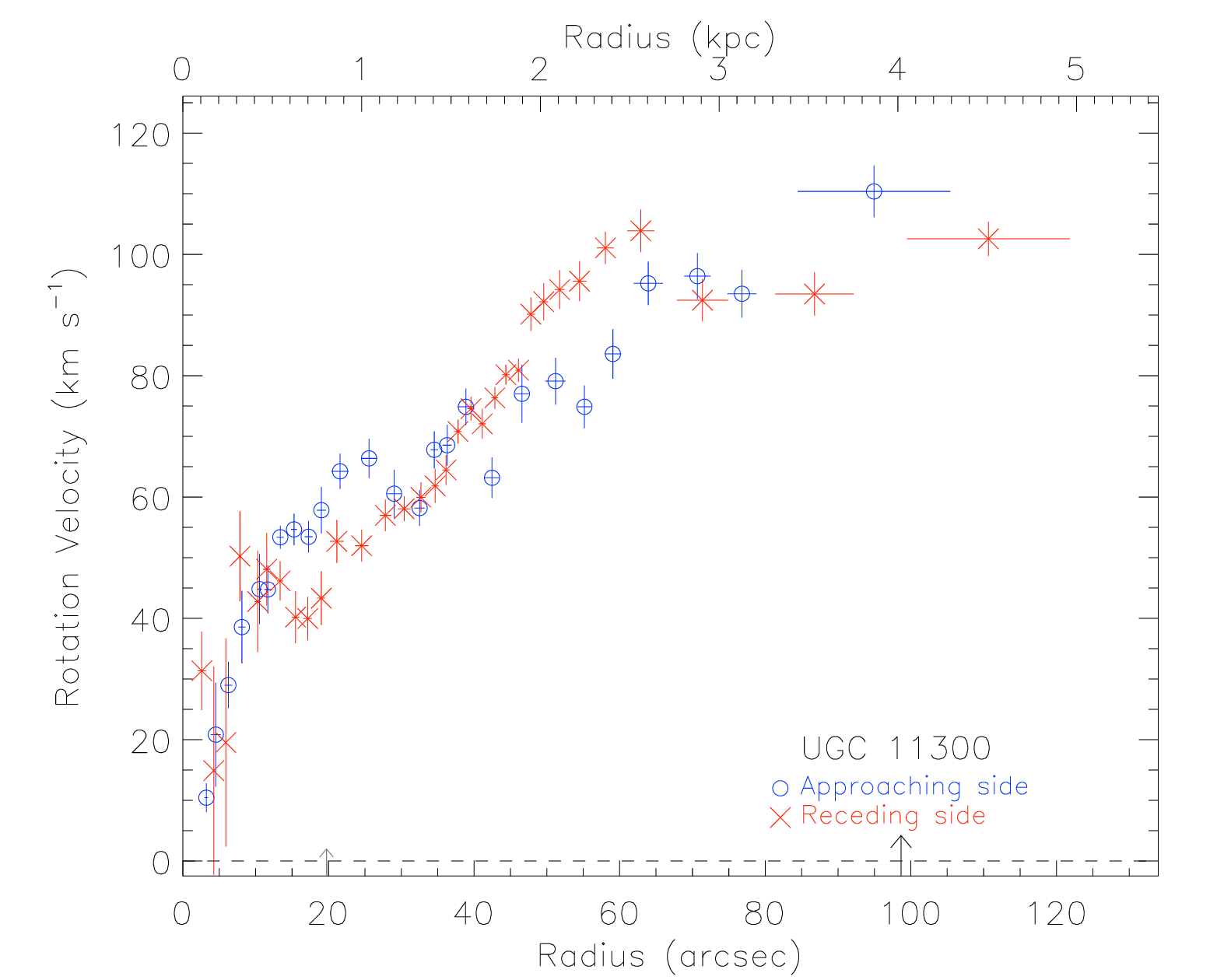}
   \includegraphics[width=8cm]{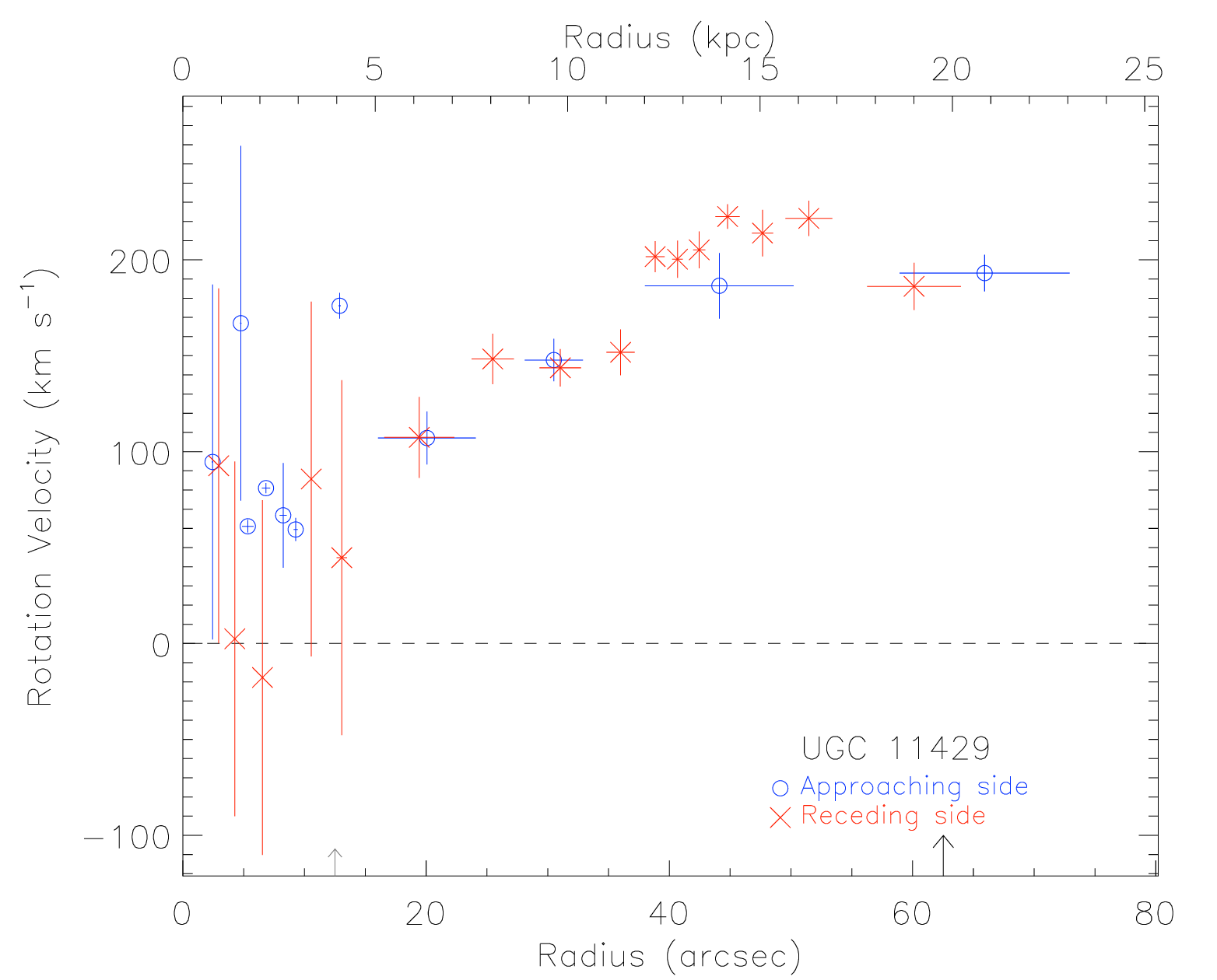}
   \includegraphics[width=8cm]{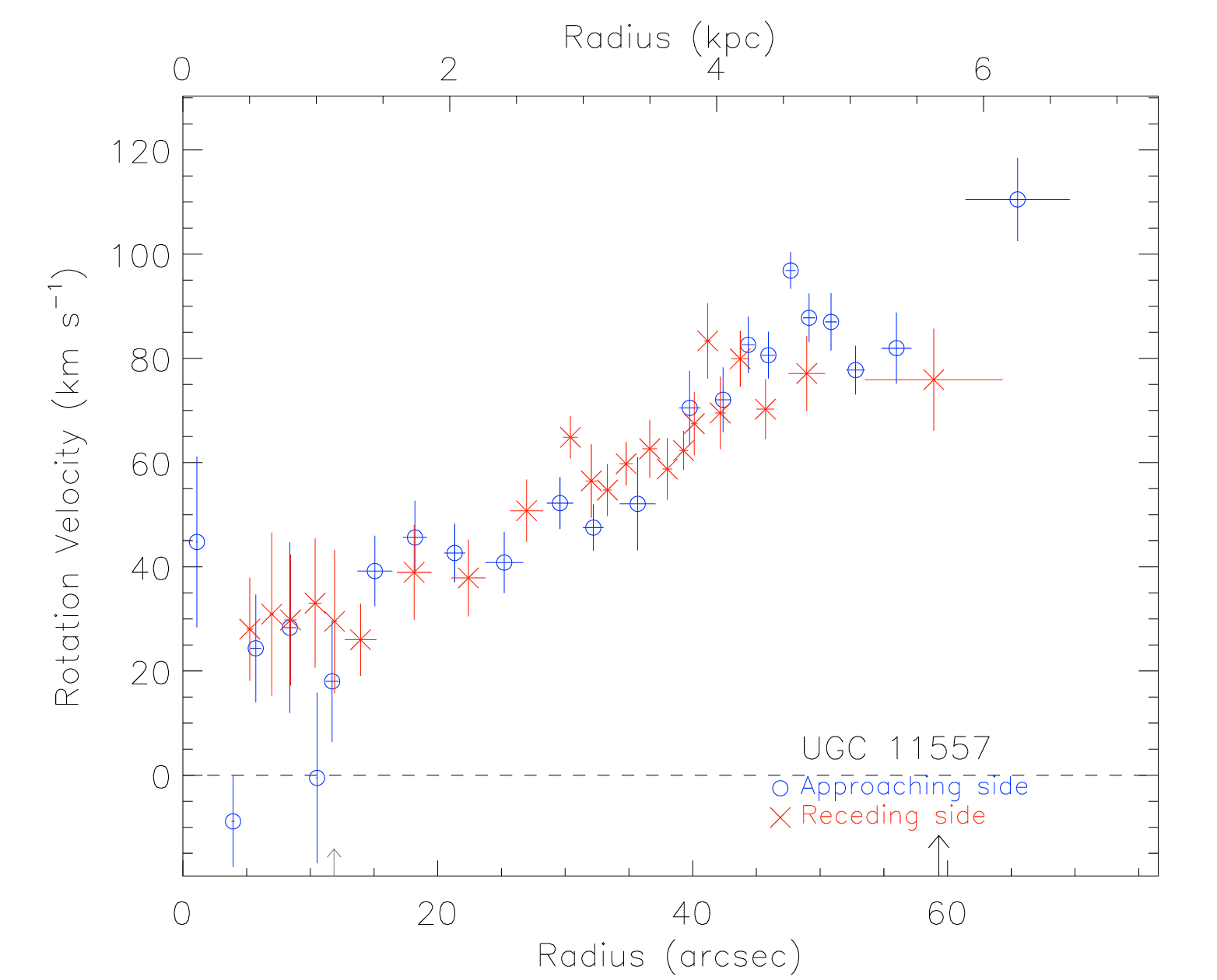}
   \includegraphics[width=8cm]{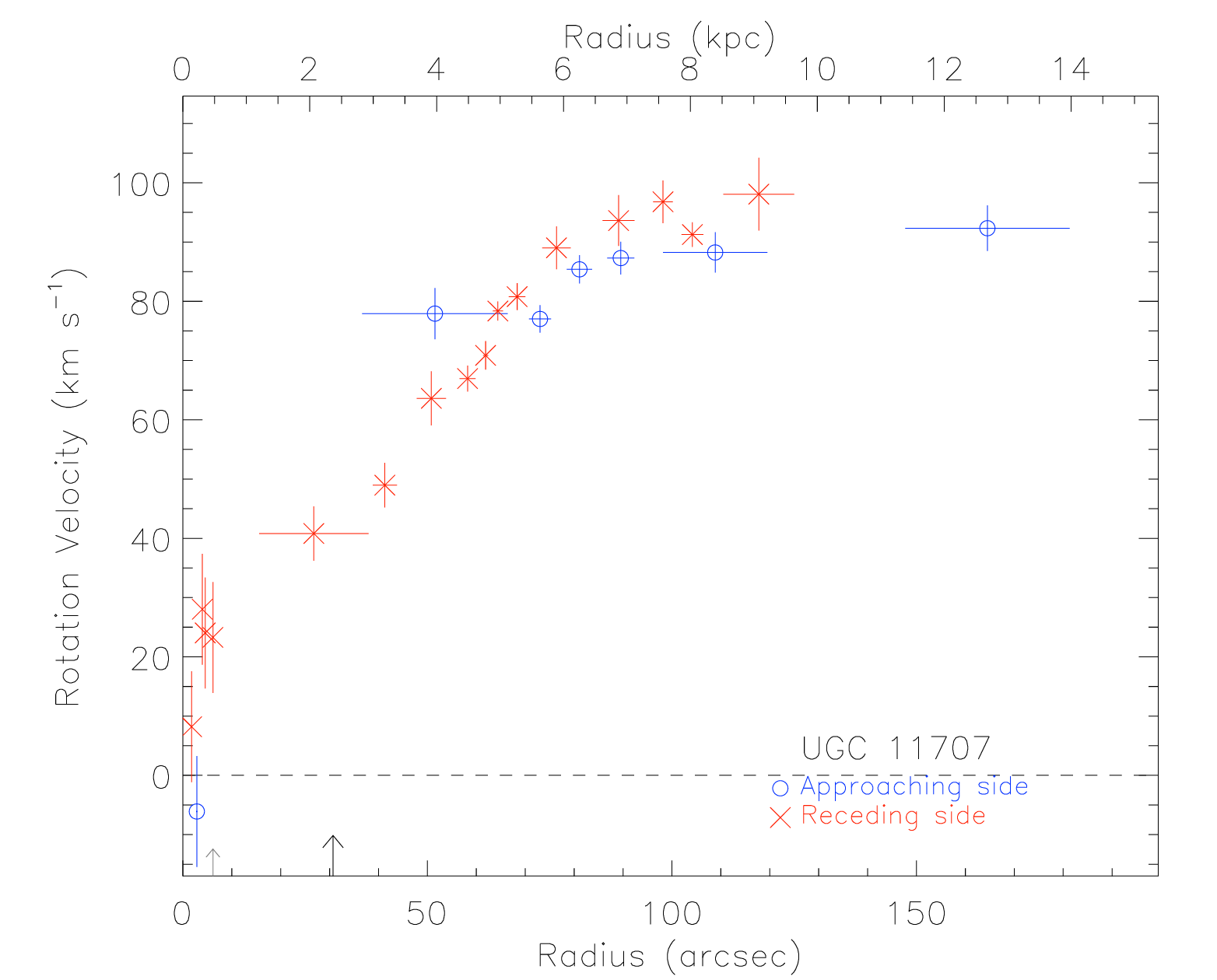}
   \includegraphics[width=8cm]{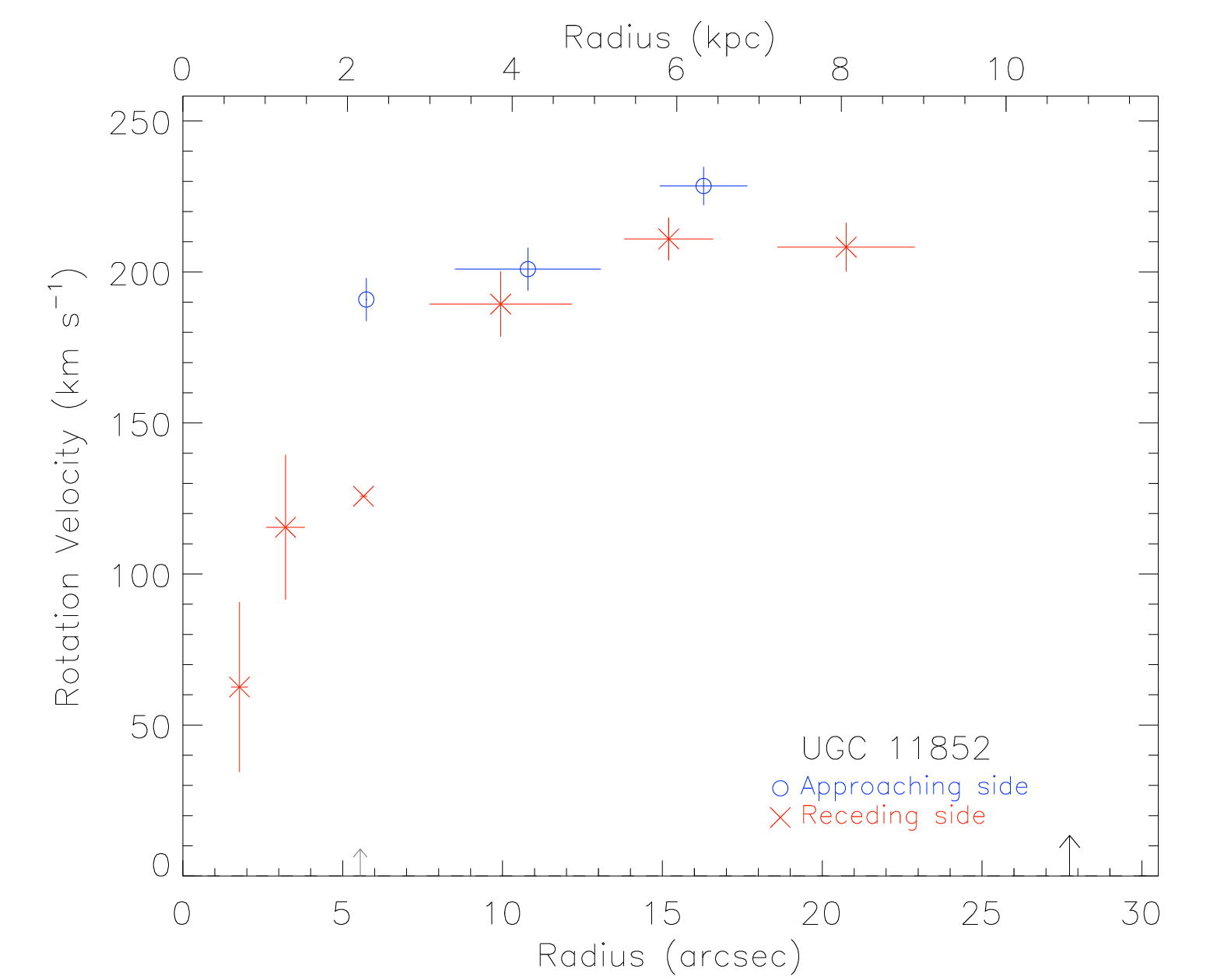}
   \includegraphics[width=8cm]{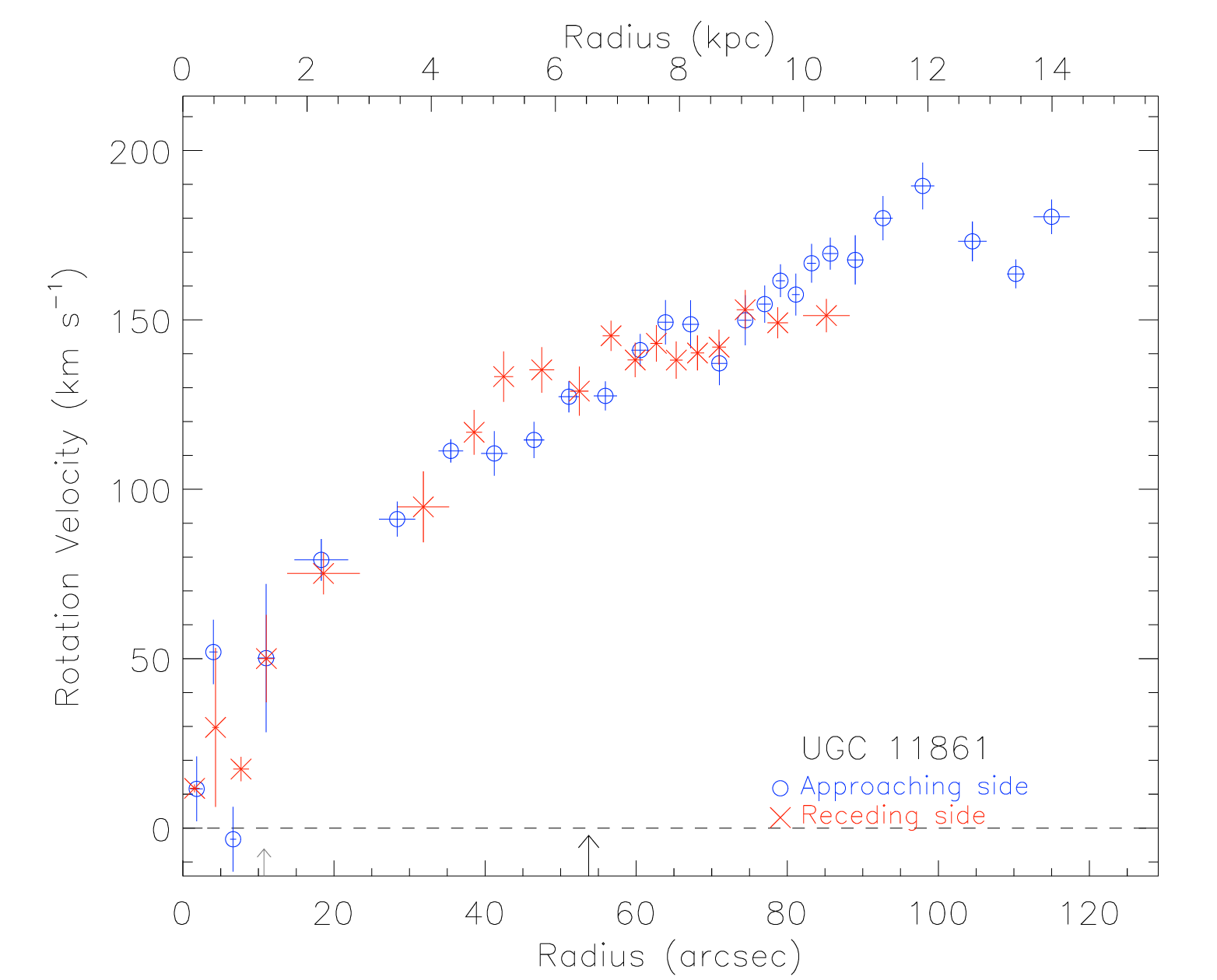}
\end{center}
\caption{From top left to bottom right: \ha~\RC~of UGC 11300, UGC 11429, UGC 11557, UGC 11707, UGC 11852, and UGC 11861.
}
\end{minipage}
\end{figure*}
\clearpage
\begin{figure*}
\begin{minipage}{180mm}
\begin{center}
   \includegraphics[width=8cm]{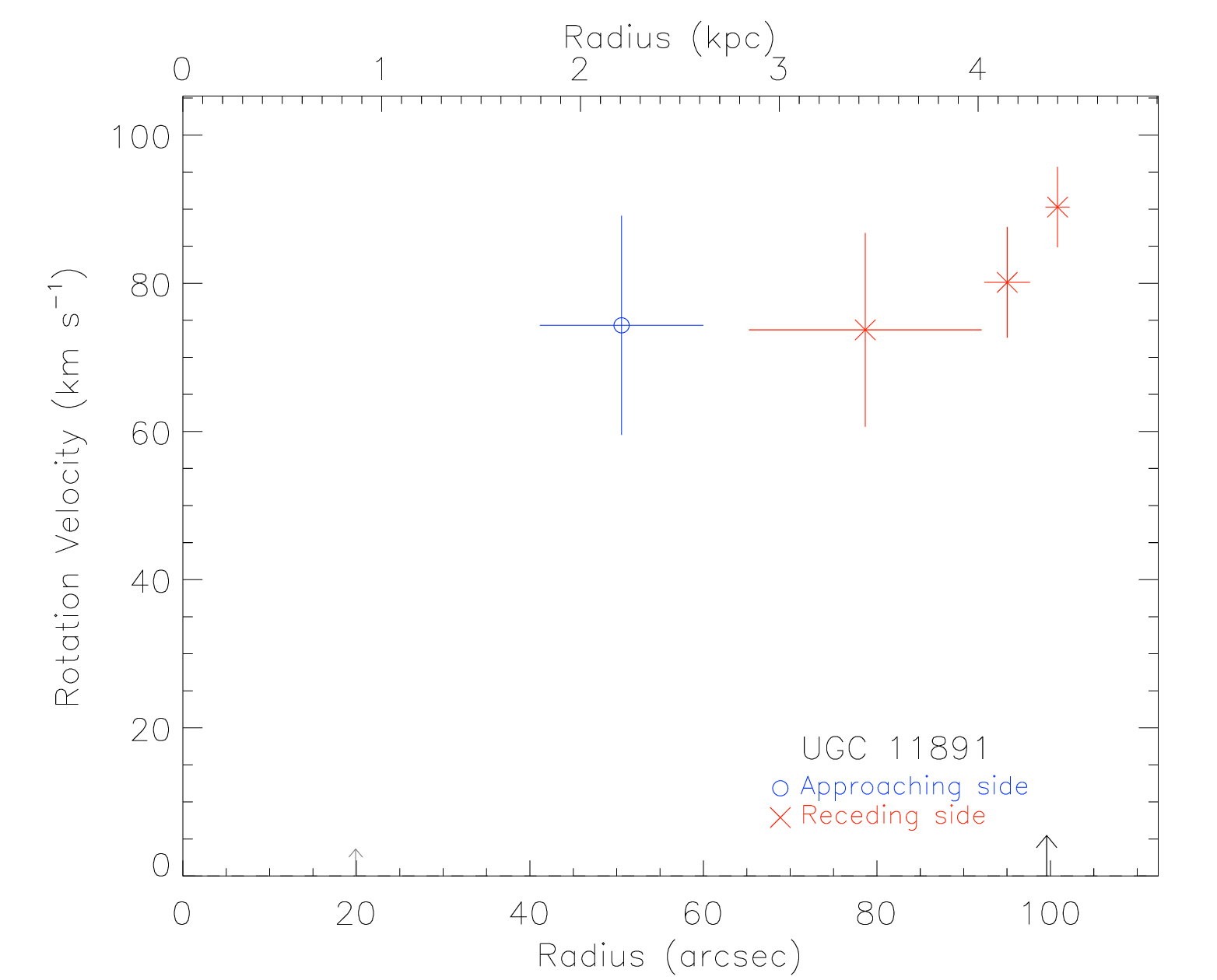}
   \includegraphics[width=8cm]{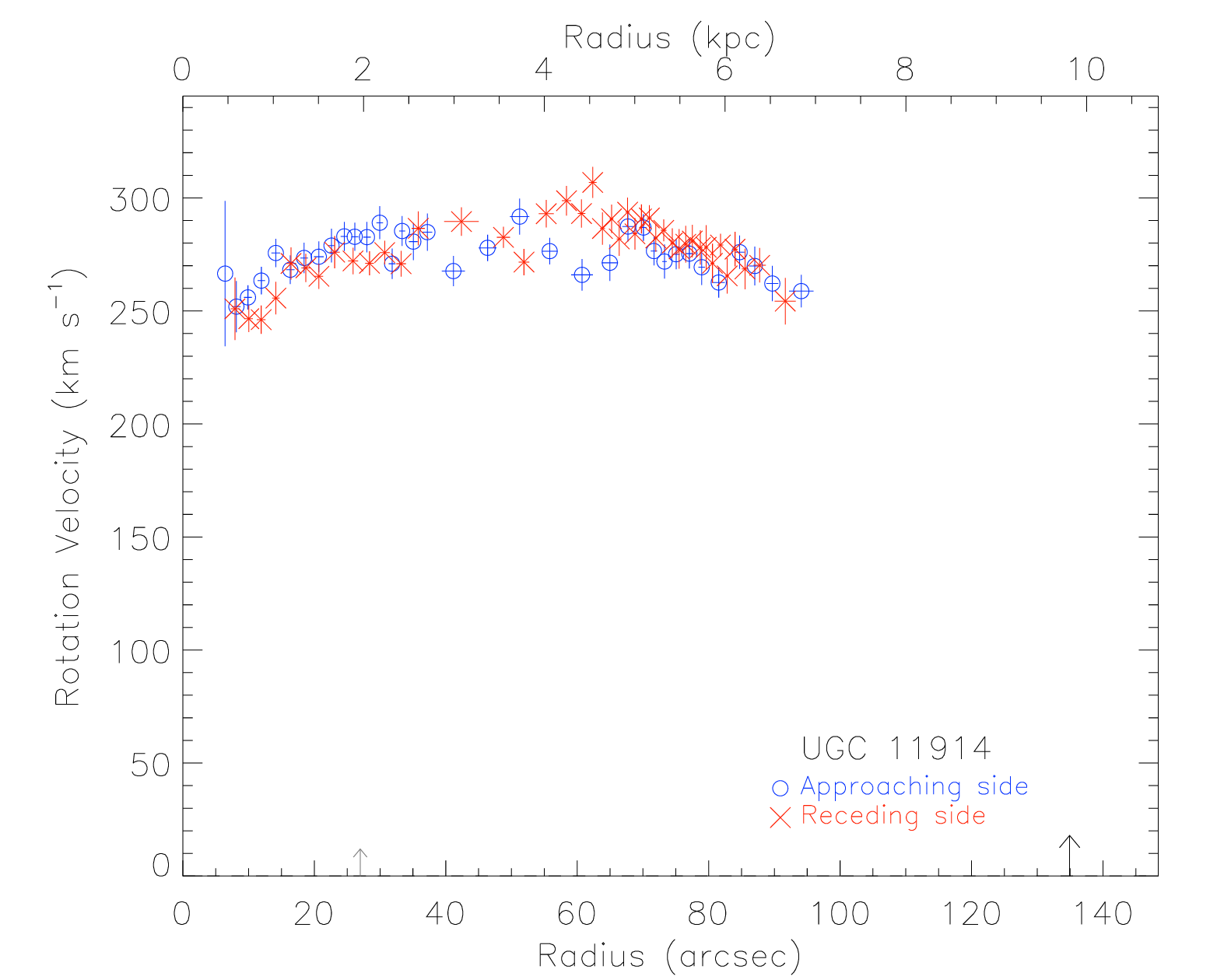}
   \includegraphics[width=8cm]{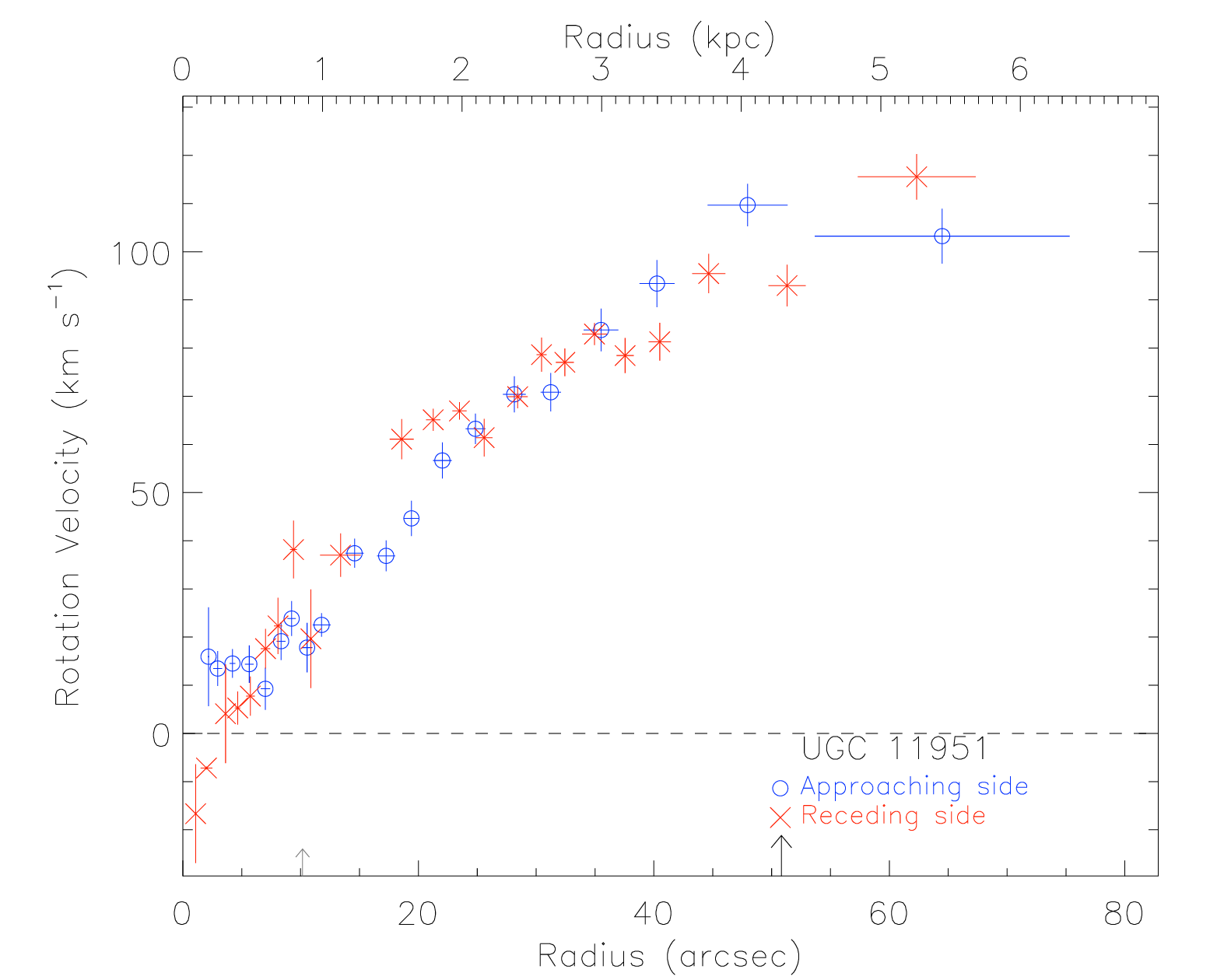}
   \includegraphics[width=8cm]{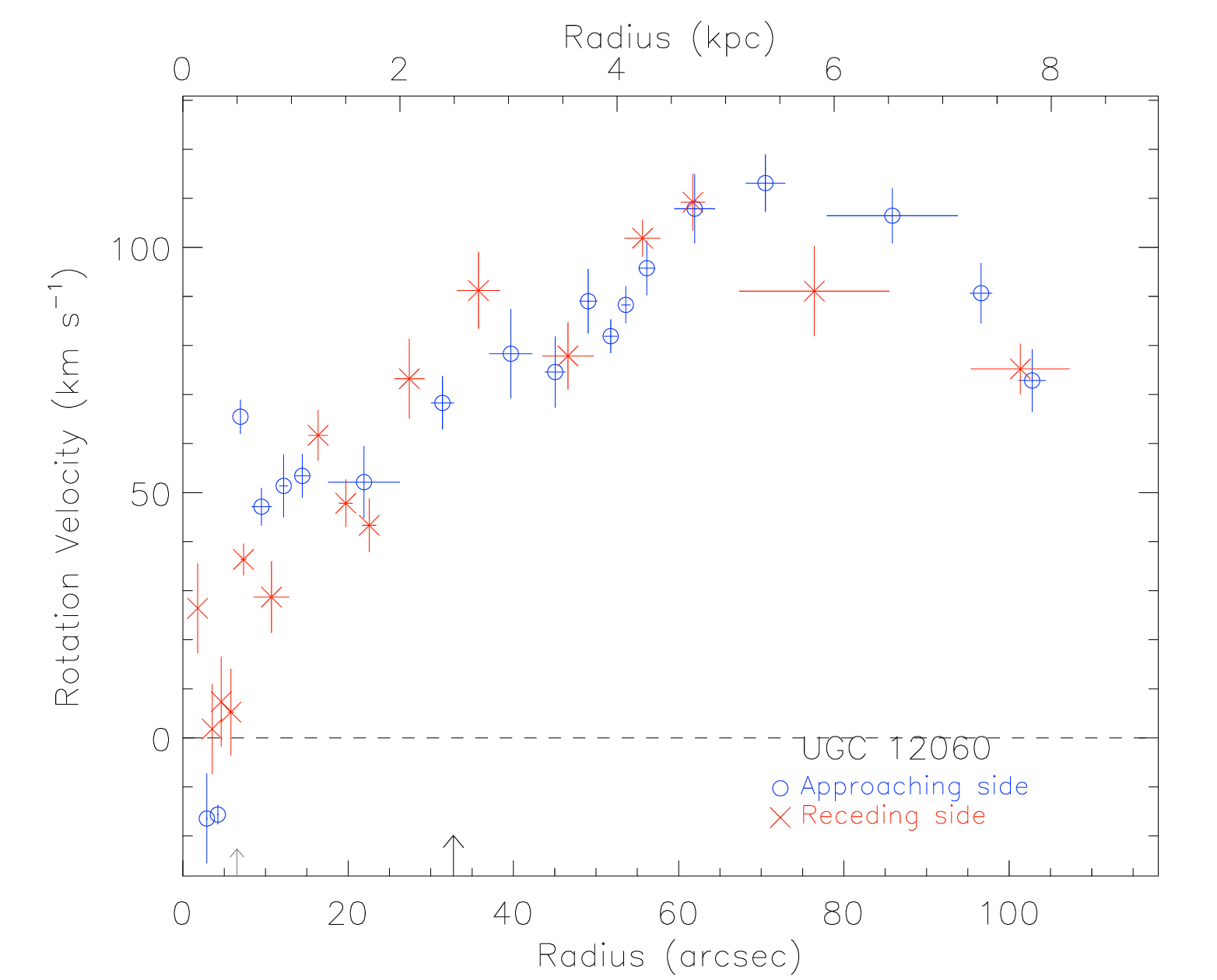}
   \includegraphics[width=8cm]{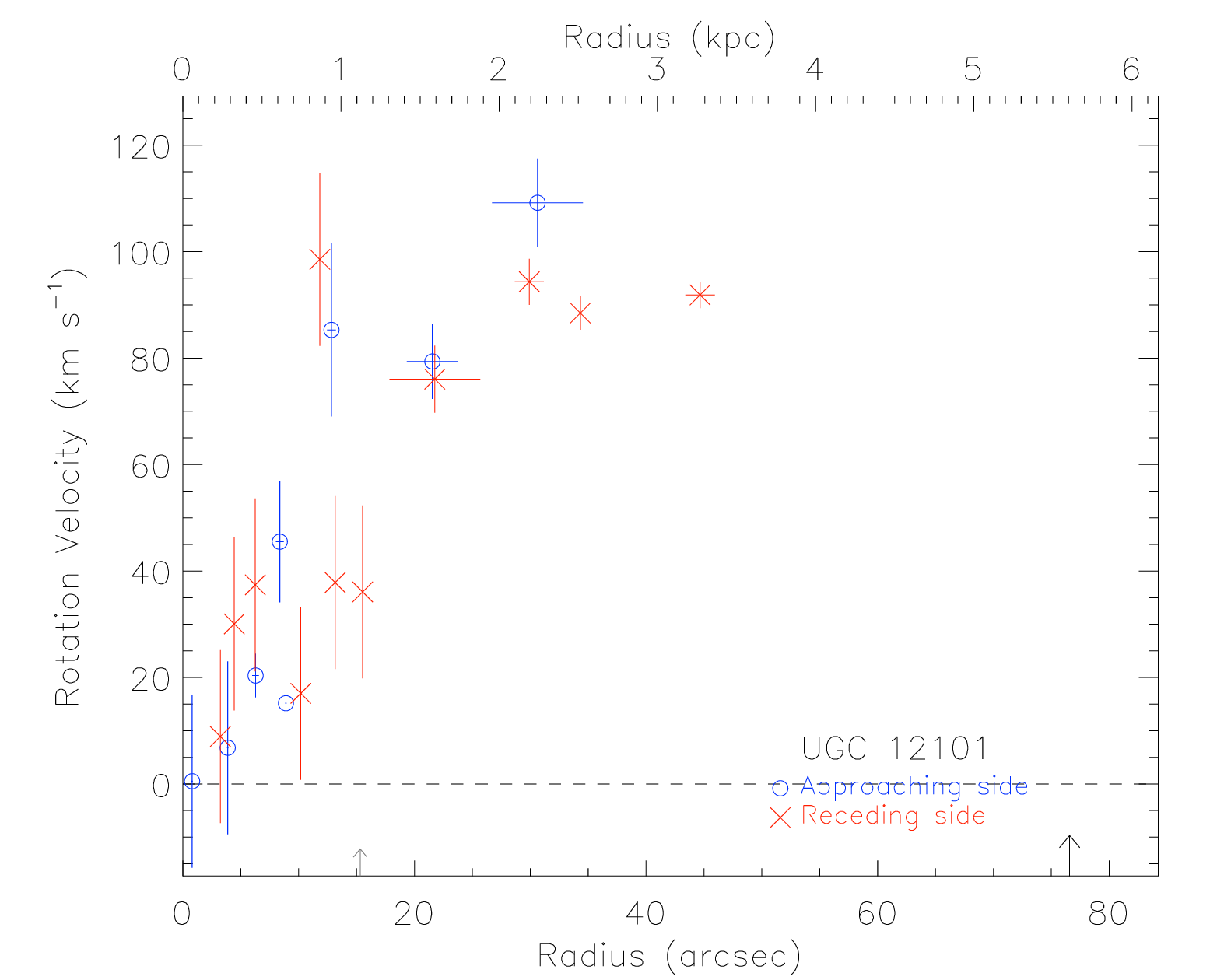}
   \includegraphics[width=8cm]{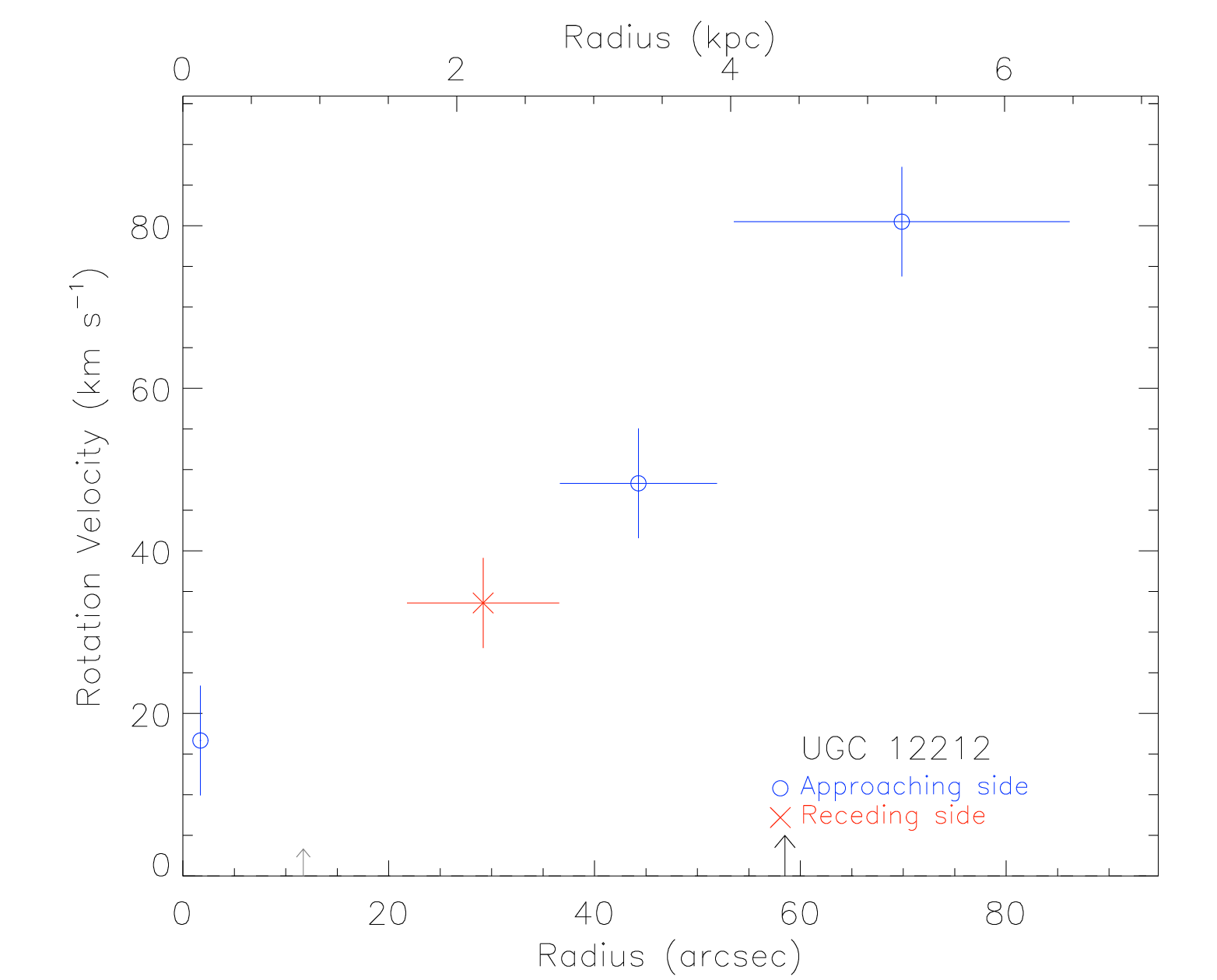}
\end{center}
\caption{From top left to bottom right: \ha~\RC~of UGC 11891, UGC 11914, UGC 11951, UGC 12060, UGC 12101, and UGC 12212.
}
\end{minipage}
\end{figure*}
\clearpage
\begin{figure*}
\begin{minipage}{180mm}
\begin{center}
   \includegraphics[width=8cm]{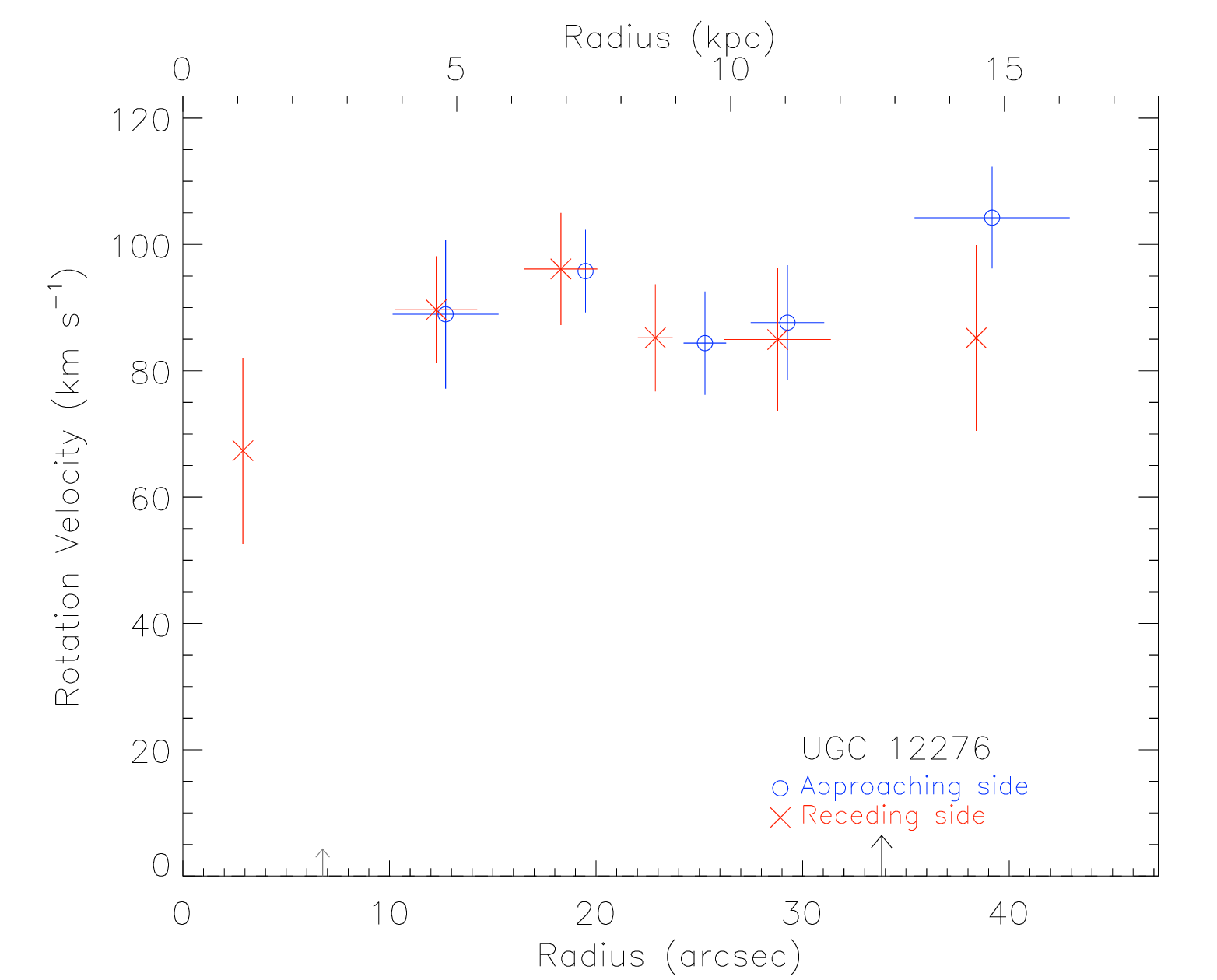}
   \includegraphics[width=8cm]{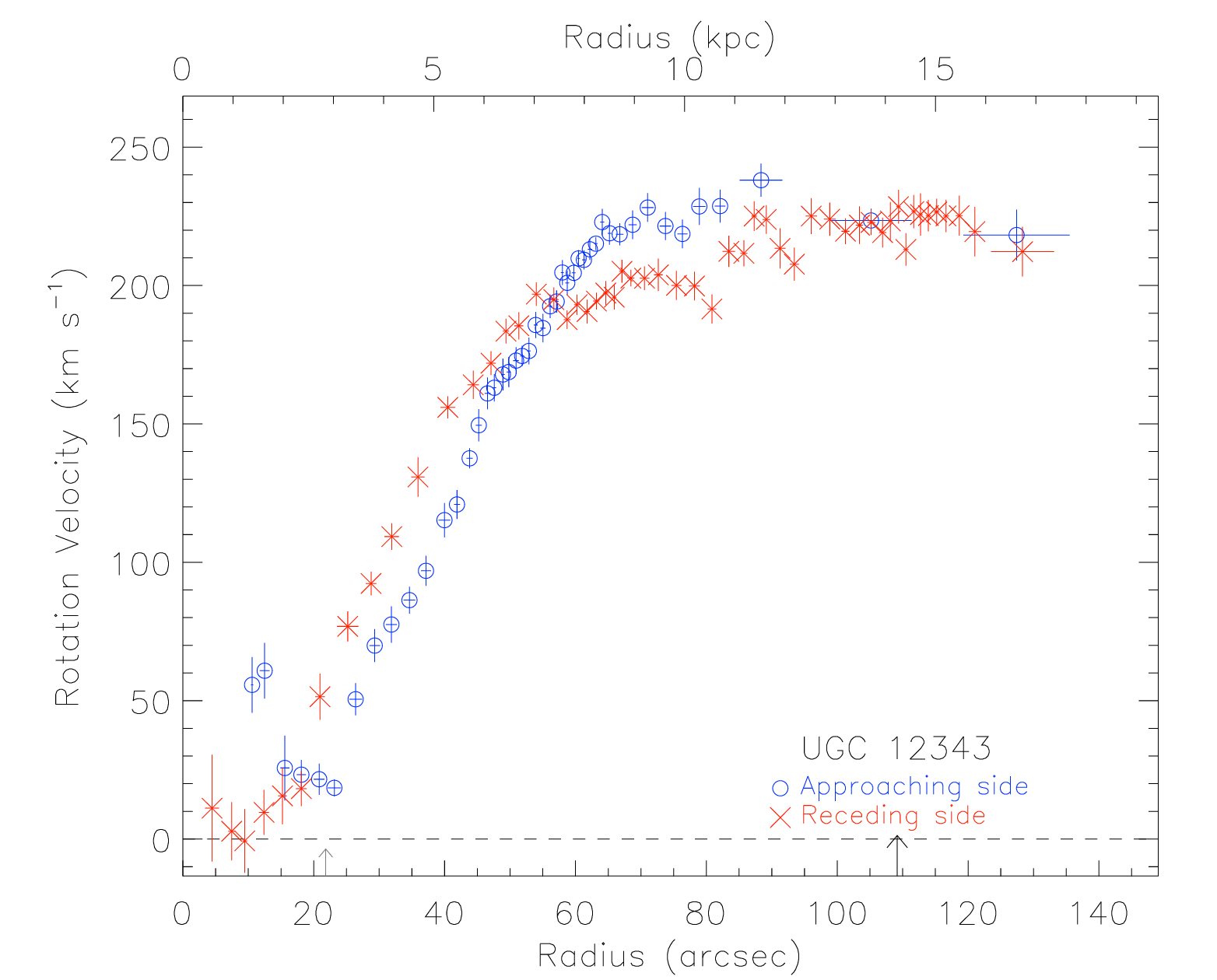}
   \includegraphics[width=8cm]{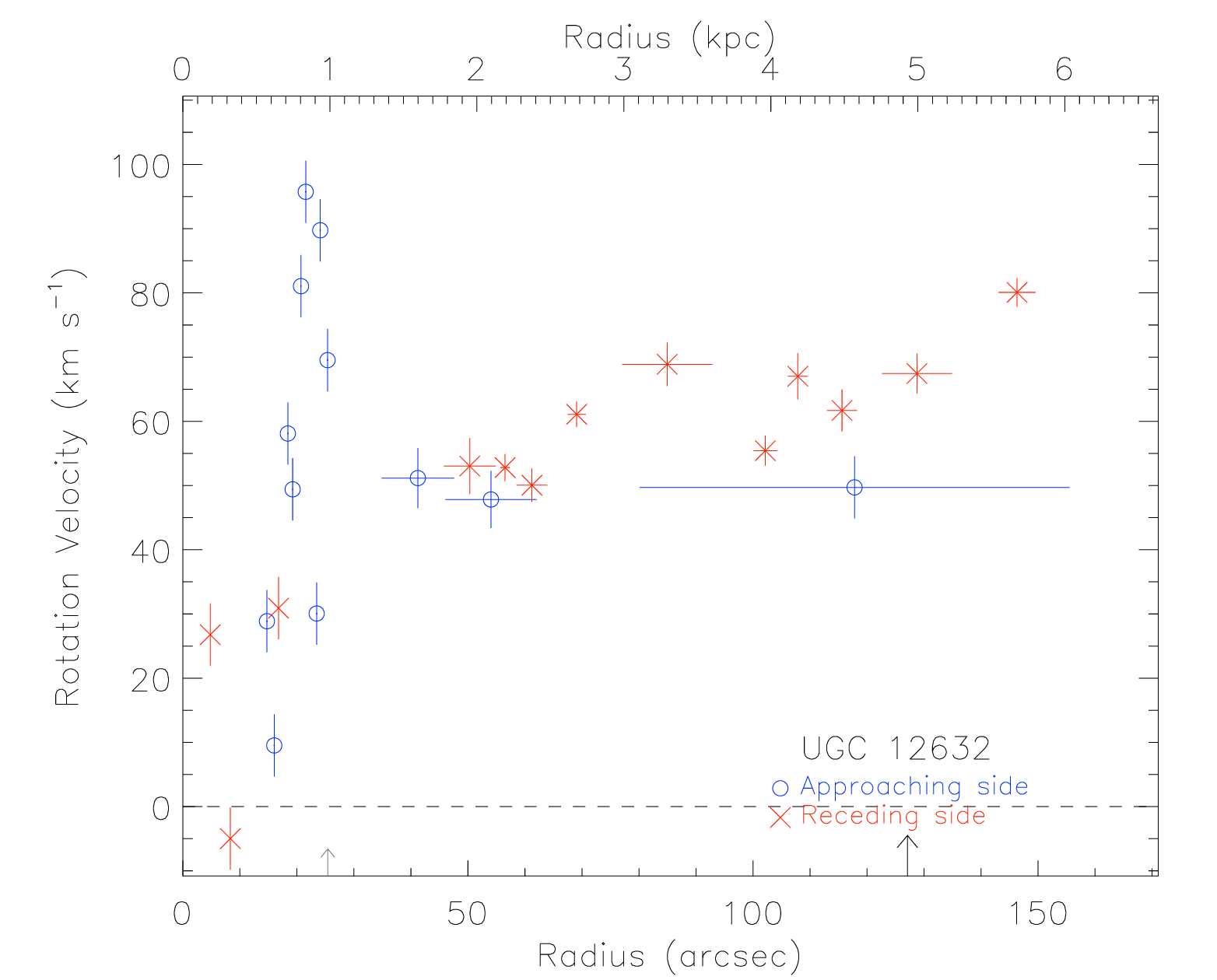}
   \includegraphics[width=8cm]{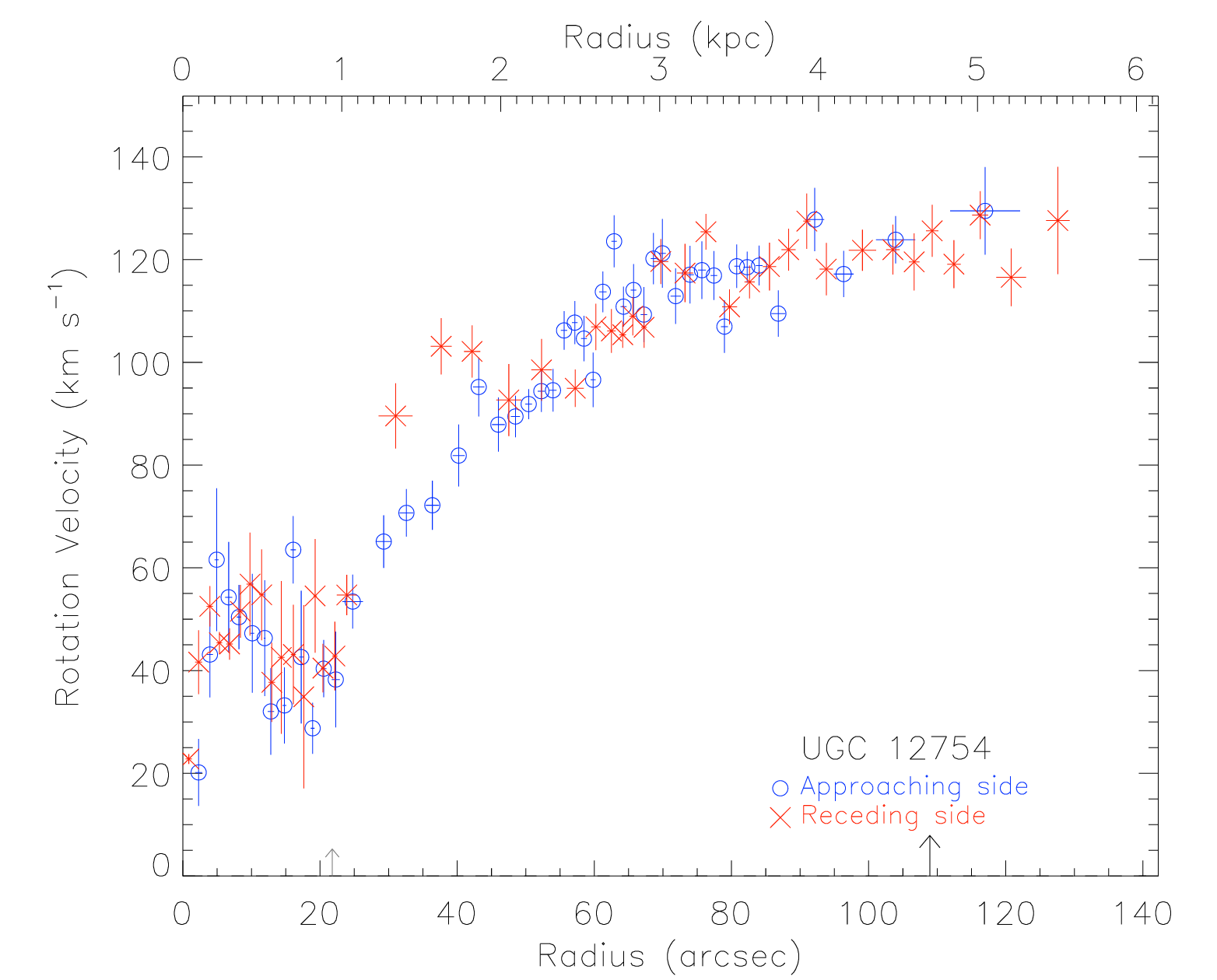}
\end{center}
\caption{From top left to bottom right: \ha~\RC~of UGC 12276, UGC 12343, UGC 12632, UGC 12754.
}
\end{minipage}
\end{figure*}
\clearpage

\clearpage
\section{Rotation curves tables}
\label{rc_tables}
\begin{table*}
\caption{UGC 00508 \RC}
\begin{tabular}{cccccccc}
\noalign{\medskip} \hline
           r (kpc)  &  $\sigma_r$ (kpc)  &             r (")  &    $\sigma_r$ (")  &          v (\kms)  & $\sigma_v$ (\kms)  &            N bins  &              side \\
(1)&(2)&(3)&(4)&(5)&(6)&(7)&(8)\\
\hline
            0.28  &            0.00  &             0.9  &             0.0  &             307  &             124  &               1  &               a \\
            0.28  &            0.00  &             0.9  &             0.0  &             364  &             124  &               1  &               r \\
            0.97  &            0.11  &             3.1  &             0.4  &             269  &              29  &               4  &               a \\
            1.28  &            0.00  &             4.1  &             0.0  &             390  &             124  &               1  &               r \\
            1.56  &            0.00  &             5.1  &             0.0  &             439  &             124  &               1  &               a \\
            2.14  &            0.24  &             6.9  &             0.8  &             329  &              42  &               2  &               a \\
            2.61  &            0.20  &             8.4  &             0.7  &             463  &              49  &               3  &               a \\
            2.98  &            0.00  &             9.6  &             0.0  &             312  &             124  &               1  &               r \\
            3.12  &            0.00  &            10.1  &             0.0  &             307  &             124  &               1  &               a \\
            3.71  &            0.10  &            12.0  &             0.3  &             395  &              22  &               3  &               a \\
            4.39  &            0.00  &            14.2  &             0.0  &             410  &             124  &               1  &               r \\
            4.61  &            0.03  &            14.9  &             0.1  &             479  &              42  &               2  &               a \\
            4.73  &            0.00  &            15.3  &             0.0  &             323  &             124  &               1  &               a \\
            5.51  &            0.03  &            17.8  &             0.1  &             493  &              72  &               2  &               a \\
            5.75  &            0.00  &            18.6  &             0.0  &             466  &             124  &               2  &               r \\
            7.78  &            1.14  &            25.2  &             3.7  &             454  &              18  &              23  &               a \\
            7.81  &            1.02  &            25.2  &             3.3  &             490  &              19  &              23  &               r \\
            9.69  &            0.39  &            31.3  &             1.3  &             465  &              15  &              23  &               r \\
           10.14  &            0.38  &            32.8  &             1.2  &             471  &              13  &              23  &               a \\
           10.64  &            0.18  &            34.4  &             0.6  &             491  &              12  &              23  &               r \\
           11.12  &            0.29  &            35.9  &             0.9  &             468  &              18  &              23  &               a \\
           11.20  &            0.16  &            36.2  &             0.5  &             488  &              10  &              23  &               r \\
           11.66  &            0.19  &            37.7  &             0.6  &             496  &              13  &              23  &               r \\
           12.09  &            0.27  &            39.1  &             0.9  &             514  &              14  &              23  &               a \\
           12.48  &            0.27  &            40.3  &             0.9  &             499  &              13  &              23  &               r \\
           13.30  &            0.50  &            43.0  &             1.6  &             499  &              16  &              23  &               a \\
           13.68  &            0.37  &            44.2  &             1.2  &             497  &              13  &              23  &               r \\
           14.82  &            0.32  &            47.9  &             1.0  &             530  &              14  &              23  &               r \\
           14.85  &            0.48  &            48.0  &             1.6  &             501  &              14  &              23  &               a \\
           15.76  &            0.27  &            50.9  &             0.9  &             526  &              14  &              23  &               r \\
           16.37  &            0.38  &            52.9  &             1.2  &             540  &              13  &              23  &               a \\
           16.90  &            0.43  &            54.6  &             1.4  &             565  &              12  &              23  &               r \\
           17.47  &            0.36  &            56.5  &             1.2  &             526  &              17  &              23  &               a \\
           18.54  &            0.57  &            59.9  &             1.8  &             567  &              12  &              23  &               r \\
           19.83  &            1.82  &            64.1  &             5.9  &             553  &              19  &              17  &               a \\
           20.11  &            0.34  &            65.0  &             1.1  &             541  &              14  &              23  &               r \\
           21.11  &            0.24  &            68.3  &             0.8  &             553  &              15  &              23  &               r \\
           22.24  &            0.67  &            71.9  &             2.2  &             514  &              17  &              21  &               r \\
\hline
\label{rclabel_ugc00508}
\end{tabular}
\\(1), (3) Galactic radius. (2), (4) Dispersion around the galactic radius. (5) Rotation velocity. (6) Dispersion on the rotation velocity. (7) Number of velocity bins. (8) Receding -- r -- or approaching -- a -- side.
\end{table*}

\label{lastpage}

\end{document}